\documentclass[aps,
prb,
reprint,
preprintnumbers,
superscriptaddress,
amssymb,
amsmath,
onecolumn
]{revtex4-2}

\usepackage[T2A,T1]{fontenc}

\usepackage{bm}
\usepackage{mathtools}
\usepackage{bigints}

\usepackage[cal=boondoxo,frak=boondox]{mathalfa}

\usepackage[table,x11names]{xcolor}
\usepackage{graphicx}
\graphicspath{{./}{./figs/}}

\usepackage{savesym}
\usepackage{enumitem}
\usepackage{hhline}
\usepackage{siunitx}
\usepackage{etoolbox}
\makeatletter
\patchcmd{\l@section}{1.5em}{2.6em}{}{}
\patchcmd{\l@subsection}{1.5em}{2.6em}{}{}
\patchcmd{\l@subsection}{2.4em}{3.4em}{}{}
\patchcmd{\l@subsubsection}{3.8em}{5.2em}{}{}
\patchcmd{\l@subsubsection}{3.2em}{4.2em}{}{}
\makeatother

\usepackage[breaklinks=true,
unicode=true,
urlcolor = RoyalBlue4,
colorlinks = true,
citecolor = RoyalBlue4,
linkcolor = blue
]{hyperref}

\renewcommand{\vec}[1]{\bm{#1}}

\makeatletter
\newcommand{\oiint}{{\mathchoice
		{\mkern-1.5mu\raisebox{0.05ex}{$\displaystyle\bigcirc$}\mkern-20.5mu\iint}%
		{\mkern-1.5mu\raisebox{0.05ex}{$\textstyle\bigcirc$}\mkern-15.5mu\iint}%
		{\mkern-1.5mu\raisebox{0.05ex}{$\scriptstyle\bigcirc$}\mkern-13mu\iint}%
		{\mkern-1.5mu\raisebox{0.05ex}{$\scriptscriptstyle\bigcirc$}\mkern-11.5mu\iint}}}
\makeatother

\begin{document}
	
	\preprint{Chapter pre-print}
	
	\title{Curvilinear nanomagnets of non-trivial geometric topology}
	
	\author{Vagson L. Carvalho-Santos}
	\email{vagson.santos@ufv.br}
	\affiliation{Departamento de Física, Universidade Federal de Viçosa, Av. PH Rolfs s/n, 36570-900 Viçosa, Brazil}
	
	\author{Luke A. Turnbull}
	\email{luke.turnbull@diamond.ac.uk}
	\affiliation{Diamond Light Source, Harwell Science and Innovation Campus, Didcot, 0X11 ODE, United Kingdom}
	
	\author{Na\"{e}mi Leo}
	\email{n.leo@lboro.ac.uk}
	\affiliation{Department of Physics, Loughborough University, Loughborough, LE11 3TU, United Kingdom}
		
	\author{Oleksandr~V.~Pylypovskyi}
	\email{o.pylypovskyi@hzdr.de}
	\affiliation{Helmholtz-Zentrum Dresden-Rossendorf e.V., Institute of Ion Beam Physics and Materials Research, 01328 Dresden, Germany}
	\affiliation{Kyiv Academic University, Kyiv 03142, Ukraine}
	
	\begin{abstract}
		In this Chapter, we focus on the question of how the global geometric characteristics given by the topology of the sample's shape are reflected in the properties of the magnetization field of the given curvilinear nanoarchitecture. We start with a brief introduction to differential geometry and topology, giving their connection to the specific properties of the vector field represented by the magnetization field defined at the specific manifold. Then we discuss modern fabrication and characterization techniques that make analysis of a sample's topology not only an interesting theoretical concept but also a directly accessible and practically relevant topic. The main attention of the Chapter is dedicated to the theoretical and experimental analysis of nanostructrures with the topology different from the extended films beyond the in-depth studied topic of tubes of the constant radius. In particular, toroidal, spherical, and non-orientable manifolds are considered.
	\end{abstract}
		
	\maketitle
	
	\tableofcontents
	
	\newpage
	\section{Introduction}
	
	The famed hairy ball theorem states, in simple terms, that it is impossible to smoothly comb the spikes of a hedgehog flat everywhere; at least one vortex, with some needles poking out, must form. This cartoon depicts the mathematical relation between global properties of a given manifold $\tilde{M}$ (here a sphere), and the global properties of the vector field $\vec{m}$ defined on this manifold $\tilde{M}$ (here represented by the hedgehogs needles, which are assumed to be of equal and unchangeable length). However, the importance of such relations extends well beyond mathematics into the natural world, where both stable and transient vortex configurations are ubiquitous in nature. Examples range from the macroscale, such as cyclones and anticyclones in the atmospheres of Earth and other planets, down to the nanoscale, where vectorial fields such as the magnetization can be represented by hedgehog-like spin densities, which are in the focus of this chapter. The following discussion is dedicated to the analysis of the role of the topology of a magnetic sample at micro- and nanoscale.
	
	From a historical perspective, attention to special geometries in magnetism can be traced back to the 1940s, particularly in the context of Ising models. Analytical results for topologically non-trivial cylindrical $M\times N$ ferromagnetic lattices had been found by Onsager~\cite{Onsager44}. A few years later, the partition function for a two-periodic square lattice, resembling the topology of a torus, was found~\cite{Kaufman49}. Towards the 1960s, magnetic reversal modes in micromagnetic structures such as infinite wires were analysed~\cite{Aharoni58,Aharoni59,Brown63}, and similar questions were raised in the 1970s for magnetically soft toroidal shells~\cite{Arrott74}. Topologically non-trivial samples, including non-planar ones (e.g., tubes), came into focus again in the 1990s in the context of Heisenberg models and mechanically soft samples that can react to the spatial distribution of magnetization~\cite{VillainGuillot94,Villain-Guillot95,Saxena97,Benoit98,Saxena98,Dandoloff02}. These initial studies integrated elastic and anisotropy interactions to investigate the role of solitonic magnetic configurations in determining the deformation of elastic shells. In the following years, major efforts focused on magnetic nanowires and tubes considering magnetic states~\cite{Escrig07,Yershov20,Fratta23,Mancilla-Almonacid20} and their dynamics, including spin wave propagation~\cite{Popov08,Gonzalez10,Yan11,Otalora16,Korber22} and resonant modes~\cite{Saavedra22,Yang16a,Saavedra25a}, and domain wall motion~\cite{Usov07,Yan10,Otalora13,Nunes26}. Notable findings that stimulated this field are related to the discovery of chiral symmetry breaking, which is intimately connected to curved geometries~\cite{Landeros10,Otalora12a,Otalora16}. 
	
	A direct link between the presence of topological defects in the sample's shape and topological \emph{magnetic} textures has been directly discussed by Saxena and Dandoloff~\cite{Saxena02}. Furthermore, it has been discussed within  the framework of the anisotropic Heisenberg model~\cite{Freitas05, Milagre07a, Belo07, Carvalho-Santos08, Carvalho-Santos12, Carvalho-Santos13a, Vilas-Boas15, CarvalhoSantos15}. The modern theory of curvilinear magnetism got a major boost in late 2000s and beginning of 2010s with the idea of the functionalization of magnetic nanostructures extended into the third dimension. This concept engages all real and spin space degrees of freedom to enhance functionalities of devices, such as magnetic memory and logic devices~\cite{Parkin08,Streubel16,Fernandez-Pacheco17}. Over the years, a profound understanding of geometric symmetry breaking~\cite{Hertel13a} and the local properties of magnetic textures on curved manifolds driven has been established~\cite{Streubel16,Fernandez-Pacheco17,Fischer20, Makarov22a, Sheka21b, Sheka22, Gubbiotti24}. 
	
	In this Chapter we focus on the question of how the global geometric characteristics given by the topology of a sample's shape are reflected in the properties of the magnetization field of the given curvilinear nanoarchitecture. The attention is mainly dedicated to the static properties of magnetic textures defined on non-developable manifolds. For the particular case of domain wall dynamics in tubular geometries, we refer the reader to review~\cite{Landeros22}.
	
	The Chapter is organized as follows. In Sec.~\ref{sec:geom_n_topo} we introduce the basic definitions of differential geometry and topology, illustrating them with examples on curvilinear magnets. The concept of a topological soliton in different dimensions~(Sec.~\ref{sec:top-defects}) is given, along with the discussion of the influence of the topology of the geometry that hosts them~(Sec.~\ref{sec:topo-theorems}). Sec.~\ref{sec:fabrication} describes the key techniques for fabrication of magnetic nanostructures with complex shapes and non-trivial topology considering both additive and subtractive fabrication methods. Experimental characterization techniques are introduced in Sec.~\ref{sec:measurement}. There, scanning, X-ray and electron microscopy are discussed with respect to their advantages and disadvantages regarding 3D shape of the sample under investigation. At the end of Sec.~\ref{sec:measurement}, micromagnetic simulations (\emph{in-silico} experiment) are introduced. 
	
	The Chapter continues with the illustrations of the previously introduced concepts, starting from the genus-1 (zero Euler characteristic) case corresponding to the geometries with one hole (Sec.~\ref{sec:toroidal}) that have a curved generatrix like the torus representing the case of a compact geometry or a rotating hyperboloid. After discussion of the fundamentals of XY model and 3D spins applied to the axially symmetric geometries (Sec.~\ref{sec:heisenberg}), we focus on toroidal geometries (Sec.~\ref{sec:Topology-Tori}) considering their statics analytically and discussing the very first experimental developments. The analytical analysis and experimental insights for samples of spherical topology (positive Euler characteristic) as well as shapes of $N$-tori-topology (negative Euler characteristic, this case includes multipodes of complex shape and gyroids) are given in Sec.~\ref{sec:finite-euler}. At the end, we discuss with the specific case of statics and linear dynamics in non-orientable M\"{o}bius rings~(Sec.~\ref{sec:nonorientable}). The Chapter ends with Conclusions and Outlook (Sec.~\ref{sec:conc}).
	
	\section{Geometry and Topology of Curvilinear Magnets}
	\label{sec:geom_n_topo}
	
	\subsection{Differential Geometry as Local Framework of Curvilinear Magnetism} 
	\label{sec:diffgeo}
	
	A thin curvilinear shell can be described with a curved surface $\vec{\varsigma}(\eta_1,\eta_2)$ with $\eta_{1,2}$ being local coordinates defined on $\vec{\varsigma}$. To describe its differential properties, we introduce two tangential vectors $\vec{g}_i = \partial_{\eta_i} \vec{\varsigma}$, $i = 1,2$, and the unit vector of the surface normal as $\vec{n} = \vec{g}_1 \times \vec{g}_2 / |\vec{g}_1 \times \vec{g}_2|$. Here, and in the following, we will use the notation $\partial_\bullet := \partial/\partial \bullet$. 	The distances along $\vec{\varsigma}$ are characterized by the so-called first fundamental form 
	\begin{equation}\label{eq:1-st-fund-form}
		g_{ij} = \vec{g}_i \cdot \vec{g}_j,\quad i,j=\overline{1,2},
	\end{equation}
	and the line element is 
	\begin{equation}
		dl^2 = g_{ij}\, d\eta_id\eta_j,
	\end{equation}
	where the Einstein summation rule is used. The second fundamental form, also known as the Weingarten map, is
	\begin{equation}\label{eq:2-nd-fund-form}
		H_{ij} = \dfrac{\vec{n}\cdot \partial_j \vec{g}_i}{\sqrt{g_{ii}g_{jj}}},\quad i,j=\overline{1,2}.
	\end{equation} 
	
	The coordinates $\eta_{1,2}$ can be chosen in such a way that the second fundamental form is represented by a diagonal matrix. In this case, $\eta_{1}$ and $\eta_{2}$ are orthogonal and  $\|H_{ij}\| = \mathrm{diag}(\kappa_1,\kappa_2)$ is a diagonal matrix, with $\kappa_{1,2}$ called the \emph{principal curvatures}, while the orthonormal vectors $\vec{e}_{1,2} = \vec{g}_{1,2}/|\vec{g}_{1,2}|$ provide the so-called principal directions. Taken together, the set of vectors $\{\vec{e}_1, \vec{e}_2, \vec{n}\}$ is a Darboux three-frame that serves as a local reference frame for the curvilinear shell.
	
	The two principal curvatures can be combined into the mean curvature $\tilde{H}$ and Gaussian curvature $\tilde{G}$,\footnote{There is an alternative definition of the mean curvature as the arithmetic mean of the principal curvatures $\tilde{H}' = (\kappa_1 + \kappa_2)/2$.} respectively, as
	\begin{equation}\label{eq:mean-gauss-curv}
		\tilde{H} = \mathrm{Tr}\,\|H_{ij}\| = \kappa_1 + \kappa_2,\quad 
		\tilde{G} = \mathrm{det}\,\|H_{ij}\| = \kappa_1\kappa_2.
	\end{equation}
	The principal curvatures $\kappa_1, \kappa_2$ as well as the mean and Gauss curvatures $\tilde{H}$ and $\tilde{G}$, respectively, are \emph{invariants} of the surface, and do not depend on the chosen reference frame. 
	This makes them unavoidable quantities for the description of the curvilinear physical effects. 
	
	A vanishing mean curvature $\tilde{H}=0$ either is observed for a flat surface (with $\kappa_1=\kappa_2=0$) or for the case of so-called \emph{minimal surfaces} $\vec{\varsigma}_\text{min}$ which are characterized by opposite principal curvatures $\kappa_1=-\kappa_2$ and thus take on complex saddle shapes. Mean curvatures appear as the solutions of variational problems of minimal surface areas for a given boundary conditions~\cite{Schoen70}, like for soap bubbles or for self-assembly of biological lamella and gyroid structures~\cite{Saranathan10}.
	
	If the Gaussian curvature vanishes, $\tilde{G} \equiv 0$, one of the principal curvatures $\kappa_i$ is also zero, and the surface $\vec{\varsigma}$ is called \emph{developable}, which are a special case of \emph{ruled surfaces}, i.e., surfaces created by the smooth motion of a straight line in 3D space. Examples of developable surfaces include cylinders and cones. For a comprehensive theory of differential geometry of manifolds in 3D space, we refer the interested reader to any of the following references:~\cite{Dubrovin84,Dubrovin85,Kuhnel15,doCarmo16,Govindarajan24}.
	
	With the defined surface shape of $\vec{\varsigma}$, one can construct a finite-volume magnetic shell with thickness $h$ as follows:
	\begin{equation}
		\vec{r}(\eta_1, \eta_2, \eta_3) = \vec{\varsigma}(\eta_1, \eta_2) + \eta_3 \vec{n},\quad \eta_3 \in [-h/2, h/2],
	\end{equation}
	where $\eta_3$ is the coordinate along the shell interior. 	Generally, lower and upper bounds for $\eta_3$ could be different to $\pm h$ resulting in a shell of varying thickness, with specific implications to the permittable magnetic states~\cite{Yershov23}.

	\subsection{Linking between Spatial Geometry and Non-Homogeneous Spin States}
	
	The link between the geometry and magnetic state can appear because of several reasons including non-local phenomena, strain and interfacial effects. In particular, in magnetically soft materials, the dominating energy contribution (next to nanoscale exchange) is the long-range magnetostatic interaction, leading to the shape anisotropy for which it is energetically favourable that the spin state tracks the sample surface to minimize surface magnetostatic charges. For a thin enough shell, the shape anisotropy direction, i.e., the local hard axis of magnetization, points along the surface normal, and the energy density $w_\text{ani}^\text{shell}$ in CGS units~\cite{Carbou01,Kohn05a,DiFratta20,Fratta20} is given by 
	\begin{equation}\label{eq:shape-ani-shell}
		w_\text{ani}^\text{shell} = 2\pi M_S^2 (\vec{m}\cdot \vec{n})^2
	\end{equation}
	where $\vec{m} = \vec{M}/M_S$ is the unit vector along the magnetization $\vec{M}$ and $M_S$ is the saturation magnetization. This kind of anisotropy appears from the necessity to minimize the surface magnetostatic charges, see also Sec.~\ref{sec:insurf-circulation} and~\ref{sec:wireframes}. In magnetic wireframes, the anisotropy in Eq.~\eqref{eq:shape-ani-shell} is also combined with the easy-axis contribution along the narrow enough segments of the wire segment~\cite{Slastikov12,Gaididei17}. 
	
	In magnetically-hard perpendicular multilayers, the easy axis of anisotropy is developed due to interfacial interactions from several thin asymmetric layers, e.g., Co sandwidhed with heavy metals like Pt or Pd. Even if the interface is not flat, the local anisotropy axis follows its normal~\cite{Smith11,Raftrey25}. 
	
	There are also special cases of the link between the geometry and magnetic state that are not discussed in this Chapter: (i)~transport phenomena reflected in the trajectories of charge carriers following the geometry of the sample~\cite{Yershov16,Makushko24} and (ii)~mechanically soft magnets~\cite{Gaididei19,Yershov19b,MirandaSilva22}.
	
	Once the fundamental forms~\eqref{eq:1-st-fund-form} and~\eqref{eq:2-nd-fund-form} for a magnetic shell of shape $\vec{\varsigma}$ are known, one can understand properties of the magnetic state in terms of the mapping the complex 3D shape onto its local reference frame $\{\eta_1, \eta_2, \eta_3\}$ as well as mapping the components of all magnetisation vectors to the Darboux three-frame as $m_i \equiv \vec{m}\cdot \vec{e}_i$, $i=1,2$ and $m_3 \equiv \vec{m}\cdot\vec{n}$. 	The power of this approach is given by restoring the local invariance of the geometry-tracking interaction. For example, the hard-axis shape anisotropy~\eqref{eq:shape-ani-shell} simplifies to be dependent on the one component of the magnetization vector only, $w_\text{ani}^\text{shell} \propto m_3^2$. The major changes in the energy functional that capture properties arising from the geometry come through terms that contain spatial derivatives. A unified formulation of the exchange energy, which is valid for both planar and curved geometries, is given by the covariant derivatives~\cite{Sheka20a,Sheka22a}:
	\begin{equation}
		\begin{aligned}
			\eth_i m_j & = \partial_i m_j + \epsilon_{jk} (\vec{e}_1 \cdot \partial_i \vec{e}_2) m_k,\\
			\eth_i m_3 & \equiv \partial_i m_3,\quad i,j=1,2,
		\end{aligned}
	\end{equation}
	where $\epsilon_{jk}$ is the totally antisymmetric tensor. The use of covariant derivatives allows us to present the exchange energy with exchange stiffness $A$ as a sum of three terms\footnote{In these expressions, the divisor $\sqrt{g_{ii}}$ could be ``eliminated'' with the use of language of covariant and contravariant vectors that are commonly denoted with sub- and superscripts, respectively~\cite{Sheka22a}.}
	\begin{equation}\label{eq:exchange-curv}
		\begin{aligned}
			w_\text{x} & = A \sum_{\alpha = x,y,z} (\partial_\alpha \vec{m})^2 = w_\text{x}^0 + w_\text{x}^\text{ani} + w_\text{x}^\text{ch},\\
			w_\text{x}^0 & = \dfrac{A}{g_{ii}} (\eth_i m_j)(\eth_i m_j),\\
			w_\text{x}^\text{ani} & = A \left[\kappa_1^2 m_1^2 + \kappa_2^2 m_2^2 + (\kappa_1^2+\kappa_2^2)m_3^2 \right],\\
			w_\text{x}^\text{ch} & = \dfrac{2A}{\sqrt{g_{ii}}}\kappa_i \mathrm{L}_{ij}^{(i)},\quad \mathrm{L}_{ij}^{(k)} = m_i\eth_k m_j - m_j \eth_k m_i.
		\end{aligned}
	\end{equation}
	Here, $w_\text{x}^0$ is the only non-vanishing term in the case of a planar extended film, representing an energy penalty due to non-collinear magnetic textures. The second term, $w_\text{x}^\text{ani}$ is the geometry-driven biaxial anisotropy that scales as the square of principal curvatures. The last, chiral, term $w_\text{x}^\text{ch}$ is linear in the principal curvatures and consists of two Lifshitz invariants $\mathrm{L}_{ij}^{(i)}$ representing symmetry of the geometry-driven Dzyaloshinskii--Moriya interaction~(DMI). In particular, this chiral contribution $w_\text{x}^\text{ch}$ is responsible for the stabilization of skyrmionic textures purely by geometric features~\cite{Kravchuk16,Pylypovskyi18}. Similar transformations could be performed with the intrinsic DMI~\cite{Sheka22a} and volume magnetostatic charges~\cite{Sheka20a}. 
	
	\subsection{Topological Properties of Curved Geometries}
	\label{sec:topology}

	Topology as a field of mathematics, in comparison with differential geometry, provides a more general overview of the properties of certain spaces. While for a detailed mathematical background and applications we refer to the dedicated publications~\cite{Shankar77,Mermin79,Kamien02,Alexander12,Nash13,Govindarajan24,Balakrishnan24}, in this Section we focus on the major definitions important for the understanding of the physics of curvilinear magnets.
	
	\subsubsection{Topological Spaces and Homotopy Classes}
	
	A topological space is defined by a mathematical set $X$ and a collection of its subsets denoted $\bar{Y} = \{ X_i \}$, which together  satisfy several conditions: (i)~The empty set and all $X$ should be in $\bar{Y}$, (ii)~any union of subsets of $X_i$ are also contained in $\bar{Y}$, and (iii)~any intersection of subsets of $X_i$ is also in $\bar{Y}$. In this case one can say that $X$ is a topological space with the topology provided by $\bar{Y}$, or, in other words, this definition establishes a ``connection'' between the collection of $X_i$ (also called the open sets), which is then called~$\bar{Y}$.
	
	The relation between two topological spaces $X$ and $Y$ can be given by a function $f$ that maps $X$ to $Y$, via $f: X \to Y$. If $f$ as well as its inverse function $f^{-1}: Y \to X$ (which should exist) are continuous, this mapping is called a homeomorphism, which establishes a connection between formally distinct sets. 
	For example, a torus is homeomorphic to a mug or a ring, or any other geometric shape that has one hole, but it also is  topologically equivalent to the direct product of two rings $\mathbb{S}^1 \times \mathbb{S}^1$ or a square with periodic boundary conditions along both axes (Clifford torus).

	There are several properties of topological spaces that are easy to illustrate with geometric examples. 
	
	A topological space is called \emph{closed} if it contains its boundary (for example, for a finite-size film its boundary line is part of the space), and \emph{unbound} if the opposite is true (such as for an infinite strip or extended thin film). 
	
	If the space is bounded and closed, it is called a \emph{compact} space. In other words, a compact space or manifold has a finite size and includes its boundary if it exists (a sphere has no boundary but is a compact manifold). 
	
	A topological space is called \emph{simply connected} any loop in the space can be shrunk to a single point, i.e, the loop length can be continuously reduced to zero, as is possible on a sphere. In contrast, complex connectivity exist in when these loops cannot disappear by continuous deformations, with the torus as an example, where loops along the poloidal direction have a minimum length due to the hole in the structure.
	
	Homeomorphism keeps the so-called topological invariants of a system unchanged, which includes the dimensionality of the topological space, as well as its connectedness and compactness. Topological invariants that arise due to the specific spatial variation of an order parameter, like the magnetization $\vec{m}$, may lead to physical quantities that could either be directly measured or inferred from experiments.
	
	Importantly, all spaces with the same topological invariants share a so-called homo\-topy class $\pi(\bullet)$, which are a crucial tool to distinguish objects with different topological properties. 
	
	\subsubsection{Topological Magnetic Defects}
	\label{sec:top-defects}
	
	Now let us consider properties of topological spaces formed by an order parameter, using magnetically ordered systems as a convenient example~\cite{Gobel19b}. If a planar thin magnetic film has a strong in-plane anisotropy, the out-of-plane magnetization component of $\vec{m}$, $m_z$, is negligible, and the magnetic state can be described by the in-plane azimuthal angle $\Phi$, with the respective topological space being a circle, $\mathbb{S}^1$: 
	\begin{equation}\label{eq:azimuthal-angle-m}
		\Phi = \arctan \dfrac{m_y}{m_x} \in [0, 2\pi).
	\end{equation}
	
	\begin{figure}[tb]
	    \centering
	    \includegraphics[width=0.95\linewidth]{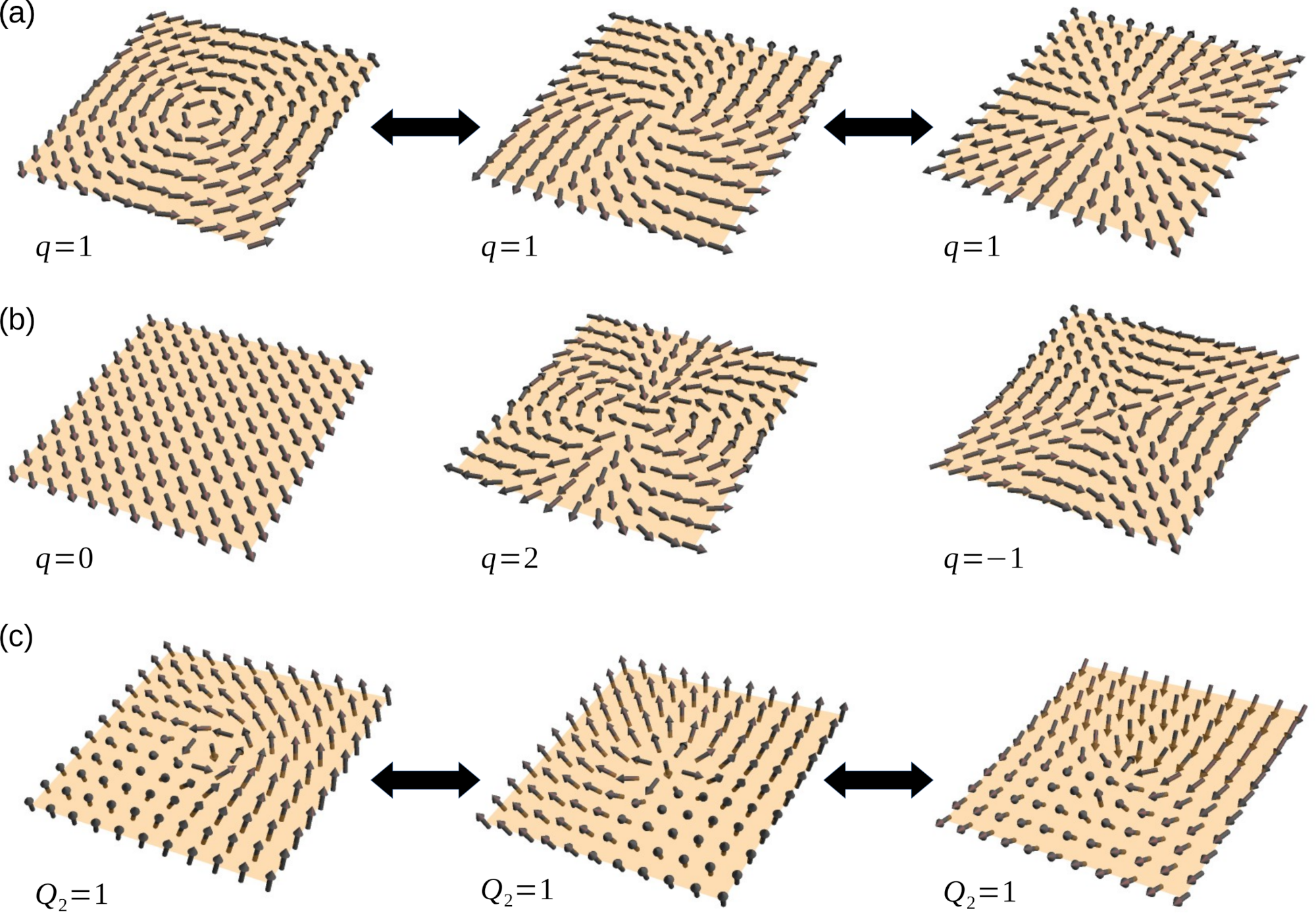} % SkBimVortex
	    \caption{Vector fields of topological textures with their respective topological charges. 
	    (a)~Vortices with three distinct helicities: azimuthal, spiral, and radial. These textures can be transformed into each other as they belong to the same $\pi_1(\mathbb{S}^1)$ homotopy group and carry the same topological charge \( q = +1 \). 
	    (b)~Other textures analyzed under the perspective of $\pi_1(\mathbb{S}^1)$ homotopy group: monodomain (\( q = 0 \)), double rotation (\( q = +2 \)), and antivortex state (\( q = -1 \)). 	    (c)~Topologically equivalent skyrmionic textures belonging to the $\pi_2(\mathbb{S}^1)$ homotopy group, with Pontryagin index \( Q_2 = +1 \) (i.e., with spins spanning the entire solid angle of $4\pi$): Bloch-type skyrmion, Néel-type skyrmion, and bimeron.
	    }
	    \label{fig:TopologicalTextures}
	\end{figure}
	
	To describe the topological quantities of a spin texture on the film, one can choose contour in the plane and calculate the variation of $\Phi$ over it. There are several possibilities of magnetic texture $\vec{m}(\vec{r})$ at this contour $\tilde{c}$ to be mapped on $\mathbb{S}^1$, and counting the windings allows us to introduce the so-called first homotopy group $\pi_1(\mathbb{S}^1)$ characterizing all such mappings. Here, the subscript $1$ indicates that the homotopy group considers 1D loops for the topological order parameter space given by the symbol in the brackets (here the circle $\mathbb{S}^1$). 
	
	Any \emph{net phase} of $\Phi$ over a closed loop will either come up to zero winding, i.e., within the order parameter space the loop can be shrunk to a point, or make one or several windings around $\mathbb{S}^1$. The additive property of windings makes $\pi_1(\mathbb{S}^1)$ isomorphic (i.e., structurally equivalent) to the additive group of integers, $\pi_1(\mathbb{S}^1) = \mathbb{Z}$. The number of windings $q$ for a given spin texture is a topological invariant, and often referred to as the topological index, topological charge or vorticity $q$. Magnetic configurations with $q = +1$ ($q = -1$) are commonly referred to as merons or vortices (antimerons or antivortices). The topological charge $q$ gives the net vorticity $\sum_i q_i$ of all merons (with $q_i=n/2$, with $n$ integer) enclosed within the contour $\tilde{c}$. 
	
	For the example of a magnetic thin film with an in-plane magnetic texture, Fig.~\ref{fig:TopologicalTextures}(a) shows azimuthal, spiral and radial vortices, which are spin textures belonging to the same $\pi_1(\mathbb{S}^1)$ homotopy group with $q=1$, and which can be continuously transformed into each other (as indicated by the black arrows). Fig.~\ref{fig:TopologicalTextures}(b) depicts the same considerations for other spin textures, including the topologically trivial monodomain state with $q=0$ (topologically equivalent to a vortex-antivortex pair), a double spin rotation with $q=2$, or a single antivortex with $q=-1$. For a one-dimensional domain wall, the sign of the topological charge $q=\pm 1$ indicates whether it is a kink or anti-kink wall.
	
	The value of the topological invariant $q$ thus allows us to quantify to the presence of specific magnetic textures or defects in the order parameter field.
	
	Extending to three-dimensional spin parameter space, but preserving the length of the magnetic moment vector $\vec{m}$, the relevant topological space to consider is the surface of a sphere $\mathbb{S}^2$. Since any loop on a sphere surface can be shrunk into a single point, the homotopy group $\pi_1(\mathbb{S}^2) = 0$ consists only of the identity element, and no topologically non-trivial one-dimensional spin textures exist in this case. 
	
	Non-trivial spin textures can be found from the second homotopy group $\pi_2(\mathbb{S}^2)$, which considers two-dimensional mappings in magnetic parameter space. For example, the easy-plane ground state is mapped to a ring around the equator of $\mathbb{S}^2$. For a magnetic vortex, the finite out-of-plane core magnetization can be mapped to the northern or southern hemisphere of $\mathbb{S}^2$; leading to a topological charge commonly referred to as \emph{polarity} $p = \pm 1$ denoting the direction of the vortex core moment~\cite{Mertens00}. The intrinsic topological charge of the second homotopy group $\pi_2(\mathbb{S}^2)$ is the so-called Pontryagin or skyrmion number
	\begin{equation}\label{eq:pontryagin}
		Q_2 = \dfrac{1}{8\pi} \int \mathcal{J} dxdy 
	\end{equation}
	where
	\begin{equation}\label{eq:gyrovector}
		\vec{\mathcal{J}} = \mathcal{J}\vec{n},\quad \mathcal{J}  = \vec{m}\cdot [\partial_x \vec{m} \times \partial_y \vec{m}]
	\end{equation}
	is the so-called gyrocoupling vector (in the case of a planar geometry $\vec{n} \equiv \vec{\hat{z}}$). 
	
	The skyrmion number $Q_2$ provides the number of coverings of the whole unit sphere $\mathbb{S}^2$, respectively the solid angle spanned by the spin texture. 	For the aforementioned case of magnetic merons, $Q_2$ can be half-integer, e.g., $Q_2 = pq/2$ for a single vortex with $p=\pm 1$ and $q=1$, as a vortex spans only half of a sphere as described above~\cite{Kosevich90}. 	In~contrast, in easy-axis systems that support the formation of skyrmions, $Q_2$ is an integer number. The same holds true for so-called bimerons (in-plane skyrmions), i.e., stable pairs of merons like vortex-antivortex pairs~\cite{Kharkov17,Cai25,Araujo20,Gobel19b}. Because skyrmions and bimerons belong to the same homotopy group with integer $Q_2$, they are topologically equivalent. This equivalence permits their continuous mutual deformation without changing their topological charge~\cite{Castro23,Ohara22,deSouza24}. Recently, it has been shown that both bimerons and skyrmions can be described by the free Abelian group $\mathbb{Z}\times \mathbb{Z}$ as the pair of integers~\cite{Rybakov25}.
	
	Similarly to the triviality of $\pi_1(\mathbb{S}^2)$, essentially two-dimensional magnetic textures such as skyrmions are topologically trivial covered by 3D spin space. In a finite-thickness samples, merons and skyrmions form the so-called Bloch lines and skyrmion strings as 3D magnetic textures.
	
	The third homotopy group $\pi_3(\mathbb{S}^2)$ that \emph{maps 3D magnetic textures on a sphere}~-- is characterized by the Hopf index~\cite{Whitehead47}
	\begin{equation}\label{eq:hopf}
		Q_3 = - \dfrac{1}{8\pi^2} \int \vec{B}^{e}\cdot \vec{A} \, dxdydz,
	\end{equation}
	where $B^{e}_x = \vec{m} \cdot [\partial_y \vec{m} \times \partial_z \vec{m}]$, and $B^{e}_y$ and $B^{e}_z$ are obtained via cyclic permutations of the indices. The components $B_i$ represent the emergent magnetic field associated to the vector potential $\vec{A}$, such that $\nabla \times \vec{A}= \vec{B}^{e}$. The Hopf index is a non-zero integer for 3D magnetic textures called Hopfions~\cite{Kent21,Balakrishnan23,Zheng23,Knapman25} discussed in Sec.~\ref{sec:Hopfions}.
	
	These topological considerations can be applied to magnetization in ferro- and ferrimagnets as well as in antiferromagnets~\cite{Kosevich90,Barker16}, where the presence of compensated couplings can give rise to further non-trivial spin textures. In a bulk antiferromagnet with a N\'{e}el order parameter, magnetic defects are similar to those in nematics. As it has been pointed out by Kovalev~\cite{Kovalev77} and Dzyaloshinskii~\cite{Dzyaloshinskiui77}, dislocations in the crystal structure give rise to the formation of antiferromagnetic domains. Later, these ideas have been developed to include magnetoelasticity~\cite{Dudko98}, and the structure of antiferromagnetic vortices has been analyzed~\cite{Dudko00,Ivanov01b,Galkina20}. We note that in particular cases, the presence of crystallographic defects is of importance for ferromagnets as well~\cite{Kovalev18,Azhar22}. 
	
	For dynamical excitations, the position of topological solitons can be tracked via different strategies. As merons and skyrmions have a single point where $m_z(\vec{r}) = \pm 1$, the first, and simplest, way to localize the soliton position $\vec{r}_c = \{x_c, y_c\}$ is to find the coordinate where
	\begin{equation}\label{eq:position-mz}
		m_z(\vec{r}_c) = \pm 1.
	\end{equation}
	
	A second widely-used approach to find the soliton position in an effectively two-dimensional magnetic texture defined in the $xy$ plane is to find the maximum the first moment of the out-of-plane magnetization via
	\begin{equation}\label{eq:position-moment}
		\vec{r}_c(z) = \dfrac{\iint \vec{r} \left[1 - m_z(\vec{r})\right]dx\,dy}{\iint\left[1 - m_z(\vec{r})\right]dx\,dy} 
	\end{equation}
	
	The third approach to localize a soliton was proposed in~\cite{Kravchuk23}, and relies on finding the maximum of the gyrovector density $\vec{\Omega}$ (also called topological charge flux density) as follows:
	\begin{equation}\label{eq:position-gyro}
		\vec{r}_c(z) = \dfrac{\iint \vec{r} \Omega_z(\vec{r})dxdy}{\iint \Omega_z(\vec{r})dxdy}
	\end{equation}
	with 
	\begin{equation}\label{eq:gyro-density}
		\vec{\Omega}= \dfrac{1}{8\pi} \vec{e}_i \epsilon_{ijk} \vec{m} \cdot \left[\partial_j \vec{m} \times \partial_k \vec{m}\right],
		\qquad \text{with} \quad i,j,k=x,y,z,
	\end{equation}
	c.f. Eq.~\eqref{eq:gyrovector}. 
	
	Using the different approaches Eqs.~\eqref{eq:position-mz}-\eqref{eq:position-gyro} for tracking the position of excited solitons results in slightly different measured dynamical behaviours, see discussion in~\cite{Kravchuk23} and references therein. Among the three strategies, the gyrovector-based approach of Eq.\eqref{eq:position-gyro} can provide the smoothest trajectory of the moving soliton in the absence of higher-order spurious dynamic modes. Similar techniques are applied to find topologically interesting features in 3D magnetic textures, including identification of local maxima in $\vec{\Omega}$~\cite{Donnelly20b} and building moments of $\vec{r}$ based on the $\tilde{\Omega} = |\vec{\Omega}|/\max |\vec{\Omega}|$~\cite{Volkov23a}.
	
	\subsubsection{Relating Topological Properties of Real Space and Order Parameter Spaces}
	\label{sec:topo-theorems}
	
	A remarkable property of the Gauss curvature, which is an intrinsic property of any given surface $\vec{\varsigma}$ defined in Eq.~\eqref{eq:mean-gauss-curv}, is established by the Gauss-Bonnet theorem. For the case of a compact surface $\vec{\varsigma}$ (i.e., which is bound and closed) this theorem reads as follows, with the integration being performed over the area of $\vec{\varsigma}$:
	\begin{equation}\label{eq:gauss-bonnet}
		\oiint \tilde{G}(\vec{r}) dS = 2\pi.
	\end{equation}
	Taking into account that the length of the gyrocoupling vector $\mathcal{J}$ [Eq.~\eqref{eq:gyrovector}] for a magnetic texture $\vec{m}(\vec{r}) = \vec{n}$ with $\vec{n}$ being the normal to the surface $\vec{\varsigma}$  is $\mathcal{J} = \mp \tilde{G}$,\footnote{One can show that the same definition $\mathcal{J} = \mp \tilde{G}$ is valid a very thin curved magnetic shell with the replacement $\partial_x \to g_{11}^{-1/2}\partial_{\eta_1}$, $\partial_y \to g_{22}^{-1/2}\partial_{\eta_2}$~\cite{Kravchuk16}.} one can conclude that the locally uniform magnetic texture is formally topologically-non-trivial by the realization of a finite integer map onto the order parameter space~$\mathbb{S}^2$. 
	
	In other words, the topology of a magnetic shell (or surface) $\vec{\varsigma}$ provides a shift of the topological charge of the magnetic texture living on this shell~\cite{Kravchuk16}
	\begin{equation}\label{eq:Qgeom}
		Q_\text{geom} = \mp (1-g),
	\end{equation}
	where $g$ is the genus of $\vec{\varsigma}$, respectively the number of holes in $\vec{\varsigma}$, so that $Q_\text{geom}^\text{sphere} = \pm 1$ and $Q_\text{geom}^\text{torus} = 0$, respectively. As consequence, for the case of a normally-magnetised sphere (radial easy-axis anisotropy) a single skyrmion with $Q_2=+1$ on a torus is always a topologically non-trivial state, while on a sphere it can be a topologically trivial magnetic texture. The latter can be easily understood from a viewpoint that in real space $\mathbb{R}^3$, skyrmions on a spherical shell can be continuously deformed into a uniform state $\vec{m}\parallel\vec{\hat{z}}$~\cite{Kravchuk16}.
	
	The Gauss-Bonnet theorem~\eqref{eq:gauss-bonnet} is also linked to the \emph{Euler characteristic} $\chi_\text{eu}$, a topological invariant that classifies surfaces according to their number of holes, via
	\begin{equation}\label{eq:Euler}
		\chi_\text{eu} = \dfrac{1}{2\pi} \oiint \tilde{G}(\vec{r}) \, dS = 2(1-g).
	\end{equation} 
	As examples, the Euler characteristic of a sphere is $\chi_\text{eu}^\text{sphere}=2$, and that of a torus $\chi_\text{eu}^\text{torus}=0$.

	Let us consider a single-valued scalar field defined on $\vec{\varsigma}$ 
	\begin{equation}
		\varpi(\eta_1,\eta_2) = \arctan \dfrac{\eta_2}{\eta_1}.
	\end{equation}
	It is analogous to the field of azimuthal angle of the magnetization [see Eq.~\eqref{eq:azimuthal-angle-m}], thus, its topological defect with charge $q$ is equivalent to the presence of the vortex or antivortex. The Poincar\'{e}--Brouwer theorem states that the total topological charge of $\varpi$ equals the Euler characteristic,
	\begin{equation}
		\sum_i q_i = \chi_\text{eu},
	\end{equation}
	thus, uniquely determined by the topology of $\vec{\varsigma}$.

	\section{Fabrication of Complex-Shaped Magnetic Nanostructures}
	\label{sec:fabrication}
	
	A great deal of experimental progress in magnetism has stemmed from advances in nanofabrication, that increasingly enable the realization of intricate 3D nanostructured geometries. This aspect is key as fabrication processes can largely determine the magnetic properties of nanostructures, and the measurement of such elements can require tight conditions are satisfied, such as surface smoothness for reflection-based measurements, or sample and substrate transparency for transmission-based measurements. 
    
	Here, we discuss two complementary routes: additive methods, in which material is deposited or grown into the target geometry, and subtractive methods, in which structures are sculpted from a larger parent form. Although presented separately for clarity, these approaches are not mutually exclusive; in practice they can be combined in numerous ways to achieve new geometries, heterostructures and to tune magnetic properties. 
	
    \subsection{Additive Fabrication Methods}
    \label{subsec:additive_methods}
    
	\begin{figure}[tb]
	\centering
		\includegraphics{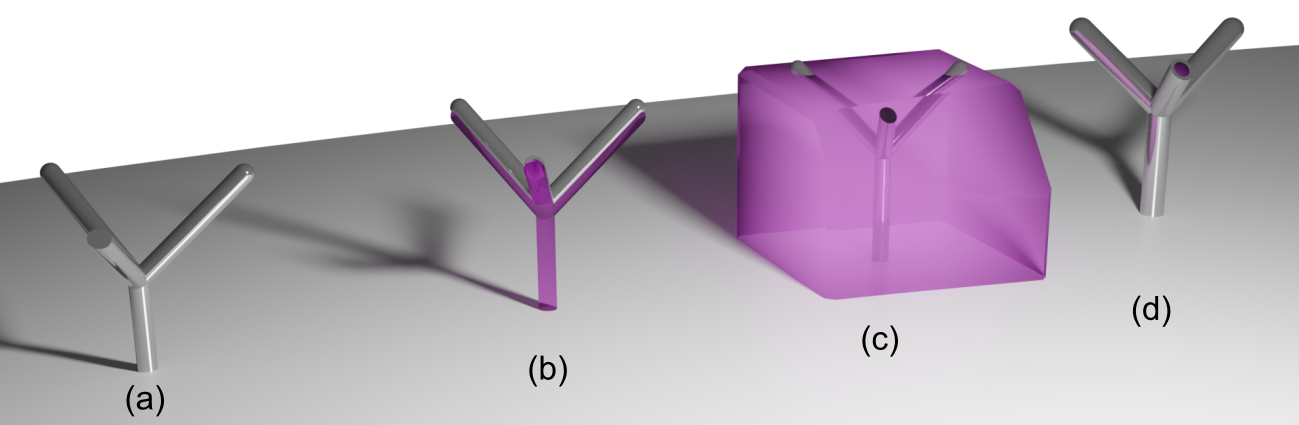} % additive-manufacturing-v2
		\caption{\textbf{Comparison of additive manufacturing methods}, with magnetic material shown in metallic gray, and non-magnetic scaffold material in pink.
		(a)~Direct writing of 3D magnetic nanostructures.
		\mbox{(b--d)}~Using non-magnetic scaffolds:
		(b)~Directional coating, potentially leading to disconnected segments, 
		(c)~bottom-up deposition into a template (e.g., via electrodeposition) leading to connected networks of solid wires, 
		(d)~conformal coatings leading to tubular geometries. The fabrication methods lead to different connectivity and cross-section of magnetic material. 
		}
		\label{fig:additive_manufacturing}
	\end{figure}
    
	Additive manufacturing of magnetic nanostructures comes in one of two main flavours: directly writing the desired magnetic structure or the creation of complex non-magnetic scaffolds which subsequently are either covered or filled with a magnetic material. These methods are schematically illustrated in Fig.~\ref{fig:additive_manufacturing}, and it is worth pointing out that the \emph{same} base structure generally will lead to \emph{different} topological properties for each chosen fabrication method (direct writing, directional coating, conformal coating or volumetric filling), as effective dimension, geometric inter\-connectivity, and anisotropy of the final magnetic structure are distinct. Availability of different fabrication methods therefore enables to create a multitude of interesting target geometries and topologies. 
	
	In addition to the geometry-defined magnetic properties from local curvature, connectivity, and effective dimension, bulk properties like saturation magnetisation, crystalline anisotropies, or damping, will also affect the overall magnetic response. Most interesting spin structures occur when the relevant intrinsic magnetic length scales, like the exchange lengths, are comparable to geometry dimensions (including wire diameter, lengths or curvature)~\cite{Guimaraes17,Hubert09,Stefanita12}. If these length scales are too disparate, certain assumptions made for an analytical predictive description (e.g., assuming local macrospin behaviour, dominant shape anisotropy, or continuously-varying spin textures) might no longer be justified.
    
    \subsubsection{Direct Printing of 3D Magnetic Nanostructures}
    \label{sec:additive_FEBID}
    
    Focused electron- or ion-beam induced deposition (FEBID/FIBID) allows to create complex free-standing metallic nanostructures [Fig.~\ref{fig:additive_manufacturing}(a)] from  magnetic materials like Co or CoFe~\cite{Teresa16,Fernandez-Pacheco20,Barth20}. FEBID uses a focused electron beam of a scanning electron microscope (respectively, the ion beam of a dual-beam FIB/SEM machine) to decompose metal-containing organic precursor molecules that are deposited via a gas injection system on the substrate surface. Using precisely-placed individual nanoscale metal deposits, complex three-dimensional geometries can be created with layer-by-layer growth, with open-source codes available to translate standard 3D stl files to printing instructions~\cite{Skoric20,f3ast_github}. Examples of geometries grown by FEBID encompass nanostructures with controlled local curvature~\cite{Donnelly22,RuizGomez25}, geometrical chirality~\cite{Sanz-Hernandez20,Fullerton25a,Gokita26}, and complex interconnectivity~\cite{Meng21a,Volkov24a}. 
    
    The quality of magnetic properties depends on the FEBID growth conditions, like beam current and gas pressure~\cite{SanzHernandez17,Skoric20,Fernandez-Pacheco20}, and often poly\-crystalline or amorphous growth is favoured so that the geometry-induced shape anisotropy dominates the magnetic response. Material properties can be improved by post-growth using thermal or electron-beam-induced annealing procedures.
    
    In addition to FEBID, other additive metal deposition techniques exist, but these usually target larger feature sizes in the micrometre regime~\cite{Hirt17}, and are thus no suitable for the study of curvilinear magnetic nanostructures.
    
    \subsubsection{Non-Magnetic 3D Scaffolds}
    
    Another route to complex three-dimensional magnetic nanostructures from additive manufacturing is to create non-magnetic scaffolds (or templates) which are subsequently decorated with a magnetic material. Schematically this approach is shown in Fig.~\ref{fig:additive_manufacturing}(b--d), where non-magnetic scaffolds are (b)~coated by directional deposition techniques, (c)~filled via electrodeposition, or (d)~are covered with a conformal coating.
    
    Separating the scaffold fabrication and growth of magnetic material offers two key advantages over direct growth:  First, due to a variety of available fabrication techniques it is possible to create scaffolds which combine both large and small feature sizes over areas and volumes that are not feasible with direct-write FEBID. Second, it is possible to grow a variety of high-quality materials onto or into the created scaffolds~\cite{Askey25}, including magnetic compounds for which no FEBID-precursors yet exists, as well as magnetic multi-layers with tailored anisotropies.

	Like in other additive manufacturing techniques, there are some limitations to creating arbitrary scaffolds as the different components of the geometry will need to be connected to a substrate, and often to each other as well. Scaffold materials can range from metals to polymers, and their mechanical, chemical, and thermal stability might be key consideration for the subsequent fabrication steps that are used to add the desired magnetic material. 
    
    \paragraph{Creating Non-Magnetic 3D Scaffolds}
    
    One method to create scaffolds is, again, FEBID using a non-magnetic precursor, most often commonly available Pt, as it is used to lift-off TEM grids prepared in FIB/SEM dual beam machines. Notable examples of FEBID-grown non-magentic scaffolds are shown in~\cite{Sanz-Hernandez18,Skoric22,Meng21a}.
    
    Another technique is two-photon polymerisation~\cite{Cui08,Hayat11,Zheng16,Williams18}, where a focused laser beam is used to expose small volumes of a liquid resist vat, resulting in complex micro-metre sized shapes including magnetic Buckyballs~\cite{Donnelly15} and extended diamond lattices as 3D extensions of planar artificial spin ices~\cite{May19,May21,Sahoo21}, as well as other artificial spin systems~\cite{Pip22,Berchialla24}, see also Sec.~\ref{sec:wireframes}. Typically, the resolution for two-photon polymerisation is in the order of a few hundred \unit{nm} with asymmetric voxels (being more elongated along the beam direction than transverse to it)~\cite{Askey25}, while post-growth treatments allow to create isolated or sub-\SI{100}{nm} features~\cite{vandenBerg23,Askey24}. % pyrolysis, beam scanning

	Self-assembly techniques offer the promise of creating nanoscale with extended volumes by repeating unit cells of a base structure.  With respect to topology, self-assembled lattices feature internal voids bridged with different interconnectivity, which affects both the anisotropic magnetic response and can also promote magnetic spin textures governed by geometric frustration. Examples for self-assembled structures include inverse opals created by regular packing of colloidal particles or polystyrene spheres~\cite{Grigoriev09a,Zeissler13,Dubitskiy17,Rana23}, anodised alumina templates that allow growth of high-aspect ratio nanowires as thin as \SI{10}{nm} and lengths of a few micrometer over large (\unit{cm^2}) areas~\cite{Vazquez05,Hauet14} with cross-connections by pulsed fields during the template growth~\cite{Martin14}, or the creation triply-periodic minimal-surface gyroid structures from di- and tri-block co-polymers~\cite{Cui08,Dolan15} (properties of nanomagnetic gyroids will be discussed in Sec.~\ref{sec:gyroids}).

	A subtractive method to create scaffolds with non-trivial, non-periodic connectivity uses ion bombardment from different directions to drill random holes. This method creates scaffolds for sparsely linked high-aspect ratio nano-wires, for which the density, directionality and average inter-connectivity can be varied by ion dose and ion beam direction. Using electro\-deposition to fill the channels and subsequent studies allow to explore magnetic properties, in particular magnetoresistance, in randomly-connected networks~\cite{Bhattacharya22,Burks20}.

	\paragraph{Coating Scaffolds with Magnetic Thin Films}
	
	Standard thin-film growth techniques including sputtering, thermal deposition or electron-beam induced deposition, can be easily used to grow thin magnetic films on a target geometry. A key advantage of this approach is that it allows to cover scaffolds and materials with thin films of tailored magnetic properties, including perpendicular magnetic anisotropy or low-damping magnonic materials.  As any of these methods creates directional coatings, any surface which is turned away from the plume (or shadowed by the scaffold) will encounter less (or no) deposition. As a result, and dependent on the scaffold shape, the deposited film might be fully or partially connected to typically form a so-called 2.5D magnetic surface [see Fig.~\ref{fig:additive_manufacturing}(b)], with half-spherical nanocaps~\cite{Sapozhnikov12,Volkov23a} and three-dimensional domain racetracks~\cite{Meng21a,Gu22} being prominent magnetic examples.  
	
	Recent progress in atomic layer deposition or other electroless deposition techniques now also allow to create truly-3D interconnected magnetic geometries by conformal growth on free-standing scaffolds [Fig.~\ref{fig:additive_manufacturing}(c)], with materials like permalloy, nickel~\cite{Pip20,Xu25b} or YIG being very attractive especially for dynamic studies of domain wall motion in tubular geometries. 
    
    \paragraph{Filling Scaffolds with Magnetic Materials}
    
    A third approach to creating 3D magnetic structures is to fill voids of a nanoscale scaffold with a bulk-like magnetic material to create an interconnected negative of the scaffold itself, as illustrated in Fig.~\ref{fig:additive_manufacturing}(d). The key method here is electro-deposition or electroplating, by which the scaffold is grown on a conducting surface, and then placed in a liquid bath. Material composition and growth rate can be controlled by the materials in solution as well as details of the pulsed voltage application~\cite{Cui08,Scherer14,Askey25}.
    
    \subsection{Subtractive Fabrication Methods}
    \label{subsec:subtractive_methods}
    
    In many solid-state systems, material properties are constrained by crystal symmetry~\cite{Birss66,Nye85}. In particular, the magnetocrystalline anisotropy as well as the Dzyaloshinskii–Moriya interaction~\cite{Dzyaloshinsky58,Moriya60} are rooted in the crystal structure, and are key for the emergence of topologically non-trivial bulk spin textures such as skyrmions~\cite{Bogdanov89,Muhlbauer09,Chacon18} and hopfions~\cite{Sutcliffe18,Zheng23}. The ability to realize long-range crystallinity in nano\-structured geometries is therefore essential for combining key intrinsic bulk interactions with the emergent effects due to non-trivial sample geometries~\cite{Volkov18}.  However, additive fabrication techniques, while offering considerable freedom in producing 3D geometries, typically involve the deposition of amorphous or nanocrystalline magnetic materials instead. For example, FEBID precursor chemistry typically yields amorphous films when the metal tends to form carbides, or metallic nanocrystallites when the metal is immiscible with carbon~\cite{Huth12}.

    Consequently, additive methods alone typically are an ill-suited solution for realizing curvilinear and topologically non-trivial geometries across a wide range of \emph{crystalline} materials.  Instead, a complementary set of subtractive, or destructive, methods allow one to sculpt 3D geometries from larger parent systems (grown by conventional single-crystal growth techniques or from  patterned heterostructures), thereby preserving crystallinity while enabling complex geometries. Importantly, subtractive fabrication approaches are not confined to magnetism~\cite{Moll18} only. Comparable strategies are increasingly relevant to other correlated electron systems, such as superconductors (where geometric confinement and topological structuring are well known to generate novel emergent phenomena) or ferroelectrics (where confinement and surface charge distributions strongly influence domain ordering and stability~\cite{Lukyanchuk20}). 
    
    \subsubsection{Chemical Etching} %%%%%%%%%%%%%%%%%%%%%%%%%%%%%%%%%%%%%%%%
    
    Etching techniques represent a well-established family of subtractive fabrication methods originating in the integrated circuit industry~\cite{Plummer23_9} and now widely applied to magnetic thin films  and heterostructures~\cite{Chen17c,Kar23}. In essence, material is selectively removed by chemical or physical processes, enabling the definition of patterns such as nanodots, disks, and rings, or serving in the fabrication of scaffolds onto which additive processes can take place.
    
    \emph{Wet chemical etching} removes material through selective chemical reactions between the etchant and the target film. High selectivity from the chemical reactivity means that the target material is removed at a significantly higher rate than any masked or surrounding layers, allowing patterned geometries to be transferred without damaging the surrounding material. For magnetic garnets such as Yttrium iron garnet (YIG), phosphoric acid wet etching has been used to fabricate microstructured films for magnonics, including one-dimensional magnonic crystals that scatter backward spin waves~\cite{Chumak08}. 
    Such studies demonstrate that wet etching can produce functional spintronic devices, but the etching process is isotropic, typically resulting in sloped walls and limiting lateral feature resolution to the micron scale~\cite{Zhu17}.
    
    \emph{Dry plasma etching methods}, especially reactive ion etching (RIE) and inductively coupled plasma RIE (ICP-RIE), combine reactive chemistry with directional ion bombardment, enabling geometric profiles with near-vertical sidewalls~\cite{Coburn79}. Plasma etching techniques have been applied in a range of magnetic materials, including IrMn layers in magnetic tunnel junction stacks~\cite{Lee12}, Heusler alloys such as Co$_{2}$MnSi~\cite{Garay15}, and other spintronic devices where sub-\SI{100}{nm} features are required~\cite{Chen17c}. The main trade-off in reactive ion etching is between process damage and chemical residues, requiring a careful choice of gas chemistry and masking strategy. 
    
    Such chemical etching provided some of the earliest demonstrations for creating topologically non-trivial magnetic geometries, most notably ferromagnetic rings in which vortex and onion states can be stabilized~\cite{Klaui03} (see Sec.~\ref{sec:toroidal}). These pioneering studies showed how subtractive methods could impose non-trivial topologies on thin films, but the approach remains fundamentally mask-defined and line-of-sight-limited, restricting structures to planar or quasi-2D magnetic stacks.

    \subsubsection{Laser Ablation}%%%%%%%%%%%%%%%%%%%%%%%%%%%%%%%%%%%%%%%%%
    
    Laser writing is a versatile tool in which focused light pulses can be harnessed to grow, ablate, or transfer material, as well as to locally modify material properties~\cite{Malinauskas16}. Ultrashort pulses on the femtosecond–picosecond (fs–ps) timescale drive multiphoton absorption processes, enabling high-resolution modification of the host material. In contrast, continuous-wave or longer-pulse irradiation causes heating, resulting in targeted thermal ablation~\cite{Gattass08} or local modification of magnetic properties~\cite{Riddiford25}. The method is inherently maskless and flexible, has been exploited in a range of rapid-prototyping contexts~\cite{Sugioka14}, and now finds increasing use in magnetism. 
    
    Furthermore, laser-based methods can serve a dual role in defining magnetic nanostructures as well as actively driving ultra-fast spin dynamics, enabling direct optical control of magnetic order.  For example, targeted laser patterning can define conduits for domain walls, imprint spatially periodic anisotropies to realize magnonic crystals, and nucleate topological spin textures such as skyrmions~\cite{Finazzi13}. In ferromagnetic and ferrimagnetic films, laser irradiation can also trigger all-optical magnetization switching, where circularly polarized light can induce helicity-dependent switching~\cite{Mangin14}. 
    
    \subsubsection{Focused Ion Beam Milling}%%%%%%%%%%%%%%%%%%%%%%
    
\begin{figure}[tb]
        \centering
        \includegraphics[width=0.95\linewidth]{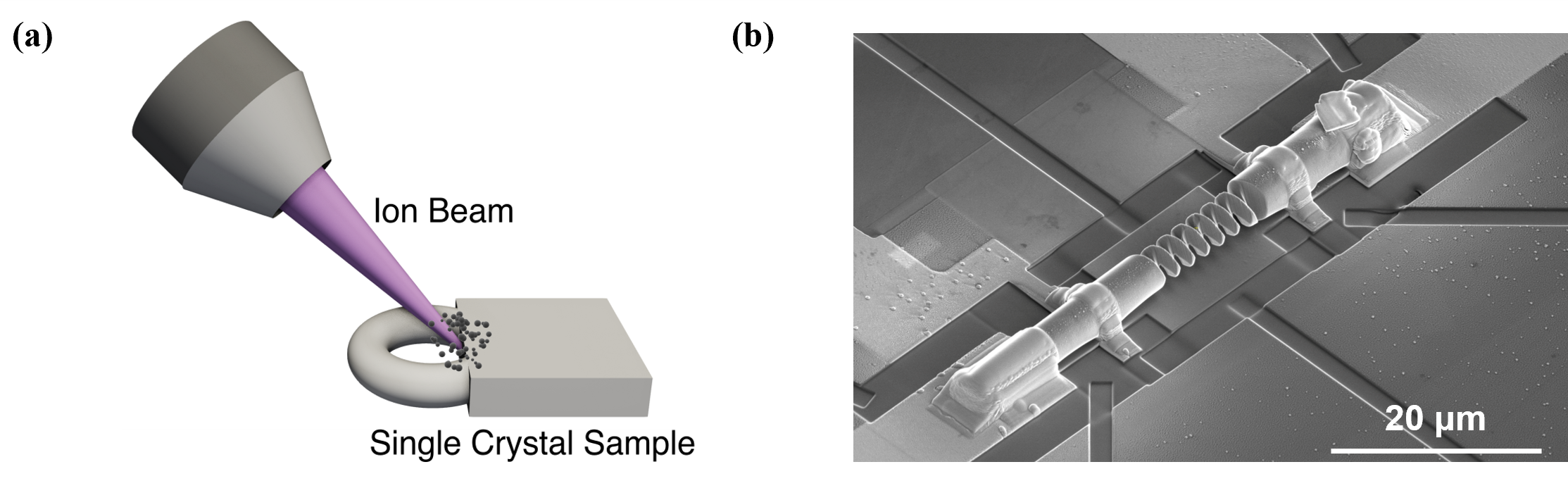} % fig_FIB
        \caption{(a) Schematic of a focused ion beam milling. b) SEM image of a single crystal Co$_{3}$Sn$_{2}$S$_{2}$ helix prepared by FIB milling on a substrate with electrical contacts for transport measurements. Image courtesy of M.~T.~Birch.
        }
        \label{Fig:FIB}
    \end{figure}

    Focused ion beam (FIB) milling is a fundamental tool for nanoscale subtractive shaping. In this technique, a tightly focused beam of energetic ions, most commonly Ga$^+$, impinges on the sample surface and removes material through sputtering with nanometric precision~\cite{Giannuzzi05,Hoflich23}. FIB is widely utilized for preparing electron microscopy lamellae, and although it is a line-of-sight method, multi-axis stages allow the sample to be tilted relative to the beam, enabling the selective milling of patterns at a variety of incidence angles, thereby enabling the sculpting of intricate topological geometries, see Fig.~\ref{Fig:FIB}(a).  Recent examples include single-crystal tori and helices~\cite{Turnbull25,Birch26} such as the helix formed by FIB milling of single crystal Co$_{3}$Sn$_{2}$S$_{2}$ shown in Fig.~\ref{Fig:FIB}(b).

    An important consideration in FIB milling is the formation of a damaged or amorphous surface layer. Ion implantation and knock-on collisions can suppress crystalline order, alter magnetic anisotropy, and reduce the local magnetic moment. These effects may be undesirable, and can be limited by controlling process parameters such as the incident milling angle and ion acceleration voltage. In addition, post-patterning treatments, such as low-dose polishing with lighter ions or post-annealing, can help to reestablish crystalline order.
    
    At the same time, the associated damage from FIB processing can be deliberately harnessed as a functional design tool. A notable example is the induction of a soft ferromagnetic layer with suppressed DMI around the edge of a helimagnetic lamella, thereby providing a practical nucleation pathway for magnetic Hopfions from its edges~\cite{Zheng23}.
    
    The choice of ion source further expands the scope of FIB processing. Light He$^+$ ions generate smaller interaction volumes and enable finer patterning~\cite{Aramesh18}, at the expense of longer processing times, whereas heavier Xe$^+$ ions in plasma FIBs provide the higher throughput required for large-scale structuring, at the expense of resolution~\cite{Liu20c}. The growing palette of available ion species thus allows researchers to tune resolution, throughput, and implantation effects according to the material system under study.
    
    Beyond geometric sculpting, ion beams can also serve as a powerful tool to locally tailor magnetic anisotropy and related material properties. In 1998 it was demonstrated that light-ion irradiation allows to accurately control the magnetic anisotropy of multilayer systems~\cite{Chappert98}.  Since then, a range of other magnetic properties -- including interlayer exchange coupling, exchange bias, magnetic damping, and saturation magnetization -- have also been shown to be modified by ion irradiation or ion implantation~\cite{Fassbender08a,Cybart16}. Together, these advances in FIB processing highlight the dual role of focused ion beams as both a subtractive nano\-fabrication technique as well as a means of function materials engineering, underscoring the potential importance in the design and realization of curvilinear and topological magnetic systems.

	\section{Experimental Characterization Techniques}
	\label{sec:measurement}%%%%%%%%%%%%%%%%%%%%%%%%%%%%%%%%%%%%%%%%%%%%%%%%%%%%%%%%

	The physical properties of magnetic systems are largely determined by the ordering of the magnetization and its response to applied stimuli, such as static and oscillating magnetic fields, electrical currents, and temperature~\cite{Buschow03}.
    
    Common experimental techniques to measure the magnetic properties and responses can be broadly classified into two categories: macroscopic probes, which examine global properties derived from that ordering~\cite{Franco21a} and microscopic probes, which evaluate the spatially resolved magnetic ordering~\cite{Donnelly22a}. A further distinction can be drawn between static and dynamic probes, sensitive to time-averaged quantities or fast responses, respectively~\cite{Mewes21, Barman21}. There exist many established magnetic measurement techniques addressing different length and time scales, however, most of these are optimized for planar or bulk samples. Therefore their application for curved 3D geometries needs to be carefully adapted to avoid artefacts and misinterpretation.
    
    With respect to topological quantities arising from complex 3D geometries, many experimental studies employ nanoscale imaging techniques~\cite{Donnelly22a,Kimel22}, as these provide direct evidence of topological features in the real-space visualization of the local magnetization field.  Many of these imaging techniques are typically associated with considerable  effort (such as experimental cycles at large-scale research facilities), and often impose stringent limitations on permitted sample geometries.
    
    In contrast, spatially averaging measurement techniques do not give direct access to local magnetic properties. However, if sufficient magnetic material is available, they still are valuable tools to give insights to averaged anisotropies, relevant spin dynamics, or indicators for which magnetic reversal processes take place in 3D nanostructures~\cite{Gokita26}. Such bulk-like measurements on 3D magnetic nanostructurses are particularly powerful if combined with three-dimensional magnetic fields~\cite{SanzHernandez23,CascalesSandoval25}.
    
    To highlight one non-imaging technique, an extension to standard hysteresis loop measurements are first-order reversal curves (FORC), which can be either measured using a standard vibrating or SQUID magnetometer~\cite{Winklhofer06,Dobrota13,Alikhani16,Muscas20} or MOKE (see also Sec.~\ref{subsec:MOKE})~\cite{Grafe14,DeBiasi17}: In addition to major switching loops, an extensive number of minor loops are measured as well, and the analysis of the dataset gives insights into cooperative magnetic ordering effects and anisotropies during the magnetic reversal process (often using supplementary micromagnetic simulations to fully understand key processes, also see Sec.~\ref{subsec:micromagnetic}). As a caveat, for a reliable FORC analysis it is critical that the minor loop is always measured from a fully saturated state, to ensure the same initial configuration throughout the dataset. Unfortunately, especially for 3D nanostructures, which often feature involved geometry-induced magnetic anisotropies, very high magnetic fields might be required, imposing additional constraints to experiments (especially MOKE setups do not enable high-field electromagnets). 
    
    \subsection{Scanning Probe Techniques}
    \label{subsec:scanning_probe}
    %%%%%%%%%%%%%%%%%%%%%%%%%%%%%%%%%%%%%%%%%%%%%%%%%%%%%%%%
    
    Scanning probe methods are among the most widely employed in nanoscale magnetism~\cite{Marchiori21}. In these approaches, a sharp tip is scanned close to the surface of a sample, and a variety of interactions between the tip and sample are exploited in order to map the magnetic state of the system with nanoscale spatial resolution. 
    
    Due to the simplicity of the measurement scheme, scanning probe microscopy is compatible with a wide range of sample environments, including applied magnetic fields and varying temperatures. As the scanning is an inherently serial process, however, throughput is usually constrained and the accessible field of view is limited to several square microns (determined by the range of the piezo actuators). 
    
\begin{figure}[tb]
        \centering
        \includegraphics[width=0.95\linewidth]{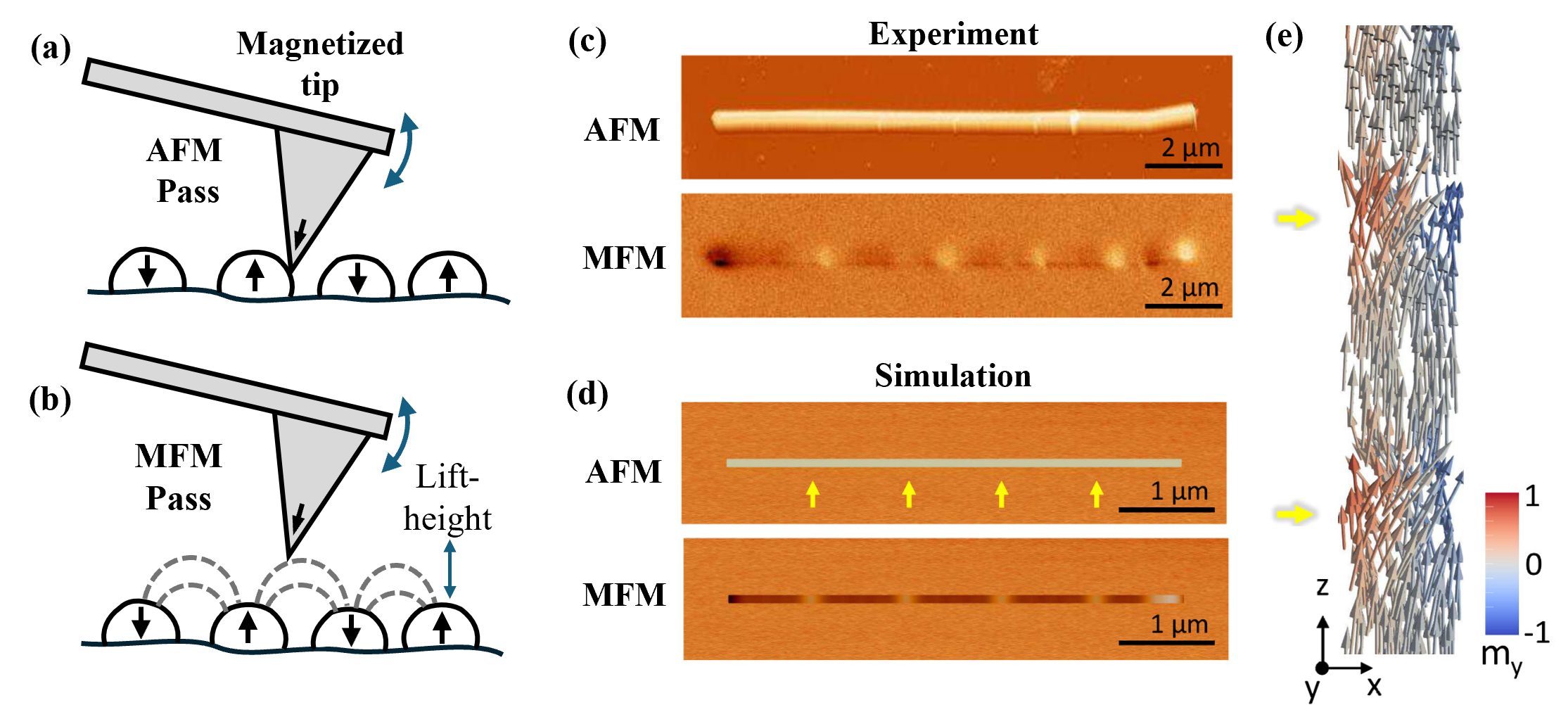} % fig_mfm
        \caption{(a,b) Schematics of the AFM and MFM passes of the probe tip during the scan. (c,d) Experimental and simulated AFM (top) and MFM (bottom) contrast for a magnetic nanowire exhibiting compositional gradients. (e) Magnetization texture predicted by micromagnetic simulation. Panels (c-–e) adapted under Creative Commons Attribution 4.0 License (https://creativecommons.org/licenses/by/4.0/) from~\cite{FernandezGonzalez25a}~\textcopyright{}~(2025) by the Authors.}
        \label{Fig:MFM}
    \end{figure}

    Magnetic Force Microscopy (MFM) has been used to image nanomagnetic structures since the 1980s~\cite{Rugar90,Kazakova19}. MFM operates in a similar fashion to Atomic Force Microscopy (AFM), but employs a magnetized tip on the end of an oscillating cantilever. In a common mode of operation, the tip first performs an AFM-type measurement to determine the surface topography of the sample, and then, in a second pass, the tip is scanned a fixed distance away from the surface and the magnetically polarized tip is used to sense the gradient of the local stray field [Fig.~\ref{Fig:MFM}(a,b)]. This magnetic tip-sample interaction is detected as a shift in the phase, amplitude, or frequency of the driven cantilever relative to its unperturbed oscillation~\cite{Kazakova19}.  The resulting stray field maps are typically qualitative, and extracting quantitative information typically requires the probe to be rigorously characterized~\cite{Haberle12,Feng22}. 
    
    MFM has been successfully used to map the magnetic ordering in a wide range of curved geometries including capped spherical particle assemblies~\cite{Ulbrich06,Baraban08}, magnetic nanowires~\cite{Jaafar11,Fernandez-Pacheco13}, and magnetic nanowire lattices~\cite{May19}. 
    Representative experimental and simulated images are presented in Fig.~\ref{Fig:MFM}(c--d), illustrating both the AFM and MFM contrast of a magnetic nanowire. The nanowire exhibits enhanced curling of the magnetization in Fe-rich regions of the sample compared to Ni-rich regions [Fig.~\ref{Fig:MFM}(e)], thereby demonstrating how compositional gradients can be used to tune magnetic properties~\cite{FernandezGonzalez25a}. Direct comparison between experimental images and simulated micromagnetic contrast, as exemplified in Figs.~\ref{Fig:MFM}(c--d) constitutes a widely adopted strategy for interpreting magnetic contrast when quantitative information on the tip is unavailable.

    \subsubsection{Other Magnetic Scanning Probe Techniques}
    
    Nitrogen vacancy (NV) magnetometry is a quantum sensing method that utilizes the localized spin of an electron in a nitrogen-vacancy defect in a diamond probe to measure static and dynamic magnetic fields with ultra-high sensitivity~\cite{Casola18} and applicable for curved films~\cite{Raftrey25}.
    Spin-Polarized Scanning Tunnelling Microscopy (SP-STM), by contrast, makes use of tunnelling currents to map magnetic ordering at an atomic level~\cite{Wiesendanger09}.
    {Scanning Brillouin Light scattering (micro-BLS) can also be employed as a magnetic probe into magnonic excitations at low wavevectors, although its spatial resolution is constrained by the optical diffraction limit~\cite{Sebastian15}.}
    
    Advanced scanning probes add to the available capabilities, particularly in terms of sensitivity and quantitative field mapping, but they remain intrinsically scanning. Furthermore, in most implementations, the magnetic stray field \emph{above} the sample surface is detected, effectively sampling a two-dimensional manifold close to the sample surface, rather than directly probing the 3D magnetization within the sample volume.

    The sensitivity to reduced components can create particular difficulty for curved and interconnected geometries where topographical artefacts and magnetic contributions from different parts of the sample may overlap in the measured signal, leading to potential ambiguities in interpretation~\cite{Grutter94}. As a result, reconstructing the underlying magnetization texture in curved and three dimensional nanostructures becomes an inverse problem that is often under-constrained without strong assumptions or prior knowledge of the underlying system~\cite{Blakely95,Xu23b,Broadway25}.
    
    \subsubsection{Experimental Challenges to Scanning Probes}
    
    When measuring 3D nanostructures, certain limitations of scanning probe methods can become evident. For example, commercially available MFM tips often have a pyramidal profile, so the extended sides of the tip can interact and collide with local topographical features, giving rise to pronounced artifacts. Strong topographic gradients, which are inherent to many curved nanostructures, can therefore couple strongly into the measured signal and obscure or dominate the magnetic contrast. 
    
    The spatial resolution of commercially-available MFM tips is typically limited to around \SI{30}{nm}, which can deteriorate significantly on rough or varying topographies~\cite{Kazakova19}. One strategy to address this challenge is the use of FEBID-grown ferromagnetic nanowires as ultra-sharp MFM tips, enhancing the resolution to around \SI{10}{nm}~\cite{Belova12}. The high aspect ratio of such tips furthermore mitigates the limitations of wider tips in probing 3D nanostructures by reducing lateral clashing and by shrinking the effective volume of the tip magnetization~\cite{Gavagnin14}. 
    
    For detecting weak signals high-moment tips are often required. This can lead to tip-induced perturbations to magnetically soft or metastable states, where the stray field of the probe modifies the configuration being imaged. Probing of antiferromagnetic ordering is also inherently restricted due to the vanishing stray fields.
    
    \subsection{Magneto-Optical Measurements}
    \label{subsec:MOKE}
    %%%%%%%%%%%%%%%%%%%%%%%%%%%%%%%%%%%%%%%%%%%%%
    
    Magneto-optical (MO) effects are phenomena in which the polarization state or intensity of incident light is modified upon reflection from, or transmission though, a magnetized material~\cite{Kimel22,Sato22a}. These effects derive from the off-diagonal elements of the dielectric tensor, and can be understood in terms of optical birefringence and dichroism, where left- and right-circularly polarized light experience different refractive indices in a magnetized sample. As a linearly-polarised light wave can be represented as the superpsition of left- and right-circularly polarised components, the magnetically-induced birefringence modifies an initially linearly polarized beam by rotating the polarization plane and inducing ellipticity.
    
    Two of the most employed magento-optical effects are the Faraday effect, observed in transmission, and the magneto-optical Kerr effect (MOKE), observed in reflection. MOKE is a particularly attractive lab-based method (i.e., not requiring access to central experimental facilities) applicable to a wide range of non-transparent samples (such as metallised magnetic films), and it is widely used to obtain ensemble-averaged information such as hysteresis loops (MOKE magnetometry), as well as to locally map magnetic domains (MOKE microscopy)~\cite{Kimel22}.  Combination with applied vector magnetic fields~\cite{CascalesSandoval25} then allows detailed studies on local anisotropies, switching behaviour, and domain propagation.
    
\begin{figure}[t!]
        \centering
        \includegraphics[width=0.95\linewidth]{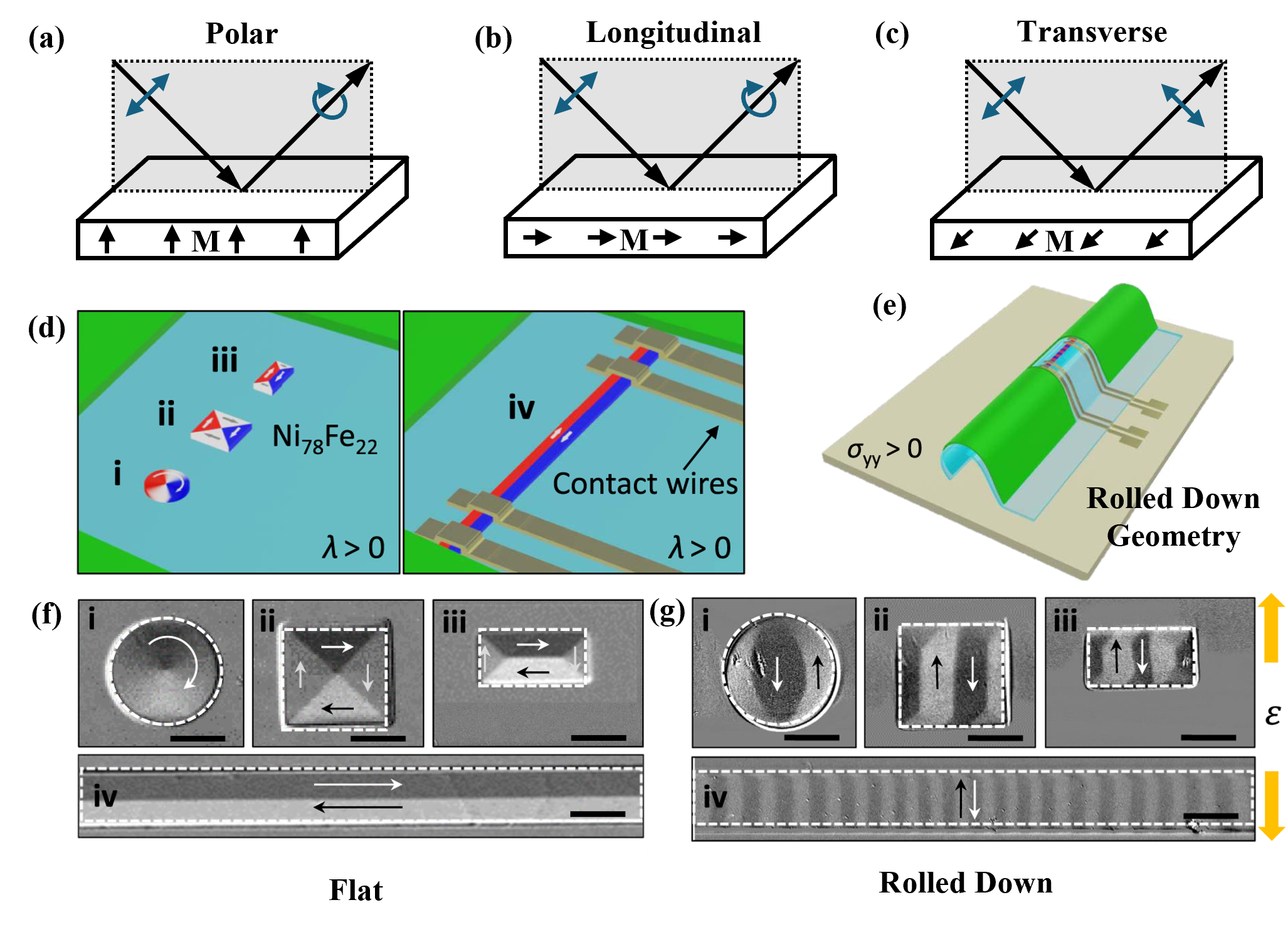} % fig_moke
        \caption{(a–c) Schematics of the polar, longitudinal, and transverse modes of MOKE, illustrating the magnetization component (M) to which each mode is sensitive. (d) Schematic representations of magnetic configurations in planar structures: i) disc, ii) square, iii) rectangle, and iv) stripe. (e) Schematic of the rolled‑down polymeric platform used to induce strain in the magnetic structures. (f) Kerr images of the planar structures, revealing archetypal vortex and Landau patterns. (g) Corresponding Kerr images of the rolled‑down magnetic structures. Yellow arrows denote the direction of tensile strain. Panels (d–g) adapted under Creative Commons Attribution 4.0 License (https://creativecommons.org/licenses/by/4.0/) from~\cite{Singh22}~\textcopyright{}~(2022) by the Authors.}
        \label{Fig:MOKE}
    \end{figure}

    The scattering geometry of the MOKE measurement determines which components of the magnetization are accessible. In planar samples such as thin films, the scattering plane is fixed, and the standard polar, longitudinal, and transverse geometries provide sensitivity to out-of-plane or in-plane magnetization components (Fig.~\ref{Fig:MOKE}). 
    On curved surfaces, however, the local surface normal and plane of incidence can shift across the object, so the local MOKE signal may be effectively sensitive to different magnetic components~\cite{Soldatov21a}. As a result, the same magnetization orientation can produce varying MOKE contrast as a function of sample position, and numerous studies report that careful analysis of the Kerr contrast generated by complex geometries is needed to eliminate spurious conclusions from geometry-induced artefacts~\cite{Soldatov21a,Richter17}.
    
    MOKE microscopy enables the imaging of magnetic domains using laboratory-based optical microscopes with polarization sensitivity, where standard objective lenses enable wide-field imaging from centimetre down to micron scales. In practice, visible range wavelengths are used, limiting the spatial resolution by diffraction. The probe depth is limited to the optical penetration depth of the light, typically a few tens of nanometres in metals, meaning Kerr imaging is broadly a surface-sensitive technique. Scattering from curved objects can furthermore distribute the reflected light to a range of angles, which can lead to signal loss if part of the reflected beam falls outside the objective lens. Preferred reflection directions, however, can also be used to detect background-free MOKE contributions from complex-shaped geometries with a dark-field MOKE approach~\cite{SanzHernandez23}.     Additionally, due to the finite depth of focus of the lenses, there can be restrictions on the extent of sample that remains in focus at any given time~\cite{Streubel14a}. This latter point, however, can be circumvented by using scanning-focus microscopy approaches~\cite{Zhao26}.
    
    MOKE microscopy has been applied to a wide array of curved systems including cylindrical microwires and thin cylinders~\cite{Richter17,Vahovsky19,Richter22,Ye13a}, rolled up nano-membranes and tubular architectures~\cite{Streubel14a,Streubel13b,Singh22}, and other curved surfaces~\cite{Wu03,Bezsmertna24}.  Notably, it was demonstrated that MOKE can directly resolve strain-induced changes to magnetic ordering in rolled-up tubular systems, where a polymeric platform was used to induce tensile strain in a series of patterned magnetic structure [Fig.~\ref{Fig:MOKE}(d,e)] giving rise to an azimuthal easy axis, as shown by the distinct domain reorientation observed in Fig.~\ref{Fig:MOKE}(f,g)~\cite{Singh22}.

    \subsection{X-ray Magnetic Imaging}
    \label{subsec:X-ray}
    %%%%%%%%%%%%%%%%%%%%%%%%%%%%%%%%%%%%%%%%%%%%%
   
    X-rays have played a pivotal role in many fields of condensed matter physics (and other sciences) since their discovery in 1895~\cite{Rontgen96}. X-rays form part of the electromagnetic spectrum, with energies ranging from $\sim$\SI{200}{eV} to $>\SI{20}{keV}$, corresponding to wavelengths ranging from a few nanometres down to the atomic scale~\cite{AlsNielsen11}. While X-rays can be produced by laboratory sources, modern X-ray science is driven by large-scale international user facilities such as synchrotron light sources and free-electron lasers (FELs), which deliver high brilliance beams with tunable energy and polarization~\cite{Jaeschke16}. As a result, X-rays are a powerful probe of nanoscale structures. 
    
    In non-resonant X-ray scattering, the electronic charge signal typically dominates the significantly weaker magnetic contribution, so for many years magnetic X-ray scattering was viewed as an impractical curiosity~\cite{AlsNielsen11}. This view changed in the 1980s and 90s with the postulation and discovery of X-ray magnetic circular dichroism (XMCD)~\cite{Thole85, vanderLaan86,Schutz87} and X-ray magnetic linear dichroism (XMLD)~\cite{Stohr98,vanderLaan09,Kuiper93}. XMCD and XMLD are \emph{resonant} spectroscopic processes in which the photon energy is tuned to an element-specific absorption edge. With an appropriate choice of absorption edge and polarization, X-rays become sensitive probes of the element-specific magnetization.  For 3d transition metals, the relevant absorption edges lie in the soft (\textless \SI{1}{keV}) X-ray range; for rare-earth elements these edges are in the hard (\textgreater \SI{1}{keV}) X-ray range~\cite{xray_data_booklet}.
    
    Whereas XMCD mainly is sensitive to a finite (element-specific) magnetization, XMLD provides sensitivity to the N\'eel vector of antiferromagnets, which is often hidden from other probes due to its vanishing net magnetization. Notably, XMCD is sensitive to the projection of the magnetization along the X-ray propagation direction, whereas XMLD depends on the relative orientation between the N\'eel vector and the linear polarization axis. As such, XMLD can distinguish between spin axes aligned parallel or perpendicular to the polarization direction, but not between antiparallel configurations~\cite{Stohr98}. Both magnetic dichroic effects also enable subtraction of non-magnetic contributions, isolating the purely magnetic signal, and even allow separation of spin and orbital contributions to the magnetic moment via the XMCD sum rules~\cite{Thole88,Thole92}. These magnetic spectroscopic contrast mechanisms can be used in conjunction with many X-ray scattering and imaging modalities, and we focus in the following on methods particularly suited to characterizing curved magnetic nanostructures.

    \subsubsection{X-ray Photoemission Electron Microscopy}
    
    X-ray Photoemission Electron Microscopy (X-PEEM) is a full-field imaging technique in which the image contrast arises from the spatially varying yield of secondary electrons generated under X-ray illumination~\cite{Locatelli25}. As such, PEEM is inherently surface sensitive as the mean free path of the liberated electrons is limited to a few manometers below the surface. Therefore, the sample does not need to be X-ray transparent and only a conductive sample surface is required, making PEEM imaging compatible with many of the complex fabrication methods described above.
    
    X-PEEM has been used extensively to image planar systems, where the X-ray beam typically illuminates from a shallow angle relative to the surface, and is therefore primarily sensitive to in-plane magnetism, when using XMCD as a contrast mechanism. Surface topography can, however, significantly affect the PEEM contrast, with concave/convex surfaces leading to decreased/increased contrast, respectively~\cite{Stohr00}.
    
\begin{figure}[tb]
        \centering
        \includegraphics[width=0.85\linewidth]{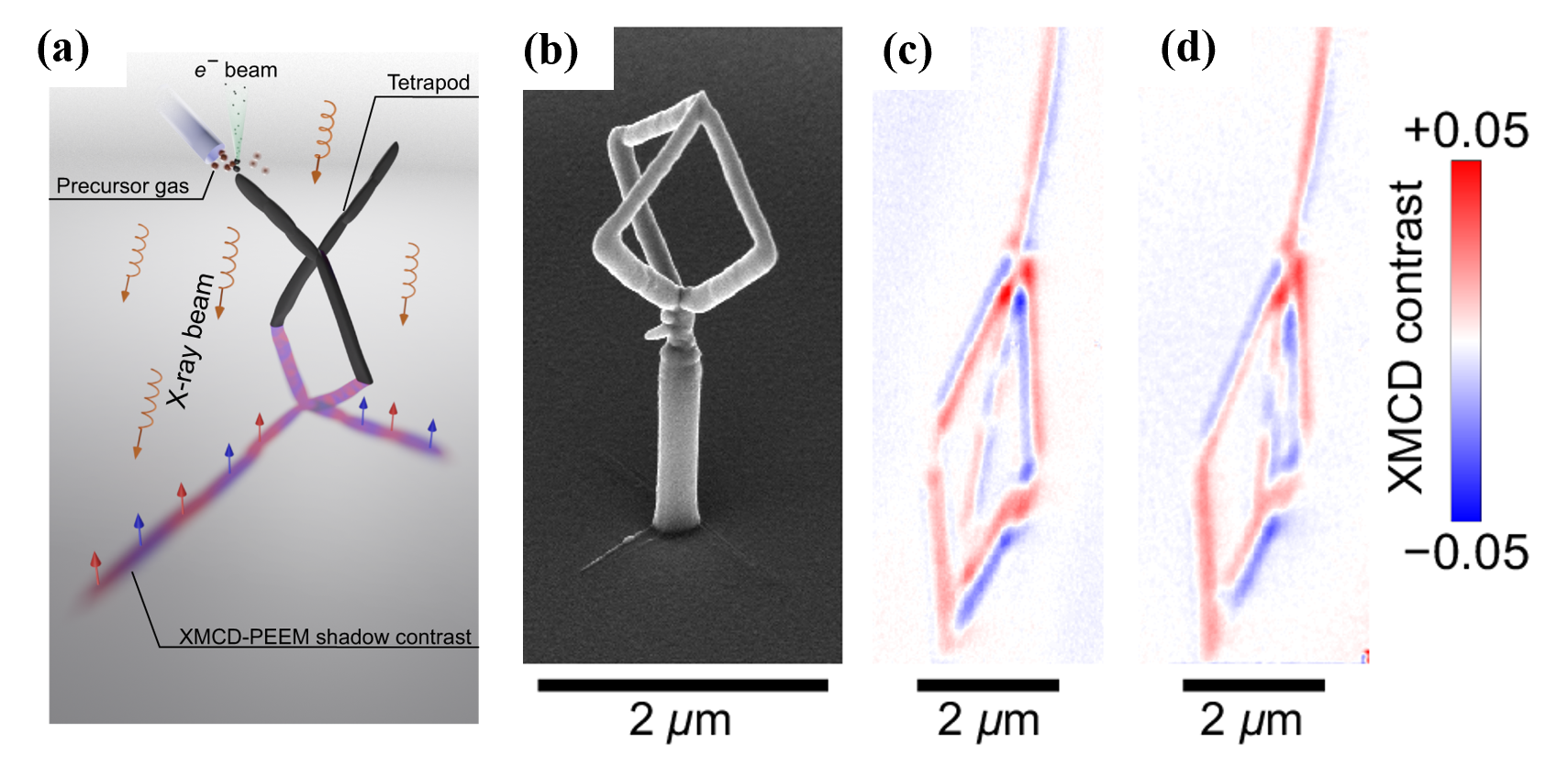} % fig_peem
        \caption{(a)~Schematic of a Co$_{3}$Fe tetrapod being imaged by shadow XMCD-PEEM. (b)~Scanning electron microscopy (SEM) of a ferromagnetic pyramid-shaped wire-frame fabricated by FEBID. (c,d)~XMCD-PEEM shadow projection providing magnetic contrast at different angles. Panels reprinted under Creative Commons Attribution 4.0 License (https://creativecommons.org/licenses/by/4.0/) from~\cite{Volkov24a}~\textcopyright{}~(2024) by the Authors.
        }
    \label{fig:PEEM}
    \end{figure}

    For raised nanostructures, so-called \emph{shadow PEEM} can be employed for magnetic imaging~\cite{Kimling11,DaCol14,Jamet15a,Volkov24a}. Due to the shallow angle of incidence of the illuminating beam, structures raised above the substrate cast a shadow across it. If the structures are X-ray transparent (as is the case for many nanometre-scaled geometries) then the magnetic contrast of the sample is also projected into this shadow due to the magnetization-dependent absorption of the X-rays. The conventional PEEM surface contrast from the sample itself can be recorded simultaneously with the shadow cast on the substrate by X-rays transmitted through the sample; this shadow encodes a two-dimensional projection of the internal magnetization and thus can substantially enhance both contrast and resolution compared to the purely surface-sensitive imaging [Fig.~\ref{fig:PEEM}(a)]. For relatively shallow 3D nanostructures the projected shadow can overlap with magnetic PEEM contrast arising from material co-deposited on the substrate underneath; this is why recent three-dimensional wireframe studies, such as those reported by Volkov~\textit{et~al.}~\cite{Volkov24a} [also see Fig.~\ref{fig:PEEM}(b--d)], deliberately elevate the magnetic network above the surface to distance the shadow and avoid mixing of undesired magnetic signals.
    
    Shadow PEEM has been used to study Bloch points in curved nanowires~\cite{RuizGomez25}, domain wall automotion in helical interconnectors~\cite{Skoric22}, vortex configuraions in nanotubes~\cite{Wyss17}, and magnetically capped rolled up nanomembranes~\cite{Streubel12b}. Another pertinent example is the study of tetrapod wireframes~\cite{Volkov24a}, with representative images shown in Fig.~\ref{fig:PEEM}(c--d). The interpretation of these images led to the experimental verification of vortex and antivortex states in samples homeomorphic to an $N$-tori, as discussed in Section~\ref{sec:wireframes}.

    %%%%%%%%%%%%%%%%%%%%%%%%%%%%%%%%%%%%%%%%%%%%%%%%%%%%%%%%%%%%%%%%%%%%%%
    \subsubsection{Scanning Transmission X-ray Microscopy}
    
    Scanning Transmission X-ray Microscopy (STXM) uses a focused, monochromatic X-ray beam with a spot size of roughly \SI{20}{nm} (generally obtained by a Fresnel Zone plate) to scan a sample, while the transmitted X-ray intensity is recorded, thereby generating a pixel-by-pixel image~\cite{Fischer01, Chao12}. The contrast reflects the integrated X-ray absorption, directly encoding the electronic and magnetic structure of the material. The magnetic signal is integrated through the sample thickness, producing a projection of the magnetization; in XMCD measurements, this corresponds to the component along the X-ray beam. While XMCD and XMLD can remove electronic charge contrast, for curved or 3D structures, projections can still appear foreshortened or overlapping. Consequently, interpreting complex magnetization configurations of non-planar structures requires careful consideration of the sample geometry. 
    
    As the name already suggests, STXM requires X-ray transparent samples, these are typically $\lesssim$~\SI{200}{nm} thick for 3d transition metals and must be mounted on electron microscopy grids or silicon nitride membranes to ensure the substrate does not absorb the X-rays. These substrates are often insulating and fragile, which can constrain fabrication strategies.
    A key advantage of STXM is the ability to measure the net magnetization through the full sample thickness, and the ability to implement time-dependent measurements to track spin excitations~\cite{Finizio22}.

    \subsubsection{Coherent Diffractive Imaging}
    
    Advances in synchrotron technology now mean that highly coherent X-ray sources such as diffraction-limited-storage-ring (DLSR) synchrotrons and FELs are increasingly accessible~\cite{Jaeschke16}. This enhanced coherence motivates the adoption of coherent diffractive imaging (CDI) methods such as holography and ptychography, both of which are primarily transmission-based techniques that computationally reconstruct the object and probe beam from coherent diffraction patterns~\cite{Chushkin25}. In holography, a reference wave interferes with the X-rays scattered by the sample, encoding both phase and amplitude information~\cite{Gabor48, GuizarSicairos08}; in ptychography scans of overlapping illuminated regions are collected, from which the object and source are reconstructed through iterative phase retrieval algorithms~\cite{Pfeiffer18,FavreNicolin20}. Both approaches decouple the achieved spatial resolution from the limits imposed by X-ray optics and the synchrotron beam diameter, and are capable of achieving single-nanometre resolution.
    
    When combined with XMCD, these methods can probe comparatively thick or three-dimensional magnetic systems. This is because the magnetic contrast is encoded in the phase of the scattered X-rays, as well as in its amplitude. In particular, the  magnetic phase signal persists further from the absorption edge where transmission is higher, allowing for thicker samples to be investigated~\cite{Neethirajan24}. 
    
    As a result, CDI techniques substantially extend the accessible sample geometries beyond those compatible with conventional absorption-based microscopy. These developments are particularly appealing in the context of 3D magnetic imaging, where multiple angular projections are combined using tomographic or laminographic reconstruction algorithms to recover the full 3D vector field of the magnetization~\cite{Donnelly17,Donnelly18,Witte20}. Such approaches have already enabled the visualization of complex 3D topological spin textures in bulk and curved nanostructures, providing access to emergent magnetic phenomena that cannot be inferred from two-dimensional projections alone~\cite{Donnelly17,Donnelly20b,DiPietroMartinez23,Fullerton25a,DiPietroMartinez25}.

    \subsection{Transmission Electron Microscopy}
    \label{subsec:EM}
    
    Electron microscopy offers a complementary path to synchrotron-based methods for nanoscale magnetic imaging. Here we focus on transmission electron microscopy (TEM)~\cite{Williams09} and scanning transmission electron microscopy (STEM)~\cite{Pennycook11}, in which  high-energy electrons ($\gtrsim$\SI{100}{keV}) transmitted through electron-transparent samples (typically up to several \SI{100}{nm} thick) are deflected and phase-shifted by the sample's magnetic flux density~\cite{Chapman99}. The resulting redistribution of intensity and phase in the transmitted electron wave encodes information on the magnetic field both within and surrounding the sample. As the electrons are sensitive to the component of the magnetic induction perpendicular to their direction of propagation, electron microscopy provides information complementary to XMCD, which probes magnetization parallel to the incident beam.
    
    Magnetic imaging can be performed in a variety of modes, including Lorentz microscopy~\cite{Loudon19}, electron holography~\cite{Kovacs18}, and differential phase contrast imaging~\cite{McVitie15}. These techniques can achieve (sub-)nanometer scale spatial resolution, and have been used to image a wide array of magnetic solitons in planar systems including vortices, skyrmions~\cite{Hassan24}, and Hopfions~\cite{Zheng23}.
    
    Electron-based measurements also benefit from coherent diffractive imaging methods mentioned above.  For example, off-axis electron holography enables the quantitative mapping of the electron phase and thus of the projected magnetic induction with high sensitivity~\cite{Kovacs18}. For this, a charged wire known as an electron biprism is used to interfere a reference wave with the electron wave that has passed through the sample, enabling reconstruction of both the amplitude and phase in the image plane. For magnetic materials this reconstruction includes the Aharonov--Bohm contribution associated with the magnetic vector potential. Off-axis electron holography has been used to prove the existence of magnetic hexapoles in interwoven helices~\cite{Fullerton25} and curved nanocaps~\cite{Dugato25}. 
    
    In such transmission-based magnetic imaging modes, the measured contrast represents a projection through the sample thickness and generally contains both magnetic and structural contributions. In curved or three-dimensional nanostructures, variations in thickness, orientation, and morphology can cause structural contrast to dominate or obscure the magnetic signal. Consequently, quantitative interpretation requires careful separation of magnetic and structural contributions, often achieved through reference measurements in field-polarized or paramagnetic states, or comparison with micromagnetic simulations.
    
    To reconstruct the full three-dimensional magnetization in complex geometries, electron microscopy can also be combined with tomographic reconstruction. In such approaches, a series of images is acquired over multiple tilt angles and used to reconstruct a volumetric maps of the magnetic flux density. These tomographic methods have been applied to unveil the 3D structure of magnetic skyrmions in FeGe nanostructures~\cite{Wolf22}, as well as to scalar-field tomography using integrated differential phase contrast~\cite{Yu24}. However, practical limitations on tilt range, sample thickness, and mechanical stability, as well as requirement for electron-transparent specimens, impose constraints on accessible geometries and length scales.
    
    \subsection{Micromagnetic Simulations}
	\label{subsec:micromagnetic}

	A complementary to the real experiments and, often even necessary technique is the \emph{in silico} experiments or different kinds of computer simulations. Mainly, they are dedicated for search of equilibrium states and analysis of dynamics by means of energy minimization or solution of the dedicated differential equations. Beyond micromagnetic approach described below, there is a wide use of atomistic simulations and accounting of coupling between magnetic and other degrees of freedom.%~\cite{Eriksson17}.}
	
	In general, micromagnetic simulations represent a paradigm shift to analytical models by confronting them with real-world complexity: irregular geometries, material defects, finite temperatures, complex topological textures, and dynamic non-equilibrium processes. By discretizing the magnetic continuum, numerical solvers do not merely approximate known solutions; they enable the discovery of unforeseen states—metastable configurations, topologically protected quasiparticles like skyrmions, and complex reversal pathways—that are often inaccessible to both theory and direct experiment. 
	
	Furthermore, from the simulated spin structure the expected measurement response can be calculated, which typically involves partial or full averaging of the local spin structure and limitations on to which magnetization components the probe is sensitive to. Simulated prediction can be used to optimize nanofabrication to avoid costly trial-and-error runs, and can be compared to the experimental results to gain further insights.
	
	The theoretical understanding of nanoscale magnetism can be determined using the Landau--Lifshitz--Gilbert (LLG) equation
	\begin{equation}\label{eq:LLG}
		\frac{\partial\vec{m}}{\partial t} = -\gamma_0 \vec{m} \times \vec{H}_\text{eff} + \alpha \vec{m} \times \frac{\partial\vec{m}}{\partial t},
	\end{equation}
	where $\alpha$ is the Gilbert Damping parameter, $\gamma_0 > 0$ is the gyromagnetic ratio, and $\vec{H}_\text{eff}$ is the effective field arising from the variational derivative of the system's total magnetic energy $E$ via $\vec{H}_\text{eff} = -(1/\mu_0 M_s) \delta E / \delta \vec{m}$. 
	
	Micromagnetic simulators numerically solve the LLG equation by calculating the effective field $\vec{H}_\text{eff}$ arising from the competition of interactions -- exchange, anisotropy, magnetostatics, and DMI -- at the nanometre scale. The landscape of micromagnetic software is diverse, with tools optimized for different priorities: computational speed, geometric flexibility, or physical comprehensiveness. They can be broadly categorized by their numerical core: the finite-difference method (FDM), favouring speed and simplicity on regular grids, and the finite-element method (FEM), prioritizing accuracy for complex shapes via irregular meshes.
	
	Among the most established FDM-based packages is the Object-Oriented MicroMagnetic Framework (OOMMF)~\cite{OOMMFb,Donahue99}. OOMMF's strength lies in its robustness, exhaustive validation, and modular Tcl-scriptable architecture. It solves the LLG equation on a Cartesian grid, integrating standard modules for exchange, anisotropy, Zeeman, and demagnetization fields. While its regular grid can be a limitation for curved geometries, OOMMF remains a vital benchmark and educational tool due to its transparency and comprehensive documentation.
	
	Pushing the boundaries of computational speed, Mumax$^3$ (Ghent University)~\cite{Vansteenkiste14} harnesses GPU acceleration to achieve performance gains orders of magnitude higher than CPU-bound codes. Also based on FDM, it combines this speed with a user-friendly interface, real-time visualization, and support for advanced physics like spin-transfer torque (STT), spin-orbit torque (SOT), and temperature effects. This makes Mumax$^3$ powerful for large-scale simulations of dynamic processes. As an independent follow-up package, Mumax+ offers similar advantages as Mumax$^3$ including GPU-acceleration, as well as allows for easier multi-physics implementation~\cite{Moreels26}.
	
	When geometric fidelity is paramount -- for instance, in nanostructures with curved edges, varying thickness, or irregular borders -- finite-element method (FEM) simulators become indispensable. MAGPAR~\cite{magpar,Scholz03} and NMag~\cite{Fischbacher07} allow for flexible meshing of arbitrary shapes, accurately capturing boundary effects and local curvatures. Similarly, codes like TetMag~\cite{Hertel23} or magnum.fe~\cite{Abert13a} and magnum.pi, to name a few, specialize in tetrahedral meshing for problems where geometry critically dictates magnetic configuration, such as in patterned nanodots or nanowires with complex cross-sections. The trade-off for this geometric precision is typically a  greater computational cost and expertise required for the generation and data analysis of the complex three-dimensional meshes.

    \section{Topological Textures in Toroidal Nanoparticles}
	\label{sec:toroidal}
	
	Among curved geometries, the torus presents a particularly compelling case, combining a non-trivial topology, which can be directly imprinted onto the magnetic subsystem to stabilize complex textures~\cite{Carvalho-Santos08,Carvalho-Santos10,Corona23,Teixeira19}, and a surface curvature that varies continuously between positive, negative, and zero values~\cite{ONeill06}. 
	This variable Gaussian curvature landscape offers a rich environment to probe how geometric gradients stabilize topological spin structures and affect their dynamics. 
	
	\subsection{Heisenberg Spins on Curved Shells}
	\label{sec:heisenberg}
	
	A magnetic system can be described by a continuous vector field representing the normalized magnetization, $\vec{m}$. This unit vector field can be parametrized using spherical coordinates in a Cartesian basis as
	\begin{equation}\label{eq:mag-vector}
		\vec{m}=\sin\Theta\cos\Phi\,\vec{\hat{x}}+\sin\Theta\sin\Phi\,\vec{\hat{y}}+\cos\Theta\,\vec{\hat{z}}.
	\end{equation}
	
	A thin magnetic shell, for which the magnetisation profile is expected to not vary across its thickness, can be effectively considered as a two-dimensional surface. The exchange energy on such a two-dimensional space can be written in terms of the Heisenberg Hamiltonian in the continuous approach as 
	\begin{eqnarray}\label{HM0}
		\mathcal{H}=J\iint\sum_{i,j=1}^2\sum_{a,b=1}^3\,\delta_{ij}h_{ab}(1+\delta_{a3}\,K)\left(\frac{\partial m^a}{\partial r_i}\right)\left(\frac{\partial m^b}{\partial r_j}\right)dxdy
	\end{eqnarray}
	where $J=Ah$ is the exchange constant, with $A$ being the exchange stiffness and $h$ the shell thickness. Also, $(r_1, r_2, r_3) \equiv (x, y, z)$, and $\delta_{ij}$ is the Kronecker delta (equal to $1$ if $i = j$ and $0$ otherwise). The metric tensor $h_{ab}$ is a function of the spin space coordinate system, which is simplified to $h_{ab} = \delta_{ab}$ in Cartesian coordinates. Here, indices $(i,j)$ correspond to physical space coordinates, while $(a,b)$ denote internal space describing the spin system.
	
	The anisotropy parameter $K$ generalizes the model: For $K = 0$, the Hamiltonian reduces to the isotropic Heisenberg case, which, combined with the constraint $|\vec{m}| = 1$, corresponds to the nonlinear sigma model. If $-1<K<0$, one obtains an easy-plane anisotropy. For $K=-1$, the so-called XY model is obtained, being equivalent to the planar rotator model under the additional constraint $m_z=0$. 
	
	Let us consider that the magnetization lies on a curved shell described by $\vec{r}=x(\eta_1,\eta_2)\,\vec{\hat{x}}+y(\eta_1,\eta_2)\,\vec{\hat{y}}+z(\eta_1,\eta_2)\,\vec{\hat{z}}$, where $\eta_1$ and $\eta_2$ are local curvilinear coordinates describing the physical space. Under this approximation, the Hamiltonian given in Eq.~\eqref{HM0} can be rewritten as 
	\begin{eqnarray}\label{HM}
		\mathcal{H}=J\iint\sum_{i,j=1}^2\sum_{a,b=1}^3\,g^{ij}h_{ab}(1+\delta_{a3}\,K)\left(\frac{\partial m^a}{\partial \eta_i}\right)\left(\frac{\partial m^b}{\partial \eta_j}\right)\sqrt{g}\,d\eta_1d\eta_2
	\end{eqnarray}
	where $\sqrt{g} = \sqrt{\det[g_{ij}]}$ defines the surface metric determinant, and $g^{ij}$ are the contra-variant components of the metric tensor satisfying $g^{ij}g_{jk} = \delta^i_k$, see also Sec.~\ref{sec:diffgeo}.  By restricting the analysis to surfaces having an orthogonal basis ($g_{ij}=0$ for $i\neq j$), and considering that the components of the magnetic unit vector are in the Cartesian basis, we obtain
	\begin{equation}\label{HM2}
		\begin{aligned}
			\mathcal{H} & = J\iint\left\{G_{21}\left[\left(\frac{\partial m_1}{\partial \eta_1}\right)^2+\left(\frac{\partial m_2}{\partial \eta_1}\right)^2+(1+K)\left(\frac{\partial m_3}{\partial \eta_1}\right)^2\right]\right.\\
			& \left.+G_{12}\left[\left(\frac{\partial m_1}{\partial \eta_2}\right)^2+\left(\frac{\partial m_2}{\partial \eta_2}\right)^2+(1+K)\left(\frac{\partial m_3}{\partial \eta_2}\right)^2\right]\right\}\,d\eta_1d\eta_2
		\end{aligned}
	\end{equation}
	where $G_{ij}=\sqrt{\frac{{g_{ii}}}{g_{jj}}}$. 
	
	Expressing the components of $\vec{m}$ in spherical coordinates (with the basis held Cartesian) simplifies the problem. Using the definitions $\Theta \equiv \Theta(\eta_1, \eta_2)$ and $\Phi \equiv \Phi(\eta_1, \eta_2)$, the three degrees of freedom ($m_1, m_2, m_3$) are reduced to two ($\Theta, \Phi$). This allows the Hamiltonian~\eqref{HM2} to be rewritten in terms of these angles:
	\begin{equation}\label{HM3}
		\begin{aligned}
			\mathcal{H} & = J\iint\left\{G_{21}\left[(1+K\sin^2\Theta)\left(\frac{\partial \Theta}{\partial \eta_1}\right)^2+\sin^2\Theta\left(\frac{\partial \Phi}{\partial \eta_1}\right)^2\right]\right.\\
			& \left.+G_{12}\left[(1+K\sin^2\Theta)\left(\frac{\partial \Theta}{\partial \eta_2}\right)^2+\sin^2\Theta\left(\frac{\partial \Phi}{\partial \eta_2}\right)^2\right]\right\}\,d\eta_1d\eta_2
		\end{aligned}
	\end{equation}
	
	For static solutions, the Hamiltonian and Lagrangian ($\mathcal{L}$) are the same (\emph{i.e.} $\mathcal{H}=\mathcal{L}$), allowing the equations of motion to be derived directly from the Euler-Lagrange equations $\frac{\partial\mathcal{L}}{\partial\Xi}-\partial_\mu\left(\frac{\partial\mathcal{L}}{\partial_\mu\Xi}\right)=0 $
	where $\Xi\equiv(\Theta,\Phi)$. Therefore, from Eq.~\eqref{HM3}, the motion equations are evaluated as
	
	\begin{subequations}\label{Euler--Lagrange}
		\begin{eqnarray}
			\frac{\sin2\Theta}{2}\left\{G_{21}\left[K(\partial_{\eta_1}\Theta)^2+\left(\partial_{\eta_1}\Phi\right)^2\right]+G_{12}\left[K(\partial_{\eta_2}\Theta)^2+\left(\partial_{\eta_2}\Phi\right)^2\right]\right\}\nonumber\\=\partial_{\eta_1}\left[(1+K\sin^2\Theta)G_{21}\,\partial_{\eta_1}\Theta\right]+\partial_{\eta_2}\left[(1+K\sin^2\Theta)G_{21}\,\partial_{\eta_2}\Theta\right]\\\nonumber\\
			\partial_{\eta_1}\left(\sin^2\Theta\,G_{21}\,\partial_{\eta_1}\Phi\right)+
			\partial_{\eta_2}\left(\sin^2\Theta\,G_{12}\,\partial_{\eta_2}\Phi\right)=0
		\end{eqnarray}
	\end{subequations}
	
	Under specific constraints imposed by the symmetry of the underlying surface, the model described above allows for the emergence of localised topological spin excitations, such as skyrmions or vortices, within curved magnetic systems. These spin textures arise due to the interplay between exchange interactions, magnetic anisotropy, and surface curvature. Notably, the characteristic length scale of these topological structures, such as their size and spatial distribution, is intrinsically governed by the underlying geometry of the magnetic shell. This geometric dependence suggests that curvature can serve as a tunable parameter to control the stability and dynamics of such excitations.

	Within the theoretical framework of the Heisenberg model on curved shells, the emergence of topological solitonic textures as excited states in systems with azimuthal symmetry can be investigated. Although Kovalchuk and Chudzi\'nski~\cite{Kovalchuk23} developed a unified framework for soliton-like solutions on curved shells using elliptic integrals across different anisotropy regimes, the following section specifically examines the cases of $K=-1$ (XY-model) and $K=0$ (isotropic model) across a range of geometries, including the cylinder~\cite{VillainGuillot94,Villain-Guillot95,Saxena97,Saxena98,Dandoloff02,Saxena98a},  sphere~\cite{Milagre07a,Kravchuk12a},  torus~\cite{Benoit98,Carvalho-Santos08,Saxena98a},  pseudosphere~\cite{Belo07}, paraboloid~\cite{Vilas-Boas15}, catenoid~\cite{Carvalho-Santos12,Carvalho-Santos13a,CarvalhoSantos15}, and hyperboloid~\cite{Carvalho-Santos12,Carvalho-Santos13a,CarvalhoSantos15,Dandoloff11}, each of which imposes distinct constraints on soliton stability.
	
	\subsubsection{Topological Spin textures in Curved Surfaces with Azimuthal Symmetry}
	
	\begin{figure}[tb]
		\centering
		\includegraphics[width=0.8\linewidth]{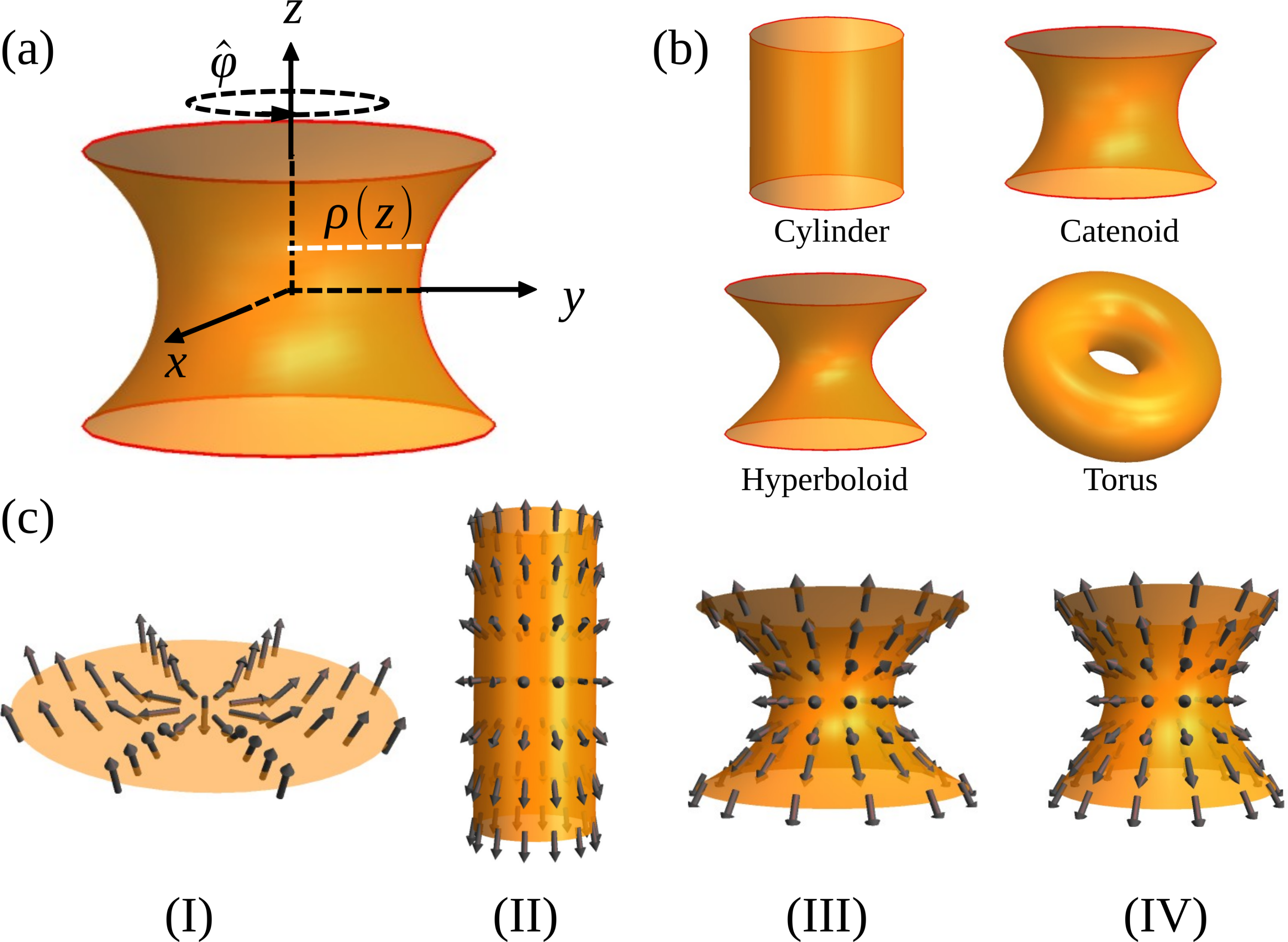} % Fig1-Vagson
		\caption{(a) Representation of the coordinate system to describe surfaces with azimuthal symmetry. (b) Example of surfaces with azimuthal. (c) I - Belavin-Polyakov soliton in a planar disk. II, III, and IV depict solitonic profiles ($K=0$) obtained  for the cylinder (left), catenoid (middle), and hyperboloid (right).}
		\label{Fig1}
	\end{figure}
	
	The anisotropic Heisenberg model has been a subject of extensive study on a range of curved manifolds, with a significant focus on azimuthally symmetric surfaces where topological spin textures can emerge. Topological solitons are predicted as excited states on infinite surfaces in the isotropic limit ($K = 0$)~\cite{Villain-Guillot95,Saxena97,Saxena02}, while fractional solitons may appear on finite surfaces~\cite{Carvalho-Santos08,Dandoloff11}. 
	
	In a cylindrical coordinate system, an azimuthally symmetric surface can be parameterized as $\vec{\vec{r}} = \rho ( z)(\cos\varphi\,\vec{\hat{x}}+\sin\varphi\,\vec{\hat{y}})+z\,\vec{\hat{z}}\,$, where $\rho(z)\equiv\rho_z$ denotes the radius of the surface at height $z$, and $\varphi$ represents the azimuthal angle, as illustrated in Fig.~\ref{Fig1}(a). The corresponding covariant metric elements are given by $g_{\varphi\varphi}=(g^{\varphi\varphi})^{-1}=\rho_z^2$ and $g_{zz}=1+\rho_z'{^2}$, where the prime denotes derivative with respect to $z$. Using this coordinate system, Eq.~\eqref{HM3} can be rewritten as
	\begin{equation}\label{HMCylinder}
		\begin{aligned}
			\mathcal{H} & = J\iint\left[(1+K\sin^2\Theta)\left({\partial_\zeta \Theta}\right)^2+\sin^2\Theta\left({\partial_\zeta \Phi}\right)^2\right.\\
			& \left.+(1+K\sin^2\Theta)\left({\partial_\varphi \Theta}\right)^2+\sin^2\Theta\left({\partial_\varphi \Phi}\right)^2\right]\,d\zeta d\varphi
		\end{aligned}
	\end{equation}
	where $d\zeta=G_{z\varphi}\,dz$. The corresponding motion equations for a spin system on a shell with azimuthal symmetry are given by
	
	\begin{subequations}
		\begin{eqnarray}
			\label{MotEq}{\sin2\Theta}\left\{K\left[(\partial_{\zeta}\Theta)^2+\partial_{\varphi}\Theta)^2\right]+\left(\partial_{\zeta}\Phi\right)^2+\left(\partial_{\varphi}\Phi\right)^2\right\}\nonumber\\=2(1+K\sin^2\Theta)\,(\partial_{\zeta}^2\Theta+\partial_\varphi^2\Theta),\\\nonumber\\
			\partial_{\zeta}\left(\sin^2\Theta\,\partial_{\zeta}\Phi\right)+
			\partial_{\varphi}\left(\sin^2\Theta\,\partial_{\varphi}\Phi\right)=0
		\end{eqnarray}
	\end{subequations}
	
	Now, by imposing boundary conditions ($\lim_{\zeta\to\infty} \Theta \equiv 0\,[\pi]$), one can use homotopy arguments to classify and characterize the obtained solutions, see also Sec.~\ref{sec:topology}. An~additional requirement is the boundary condition $\lim_{\zeta\to\infty} {\partial_\zeta\Theta} = 0$, with the decay rate ensuring the convergence of the integral in Eq.~\eqref{HMCylinder}. Under these conditions, one can identify all points at infinity and compactify the spin vector field on a surface with circular symmetry into a sphere (where the surface's endpoints become the sphere's two poles). The correspondence of this sphere with the order parameter manifold $S^2$ yields the homotopy group $\pi_2(\mathbb{S}^2) = \mathbb{Z}$. This result classifies the topological excitations on these surfaces, meaning spin distributions can be categorized into distinct classes of topologically non-trivial configurations~\cite{Saxena98a}. All configurations within a given class are topologically equivalent. 
	
	To discuss possible topological solutions, one can restrict the analysis to azimuthally symmetric solutions, where the order parameters $\Theta$ and $\Phi$ satisfy $\Phi = q\, \varphi+\phi_0 \quad \text{and} \quad {\partial_\varphi \Theta} = 0$, with $q$ integer and $\phi_0$ a phase that does not account to the energy. Under these conditions, the Hamiltonian~\eqref{HMCylinder} simplifies to
	\begin{eqnarray}\label{HMCylinderSimp}
		\mathcal{H}=2\pi J\int\left[(1+K\sin^2\Theta)\left({\partial_\zeta \Theta}\right)^2+q^2\,{\sin^2\Theta}\right]\,d\zeta\,\,
	\end{eqnarray}
	and the motion equation~\eqref{MotEq} reads
	\begin{eqnarray}\label{MotFin}
		{\sin2\Theta}\left\{K(\partial_{\zeta}\Theta)^2+q^2\right\}=2(1+K\sin^2\Theta)\,\partial_{\zeta}^2\Theta
	\end{eqnarray}
	
	\subsubsection{Isotropic Regime and Soliton-like Solutions ($K=0$)}\label{sec:Isotropic_Heisem}
	
	The simplest case to analyze is the isotropic regime (\(K=0\)). Since the energy scales with \(q^2\), we can focus on solutions with \(q = 1\). In this case, Eq.~\eqref{MotFin} reduces to $2\partial_{\zeta}^2 \Theta = \sin 2\Theta$, which is the sine-Gordon equation. Imposing the boundary conditions \(\Theta(-\infty) = 0\) and \(\Theta(\infty) = \pi\) yields a topological soliton solution representing a smooth transition between the two neighbouring vacua
	\begin{equation}
		\Theta(\zeta) = 2 \arctan\left(e^{\zeta}\right).
		\label{eq:soliton_solution}
	\end{equation}
	
	It is important to note that the soliton typically has a characteristic length scale, which depends on the geometry of the underlying manifold. In general, this characteristic length scale appears as a prefactor of the $\sin 2\Theta$ term in the sine-Gordon equation, and thus is is determined by the ratio of metric components $G_{z\varphi}$. Under the considered assumptions, this scale is absorbed into the dimensionless coordinate $\zeta=\int G_{z\varphi}\,dz$. For example, in the case of an infinite cylinder of radius $r$, the soliton's characteristic length scale would originally be proportional to $r$. However, the rescaling $z\rightarrow z/r$ to the dimensionless coordinate $\zeta$ removes any explicit dependence on $r$ in the final solution~\cite{Villain-Guillot95}. As a result, the soliton's effective width in Eq.~\eqref{eq:soliton_solution} is normalized to unity.
	
	The corresponding energy density, integrated over the domain \([\zeta_1, \zeta_2]\), is given by
	\begin{equation}
		2\pi J\int_{\zeta_1}^{\zeta_2} \left[(\partial_{\zeta} \Theta)^2 + \sin^2 \Theta \right] d\zeta = \left. -\frac{4\pi J}{e^{2\zeta} + 1} \right|_{\zeta_1}^{\zeta_2}\,
		\label{eq:energy_density}
	\end{equation}
	which reveals that the energy is localized and decays exponentially towards the boundaries of the domain. This is a hallmark of a topological soliton, whose energy is not distributed uniformly but is instead concentrated within a finite region.
	
	In the study of topological solitons, a fundamental objective is the quantitative classification of these non-trivial field configurations. This classification is achieved by calculating a topological invariant, a quantity that remains unchanged under continuous deformations of the spin vector field. To this end, one can evaluate the solitonic topological charge, defined through the (2D) topological Pontryagin invariant $Q_2$ [see Eq.~\eqref{eq:pontryagin}] (or Skyrmion number), which in angular magnetic variables is
	\begin{equation}
		Q_2 = \frac{1}{4\pi} \int \sin \Theta \, d\Theta d\Phi.
		\label{eq:soliton_charge}
	\end{equation}
	From the solution given in Eq.~\eqref{eq:soliton_solution}, one obtains $Q_2 = -\frac{1}{e^{2\zeta} + 1}$. Over the interval $\zeta\in(-\infty,\infty)$, this topological charge corresponds to a $\pi$-soliton representing the first class of the second homotopy group $\pi_2({S}^2)$ (see Sec.~\ref{sec:top-defects}), having energy $E=4\pi J$, in agreement with the saturated Bogomolnyi inequality~\cite{Bogomolnyi76}. 
	
	However, for finite surfaces, the mapping becomes incomplete, leading to a fractional soliton solution. In such cases, standard homotopy arguments cannot classify Eq.~\eqref{eq:soliton_solution} as a topological excitation. Specifically, for non-simply connected surfaces, the soliton develops a finite characteristic length scale that stabilizes its size and prevents collapse. Examples include the annulus, the truncated cone~\cite{Freitas05}, the punctured pseudosphere~\cite{Belo07}, and the torus~\cite{Carvalho-Santos08}. Furthermore, periodic soliton solutions can be constructed for rotationally symmetric surfaces~\cite{Villain-Guillot95,Dandoloff11}; however, a detailed exploration of these periodic structures falls beyond the scope of this chapter.  
	
	It is instructive to mention that when one analyzes solutions of Eq.~\eqref{HMCylinder} in a planar surface, one obtains a special type of topological soliton, the Belavin-Polyakov soliton~\cite{Belavin75}, which consists of a stable finite-energy particle-like solution where the field configuration wraps around the target sphere exactly once (or a multiple integer number of times). This particle-like metastable state can be represented by the profile $\Theta=\arccos\left[(R^2-r^2)/(R^2+r^2)\right]$ and $\Phi={\varphi}+\pi/2$, where $r=\sqrt{x^2+y^2}$. The vector field related to this magnetization pattern is shown in Fig.~\ref{Fig1}(c-I), and can be viewed as a localized kink in a field that points in different directions on a sphere. Its stability is guaranteed not by energy minimization but by its topological nature.
	
	The influence of the shell geometry on the characteristic length of the solitonic solution can be illustrated through the particular cases of a cylinder, catenoid, hyperboloid, and torus, whose shapes are shown in Fig.~\ref{Fig1}(b). Of these, the first three geometries -- cylinder, catenoid, and hyperboloid -- share similar topological and geometrical properties, allowing for a unified treatment. The torus, being a compact manifold, requires separate consideration, and is discussed in Sec.~\ref{sec:Topology-Tori}. 
	
	\paragraph{Cylinder, Catenoid, and Hyperboloid with $K=0$}
	
	The cylinder, catenoid, and hyperboloid have their geometries characterized by $\rho_z^{\text{cyl}}=r$, $\rho_z^{\text{cat}}=r \cosh(z/r)$, and $\rho_z^{\text{hyp}}=r \sqrt{1+z^2/b^2}$, where $b$ is a parameter determining the relation between the hyperboloid radius in the $xy$-plane and its height. The general parametrization for the hyperboloid results in an integral for $\zeta$ that is cumbersome and will not be discussed here. A particularly interesting case arises for a specific class of hyperboloids, described by the polar hyperbolic coordinate system (biharmonic coordinates) by setting $b=r$, for which the adopted geometrical parameterizations for the cylinder, catenoid, and polar hyperboloid yield 
	\begin{subequations}
		\begin{equation}
			\zeta^{\text{cyl}}=\zeta^{\text{cat}}=\eta, \end{equation}
		\begin{equation}
			\zeta^{\text{hyp}}=-\text{arctanh}\left[\sqrt{2} (\eta^2+1)  - \eta\sqrt{1 + 2 \eta^2}\right] + 
			\sqrt{2} \ln\left[ \sqrt{r^2 + 2 z^2}-\sqrt{2} z\right],
		\end{equation}
	\end{subequations}
	where $\eta=z/r$. Therefore, the soliton adopts a characteristic length that depends on the manifold's geometry. For the cylinder and catenoid, this length is precisely equal to the surface's characteristic radius ${r}$. Consequently, in the case of a circular cylinder or catenoid with isotropic spin coupling, there is no mismatch or geometric frustration. For the hyperboloid, however, the characteristic length becomes more intricate, as it depends on both the shape parameter $b$ (the asymptotic curvature) and the radius $r$, while the soliton hosted on a polar hyperboloid presents a characteristic length depending on, but not equal to $r$. 
	
	Despite the dependence of the dimensionless coordinate $\zeta$ on $b$, the soliton energy in the asymptotic limits $\zeta\in(-\infty,\infty)$ remains $4\pi J$, indicating no geometric frustration for any of the considered surfaces. This is because there are no mismatches in length scales, unlike for cases such as the elliptic cylinder~\cite{Saxena98a}. The vector field given by the obtained soliconic profiles is depicted in Fig.~\ref{Fig1}(c-II, III, IV). 
	
	\begin{figure}[tb]
		\centering
		\includegraphics[width=0.8\linewidth]{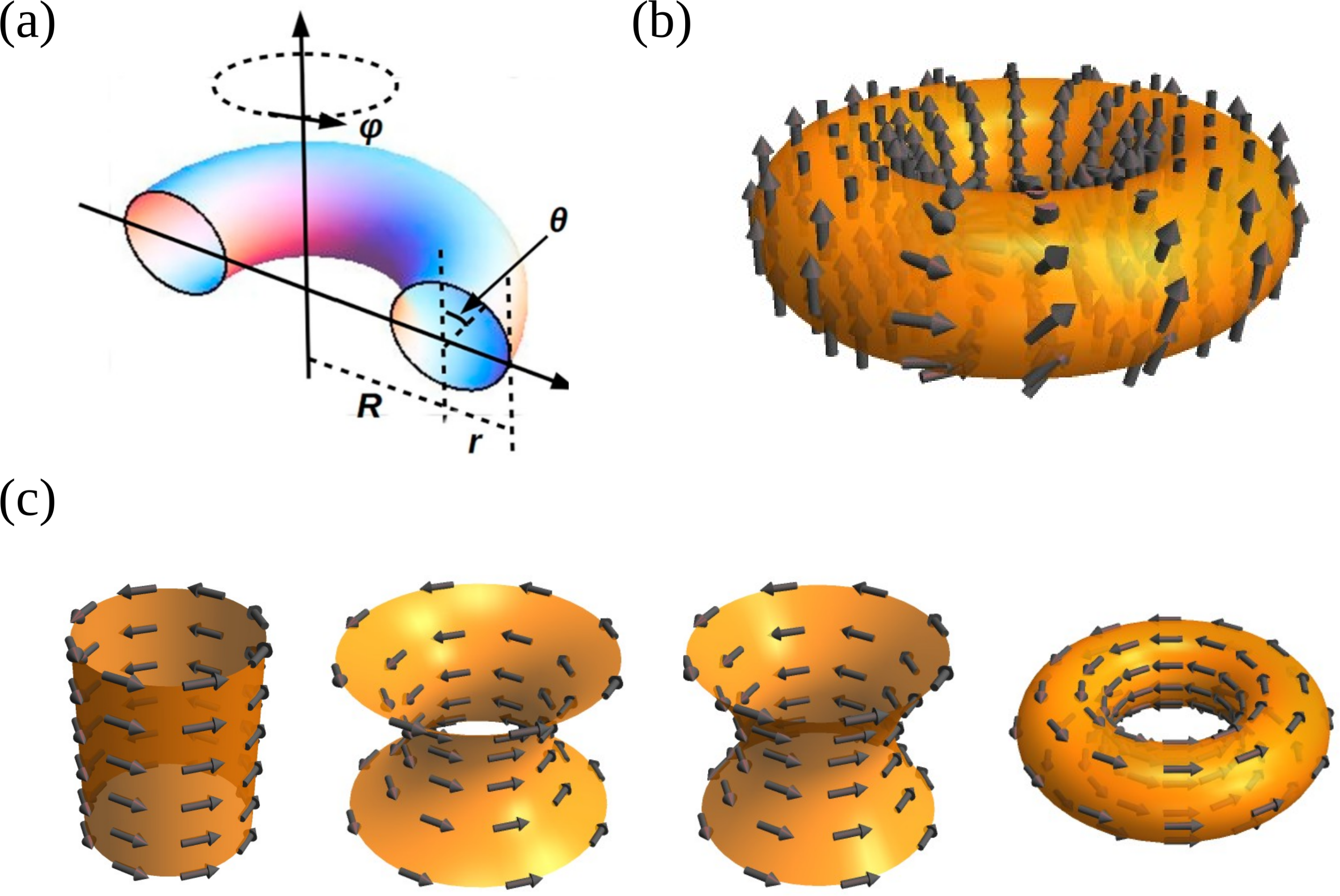} % Fig2-Vagson
		\caption{(a)~Coordinate system describing a torus surface. Reprinted from~\cite{Teixeira19},~\textcopyright{}~(2019), with permission from Elsevier. (b)~Vector field of the solitonic profile on a toroidal surface. (c)~In-surface state on the surfaces of the cylinder, catenoid, hyperboloid, and torus. For all considered cases $K=-1$.
		}
		\label{Fig2}
	\end{figure}
	
	\paragraph{Toroidal Geometry ($K=0$ and $K=-1$)}
	
	The toroidal geometry is analyzed separately due to its distinct boundary conditions and topological constraints. Unlike the catenoid, hyperboloid, and cylinder, the torus presents unique challenges, including its genus, periodic boundary conditions, and Gaussian curvature varying from positive to negative across its surface. These features imply a distinct analytical approach to determine the soliton characteristic length. 
	
	The simplest approach to finding possible solitonic solutions for the isotropic regime on a torus is using the toroidal coordinate system~\cite{Gradshteyn07}, also shown in Fig.~\ref{Fig2}(a), given by
	\begin{equation}\label{TorCoord}
		x=\frac{a\,\sinh\beta}{\cosh\beta-\sin\xi}\hat{\vec{R}}+\frac{a\,\cos\xi}{\cosh\beta-\sin\xi}\vec{\hat{z}},
	\end{equation}
	where $\hat{\vec{R}}=\cos\varphi\,\vec{\hat{x}}+\sin\varphi\,\vec{\hat{y}}$. $\varphi$ and $\xi$ vary from 0 to $2\pi$, while the constant parameters $a=\sqrt{R^2-r^2}$ and $\cosh\beta=R/r$ are both real and positive. The angle $\xi$ is related to the polar angle $\theta$ by the expression $\tan(\xi/2)=\sqrt{(R+r)/(R-r)}\,\tan(\theta/2)$. Here, $R$ and $r$ represent the toroidal and poloidal radii, respectively, as illustrated in Fig.~\ref{Fig2}(a). 
	
	In this set of coordinates, the Hamiltonian~\eqref{HM3} describing toroidally symmetric solutions, $\Theta\equiv\Theta(\varphi)$ and $\Phi\equiv\Phi(\eta)$, gets the form
	\begin{equation}
		\mathcal{H}_t=J\iint\left[\frac{(\partial_\varphi\Theta)^2}{\sinh\beta}+\sinh\beta\,\sin^2\Theta(\partial_\eta\Phi)^2\right]d\eta d\varphi
	\end{equation}
	and the set of Eqs.~\eqref{Euler--Lagrange} read
	\begin{eqnarray}
		\sin2\Theta=2(\sinh{\beta})^{-2}\,\partial_\varphi^2\Theta
	\end{eqnarray}
	which can be written as $\sin2\Theta=2\,\partial_\zeta^2\Theta$ by taking $\zeta=\varphi\sinh\beta$. 
	
	Consequently, the Heisenberg spin system on a torus admits solitonic-like solutions of the kind $\Theta=2\arctan(\text{e}^{\varphi\sinh\beta})$, whose profile is depicted in Fig.~\ref{Fig2}(b). For this solution, the topological charge is evaluated as $Q_t = \tanh(\pi \sinh \beta)$, which, in the limit $R\rightarrow\infty$ result in $Q_t\rightarrow1$. Therefore, for finite $R$, the spin-sphere mapping is incomplete on the torus surface [as depicted in Fig.~\ref{Fig2}(b), where the spins twist but fail to achieve a full $4\pi$ winding, and homotopy arguments cannot classify the solitonic solution as a topological excitation. In this case, the topology of the geometric support must be considered: While the genus obstructs a full mapping from the spin sphere to the torus (resulting in a fractional charge $|Q_2|<1$), it also ensures topological stability, preventing the fractional soliton from decaying into the ground state. Crucially, the soliton acquires a finite characteristic length scale depending on the toroidal and poloidal radii, generating a geometric frustration, which inhibits collapse and prevents its size from vanishing. Similar behaviour arises in systems with an annulus, like a truncated cone~\cite{Freitas05} or a punctured pseudo-sphere~\cite{Belo07}. 
	
	\subsubsection{XY Model and Vortex-like Solutions ($K=-1$)}
	
	Geometrically, a vortex with winding number $q$ can be understood as a magnetic configuration where the spin orientation rotates continuously around a central core region that acts as either a singular point or topological obstruction. This core prevents continuous deformation to a uniform ground state without requiring modifications to spins at arbitrary distances, making the vortex a topologically stable excitation. When modeling vortices using continuum spin descriptions, the treatment is limited to its outer region, as the core's microscopic spin arrangement and precise structure lie beyond analytical approaches and require numerical simulations or discrete lattice techniques for complete characterization. 
	
	The surface topology fundamentally influences the vortex structure: In non-simply connected surfaces (such as annuli or punctured disks), the inherent hole provides a natural cutoff that yields coreless solutions where topological stability emerges from boundary conditions, while in simply-connected surfaces, a singular core necessarily develops at surface self-intersection points.
	
	A vortex-like solution \emph{can} be derived from the isotropic Heisenberg Hamiltonian. However, in the previously presented framework one can also choose to work with the XY-model by setting \(K = -1\). This choice is motivated by the fact that it imposes constraints on the solutions for \(\Theta\) by confining them to the \(xy\)-plane. Keeping the assumption of rotational symmetry, Eqs.~\eqref{HMCylinderSimp} and~\eqref{MotFin} simplify to
	\begin{subequations}
		\begin{eqnarray}\label{HMCylinderAni}
			\mathcal{H}=2\pi J\int\left[(1-\sin^2\Theta)\left({\partial_\zeta \Theta}\right)^2+q^2\,{\sin^2\Theta}\right]\,d\zeta,
		\end{eqnarray}
		\begin{eqnarray}\label{MotFinAni}
			{\sin2\Theta}\left\{-(\partial_{\zeta}\Theta)^2+q^2\right\}=2(1-\sin^2\Theta)\,\partial_{\zeta}^2\Theta
		\end{eqnarray}
	\end{subequations}
	
	The simplest solutions for the above set of equations are given by \(\Theta = n\pi/2\), with $n\in\mathbb{Z}$. However, solutions with even $n$ yield the trivial spin configuration, with the magnetic moments pointing along the $z$-axis direction. Excited states emerge for odd values of $n$, which also satisfy Eq.~\eqref{MotFinAni} and give a vector field configuration where the spins remain confined to the \(xy\)-plane. The solution $\Phi=q\varphi+\phi_0$ represents a vortex with winding number $q$, formally defined, in the continuum limit, as $q=(2\pi)^{-1}\oint_C (\nabla\Phi)\cdot d\vec{l}$, where the integration is performed along a closed path \( C \) enclosing the surface. 
	
	The vortex patterns with \( q = 1 \) for the cylinder, catenoid, hyperboloid, and torus surfaces are illustrated in Fig.~\ref{Fig2}(c). From Eq.~\eqref{HMCylinderSimp} one notices that the vortex energy scales with \( q^2 \), rendering vortices with \( q > 1 \) physically unstable despite their topological character. For a rotationally symmetric surface, the vortex energy is evaluated as
	\begin{equation}
		E_{\text{v}} = 2\pi J \zeta \big\vert_{\zeta_1}^{\zeta_2}
	\end{equation}
	where it can be observed that the vortex energy depends on the dimensionless parameter $\zeta$ (introduced on page 34), which is directly related to both the surface curvature and the radius of the surface at height $z$. 
	
	\subsection{Topological Magnetic Patterns in Nanotori}
	\label{sec:Topology-Tori}
	
	Toroidal nanostructures present singular topological features due to their non-zero genus, which allows the emergence of self-closed 3D magnetic textures. Although cylindrical nanorings share the same genus-1 topology as a torus and can be continuously deformed into one, their geometrical properties differ significantly. A key distinction lies in the Gaussian curvature: On a toroidal surface it varies smoothly across it, changing from negative on the inner rim to positive on the outer rim. In contrast, the Gaussian curvature of a cylindrical nanoring is identically zero across its entire surface and is undefined at its sharp edges, leading to divergent local curvature. As a result, toroidal systems provide a unique and richer platform for exploring curvature-induced phenomena, such as chiral symmetry breaking, non-trivial magnetization ground states, and the stabilization of complex two- and three-dimensional magnetic textures, which are not accessible in their cylindrical counterparts.

	\subsubsection{In-Surface Circulation, Vortices and Antivortices in Nanotori}
	\label{sec:insurf-circulation}
	
	The analysis in Sec.~\ref{sec:heisenberg} of topological excitations in isotropic and anisotropic Heisenberg model in curved manifolds completely neglected additional magnetic energy terms. Among these, the magnetostatic energy is crucial, as it determines the (meta)stability of spin textures and governs the nucleation of distinct magnetization patterns. 
	
	Within a continuous approach, the magnetostatic energy is expressed as~\cite{Aharoni96}
	\begin{subequations}\label{eq:magnetostatics}
		\begin{equation}
			E_{\text{mag}} = \frac{\mu_0 M_S}{2} \int_V \vec{m}(\vec{r}) \cdot \nabla V_m(\vec{r}) dV ,
		\end{equation}
		where the magnetostatic potential is given by
		\begin{equation}
			V_m(\vec{r}) = -\frac{M_S}{4\pi} \int_{V'} \frac{\nabla\cdot\vec{m}}{|\vec{r} - \vec{r}'|} dV' + \frac{M_S}{4\pi} \int_{S'} \frac{\vec{m}\cdot\vec{n}}{|\vec{r} - \vec{r}'|} dS'
		\end{equation}
	\end{subequations}
	Here, $\vec{n}$ is the unit normal vector to the torus surface, $\mu_0$ is the vacuum permeability, and $V'$ and $S'$ denote the volume and surface of the magnetic body, respectively.
	
	Since the early 2000s, extensive research has been devoted to characterizing vortex states in circular nanodisks~\cite{Guslienko01b,Klaui03,Shibata03a,Hertel04,Pereira07}, nanospheres~\cite{Goll04,Kravchuk12a}, and nanorings~\cite{Beleggia06,Vaz07}. Magnetic vortices typically exhibit an out-of-plane magnetization component at the core, which prevents exchange-energy divergences. Introducing a central hole transforms the structure into a nanoring, and suppresses the vortex-core formation by causing the out-of-plane magnetization to diminish with increasing hole size~\cite{Kravchuk07}. This coreless configuration enhances stability, allowing vortices to persist in nanorings at smaller particle sizes compared to their nanodisk counterparts~\cite{Landeros06,Kravchuk07}. 
	
\begin{figure}[tb]
		\centering
		\includegraphics[width=0.8\linewidth]{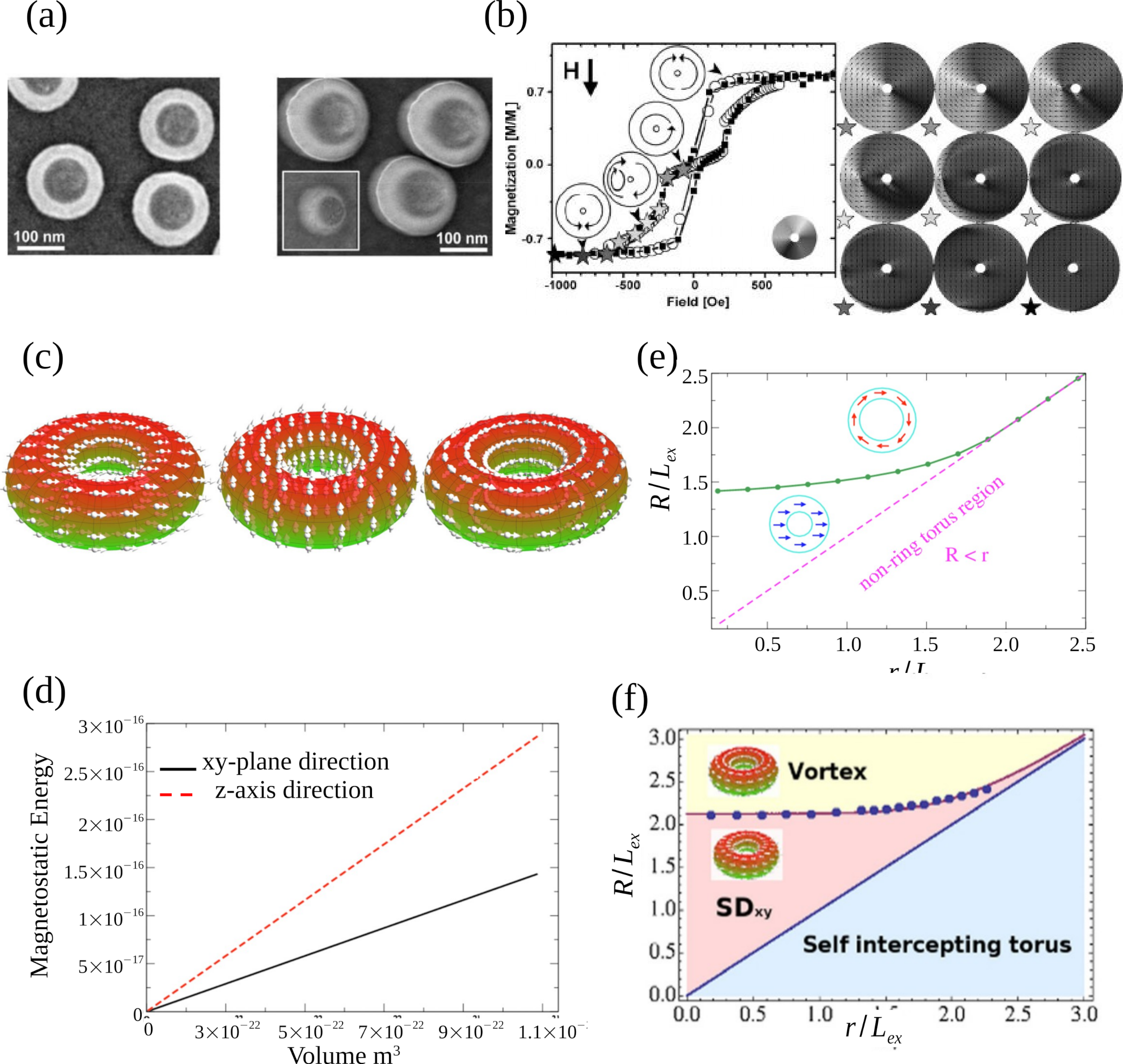} % Fig3-Vagson
		\caption{(a) Scanning electron micrographs of symmetric and asymmetric nanorings fabricated of Co nanorings obtained by sputter-depositing a 40\,nm film using a high-purity (99.995\%) target in a magnetron sputtering system equipped with in situ substrate tilting-angle adjustment. (b) Hysteresis loop of a Co nanoring array (left) and photoemission electron microscopy image (right) showing vortex states (clockwise/counterclockwise) and an onion state. (c) Magnetic states stabilized in Py nanotori: single domain along $xy$-plane (left), single domain along $z$-axis, and vortex state (right). (d) Magnetostatic energy of single domain along $xy$ and $z$ direction. (e) and (f) show the state diagram for nanotori obtained through analytical calculations and micromagnetic simulations, respectively. Panel~(a) reprinted with permission from~\cite{Zhu06a}~\textcopyright{}~(2006) by the American Physical Society. Panel~(b) reprinted from~\cite{Klaui04b}, with the permission of AIP Publishing. Panels~(c,f) are reprinted from~\cite{Vojkovic16}, with the permission of AIP Publishing. Panels~(d,e) are reprinted from~\cite{Carvalho-Santos10}, with the permission of AIP Publishing.}
		\label{Fig3}
	\end{figure}

	The possibility of nucleating and reversing the vortex circulation direction in symmetric and asymmetric mesoscopic ring magnets has been demonstrated experimentally, as e.g., shown in Figs.~\ref{Fig3}(a) and (b), through the controlled application of magnetic fields, notches, and asymmetries~\cite{Klaui01,Klaui02,Klaui03,Klaui03b,Klaui03a,Zhu06a}, enabling precise manipulation of their magnetic state for device applications.

	\paragraph{Smoothed Toroidal Nanodisks}

	The geometric constraint of a nanoring is defined by the presence of a central hole, which renders it topologically equivalent to a torus (genus 1). In such a torus-shaped nanoparticle, the curvature and continuous symmetry eliminate sharp geometric features, thereby suppressing stray fields and defects typically associated with boundaries. This smoothness significantly increases the stability of the vortex configuration~\cite{Carvalho-Santos10,Vojkovic16,Lewis20}  (see also Sec.~\ref{subsec:experiment-torus}). Consequently, vortices emerge as stable ground states in toroidal nanoparticles where the competition between exchange and magnetostatic interactions governs magnetic properties. The enhanced stability of the vortex state ($\vec{m}_{v}=-\sin\varphi\,\vec{\hat{x}}+\cos\varphi\,\vec{\hat{y}}$) in a magnetic nanotorus is demonstrated by comparing its energy to monodomain configurations aligned along both the $z$-direction ($\vec{m}_{z}=\vec{\hat{z}}$) and the $xy$-plane ($\vec{m}_{xy}=\vec{\hat{y}}$), as illustrated in Figs.~\ref{Fig3}(c,d).
	
	In the case of a purely in-surface vortex configuration, the dominant contribution to the energy arises only from the exchange interaction. Since the magnetization remains confined to the surface plane, the magnetostatic energy (typically arising from volume charges or edge effects) vanishes. This simplifies the analysis, as the magnetic energy is governed exclusively by the exchange term, which can be written as $E_{\text{x}}=A\int(\nabla{\vec{m}})^2dV$, with $A$ being the exchange stiffness. In this case, by adopting the geometric parameters presented in Fig.~\ref{Fig2}(a), one obtains 
	\begin{equation}
		E_{v}=2\pi^2 A\left[\sqrt{a^2+r^2}-a\right]
	\end{equation} 
	where the relation $a=\sqrt{R^2-r^2}$ is reiterated. 
	
	The single-domain states with in-plane and out-of-plane magnetization orientations exhibit magnetic charge distributions across the torus surface, resulting in a more complex energy landscape compared to the vortex state. While the in-plane configuration may show slight deviations from perfect parallel alignment, these variations introduce only minor energy corrections relative to the idealized single-domain state~\cite{Carvalho-Santos10}. Consequently, calculating the energy of the idealized single-domain configuration proves sufficient for analysing the vortex stability in the toroidal system. 
	
	\paragraph{Magnetostatic Energy of a Single-Domain State on a Bulk Torus}
	
	As the exchange energy vanishes for the single-domain mode, one only needs to compute the magnetostatic contribution. This task can be performed by computing the magnetostatic potential $V_m$, from which the magnetic field $\vec{H}=-\nabla V_m$ is derived. To determine the potential, one can adopt the toroidal coordinate system given in Eq.~\eqref{TorCoord}, which is highly useful because the Laplace equation admits a well-known solution within it, for which the method of separation of variables is valid (see Ref.~\cite[p.~1302]{Morse53})
	\begin{eqnarray}
		V_m(\beta,\xi,\varphi)=\sqrt{\cosh\beta-\cos\xi}\,\,B(\beta)X(\xi)F(\varphi)
	\end{eqnarray}
	where the functions $B$, $X$, and $F$ satisfy ($\chi = \cosh \beta$, and $n$, $k$ are constants). The separated variable form of the potential allows us to decompose Laplace's equation into the following independent equations
	\begin{subequations}
		\begin{equation}
			(\chi^2 - 1)\frac{d^2 B}{d\chi^2} + 2\chi\frac{dB}{d\chi} - \left[ n^2 - \frac{1}{4} + \frac{k^2}{\chi^2 - 1} \right]B = 0, 
		\end{equation}
		\begin{equation}
			\frac{d^2 X}{d\xi^2} + n^2 X = 0, 
		\end{equation}
		\begin{equation}
			\frac{d^2 F}{d\phi^2} + k^2 F = 0. 
		\end{equation}
	\end{subequations}
	The functions $X(\xi)$ and $F(\phi)$ are linear combinations of trigonometric functions, as shown below
	\begin{align}
		X_n(\xi) &= f_n \cos(n\beta) + g_n \sin(n\beta), \\
		F_k(\phi)  &= u_k \cos(k\phi) + w_k \sin(k\phi).
	\end{align}
	where $f_n$, $g_n$, $u_k$, and $w_k$ are constant coefficients to be determined from the boundary conditions established for the problem. Since $\xi$ and $\phi$ are variables ranging from $0$ to $2\pi$, the solutions are periodic. This condition implies that $k = 0,1,2,\dots$ and $n = 0,1,2,\dots$; that is, rotational invariance imposes that $k$ and $n$ are integers.
	
	The functions $B(\beta)$ consist of the associated Legendre functions of half-integer degree of the first kind, $P_{n-1/2}^k (\cosh \alpha)$, and of the second kind, $Q_{n-1/2}^k (\cosh \alpha)$, also called toroidal harmonics, whose asymptotic behavior is given by $
	\lim_{\alpha \to \infty} P_{n-1/2}^{k}(\cosh \beta) = \infty,$ and $\lim_{\alpha \to 0} Q_{n-1/2}^{k}(\cosh \beta) = \infty$. For more details on the toroidal harmonics, the reader is referred to the Refs.~\cite{Morse53}. Under this formalism, the expansion of $1/|\vec{r} - \vec{r}'|$, in terms of Green's functions, for the toroidal coordinate system, is given by 
	\begin{equation}\label{TorHarm}
		\begin{aligned}
			\frac{1}{|\vec{r} - \vec{r}'|} & = \frac{\sqrt{\Lambda\,\Lambda'}}{\pi b} 
			\sum_{n=0}^{\infty} \sum_{k=0}^{\infty} (-1)^k \epsilon_n \epsilon_k \cos[n(\xi - \xi')] \cos[k(\phi - \phi')] \\
			& \times \frac{\Gamma\left(n - k + \frac{1}{2}\right)}{\Gamma\left(n + k + \frac{1}{2}\right)} P_{n-1/2}^k (\cosh \beta_<) Q_{n-1/2}^k (\cosh \beta_>),
		\end{aligned}
	\end{equation}
	where $\Lambda=\cosh \beta - \cos \xi\,$, $\Lambda'= \cosh \beta' - \cos \xi'$, $\epsilon_n = (2 - \delta_{n,0})$, $\epsilon_k = (2 - \delta_{k,0})$, and the vector $\vec{r}'$ is parameterized by $(\alpha', \beta', \phi')$ (see Ref.~\cite[p.~1304]{Morse53}). The fact that $P_{n-1/2}^k(\cosh \alpha)$ and $Q_{n-1/2}^k(\cosh \alpha)$ are related to the signs $<$ and $>$, respectively, reflects the asymptotic behavior of these functions.
	
	Using the 3D Green’s function given in Eq.~\eqref{TorHarm}, the magnetostatic energy for the single domain configuration along the $z$-axis and $xy$-plane is given respectively by~\cite{Carvalho-Santos10}
	\begin{align}\label{EMZ}
		E_m^{z}&=\frac{128\mu_0 M_S^2\,a^3}{9}\sinh^2\beta_0\sum_ {n=1}^{\infty}n^2\,P_{n-1/2}Q_{n-1/2}(Q^{1}_{n-1/2})^2\\\nonumber\\\label{EMXY}
		E_m^{xy}&=-\frac{32\mu_0 M_S^2\,a^3}{9}\sum_ {n=1}^{\infty}\epsilon_n\,C_n^2\,P^1_{n-1/2}Q^{-1}_{n-1/2}
	\end{align}
	where the superscripts $k = 0$ in the Legendre functions have been omitted and $P_\nu^\mu(\cosh\beta_0)\equiv P_\nu^\mu$ and $Q_\nu^\mu(\cosh\beta_0)\equiv Q_\nu^\mu$, where $\cosh \beta_0 = R / r$ is the torus aspect ratio. Also, $$C_n=\cosh\beta_0\left(\frac{Q^2_{n+1/2}}{\sinh\beta_0}-nQ^1_{n-1/2}\right)-\frac{Q^2_{n-1/2}}{\sinh\beta_0}$$
	
	Although one can obtain analytical expressions for the magnetostatic energy of the single domain state in the torus, given by Eqs.~\eqref{EMZ} and~\eqref{EMXY} present a physical content that is not immediately apparent due to their complex forms. To extract relevant results, one can evaluate them numerically considering the magnetic parameters of Permalloy~\cite{Carvalho-Santos10}, whose results are summarized in Fig.~\ref{Fig3}(d). The comparison between the energies associated with the single-domain states along $z$-axis and $xy$-plane, evidences that they can be approximated by the following expressions (where $V=2\pi^2r^2R$ is the torus volume)
	\begin{eqnarray}
		E_m^z\approx \frac{\mu_0 M_S^2}{4}V\hspace{1cm}
		E_m^{xy}\approx \frac{\mu_0 M_S^2}{8}V
	\end{eqnarray}
	evidencing that the single domain configuration along the $xy$-plane costs nearly half of the energy from its counterpart along $z$. 
	
	Finally, to analyse the vortex stability in magnetic nanotori, one should compare the vortex and single-domain $xy$ states to determine the magnetization ground state. Fig.~\ref{Fig3}(e) summarizes the main findings, indicating the energetically favourable state as a function of the torus aspect ratio, $R/r$. Considering only the physically relevant geometric regime where $R>r$, one notices that the vortex state is the ground state for all tori with $r\gtrsim\SI{10}{nm}$ (i.e., $\rho=r/\ell_{\text{ex}}\gtrsim2$, with $\ell_{\text{ex}} = \sqrt{A/(4\pi M_S^2)}$ in CGS or $\ell_{\text{ex}} = \sqrt{A/(\mu_0 M_S^2)}$ in SI being the exchange length of the material). For tori with $r\lesssim\SI{10}{nm}$, the vortex still minimizes the total energy when $R\gtrsim\SI{10}{nm}$, whereas the single-domain state becomes favourable for smaller $R$. It is worth highlighting that in the region where $\rho\lesssim1$ and $\tau=R/\ell_{\text{ex}}\lesssim1.5$, the true ground state is not exactly the single domain state, but a slightly different onion-like patterns with energies at most 3.2\% lower than the single domain state~\cite{Carvalho-Santos10}. Consequently, the transition curve in this region carries an uncertainty of approximately 3.2\%. 
	
	Micromagnetic simulations, as intrudocued in Sec.~\ref{subsec:micromagnetic}, can be applied to analyze the vortex stability in magnetic nanotori. The state diagram given by the theoretical model has also been obtained through micromagnetic simulations~\cite{Vojkovic16} using the OOMMF~\cite{OOMMFb,Donahue99} package, as shown in Fig.~\ref{Fig3}(f). In this case, one can initialize the magnetization in a ferromagnetic or vortex state, relax the system by solving the LLG equation and compare the energy of the final states. As evidenced in the diagram, results from micromagnetic simulations corroborate analytical findings, where the vortex state is strongly stable in magnetic nanotori. This behavior, which has been experimentally corroborated~\cite{Lewis20}, arises from a smooth variation in curvature. This is in contrast to cylindrical nanorings, where sharp edges create discontinuities and divergences in curvature. 
	
	\paragraph{Magnetic Surface States on Hollow Nanotori}
	
	While the stabilized vortices in both bulk and hollow nanotori are typically azimuthal, one can also analyse the possibility of nucleating in-surface states with non-azimuthal orientations in very thin toroidal nanoshells. To describe these states, the magnetization can be parameterized in the local coordinate frame as $\vec{m}=\hat{\vec{e}}_1\,\sin\Theta\cos\Phi+\hat{\vec{e}}_2\,\sin\Theta\cos\Phi+\vec{n}\,\cos\Theta$, where $\vec{n}$ is the unit vector of the normal to the surface. To determine the exchange energy using a curvilinear basis, one can use the theoretical formalism developed by Gaididei~\textit{et~al.}~\cite{Gaididei14}, see Sec.~\ref{sec:diffgeo}.
	
	In the Cartesian frame of reference, the exchange energy density is given by $\varepsilon_\text{ex} = (\nabla m_i) (\nabla m_i)$, where $m_i$ are the Cartesian components of the magnetization vector. These components are expressed in the local curvilinear basis as $m_i = m_\alpha (\vec{e}_\alpha \cdot \vec{\hat{x}}_i) + m_n (\vec{n} \cdot \vec{\hat{x}}_i)$, where $m_\alpha$ and $m_n$ are the projections onto the local tangential and normal directions, respectively. To evaluate the energy density, this expression for $m_i$ is substituted into $E_{\text{x}}$, and the gradient operator is applied in its curvilinear form, $\nabla \equiv (g_{\alpha\alpha})^{-1/2} \vec{e}_\alpha \partial_\alpha$. After an algebraic manipulation, one obtains that the exchange energy is given by~\cite{Gaididei14}
	\begin{eqnarray}\label{Gaididei-Exchange}
		E_\text{ex}=Ah\int\left\{\left[\nabla\Theta-\vec{\Gamma}(\Phi)\right]^2+\left[\sin\Theta(\nabla\Phi-\vec{\Omega})-\cos\Theta\frac{\partial\vec{\Gamma}(\Phi)}{\partial\Phi}\right]^2\right\}dS
	\end{eqnarray}
	where $dS=\sqrt{\mathrm{det}\,\|g_{\alpha\beta}\|}d\xi_1 d\xi_2$ the vector $\vec{\Omega} = (\varpi_1/\sqrt{g_{11}}, \varpi_2/\sqrt{g_{22}})$ is a modified spin connection with $\varpi_\alpha = \vec{e}_1 \cdot \partial_\alpha \vec{e}_2$, and the vector $\vec{\Gamma}(\Phi)$ is determined by $
	\vec{\Gamma}(\Phi) = |H_{\alpha\beta}| \vec{\tau}(\Phi)$, with $\vec{\tau}(\Phi) = (\cos\Phi, \sin\Phi)$, see also Eq.~\eqref{eq:exchange-curv}.
	
	\begin{figure}[tb]
		\centering
		\includegraphics[width=0.8\linewidth]{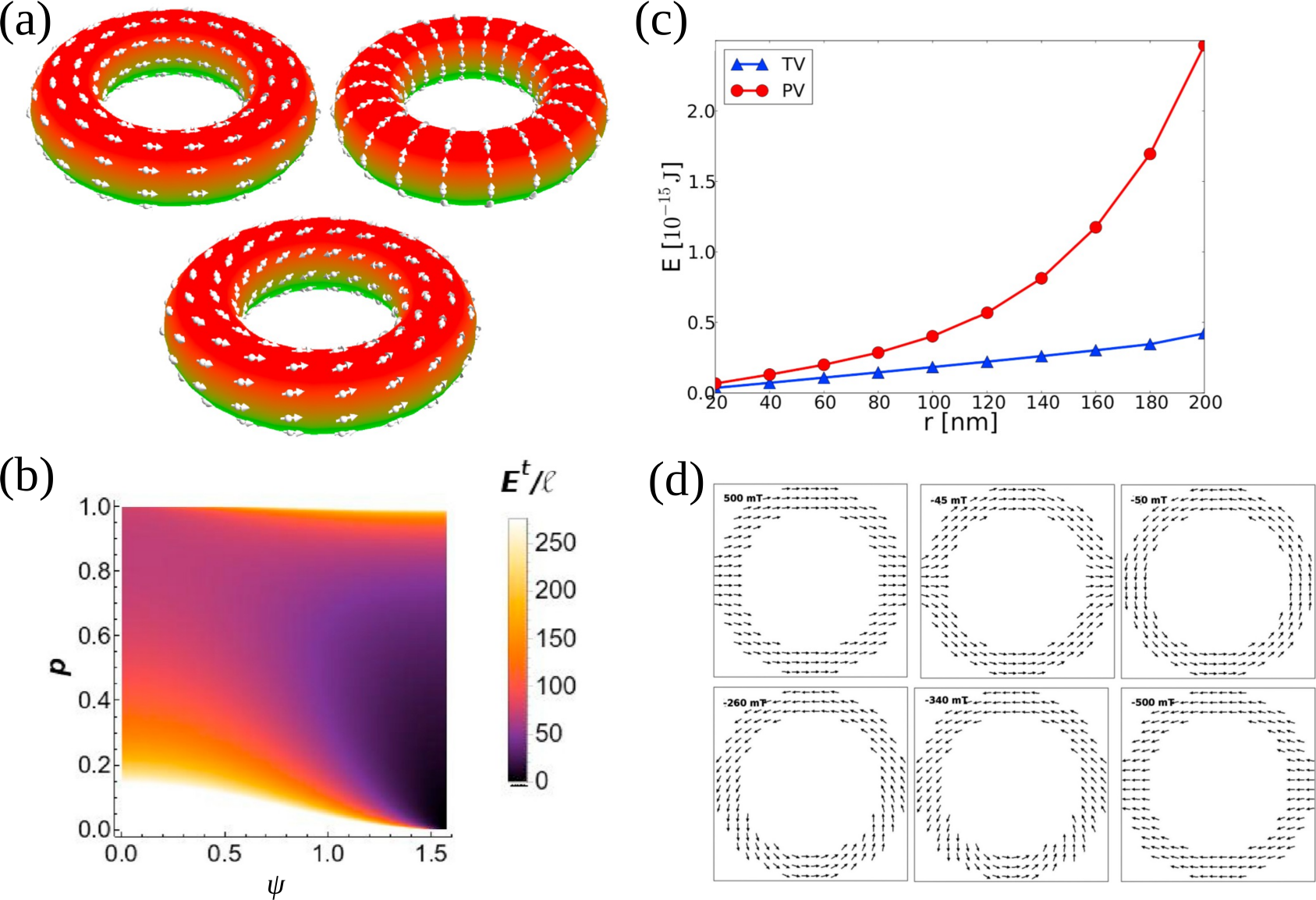} % Fig4-Vagson
		\caption{(a)~In-surface states in a toroidal shell for $\psi=0$, $\psi=\pi/2$, and $\psi=\pi/4$. (b)~Color map of exchange energy of an in-surface state as a function of $p$. Dark colors represents lower values of energy. (c)~Total energy of a TV (blue) and a PV (red) states for $R = 200$\,nm as a function~of~$r$. (d)~Snapshot of vortex-mediated reversal process in a toroidal nanomagnet with $R=26$\,nm and $r=4$\,nm. Panels~(a--c) are reprinted from~\cite{Teixeira19}~\textcopyright{}~(2019) with permission from Elsevier. Panel~(d) is reprinted from~\cite{Vojkovic16}, with the permission of AIP Publishing.}
		\label{Fig4}
	\end{figure}
	
	To describe an in-surface magnetization configuration within a nanotorus, the magnetization vector can be represented in the local curvilinear orthonormal basis. For a uniform state where all magnetic moments are aligned in a single direction tangent to the surface, this is given by $\vec{m} = \hat{\vec{\theta}} \cos\psi + \hat{\vec{\varphi}} \sin\psi$, where the unit vectors $\hat{\vec{\theta}}$ and $\hat{\vec{\varphi}}$ correspond to the toroidal and poloidal directions, respectively, and $\psi$ is a constant angle determining the specific orientation of the moments within the surface. One can refer to the resulting magnetization profiles for an arbitrary $\psi$ as a tokamak-like (TK) configuration. 
	
	Specific cases include the toroidal vortex (TV) for $\psi=\pi/2$ and the poloidal vortex (PV) for $\psi=0$. The magnetization profiles for the TV, PV, and TK (with $\psi=\pi/4$) configurations are shown in Fig.~\ref{Fig4}(a). In this case, using the basis $\vec{r}=(R+r\sin\theta)\hat{\vec{R}}+r\cos\theta\,\vec{\hat{z}}$ to describe a torus, one obtains that $\vec{\Gamma}=-\hat{\vec{\theta}}\,r^{-1}\cos\psi-\hat{\vec{\varphi}}\,(R+r\sin\theta)^{-1}\,\sin\theta\sin\psi$ and $\vec{\Omega}=-\hat{\vec{\varphi}}\,(R+r\sin\theta)^{-1}\cos\theta$. In this context, the exchange energy density in Eq.~\eqref{Gaididei-Exchange} is simplified to $\varepsilon_\text{ex}=\vec{\Gamma}^2+\vec{\Omega}^2$~\cite{Teixeira19}, and the exchange energy of the in-surface state is given by 
	\begin{eqnarray}
		E_{\text{x}}=4\pi^2AL\left[\frac{p}{1+\sqrt{1-p^2}}+\frac{p\,\sin^2\psi}{1-p^2+\sqrt{1-p^2}}+\frac{\cos^2\psi}{p}\right]\,
	\end{eqnarray}
	where $p=r/R$. 
	
	Results for $E_{\text{x}}/\ell_{\text{ex}}$ as a function of the aspect ratio $p$ are presented in Fig.~\ref{Fig4}(b). One can notice that the exchange energy of the TK state is higher than that of either the TV or PV state for any value of $p$. A key finding is that the TV state minimizes the exchange energy for $p<0.87$. However, at the specific aspect ratio $p\approx0.87$, the exchange energy is minimized for any value of $\psi$, meaning that any in-surface orientation of the magnetization yields an equivalent energy. Consequently, TK states can minimize the exchange energy only at this singular point. For $p>0.87$, the PV state becomes the lowest-energy configuration, with its exchange energy dropping below that of the TV state.
	
	It is worth highlighting that minimizing the exchange energy alone, however, does not guarantee the minimization of the total energy. Indeed, unlike the toroidal vortex (TV) state, the poloidal vortex (PV) state presents a magnetostatic energy contribution arising from volumetric magnetic charges, formally defined as $\rho_m = -\nabla \cdot \vec{m}$, which persists despite the PV state having no surface magnetic charges. Therefore, this volumetric magnetostatic contribution should be considered to determine the in-surface magnetization profile that minimizes the total energy. 
	
	Micromagnetic simulations using the OOMMF code~\cite{OOMMFb,Donahue99} have been performed to calculate the total energy for each state in permalloy nanotori. These calculations explicitly treat the magnetostatic interaction rather than approximating it as an in-surface anisotropy, ensuring the total energy is given by the direct sum of the exchange and dipolar energies. Results presented in Fig.~\ref{Fig4}(c) allow one to notice that although the PV configuration exhibits the lowest exchange energy for $r/R > 0.87$, its total energy is higher due to this dominant dipolar contribution. Consequently, when purely in-surface states are considered, the TV state remains the lowest-energy configuration for \emph{any} aspect ratio in a toroidal nanoparticle.

	\paragraph{Magnetisation Reversal in Bulk Nanotori}
		
	In addition to stabilized magnetic patterns, the magnetization reversal mechanisms in magnetic nanotori allow the emergence of a rich variety of two- and three-dimensional topological metastable states depending on the torus aspect ratio and the direction of the applied magnetic field, arising from the smooth curvature variations across the nanotorus surface. In the following, some key results on the nucleation mechanisms and stability of these topological textures will be presented, highlighting their dependence on geometric and field parameters.
	
	The simplest reversal mechanism occurs for bulk nanotori with small values of $R$ under the action of an in-plane magnetic field. In this case, micromagnetic simulations considering Permalloy nanotori, characterized by an exchange stiffness $A=1.3\times10^{-11}$~J/m and a magnetization saturation of $M_S=860$~kA/m, have revealed that two main mechanisms characterize the hysteresis loop. For structures with $R\lesssim16$~nm, the hysteresis loops are characteristically square, indicative of a rapid and coherent rotation of the magnetization~\cite{Vojkovic16}. Since the magnetostatic and exchange energies remain constant during this rotation, the driving force for the abrupt switch is the minimization of Zeeman energy. Notably, the coercivity in this regime depends only on $R$~\cite{Vojkovic16}. Larger $R$ values, corresponding to a greater magnetic volume, require a faster rotation to overcome the energy barrier, resulting in a lower coercive field. 
	
	In contrast, for nanotori with $R\gtrsim18$~nm, the hysteresis loops exhibit a pronounced neck, revealing a more complex reversal process~\cite{Vojkovic16}. A neck-like shape is a signature of a non-uniform magnetization reversal mediated by the nucleation of a vortex. As detailed in the snapshots shown in Fig.~\ref{Fig4}(c), the reversal begins with the formation of an onion state, from which a vortex nucleates. The vortex progressively aligns more moments with the field, resulting in a gradual decrease in magnetization. The stability of this vortex state across a wide range of parameters is responsible for the consistent appearance of these necked loops for $R \gtrsim 18$~nm. Furthermore, in this vortex-mediated regime, the field required for the final annihilation of the vortex structure shows a strong dependence on $r$. 
	
	\begin{figure}[tb]
		\centering
		\includegraphics[width=0.8\linewidth]{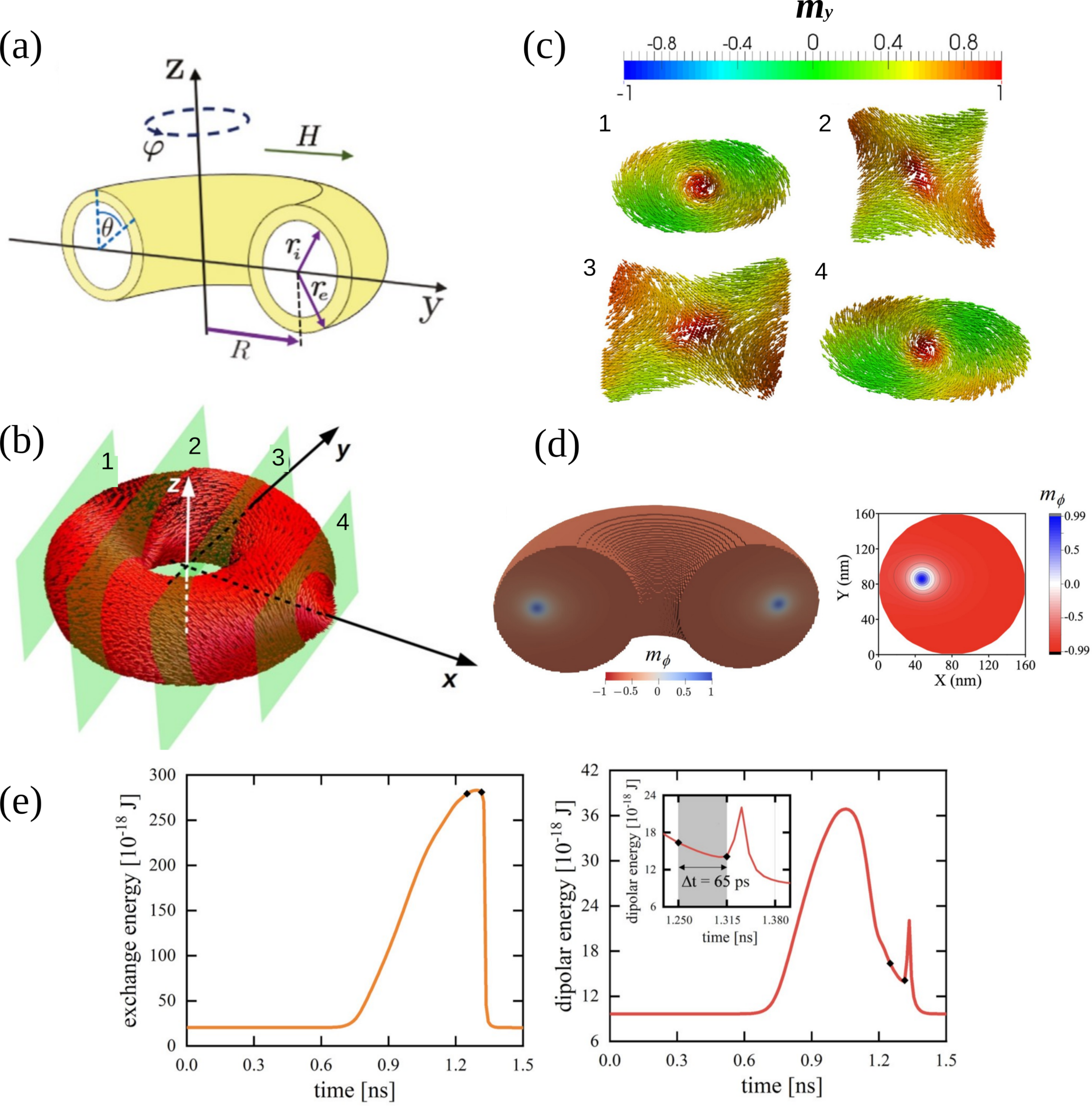} % Fig5-Vagson
		\caption{(a)~Geometric parameters of a toroidal tube with thickness $r_e-r_i$. Remanent state during the reversal process of a nanotorus where the nucleation of a vortex at the external border is observed. (b)~A~ferromagnetic torus and (c)~snapshots of the magnetization configuration in the planes 1, 2, 3, and 4. (d) Snapshot of the azimuthal component of the magnetization in the toroidal cross-section during the reversal under the action of an Oersted magnetic field. (d)~Cross-section of a torus and~(e)~magnetic energy as a function of time for the reversal process for this geometry. Panel~(a) is reprinted from~\cite{Saavedra22},~\textcopyright{}~(2022), with permission from Elsevier. Panels~(b,c) are reprinted from~\cite{Vojkovic17}, with the permission of AIP Publishing. Panels~(d,e) are reprinted with permission from~\cite{Castillo-Sepulveda21}~\textcopyright{}~(2021) by the American Physical Society.}
		\label{Fig5}
	\end{figure}
	
	\paragraph{Reversal of a Hollow Torus under an In-Plane Field}
	
	Through micromagnetic simulations using OOMMF package, Vojkovic~\textit{et~al.}~\cite{Vojkovic17} investigated the magnetization reversal process in a Permalloy nanotorus under an in-plane magnetic field by modeling the structure as a thin hollow torus with internal and external tube radii denoted by $r_i$ and $r_e$, respectively [see Fig.~\ref{Fig5}(a)]. Their model adopted a toroidal radius of $R = 52$\,nm, with poloidal radii of $r_i = 20$\,nm and $r_e = 26$\,nm, resulting in a torus with a uniform thickness of 6\,nm. In the simulations, the magnetization was first saturated by a strong external field ($\vec{H}=H\vec{\hat{y}}$) to establish a single-domain initial state aligned with the field's in-plane direction. The reversal behavior was then obtained by progressively reducing the magnetic field to zero, allowing the system to relax into its remanent state. The magnetic configuration obtained from this procedure is presented in Fig.~\ref{Fig5}(b), where one notices the appearance of a vortex in the torus region with positive Gaussian curvature. To analyze the magnetization at the negative Gaussian curvature, one can analyze four cross-sectional cutoffs at the planes highlighted in green color in Fig.~\ref{Fig5}(b). This procedure yields the emergence of a vortex-antivortex (VA) pair within the nanotorus. Specifically, two vortices with opposite chiralities appear at the external borders (positive curvature), while two antivortices, also with opposing chiralities, are observed at the internal borders (negative curvature), as depicted in Fig.~\ref{Fig5}(c).
	
	These vortex-antivortex (VA) pairs can be interpreted as minimal versions of a cross-tie domain wall~\cite{McCord09,Middelhoek63}, which separates two distinct magnetic domains formed along the poloidal angle: one domain magnetized along $-\hat{\vec{\varphi}}$ (for $-\pi/6 \lesssim h \lesssim \pi/6$) and another along $+\hat{\vec{\varphi}}$ (for $5\pi/6 \lesssim h \lesssim 7\pi/6$). The formation of these domains in the upper and lower regions of the toroidal nanoparticle results from the interplay between shape anisotropy (magnetostatic) and exchange interactions. Equivalently, this can be understood through the interplay of dipolar, DMI-like, and curvature-induced effective anisotropy interactions~\cite{Gaididei14}. The nucleation of this VA pair can be understood by expressing the magnetization as
	\begin{equation}\label{eq:mag-vec-curv}
		\vec{m}=\sin\Theta\cos\Phi \vec{e}_1 +\sin\Theta\sin\Phi\vec{\hat{\varphi}}+\cos\Theta\vec{\hat{n}},
	\end{equation}
	where $\vec{e}_1 = \vec{\hat{\theta}}$ and $\vec{e}_2 = \vec{\hat{\varphi}}$. By modeling the magnetostatic energy as an easy-tangential anisotropy and considering a tangential magnetization configuration ($\Theta=\pi/2$), the exchange energy density given in Eq.~\eqref{Gaididei-Exchange} can be decomposed into three components: (i) a ``standard'' exchange interaction $\varepsilon_\text{ex}^1 = A(\nabla\Phi)^2$, which promotes a homogeneous spatial distribution of the magnetization and is minimized for constant $\Phi$; (ii) an effective anisotropy given by $\varepsilon_\text{ex}^2 =A \vec{\Gamma}^2$; and (iii) an effective DMI-like interaction $\varepsilon_\text{ex}^3=-2A(\nabla\Phi\cdot\vec{\Omega})$, which is minimized when the magnetization displays inhomogeneous textures. Within this framework, one obtains:
	
	\begin{subequations}\label{Effective-Torus}
		\begin{align}\label{Effective-Torus1}
			\varepsilon_\text{ex}^2&=A\left[\frac{1}{r^2}\cos^2\Phi+\frac{\sin^2\theta}{(R+r\sin\theta)^2}\sin^2\Phi\right],\\
			% \end{eqnarray}
		% \begin{eqnarray}
			\label{Effective-Torus2}\varepsilon_\text{ex}^3&=\frac{2A\cos\theta}{(R+r\sin\theta)^2}\partial_\varphi\Phi.
		\end{align}
	\end{subequations}
	
	The analysis of Eqs.~\eqref{Effective-Torus} reveals that for $\theta = 0$ or $\theta = \pi$, $\varepsilon_\text{ex}^3 \sim 2/R^2$ and $\varepsilon_\text{ex}^2 \sim 1/r^2$. Given that $R = 2r$, $\varepsilon_\text{ex}^2$ dominates, the in-surface solution $\Phi = \pi/2$ more favorable. Conversely, for $\theta \to \pi/2$, the first term in Eq.~\eqref{Effective-Torus1} dominates and favors the solution $\Phi = 0$. Consequently, the curvature-induced DMI-like interaction ($\varepsilon_\text{ex}^3$) generates spatially inhomogeneous magnetization distributions. These inhomogeneous states lead to the formation of the upper and lower domains observed in Fig.~\ref{Fig5}(b). The main consequence of these two domains is the formation of the VA pair, where the curvature-induced chiral states determine the specific regions where the vortex and antivortex appear. By adopting the ansatz $\Phi = \pm \arctan\left[ (r \cos \theta)/((R + r \sin \theta) \sin \varphi) \right]$ to describe vortex ($+$) or antivortex ($-$) configurations, the energy term given in Eq.~\eqref{Effective-Torus2} is minimized for $\partial_\varphi \Phi < 0$ (vortex) around $\theta = \pi/2$ and $\partial_\varphi \Phi > 0$ (antivortex) around $\theta = -\pi/2$. Consequently, the curvature-induced DMI-like interaction is responsible for the nucleation of a vortex at the torus region with positive Gaussian curvature and an antivortex at the negative one. This finding is directly associated with the curvature-induced magnetochiral effects~\cite{Hertel13a,Gaididei14}, responsible for the enhancing of the stability of topological textures in dome-shaped nanoparticles~\cite{CarvalhoSantos15a,Carvalho-Santos20,Bezsmertna24,Raftrey25} and a winding number selection of merons~\cite{Elias19,MirandaSilva22,Volkov24a}, a mechanism whereby the Gaussian curvature of the magnetic body determines the topological charge of the hosted meron.
	
	\subsubsection{Three-Dimensional Solitons in Bulk Nanotori}
	\label{sec:Hopfions}
	
	Driven by advances in the synthesis and characterization of magnetic meso- and nanoparticles with diverse geometries, the field of three-dimensional (3D) nanomagnetism has emerged as a new and exciting area of research~\cite{Fernandez-Pacheco17,Makarov22,Gubbiotti24}. Within it, 3D topological magnetic textures can emerge as stable or metastable states within magnetic bodies, which are unattaimable in planar or quasi-one dimensional structures. 
	
\begin{figure}[tb]
		\centering
		\includegraphics[width=0.8\linewidth]{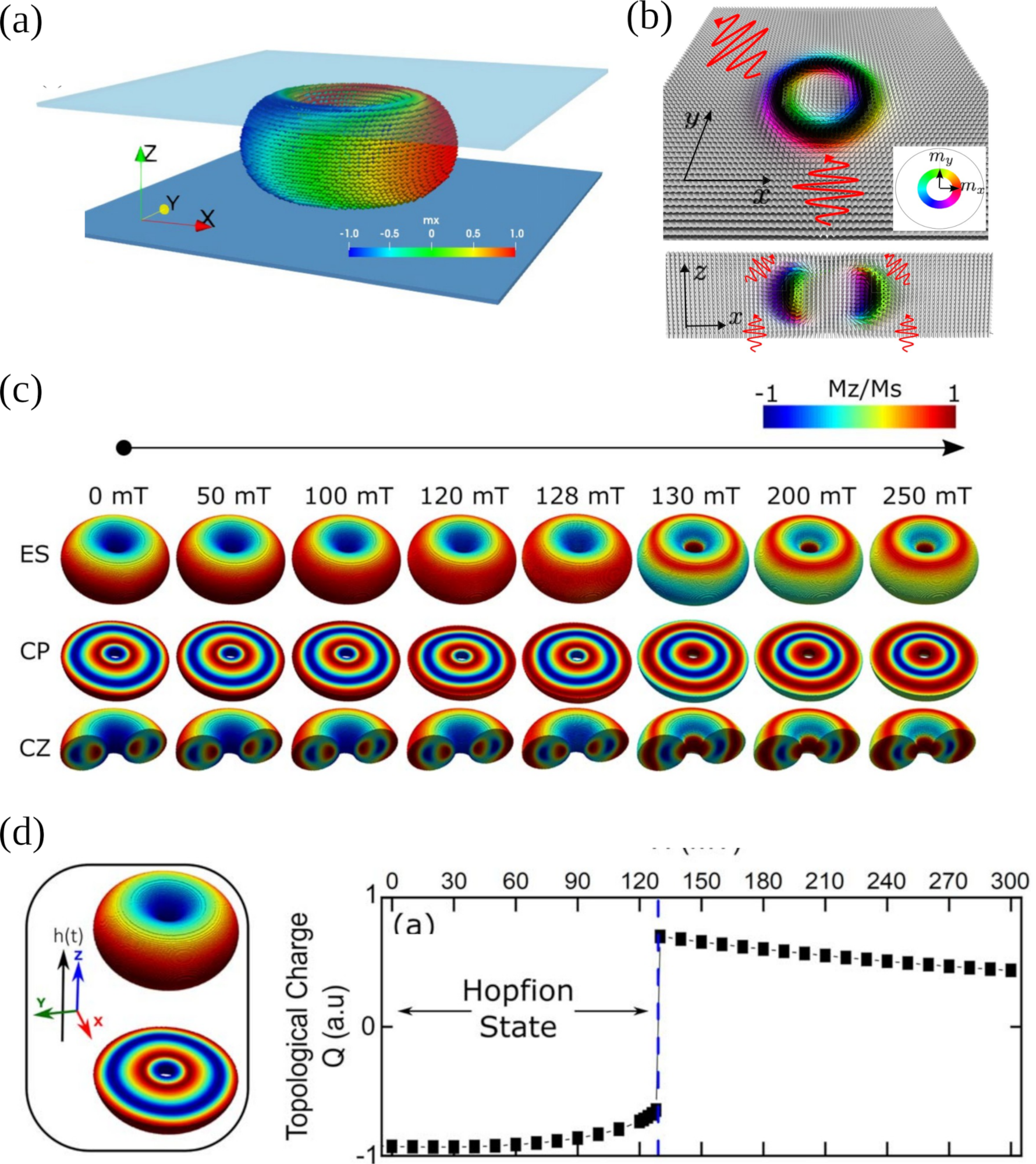} % Fig6-Vagson
		\caption{(a)~Schematic view of the Hopfion pattern exhibiting its toroidal symmetry. (b)~Schematic representation of spin wave scattering on Hopfions: Top: $xy$-plane cross-section. Bottom: $xz$-plane cross-section. Colour code indicates the $\vec{m}$-projection in the $xy$-plane. Red wavy lines denote spin wave Hopfion interactions. Spin waves in the $xy$-plane undergo skew scattering, while for those along the symmetry axis a lens-like effect occurs. (c)~Stabilized states in a nanotorus with DM interaction under the action of a uniaxial magnetic field. $z$-component of the magnetization with their respective values of the magnetic field. (d)~Topological charge of the 2D skyrmion appearing in the central plane for the case $H=0$. Panel~(a) is reprinted with permission from~\cite{Li22d}~\textcopyright{}~(2022) by the American Physical Society. Panel~(a) is reprinted with permission from~\cite{Saji23}~\textcopyright{}~(2023) by the American Physical Society. Panels~(c,d) are reprinted with permission from~\cite{Corona23}~\textcopyright{}~(2023) by IOP Publishing Ltd..}
		\label{Fig6}
	\end{figure}

	A particular class of 3D magnetization patterns of interest is the \emph{Hopfion}, a soliton characterized by a knotted magnetization field. Hopfions can be considered as a 3D skyrmion (bimeron) tube closed in itself, corresponding to a $2\pi$ rotation of the magnetization vector between the central domain (where $\vec{m}(\vec{r} = 0) = -\vec{\hat{z}}$) and the domain at infinity (where $\vec{m}(\vec{r}\rightarrow\infty) = \vec{\hat{z}}$) as depicted in Fig.~\ref{Fig6}(a) (see also Sec.~\ref{sec:top-defects}). As localized magnetic solitons, Hopfions present a magnetization field $\vec{m}$ that approaches a constant value $\vec{m}_0 = \vec{m}(\vec{r} \rightarrow \infty)$ at $|\vec{r}|\rightarrow \infty$. This far-field condition allows the compactification of the spin space, ensuring the value of the Hopf number $Q_3$, defined in Eq. \eqref{eq:hopf}, for a Hopfion is an integer number in infinite samples. 
	
	Following more than twenty years of theoretical investigation~\cite{Kosevich90,Faddeev97}, Hopfions have recently experienced a surge of activity focused on their nucleation and stabilization mechanisms, and mathematical representation~\cite{Sutcliffe17,Sutcliffe18,Rybakov22,Guslienko25}, with their experimental observation having been claimed only very recently~\cite{Kent21,Zheng23}. 
 
	The rotational symmetry of Hopfions endows them with a toroidal moment, defined as $\vec{t} = \int dV[\vec{r} \times \vec{B}] / 2 = t_z\vec{\hat{z}}$, where the $z$-axis is the symmetry axis. This toroidal moment simultaneously breaks spatial inversion symmetry and time reversal on a macroscopic scale, and thus impacts spin wave propagation~\cite{Brevis25}, leading to a lens-like effect~\cite{Saji23}, as schematically shown in Fig.~\ref{Fig6}(b). Consequently, the scattering of spin waves by Hopfions excites a wide spectrum of radial, azimuthal, and hybridized modes in confined systems~\cite{Bo21,Raftrey21}. This response makes spin wave scattering a probe for detecting these topological textures in magnetic nanoparticles~\cite{Wu25}.
	
	The intrinsic geometrical properties of nanotoroidal systems makes them excellent candidates for nucleating, stabilizing nanoscale 3D topological solitons, including Hopfions~\cite{Sutcliffe17,Liu18a,Gobel20a,Liu20a,Rybakov22,Liu22d,Khodzhaev22,Saji23}, torons~\cite{Zheng17a,Ackerman17,Tai20,Hu21,Li22d,Saavedra24a,Kuchkin25}, and vortex rings~\cite{Donnelly20b,Liu24d}, as well as their precise manipulation. This prospect is bolstered by preliminary theoretical investigations, which indicate the viability of toroidal systems for hosting complex 3D spin structures~\cite{Carvalho-Santos20,Corona23}, as discussed for the following scenario.
	
	\subparagraph{Reversal under an Oersted Magnetic Field on Bulk Torus} 
	
	The analysis of the reversal process in toroidal nanoparticles with large $R$ under the action of an Oersted magnetic field has been performed by Castillo-Sep\'ulveda~\textit{et~al.}~\cite{Carvalho-Santos20} for a permalloy system. Through micromagnetic simulations using the Mumax3 code~\cite{Vansteenkiste14}, the authors have solved the LLG equation \eqref{eq:LLG} for the  magnetization dynamics during the reversal process. Given the absence of chiral interactions and magnetocrystalline anisotropy, the total magnetic energy density comprises only the exchange, magnetostatic, and Zeeman contributions, defined as $E = E_{\text{ex}} + E_\text{m} + E_\text{Z}$. 
	
	The external Oersted field, $\vec{H}$, is radially non-homogeneous, with its magnitude decreasing as a function of the cylindrical radial distance. The experimental realization of this azimuthal magnetic field requires positioning a straight current-carrying conductor exactly along the torus' symmetry axis, such that it passes through the centre of curvature defined by the major radius. This specific geometric arrangement ensures that the generated field lines are purely azimuthal relative to the toroidal surface.

	The reversal process occurs through one of four distinct mechanisms dependent on the dimensionless geometry variable $\tilde{B}=r_i/r_e$ [See Fig.~\ref{Fig5}(a)].
	
	The first mechanism consists of a skyrmion tube closed in on itself due to the toroidal topology and geometry, characterizing a Hopfion-like state. It is worth noticing that this Hopfion nucleation occurs even without an intrinsic DMI, resulting from the competition between exchange and dipolar interactions, akin to skyrmion nucleation in nanodots without DM interaction~\cite{VidalSilva19}. 
	The magnetization pattern of the torus cross-section reveals a 2D skyrmion, as evidenced by the azimuthal component of the magnetization shown in Fig.~\ref{Fig5}(d). Changes in the tube thickness yield the nucleation of other three modes~\cite{Castillo-Sepulveda21}: the pinned-Hopfion mechanism, where reversal is mediated by a Hopfion with a non-integer topological index pinned at the central hole; a quasi-coherent fast reversal, mediated by a quasi-tangential poloidal vortex where moments rotate around the poloidal direction; and the nucleation of two transversal radial domain walls that propagate along the torus cross-section. These four processes involve the appearance of a cross-vortex component forming intricate 3D emergent spin structures. 
	
	The behaviour of the magnetic energy during the Hopfion-mediated reversal process is depicted in Fig.~\ref{Fig5}(e) for a torus with $r_i=\SI{80}{nm}$ and $\beta = 0.0$; the nucleation of the Hopfion texture increases the exchange energy due to the formation of a swirling spin structure. Concurrently, the magnetostatic energy exhibits a non-monotonic behaviour: it initially increases due to geometry-induced non-local chiral interactions, which give place to surface and volumetric magnetic charges at the onset of nucleation~\cite{Sheka20a}. However, it undergoes an abrupt drop just before the Hopfion is fully formed. This drop to the swirling texture minimizes surface magnetic charges in the region between the black dots. Despite this reduction, the dipolar energy of the Hopfion state remains higher than that of an in-surface magnetization configuration. Therefore, the nucleated Hopfions are transient states that only appear for a large poloidal radius, which provides the necessary space to nucleate the swirling texture without an excessive exchange energy cost. Finally, they observe that the radially decreasing external field displaces the Hopfion center toward the torus border. Its subsequent annihilation is characterized by an abrupt reduction in exchange energy and a concurrent bump in the dipolar term, a signature of a topological phase transition as the system decays back to the vortex state.
	
	Although the torus geometry allows the nucleation of Hopfions in systems without chiral interactions, these structures emerge only as transient states with a lifetime on the order of nanoseconds~\cite{Castillo-Sepulveda21}. The stabilization of Hopfions as stable 3D textures requires a chiral interaction, which promotes twists in the magnetization throughout the nanotorus cross-section. Furthermore, because the shape anisotropy inherent to the torus shares the same symmetry as the Hopfion, additional interfacial anisotropies are not necessary for their nucleation in magnetic nanotori~\cite{Corona23,Saavedra24b}, unlike those required for their planar counterparts~\cite{Tai18}.  
	
	In a nanotorus, a stable Hopfion state can be realized by considering systems with bulk Dzyaloshinskii-Moriya interaction (DMI). In these materials, the lack of structural inversion symmetry gives rise to a chiral energy term. The energy density associated with the bulk DMI can be mathematically expressed as $\varepsilon_{\mathrm{DMI}} = D \, \mathbf{m} \cdot ( \nabla \times \mathbf{m} )$,  where $D$ is the DMI constant. The stability of the resulting Hopfion state is then assessed by analysing the evolution of the static magnetic configuration as a function of an external uniaxial magnetic field, $\vec{H} = H\vec{\hat{z}}$. To nucleate a stable Hopfion and analyze its response to an external magnetic field, one can perform micromagnetic simulations inserting the transient Hopfion obtained in Ref.~\cite{Carvalho-Santos20} and shown in Fig.~\ref{Fig5}(d). By numerically solving the LLG equation using the Mumax3 code, the relaxed states as a function of the external field values ranging from 0 to 250\,mT can be obtained. A selection of the resulting magnetization configurations is depicted in Fig.~\ref{Fig6}(c), where one can see the $z$-component of the magnetization on the nanotorus external surface (ES), its central $xy$-plane cross-section (CP), and its central $z$-axis cross-section (CZ), presented in the upper, middle, and bottom rows, respectively. The stabilization of two distinct magnetization profiles is observed, separated by a threshold magnetic field of $H = 128$\,mT. Below this value, the magnetization on the external surface is predominantly aligned along the $z$-axis. In contrast, above this threshold, the surface magnetization lies primarily in the $xy$-plane. Analysis of the central plane (CP) cross-section reveals that the intrinsic DMI stabilizes chiral structures where the magnetization rotates periodically from the inner to the outer external border of the nanoring. Since the characteristic size of chiral magnetic textures is governed by the interplay of competing interactions in the system, the observed states exhibit magnetization rotations of either $3\pi$ radians (for $H \leq 128$\,mT) or $2\pi$ radians (for $H > 128$\,mT) along the radial direction of the central plane (CP). Analysis of the central $z$-axis cross section (CZ) provides a complete three-dimensional description of these configurations. For magnetic fields $H \leq 128$\,mT, one identifies the nucleation of a 2D skyrmion in the torus cross-section. As this pattern forms a closed loop along the nanoring, it constitutes a Hopfion configuration. Conversely, for $H > 128$\,mT, the nucleation of a linear helical state (LHS) is observed, as in magnetic nanospheres~\cite{Pathak21}.
	
	Although 3D topological configurations are characterized by the Hopf index~\cite{Wilczek83}, the topology of the stabilized states can be determined by the skyrmion number of the 2D structure observed in the central plane (CP). Fig.~\ref{Fig6}(d) reveals an abrupt change in the topological charge at the critical field of $128$\,mT. For $H \leq 128$\,mT, the topological charge is $Q_2 \approx -1$, while for $H > 128$\,mT, a non-integer value is obtained. This non-integer value indicates that, from a topological perspective, this configuration is equivalent to a ferromagnetic state. More specifically, the skyrmion number of the central plane (CP) varies continuously with the applied field, signalling the existence of incomplete topological spin structures. In the range of 60 to 128\,mT, the magnetic field orients the magnetic moments near the torus centre along its direction, thereby progressively reducing the absolute value of the topological charge. At higher fields, the Hopfion state gives way to a helicoidal state. As the field strength increases further, this helicoidal state exhibits a shrinking region where the magnetization remains antiparallel to the field.

	\subsection{Experimental Studies of Toroidal Nanoparticles}
    \label{subsec:experiment-torus}
    
    \subsubsection{Planar Nanorings}
    
\begin{figure}[tb]
        \centering
        \includegraphics[width=0.95\linewidth]{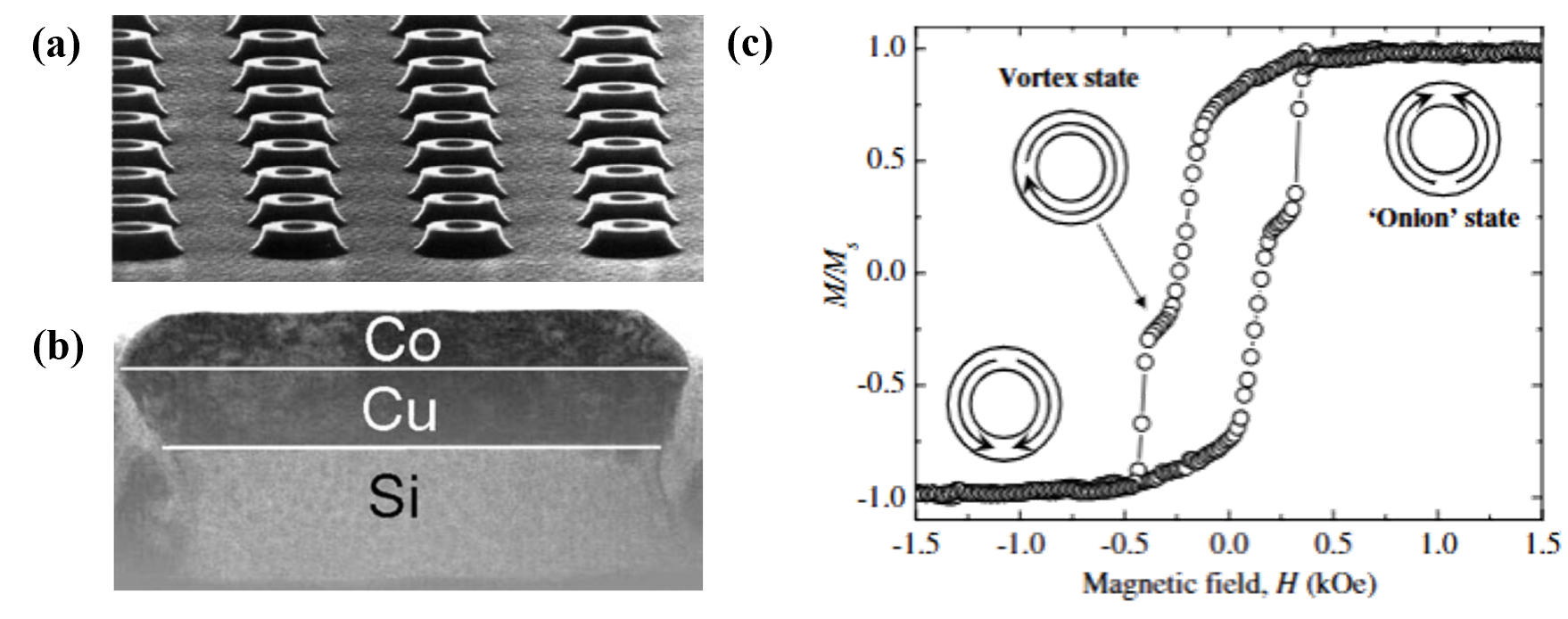} % fig_rings
         \caption{(a,b) SEM images of an array of nanorings and the corresponding material layer. Panels (a,b) reprinted with permission from~\cite{Klaui03}~\textcopyright{}~(2003) by IOP Publishing Ltd. (c) Characteristic hysteresis loop for magnetic nanorings, indicating the switching between vortex and onion states. Panel~(c) reproduced with permission from~\cite{Vaz07}~\textcopyright{}~(2007) by IOP Publishing Ltd.}
        \label{Fig:rings}
    \end{figure}

    Some of the first experimental signatures of spin textures in genus-1 geometries were measured in 2001 from mesoscopic ferromagnetic planar rings~\cite{Rothman01}.
    %, where Rothman~\textit{et~al.} studied micron-scale rings using a combination of magnetometry measurements and micromagnetic simulations . 
    These rings with inner and outer diameters of \SI{1.3}{\micro m} and \SI{1.6}{\micro m}, respectively, were fabricated by a multi-step lithography process and are shown in Figs.\ref{Fig:rings}(a,b). The ring shapes were patterned into a resist layer using electron-beam lithography (EBL), which served as a mask to transfer the ring imprint into a silicon substrate via etching; Cu and Co layers (the latter about \SI{34}{nm} thick) were then deposited onto the textured substrate by molecular-beam-epitaxy (MBE).
    
	To characterize the magnetic behavior of the nanorings, the authors used MOKE magnetometry under varying  in-plane magnetic fields to generated averaged $M(H)$ hysteresis loop. Regardless of the deposition of a magnetic film across the entire substrate, the authors exploited the difference in the polarization of light reflected from the etched trenches and upper surfaces of the nanostructures, arising from interference effects, to suppress the trench contribution~\cite{Nee96}.
    
    Steps in the observed hysteresis loop shown in Fig.~\ref{Fig:rings}(c), can be interpreted in conjunction with micromagnetic simulations, and enabled the identification of the characteristic onion and flux-closure vortex states discussed in Sec.~\ref{sec:insurf-circulation}. Starting from positive saturation, the magnetization decreases only marginally upon reducing the applied field to zero, indicating the net moment associated with the onion state (Fig.~\ref{Fig:rings}c). As the field is reversed, the magnetization exhibits a discrete step to a state with near-zero remanence, marking the transition to the vortex state. With further sweeping of the field, the onion state is restabilized with the magnetization reversed. 
    
    Micromagnetic simulations associated these hysteretic features with transitions between dinstinct states mediated by domain walls~\cite{Klaui04}. Yoo~\textit{et~al.} systematically extended these observations in 2003, constructing a switching-field phase diagram for Co nanorings~\cite{Yoo03}. By varying ring diameter, width, and thickness, and sweeping the field along different axes, they identified regions where qualitatively distinct reversal pathways of domain wall nucleation, motion, and annihilation occur~\cite{Vaz07}. These observations highlight that while topologically non-trivial sample geometries enable characteristic ordering phenomena, the absolute dimensions of the system still dictate the balance of the relevant micromagnetic energy scales and thus the realizable regimes.
    
    A key experimental advance was the transition from ensemble-averaged to single device measurements, capable of detecting localized magnetization configurations in individual nanostructures. In a notable study from 2005, Kl\"aui~\textit{et~al.} employed lithographic processes to electrically contact individual permalloy rings at multiple points and exploited anisotropic magnetoresistance (AMR), where the electrical resistance of a conductor depends on the orientation of the local magnetic field, as a direct measure of the domain wall position~\cite{Klaui05b}.
    The magnetic-field dependent resistance exhibited quantized and reproducible steps, which the authors correlated with the nucleation, de-pinning and propagation of domain walls around the ring.  Furthermore, by introducing controlled asymmetries and local features, such as constrictions, notches, and deliberately asymmetric rings, they demonstrated that the domain wall de-pinning and switching fields could be geometrically engineered, establishing these planar samples as a rich environment for the exploration of magnetic phenomena in a precisely defined and tunable system architecture~\cite{Klaui01,Klaui02}. More recently in 2023, networks of nanorings have been used to enable non-conventional nanomagnetic computation~\cite{Vidamour23}, making use of the interesting dynamic response. 
    
    \begin{figure}[tb]
        \centering
        \includegraphics[width=0.95\linewidth]{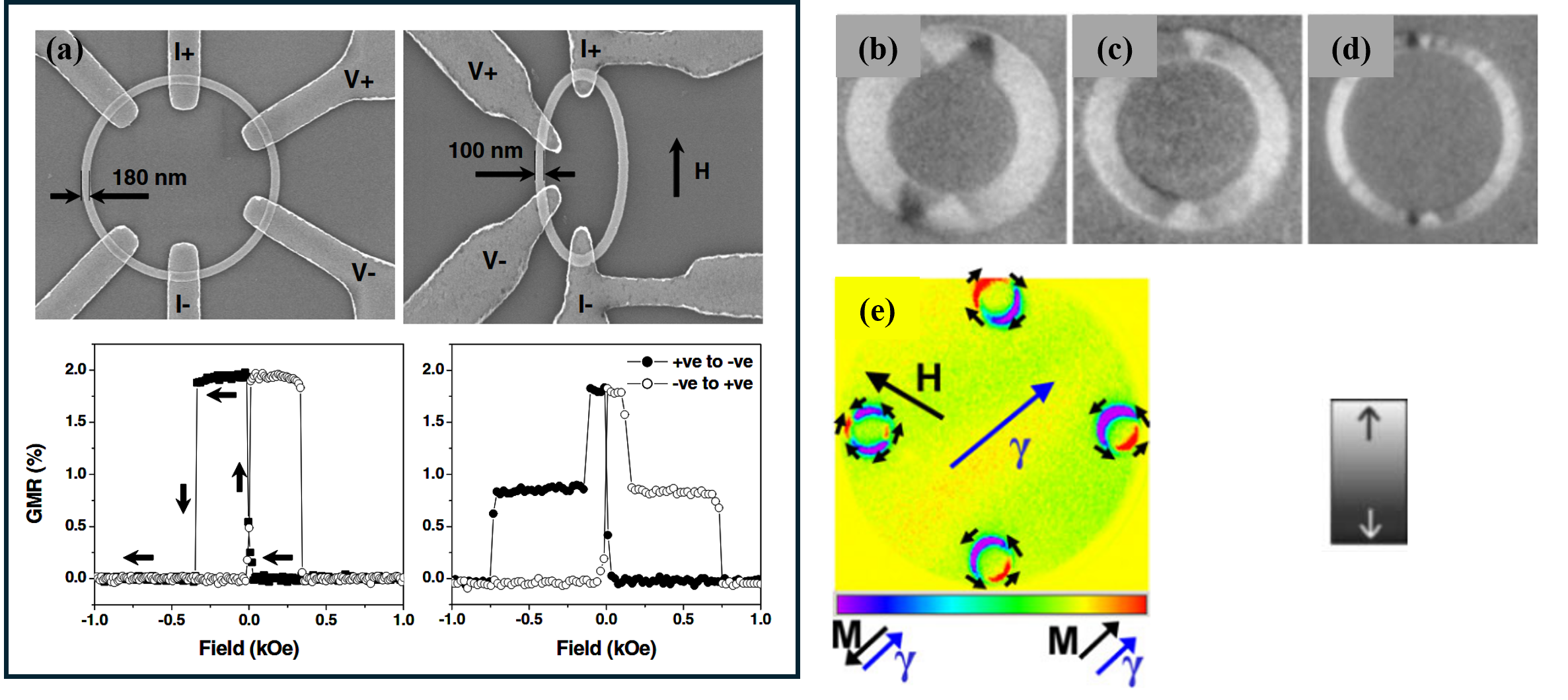} % fig_GMR
        \caption{(a)~SEM images of circular and elliptical pseudo-spin-vales, and their respective magnetoresistive curves. The outer ring diameter and majox axis of the elliptical ring are \SI{5}{\micro m}.  (b--e)~Example images of PEEM images of onion and vortex states in ferromagnetic rings. Panel~(a) reprinted with permission from~\cite{Vaz07}~\textcopyright{}~(2007) by IOP Publishing Ltd. Panels~(b--d) reprinted with permission from~\cite{Laufenberg06}~\textcopyright{}~(2003) by AIP Publishing. Panel~(e) reprinted with permission from~\cite{Klaui03}~\textcopyright{}~(2003) by IOP Publishing Ltd.
        }
        \label{Fig:GMR}
    \end{figure}
    
    A natural progression of this work was the investigation of multilayer ring structures incorporating giant magnetoresistance, such as pseudo-spin-valve systems. In these systems, a soft ferromagnetic layer (NiFe) and a hard ferromagnetic layer (Co) are separated by a non-magnetic Cu spacer to suppress direct interlayer exchange coupling. Nevertheless, strong magnetostatic interactions persist due to the stray fields associated with the onion-state, leading to a pronounced influence of the hard layer on the reversal processes of the soft layer.
    
    Magnetoresistance measurements on circular and elliptical pseudo-spin-valves shown in Fig.~\ref{Fig:GMR}(a) reveal that resistance changes are governed by domain wall motion that presents as a flat baseline with abrupt jumps as the walls nucleate and move through the ring~\cite{Vaz07}. In the case of the elliptical ring, plateaus in the magnetoresistive characteristics corresponds to the switching of the Co layer to a vortex state before reversing to an onion state. 
    
    These transport experiments were soon complemented by direct, real-space imaging of the spin textures themselves.
    Using XMCD-XPEEM nanoscale spatial resolution maps of the magnetization in individual Co and permalloy rings, unambiguously show the onion and flux-closure vortex states [Fig.~\ref{Fig:GMR}(b--e)]~\cite{Klaui03c, Laufenberg06}. Systematic imaging across a range of ring widths and thickness led to an experimental phase diagram delineating the regimes where different types of domain walls are stabilized by the balance of exchange and magnetostatic energies~\cite{Klaui04a}. In wider annular structures, the binary onion–vortex picture breaks down: multi-vortex, cross-tie–like and more complex flux-closure patterns emerge across the ring width~\cite{Klaui03c}. 
    
    Beyond static configurations, planar rings have also served as model systems for exploring magnetization dynamics in a closed, curved track. Current-driven experiments on contacted permalloy rings demonstrated that spin-transfer torque (STT) can displace domain walls around the circumference in a controlled and reproducible fashion, with individual pulses shifting walls by well-defined arc lengths~\cite{Klaui03d,Klaui05}. Such controlled domain-wall motion underpins many of the advanced spintronic device architectures currently being explored. 
    
    Furthermore, time-resolved Kerr microscopy measurements were used to map the eigenmode spectra of the archetypal vortex and onion states, revealing spin wave modes strongly localized at the domain wall configurations. In their analysis~\cite{Neudecker06}, Neudecker~\textit{et~al.} note that modes with an even number of nodes were not observed in experiment or simulations, a selection effect intrinsic to the excitation geometry that only excites modes with an odd number of half-wavelengths in the ring. The authors highlight that despite their absence in MOKE measurements, even-node modes should remain accessible using alternative techniques such as Brillouin light scattering, where the excitation conditions differ. 
    
        \begin{figure}[tb]
        \centering
        \includegraphics[width=0.95\linewidth]{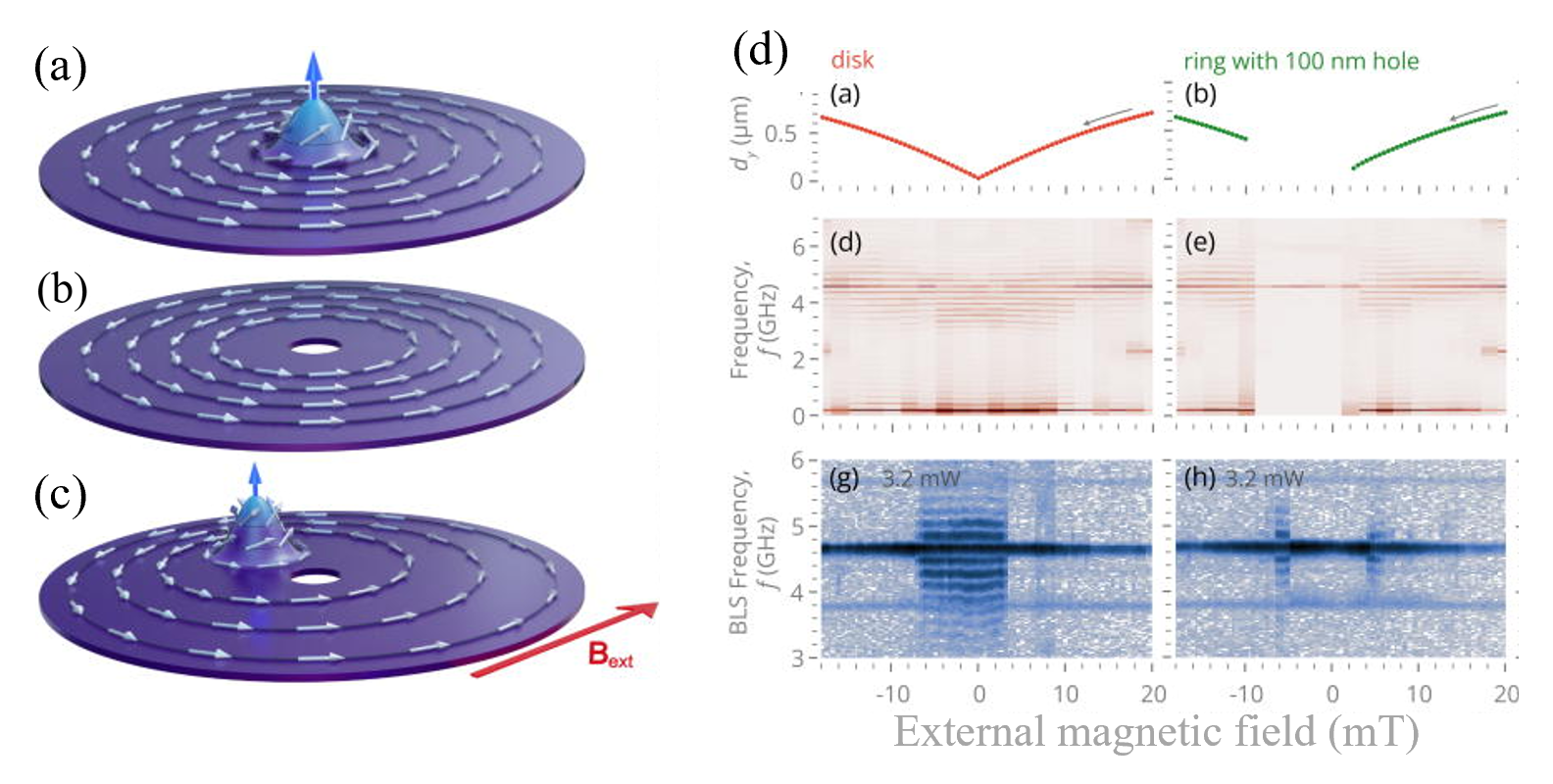} % fig_BLS
        \caption{(a) Schematic illustration of a static magnetic vortex in a nanodisc. (b) Schematic illustration of the texture in a magnetic nanoring, where only the magnetization in the plane of the ring remains. (c) Schematic of nanoring under the application of an external in-plane magnetic field that nucleates a vortex core. (d) The top panels showing the displacement ,d$_{y}$, of the vortex core from the centre of the disc as a function of external in-plane magnetic field for a disc and ring, respectively. The middle and bottom panels show micromagnetic simulations and measured BLS results, respectively, illustrating the formation of the frequency comb in the different geometries when exciting at 4.6 GHz. Reproduced under Creative Commons Attribution 4.0 License (https://creativecommons.org/licenses/by/4.0/) from~\cite{Heins26a}~\textcopyright{}~(2026) by the Authors.
}
        \label{Fig:BLS}
    \end{figure}
    
    Building on this distinction between different excitation pathways, more recent work has highlighted the role of measurement techniques in accessing otherwise hidden spectral features. In particular, Heins~\textit{et~al.}~\cite{Heins26a} employed Brillouin light scattering microscopy to investigate the emergence of magnon frequency combs in both magnetic discs and rings (Fig.~\ref{Fig:BLS}). They show that in disc geometries, the coupling between driven magnon modes and the vortex core gyration leads to the formation of Floquet states and, consequently, evenly spaced frequency combs (Fig.~\ref{Fig:BLS}d). In contrast, for ring structures, where the vortex core is absent due to the central hole, this mechanism is suppressed, and no comb formation is observed, even under strong excition. Crucially, the authors demostrate that reintroducing an offset vortex core in the disc via an external in-plane magnetic field (Fig.~\ref{Fig:BLS}c) restores the nonlinear couple and enables comb generation, thereby providing a means to actively control the accessible magnon spectrum through both non-trivial geometry and excitation conditions.
    
    Taken together, the transport, imaging, and dynamical studies establish planar nanorings and nanodiscs as well-characterized genus-1 systems in which static textures, engineered energy landscapes, and curvature-modified spin dynamics can all be probed within a reliable lithographic framework. Crucially, these systems remain geometrically flat: curvature is confined to the azimuthal direction in the plane of the film, and the ring thickness is small enough that its magnetization is effectively uniform through the film.

    \subsubsection{Towards 3D Tori}
    
\begin{figure}[tb]
        \centering
        \includegraphics[width=0.95\linewidth]{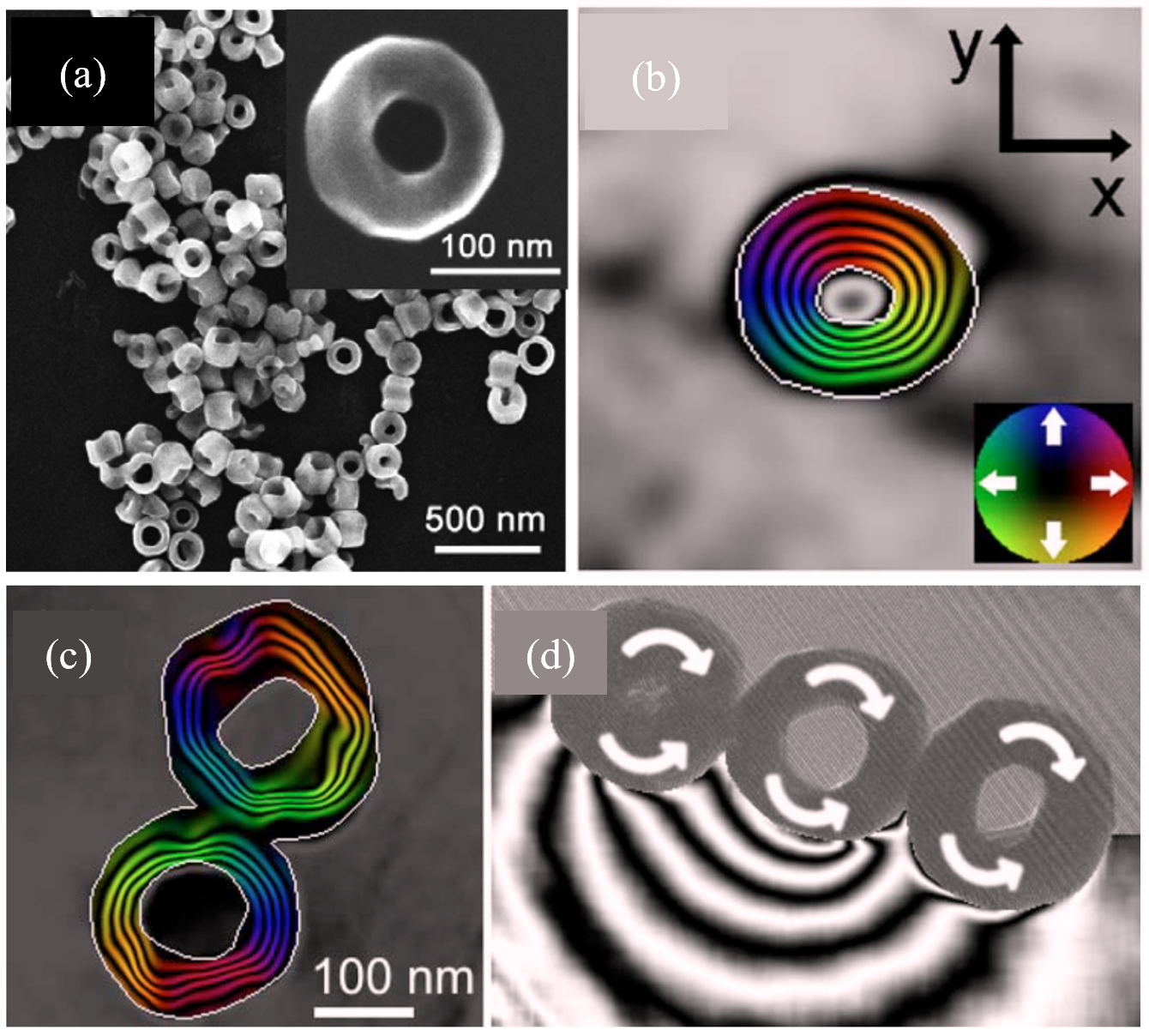} % fig_ebeam_nanorings
	    \caption{(a)~SEM image of an array of toroidal magnetite nanoparticles. (b,c)~Magnetic induction maps of a single and double torus, respectively. The orientation of the magnetization is indicated by the colour wheel. (d)~Stray field of an onion state from a series of three nanoparticles. Panel~(a) is adapted with permission from~\cite{Jia08}.~\textcopyright{}~2008 American Chemical Society. Panels~(b--d) reprinted from~\cite{Eltschka08} with the permission of AIP Publishing.}
        \label{Fig:tem_rings}
    \end{figure}

    A significant advance towards truly 3D tori came from the chemical synthesis of iron-oxide nanorings. These are single crystal, ring-shaped particles with diameters of approximately \SI{100}{nm} to \SI{200}{nm}, produced by a solution-based hydrothermal method~\cite{Jia08}. 
    Using a double anion-assisted approach directly yields a toroidal geometry for hematite ($\alpha$-Fe$_{2}$O$_{3}$) nanoparticles, which can subsequently be reduced to ferromagnetic magnetite (Fe$_3$O$_4$) and further oxidized to maghemite ($\gamma$-Fe$_{2}$O$_{3}$). This process yields ferromagnetic particles like those shown in Fig.~\ref{Fig:tem_rings}(a), while preserving the morphology and topology of the initially synthesized geometry. Compared to lithographically defined planar rings, these structures possess an appreciable three-dimensional cross-section, allowing the magnetization to vary throughout the full volume of the ring.
    
    Building on this synthesis, Eltschka~\textit{et~al.} used off-axis electron holography combined with electron tomographic methods to correlate the 3D ring geometry with its magnetization configuration~\cite{Eltschka08}.
    For isolated, nominally uniform rings, the ground state was the flux-closure vortex, revealed as concentric contours of magnetic induction [Fig.~\ref{Fig:tem_rings}(b)], reflecting the 2D planar analogue. The same synthesis methods can also produce neighboring rings which are in direct contact, enabling exchange-coupled higher genus tori to be realized. In such genus-2 tori, counter-rotating vortex pairs are favored due to the orientation of the magnetization at the shared interface [Fig.~\ref{Fig:tem_rings}(c)]. Linear series of three or more rings were also isolated and, under application of an external magnetic fields with an in-plane component, stabilized extended onion-like textures in which the magnetization follows the perimeter of the rings and produces large dipole-like stray field at the ends of the chain, visible in electron holography [Fig.~\ref{Fig:tem_rings}(d)]. Finally, in rings with a more pronounced 3D cross section-section, where the structure tends towards a tube-like shape, the magnetization escapes the plane, producing multiple coexisting counter-rotating vortices. 
    
    Lewis~\textit{et~al.} provided a 3D treatment of similar toroidal nanoparticles by combining electron tomography with micromagnetic simulations~\cite{Lewis20}. In their study, the authors performed High-Angle Annular Dark-Field Scanning Transmission Electron Microscopy (HAADF-STEM) tomography in order to reconstruct the 3D volumes of a cluster of nanoparticles. They then isolated isosurfaces representing the morphology of individual rings and used the resulting voxel meshes directly as input for micromagnetic simulations. By comparing their measured structures with ideal toroidal geometries they showed that deviation from a perfect torus structure leads to an increasing net magnetization, indicating a lower quality of the vortex state and modifying the energy barrier, which in some cases even facilitate transitions between the vortex and onion states. An approach similar to that of Wolf~\textit{et~al.}, based on full vector-field electron tomography, could, in the future, be applied to these systems to directly reconstruct the three-dimensional magnetization field inside individual toroidal nanoparticles, going beyond shape-constrained simulations to true 3D magnetization imaging~\cite{Wolf22}.

    \subsubsection{Bulk Chiral Magnets with Toroidal Geometries}
    
    Finding spontaneously-forming  iron-oxide nanorings was serendipitous, however the vast majority of crystalline magnetic materials does not naturally adopt toroidal morphologies and must instead be deliberately sculpted into a desired target geometry. As discussed in Sec.~\ref{subsec:subtractive_methods}, recent advances in FIB patterning methodologies have enabled early experimental demonstrations of single-crystal systems shaped into well-defined three-dimensional architectures, including nano-tori milled directly from a chiral helimagnetic system~\cite{Turnbull25}.
    
    In chiral helimagnets, the bulk DMI favours a long-wavelength helical winding of the magnetization as the ground state~\cite{Lebech89} (typical periodicities are a few hundred \unit{nm}). This intrinsic twisting introduces new possibilities when such materials are nanopatterned into toroidal geometries. If a helimagnet is sculpted into a torus, the helical modulation is geometrically constrained to a closed path around the toroidal circumference, particularly if the minor radius is comparable to the helical pitch. Here self-closure in the non-trivial geometry pertains to the helical wavevector rather than that of the magnetization itself. As a result, the system must either adopt an integer number of windings around the ring to retain a smoothly varying texture, or must introduce topological defects that accommodate the mismatch between the imposed geometry and the preferred helical period.
    
    It also has been noted that the combination of confinement and curvature in such chiral nanorings leads to two intertwined surface helices running around the torus, as a result of the additional symmetry breaking at the surface when compared to the helix within the volume. Resonant soft X-ray imaging combined with micromagnetic simulations revealed that the aforementioned defects likely correspond to super-coiling defects in the surface helices, akin to tendril perversions in climbing plants or circular DNA~\cite{Turnbull25}.
    
    Conceptually, these toroidal helimagnets mark a transition from shape-stabilized vortices in soft ferromagnetic tori to more complex exchange-interaction-driven, multi-winding textures in crystalline magnets. In soft feromagnets, curvature and topology mainly provide a low-energy route to flux closure for an otherwise isotropic ferromagnet, yielding a single-winding toroidal vortex or onion stated governed by the balance of exchange and dipolar energies. In bulk helimagnets, the ground state is set by an intrinsic helical winding and the toroidal geometry acts as a topological constraint that forces the helix to close on itself, imprinting non-trivial topology into the magnetization via the sample geometry. 
    
    The progression from planar ferromagnetic rings through iron-oxide nanotori to sculpted toroidal helimagnets illustrates how increasing control over both material and geometry opens access to an expanding range of orderings in magnetic toroidal geometries, from simple vortices to multi-winding configurations.

	\section{Spin Textures in Geometries with non-zero Euler Characteristic}
	\label{sec:finite-euler}
	
	In this Section, we consider the case of magnetic samples that either have no holes, i.e., homeomorphic to a sphere, or have more than one hole. Their distinction from a torus is well pronounced for the case of a magnetically soft sample: while torus can be magnetized in a vortex manner keeping high local uniformity of its magnetic state, sphere or any geometry with more than one hole necessarily contains at least several merons.
	
	\subsection{Magnetization Distributions in Spherical Shells}
	\label{sec:spheres}
	
	\subsubsection{3D Onion and Whirligig States in Magnetically Soft Spheres}
	
	A spherical shell of radius $R$ (here and below, we consider the average of inner and outer radii) and thickness $h$ can be understood as a compact surface of radius $R$ if the magnetization texture along the radial direction is uniform. It has a zero genus $g = 0$, thus the Euler characteristic $\chi_\text{eu} = 2$, imposing the total vorticity of magnetic texture to be equal to two as well. Let us start from the case of a magnetically soft spherical shell, where this can be reflected in the presence of two vortices at the opposite sphere's poles.
	
	A spherical surface can be parameterized as 
	\begin{equation}
		\vec{\varsigma}(\vartheta,\varphi) = R \{ \sin\vartheta\cos\varphi, \sin\vartheta\sin\varphi, \cos\vartheta \},
	\end{equation}
	where $\vartheta \equiv \eta_1$ and $\varphi \equiv \eta_2$. In the magnetic energy, we account for the exchange~\eqref{Gaididei-Exchange} and magnetostatics~\eqref{eq:magnetostatics}. A high symmetry of the system allows limiting consideration to the magnetic textures of axial symmetry~\cite{Kravchuk12a,Sloika17}, i.e.
	\begin{equation}
		\vec{m} = \sin\Theta\cos\Phi \vec{e}_\vartheta + \sin\theta\sin\Phi \vec{e}_\varphi + \cos\Theta\vec{n}
	\end{equation}
	with $\Theta = \Theta(\vartheta)$ and $\Phi = \pi - \Phi_0$.
	
	\begin{figure}[tb]
		\centering
		\includegraphics[width=\linewidth]{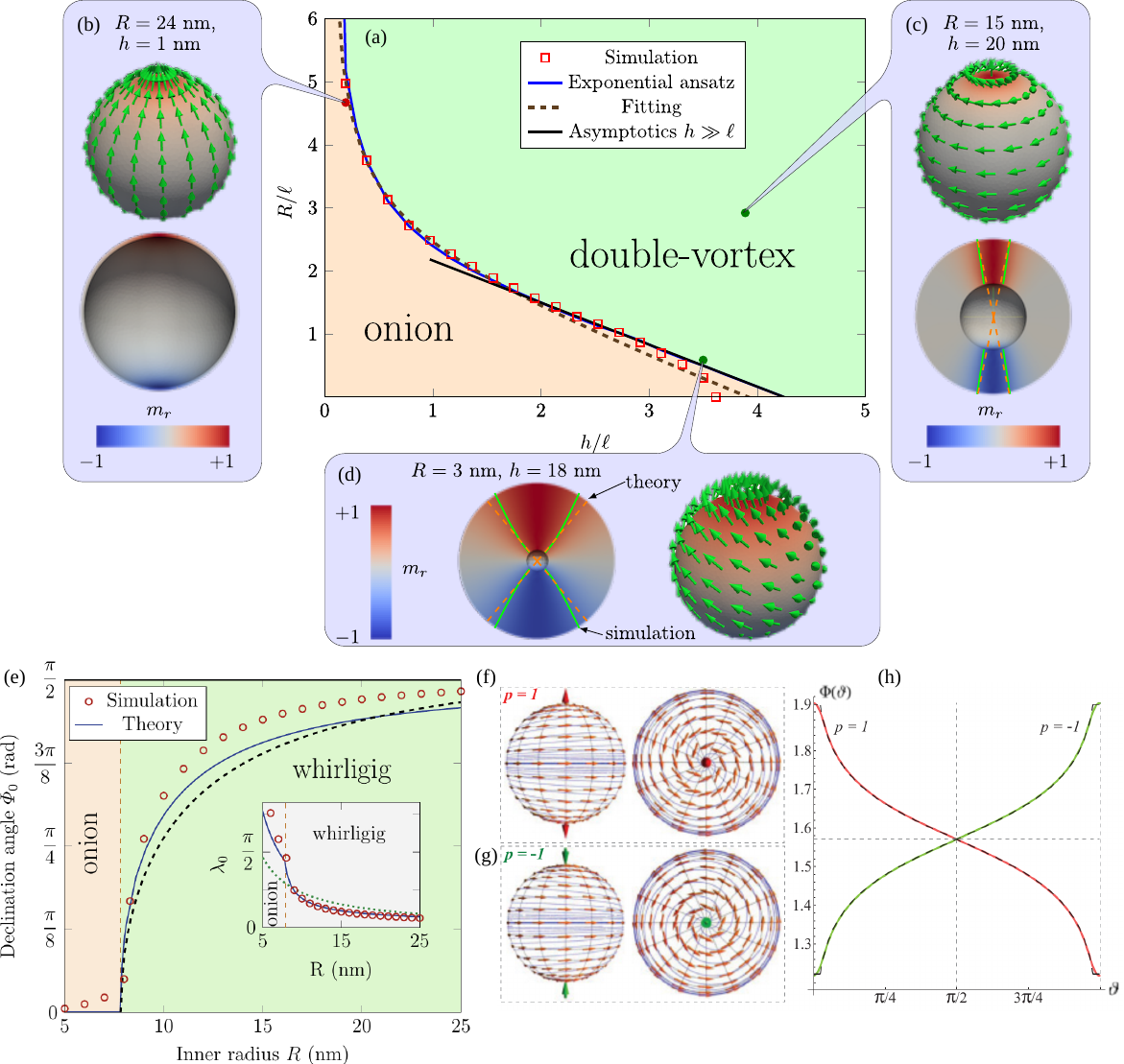} % sphere-soft
		\caption{\textbf{Magnetic states of a magnetically soft spherical shell.} (a)~Diagram of equilibrium states distinguishing 3D onion and whiligig (double-vortex) states depending on sphere thickness $h$ and radius $R$. Exemplary (b)~3D onion and (c,d)~whirligig states. (e)~Change $\Phi_0$ in the magnetic azimuthal angle depending on the sphere radius $R$. The inset shows the equilibrium value of the parameter~$\lambda$ defined in Eq.~\eqref{eq:sphere-soft-ansatz}~\cite{Sloika17}. (f,g)~Polarity-circulation coupling in whirligig state with (f)~outward and (g)~inward vortex polarities. (h)~The spatial distribution of the magnetic azimuthal angle $\Phi$ along the sphere's meridian~\cite{Kravchuk12a}. Panels~(a--e) reprinted from~\cite{Sloika17},~\textcopyright{}~(2017), with permission from Elsevier. Panels~(f--h) reprinted with permission from~\cite{Kravchuk12a}~\textcopyright{}~(2012) by the American Physical Society.
		}
		\label{fig:sphere-soft}
	\end{figure}
	
	If a sphere is small enough, the exchange energy is dominant, leading to the so-called 3D onion state, which is approaching the uniform magnetization distribution tilted towards the surface of the shell [Fig.~\ref{fig:sphere-soft}(a,b)]. Around the sphere poles, it can be understood as two nearly radial vortices of the same polarity with $\Phi_0 \approx 0$. The numerical analysis shows that for a fixed $h$ and larger $R$, there is a phase transition to the state resembling two vortices with a finite $\Phi_0$~\cite{Sloika17} [Fig.~\ref{fig:sphere-soft}(c--e)]. The out-of-surface magnetization can be described with the Ansatz~\cite{Sloika17}
	\begin{equation}\label{eq:sphere-soft-ansatz}
		\begin{aligned}
			\Theta & = \begin{cases}
				\pi/2 - f(\vartheta,\lambda), & \vartheta \in [0,\pi/2),\\
				\pi/2, & \vartheta = \pi/2,\\
				\pi/2 + f(\vartheta,\lambda), & \vartheta \in (\pi/2,\pi],\\
			\end{cases}\\
			f(\vartheta,\lambda) & = \dfrac{\pi}{2} \dfrac{e^{-\vartheta/\lambda} - e^{(\vartheta-\pi)/\lambda}}{1-e^{-\pi/\lambda}},
		\end{aligned}
	\end{equation} 
	where the parameter $\lambda$ is responsible for the vortex core size. Then, the energy of the magnetic state can be presented as~\cite{Sloika17}
	\begin{equation}\label{eq:energy-ms-sphere}
		E = E_\text{on} + \mathcal{E}_1 \sin^2 \dfrac{\Phi_0}{2} + \mathcal{E}_2 \sin^4 \dfrac{\Phi_0}{2}.
	\end{equation}
	Here, $E_\text{on}$ is the energy of a pure 3D onion state with $\Phi_0 = 0$ and $\mathcal{E}_{1,2}$ are certain coefficients. All these three quantities depend on the ratio $h/R$ and $\lambda$. In particular, $\mathcal{E}_2 > 0$ and does not depend on material parameters originating from the volume magnetostatic charges. Together, $\lambda$ and $\Phi_0$ are the variational parameters that can be determined for the given geometry. 
	
	Depending on the value of $\mathcal{E}_1$, there are several possibilities to get the minimal energy $E$ in~\eqref{eq:energy-ms-sphere}. For $\mathcal{E}_1 > 0$ and $\mathcal{E}_1 < -2\mathcal{E}_2$ the ground state of the model corresponds to the 3D onion state with the magnetization strictly following the sphere's meridians. Otherwise, a finite $\Phi_0 = \pm 2\arcsin \sqrt{-\mathcal{E}_1/(2\mathcal{E}_2})$ appears. This analytical model describes the second-order phase transition between the 3D onion and the so-called whirligig state shown in Fig.~\ref{fig:sphere-soft}(d) that is the pair of vortices on a sphere with $\Phi$ substantially different from $\pi/2$ as it is common for the planar case [Fig.~\ref{fig:sphere-soft}(e)]. Considering the magnetic state in micromagnetic simulations, the transition between these states is slightly smoothed because of the finite $\Phi_0 \ll 1$ even in the 3D onion state~\cite{Sloika17}.
	
	The transition between 3D onion and whirligig states can be analyzed considering the case of small $\Phi_0$, at which the energy~\eqref{eq:energy-ms-sphere} is~\cite{Sloika17}
	\begin{equation}
		E = E_\text{on} + \mathcal{E}_1 \dfrac{\Phi_0^2}{4} + \left( \mathcal{E}_2 - \dfrac{\mathcal{E}_1}{3} \right) \dfrac{\Phi_0^4}{16} + \mathcal{O}\left(\Phi_0^6\right).
	\end{equation}
	Note that at the transition point $\mathcal{E}_1 = 0$ and the energy is smooth with respect to $\lambda$. Analytically, one can access two limiting cases that provide the critical radius of the sphere that separates two ground states~\cite{Sloika17}:
	\begin{equation}
		\dfrac{R_c}{\ell_{\text{ex}}} = \begin{cases}
			2\sqrt{2} - \dfrac{2}{3} \dfrac{h}{\ell_{\text{ex}}}, & R \ll \ell_{\text{ex}},\\
			1.59 \left( \dfrac{\ell_{\text{ex}}}{h} \right)^{2/3}, & R \gg \ell_{\text{ex}},
		\end{cases}
	\end{equation}
	where the last equality holds approximately.
	
	If the sphere is large and thin enough, the magnetostatic energy can be replaced by the hard-normal anisotropy~\eqref{eq:shape-ani-shell}. This case allows an analytical analysis of the generalized whirligig state [Fig.~\ref{fig:sphere-soft}(c,d)] allowing vortices at the poles to have arbitrary polarities with the focus on the spatial distribution of the azimuthal angle $\Phi = \Phi(\vartheta)$. The out-of-surface magnetization can be approximated with the Ansatz~\cite{Kravchuk12a}
	\begin{equation}
		\cos\Theta = p_1 \exp \left( - \dfrac{\vartheta^2}{2\vartheta_c^2} \right) + p_2 \exp \left( - \dfrac{(\pi-\vartheta)^2}{2\vartheta_c^2} \right),
	\end{equation}
	where $p_{1,2} = \pm 1$ are the polarities of the vortices at the north and the south poles, and $\vartheta_c$ characterizes the polar size of the vortex core. If $p_1 = -p_2$, the magnetic energy reaches the local minimum at $\Phi = 0$ or $\Phi = \pi$ realizing the 3D onion state discussed above and shown in Fig.~\ref{fig:sphere-soft}(b). For the same polarities $p_1 = p_2 = p$, i.e. both vortex cores directed out of sphere, the spatial distribution of the magnetic azimuthal angle is~\cite{Kravchuk12a}
	\begin{equation}
		\Phi(\vartheta) \approx \pm \dfrac{\pi}{2} \left( 1 - p\vartheta_c x \ln \tan \dfrac{\vartheta}{2} \right),
	\end{equation}
	where $x$ is the solution of $x = \cos (0.5 x \vartheta_c u_c \pi)$ and $u_c \approx \ln \cot (\vartheta_c/2)$. This is the illustration of how the polarity-circulation coupling in the meronic distribution on the curved surfaces arises in the presence of the out-of-surface magnetic component, see Fig.~\ref{fig:sphere-soft}(f--h). Later this effect has been related to the consequence of the geometry-induced DMI~\cite{Sloika22} that also alters the vortex dynamics on spherical caps~\cite{Sloika25}. 
	
	\subsubsection{Skyrmionic States on Spheres with Radial Easy Axis}
	
	Spherical shells with the easy axis of magnetization along the radial direction $\vec{n}$ are known to arise at the nanoscale due to specific fabrication procedures like the Kirkendall effect~\cite{Sayed18a}. In analytical analysis, they possess a high symmetry, allowing a radially symmetric ``hedgehog'' state and various axially symmetric states. Taking into account that for small enough $h$, the magnetostatics~\eqref{eq:magnetostatics} scales the out-of-surface anisotropy only, in the following, we consider the model of a ferromagnetic spherical shell taking into account the exchange~\eqref{Gaididei-Exchange}, anisotropy energy $E_\text{an} = -K \int m_n^2 d\vec{r}$ with $K > 0$ being the anisotropy coefficient, and the DMI originating from the shell's material of the interfacial symmetry
	\begin{equation}\label{eq:DMI-at-curved}
		E_\textsc{dm} = D \int \left[ m_n \nabla \cdot \vec{m} - (\vec{m}\cdot \nabla) m_n \right] d\vec{r},
	\end{equation}
	where $D$ is the DMI constant. 
	
	The radially symmetric magnetic textures with $\Phi = 0$ or $\Phi = \pi$ are the solutions of the equation~\cite{Kravchuk16}
	\begin{equation}\label{eq:spherical-radial}
		\begin{aligned}
			\partial_\vartheta^2\Theta + \cot \vartheta \partial_\vartheta \Theta - \sin\Theta\cos\Theta \left( \dfrac{\cos2\vartheta}{\sin^2\vartheta} \right. & \left.+ \dfrac{R^2}{\ell^2} - \dfrac{4D}{D_c} \right) \\ 
			& + 2\left(1 + \dfrac{D}{D_c}\right)\cot\vartheta\sin^2\Theta = 0.
		\end{aligned}
	\end{equation}
	Here, the magnetic length $\ell = \sqrt{A/K}$, the critical DMI is $D_c = 2A/R$ and the explicit value of $\Phi$ is accounted for the functional dependency of the polar magnetic angle $\Theta \in \mathbb{R}$. In particular, the boundary conditions on $\Theta$ can be of two kinds: either $\Theta(0) = 0$ with $\Theta(\pi) = (n - 1)\pi$, or $\Theta(0) = \pi$ with $\Theta(\pi) = n\pi$. The number $n \in \mathbb{Z}$ determines the helicity of the texture. 
	
\begin{figure}[tb]
		\centering
		\includegraphics[width=\linewidth]{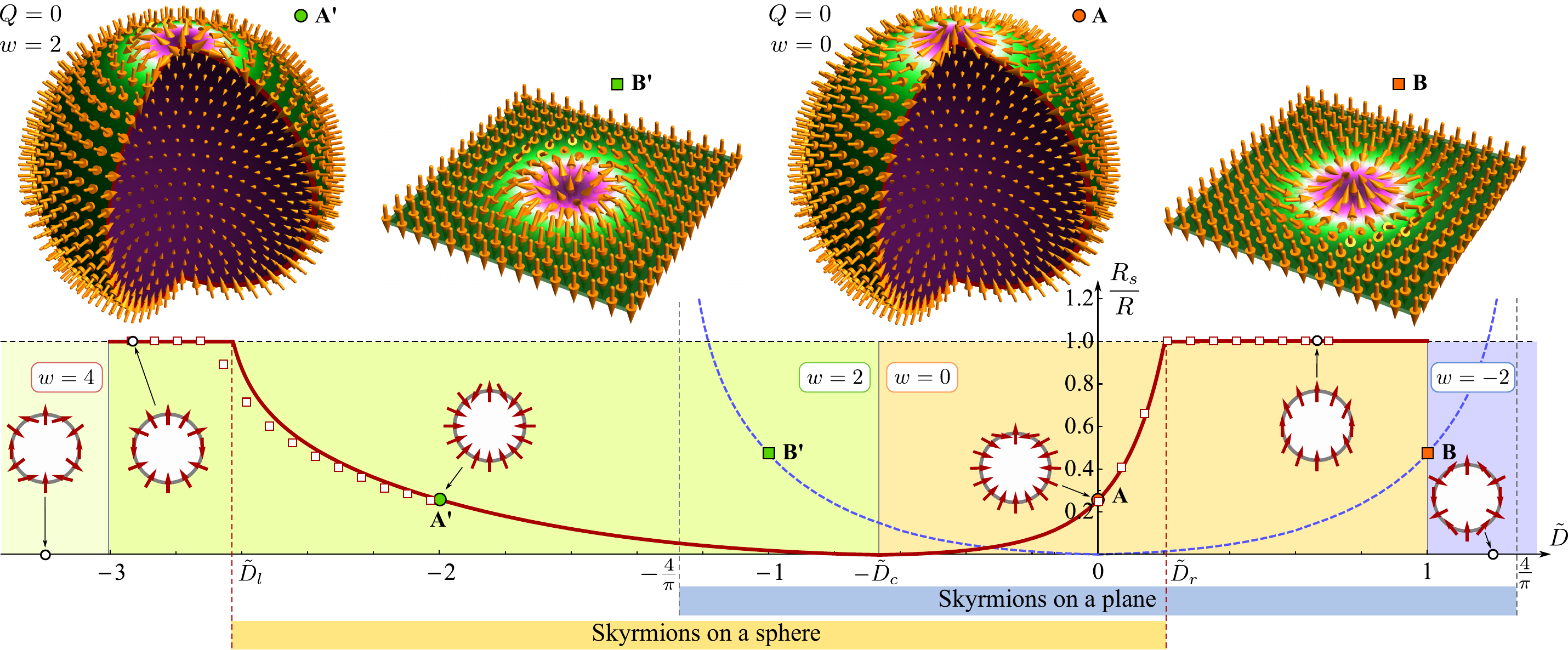} % sphere-ea
		\caption{\textbf{Skyrmionic states in spherical shells with the radial easy axis of anisotropy that are topologically trivial due to the sphere's topology.} (A,A')~Skyrmions with the different helicity numbers $w$ [$n$ in the text  after Eq.~\eqref{eq:spherical-radial}] and (B,B') their planar counterparts. The bottom diagram shows skyrmion radius as a function of DMI strength $\tilde{D} = D/\sqrt{AK}$ for the skyrmions on a plane (blue dashed line) and on a sphere where the geometry-induced DMI influences the magnetic state. Insets shows specific solutions. Reprinted with permission from~\cite{Kravchuk16}~\textcopyright{}~(2016) by the American Physical Society.
		}
		\label{fig:sphere-ea}
	\end{figure}

	Let us consider the case of intrinsically achiral material with $D = 0$. Similarly to the case of the magnetically soft shell, if its radius is small enough ($R \le R_c \approx 1.8\ell$), the dominating energy is the exchange. It forms a quasi-uniform magnetization distribution in the real space, which corresponds to a skyrmion of radius $R_\text{sc} = R$ in the local reference frame. Unlike the 3D onion state [Fig.~\ref{fig:sphere-soft}(b) and inset for $\tilde{D} \lesssim 1$ in Fig.~\ref{fig:sphere-ea}], the magnetization outside the poles and equator is tilted towards the normal $\vec{n}$ from $\vec{\hat{z}}$ direction. For larger $R$ the anisotropy energy becomes dominant, and the ground state becomes a hedgehog~\cite{Kravchuk16}. 
	
	A skyrmion can exist on the background of the hedgehog state being stabilized purely by the curvature-induced DMI, see inset A in Fig.~\ref{fig:sphere-ea}. For $R > R_c $, the equilibrium skyrmion radius $R_\text{sc}$ rapidly decays with larger $R$. The explicit accounting of magnetostatics in the numerical model changes the slope of the function $R_\text{sc}(R)$. Collapse of the curvature-stabilized skyrmions for thin shells has been reported at $R \approx 3.5\ell$~\cite{Kravchuk16}.
	
	Accounting for the intrinsic DMI~\eqref{eq:DMI-at-curved} changes the diagram of equilibrium states, adding higher-helicity states~\cite{Kravchuk16,Guo24}. A positive DMI constant $D$ suppresses the area of the stable hedgehog state. For larger $D$, it is possible to stabilize textures with the helicity number $n < 0$ as the lowest-energy ones in the class of the axially symmetric textures. At negative $D$, it is possible to compensate the curvature-induced DMI. This interplay between the intrinsic and curvature-induced DMIs increases the region of stability of the hedgehog state and introduces the lowest-energy states with positive helicity numbers. These states are summarized in Fig.~\ref{fig:sphere-ea}.

	\subsection{Wireframes}
	\label{sec:wireframes}
	
	\begin{figure}[tb]
	\centering
	\includegraphics[width=\linewidth]{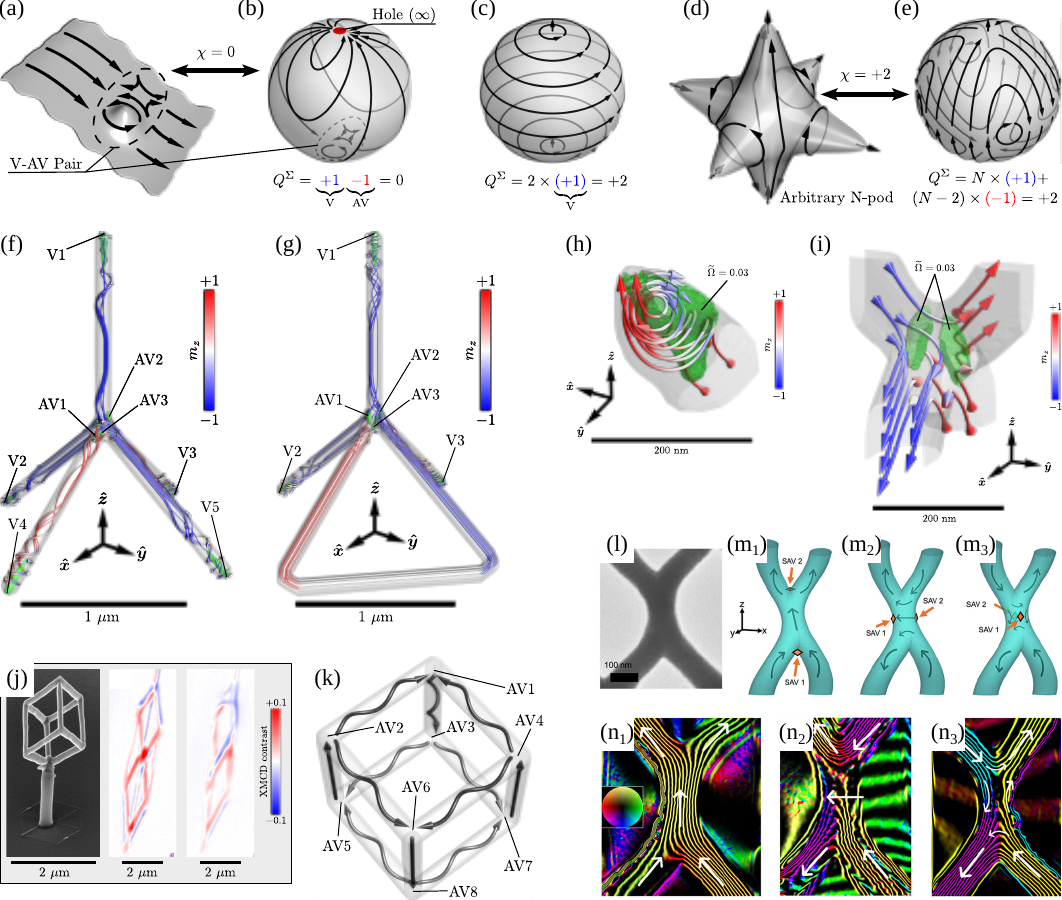} % wireframes
	\caption{\textbf{Magnetic samples with topologies homeomorphic to a sphere and $N$-tori.} (a)~An extended ferromagnetic shell can locally contain vortices~(V) and antivortices~(AV) keeping zero total vorticity in the ground state, and (b)~is homeomorphic to a sphere with a hole. (c)~Complete sphere keeps the total vorticity equal to 2 and is homeomorphic to (d,e) more complex shapes without holes that can stabilize more vortices and antivortices. (f)~Micromagnetic models of pentapods homeomorphic either to a sphere (vortices~V1\ldots V5 and antivortices AV1\ldots AV3) (g)~or a torus (vortices V1\ldots V3 and antivortices AV1\ldots AV3). (h)~Surface vortex and (i)~antivortex states in simulations of the experimental geometry. Isosurfaces of the topological charge flux density $\tilde{\Omega}$ [see Eq.~\eqref{eq:gyro-density}] are indicated. (j)~Scanning electron microscopy (SEM) of a ferromagnetic cube-shaped wireframe fabricated by FEBID and its XMCD-PEEM shadow contrast providing magnetic contrast. (k)~Schematics of positions of antivortices for the sample shown in panel~(j). (l)~Transmission electron microscopy (TEM) of an individual wire junction. (m$_{1\ldots3}$)~Schematics of ``two in/two out'' magnetic states and (n$_{1\ldots3}$) respective magnetic induction contour images obtained by electron holography. Panels~(a--k) reprinted under Creative Commons Attribution 4.0 License (https://creativecommons.org/licenses/by/4.0/) from~\cite{Volkov24a}~\textcopyright{}~(2024) by the Authors. Panels~(l--n) reprinted under Creative Commons Attribution 4.0 License (https://creativecommons.org/licenses/by/4.0/) from~\cite{Fullerton25b}~\textcopyright{}~(2025) by the Authors.
	}
	\label{fig:wireframes}
	\end{figure}
	
	Shapes can have a complex geometry but still be  homeomorphic to a sphere. In this case, its topological features in the magnetization field will be equivalent to that of a sphere, but pinned by the local geometric features, see Fig.~\ref{fig:wireframes}(a--e) and Fig.~\ref{fig:PEEM}.  
	
	To illustrated this point, we start from the discussion of multipod geometries of interconnected nanowire segments which can be fabricated using FEBID~\cite{Fernandez-Pacheco20,Hoflich23}. Let us consider a tetrapod consisting of four \SI{1}{\micro m}  long and \SI{100}{nm} wide segments joined in two V-shaped parts~\cite{Volkov24a}, as shown in Fig.~\ref{fig:PEEM}(a). 	A common feature of magnetic FEBID nanowires is the dominating shape anisotropy leading to the mainly in-surface magnetization distribution with multiple merons. 
	
	The total vorticity of the magnetic texture on the nanostructure surface should be equal to two, $\chi_\text{eu}=+2$ (see Sec.~\ref{sec:topo-theorems}). The shape anisotropy forces each of the nanowire segments to be uniformly magnetized along its long direction, which in practice is only strictly enforced in the middle part of the nanowire. At the free ends of the tetrapod, magnetostatic interactions lead to the formation of four vortices. In comparison with the vortices in nanodots, these vortices represent the specific surface state because their Bloch line naturally transforms into the uniform magnetization in the segment interior, similarly to the flower states in nanowires~\cite{Ternovsky01,Hertel04a}, as shown in Fig.~\ref{fig:wireframes}(h).	The topological requirement for the net vorticity to be equal two forces the formation of an antivortex, the energetically favourable position of which is in the central joint of all four segments of the tetrapod [Fig.~\ref{fig:wireframes}(i)]. Naturally, a single bulk antivortex forms two antivortex textures at the opposite sides of the central region of the tetrapod, completing the total vorticity of all merons at the tetrapod surface. 

	Properties of the antivortex in the tetrapod are determined by the mutual orientation and size of its segments. If all the segments are equivalent in diameter, independently of the relative angle between the top and bottom V-shaped parts of the tetrapod, the antivortex forms a well-defined Bloch line through the interior of the geometry. In contrast, a broken symmetry of wire segments between the V-shaped parts allows the formation of surface antivortex states if their planes are placed under 90$^\circ$ [Fig.~\ref{fig:wireframes}(i)]. The surface antivortices keep a well-defined Bloch line only in the vicinity of the surface of the sample [green pockets in Fig.~\ref{fig:wireframes}(i)], while in the interior the winding number of magnetization around any closed loop vanishes~\cite{Volkov24a}. 
	
	The amount of surface equilibrium vortices on a multipod surface can be easily defined by the amount of line segments. For example, a large enough pentapod necessarily hosts five vortices, see Fig.~\ref{fig:wireframes}(f). The topological requirements impose the presence of three antivortices at the surface: in equilibrium, all of them are located in the joint part of the pentapod. In the case of an odd number of vortices, at least one antivortex is purely a surface one with the Bloch line localized in the vicinity of the sample's surface. 
	
	The idea of X-shaped tetrapod can be easily extended to more complex geometries, e.g. a pentapod~[Fig.~\ref{fig:wireframes}(f)]. Then, one of the antivortices located in its central case is necessarily the surface state under any geometric conditions.  The pentapod geometry can be straightforwardly developed into a toroidal one by joining two segments together, as shown in Fig.~\ref{fig:wireframes}(g). This construction keeps the three vortices and three antivortices in the ground magnetic state, while the total vorticity of the surface magnetization is zero. Joining other segments of the multipod together allows us to construct a topology homeomorphic to a torus with multiple holes [Fig.~\ref{fig:wireframes}(j,k)]. The resulting total vorticity of magnetic texture becomes negative alongside the Euler characteristic, making antivortices dominant over vortices, to the point where antivortices are the only merons present in the sample~\cite{Volkov24a}, see Fig.~\ref{fig:wireframes}(j,k). 
	
	High-genus nanostructures designed, e.g., as a series of interconnected helices with spatially variable wire diameter therefore enable to stabilize a broad range of magnetic textures, including domain walls pinned in different geometric sites, see Fig.~\ref{fig:wireframes}(l,m,n). The specific position of the given magnetic texture is associated with the diameter of the wires at the interconnection, and the direction of the applied magnetic field~\cite{Fullerton25b}.
	
	\paragraph{Buckyballs}
	
\begin{figure}[t!]
		\centering
		\includegraphics[width=\linewidth]{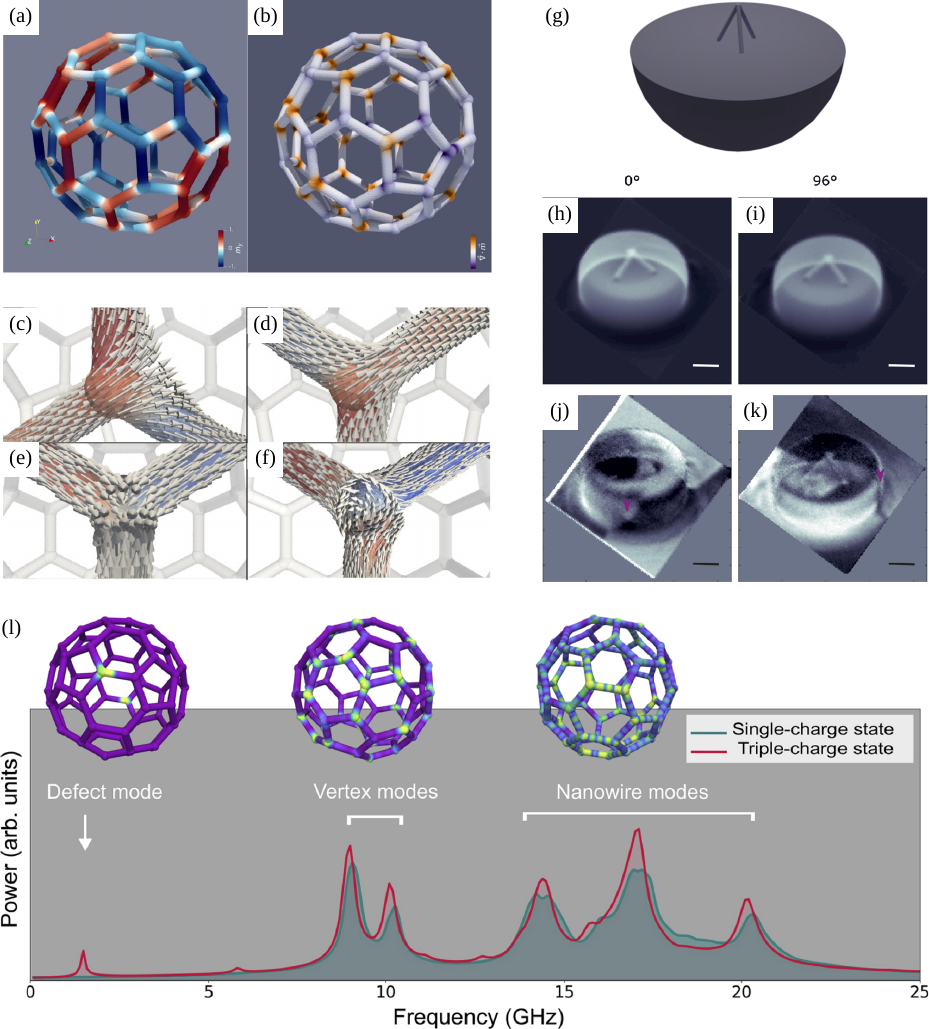} % buckyballs
		\caption{\textbf{Magnetic states in buckyballs.} 
			(a)~A buckyball with \SI{100}{nm} long segments with the $m_y$ component of magnetization highlighted. 
			(b)~Colormap over the buckyball shown in panel~(a) showing the local value of $\nabla\cdot\vec{m}$. 
			(c)~Examples of junctions with the states ``one in, two out'', 
			(d)~``two in, one out'', and 
			(e,f)~two possible ``three in'' configurations~\cite{Cheenikundil21}. 
			(g)~Schematics of a single tripod that mimics part of a buckyball, on a hemispherical support prepared with two-photon lithography. 
			(h,i)~Projections of the electronic contrast of the experimental geometry for different angles around the laminography rotation axis and 
			(j,k)~respective XMCD contrast indicating magnetization projection along the vertical projections. The purple arrow indicates a domain wall in the pillar side~\cite{Pip22}. The scale bar corresponds to \SI{1}{\micro m}. 
			(l)~Power spectrum of spin modes in a buckyball made of nanowires for the cases of the ice-rule-obeying configurations (single-charge state) and with violation of them (triple-charge state)~\cite{Cheenikundil22}. 
			Panels~(a--f) reprinted from~\cite{Cheenikundil21}, with the permission of AIP Publishing. Panels~(g--k) reprinted under Creative Commons Attribution 4.0 License (https://creativecommons.org/licenses/by/4.0/) from~\cite{Pip22}~\textcopyright{}~(2022) by the Authors. Panel~(l) reprinted under Creative Commons Attribution 4.0 License (https://creativecommons.org/licenses/by/4.0/) from~\cite{Cheenikundil22}~\textcopyright{}~(2022) by the Authors.
		}
		\label{fig:buckyballs}
	\end{figure}

	A special case of high symmetry is represented by a buckyball C$_{60}$ structure [Fig.~\ref{fig:buckyballs}(a,b)]. A buckyball consists of 60 vertices and 90 straight interconnections, giving the Euler characteristic $\chi_\text{eu} = -60$ manifesting the dominating number of antivortices. 
	
	Depending on the length and diameter of the interconnections, they may support transversal or Bloch-point domain walls as metastable states, while in general the wire segments tend towards a uniform magnetization, as has been reported for the multipodes. 
	
	The low-energy magnetic states of a buckyball can be understood in terms of spin ice systems used to analyse effects of geometric frustration on the magnetic response~\cite{Cheenikundil21,Pip22}. At the vertices represented by Y-shaped junctions, common spin state states correspond to ``one in, two out'' or ``two in, one out'' configurations shown in Fig.~\ref{fig:buckyballs}(c,d); these ``single-charged'' local states minimize (but not fully remove) the local magnetic frustration inherent to the system. In this case, a topological defect in the magnetization field forms at the side surface of the junction, and almost lying in the plane of the junction itself. 
	%Such states violate common spin-ice rules of charge formation. 
	Occasionally, high-energy ``three in'' or ``three out'' vortices can be observed~\cite{Cheenikundil21}, see Fig.~\ref{fig:buckyballs}(e,f), but these are absent in the globally demagnetized state formed by a combination of 30 ``two in, one out'' and 30 ``one in, two two out'' junction states. 
	
	The emergent local single or triple magnetostatic charges  at the junctions affect the magnon dynamics profoundly. Namely, the absorption power spectrum demonstrates clearly distinguishable peaks corresponding to modes at triple in/out junctions, as well as modes associated with ice-rule-obeying structures and interconnections~\cite{Cheenikundil22} [Fig.~\ref{fig:buckyballs}(l)]. Experiments with 3D spin-ice building blocks demonstrate an additional possibility to tailor their state by accounting for the substrate covered by a magnetic material~\cite{Pip22}, see Fig.~\ref{fig:buckyballs}(g-k) demonstrating a complex magnetic texture shared between tripod and substrate.
	
	\paragraph{Topological Stray Fields}
	
	More complex-shaped interconnectors and application of external fields make it possible to tun magnetic textures and even design of the surrounding stray fields~\cite{Donnelly22a,Volkov24a,Fullerton25}. For example, the complex topology of stray fields around non-touching interwoven helix-shaped nanowires can lead to coupled domain-wall pairs as the geometric helical twist of the wires leads to a cross-tie-like topology of stray fields~\cite{Donnelly22a,MoralesFernandez26}. 
	
	Complex stray fields also surround merons in multipods~\cite{Volkov24a}. Compositions of interconnected helices allow shaping of $B$-field lines within holes in the magnetic material. Fine-tuning the spatial distribution, including that of antivortex and quadrupolar stray field lines, is possible by the appropriate initialization of domain walls and (anti)vortices at the geometric interconnections~\cite{Fullerton25}.
	
	\clearpage
	
	\subsection{Nanoscale Magnetic Gyroids}
	\label{sec:gyroids}

	\begin{figure}[t!]
		\centering
		\includegraphics{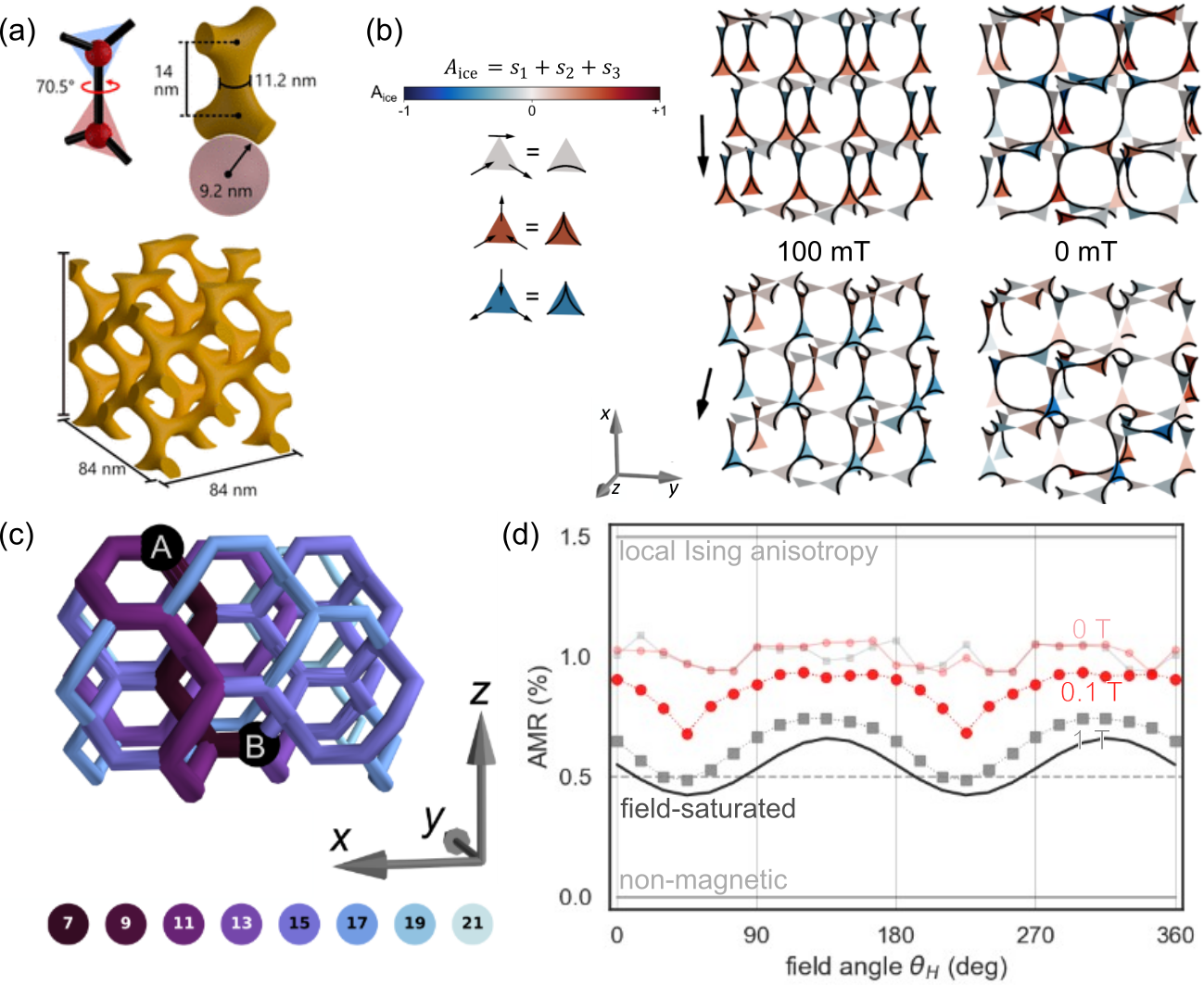} % gyroids-v1
		\caption{\textbf{Properties of extended gyroids lattices.}
			(a)~Base elements of corner-sharing triangles with a chiral twist (top) form an extended gyroid lattice (bottom).
			Reprinted (adapted) with permission from~\cite{Llandro20}.~\textcopyright{}~(2020) American Chemical Society.
			(b)~Magnetic lattice flux defined by ice-rule obeying two-in-one out (red), one-in-two out (blue) or one-in-one-out (gray) plaquettes. For applied fields of \SI{100}{mT} (left row) regular flux patterns form, whereas at remanence (right row) complex, frustrated low-energy configurations can form.
			(c)~Parallel conduction channels between nodes $A$ and $B$, coloured by legnth of connecting segments.
			(d)~Anisotropic magnetoresistance between $A$ and $B$, for theoreitical limiting cases and from simulations at different fields (data points).
			Panels~(b,d) reprinted with permission from~\cite{Koshikawa23}.~\textcopyright{}~(2023) by the American Physical Society.
			}
		\label{fig:gyroids}
	\end{figure}

	Above, we considered the topological properties of wireframe building blocks and self-contained structures like the buckyball constructed from straight wire segments. In the following we consider extended interconnected magnetic networks, with the gyroid as a particularly interesting example case.
		
	Gyroid structures are a special case of an ``infinite periodic minimal surface without self-intersections'' first postulated by Alan Schoen in 1970~\cite{Schoen70}. Decades later, it was found that such complex lattices can be realized on the nanoscale by self-assembly of di- and tri-block co-polymers~\cite{Dolan15,Wu15a,Dolan18} and are even found in wings of some butterfly species where they act as biological photonic crystals~\cite{Saranathan10}.
	
	Gyroid structures can be grown in two variants -- single gyroids as well as double gyroids with two interconnected but physically separate sub-lattices. As shown in Fig.~\ref{fig:gyroids}, nanoscale single gyroid lattices are built from a an inherently chiral periodic network of chiral triple junctions (represented by triangular plaquettes), mathematically also known as \emph{srs} net.
	
	Electron holography allowed to experimentally image the magnetic state of small-scale single and double Ni gyroid samples~\cite{Llandro20}, which demonstrated that the local moment prefers to point along the direction of the individual struts, no high-energy all-in or all-out vertex configurations were observed at remanence. Finite-element micromagnetic simulations gave further insights to the local anisotropy at remanence~\cite{Koshikawa23}, finding that individual struts can be described by a macrospin picture where all local moments aligning in a parallel fashion. However, the local orientation of a strut ``macrospin'' moment
	is complex, and heavily influenced by the magnetic continuity in the gyrating channels as well as by curvature-induced DMI~\cite{Streubel16,Volkov18}. 
	
	Regardless of the complex local anisotropy 	in nanoscale gyroids, an analysis of the scalar spin chirality $\Omega_s$~\cite{Taguchi01,Park08} 
	\begin{equation}
	    \Omega_s = \frac{1}{3} \left( s_1^Is_2^I + s_2^Is_3^I + s_3^Is_1^I \right),
	    \label{eq:spin_chirality}
	\end{equation}
	showed that the triangular plaquettes with moments $s_i^I$ (with $i=\overline{1,3}$, with $I$ denoting its orientation in the local reference frame) typically form ice-rule-obeying spin configurations that collectively minimize the geometrical frustration inherent to the triple-connected lattice~\cite{Wannier50,Diep13,Lacroix11,Skjaervo19}. These spin-ice states correspond to two-in-one-out, one-out-two-in, or, here additionally, one-in-one-out configurations; combining these three options it is possible to define a lattice-spanning network of magnetic flux, as shown in Fig.~\ref{fig:gyroids}(b) for applied fields (top row) and at remanence (bottom row).
	
	Importantly, the 3D connectivity of the twisting neighbouring triangular plaquettes naturally supports \emph{non-coplanar} spin structures (i.e., especially in the one-in-one-out configuration coloured in gray). This non-coplanarity allows for an emergent vector spin chirality~\cite{Park08}, which is relevant for non-reciprocal magnetotransport and spin-wave propagation~\cite{Rikken01,Seki16a,Tokura18,Barman20,Atzori21}. It is worth noting that the generalized skyrmion number, respectively the emergent magnetic field, are the equivalent topological measures in a continuous micromagnetic picture.
	
	Finally, the interplay between the spatial connectivity and frustrated magnetic order in gyroids leads to a complex magneto-transport response. For the simplest case of longitudinal anisotropic magnetoresistance (AMR)~\cite{Koshikawa23}, the local resistance of each strut is modulated by the relative alignment between current flow and macrospin direction, $\rho_\text{AMR}\propto \vec{j}_i\cdot\vec{m}_i\propto m_i^\parallel$. As shown in Fig.~\ref{fig:gyroids}(c), a multitude of paths between two chosen nodes of varying lengths exist, which form parallel conduction channels. Applying Kirchhoff's law and the piece-wise modulated local resistances then allows to theoretically calculated the expected AMR response. Due to the coexisting three-dimensional conduction channels the resulting AMR behaviour with varying field direction and amplitude is decidedly distinct from the bulk case~\cite{Pignard00}, or even two-dimensional wire networks~\cite{Wang05b,Branford12,Le17}. For example, typical limiting cases for saturated and remanent spin order feature intermediate or maximum response in gyroids, see Fig.~\ref{fig:gyroids}(d), compared to the maximum and minimum response expected in bulk, respectively (the reason for this discrepancy are small spin canting in the order of the geometric DMI which persist at high fields)~\cite{Koshikawa23}.	Furthermore, even small moment deviations from the local current direction might lead to noticeable difference to theoretically expected limits (e.g., compare the gray squares from micromagnetic simulations to the black line as limiting response for the calculated finite-sized case).
	
	While Ref.~\cite{Koshikawa23} considered AMR contributions in a finite-size gyroid network only, more interesting magnetotransport responses related to the non-zero vector spin chirality should be appreciable by transverse magnetoresistance measurements~\cite{Meng21}.

	\section{Non-Orientable Surfaces: M\"{o}bius Rings}
	\label{sec:nonorientable}
	
	\subsection{Geometry of a M\"{o}bius Ring}
	
\begin{figure}[tb]
		\centering
		\includegraphics[width=0.7\linewidth]{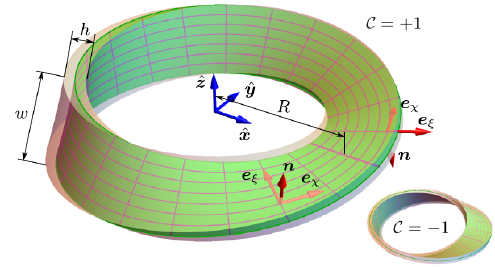} % moebius-geom
		\caption{\textbf{Geometry of a M\"{o}bius ring.} M\"{o}bius ring of opposite chiralities $C=\pm1$ and the local reference frame defined by Eq.~(\ref{eq:moebius_coordinates}).
		Reprinted under Creative Commons Attribution 3.0 License (https://creativecommons.org/licenses/by/3.0/) from~\cite{Pylypovskyi15b}~\textcopyright{}~(2015) by the Authors.}
		\label{fig:moebius-geom}
	\end{figure}

	The surface $\vec{\varsigma}$ of a M\"{o}bius ring can be parameterized using the azimuthal angle $\eta_1 = \chi$ around the ring and the transversal coordinate $\eta_2 = \xi$, as shown in Fig.~\ref{fig:moebius-geom}:
	\begin{equation}
		\begin{aligned}
			\vec{\varsigma} & = \left\{ \left(R + \xi \cos\dfrac{\chi}{2}\right)\cos\chi, \left(R + \xi \cos\dfrac{\chi}{2}\right)\sin\chi, C \xi \sin\dfrac{\chi}{2} \right\},\\
			& \text{with}\qquad 0 \le \chi < 2\pi,\quad \xi \in \left[ -\dfrac{w}{2}, \dfrac{w}{2} \right],
		\end{aligned}
		\label{eq:moebius_coordinates}
	\end{equation}
	Here, $R$ is the ring's radius, $w$ is the width of the band, and $C = \pm 1$ is the chirality defining the ring's twist orientation along the azimuthal direction (counter-clockwise or clockwise, respectively). A shell of the finite thickness $h$ is given by the extension of the surface $\vec{\varsigma}$ along its normal $\vec{n} = \vec{e}_\chi \times \vec{e}_\xi$, and self-intersections of the geometry are absent if $h \ll R - w/2$. 
	
	The peculiar topological property of the M\"{o}bius ring is its non-orientability: the normal field $\vec{n}$ is necessarily discontinuous by $\chi$. Being mapped to a 2D rectangle $\mathcal{R} = \{(x,y): x \in [0,2\pi R],\; y \in [-w/2,w/2]\} $, the M\"{o}bius-ring topology is reflected in antiperiodic (also called twisted periodic) boundary conditions for any scalar function $f(\bullet)$ on $\mathcal{R}$, i.e., $f(x,y) = f(x+2\pi R,-y)$.
	
	\subsection{Static Magnetic States on a M\"{o}bius Ring}
	
	Let us start from the critical phenomena, in particular, finite-size scaling effects, that directly reflect sample topology and boundary conditions for the order parameter. Comparing to toroidal systems, the case of a M\"{o}bius ring is dependent on the number of lattice sites along $\vec{e}_\xi$~\cite{Lu01}. The finite-size scaling for the magnetization $m$ at a given temperature $T$ can be characterized by the so-called Binder parameter 
	\begin{equation}
		b_{M\times N}(T) = \dfrac{1}{2} \left( 3 - \dfrac{\langle m^4\rangle}{\langle m^2\rangle^2}\right),
	\end{equation}
	where $\langle \bullet \rangle$ means averaging Monte Carlo steps and the subscript $M\times N$ indicates the size of the discretised lattice. The value of the Binder parameter depends on the system geometry, type of boundary conditions, and system size. There is a scaling of the Ising model that collapses curves $b(T)$ obtained for different $M\times N$ lattice sizes for the specific topology. Comparing different topologies at the given temperature, the Binder parameters are ordered as follows:
	\begin{equation}
		b_\text{plane} < b_\text{cylinder} < b_\text{torus} < b_\text{M\"{o}bius}.
	\end{equation}
	This behavior reflects the dependence of the number and size of percolating clusters that is influenced by the structure of the boundary conditions of the lattice~\cite{Kaneda01}. Further studies indicate that the finite-size scaling of the magnetic susceptibility is also directly dependent on the topology of the geometry, making the M\"{o}bius ring distinct from the torus and projective plane\footnote{Projective plane is the generalization of a usual plane that contains a line (boundary) at the infinity instead of the single point associated with all infinitely distant points.}~\cite{Vasilyev19}.
	
	In the following, we focus on the model that includes Heisenberg exchange, uniaxial anisotropy, dipolar interaction, and interaction with external magnetic fields. We start our consideration of static spin states on a M\"{o}bius ring from the case when its radius is much larger than the magnetic length, $R \gg \ell = \sqrt{A/K_\text{eff}}$, where $K_\text{eff} = K - 4\pi M_S^2$ in CGS, $K$ is the coefficient of magnetocrystalline anisotropy and the total effective anisotropy includes the shape anisotropy as well. The ratio between the dipolar interaction (magnetostatics) and the uniaxial out-of-plane anisotropy can be characterized by the so-called quality factor $Q = K/(2\pi M_S)$. The major contribution of the magnetocrystalline anisotropy to the magnetic state is realized for $Q \gtrsim 1$, otherwise, the shape anisotropy is dominating.
	
\begin{figure}[tb]
		\centering
		\includegraphics[width=\linewidth]{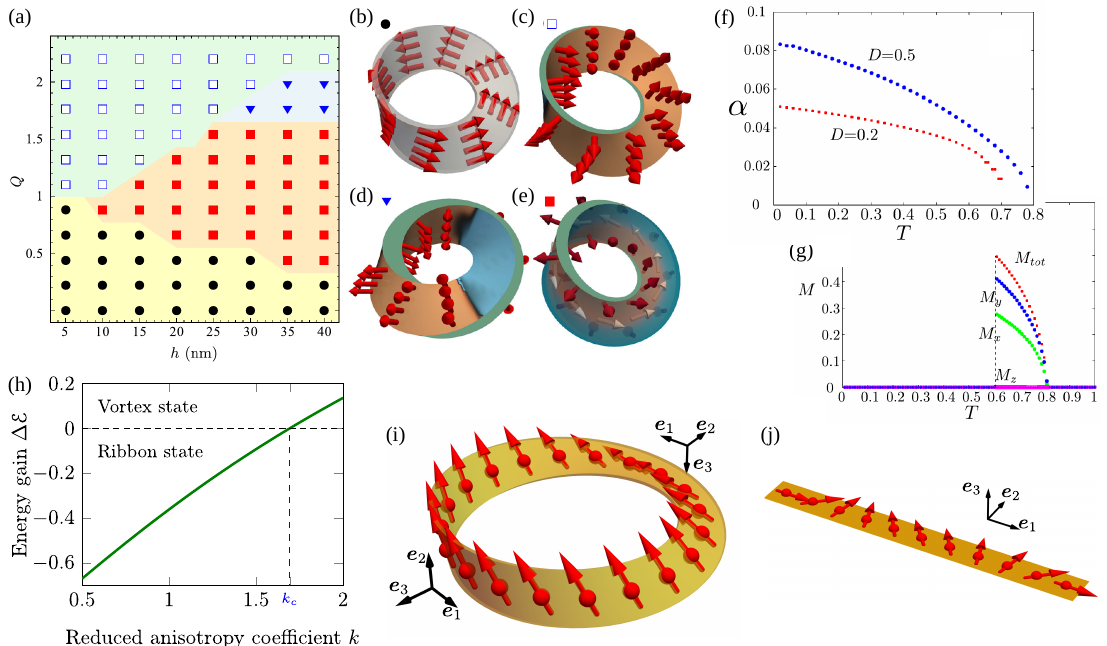} % moebius-static
		\caption{\textbf{Static states of a ferromagnetic M\"{o}bius ring.} (a)~Diagram of ground states at low temperatures for different quality factors~$Q$ and sample thickness~$h$: (b)~vortex, (c)~transversal wall, (d)~multidomain state, (e)~longitudinal wall~\cite{Pylypovskyi15b}. (f)~Diagram of ground states at finite temperature~$T$ depending on the out-of-surface anisotropy coefficient~$D$ and in-surface exchange anisotropy~$\alpha$. The regions above (below) dotted lines correspond to the lowest-energy transversal (longitudinal) domain wall~\cite{Yoneya08}. (g)~Transformation of the longitudinal domain wall (zero net magnetization $M$) into transversal wall (finite magnetization $M$) with temperature~\cite{Yoneya08}. (h)~Difference between vortex and ribbon states as a function of effective anisotropy $k$~\cite{Gaididei17}. (i)~Real-space and (j)~untwisted view of the ribbon state in a narrow M\"{o}bius ring. Panels~(a--e) are reprinted under Creative Commons Attribution 3.0 License (https://creativecommons.org/licenses/by/3.0/) from~\cite{Pylypovskyi15b}~\textcopyright{}~(2015) by the Authors. Panels~(f,g) reprinted with permission from~\cite{Yoneya08}~\textcopyright{}~(2008) by the American Physical Society.  Panels~(h--j) are adapted from~\cite{Gaididei17},~\textcopyright{}~IOP Publishing. Reproduced with permission. All rights reserved.
		}
		\label{fig:moebius-static}
	\end{figure}

	In the case of a magnetically soft material (i.e., $K\rightarrow 0$, and thus $Q\rightarrow 0$), the ground state of a M\"{o}bius ring is a vortex, similar to the planar rings~\cite{Pylypovskyi15b}, see Fig.~\ref{fig:moebius-static}(a,b). Experimentally, M\"{o}bius rings from magnetically soft materials can be fabricated by means of FEBID, see Sec.~\ref{sec:additive_FEBID}. 
	
	The non-trivial geometrical topology of the M\"{o}bius ring becomes pronounced for larger $Q$ values, when the easy axis follows the surface normal~$\vec{n}$. For a large enough $Q$, the ground state of the sample is represented by a domain wall that is topologically protected because of the inherent discontinuity in the normal field. There are two possible configurations of such domain walls: with the domain wall plane either along $\vec{e}_\xi$ [Fig.~\ref{fig:moebius-static}(c,d)], or along $\vec{e}_\chi$~\cite{Yoneya08} [Fig.~\ref{fig:moebius-static}(e)]. Here, we refer to them as the transversal and longitudinal walls, respectively. These domain wall states differ by the total macroscopic moment of the sample: the longitudinal-wall state is fully compensated, while the presence of a transversal wall leads to a sizeable moment perpendicular to the M\"{o}bius ring axis $\vec{\hat{z}}$. 
	
	Which one of the two domain walls is the lowest-energy state at low temperatures is determined either by accounting for in-plane anisotropy~\cite{Yoneya08}, or by magnetostatics that become decisive for a small but finite thickness $h$ of the ring~\cite{Pylypovskyi15b}. For a dominating effective in-plane anisotropy ($Q<1$), the transversal domain wall has a lower energy compared to the longitudinal wall, see Fig.~\ref{fig:moebius-static}(f). Furthermore, there is a critical temperature at which the longitudinal domain wall loses its thermodynamic stability and transforms into the transversal one, leading to abrupt changes of the total magnetic moment of the system, see Fig.~\ref{fig:moebius-static}(g)~\cite{Yoneya08}.
	
	The detailed micromagnetic study of the properties of both types of topologically protected walls can be completed within the framework of curvilinear magnetism. Here, we rely on the angular parameterization of the magnetization vector $\vec{m}$ given by Eq.~\eqref{eq:mag-vec-curv} with $\vec{e}_1 = \vec{\hat{\chi}}$ and $\vec{e}_2 = \vec{\hat{\xi}}$. Comparison with micromagnetic simulations justifies the following Ansatz for the transversal domain wall:
	\begin{equation}\label{eq:ansatz-transv}
		\Theta_\text{tr} = 2\arctan e^{p(\chi - X)/\chi_0},\quad \Phi_\text{tr} = \bar{C} p \dfrac{\pi}{2},
	\end{equation}
	where $p = \pm 1$ is the domain wall polarity (i.e. kink or antikink), $X$ and $\chi_0$ are its azimuthal coordinate and width, respectively, and $\bar{C} = \pm 1$ is the magnetic chirality of the wall. Then, the total energy of a narrow enough ring is
	\begin{equation}\label{eq:transwersal-dw-energy}
		\dfrac{E_\text{tr}}{2Ah} \approx \dfrac{w}{R} \left[ \dfrac{1}{\chi_0} + \bar{C}C \dfrac{\pi}{2} + \chi_0 \left(\cos X - \dfrac{1}{4}\right)\right] + \chi_0 \dfrac{Rw}{\ell^2} + \text{const}.
	\end{equation}
	This expression reaches a minimum for $X = \pi$ (i.e, ``vertical'' part of the M\"{o}bius ring) and $\chi_0 = \ell/R$. Remarkably, the coupling between chiralities of the ring geometry $C$ and domain wall chirality $\bar{C}$ appears: for the given definitions, they should be of the opposite rotation directions with the energy barrier between states equal to $2\pi Ahw/R$. The same analysis of the longitudinal wall with the Ansatz
	\begin{equation}\label{eq:ansatz-longitudinal}
		\Theta_\text{lon} = 2\arctan e^{p \xi/\xi_0},\quad \Phi_\text{lon} = \pi \dfrac{\bar{C} + p}{2},
	\end{equation}
	with $\chi_0$ being the domain wall width approximately equal to $\ell$ in the equilibrium, shows the presence of the same coupling between magnetic and geometric chiralities~\cite{Pylypovskyi15b}.
	
	Accounting for the finite thickness of the M\"{o}bius ring with magnetostatics for different values of the quality factor $Q$ and thickness $h$ shows that for $h \lesssim \ell$, the transversal wall becomes energetically favorable over the vortex in the plane of the ring for $Q > 1$, see Fig.~\ref{fig:moebius-static}(a). With increasing thickness of the M\"{o}bius ring, a finite interval around $Q = 1$ appears where the longitudinal wall is the ground state. With further increase of the role of magnetostatics, multiple walls appear due to spontaneous split into magnetic domains around the M\"{o}bius ring~\cite{Pylypovskyi15b}.
	
	The case of a narrow and small-radius M\"{o}bius ring asymptotically approaches a flat ring on the macroscale. However, the geometry-induced shape anisotropy of the exchange origin forces the system to ``remember'' its true underlying topology. Namely, such an object becomes equivalent to a closed nanowire with antiperiodic boundary conditions on the magnetic angles, and spatially dependent curvature and torsion
	\begin{equation}\label{eq:eff-curv-tors-narrow-moebius}
		\kappa_\text{eff} = - \dfrac{2\cos\frac{\chi}{2}}{R} \dfrac{1 + 2\sin^2\frac{\chi}{2}}{1 + 4\sin^2\frac{\chi}{2}},\quad \tau_\text{eff} = - \dfrac{C}{2R} \sqrt{1 + 4\sin^2\frac{\chi}{2}}.
	\end{equation} 
	In case of $Q \ll 1$, this effective nanowire geometry supports two equilibrium solutions, see Fig.~\ref{fig:moebius-static}(h). For a strong enough shape anisotropy, the lowest-energy vortex state is analogous to the vortex in magnetically soft large M\"{o}bius rings has, and suppresses effects of curvature beyond those stemming from the ring topology. If local exchange effects are dominating, a locally inhomogeneous magnetization distribution forms which is tilted at the ``vertical'' part of the M\"{o}bius ring near $\chi = \pi$~\cite{Gaididei17}, see the ribbon state in Fig.~\ref{fig:moebius-static}(h) and Fig.~\ref{fig:moebius-static}(i,j).

	To properly characterize the geometric and topological signatures of magnetic textures on non-orientable manifolds, one should evaluate their emergent gauge fields and define the corresponding topological charge. Since the manifold itself lacks a global orientation, one can construct its orientable double cover,  \(\widetilde{\mathcal{M}}\), which is periodic and provides a well-defined setting for integration~\cite{Hatcher02}. Geometrically, this construction ``unwinds'' the non-orientable substrate, transforming a M\"{o}bius ring into an ordinary  cylinder by aligning two copies that cancel the intrinsic directional reversal.

    The emergent magnetic field, defined in Eq. \eqref{eq:hopf}, arising from the magnetization texture induces, on a two-dimensional surface, the curvature two-form
    \[
        \mathcal{F} = \mathbf{m} \cdot \left( \partial_{\eta_{1}}\mathbf{m} \times \partial_{\eta_{2}} \mathbf{m} \right) \, \sqrt{g}\,d\eta_{1}  d\eta_{2},
    \]  
    where \(\eta_{1}, \eta_{2}\) denote local curvilinear coordinates. For the non-orientable base manifold \(\mathcal{M}\), the topological charge is naturally defined via the projection \(\Pi: \widetilde{\mathcal{M}} \to \mathcal{M}\) and the pullback \(\widetilde{\mathcal{F}}\) of the curvature form to the double cover. Geometrically, \(\widetilde{\mathcal{M}}\) corresponds to the \(\mathbb{Z}_2\)-bundle of frame orientations over \(\mathcal{M}\). This is achieved by lifting the curvature form \(\mathcal{F}\) from \(\mathcal{M}\) to \(\widetilde{\mathcal{M}}\) via the projection map \(\Pi\), meaning that for tangent vectors on the double cover, \(\widetilde{\mathcal{F}}\) measures the same curvature as \(\mathcal{F}\) does on their projected images on the base manifold. In this framework, the topological charge for a non-orientable surface can be expressed directly on \(\mathcal{M}\) as  
    \[
        \mathcal{Q}_{\mathrm{top}} = \frac{1}{8\pi} \int_{\widetilde{\mathcal{M}}} \widetilde{\mathcal{F}},
    \label{NOCharge}
    \]
    where \(\widetilde{\mathcal{F}}\) acts as the effective topological charge density. The prefactor \(1/8\pi\) accounts for the double covering and ensures that \(\mathcal{Q}_{\mathrm{top}}\) reduces to the conventional normalization \(1/4\pi\) on orientable manifolds. This formulation incorporates the local orientation reversal inherent to non-orientable geometries, remains valid for smooth magnetization textures whose characteristic scales exceed the system’s microscopic width, and naturally reverts to the standard definition when the underlying manifold is orientable~\cite{Saji25a}.
	
	\subsection{Spin Wave Dynamics on a M\"{o}bius Ring}
	
	Up to now, the spin wave dynamics on a nanoscale M\"{o}bius ring have been mainly investigated for the magnetically soft case, i.e., $Q = 0$, within its vortex ground state. Typically observed modes correspond to azimuthally traveling waves, which can be classified by the number of wave periods along circumference $n$, with $2n \in \mathbb{Z}$, as well as the number of nodal lines along the transversal direction and band thickness. We stress here that the M\"{o}bius topology permits half-integer $n$ that realizes in the case of odd number of transversal and cross-sectional nodes in the given mode~\cite{Korber23}.
	
\begin{figure}[tb]
	    \centering
	    \includegraphics[width=\linewidth]{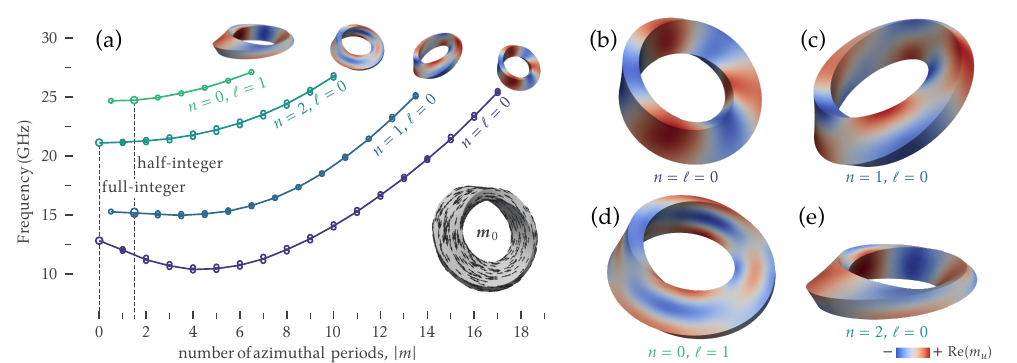} % moebius-spectrum
	    \caption{\textbf{Magnon modes on a Permalloy M\"{o}bius ring of radius 100\,nm, width 60\,nm and thickness 25\,nm.} (a)~Frequencies depending on the mode number. (b--e) Exemplary profiles of modes shown by the real part of the excitation. Reprinted under Creative Commons Attribution 4.0 License (https://creativecommons.org/licenses/by/4.0/) from~\cite{Korber23}~\textcopyright{}~(2023) by the Author.}
	    \label{fig:moebius-spectrum}
	\end{figure}

	Similarly to tubes and thin films, the dispersion of these spin waves reaches a minimum for low numbers of the transversal and cross-sectional nodal lines. Many of them can be arranged into degenerate doublets, supplemented by the observation of a symmetry-induced splitting of certain azimuthal modes, see Fig.~\ref{fig:moebius-spectrum}. 
	
	It is natural to compare the spin wave dynamics in the M\"{o}bius ring with that on an orientable ring. For a planar, orientable ring the standing modes are determined by the boundary conditions, and it can be spatially deformed to mimic the general shape of the M\"{o}bius ring but without twist~\cite{Thonikkadavan26}. This spatial deformation of the planar ring leads to the appearance of excitation doublets that could be understood by a formation of standing waves by a superposition of two counter-propagating waves. These modes have the same frequency, while their phase is different due to spatially inhomogeneous torsional distortion of geometry and respective distribution of mode maxima and minima~\cite{Thonikkadavan26}.
	
	The lowest-frequency mode in the M\"{o}bius ring has a full correspondence to the same mode of the orientable ring. However, higher modes, while still grouped into doublets, develop a qualitative distinction as they are propagating. Within the doublets, counter-rotating modes have a slightly different frequencies, below and above the corresponding mode in the orientable ring. This behavior has been numerically tested on the rings with multiple twists and associated with the geometric phase originating from the helical twist of the geometry~\cite{Thonikkadavan26}.
	
	An analytical analysis of the spin wave modes in complex geometries has been done under the assumption of narrow stripes~\cite{Gaididei17}. The shape anisotropy expressed in terms of the effective curvature and torsion [c.f.~Eq.~\eqref{eq:eff-curv-tors-narrow-moebius}] generally leads to the non-reciprocal propagation of the spin waves due to the geometry-driven DMI. More precisely, rotation of the polarization plane of the spin wave due to geometry can be understood as the Berry phase accumulated along the propagated distance~\cite{dAquino25}. Recently, this framework has been applied to the case of ultrafast magnetization dynamics described by the so-called inertial LLG~\eqref{eq:LLG} as follows~\cite{Ciornei11,Mondal17,Neeraj21,Quarenta24}
	\begin{equation}
		\partial_t \vec{m} = - \gamma_0 \vec{m} \times \left( \vec{H}_\text{eff} - \dfrac{\alpha}{\gamma_0} \partial_t \vec{m} - \dfrac{\mathcal{I}}{\gamma_0} \partial_t^2 \vec{m} \right),
	\end{equation}
	where the mass-like coefficient $\mathcal{I}$ is responsible for the inertial dynamics~\cite{ReyesOsorio25}. Frequencies of the fundamental nutation and precession modes are dependent on the curvature and torsion of the geometry and further scaled by $\mathcal{I}$ (which furthermore is potentially time-dependent). In particular, the nutation frequency at zero wave vector $k = 0$ grows as $1/\mathcal{I}$. At finite $k$, the M\"{o}bius ring geometry imposes the quantization
	\begin{equation}
		k = \dfrac{1 + 2\tilde{n}}{2R},\quad \tilde{n} \in \mathbb{Z}.
	\end{equation}
	In this way, the predicted fundamental frequencies in a narrow Permalloy ribbon are mapped to dozens of GHz and sub-THz ranges for the precession and nutation of magnetization, respectively~\cite{dAquino25}.

	\section{Conclusions and Outlook}
	\label{sec:conc}
	
	To summarise, we review major developments in curvilinear magnetism, focusing on the role of the non-trivial topology of a sample's shape. The specific point of topology is that it provides us certain knowledge of the global state of the magnetization field of the sample even if the local magnetic interactions (e.g., exchange and surface anisotropy) are dominating. For a long time, the major focus in the community was on nanotubes: they represent a simple enough playground for a theoretical analysis, capturing many features of curved geometry and non-trivial topology. Among the key exciting features of nanotubes one can mention discovery of the interplay between tubular geometry and unusual behavior of magnetization, such as dependency of the vortex wall propagation velocity on its chirality~\cite{Landeros10,Otalora12a}, suppression of Walker breakdown~\cite{Yan11} and non-reciprocal propagation of spin waves~\cite{Otalora16,Korber21a} to name a few. Referring the reader to the focused reviews on magnetic nanotubes~\cite{Dhungana14,Otalora19,Proenca20,Landeros22}, here, we address more complex geometries including axially symmetric hyperbolic geometries, tori, spheres, as well as samples of low symmetry, such as wireframes, nanogyroids, and the M\"{o}bius ring as an example of non-orientable surface. 
	
	Tubular geometries of toroidal topology (Sec.~\ref{sec:toroidal}) demonstrate a complex interplay between the local curvatures and shape of topologically non-trivial magnetic solitons that wrap around the sample, thus, cannot collapse (Sec.~\ref{sec:heisenberg}). Large enough samples of compact geometries like a ring-shaped torus, commonly keep the locally uniform vortex ground state, but under external perturbations reveal appearance of vortex-antivortex pairs driven by the sign of local mean curvature (Sec.~\ref{sec:Topology-Tori} and~\ref{subsec:experiment-torus}). The ground states become more complex for more specific nanostructures. Indeed, bulk nanotori with large aspect ratio between the toroidal and poloidal radii allow the nucleation of metastable states such as Hopfion-like patterns~\ref{sec:Hopfions}. Still the continuity of properties is satisfied by the fact that even local pinning of merons keeps the total balance of their vorticity over the sample that is determined by the topology via genus of the specific sample (Sec.~\ref{sec:spheres}). Furthermore, magnetic wireframes (Sec.~\ref{sec:wireframes}) and nanogyroids (Sec.~\ref{sec:gyroids}) offer a room for the direct analysis of (artificial) spin-ice systems. A M\"{o}bius ring is the specific example of the natural antiperiodic boundary conditions. They lead to the domain wall states protected by sample topology as well as coupling of geometric and magnetic chiralities (Sec.~\ref{sec:nonorientable}).
	
	After a long time of purely analytical and numerical research, recent technological advances provide to experiemntally validate theoretical predictions. In particular, direct writing (Sec.~\ref{subsec:additive_methods}) with focused ion/electron beams or covering of non-magnetic templates by a magnetic material allow fabrication of complex, magnetically soft or hard samples. In contrast, subtractive techniques (Sec.~\ref{subsec:subtractive_methods}) make it possible to shape highly crystalline materials into complex geometries. The magnetic state of complex nanoarchitectures can be accessed with a variety of experimental methods, including 2D imaging and 3D tomographic reconsutrctions using x-rays or electrons as probe, as well as with local scanning techniques or magneto-optical methods (Sec.~\ref{sec:measurement}). Remarkably, magnetic states in 3D nanotori~\cite{Eltschka08,Lewis20,Turnbull25}, wireframes~\cite{Fernandez-Pacheco20,Gubbiotti24} and nanogyroids~\cite{Llandro20} are analyzed not only theoretically, but also experimentally.
	
	We anticipate that many as-yet-unanswered questions related to the influence of a samples' topologogy on its magnetic configuartion will be addressed in the future. High-genus magnetically soft samples can be considered as a convenient platform to study phenomena related to antivortices that otherwise is difficult to access in planar samples~\cite{Shigeto02,Wang07,Gliga08,Li14}. Topologically-protected stabilization of magnetic textures offers promising possibilities for unconventional computing due to complexity and high stability of magnetic states in complex curvilinear geometries. For example, geometrically-designed states with multiple merons can be of advantage for neuromorphic and reservoir computing by expansion of the parameter space in their responses on external fields or currents~\cite{Tey22}. Such activities may require novel computational approaches for magnetism~\cite{Muller19,Kovacs22,Abert25}. However, at the moment we still have only a limited understanding regarding the dynamics of solitons and spin wave excitations in complex topologies. After the recent works on eigenmodes in buckyball~\cite{Cheenikundil21,Cheenikundil22,Pip22} and M\"{o}bius ring~\cite{dAquino25,Thonikkadavan26} topologies, it is relevant to develop a generalized framework accounting for topological features and mapping them into magnetization dynamics, including peculiar surface magnetostatic modes in geometries with non-trivial geometry. 
	
	In contrast to ferromagnets, there is limited understanding of other types of magnetic ordering with respect to the manifolds at which they are found. In particular, curvilinear antiferromagnets are mainly studied for the cases of spin chains~\cite{Pylypovskyi20,Borysenko22,Yershov22a,DasGupta22} and generalized cylinders~\cite{Yershov22,Yershov25}. At the same time, even bulk antiferromagnets with crystal defects introduces topological features in their N\'{e}el ordering~\cite{Galkina20}. Similarly, it would be of relevance to analyze phenomena in ferrimagnets~\cite{Ivanov19} and superconductors~\cite{Salamone21,JohanSkarpeid24,Heinrich25}. 
	
	Beyond large-scale collective phenomena, well captured by the physics of Landau--Lifshitz equation, modification of the material parameters under shaping of the sample requires better understanding. This includes changes in exchange couplings and spin-orbit phenomena in curvilinear magnets~\cite{Edstrom22,Gao26}, as well as topology of electronic bands~\cite{Kosevich04a} and their spin polarization. At the very nanoscale, unconventional topology may also appear with complex-shaped molecular magnets~\cite{Cador04} also carrying magnetoelectric multipoles~\cite{Murray22}, therefore, offering additional responses to be understood. 
	
	From an experimental perspective, the exploration of magnetism in non-trivial topologies poses both significant challenges and opportunities. Advances in three-dimensional nanofabrication techniques enable the realization of curvilinear magnetic architectures in a growing range of materials with increasing precision. Complementarily, progress in magnetic measurement methods, both with increasing access to large-scale experimental facilities as well as advances for table-top in-lab equipment provide access to spatially resolved characterization of nanoscale magnetization textures and their dynamics in complex nanoscale geometries. While we discussed several pioneering results in this chapter, more experimental studies in close interplay with theory and simulation will be essential to validate the predicted static and dynamic topological effects, uncover emergent phenomena, and guide the design of functional devices.

	\section*{Conflict of Interest}
	
	The authors declare no conflict of interest.
	
	\section*{Acknowledgements}
	
	O.V.P. thanks Dr.~Filipp~N.~Rybakov (Uppsala University) and Dr.~Lukas~K\"{o}rber (Radboud University Nijmegen) for fruitful discussions. The activities that led to the results reported in this chapter were financed via numerous third party projects, whose support we greatly acknowledge. Among the most recent funding,  O.V.P acknowledges ERC grant 3DmultiFerro (project number: 101141331). 
	V.L.C.-S. thanks the Brazilian agencies CNPq (Grant No. 300324/2026-0), INCT of Spintronics and Advanced Magnetic Nanostructures (INCT-SpinNanoMag - CNPq 406836/2022-1), and Rede Mineira de Nanomagnetismo/FAPEMIG. 
	N.L.\ acknowledges funding from UK Research and Innovation (UKRI) via the Future Leader Fellowship MR/X033910/1 (LIONESS).
	L.A.T. thanks the Alexander von Humboldt Foundation for support and Dr. Max Birch (RIKEN CEMS, Tokyo).


\begin{thebibliography}{447}%
		\makeatletter
		\providecommand \@ifxundefined [1]{%
			\@ifx{#1\undefined}
		}%
		\providecommand \@ifnum [1]{%
			\ifnum #1\expandafter \@firstoftwo
			\else \expandafter \@secondoftwo
			\fi
		}%
		\providecommand \@ifx [1]{%
			\ifx #1\expandafter \@firstoftwo
			\else \expandafter \@secondoftwo
			\fi
		}%
		\providecommand \natexlab [1]{#1}%
		\providecommand \enquote  [1]{``#1''}%
		\providecommand \bibnamefont  [1]{#1}%
		\providecommand \bibfnamefont [1]{#1}%
		\providecommand \citenamefont [1]{#1}%
		\providecommand \href@noop [0]{\@secondoftwo}%
		\providecommand \href [0]{\begingroup \@sanitize@url \@href}%
		\providecommand \@href[1]{\@@startlink{#1}\@@href}%
		\providecommand \@@href[1]{\endgroup#1\@@endlink}%
		\providecommand \@sanitize@url [0]{\catcode `\\12\catcode `\$12\catcode
			`\&12\catcode `\#12\catcode `\^12\catcode `\_12\catcode `\%12\relax}%
		\providecommand \@@startlink[1]{}%
		\providecommand \@@endlink[0]{}%
		\providecommand \url  [0]{\begingroup\@sanitize@url \@url }%
		\providecommand \@url [1]{\endgroup\@href {#1}{\urlprefix }}%
		\providecommand \urlprefix  [0]{URL }%
		\providecommand \Eprint [0]{\href }%
		\providecommand \doibase [0]{https://doi.org/}%
		\providecommand \selectlanguage [0]{\@gobble}%
		\providecommand \bibinfo  [0]{\@secondoftwo}%
		\providecommand \bibfield  [0]{\@secondoftwo}%
		\providecommand \translation [1]{[#1]}%
		\providecommand \BibitemOpen [0]{}%
		\providecommand \bibitemStop [0]{}%
		\providecommand \bibitemNoStop [0]{.\EOS\space}%
		\providecommand \EOS [0]{\spacefactor3000\relax}%
		\providecommand \BibitemShut  [1]{\csname bibitem#1\endcsname}%
		\let\auto@bib@innerbib\@empty
		%</preamble>
		\bibitem [{\citenamefont {Onsager}(1944)}]{Onsager44}%
		\BibitemOpen
		\bibfield  {author} {\bibinfo {author} {\bibfnamefont {L.}~\bibnamefont
				{Onsager}},\ }\bibfield  {title} {\bibinfo {title} {Crystal statistics. i. a
				two-dimensional model with an order-disorder transition},\ }\href
		{https://doi.org/10.1103/physrev.65.117} {\bibfield  {journal} {\bibinfo
				{journal} {Physical Review}\ }\textbf {\bibinfo {volume} {65}},\ \bibinfo
			{pages} {117} (\bibinfo {year} {1944})}\BibitemShut {NoStop}%
		\bibitem [{\citenamefont {Kaufman}(1949)}]{Kaufman49}%
		\BibitemOpen
		\bibfield  {author} {\bibinfo {author} {\bibfnamefont {B.}~\bibnamefont
				{Kaufman}},\ }\bibfield  {title} {\bibinfo {title} {Crystal statistics. ii.
				partition function evaluated by spinor analysis},\ }\href
		{https://doi.org/10.1103/physrev.76.1232} {\bibfield  {journal} {\bibinfo
				{journal} {Physical Review}\ }\textbf {\bibinfo {volume} {76}},\ \bibinfo
			{pages} {1232} (\bibinfo {year} {1949})}\BibitemShut {NoStop}%
		\bibitem [{\citenamefont {Aharoni}\ and\ \citenamefont
			{Shtrikman}(1958)}]{Aharoni58}%
		\BibitemOpen
		\bibfield  {author} {\bibinfo {author} {\bibfnamefont {A.}~\bibnamefont
				{Aharoni}}\ and\ \bibinfo {author} {\bibfnamefont {S.}~\bibnamefont
				{Shtrikman}},\ }\bibfield  {title} {\bibinfo {title} {Magnetization curve of
				the infinite cylinder},\ }\href {https://doi.org/10.1103/physrev.109.1522}
		{\bibfield  {journal} {\bibinfo  {journal} {Physical Review}\ }\textbf
			{\bibinfo {volume} {109}},\ \bibinfo {pages} {1522} (\bibinfo {year}
			{1958})}\BibitemShut {NoStop}%
		\bibitem [{\citenamefont {Aharoni}(1959)}]{Aharoni59}%
		\BibitemOpen
		\bibfield  {author} {\bibinfo {author} {\bibfnamefont {A.}~\bibnamefont
				{Aharoni}},\ }\bibfield  {title} {\bibinfo {title} {Some recent developments
				in micromagnetics at the {W}eizmann institute of science},\ }\href
		{https://doi.org/10.1063/1.2185971} {\bibfield  {journal} {\bibinfo
				{journal} {Journal of Applied Physics}\ }\textbf {\bibinfo {volume} {30}},\
			\bibinfo {pages} {S70} (\bibinfo {year} {1959})}\BibitemShut {NoStop}%
		\bibitem [{\citenamefont {Brown}(1963)}]{Brown63}%
		\BibitemOpen
		\bibfield  {author} {\bibinfo {author} {\bibfnamefont {W.~F.}\ \bibnamefont
				{Brown}, \bibfnamefont {Jr.}},\ }\href@noop {} {\emph {\bibinfo {title}
				{Micromagnetics}}}\ (\bibinfo  {publisher} {Wiley, New York},\ \bibinfo
		{year} {1963})\BibitemShut {NoStop}%
		\bibitem [{\citenamefont {Arrott}\ \emph {et~al.}(1974)\citenamefont {Arrott},
			\citenamefont {Heinrich},\ and\ \citenamefont {Bloomberg}}]{Arrott74}%
		\BibitemOpen
		\bibfield  {author} {\bibinfo {author} {\bibfnamefont {A.}~\bibnamefont
				{Arrott}}, \bibinfo {author} {\bibfnamefont {B.}~\bibnamefont {Heinrich}},\
			and\ \bibinfo {author} {\bibfnamefont {D.}~\bibnamefont {Bloomberg}},\
		}\bibfield  {title} {\bibinfo {title} {Micromagnetics of magnetization
				processes in toroidal geometries},\ }\href
		{https://doi.org/10.1109/tmag.1974.1058423} {\bibfield  {journal} {\bibinfo
				{journal} {{IEEE} Transactions on Magnetics}\ }\textbf {\bibinfo {volume}
				{10}},\ \bibinfo {pages} {950} (\bibinfo {year} {1974})}\BibitemShut
		{NoStop}%
		\bibitem [{\citenamefont {Villain-Guillot}\ \emph {et~al.}(1994)\citenamefont
			{Villain-Guillot}, \citenamefont {Dandoloff},\ and\ \citenamefont
			{Saxena}}]{VillainGuillot94}%
		\BibitemOpen
		\bibfield  {author} {\bibinfo {author} {\bibfnamefont {S.}~\bibnamefont
				{Villain-Guillot}}, \bibinfo {author} {\bibfnamefont {R.}~\bibnamefont
				{Dandoloff}},\ and\ \bibinfo {author} {\bibfnamefont {A.}~\bibnamefont
				{Saxena}},\ }\bibfield  {title} {\bibinfo {title} {Heisenberg spins on an
				infinite cylinder: a geometrical effect of anisotropy},\ }\href
		{https://doi.org/10.1016/0375-9601(94)90473-1} {\bibfield  {journal}
			{\bibinfo  {journal} {Physics Letters A}\ }\textbf {\bibinfo {volume}
				{188}},\ \bibinfo {pages} {343} (\bibinfo {year} {1994})}\BibitemShut
		{NoStop}%
		\bibitem [{\citenamefont {Villain-Guillot}\ \emph {et~al.}(1995)\citenamefont
			{Villain-Guillot}, \citenamefont {Dandoloff}, \citenamefont {Saxena},\ and\
			\citenamefont {Bishop}}]{Villain-Guillot95}%
		\BibitemOpen
		\bibfield  {author} {\bibinfo {author} {\bibfnamefont {S.}~\bibnamefont
				{Villain-Guillot}}, \bibinfo {author} {\bibfnamefont {R.}~\bibnamefont
				{Dandoloff}}, \bibinfo {author} {\bibfnamefont {A.}~\bibnamefont {Saxena}},\
			and\ \bibinfo {author} {\bibfnamefont {A.~R.}\ \bibnamefont {Bishop}},\
		}\bibfield  {title} {\bibinfo {title} {Topological solitons and geometrical
				frustration},\ }\href {https://doi.org/10.1103/PhysRevB.52.6712} {\bibfield
			{journal} {\bibinfo  {journal} {Physical Review B}\ }\textbf {\bibinfo
				{volume} {52}},\ \bibinfo {pages} {6712} (\bibinfo {year}
			{1995})}\BibitemShut {NoStop}%
		\bibitem [{\citenamefont {Saxena}\ and\ \citenamefont
			{Dandoloff}(1997)}]{Saxena97}%
		\BibitemOpen
		\bibfield  {author} {\bibinfo {author} {\bibfnamefont {A.}~\bibnamefont
				{Saxena}}\ and\ \bibinfo {author} {\bibfnamefont {R.}~\bibnamefont
				{Dandoloff}},\ }\bibfield  {title} {\bibinfo {title} {Curvature-induced
				geometrical frustration in magnetic systems},\ }\href
		{https://doi.org/10.1103/physrevb.55.11049} {\bibfield  {journal} {\bibinfo
				{journal} {Physical Review B}\ }\textbf {\bibinfo {volume} {55}},\ \bibinfo
			{pages} {11049} (\bibinfo {year} {1997})}\BibitemShut {NoStop}%
		\bibitem [{\citenamefont {Benoit}\ and\ \citenamefont
			{Dandoloff}(1998)}]{Benoit98}%
		\BibitemOpen
		\bibfield  {author} {\bibinfo {author} {\bibfnamefont {J.}~\bibnamefont
				{Benoit}}\ and\ \bibinfo {author} {\bibfnamefont {R.}~\bibnamefont
				{Dandoloff}},\ }\bibfield  {title} {\bibinfo {title} {Heisenberg spins on an
				elastic torus section},\ }\href
		{https://doi.org/10.1016/s0375-9601(98)00657-4} {\bibfield  {journal}
			{\bibinfo  {journal} {Physics Letters A}\ }\textbf {\bibinfo {volume}
				{248}},\ \bibinfo {pages} {439} (\bibinfo {year} {1998})}\BibitemShut
		{NoStop}%
		\bibitem [{\citenamefont {Saxena}\ and\ \citenamefont
			{Dandoloff}(1998)}]{Saxena98}%
		\BibitemOpen
		\bibfield  {author} {\bibinfo {author} {\bibfnamefont {A.}~\bibnamefont
				{Saxena}}\ and\ \bibinfo {author} {\bibfnamefont {R.}~\bibnamefont
				{Dandoloff}},\ }\bibfield  {title} {\bibinfo {title} {Heisenberg spins on a
				cylinder in an axial magnetic field},\ }\href
		{https://doi.org/10.1103/physrevb.58.r563} {\bibfield  {journal} {\bibinfo
				{journal} {Physical Review B}\ }\textbf {\bibinfo {volume} {58}},\ \bibinfo
			{pages} {R563} (\bibinfo {year} {1998})}\BibitemShut {NoStop}%
		\bibitem [{\citenamefont {Dandoloff}\ and\ \citenamefont
			{Saxena}(2002)}]{Dandoloff02}%
		\BibitemOpen
		\bibfield  {author} {\bibinfo {author} {\bibfnamefont {R.}~\bibnamefont
				{Dandoloff}}\ and\ \bibinfo {author} {\bibfnamefont {A.}~\bibnamefont
				{Saxena}},\ }\bibfield  {title} {\bibinfo {title} {Skyrmions on an elastic
				cylinder},\ }\href {https://doi.org/10.1140/epjb/e2002-00298-2} {\bibfield
			{journal} {\bibinfo  {journal} {The European Physical Journal B - Condensed
					Matter}\ }\textbf {\bibinfo {volume} {29}},\ \bibinfo {pages} {265} (\bibinfo
			{year} {2002})}\BibitemShut {NoStop}%
		\bibitem [{\citenamefont {Escrig}\ \emph {et~al.}(2007)\citenamefont {Escrig},
			\citenamefont {Landeros}, \citenamefont {D.}, \citenamefont {Vogel},\ and\
			\citenamefont {Vargas}}]{Escrig07}%
		\BibitemOpen
		\bibfield  {author} {\bibinfo {author} {\bibfnamefont {J.}~\bibnamefont
				{Escrig}}, \bibinfo {author} {\bibfnamefont {P.}~\bibnamefont {Landeros}},
			\bibinfo {author} {\bibfnamefont {A.}~\bibnamefont {D.}}, \bibinfo {author}
			{\bibfnamefont {E.}~\bibnamefont {Vogel}},\ and\ \bibinfo {author}
			{\bibfnamefont {P.}~\bibnamefont {Vargas}},\ }\bibfield  {title} {\bibinfo
			{title} {Phase diagrams of magnetic nanotubes},\ }\href
		{https://doi.org/10.1016/j.jmmm.2006.05.019} {\bibfield  {journal} {\bibinfo
				{journal} {Journal of Magnetism and Magnetic Materials}\ }\textbf {\bibinfo
				{volume} {308}},\ \bibinfo {pages} {233} (\bibinfo {year}
			{2007})}\BibitemShut {NoStop}%
		\bibitem [{\citenamefont {Yershov}\ \emph {et~al.}(2020)\citenamefont
			{Yershov}, \citenamefont {Kravchuk}, \citenamefont {Sheka},\ and\
			\citenamefont {Roessler}}]{Yershov20}%
		\BibitemOpen
		\bibfield  {author} {\bibinfo {author} {\bibfnamefont {K.}~\bibnamefont
				{Yershov}}, \bibinfo {author} {\bibfnamefont {V.}~\bibnamefont {Kravchuk}},
			\bibinfo {author} {\bibfnamefont {D.}~\bibnamefont {Sheka}},\ and\ \bibinfo
			{author} {\bibfnamefont {U.}~\bibnamefont {Roessler}},\ }\bibfield  {title}
		{\bibinfo {title} {Curvature effects on phase transitions in chiral
				magnets},\ }\href {https://doi.org/10.21468/scipostphys.9.4.043} {\bibfield
			{journal} {\bibinfo  {journal} {{SciPost} Physics}\ }\textbf {\bibinfo
				{volume} {9}},\ \bibinfo {pages} {043} (\bibinfo {year} {2020})}\BibitemShut
		{NoStop}%
		\bibitem [{\citenamefont {Fratta}\ \emph {et~al.}(2023)\citenamefont {Fratta},
			\citenamefont {Fiorenza},\ and\ \citenamefont {Slastikov}}]{Fratta23}%
		\BibitemOpen
		\bibfield  {author} {\bibinfo {author} {\bibfnamefont {G.~D.}\ \bibnamefont
				{Fratta}}, \bibinfo {author} {\bibfnamefont {A.}~\bibnamefont {Fiorenza}},\
			and\ \bibinfo {author} {\bibfnamefont {V.}~\bibnamefont {Slastikov}},\
		}\bibfield  {title} {\bibinfo {title} {On symmetry of energy minimizing
				harmonic-type maps on cylindrical surfaces},\ }\href
		{https://doi.org/10.3934/mine.2023056} {\bibfield  {journal} {\bibinfo
				{journal} {Mathematics in Engineering}\ }\textbf {\bibinfo {volume} {5}},\
			\bibinfo {pages} {1} (\bibinfo {year} {2023})}\BibitemShut {NoStop}%
		\bibitem [{\citenamefont {Mancilla-Almonacid}\ \emph
			{et~al.}(2020)\citenamefont {Mancilla-Almonacid}, \citenamefont {Castro},
			\citenamefont {Fonseca}, \citenamefont {Altbir}, \citenamefont {Allende},\
			and\ \citenamefont {Carvalho-Santos}}]{Mancilla-Almonacid20}%
		\BibitemOpen
		\bibfield  {author} {\bibinfo {author} {\bibfnamefont {D.}~\bibnamefont
				{Mancilla-Almonacid}}, \bibinfo {author} {\bibfnamefont {M.}~\bibnamefont
				{Castro}}, \bibinfo {author} {\bibfnamefont {J.}~\bibnamefont {Fonseca}},
			\bibinfo {author} {\bibfnamefont {D.}~\bibnamefont {Altbir}}, \bibinfo
			{author} {\bibfnamefont {S.}~\bibnamefont {Allende}},\ and\ \bibinfo {author}
			{\bibfnamefont {V.}~\bibnamefont {Carvalho-Santos}},\ }\bibfield  {title}
		{\bibinfo {title} {Magnetic ground states for bent nanotubes},\ }\href
		{https://doi.org/10.1016/j.jmmm.2020.166754} {\bibfield  {journal} {\bibinfo
				{journal} {Journal of Magnetism and Magnetic Materials}\ }\textbf {\bibinfo
				{volume} {507}},\ \bibinfo {pages} {166754} (\bibinfo {year}
			{2020})}\BibitemShut {NoStop}%
		\bibitem [{\citenamefont {Popov}\ and\ \citenamefont
			{Zavislyak}(2008)}]{Popov08}%
		\BibitemOpen
		\bibfield  {author} {\bibinfo {author} {\bibfnamefont {M.}~\bibnamefont
				{Popov}}\ and\ \bibinfo {author} {\bibfnamefont {I.}~\bibnamefont
				{Zavislyak}},\ }\bibfield  {title} {\bibinfo {title} {Modeling of
				magnetostatic surface oscillations in elliptic nanotubes},\ }\href
		{http://archive.ujp.bitp.kiev.ua/index.php?item=j&id=107} {\bibfield
			{journal} {\bibinfo  {journal} {Ukr. J. Phys}\ }\textbf {\bibinfo {volume}
				{53}},\ \bibinfo {pages} {702} (\bibinfo {year} {2008})}\BibitemShut
		{NoStop}%
		\bibitem [{\citenamefont {Gonz\'{a}lez}\ \emph {et~al.}(2010)\citenamefont
			{Gonz\'{a}lez}, \citenamefont {Landeros},\ and\ \citenamefont {\'{A}lvaro
				S.~N\'{u}\~{n}ez}}]{Gonzalez10}%
		\BibitemOpen
		\bibfield  {author} {\bibinfo {author} {\bibfnamefont {A.}~\bibnamefont
				{Gonz\'{a}lez}}, \bibinfo {author} {\bibfnamefont {P.}~\bibnamefont
				{Landeros}},\ and\ \bibinfo {author} {\bibnamefont {\'{A}lvaro
					S.~N\'{u}\~{n}ez}},\ }\bibfield  {title} {\bibinfo {title} {Spin wave
				spectrum of magnetic nanotubes},\ }\href
		{https://doi.org/http://dx.doi.org/10.1016/j.jmmm.2009.10.010} {\bibfield
			{journal} {\bibinfo  {journal} {Journal of Magnetism and Magnetic Materials}\
			}\textbf {\bibinfo {volume} {322}},\ \bibinfo {pages} {530} (\bibinfo {year}
			{2010})}\BibitemShut {NoStop}%
		\bibitem [{\citenamefont {Yan}\ \emph {et~al.}(2011)\citenamefont {Yan},
			\citenamefont {Andreas}, \citenamefont {K{\'a}kay}, \citenamefont
			{Garc{\'i}a-S{\'a}nchez},\ and\ \citenamefont {Hertel}}]{Yan11}%
		\BibitemOpen
		\bibfield  {author} {\bibinfo {author} {\bibfnamefont {M.}~\bibnamefont
				{Yan}}, \bibinfo {author} {\bibfnamefont {C.}~\bibnamefont {Andreas}},
			\bibinfo {author} {\bibfnamefont {A.}~\bibnamefont {K{\'a}kay}}, \bibinfo
			{author} {\bibfnamefont {F.}~\bibnamefont {Garc{\'i}a-S{\'a}nchez}},\ and\
			\bibinfo {author} {\bibfnamefont {R.}~\bibnamefont {Hertel}},\ }\bibfield
		{title} {\bibinfo {title} {Fast domain wall dynamics in magnetic nanotubes:
				Suppression of {W}alker breakdown and {C}herenkov-like spin wave emission},\
		}\href {https://doi.org/10.1063/1.3643037} {\bibfield  {journal} {\bibinfo
				{journal} {Applied Physics Letters}\ }\textbf {\bibinfo {volume} {99}},\
			\bibinfo {pages} {122505} (\bibinfo {year} {2011})}\BibitemShut {NoStop}%
		\bibitem [{\citenamefont {Ot\'alora}\ \emph {et~al.}(2016)\citenamefont
			{Ot\'alora}, \citenamefont {Yan}, \citenamefont {Schultheiss}, \citenamefont
			{Hertel},\ and\ \citenamefont {K\'akay}}]{Otalora16}%
		\BibitemOpen
		\bibfield  {author} {\bibinfo {author} {\bibfnamefont {J.~A.}\ \bibnamefont
				{Ot\'alora}}, \bibinfo {author} {\bibfnamefont {M.}~\bibnamefont {Yan}},
			\bibinfo {author} {\bibfnamefont {H.}~\bibnamefont {Schultheiss}}, \bibinfo
			{author} {\bibfnamefont {R.}~\bibnamefont {Hertel}},\ and\ \bibinfo {author}
			{\bibfnamefont {A.}~\bibnamefont {K\'akay}},\ }\bibfield  {title} {\bibinfo
			{title} {Curvature-induced asymmetric spin-wave dispersion},\ }\href
		{https://doi.org/10.1103/PhysRevLett.117.227203} {\bibfield  {journal}
			{\bibinfo  {journal} {Physical Review Letters}\ }\textbf {\bibinfo {volume}
				{117}},\ \bibinfo {pages} {227203} (\bibinfo {year} {2016})}\BibitemShut
		{NoStop}%
		\bibitem [{\citenamefont {K\"{o}rber}\ \emph {et~al.}(2022)\citenamefont
			{K\"{o}rber}, \citenamefont {Verba}, \citenamefont {Ot\'{a}lora},
			\citenamefont {Kravchuk}, \citenamefont {Lindner}, \citenamefont
			{Fassbender},\ and\ \citenamefont {K\'{a}kay}}]{Korber22}%
		\BibitemOpen
		\bibfield  {author} {\bibinfo {author} {\bibfnamefont {L.}~\bibnamefont
				{K\"{o}rber}}, \bibinfo {author} {\bibfnamefont {R.}~\bibnamefont {Verba}},
			\bibinfo {author} {\bibfnamefont {J.~A.}\ \bibnamefont {Ot\'{a}lora}},
			\bibinfo {author} {\bibfnamefont {V.}~\bibnamefont {Kravchuk}}, \bibinfo
			{author} {\bibfnamefont {J.}~\bibnamefont {Lindner}}, \bibinfo {author}
			{\bibfnamefont {J.}~\bibnamefont {Fassbender}},\ and\ \bibinfo {author}
			{\bibfnamefont {A.}~\bibnamefont {K\'{a}kay}},\ }\bibfield  {title} {\bibinfo
			{title} {Curvilinear spin-wave dynamics beyond the thin-shell approximation:
				Magnetic nanotubes as a case study},\ }\href
		{https://doi.org/10.1103/physrevb.106.014405} {\bibfield  {journal} {\bibinfo
				{journal} {Physical Review B}\ }\textbf {\bibinfo {volume} {106}},\ \bibinfo
			{pages} {014405} (\bibinfo {year} {2022})}\BibitemShut {NoStop}%
		\bibitem [{\citenamefont {Saavedra}\ \emph {et~al.}(2022)\citenamefont
			{Saavedra}, \citenamefont {Castillo-Sep\'{u}lveda}, \citenamefont {Corona},
			\citenamefont {Altbir}, \citenamefont {Escrig},\ and\ \citenamefont
			{Carvalho-Santos}}]{Saavedra22}%
		\BibitemOpen
		\bibfield  {author} {\bibinfo {author} {\bibfnamefont {E.}~\bibnamefont
				{Saavedra}}, \bibinfo {author} {\bibfnamefont {S.}~\bibnamefont
				{Castillo-Sep\'{u}lveda}}, \bibinfo {author} {\bibfnamefont {R.}~\bibnamefont
				{Corona}}, \bibinfo {author} {\bibfnamefont {D.}~\bibnamefont {Altbir}},
			\bibinfo {author} {\bibfnamefont {J.}~\bibnamefont {Escrig}},\ and\ \bibinfo
			{author} {\bibfnamefont {V.}~\bibnamefont {Carvalho-Santos}},\ }\bibfield
		{title} {\bibinfo {title} {Influence of curvature on the dynamical
				susceptibility of bent nanotubes},\ }\href
		{https://doi.org/10.1016/j.rinp.2022.105290} {\bibfield  {journal} {\bibinfo
				{journal} {Results in Physics}\ }\textbf {\bibinfo {volume} {35}},\ \bibinfo
			{pages} {105290} (\bibinfo {year} {2022})}\BibitemShut {NoStop}%
		\bibitem [{\citenamefont {Yang}\ \emph {et~al.}(2016)\citenamefont {Yang},
			\citenamefont {Kim}, \citenamefont {Kim}, \citenamefont {Cho}, \citenamefont
			{Lee},\ and\ \citenamefont {Kim}}]{Yang16a}%
		\BibitemOpen
		\bibfield  {author} {\bibinfo {author} {\bibfnamefont {J.}~\bibnamefont
				{Yang}}, \bibinfo {author} {\bibfnamefont {J.}~\bibnamefont {Kim}}, \bibinfo
			{author} {\bibfnamefont {B.}~\bibnamefont {Kim}}, \bibinfo {author}
			{\bibfnamefont {Y.-J.}\ \bibnamefont {Cho}}, \bibinfo {author} {\bibfnamefont
				{J.-H.}\ \bibnamefont {Lee}},\ and\ \bibinfo {author} {\bibfnamefont {S.-K.}\
				\bibnamefont {Kim}},\ }\bibfield  {title} {\bibinfo {title} {Characteristic
				dynamic modes and domain-wall motion in magnetic nanotubes excited by
				resonant rotating magnetic fields},\ }\href
		{https://doi.org/10.1103/physrevb.94.024425} {\bibfield  {journal} {\bibinfo
				{journal} {Physical Review B}\ }\textbf {\bibinfo {volume} {94}},\ \bibinfo
			{pages} {024425} (\bibinfo {year} {2016})}\BibitemShut {NoStop}%
		\bibitem [{\citenamefont {Saavedra}\ \emph {et~al.}(2025)\citenamefont
			{Saavedra}, \citenamefont {Saji}, \citenamefont {Laroze}, \citenamefont
			{Allende},\ and\ \citenamefont {Castro}}]{Saavedra25a}%
		\BibitemOpen
		\bibfield  {author} {\bibinfo {author} {\bibfnamefont {E.}~\bibnamefont
				{Saavedra}}, \bibinfo {author} {\bibfnamefont {C.}~\bibnamefont {Saji}},
			\bibinfo {author} {\bibfnamefont {D.}~\bibnamefont {Laroze}}, \bibinfo
			{author} {\bibfnamefont {S.}~\bibnamefont {Allende}},\ and\ \bibinfo {author}
			{\bibfnamefont {M.~A.}\ \bibnamefont {Castro}},\ }\bibfield  {title}
		{\bibinfo {title} {Curvature-dependent spin-wave resonance of magnetic
				bimerons in cylindrical nanotubes},\ }\href
		{https://doi.org/10.1063/5.0300885} {\bibfield  {journal} {\bibinfo
				{journal} {Applied Physics Letters}\ }\textbf {\bibinfo {volume} {127}},\
			\bibinfo {pages} {252406} (\bibinfo {year} {2025})}\BibitemShut {NoStop}%
		\bibitem [{\citenamefont {Usov}\ \emph {et~al.}(2007)\citenamefont {Usov},
			\citenamefont {Zhukov},\ and\ \citenamefont {Gonzalez}}]{Usov07}%
		\BibitemOpen
		\bibfield  {author} {\bibinfo {author} {\bibfnamefont {N.}~\bibnamefont
				{Usov}}, \bibinfo {author} {\bibfnamefont {A.}~\bibnamefont {Zhukov}},\ and\
			\bibinfo {author} {\bibfnamefont {J.}~\bibnamefont {Gonzalez}},\ }\bibfield
		{title} {\bibinfo {title} {Domain walls and magnetization reversal process in
				soft magnetic nanowires and nanotubes},\ }\href
		{https://doi.org/10.1016/j.jmmm.2007.02.138} {\bibfield  {journal} {\bibinfo
				{journal} {Journal of Magnetism and Magnetic Materials}\ }\textbf {\bibinfo
				{volume} {316}},\ \bibinfo {pages} {255} (\bibinfo {year}
			{2007})}\BibitemShut {NoStop}%
		\bibitem [{\citenamefont {Yan}\ \emph {et~al.}(2010)\citenamefont {Yan},
			\citenamefont {K\'akay}, \citenamefont {Gliga},\ and\ \citenamefont
			{Hertel}}]{Yan10}%
		\BibitemOpen
		\bibfield  {author} {\bibinfo {author} {\bibfnamefont {M.}~\bibnamefont
				{Yan}}, \bibinfo {author} {\bibfnamefont {A.}~\bibnamefont {K\'akay}},
			\bibinfo {author} {\bibfnamefont {S.}~\bibnamefont {Gliga}},\ and\ \bibinfo
			{author} {\bibfnamefont {R.}~\bibnamefont {Hertel}},\ }\bibfield  {title}
		{\bibinfo {title} {Beating the {Walker} limit with massless domain walls in
				cylindrical nanowires},\ }\href
		{https://doi.org/10.1103/PhysRevLett.104.057201} {\bibfield  {journal}
			{\bibinfo  {journal} {Physical Review Letters}\ }\textbf {\bibinfo {volume}
				{104}},\ \bibinfo {pages} {057201} (\bibinfo {year} {2010})}\BibitemShut
		{NoStop}%
		\bibitem [{\citenamefont {Ot\'{a}lora}\ \emph {et~al.}(2013)\citenamefont
			{Ot\'{a}lora}, \citenamefont {L\'{o}pez-L\'{o}pez}, \citenamefont {Landeros},
			\citenamefont {Vargas},\ and\ \citenamefont {A.S.}}]{Otalora13}%
		\BibitemOpen
		\bibfield  {author} {\bibinfo {author} {\bibfnamefont {J.}~\bibnamefont
				{Ot\'{a}lora}}, \bibinfo {author} {\bibfnamefont {J.}~\bibnamefont
				{L\'{o}pez-L\'{o}pez}}, \bibinfo {author} {\bibfnamefont {P.}~\bibnamefont
				{Landeros}}, \bibinfo {author} {\bibfnamefont {P.}~\bibnamefont {Vargas}},\
			and\ \bibinfo {author} {\bibfnamefont {N.~n.}\ \bibnamefont {A.S.}},\
		}\bibfield  {title} {\bibinfo {title} {Breaking of chiral symmetry in vortex
				domain wall propagation in ferromagnetic nanotubes},\ }\href
		{https://doi.org/10.1016/j.jmmm.2013.04.014} {\bibfield  {journal} {\bibinfo
				{journal} {Journal of Magnetism and Magnetic Materials}\ }\textbf {\bibinfo
				{volume} {341}},\ \bibinfo {pages} {86} (\bibinfo {year} {2013})}\BibitemShut
		{NoStop}%
		\bibitem [{\citenamefont {Nunes}\ \emph {et~al.}(2026)\citenamefont {Nunes},
			\citenamefont {Castillo-Sepulveda}, \citenamefont {Costilla}, \citenamefont
			{de~Souza}, \citenamefont {Corona}, \citenamefont {Altbir},\ and\
			\citenamefont {Carvalho-Santos}}]{Nunes26}%
		\BibitemOpen
		\bibfield  {author} {\bibinfo {author} {\bibfnamefont {J.~V.}\ \bibnamefont
				{Nunes}}, \bibinfo {author} {\bibfnamefont {S.}~\bibnamefont
				{Castillo-Sepulveda}}, \bibinfo {author} {\bibfnamefont {J.~I.}\ \bibnamefont
				{Costilla}}, \bibinfo {author} {\bibfnamefont {S.~F.}\ \bibnamefont
				{de~Souza}}, \bibinfo {author} {\bibfnamefont {R.~M.}\ \bibnamefont
				{Corona}}, \bibinfo {author} {\bibfnamefont {D.}~\bibnamefont {Altbir}},\
			and\ \bibinfo {author} {\bibfnamefont {V.~L.}\ \bibnamefont
				{Carvalho-Santos}},\ }\bibfield  {title} {\bibinfo {title} {Curvature-induced
				effects on the vortex domain wall properties in bent nanotubes},\ }\href
		{https://doi.org/10.1038/s44306-026-00127-1} {\bibfield  {journal} {\bibinfo
				{journal} {npj Spintronics}\ }\textbf {\bibinfo {volume} {4}},\ \bibinfo
			{pages} {7} (\bibinfo {year} {2026})}\BibitemShut {NoStop}%
		\bibitem [{\citenamefont {Landeros}\ and\ \citenamefont
			{N\'{u}{\~n}ez}(2010)}]{Landeros10}%
		\BibitemOpen
		\bibfield  {author} {\bibinfo {author} {\bibfnamefont {P.}~\bibnamefont
				{Landeros}}\ and\ \bibinfo {author} {\bibfnamefont {A.~S.}\ \bibnamefont
				{N\'{u}{\~n}ez}},\ }\bibfield  {title} {\bibinfo {title} {Domain wall motion
				on magnetic nanotubes},\ }\href {https://doi.org/10.1063/1.3466747}
		{\bibfield  {journal} {\bibinfo  {journal} {Journal of Applied Physics}\
			}\textbf {\bibinfo {volume} {108}},\ \bibinfo {pages} {033917} (\bibinfo
			{year} {2010})}\BibitemShut {NoStop}%
		\bibitem [{\citenamefont {Ot\'{a}lora}\ \emph {et~al.}(2012)\citenamefont
			{Ot\'{a}lora}, \citenamefont {L\'{o}pez-L\'{o}pez}, \citenamefont {Vargas},\
			and\ \citenamefont {Landeros}}]{Otalora12a}%
		\BibitemOpen
		\bibfield  {author} {\bibinfo {author} {\bibfnamefont {J.}~\bibnamefont
				{Ot\'{a}lora}}, \bibinfo {author} {\bibfnamefont {J.}~\bibnamefont
				{L\'{o}pez-L\'{o}pez}}, \bibinfo {author} {\bibfnamefont {P.}~\bibnamefont
				{Vargas}},\ and\ \bibinfo {author} {\bibfnamefont {P.}~\bibnamefont
				{Landeros}},\ }\bibfield  {title} {\bibinfo {title} {Chirality switching and
				propagation control of a vortex domain wall in ferromagnetic nanotubes},\
		}\href {https://doi.org/10.1063/1.3687154} {\bibfield  {journal} {\bibinfo
				{journal} {Applied Physics Letters}\ }\textbf {\bibinfo {volume} {100}},\
			\bibinfo {pages} {072407} (\bibinfo {year} {2012})}\BibitemShut {NoStop}%
		\bibitem [{\citenamefont {Saxena}\ and\ \citenamefont
			{Dandoloff}(2002)}]{Saxena02}%
		\BibitemOpen
		\bibfield  {author} {\bibinfo {author} {\bibfnamefont {A.}~\bibnamefont
				{Saxena}}\ and\ \bibinfo {author} {\bibfnamefont {R.}~\bibnamefont
				{Dandoloff}},\ }\bibfield  {title} {\bibinfo {title} {Stabilization of
				half-skyrmions: {H}eisenberg spins on a non-simply connected manifold},\
		}\href {https://doi.org/10.1103/PhysRevB.66.104414} {\bibfield  {journal}
			{\bibinfo  {journal} {Physical Review B}\ }\textbf {\bibinfo {volume} {66}},\
			\bibinfo {pages} {104414} (\bibinfo {year} {2002})}\BibitemShut {NoStop}%
		\bibitem [{\citenamefont {Freitas}\ \emph {et~al.}(2005)\citenamefont
			{Freitas}, \citenamefont {Moura-Melo},\ and\ \citenamefont
			{Pereira}}]{Freitas05}%
		\BibitemOpen
		\bibfield  {author} {\bibinfo {author} {\bibfnamefont {W.~A.}\ \bibnamefont
				{Freitas}}, \bibinfo {author} {\bibfnamefont {W.~A.}\ \bibnamefont
				{Moura-Melo}},\ and\ \bibinfo {author} {\bibfnamefont {A.}~\bibnamefont
				{Pereira}},\ }\bibfield  {title} {\bibinfo {title} {Heisenberg spins on a
				cone: an interplay between geometry and magnetism},\ }\href
		{https://doi.org/10.1016/j.physleta.2004.12.039} {\bibfield  {journal}
			{\bibinfo  {journal} {Physics Letters A}\ }\textbf {\bibinfo {volume}
				{336}},\ \bibinfo {pages} {412} (\bibinfo {year} {2005})}\BibitemShut
		{NoStop}%
		\bibitem [{\citenamefont {Milagre}\ and\ \citenamefont
			{Moura-Melo}(2007)}]{Milagre07a}%
		\BibitemOpen
		\bibfield  {author} {\bibinfo {author} {\bibfnamefont {G.}~\bibnamefont
				{Milagre}}\ and\ \bibinfo {author} {\bibfnamefont {W.~A.}\ \bibnamefont
				{Moura-Melo}},\ }\bibfield  {title} {\bibinfo {title} {Magnetic vortex-like
				excitations on a sphere},\ }\href
		{https://doi.org/10.1016/j.physleta.2007.04.002} {\bibfield  {journal}
			{\bibinfo  {journal} {Physics Letters A}\ }\textbf {\bibinfo {volume}
				{368}},\ \bibinfo {pages} {155} (\bibinfo {year} {2007})}\BibitemShut
		{NoStop}%
		\bibitem [{\citenamefont {Belo}\ \emph {et~al.}(2007)\citenamefont {Belo},
			\citenamefont {Oliveira-Neto}, \citenamefont {Moura-Melo}, \citenamefont
			{Pereira},\ and\ \citenamefont {Ercolessi}}]{Belo07}%
		\BibitemOpen
		\bibfield  {author} {\bibinfo {author} {\bibfnamefont {L.}~\bibnamefont
				{Belo}}, \bibinfo {author} {\bibfnamefont {N.}~\bibnamefont {Oliveira-Neto}},
			\bibinfo {author} {\bibfnamefont {W.}~\bibnamefont {Moura-Melo}}, \bibinfo
			{author} {\bibfnamefont {A.}~\bibnamefont {Pereira}},\ and\ \bibinfo {author}
			{\bibfnamefont {E.}~\bibnamefont {Ercolessi}},\ }\bibfield  {title} {\bibinfo
			{title} {Heisenberg model on a space with negative curvature: Topological
				spin textures on the pseudosphere},\ }\href
		{https://doi.org/10.1016/j.physleta.2007.01.044} {\bibfield  {journal}
			{\bibinfo  {journal} {Physics Letters A}\ }\textbf {\bibinfo {volume}
				{365}},\ \bibinfo {pages} {463} (\bibinfo {year} {2007})}\BibitemShut
		{NoStop}%
		\bibitem [{\citenamefont {Carvalho-Santos}\ \emph {et~al.}(2008)\citenamefont
			{Carvalho-Santos}, \citenamefont {Moura}, \citenamefont {Moura-Melo},\ and\
			\citenamefont {Pereira}}]{Carvalho-Santos08}%
		\BibitemOpen
		\bibfield  {author} {\bibinfo {author} {\bibfnamefont {V.~L.}\ \bibnamefont
				{Carvalho-Santos}}, \bibinfo {author} {\bibfnamefont {A.~R.}\ \bibnamefont
				{Moura}}, \bibinfo {author} {\bibfnamefont {W.~A.}\ \bibnamefont
				{Moura-Melo}},\ and\ \bibinfo {author} {\bibfnamefont {A.~R.}\ \bibnamefont
				{Pereira}},\ }\bibfield  {title} {\bibinfo {title} {Topological spin
				excitations on a rigid torus},\ }\href
		{https://doi.org/10.1103/PhysRevB.77.134450} {\bibfield  {journal} {\bibinfo
				{journal} {Physical Review B}\ }\textbf {\bibinfo {volume} {77}},\ \bibinfo
			{pages} {134450} (\bibinfo {year} {2008})}\BibitemShut {NoStop}%
		\bibitem [{\citenamefont {Carvalho-Santos}\ and\ \citenamefont
			{Dandoloff}(2012)}]{Carvalho-Santos12}%
		\BibitemOpen
		\bibfield  {author} {\bibinfo {author} {\bibfnamefont {V.~L.}\ \bibnamefont
				{Carvalho-Santos}}\ and\ \bibinfo {author} {\bibfnamefont {R.}~\bibnamefont
				{Dandoloff}},\ }\bibfield  {title} {\bibinfo {title} {Coupling between
				magnetic field and curvature in heisenberg spins on surfaces with rotational
				symmetry},\ }\href {https://doi.org/10.1016/j.physleta.2012.10.027}
		{\bibfield  {journal} {\bibinfo  {journal} {Physics Letters A}\ }\textbf
			{\bibinfo {volume} {376}},\ \bibinfo {pages} {3551} (\bibinfo {year}
			{2012})}\BibitemShut {NoStop}%
		\bibitem [{\citenamefont {Carvalho-Santos}\ and\ \citenamefont
			{Dandoloff}(2013)}]{Carvalho-Santos13a}%
		\BibitemOpen
		\bibfield  {author} {\bibinfo {author} {\bibfnamefont {V.~L.}\ \bibnamefont
				{Carvalho-Santos}}\ and\ \bibinfo {author} {\bibfnamefont {R.}~\bibnamefont
				{Dandoloff}},\ }\bibfield  {title} {\bibinfo {title} {Topological spin
				excitations induced by an external magnetic field coupled to a surface with
				rotational symmetry},\ }\href {https://doi.org/10.1007/s13538-013-0126-1}
		{\bibfield  {journal} {\bibinfo  {journal} {Brazilian Journal of Physics}\
			}\textbf {\bibinfo {volume} {43}},\ \bibinfo {pages} {130} (\bibinfo {year}
			{2013})}\BibitemShut {NoStop}%
		\bibitem [{\citenamefont {Vilas-Boas}\ \emph {et~al.}(2015)\citenamefont
			{Vilas-Boas}, \citenamefont {Elias}, \citenamefont {Altbir}, \citenamefont
			{Fonseca},\ and\ \citenamefont {Carvalho-Santos}}]{Vilas-Boas15}%
		\BibitemOpen
		\bibfield  {author} {\bibinfo {author} {\bibfnamefont {P.~S.}\ \bibnamefont
				{Vilas-Boas}}, \bibinfo {author} {\bibfnamefont {R.~G.}\ \bibnamefont
				{Elias}}, \bibinfo {author} {\bibfnamefont {D.}~\bibnamefont {Altbir}},
			\bibinfo {author} {\bibfnamefont {J.~M.}\ \bibnamefont {Fonseca}},\ and\
			\bibinfo {author} {\bibfnamefont {V.~L.}\ \bibnamefont {Carvalho-Santos}},\
		}\bibfield  {title} {\bibinfo {title} {Topological magnetic solitons on a
				paraboloidal shell},\ }\href {https://doi.org/10.1016/j.physleta.2014.10.025}
		{\bibfield  {journal} {\bibinfo  {journal} {Physics Letters A}\ }\textbf
			{\bibinfo {volume} {379}},\ \bibinfo {pages} {47} (\bibinfo {year}
			{2015})}\BibitemShut {NoStop}%
		\bibitem [{\citenamefont {Carvalho-Santos}\ \emph
			{et~al.}(2015{\natexlab{a}})\citenamefont {Carvalho-Santos}, \citenamefont
			{Elias}, \citenamefont {Altbir},\ and\ \citenamefont
			{Fonseca}}]{CarvalhoSantos15}%
		\BibitemOpen
		\bibfield  {author} {\bibinfo {author} {\bibfnamefont {V.}~\bibnamefont
				{Carvalho-Santos}}, \bibinfo {author} {\bibfnamefont {R.}~\bibnamefont
				{Elias}}, \bibinfo {author} {\bibfnamefont {D.}~\bibnamefont {Altbir}},\ and\
			\bibinfo {author} {\bibfnamefont {J.}~\bibnamefont {Fonseca}},\ }\bibfield
		{title} {\bibinfo {title} {Stability of skyrmions on curved surfaces in the
				presence of a magnetic field},\ }\href
		{https://doi.org/10.1016/j.jmmm.2015.04.078} {\bibfield  {journal} {\bibinfo
				{journal} {Journal of Magnetism and Magnetic Materials}\ }\textbf {\bibinfo
				{volume} {391}},\ \bibinfo {pages} {179} (\bibinfo {year}
			{2015}{\natexlab{a}})}\BibitemShut {NoStop}%
		\bibitem [{\citenamefont {Parkin}\ \emph {et~al.}(2008)\citenamefont {Parkin},
			\citenamefont {Hayashi},\ and\ \citenamefont {Thomas}}]{Parkin08}%
		\BibitemOpen
		\bibfield  {author} {\bibinfo {author} {\bibfnamefont {S.~S.~P.}\
				\bibnamefont {Parkin}}, \bibinfo {author} {\bibfnamefont {M.}~\bibnamefont
				{Hayashi}},\ and\ \bibinfo {author} {\bibfnamefont {L.}~\bibnamefont
				{Thomas}},\ }\bibfield  {title} {\bibinfo {title} {{M}agnetic domain-wall
				racetrack memory},\ }\href {https://doi.org/10.1126/science.1145799}
		{\bibfield  {journal} {\bibinfo  {journal} {Science}\ }\textbf {\bibinfo
				{volume} {320}},\ \bibinfo {pages} {190} (\bibinfo {year}
			{2008})}\BibitemShut {NoStop}%
		\bibitem [{\citenamefont {Streubel}\ \emph {et~al.}(2016)\citenamefont
			{Streubel}, \citenamefont {Fischer}, \citenamefont {Kronast}, \citenamefont
			{Kravchuk}, \citenamefont {Sheka}, \citenamefont {Gaididei}, \citenamefont
			{Schmidt},\ and\ \citenamefont {Makarov}}]{Streubel16}%
		\BibitemOpen
		\bibfield  {author} {\bibinfo {author} {\bibfnamefont {R.}~\bibnamefont
				{Streubel}}, \bibinfo {author} {\bibfnamefont {P.}~\bibnamefont {Fischer}},
			\bibinfo {author} {\bibfnamefont {F.}~\bibnamefont {Kronast}}, \bibinfo
			{author} {\bibfnamefont {V.~P.}\ \bibnamefont {Kravchuk}}, \bibinfo {author}
			{\bibfnamefont {D.~D.}\ \bibnamefont {Sheka}}, \bibinfo {author}
			{\bibfnamefont {Y.}~\bibnamefont {Gaididei}}, \bibinfo {author}
			{\bibfnamefont {O.~G.}\ \bibnamefont {Schmidt}},\ and\ \bibinfo {author}
			{\bibfnamefont {D.}~\bibnamefont {Makarov}},\ }\bibfield  {title} {\bibinfo
			{title} {Magnetism in curved geometries ({T}opical {R}eview)},\ }\href
		{https://doi.org/10.1088/0022-3727/49/36/363001} {\bibfield  {journal}
			{\bibinfo  {journal} {Journal of Physics D: Applied Physics}\ }\textbf
			{\bibinfo {volume} {49}},\ \bibinfo {pages} {363001} (\bibinfo {year}
			{2016})}\BibitemShut {NoStop}%
		\bibitem [{\citenamefont {Fern{\'a}ndez-Pacheco}\ \emph
			{et~al.}(2017)\citenamefont {Fern{\'a}ndez-Pacheco}, \citenamefont
			{Streubel}, \citenamefont {Fruchart}, \citenamefont {Hertel}, \citenamefont
			{Fischer},\ and\ \citenamefont {Cowburn}}]{Fernandez-Pacheco17}%
		\BibitemOpen
		\bibfield  {author} {\bibinfo {author} {\bibfnamefont {A.}~\bibnamefont
				{Fern{\'a}ndez-Pacheco}}, \bibinfo {author} {\bibfnamefont {R.}~\bibnamefont
				{Streubel}}, \bibinfo {author} {\bibfnamefont {O.}~\bibnamefont {Fruchart}},
			\bibinfo {author} {\bibfnamefont {R.}~\bibnamefont {Hertel}}, \bibinfo
			{author} {\bibfnamefont {P.}~\bibnamefont {Fischer}},\ and\ \bibinfo {author}
			{\bibfnamefont {R.~P.}\ \bibnamefont {Cowburn}},\ }\bibfield  {title}
		{\bibinfo {title} {Three-dimensional nanomagnetism},\ }\href
		{https://doi.org/10.1038/ncomms15756} {\bibfield  {journal} {\bibinfo
				{journal} {Nature Communications}\ }\textbf {\bibinfo {volume} {8}},\
			\bibinfo {pages} {15756} (\bibinfo {year} {2017})}\BibitemShut {NoStop}%
		\bibitem [{\citenamefont {Hertel}(2013)}]{Hertel13a}%
		\BibitemOpen
		\bibfield  {author} {\bibinfo {author} {\bibfnamefont {R.}~\bibnamefont
				{Hertel}},\ }\bibfield  {title} {\bibinfo {title} {Curvature--induced
				magnetochirality},\ }\href {https://doi.org/10.1142/s2010324713400092}
		{\bibfield  {journal} {\bibinfo  {journal} {SPIN}\ }\textbf {\bibinfo
				{volume} {03}},\ \bibinfo {pages} {1340009} (\bibinfo {year}
			{2013})}\BibitemShut {NoStop}%
		\bibitem [{\citenamefont {Fischer}\ \emph {et~al.}(2020)\citenamefont
			{Fischer}, \citenamefont {Sanz-Hern{\'{a}}ndez}, \citenamefont {Streubel},\
			and\ \citenamefont {Fern{\'{a}}ndez-Pacheco}}]{Fischer20}%
		\BibitemOpen
		\bibfield  {author} {\bibinfo {author} {\bibfnamefont {P.}~\bibnamefont
				{Fischer}}, \bibinfo {author} {\bibfnamefont {D.}~\bibnamefont
				{Sanz-Hern{\'{a}}ndez}}, \bibinfo {author} {\bibfnamefont {R.}~\bibnamefont
				{Streubel}},\ and\ \bibinfo {author} {\bibfnamefont {A.}~\bibnamefont
				{Fern{\'{a}}ndez-Pacheco}},\ }\bibfield  {title} {\bibinfo {title} {Launching
				a new dimension with 3{D} magnetic nanostructures},\ }\href
		{https://doi.org/10.1063/1.5134474} {\bibfield  {journal} {\bibinfo
				{journal} {{APL} Materials}\ }\textbf {\bibinfo {volume} {8}},\ \bibinfo
			{pages} {010701} (\bibinfo {year} {2020})}\BibitemShut {NoStop}%
		\bibitem [{\citenamefont {Makarov}\ and\ \citenamefont
			{Sheka}(2022)}]{Makarov22a}%
		\BibitemOpen
		\bibinfo {editor} {\bibfnamefont {D.}~\bibnamefont {Makarov}}\ and\ \bibinfo
		{editor} {\bibfnamefont {D.~D.}\ \bibnamefont {Sheka}},\ eds.,\ \href
		{https://doi.org/10.1007/978-3-031-09086-8} {\emph {\bibinfo {title}
				{Curvilinear Micromagnetism}}}\ (\bibinfo  {publisher} {Springer
			International Publishing},\ \bibinfo {address} {Springer Cham},\ \bibinfo
		{year} {2022})\BibitemShut {NoStop}%
		\bibitem [{\citenamefont {Sheka}(2021)}]{Sheka21b}%
		\BibitemOpen
		\bibfield  {author} {\bibinfo {author} {\bibfnamefont {D.~D.}\ \bibnamefont
				{Sheka}},\ }\bibfield  {title} {\bibinfo {title} {A perspective on
				curvilinear magnetism},\ }\href {https://doi.org/10.1063/5.0048891}
		{\bibfield  {journal} {\bibinfo  {journal} {Applied Physics Letters}\
			}\textbf {\bibinfo {volume} {118}},\ \bibinfo {pages} {230502} (\bibinfo
			{year} {2021})}\BibitemShut {NoStop}%
		\bibitem [{\citenamefont {Sheka}\ \emph
			{et~al.}(2022{\natexlab{a}})\citenamefont {Sheka}, \citenamefont
			{Pylypovskyi}, \citenamefont {Volkov}, \citenamefont {Yershov}, \citenamefont
			{Kravchuk},\ and\ \citenamefont {Makarov}}]{Sheka22}%
		\BibitemOpen
		\bibfield  {author} {\bibinfo {author} {\bibfnamefont {D.~D.}\ \bibnamefont
				{Sheka}}, \bibinfo {author} {\bibfnamefont {O.~V.}\ \bibnamefont
				{Pylypovskyi}}, \bibinfo {author} {\bibfnamefont {O.~M.}\ \bibnamefont
				{Volkov}}, \bibinfo {author} {\bibfnamefont {K.~V.}\ \bibnamefont {Yershov}},
			\bibinfo {author} {\bibfnamefont {V.~P.}\ \bibnamefont {Kravchuk}},\ and\
			\bibinfo {author} {\bibfnamefont {D.}~\bibnamefont {Makarov}},\ }\bibfield
		{title} {\bibinfo {title} {Fundamentals of curvilinear ferromagnetism:
				Statics and dynamics of geometrically curved wires and narrow ribbons},\
		}\href {https://doi.org/10.1002/smll.202105219} {\bibfield  {journal}
			{\bibinfo  {journal} {Small}\ }\textbf {\bibinfo {volume} {18}},\ \bibinfo
			{pages} {2105219} (\bibinfo {year} {2022}{\natexlab{a}})}\BibitemShut
		{NoStop}%
		\bibitem [{\citenamefont {Gubbiotti}\ \emph {et~al.}(2024)\citenamefont
			{Gubbiotti}, \citenamefont {Barman}, \citenamefont {Ladak}, \citenamefont
			{Bran}, \citenamefont {Grundler}, \citenamefont {Huth}, \citenamefont
			{Plank}, \citenamefont {Schmidt}, \citenamefont {van Dijken}, \citenamefont
			{Streubel}, \citenamefont {Dobrovolskiy}, \citenamefont {Scagnoli},
			\citenamefont {Heyderman}, \citenamefont {Donnelly}, \citenamefont {Hellwig},
			\citenamefont {Fallarino}, \citenamefont {Jungfleisch}, \citenamefont
			{Farhan}, \citenamefont {Maccaferri}, \citenamefont {Vavassori},
			\citenamefont {Fischer}, \citenamefont {Tomasello}, \citenamefont
			{Finocchio}, \citenamefont {Clérac}, \citenamefont {Sessoli}, \citenamefont
			{Makarov}, \citenamefont {Sheka}, \citenamefont {Krawczyk}, \citenamefont
			{Gallardo}, \citenamefont {Landeros}, \citenamefont {d'Aquino}, \citenamefont
			{hertel}, \citenamefont {Pirro}, \citenamefont {Ciubotaru}, \citenamefont
			{Becherer}, \citenamefont {Gartside}, \citenamefont {Ono}, \citenamefont
			{Bortolotti},\ and\ \citenamefont {Fern\`{a}ndez-Pacheco}}]{Gubbiotti24}%
		\BibitemOpen
		\bibfield  {author} {\bibinfo {author} {\bibfnamefont {G.}~\bibnamefont
				{Gubbiotti}}, \bibinfo {author} {\bibfnamefont {A.}~\bibnamefont {Barman}},
			\bibinfo {author} {\bibfnamefont {S.}~\bibnamefont {Ladak}}, \bibinfo
			{author} {\bibfnamefont {C.}~\bibnamefont {Bran}}, \bibinfo {author}
			{\bibfnamefont {D.}~\bibnamefont {Grundler}}, \bibinfo {author}
			{\bibfnamefont {M.}~\bibnamefont {Huth}}, \bibinfo {author} {\bibfnamefont
				{H.}~\bibnamefont {Plank}}, \bibinfo {author} {\bibfnamefont
				{G.}~\bibnamefont {Schmidt}}, \bibinfo {author} {\bibfnamefont
				{S.}~\bibnamefont {van Dijken}}, \bibinfo {author} {\bibfnamefont
				{R.}~\bibnamefont {Streubel}}, \bibinfo {author} {\bibfnamefont {O.~V.}\
				\bibnamefont {Dobrovolskiy}}, \bibinfo {author} {\bibfnamefont
				{V.}~\bibnamefont {Scagnoli}}, \bibinfo {author} {\bibfnamefont {L.~J.}\
				\bibnamefont {Heyderman}}, \bibinfo {author} {\bibfnamefont {C.}~\bibnamefont
				{Donnelly}}, \bibinfo {author} {\bibfnamefont {O.}~\bibnamefont {Hellwig}},
			\bibinfo {author} {\bibfnamefont {L.}~\bibnamefont {Fallarino}}, \bibinfo
			{author} {\bibfnamefont {M.~B.}\ \bibnamefont {Jungfleisch}}, \bibinfo
			{author} {\bibfnamefont {A.}~\bibnamefont {Farhan}}, \bibinfo {author}
			{\bibfnamefont {N.}~\bibnamefont {Maccaferri}}, \bibinfo {author}
			{\bibfnamefont {P.}~\bibnamefont {Vavassori}}, \bibinfo {author}
			{\bibfnamefont {P.}~\bibnamefont {Fischer}}, \bibinfo {author} {\bibfnamefont
				{R.}~\bibnamefont {Tomasello}}, \bibinfo {author} {\bibfnamefont
				{G.}~\bibnamefont {Finocchio}}, \bibinfo {author} {\bibfnamefont
				{R.}~\bibnamefont {Clérac}}, \bibinfo {author} {\bibfnamefont
				{R.}~\bibnamefont {Sessoli}}, \bibinfo {author} {\bibfnamefont
				{D.}~\bibnamefont {Makarov}}, \bibinfo {author} {\bibfnamefont
				{D.}~\bibnamefont {Sheka}}, \bibinfo {author} {\bibfnamefont
				{M.}~\bibnamefont {Krawczyk}}, \bibinfo {author} {\bibfnamefont {R.~A.}\
				\bibnamefont {Gallardo}}, \bibinfo {author} {\bibfnamefont {P.}~\bibnamefont
				{Landeros}}, \bibinfo {author} {\bibfnamefont {M.}~\bibnamefont {d'Aquino}},
			\bibinfo {author} {\bibfnamefont {R.}~\bibnamefont {hertel}}, \bibinfo
			{author} {\bibfnamefont {P.}~\bibnamefont {Pirro}}, \bibinfo {author}
			{\bibfnamefont {F.}~\bibnamefont {Ciubotaru}}, \bibinfo {author}
			{\bibfnamefont {M.}~\bibnamefont {Becherer}}, \bibinfo {author}
			{\bibfnamefont {J.}~\bibnamefont {Gartside}}, \bibinfo {author}
			{\bibfnamefont {T.}~\bibnamefont {Ono}}, \bibinfo {author} {\bibfnamefont
				{P.}~\bibnamefont {Bortolotti}},\ and\ \bibinfo {author} {\bibfnamefont
				{A.}~\bibnamefont {Fern\`{a}ndez-Pacheco}},\ }\bibfield  {title} {\bibinfo
			{title} {2025 roadmap on {3D} nano-magnetism},\ }\href
		{https://doi.org/10.1088/1361-648x/ad9655} {\bibfield  {journal} {\bibinfo
				{journal} {Journal of Physics: Condensed Matter}\ }\textbf {\bibinfo {volume}
				{37}},\ \bibinfo {pages} {143502} (\bibinfo {year} {2024})}\BibitemShut
		{NoStop}%
		\bibitem [{\citenamefont {Landeros}\ \emph {et~al.}(2022)\citenamefont
			{Landeros}, \citenamefont {Ot\'{a}lora}, \citenamefont {Streubel},\ and\
			\citenamefont {K\'{a}kay}}]{Landeros22}%
		\BibitemOpen
		\bibfield  {author} {\bibinfo {author} {\bibfnamefont {P.}~\bibnamefont
				{Landeros}}, \bibinfo {author} {\bibfnamefont {J.~A.}\ \bibnamefont
				{Ot\'{a}lora}}, \bibinfo {author} {\bibfnamefont {R.}~\bibnamefont
				{Streubel}},\ and\ \bibinfo {author} {\bibfnamefont {A.}~\bibnamefont
				{K\'{a}kay}},\ }\bibinfo {title} {Tubular geometries},\ in\ \href
		{https://doi.org/10.1007/978-3-031-09086-8_4} {\emph {\bibinfo {booktitle}
				{Topics in Applied Physics}}}\ (\bibinfo  {publisher} {Springer International
			Publishing},\ \bibinfo {address} {Cham},\ \bibinfo {year} {2022})\ pp.\
		\bibinfo {pages} {163--213}\BibitemShut {NoStop}%
		\bibitem [{Note1()}]{Note1}%
		\BibitemOpen
		\bibinfo {note} {There is an alternative definition of the mean curvature as
			the arithmetic mean of the principal curvatures $\protect \tilde{H}' =
			(\kappa _1 + \kappa _2)/2$.}\BibitemShut {Stop}%
		\bibitem [{\citenamefont {Schoen}(1970)}]{Schoen70}%
		\BibitemOpen
		\bibfield  {author} {\bibinfo {author} {\bibfnamefont {A.~H.}\ \bibnamefont
				{Schoen}},\ }\href@noop {} {\emph {\bibinfo {title} {Infinite Periodic
					Minimal Surfaces Without Self-intersections}}}\ (\bibinfo  {publisher}
		{National Aeronautics and Space Administration},\ \bibinfo {year}
		{1970})\BibitemShut {NoStop}%
		\bibitem [{\citenamefont {Saranathan}\ \emph {et~al.}(2010)\citenamefont
			{Saranathan}, \citenamefont {Osuji}, \citenamefont {Mochrie}, \citenamefont
			{Noh}, \citenamefont {Narayanan}, \citenamefont {Sandy}, \citenamefont
			{Dufresne},\ and\ \citenamefont {Prum}}]{Saranathan10}%
		\BibitemOpen
		\bibfield  {author} {\bibinfo {author} {\bibfnamefont {V.}~\bibnamefont
				{Saranathan}}, \bibinfo {author} {\bibfnamefont {C.~O.}\ \bibnamefont
				{Osuji}}, \bibinfo {author} {\bibfnamefont {S.~G.~J.}\ \bibnamefont
				{Mochrie}}, \bibinfo {author} {\bibfnamefont {H.}~\bibnamefont {Noh}},
			\bibinfo {author} {\bibfnamefont {S.}~\bibnamefont {Narayanan}}, \bibinfo
			{author} {\bibfnamefont {A.}~\bibnamefont {Sandy}}, \bibinfo {author}
			{\bibfnamefont {E.~R.}\ \bibnamefont {Dufresne}},\ and\ \bibinfo {author}
			{\bibfnamefont {R.~O.}\ \bibnamefont {Prum}},\ }\bibfield  {title} {\bibinfo
			{title} {Structure, function, and self-assembly of single network gyroid
				{$(I4_132)$} photonic crystals in butterfly wing scales},\ }\href
		{https://doi.org/10.1073/pnas.0909616107} {\bibfield  {journal} {\bibinfo
				{journal} {Proceedings of the National Academy of Sciences}\ }\textbf
			{\bibinfo {volume} {107}},\ \bibinfo {pages} {11676} (\bibinfo {year}
			{2010})}\BibitemShut {NoStop}%
		\bibitem [{\citenamefont {Dubrovin}\ \emph {et~al.}(1984)\citenamefont
			{Dubrovin}, \citenamefont {Fomenko},\ and\ \citenamefont
			{Novikov}}]{Dubrovin84}%
		\BibitemOpen
		\bibfield  {author} {\bibinfo {author} {\bibfnamefont {B.~A.}\ \bibnamefont
				{Dubrovin}}, \bibinfo {author} {\bibfnamefont {A.~T.}\ \bibnamefont
				{Fomenko}},\ and\ \bibinfo {author} {\bibfnamefont {S.~P.}\ \bibnamefont
				{Novikov}},\ }\href@noop {} {\emph {\bibinfo {title} {Modern Geometry --
					Methods and Applications: Part {I}. {T}he Geometry of Surfaces,
					Transformation Groups, and Fields}}},\ \bibinfo {edition} {2nd}\ ed.,\
		\bibinfo {series} {Graduate Texts in Mathematics 93}, Vol.~\bibinfo {volume}
		{1}\ (\bibinfo  {publisher} {Springer New York},\ \bibinfo {year}
		{1984})\BibitemShut {NoStop}%
		\bibitem [{\citenamefont {Dubrovin}\ \emph {et~al.}(1985)\citenamefont
			{Dubrovin}, \citenamefont {Fomenko},\ and\ \citenamefont
			{Novikov}}]{Dubrovin85}%
		\BibitemOpen
		\bibfield  {author} {\bibinfo {author} {\bibfnamefont {B.}~\bibnamefont
				{Dubrovin}}, \bibinfo {author} {\bibfnamefont {A.}~\bibnamefont {Fomenko}},\
			and\ \bibinfo {author} {\bibfnamefont {S.}~\bibnamefont {Novikov}},\
		}\href@noop {} {\emph {\bibinfo {title} {Modern Geometry - Methods and
					Applications: Part {II}: The Geometry and Topology of Manifolds}}},\ GTM093\
		(\bibinfo  {publisher} {Springer},\ \bibinfo {year} {1985})\BibitemShut
		{NoStop}%
		\bibitem [{\citenamefont {K\"{u}hnel}(2015)}]{Kuhnel15}%
		\BibitemOpen
		\bibfield  {author} {\bibinfo {author} {\bibfnamefont {W.}~\bibnamefont
				{K\"{u}hnel}},\ }\href {https://doi.org/10.1090/stml/077} {\emph {\bibinfo
				{title} {Differential Geometry: Curves -- Surfaces -- Manifolds}}}\ (\bibinfo
		{publisher} {American Mathematical Society},\ \bibinfo {year}
		{2015})\BibitemShut {NoStop}%
		\bibitem [{\citenamefont {do~Carmo}(2016)}]{doCarmo16}%
		\BibitemOpen
		\bibfield  {author} {\bibinfo {author} {\bibfnamefont {M.~P.}\ \bibnamefont
				{do~Carmo}},\ }\href@noop {} {\emph {\bibinfo {title} {Differential Geometry
					of Curves and Surfaces}}}\ (\bibinfo  {publisher} {Dover Publications Inc.},\
		\bibinfo {year} {2016})\BibitemShut {NoStop}%
		\bibitem [{\citenamefont {Govindarajan}\ and\ \citenamefont
			{Ramadevi}(2024)}]{Govindarajan24}%
		\BibitemOpen
		\bibfield  {author} {\bibinfo {author} {\bibfnamefont {T.~R.}\ \bibnamefont
				{Govindarajan}}\ and\ \bibinfo {author} {\bibfnamefont {P.}~\bibnamefont
				{Ramadevi}},\ }\href {https://doi.org/10.1007/978-3-031-59501-1} {\emph
			{\bibinfo {title} {Geometry and Topology of Low Dimensional Systems}}}\
		(\bibinfo  {publisher} {Springer International Publishing},\ \bibinfo
		{address} {Cham},\ \bibinfo {year} {2024})\BibitemShut {NoStop}%
		\bibitem [{\citenamefont {Yershov}\ and\ \citenamefont
			{Sheka}(2023)}]{Yershov23}%
		\BibitemOpen
		\bibfield  {author} {\bibinfo {author} {\bibfnamefont {K.~V.}\ \bibnamefont
				{Yershov}}\ and\ \bibinfo {author} {\bibfnamefont {D.~D.}\ \bibnamefont
				{Sheka}},\ }\bibfield  {title} {\bibinfo {title} {Control of magnetic
				response in curved stripes by tailoring the cross section},\ }\href
		{https://doi.org/10.1103/physrevb.107.l100415} {\bibfield  {journal}
			{\bibinfo  {journal} {Physical Review B}\ }\textbf {\bibinfo {volume}
				{107}},\ \bibinfo {pages} {L100415} (\bibinfo {year} {2023})}\BibitemShut
		{NoStop}%
		\bibitem [{\citenamefont {Carbou}(2001)}]{Carbou01}%
		\BibitemOpen
		\bibfield  {author} {\bibinfo {author} {\bibfnamefont {G.}~\bibnamefont
				{Carbou}},\ }\bibfield  {title} {\bibinfo {title} {Thin layers in
				micromagnetism},\ }\href {https://doi.org/10.1142/S0218202501001458}
		{\bibfield  {journal} {\bibinfo  {journal} {Mathematical Models and Methods
					in Applied Sciences (M3AS)}\ }\textbf {\bibinfo {volume} {11}},\ \bibinfo
			{pages} {1529} (\bibinfo {year} {2001})}\BibitemShut {NoStop}%
		\bibitem [{\citenamefont {Kohn}\ and\ \citenamefont
			{Slastikov}(2005)}]{Kohn05a}%
		\BibitemOpen
		\bibfield  {author} {\bibinfo {author} {\bibfnamefont {R.~V.}\ \bibnamefont
				{Kohn}}\ and\ \bibinfo {author} {\bibfnamefont {V.~V.}\ \bibnamefont
				{Slastikov}},\ }\bibfield  {title} {\bibinfo {title} {Another thin-film limit
				of micromagnetics},\ }\href {https://doi.org/10.1007/s00205-005-0372-7}
		{\bibfield  {journal} {\bibinfo  {journal} {Archive for Rational Mechanics
					and Analysis}\ }\textbf {\bibinfo {volume} {178}},\ \bibinfo {pages} {227}
			(\bibinfo {year} {2005})}\BibitemShut {NoStop}%
		\bibitem [{\citenamefont {Fratta}(2020)}]{DiFratta20}%
		\BibitemOpen
		\bibfield  {author} {\bibinfo {author} {\bibfnamefont {G.~D.}\ \bibnamefont
				{Fratta}},\ }\bibfield  {title} {\bibinfo {title} {Micromagnetics of curved
				thin films},\ }\href {https://doi.org/10.1007/s00033-020-01336-2} {\bibfield
			{journal} {\bibinfo  {journal} {Zeitschrift f{\"{u}}r angewandte Mathematik
					und Physik}\ }\textbf {\bibinfo {volume} {71}},\ \bibinfo {pages} {111}
			(\bibinfo {year} {2020})}\BibitemShut {NoStop}%
		\bibitem [{\citenamefont {Fratta}\ \emph {et~al.}(2020)\citenamefont {Fratta},
			\citenamefont {Muratov}, \citenamefont {Rybakov},\ and\ \citenamefont
			{Slastikov}}]{Fratta20}%
		\BibitemOpen
		\bibfield  {author} {\bibinfo {author} {\bibfnamefont {G.~D.}\ \bibnamefont
				{Fratta}}, \bibinfo {author} {\bibfnamefont {C.~B.}\ \bibnamefont {Muratov}},
			\bibinfo {author} {\bibfnamefont {F.~N.}\ \bibnamefont {Rybakov}},\ and\
			\bibinfo {author} {\bibfnamefont {V.~V.}\ \bibnamefont {Slastikov}},\
		}\bibfield  {title} {\bibinfo {title} {Variational principles of
				micromagnetics revisited},\ }\href {https://doi.org/10.1137/19m1261365}
		{\bibfield  {journal} {\bibinfo  {journal} {{SIAM} Journal on Mathematical
					Analysis}\ }\textbf {\bibinfo {volume} {52}},\ \bibinfo {pages} {3580}
			(\bibinfo {year} {2020})}\BibitemShut {NoStop}%
		\bibitem [{\citenamefont {Slastikov}\ and\ \citenamefont
			{Sonnenberg}(2012)}]{Slastikov12}%
		\BibitemOpen
		\bibfield  {author} {\bibinfo {author} {\bibfnamefont {V.~V.}\ \bibnamefont
				{Slastikov}}\ and\ \bibinfo {author} {\bibfnamefont {C.}~\bibnamefont
				{Sonnenberg}},\ }\bibfield  {title} {\bibinfo {title} {Reduced models for
				ferromagnetic nanowires},\ }\href {https://doi.org/10.1093/imamat/hxr019}
		{\bibfield  {journal} {\bibinfo  {journal} {IMA Journal of Applied
					Mathematics}\ }\textbf {\bibinfo {volume} {77}},\ \bibinfo {pages} {220}
			(\bibinfo {year} {2012})}\BibitemShut {NoStop}%
		\bibitem [{\citenamefont {Gaididei}\ \emph {et~al.}(2017)\citenamefont
			{Gaididei}, \citenamefont {Goussev}, \citenamefont {Kravchuk}, \citenamefont
			{Pylypovskyi}, \citenamefont {Robbins}, \citenamefont {Sheka}, \citenamefont
			{Slastikov},\ and\ \citenamefont {Vasylkevych}}]{Gaididei17}%
		\BibitemOpen
		\bibfield  {author} {\bibinfo {author} {\bibfnamefont {Y.}~\bibnamefont
				{Gaididei}}, \bibinfo {author} {\bibfnamefont {A.}~\bibnamefont {Goussev}},
			\bibinfo {author} {\bibfnamefont {V.~P.}\ \bibnamefont {Kravchuk}}, \bibinfo
			{author} {\bibfnamefont {O.~V.}\ \bibnamefont {Pylypovskyi}}, \bibinfo
			{author} {\bibfnamefont {J.~M.}\ \bibnamefont {Robbins}}, \bibinfo {author}
			{\bibfnamefont {D.}~\bibnamefont {Sheka}}, \bibinfo {author} {\bibfnamefont
				{V.}~\bibnamefont {Slastikov}},\ and\ \bibinfo {author} {\bibfnamefont
				{S.}~\bibnamefont {Vasylkevych}},\ }\bibfield  {title} {\bibinfo {title}
			{Magnetization in narrow ribbons: curvature effects},\ }\href
		{https://doi.org/10.1088/1751-8121/aa8179} {\bibfield  {journal} {\bibinfo
				{journal} {Journal of Physics A: Mathematical and Theoretical}\ }\textbf
			{\bibinfo {volume} {50}},\ \bibinfo {pages} {385401} (\bibinfo {year}
			{2017})}\BibitemShut {NoStop}%
		\bibitem [{\citenamefont {Smith}\ \emph {et~al.}(2011)\citenamefont {Smith},
			\citenamefont {Makarov}, \citenamefont {Sanchez}, \citenamefont {Fomin},\
			and\ \citenamefont {Schmidt}}]{Smith11}%
		\BibitemOpen
		\bibfield  {author} {\bibinfo {author} {\bibfnamefont {E.~J.}\ \bibnamefont
				{Smith}}, \bibinfo {author} {\bibfnamefont {D.}~\bibnamefont {Makarov}},
			\bibinfo {author} {\bibfnamefont {S.}~\bibnamefont {Sanchez}}, \bibinfo
			{author} {\bibfnamefont {V.~M.}\ \bibnamefont {Fomin}},\ and\ \bibinfo
			{author} {\bibfnamefont {O.~G.}\ \bibnamefont {Schmidt}},\ }\bibfield
		{title} {\bibinfo {title} {Magnetic microhelix coil structures},\ }\href
		{https://doi.org/10.1103/PhysRevLett.107.097204} {\bibfield  {journal}
			{\bibinfo  {journal} {Physical Review Letters}\ }\textbf {\bibinfo {volume}
				{107}},\ \bibinfo {pages} {097204} (\bibinfo {year} {2011})}\BibitemShut
		{NoStop}%
		\bibitem [{\citenamefont {Raftrey}\ \emph {et~al.}(2025)\citenamefont
			{Raftrey}, \citenamefont {Bhattacharya}, \citenamefont {Langton},
			\citenamefont {Fugetta}, \citenamefont {Satapathy}, \citenamefont
			{Bezsmertna}, \citenamefont {Sorrentino}, \citenamefont {Makarov},
			\citenamefont {Yin}, \citenamefont {Fischer},\ and\ \citenamefont
			{Liu}}]{Raftrey25}%
		\BibitemOpen
		\bibfield  {author} {\bibinfo {author} {\bibfnamefont {D.}~\bibnamefont
				{Raftrey}}, \bibinfo {author} {\bibfnamefont {D.}~\bibnamefont
				{Bhattacharya}}, \bibinfo {author} {\bibfnamefont {C.}~\bibnamefont
				{Langton}}, \bibinfo {author} {\bibfnamefont {B.~J.}\ \bibnamefont
				{Fugetta}}, \bibinfo {author} {\bibfnamefont {S.}~\bibnamefont {Satapathy}},
			\bibinfo {author} {\bibfnamefont {O.}~\bibnamefont {Bezsmertna}}, \bibinfo
			{author} {\bibfnamefont {A.}~\bibnamefont {Sorrentino}}, \bibinfo {author}
			{\bibfnamefont {D.}~\bibnamefont {Makarov}}, \bibinfo {author} {\bibfnamefont
				{G.}~\bibnamefont {Yin}}, \bibinfo {author} {\bibfnamefont {P.}~\bibnamefont
				{Fischer}},\ and\ \bibinfo {author} {\bibfnamefont {K.}~\bibnamefont {Liu}},\
		}\bibfield  {title} {\bibinfo {title} {Curvature induced modifications of
				chirality and magnetic configuration in perpendicular films},\ }\href
		{https://doi.org/10.1021/acsnano.5c08926} {\bibfield  {journal} {\bibinfo
				{journal} {ACS Nano}\ }\textbf {\bibinfo {volume} {19}},\ \bibinfo {pages}
			{31609} (\bibinfo {year} {2025})}\BibitemShut {NoStop}%
		\bibitem [{\citenamefont {Yershov}\ \emph {et~al.}(2016)\citenamefont
			{Yershov}, \citenamefont {Kravchuk}, \citenamefont {Sheka},\ and\
			\citenamefont {Gaididei}}]{Yershov16}%
		\BibitemOpen
		\bibfield  {author} {\bibinfo {author} {\bibfnamefont {K.~V.}\ \bibnamefont
				{Yershov}}, \bibinfo {author} {\bibfnamefont {V.~P.}\ \bibnamefont
				{Kravchuk}}, \bibinfo {author} {\bibfnamefont {D.~D.}\ \bibnamefont
				{Sheka}},\ and\ \bibinfo {author} {\bibfnamefont {Y.}~\bibnamefont
				{Gaididei}},\ }\bibfield  {title} {\bibinfo {title} {Curvature and torsion
				effects in spin-current driven domain wall motion},\ }\href
		{https://doi.org/10.1103/PhysRevB.93.094418} {\bibfield  {journal} {\bibinfo
				{journal} {Physical Review B}\ }\textbf {\bibinfo {volume} {93}},\ \bibinfo
			{pages} {094418} (\bibinfo {year} {2016})}\BibitemShut {NoStop}%
		\bibitem [{\citenamefont {Makushko}\ \emph {et~al.}(2024)\citenamefont
			{Makushko}, \citenamefont {Kovalev}, \citenamefont {Zabila}, \citenamefont
			{Ilyakov}, \citenamefont {Ponomaryov}, \citenamefont {Arshad}, \citenamefont
			{Prajapati}, \citenamefont {de~Oliveira}, \citenamefont {Deinert},
			\citenamefont {Chekhonin}, \citenamefont {Veremchuk}, \citenamefont {Kosub},
			\citenamefont {Skourski}, \citenamefont {Ganss}, \citenamefont {Makarov},\
			and\ \citenamefont {Ortix}}]{Makushko24}%
		\BibitemOpen
		\bibfield  {author} {\bibinfo {author} {\bibfnamefont {P.}~\bibnamefont
				{Makushko}}, \bibinfo {author} {\bibfnamefont {S.}~\bibnamefont {Kovalev}},
			\bibinfo {author} {\bibfnamefont {Y.}~\bibnamefont {Zabila}}, \bibinfo
			{author} {\bibfnamefont {I.}~\bibnamefont {Ilyakov}}, \bibinfo {author}
			{\bibfnamefont {A.}~\bibnamefont {Ponomaryov}}, \bibinfo {author}
			{\bibfnamefont {A.}~\bibnamefont {Arshad}}, \bibinfo {author} {\bibfnamefont
				{G.~L.}\ \bibnamefont {Prajapati}}, \bibinfo {author} {\bibfnamefont {T.~V.
					A.~G.}\ \bibnamefont {de~Oliveira}}, \bibinfo {author} {\bibfnamefont
				{J.-C.}\ \bibnamefont {Deinert}}, \bibinfo {author} {\bibfnamefont
				{P.}~\bibnamefont {Chekhonin}}, \bibinfo {author} {\bibfnamefont
				{I.}~\bibnamefont {Veremchuk}}, \bibinfo {author} {\bibfnamefont
				{T.}~\bibnamefont {Kosub}}, \bibinfo {author} {\bibfnamefont
				{Y.}~\bibnamefont {Skourski}}, \bibinfo {author} {\bibfnamefont
				{F.}~\bibnamefont {Ganss}}, \bibinfo {author} {\bibfnamefont
				{D.}~\bibnamefont {Makarov}},\ and\ \bibinfo {author} {\bibfnamefont
				{C.}~\bibnamefont {Ortix}},\ }\bibfield  {title} {\bibinfo {title} {A tunable
				room-temperature nonlinear {H}all effect in elemental bismuth thin films},\
		}\href {https://doi.org/10.1038/s41928-024-01118-y} {\bibfield  {journal}
			{\bibinfo  {journal} {Nature Electronics}\ }\textbf {\bibinfo {volume} {7}},\
			\bibinfo {pages} {207} (\bibinfo {year} {2024})}\BibitemShut {NoStop}%
		\bibitem [{\citenamefont {Gaididei}\ \emph {et~al.}(2019)\citenamefont
			{Gaididei}, \citenamefont {Yershov}, \citenamefont {Sheka}, \citenamefont
			{Kravchuk},\ and\ \citenamefont {Saxena}}]{Gaididei19}%
		\BibitemOpen
		\bibfield  {author} {\bibinfo {author} {\bibfnamefont {Y.}~\bibnamefont
				{Gaididei}}, \bibinfo {author} {\bibfnamefont {K.~V.}\ \bibnamefont
				{Yershov}}, \bibinfo {author} {\bibfnamefont {D.~D.}\ \bibnamefont {Sheka}},
			\bibinfo {author} {\bibfnamefont {V.~P.}\ \bibnamefont {Kravchuk}},\ and\
			\bibinfo {author} {\bibfnamefont {A.}~\bibnamefont {Saxena}},\ }\bibfield
		{title} {\bibinfo {title} {Magnetization-induced shape transformations in
				flexible ferromagnetic rings},\ }\href
		{https://doi.org/10.1103/PhysRevB.99.014404} {\bibfield  {journal} {\bibinfo
				{journal} {Physical Review B}\ }\textbf {\bibinfo {volume} {99}},\ \bibinfo
			{pages} {014404} (\bibinfo {year} {2019})}\BibitemShut {NoStop}%
		\bibitem [{\citenamefont {Yershov}\ \emph {et~al.}(2019)\citenamefont
			{Yershov}, \citenamefont {Kravchuk}, \citenamefont {Sheka}, \citenamefont
			{van~den Brink},\ and\ \citenamefont {Gaididei}}]{Yershov19b}%
		\BibitemOpen
		\bibfield  {author} {\bibinfo {author} {\bibfnamefont {K.~V.}\ \bibnamefont
				{Yershov}}, \bibinfo {author} {\bibfnamefont {V.~P.}\ \bibnamefont
				{Kravchuk}}, \bibinfo {author} {\bibfnamefont {D.~D.}\ \bibnamefont {Sheka}},
			\bibinfo {author} {\bibfnamefont {J.}~\bibnamefont {van~den Brink}},\ and\
			\bibinfo {author} {\bibfnamefont {Y.}~\bibnamefont {Gaididei}},\ }\bibfield
		{title} {\bibinfo {title} {{Spontaneous deformation of flexible ferromagnetic
					ribbons induced by Dzyaloshinskii-Moriya interaction}},\ }\href
		{https://doi.org/10.1103/PhysRevB.100.140407} {\bibfield  {journal} {\bibinfo
				{journal} {Physical Review B}\ }\textbf {\bibinfo {volume} {100}},\ \bibinfo
			{pages} {140407(R)} (\bibinfo {year} {2019})}\BibitemShut {NoStop}%
		\bibitem [{\citenamefont {Miranda-Silva}\ \emph {et~al.}(2022)\citenamefont
			{Miranda-Silva}, \citenamefont {Taveira}, \citenamefont {Teixeira},
			\citenamefont {Fonseca}, \citenamefont {Rodrigues}, \citenamefont
			{El\'{i}as}, \citenamefont {Riveros}, \citenamefont {Vidal-Silva},\ and\
			\citenamefont {Carvalho-Santos}}]{MirandaSilva22}%
		\BibitemOpen
		\bibfield  {author} {\bibinfo {author} {\bibfnamefont {B.}~\bibnamefont
				{Miranda-Silva}}, \bibinfo {author} {\bibfnamefont {P.~H.~C.}\ \bibnamefont
				{Taveira}}, \bibinfo {author} {\bibfnamefont {A.~W.}\ \bibnamefont
				{Teixeira}}, \bibinfo {author} {\bibfnamefont {J.~M.}\ \bibnamefont
				{Fonseca}}, \bibinfo {author} {\bibfnamefont {L.~N.}\ \bibnamefont
				{Rodrigues}}, \bibinfo {author} {\bibfnamefont {R.~G.}\ \bibnamefont
				{El\'{i}as}}, \bibinfo {author} {\bibfnamefont {A.}~\bibnamefont {Riveros}},
			\bibinfo {author} {\bibfnamefont {N.}~\bibnamefont {Vidal-Silva}},\ and\
			\bibinfo {author} {\bibfnamefont {V.~L.}\ \bibnamefont {Carvalho-Santos}},\
		}\bibfield  {title} {\bibinfo {title} {Manipulating the shape of flexible
				magnetic nanodisks with meronlike magnetic states},\ }\href
		{https://doi.org/10.1103/physrevb.105.104430} {\bibfield  {journal} {\bibinfo
				{journal} {Physical Review B}\ }\textbf {\bibinfo {volume} {105}},\ \bibinfo
			{pages} {104430} (\bibinfo {year} {2022})}\BibitemShut {NoStop}%
		\bibitem [{\citenamefont {Sheka}\ \emph {et~al.}(2020)\citenamefont {Sheka},
			\citenamefont {Pylypovskyi}, \citenamefont {Landeros}, \citenamefont
			{Gaididei}, \citenamefont {K\'{a}kay},\ and\ \citenamefont
			{Makarov}}]{Sheka20a}%
		\BibitemOpen
		\bibfield  {author} {\bibinfo {author} {\bibfnamefont {D.~D.}\ \bibnamefont
				{Sheka}}, \bibinfo {author} {\bibfnamefont {O.~V.}\ \bibnamefont
				{Pylypovskyi}}, \bibinfo {author} {\bibfnamefont {P.}~\bibnamefont
				{Landeros}}, \bibinfo {author} {\bibfnamefont {Y.}~\bibnamefont {Gaididei}},
			\bibinfo {author} {\bibfnamefont {A.}~\bibnamefont {K\'{a}kay}},\ and\
			\bibinfo {author} {\bibfnamefont {D.}~\bibnamefont {Makarov}},\ }\bibfield
		{title} {\bibinfo {title} {Nonlocal chiral symmetry breaking in curvilinear
				magnetic shells},\ }\href {https://doi.org/10.1038/s42005-020-0387-2}
		{\bibfield  {journal} {\bibinfo  {journal} {Communications Physics}\ }\textbf
			{\bibinfo {volume} {3}},\ \bibinfo {pages} {128} (\bibinfo {year}
			{2020})}\BibitemShut {NoStop}%
		\bibitem [{\citenamefont {Sheka}\ \emph
			{et~al.}(2022{\natexlab{b}})\citenamefont {Sheka}, \citenamefont {Kravchuk},
			\citenamefont {Peddis}, \citenamefont {Varvaro}, \citenamefont {Krupiński},
			\citenamefont {Albrecht}, \citenamefont {Erb}, \citenamefont {Facsko},\ and\
			\citenamefont {Makarov}}]{Sheka22a}%
		\BibitemOpen
		\bibfield  {author} {\bibinfo {author} {\bibfnamefont {D.~D.}\ \bibnamefont
				{Sheka}}, \bibinfo {author} {\bibfnamefont {V.~P.}\ \bibnamefont {Kravchuk}},
			\bibinfo {author} {\bibfnamefont {D.}~\bibnamefont {Peddis}}, \bibinfo
			{author} {\bibfnamefont {G.}~\bibnamefont {Varvaro}}, \bibinfo {author}
			{\bibfnamefont {M.}~\bibnamefont {Krupiński}}, \bibinfo {author}
			{\bibfnamefont {M.}~\bibnamefont {Albrecht}}, \bibinfo {author}
			{\bibfnamefont {D.}~\bibnamefont {Erb}}, \bibinfo {author} {\bibfnamefont
				{S.}~\bibnamefont {Facsko}},\ and\ \bibinfo {author} {\bibfnamefont
				{D.}~\bibnamefont {Makarov}},\ }\bibinfo {title} {Curvilinear magnetic
			shells},\ in\ \href {https://doi.org/10.1007/978-3-031-09086-8_3} {\emph
			{\bibinfo {booktitle} {Topics in Applied Physics}}}\ (\bibinfo  {publisher}
		{Springer International Publishing},\ \bibinfo {address} {Cham},\ \bibinfo
		{year} {2022})\ pp.\ \bibinfo {pages} {83--161}\BibitemShut {NoStop}%
		\bibitem [{Note2()}]{Note2}%
		\BibitemOpen
		\bibinfo {note} {In these expressions, the divisor $\protect \sqrt {g_{ii}}$
			could be ``eliminated'' with the use of language of covariant and
			contravariant vectors that are commonly denoted with sub- and superscripts,
			respectively~\cite {Sheka22a}.}\BibitemShut {Stop}%
		\bibitem [{\citenamefont {Kravchuk}\ \emph {et~al.}(2016)\citenamefont
			{Kravchuk}, \citenamefont {R\"o\ss{}ler}, \citenamefont {Volkov},
			\citenamefont {Sheka}, \citenamefont {van~den Brink}, \citenamefont
			{Makarov}, \citenamefont {Fuchs}, \citenamefont {Fangohr},\ and\
			\citenamefont {Gaididei}}]{Kravchuk16}%
		\BibitemOpen
		\bibfield  {author} {\bibinfo {author} {\bibfnamefont {V.~P.}\ \bibnamefont
				{Kravchuk}}, \bibinfo {author} {\bibfnamefont {U.~K.}\ \bibnamefont
				{R\"o\ss{}ler}}, \bibinfo {author} {\bibfnamefont {O.~M.}\ \bibnamefont
				{Volkov}}, \bibinfo {author} {\bibfnamefont {D.~D.}\ \bibnamefont {Sheka}},
			\bibinfo {author} {\bibfnamefont {J.}~\bibnamefont {van~den Brink}}, \bibinfo
			{author} {\bibfnamefont {D.}~\bibnamefont {Makarov}}, \bibinfo {author}
			{\bibfnamefont {H.}~\bibnamefont {Fuchs}}, \bibinfo {author} {\bibfnamefont
				{H.}~\bibnamefont {Fangohr}},\ and\ \bibinfo {author} {\bibfnamefont
				{Y.}~\bibnamefont {Gaididei}},\ }\bibfield  {title} {\bibinfo {title}
			{Topologically stable magnetization states on a spherical shell:
				Curvature-stabilized skyrmions},\ }\href
		{https://doi.org/10.1103/PhysRevB.94.144402} {\bibfield  {journal} {\bibinfo
				{journal} {Physical Review B}\ }\textbf {\bibinfo {volume} {94}},\ \bibinfo
			{pages} {144402} (\bibinfo {year} {2016})}\BibitemShut {NoStop}%
		\bibitem [{\citenamefont {Pylypovskyi}\ \emph {et~al.}(2018)\citenamefont
			{Pylypovskyi}, \citenamefont {Makarov}, \citenamefont {Kravchuk},
			\citenamefont {Gaididei}, \citenamefont {Saxena},\ and\ \citenamefont
			{Sheka}}]{Pylypovskyi18}%
		\BibitemOpen
		\bibfield  {author} {\bibinfo {author} {\bibfnamefont {O.~V.}\ \bibnamefont
				{Pylypovskyi}}, \bibinfo {author} {\bibfnamefont {D.}~\bibnamefont
				{Makarov}}, \bibinfo {author} {\bibfnamefont {V.~P.}\ \bibnamefont
				{Kravchuk}}, \bibinfo {author} {\bibfnamefont {Y.}~\bibnamefont {Gaididei}},
			\bibinfo {author} {\bibfnamefont {A.}~\bibnamefont {Saxena}},\ and\ \bibinfo
			{author} {\bibfnamefont {D.~D.}\ \bibnamefont {Sheka}},\ }\bibfield  {title}
		{\bibinfo {title} {Chiral skyrmion and skyrmionium states engineered by the
				gradient of curvature},\ }\href
		{https://doi.org/10.1103/physrevapplied.10.064057} {\bibfield  {journal}
			{\bibinfo  {journal} {Physical Review Applied}\ }\textbf {\bibinfo {volume}
				{10}},\ \bibinfo {pages} {064057} (\bibinfo {year} {2018})}\BibitemShut
		{NoStop}%
		\bibitem [{\citenamefont {Shankar}(1977)}]{Shankar77}%
		\BibitemOpen
		\bibfield  {author} {\bibinfo {author} {\bibfnamefont {R.}~\bibnamefont
				{Shankar}},\ }\bibfield  {title} {\bibinfo {title} {Applications of topology
				to the study of ordered systems},\ }\href
		{https://doi.org/10.1051/jphys:0197700380110140500} {\bibfield  {journal}
			{\bibinfo  {journal} {Journal de Physique}\ }\textbf {\bibinfo {volume}
				{38}},\ \bibinfo {pages} {1405} (\bibinfo {year} {1977})}\BibitemShut
		{NoStop}%
		\bibitem [{\citenamefont {Mermin}(1979)}]{Mermin79}%
		\BibitemOpen
		\bibfield  {author} {\bibinfo {author} {\bibfnamefont {N.~D.}\ \bibnamefont
				{Mermin}},\ }\bibfield  {title} {\bibinfo {title} {The topological theory of
				defects in ordered media},\ }\href
		{https://doi.org/10.1103/revmodphys.51.591} {\bibfield  {journal} {\bibinfo
				{journal} {Reviews of Modern Physics}\ }\textbf {\bibinfo {volume} {51}},\
			\bibinfo {pages} {591} (\bibinfo {year} {1979})}\BibitemShut {NoStop}%
		\bibitem [{\citenamefont {Kamien}(2002)}]{Kamien02}%
		\BibitemOpen
		\bibfield  {author} {\bibinfo {author} {\bibfnamefont {R.}~\bibnamefont
				{Kamien}},\ }\bibfield  {title} {\bibinfo {title} {The geometry of soft
				materials: a primer},\ }\href {https://doi.org/10.1103/RevModPhys.74.953}
		{\bibfield  {journal} {\bibinfo  {journal} {Reviews of Modern Physics}\
			}\textbf {\bibinfo {volume} {74}},\ \bibinfo {pages} {953} (\bibinfo {year}
			{2002})}\BibitemShut {NoStop}%
		\bibitem [{\citenamefont {Alexander}\ \emph {et~al.}(2012)\citenamefont
			{Alexander}, \citenamefont {Chen}, \citenamefont {Matsumoto},\ and\
			\citenamefont {Kamien}}]{Alexander12}%
		\BibitemOpen
		\bibfield  {author} {\bibinfo {author} {\bibfnamefont {G.~P.}\ \bibnamefont
				{Alexander}}, \bibinfo {author} {\bibfnamefont {B.~G.-g.}\ \bibnamefont
				{Chen}}, \bibinfo {author} {\bibfnamefont {E.~A.}\ \bibnamefont
				{Matsumoto}},\ and\ \bibinfo {author} {\bibfnamefont {R.~D.}\ \bibnamefont
				{Kamien}},\ }\bibfield  {title} {\bibinfo {title} {\textit{Colloquium}:
				Disclination loops, point defects, and all that in nematic liquid crystals},\
		}\href {https://doi.org/10.1103/RevModPhys.84.497} {\bibfield  {journal}
			{\bibinfo  {journal} {Reviews of Modern Physics}\ }\textbf {\bibinfo {volume}
				{84}},\ \bibinfo {pages} {497} (\bibinfo {year} {2012})}\BibitemShut
		{NoStop}%
		\bibitem [{\citenamefont {Nash}\ and\ \citenamefont {Sen}(2013)}]{Nash13}%
		\BibitemOpen
		\bibfield  {author} {\bibinfo {author} {\bibfnamefont {C.}~\bibnamefont
				{Nash}}\ and\ \bibinfo {author} {\bibfnamefont {S.}~\bibnamefont {Sen}},\
		}\href@noop {} {\emph {\bibinfo {title} {Topology and Geometry for
					Physicists}}},\ Dover Books on Mathematics\ (\bibinfo  {publisher} {Courier
			Corporation},\ \bibinfo {year} {2013})\BibitemShut {NoStop}%
		\bibitem [{\citenamefont {Balakrishnan}\ \emph {et~al.}(2024)\citenamefont
			{Balakrishnan}, \citenamefont {Dandoloff},\ and\ \citenamefont
			{Saxena}}]{Balakrishnan24}%
		\BibitemOpen
		\bibfield  {author} {\bibinfo {author} {\bibfnamefont {R.}~\bibnamefont
				{Balakrishnan}}, \bibinfo {author} {\bibfnamefont {R.}~\bibnamefont
				{Dandoloff}},\ and\ \bibinfo {author} {\bibfnamefont {A.}~\bibnamefont
				{Saxena}},\ }\bibfield  {title} {\bibinfo {title} {Twisted curve geometry
				underlying topological invariants},\ }\href
		{https://doi.org/10.1016/j.physleta.2023.129261} {\bibfield  {journal}
			{\bibinfo  {journal} {Physics Letters A}\ }\textbf {\bibinfo {volume}
				{493}},\ \bibinfo {pages} {129261} (\bibinfo {year} {2024})}\BibitemShut
		{NoStop}%
		\bibitem [{\citenamefont {G{\"{o}}bel}\ \emph {et~al.}(2019)\citenamefont
			{G{\"{o}}bel}, \citenamefont {Mook}, \citenamefont {Henk}, \citenamefont
			{Mertig},\ and\ \citenamefont {Tretiakov}}]{Gobel19b}%
		\BibitemOpen
		\bibfield  {author} {\bibinfo {author} {\bibfnamefont {B.}~\bibnamefont
				{G{\"{o}}bel}}, \bibinfo {author} {\bibfnamefont {A.}~\bibnamefont {Mook}},
			\bibinfo {author} {\bibfnamefont {J.}~\bibnamefont {Henk}}, \bibinfo {author}
			{\bibfnamefont {I.}~\bibnamefont {Mertig}},\ and\ \bibinfo {author}
			{\bibfnamefont {O.~A.}\ \bibnamefont {Tretiakov}},\ }\bibfield  {title}
		{\bibinfo {title} {Magnetic bimerons as skyrmion analogues in in-plane
				magnets},\ }\href {https://doi.org/10.1103/physrevb.99.060407} {\bibfield
			{journal} {\bibinfo  {journal} {Physical Review B}\ }\textbf {\bibinfo
				{volume} {99}},\ \bibinfo {pages} {060407} (\bibinfo {year}
			{2019})}\BibitemShut {NoStop}%
		\bibitem [{\citenamefont {Mertens}\ and\ \citenamefont
			{Bishop}(2000)}]{Mertens00}%
		\BibitemOpen
		\bibfield  {author} {\bibinfo {author} {\bibfnamefont {F.~G.}\ \bibnamefont
				{Mertens}}\ and\ \bibinfo {author} {\bibfnamefont {A.~R.}\ \bibnamefont
				{Bishop}},\ }\bibfield  {title} {\bibinfo {title} {Dynamics of vortices in
				two--dimensional magnets},\ }in\ \href
		{https://doi.org/10.1007/3-540-46629-0_7} {\emph {\bibinfo {booktitle}
				{Nonlinear Science at the Dawn of the 21th Century}}},\ \bibinfo {editor}
		{edited by\ \bibinfo {editor} {\bibfnamefont {P.~L.}\ \bibnamefont
				{Christiansen}}, \bibinfo {editor} {\bibfnamefont {M.~P.}\ \bibnamefont
				{Soerensen}},\ and\ \bibinfo {editor} {\bibfnamefont {A.~C.}\ \bibnamefont
				{Scott}}}\ (\bibinfo  {publisher} {Springer--Verlag},\ \bibinfo {address}
		{Berlin},\ \bibinfo {year} {2000})\ pp.\ \bibinfo {pages}
		{137--170}\BibitemShut {NoStop}%
		\bibitem [{\citenamefont {Kosevich}\ \emph {et~al.}(1990)\citenamefont
			{Kosevich}, \citenamefont {Ivanov},\ and\ \citenamefont
			{Kovalev}}]{Kosevich90}%
		\BibitemOpen
		\bibfield  {author} {\bibinfo {author} {\bibfnamefont {A.~M.}\ \bibnamefont
				{Kosevich}}, \bibinfo {author} {\bibfnamefont {B.~A.}\ \bibnamefont
				{Ivanov}},\ and\ \bibinfo {author} {\bibfnamefont {A.~S.}\ \bibnamefont
				{Kovalev}},\ }\bibfield  {title} {\bibinfo {title} {Magnetic solitons},\
		}\href {https://doi.org/10.1016/0370-1573(90)90130-T} {\bibfield  {journal}
			{\bibinfo  {journal} {Physics Reports}\ }\textbf {\bibinfo {volume} {194}},\
			\bibinfo {pages} {117} (\bibinfo {year} {1990})}\BibitemShut {NoStop}%
		\bibitem [{\citenamefont {Kharkov}\ \emph {et~al.}(2017)\citenamefont
			{Kharkov}, \citenamefont {Sushkov},\ and\ \citenamefont
			{Mostovoy}}]{Kharkov17}%
		\BibitemOpen
		\bibfield  {author} {\bibinfo {author} {\bibfnamefont {Y.~A.}\ \bibnamefont
				{Kharkov}}, \bibinfo {author} {\bibfnamefont {O.~P.}\ \bibnamefont
				{Sushkov}},\ and\ \bibinfo {author} {\bibfnamefont {M.}~\bibnamefont
				{Mostovoy}},\ }\bibfield  {title} {\bibinfo {title} {Bound states of
				skyrmions and merons near the lifshitz point},\ }\href
		{https://doi.org/10.1103/physrevlett.119.207201} {\bibfield  {journal}
			{\bibinfo  {journal} {Physical Review Letters}\ }\textbf {\bibinfo {volume}
				{119}},\ \bibinfo {pages} {207201} (\bibinfo {year} {2017})}\BibitemShut
		{NoStop}%
		\bibitem [{\citenamefont {Cai}\ \emph {et~al.}(2025)\citenamefont {Cai},
			\citenamefont {Wang}, \citenamefont {Zhang}, \citenamefont {Bao},
			\citenamefont {Shen}, \citenamefont {Ren}, \citenamefont {Qiu}, \citenamefont
			{Yu}, \citenamefont {Luo}, \citenamefont {Kl\"{a}ui}, \citenamefont {Zhang},
			\citenamefont {Jaouen}, \citenamefont {van~der Laan}, \citenamefont
			{Hesjedal}, \citenamefont {Shen},\ and\ \citenamefont {Zhang}}]{Cai25}%
		\BibitemOpen
		\bibfield  {author} {\bibinfo {author} {\bibfnamefont {M.}~\bibnamefont
				{Cai}}, \bibinfo {author} {\bibfnamefont {S.}~\bibnamefont {Wang}}, \bibinfo
			{author} {\bibfnamefont {Y.}~\bibnamefont {Zhang}}, \bibinfo {author}
			{\bibfnamefont {X.}~\bibnamefont {Bao}}, \bibinfo {author} {\bibfnamefont
				{D.}~\bibnamefont {Shen}}, \bibinfo {author} {\bibfnamefont {J.}~\bibnamefont
				{Ren}}, \bibinfo {author} {\bibfnamefont {L.}~\bibnamefont {Qiu}}, \bibinfo
			{author} {\bibfnamefont {H.}~\bibnamefont {Yu}}, \bibinfo {author}
			{\bibfnamefont {Z.}~\bibnamefont {Luo}}, \bibinfo {author} {\bibfnamefont
				{M.}~\bibnamefont {Kl\"{a}ui}}, \bibinfo {author} {\bibfnamefont
				{S.}~\bibnamefont {Zhang}}, \bibinfo {author} {\bibfnamefont
				{N.}~\bibnamefont {Jaouen}}, \bibinfo {author} {\bibfnamefont
				{G.}~\bibnamefont {van~der Laan}}, \bibinfo {author} {\bibfnamefont
				{T.}~\bibnamefont {Hesjedal}}, \bibinfo {author} {\bibfnamefont
				{K.}~\bibnamefont {Shen}},\ and\ \bibinfo {author} {\bibfnamefont
				{J.}~\bibnamefont {Zhang}},\ }\bibfield  {title} {\bibinfo {title}
			{Stabilization and observation of large-area ferromagnetic bimeron lattice},\
		}\href {https://doi.org/10.1103/587m-xvgv} {\bibfield  {journal} {\bibinfo
				{journal} {Physical Review Letters}\ }\textbf {\bibinfo {volume} {135}},\
			\bibinfo {pages} {116703} (\bibinfo {year} {2025})}\BibitemShut {NoStop}%
		\bibitem [{\citenamefont {Ara\'{u}jo}\ \emph {et~al.}(2020)\citenamefont
			{Ara\'{u}jo}, \citenamefont {Lopes}, \citenamefont {Carvalho-Santos},
			\citenamefont {Pereira}, \citenamefont {Silva}, \citenamefont {Silva},\ and\
			\citenamefont {Altbir}}]{Araujo20}%
		\BibitemOpen
		\bibfield  {author} {\bibinfo {author} {\bibfnamefont {A.~S.}\ \bibnamefont
				{Ara\'{u}jo}}, \bibinfo {author} {\bibfnamefont {R.~J.~C.}\ \bibnamefont
				{Lopes}}, \bibinfo {author} {\bibfnamefont {V.~L.}\ \bibnamefont
				{Carvalho-Santos}}, \bibinfo {author} {\bibfnamefont {A.~R.}\ \bibnamefont
				{Pereira}}, \bibinfo {author} {\bibfnamefont {R.~L.}\ \bibnamefont {Silva}},
			\bibinfo {author} {\bibfnamefont {R.~C.}\ \bibnamefont {Silva}},\ and\
			\bibinfo {author} {\bibfnamefont {D.}~\bibnamefont {Altbir}},\ }\bibfield
		{title} {\bibinfo {title} {Typical skyrmions versus bimerons: A long-distance
				competition in ferromagnetic racetracks},\ }\href
		{https://doi.org/10.1103/physrevb.102.104409} {\bibfield  {journal} {\bibinfo
				{journal} {Physical Review B}\ }\textbf {\bibinfo {volume} {102}},\ \bibinfo
			{pages} {104409} (\bibinfo {year} {2020})}\BibitemShut {NoStop}%
		\bibitem [{\citenamefont {Castro}\ \emph {et~al.}(2023)\citenamefont {Castro},
			\citenamefont {Altbir}, \citenamefont {Galvez-Poblete}, \citenamefont
			{Corona}, \citenamefont {Oyarz\'{u}n}, \citenamefont {Pereira}, \citenamefont
			{Allende},\ and\ \citenamefont {Carvalho-Santos}}]{Castro23}%
		\BibitemOpen
		\bibfield  {author} {\bibinfo {author} {\bibfnamefont {M.~A.}\ \bibnamefont
				{Castro}}, \bibinfo {author} {\bibfnamefont {D.}~\bibnamefont {Altbir}},
			\bibinfo {author} {\bibfnamefont {D.}~\bibnamefont {Galvez-Poblete}},
			\bibinfo {author} {\bibfnamefont {R.~M.}\ \bibnamefont {Corona}}, \bibinfo
			{author} {\bibfnamefont {S.}~\bibnamefont {Oyarz\'{u}n}}, \bibinfo {author}
			{\bibfnamefont {A.~R.}\ \bibnamefont {Pereira}}, \bibinfo {author}
			{\bibfnamefont {S.}~\bibnamefont {Allende}},\ and\ \bibinfo {author}
			{\bibfnamefont {V.~L.}\ \bibnamefont {Carvalho-Santos}},\ }\bibfield  {title}
		{\bibinfo {title} {Skyrmion-bimeron dynamic conversion in magnetic
				nanotracks},\ }\href {https://doi.org/10.1103/physrevb.108.094436} {\bibfield
			{journal} {\bibinfo  {journal} {Physical Review B}\ }\textbf {\bibinfo
				{volume} {108}},\ \bibinfo {pages} {094436} (\bibinfo {year}
			{2023})}\BibitemShut {NoStop}%
		\bibitem [{\citenamefont {Ohara}\ \emph {et~al.}(2022)\citenamefont {Ohara},
			\citenamefont {Zhang}, \citenamefont {Chen}, \citenamefont {Kato},
			\citenamefont {Xia}, \citenamefont {Ezawa}, \citenamefont {Tretiakov},
			\citenamefont {Hou}, \citenamefont {Zhou}, \citenamefont {Zhao},
			\citenamefont {Yang},\ and\ \citenamefont {Liu}}]{Ohara22}%
		\BibitemOpen
		\bibfield  {author} {\bibinfo {author} {\bibfnamefont {K.}~\bibnamefont
				{Ohara}}, \bibinfo {author} {\bibfnamefont {X.}~\bibnamefont {Zhang}},
			\bibinfo {author} {\bibfnamefont {Y.}~\bibnamefont {Chen}}, \bibinfo {author}
			{\bibfnamefont {S.}~\bibnamefont {Kato}}, \bibinfo {author} {\bibfnamefont
				{J.}~\bibnamefont {Xia}}, \bibinfo {author} {\bibfnamefont {M.}~\bibnamefont
				{Ezawa}}, \bibinfo {author} {\bibfnamefont {O.~A.}\ \bibnamefont
				{Tretiakov}}, \bibinfo {author} {\bibfnamefont {Z.}~\bibnamefont {Hou}},
			\bibinfo {author} {\bibfnamefont {Y.}~\bibnamefont {Zhou}}, \bibinfo {author}
			{\bibfnamefont {G.}~\bibnamefont {Zhao}}, \bibinfo {author} {\bibfnamefont
				{J.}~\bibnamefont {Yang}},\ and\ \bibinfo {author} {\bibfnamefont
				{X.}~\bibnamefont {Liu}},\ }\bibfield  {title} {\bibinfo {title} {Reversible
				transformation between isolated skyrmions and bimerons},\ }\href
		{https://doi.org/10.1021/acs.nanolett.2c03106} {\bibfield  {journal}
			{\bibinfo  {journal} {Nano Letters}\ }\textbf {\bibinfo {volume} {22}},\
			\bibinfo {pages} {8559} (\bibinfo {year} {2022})}\BibitemShut {NoStop}%
		\bibitem [{\citenamefont {de~Souza}\ \emph {et~al.}(2024)\citenamefont
			{de~Souza}, \citenamefont {Teixeira}, \citenamefont {Laroze}, \citenamefont
			{Pereira}, \citenamefont {Carvalho-Santos},\ and\ \citenamefont
			{Fonseca}}]{deSouza24}%
		\BibitemOpen
		\bibfield  {author} {\bibinfo {author} {\bibfnamefont {S.~F.}\ \bibnamefont
				{de~Souza}}, \bibinfo {author} {\bibfnamefont {A.~W.}\ \bibnamefont
				{Teixeira}}, \bibinfo {author} {\bibfnamefont {D.}~\bibnamefont {Laroze}},
			\bibinfo {author} {\bibfnamefont {A.~R.}\ \bibnamefont {Pereira}}, \bibinfo
			{author} {\bibfnamefont {V.~L.}\ \bibnamefont {Carvalho-Santos}},\ and\
			\bibinfo {author} {\bibfnamefont {J.~M.}\ \bibnamefont {Fonseca}},\
		}\bibfield  {title} {\bibinfo {title} {Field-driven conversion of
				two-dimensional solitonic magnetic textures},\ }\bibfield  {journal}
		{\bibinfo  {journal} {Journal of Applied Physics}\ }\textbf {\bibinfo
			{volume} {136}},\ \href {https://doi.org/10.1063/5.0231858}
		{10.1063/5.0231858} (\bibinfo {year} {2024})\BibitemShut {NoStop}%
		\bibitem [{\citenamefont {Rybakov}\ \emph {et~al.}(2025)\citenamefont
			{Rybakov}, \citenamefont {Eriksson},\ and\ \citenamefont
			{Kiselev}}]{Rybakov25}%
		\BibitemOpen
		\bibfield  {author} {\bibinfo {author} {\bibfnamefont {F.~N.}\ \bibnamefont
				{Rybakov}}, \bibinfo {author} {\bibfnamefont {O.}~\bibnamefont {Eriksson}},\
			and\ \bibinfo {author} {\bibfnamefont {N.~S.}\ \bibnamefont {Kiselev}},\
		}\bibfield  {title} {\bibinfo {title} {Topological invariants of vortices,
				merons, skyrmions, and their combinations in continuous and discrete
				systems},\ }\href {https://doi.org/10.1103/physrevb.111.134417} {\bibfield
			{journal} {\bibinfo  {journal} {Physical Review B}\ }\textbf {\bibinfo
				{volume} {111}},\ \bibinfo {pages} {134417} (\bibinfo {year}
			{2025})}\BibitemShut {NoStop}%
		\bibitem [{\citenamefont {Whitehead}(1947)}]{Whitehead47}%
		\BibitemOpen
		\bibfield  {author} {\bibinfo {author} {\bibfnamefont {J.~H.~C.}\
				\bibnamefont {Whitehead}},\ }\bibfield  {title} {\bibinfo {title} {An
				expression of hopf{\textquotesingle}s invariant as an integral},\ }\href
		{https://doi.org/10.1073/pnas.33.5.117} {\bibfield  {journal} {\bibinfo
				{journal} {Proceedings of the National Academy of Sciences}\ }\textbf
			{\bibinfo {volume} {33}},\ \bibinfo {pages} {117} (\bibinfo {year}
			{1947})}\BibitemShut {NoStop}%
		\bibitem [{\citenamefont {Kent}\ \emph {et~al.}(2021)\citenamefont {Kent},
			\citenamefont {Reynolds}, \citenamefont {Raftrey}, \citenamefont {Campbell},
			\citenamefont {Virasawmy}, \citenamefont {Dhuey}, \citenamefont {Chopdekar},
			\citenamefont {Hierro-Rodriguez}, \citenamefont {Sorrentino}, \citenamefont
			{Pereiro}, \citenamefont {Ferrer}, \citenamefont {Hellman}, \citenamefont
			{Sutcliffe},\ and\ \citenamefont {Fischer}}]{Kent21}%
		\BibitemOpen
		\bibfield  {author} {\bibinfo {author} {\bibfnamefont {N.}~\bibnamefont
				{Kent}}, \bibinfo {author} {\bibfnamefont {N.}~\bibnamefont {Reynolds}},
			\bibinfo {author} {\bibfnamefont {D.}~\bibnamefont {Raftrey}}, \bibinfo
			{author} {\bibfnamefont {I.~T.~G.}\ \bibnamefont {Campbell}}, \bibinfo
			{author} {\bibfnamefont {S.}~\bibnamefont {Virasawmy}}, \bibinfo {author}
			{\bibfnamefont {S.}~\bibnamefont {Dhuey}}, \bibinfo {author} {\bibfnamefont
				{R.~V.}\ \bibnamefont {Chopdekar}}, \bibinfo {author} {\bibfnamefont
				{A.}~\bibnamefont {Hierro-Rodriguez}}, \bibinfo {author} {\bibfnamefont
				{A.}~\bibnamefont {Sorrentino}}, \bibinfo {author} {\bibfnamefont
				{E.}~\bibnamefont {Pereiro}}, \bibinfo {author} {\bibfnamefont
				{S.}~\bibnamefont {Ferrer}}, \bibinfo {author} {\bibfnamefont
				{F.}~\bibnamefont {Hellman}}, \bibinfo {author} {\bibfnamefont
				{P.}~\bibnamefont {Sutcliffe}},\ and\ \bibinfo {author} {\bibfnamefont
				{P.}~\bibnamefont {Fischer}},\ }\bibfield  {title} {\bibinfo {title}
			{Creation and observation of {H}opfions in magnetic multilayer systems},\
		}\href {https://doi.org/10.1038/s41467-021-21846-5} {\bibfield  {journal}
			{\bibinfo  {journal} {Nature Communications}\ }\textbf {\bibinfo {volume}
				{12}},\ \bibinfo {pages} {1562} (\bibinfo {year} {2021})}\BibitemShut
		{NoStop}%
		\bibitem [{\citenamefont {Balakrishnan}\ \emph {et~al.}(2023)\citenamefont
			{Balakrishnan}, \citenamefont {Dandoloff},\ and\ \citenamefont
			{Saxena}}]{Balakrishnan23}%
		\BibitemOpen
		\bibfield  {author} {\bibinfo {author} {\bibfnamefont {R.}~\bibnamefont
				{Balakrishnan}}, \bibinfo {author} {\bibfnamefont {R.}~\bibnamefont
				{Dandoloff}},\ and\ \bibinfo {author} {\bibfnamefont {A.}~\bibnamefont
				{Saxena}},\ }\bibfield  {title} {\bibinfo {title} {Exact hopfion vortices in
				a {3D} {H}eisenberg ferromagnet},\ }\href
		{https://doi.org/10.1016/j.physleta.2023.128975} {\bibfield  {journal}
			{\bibinfo  {journal} {Physics Letters A}\ }\textbf {\bibinfo {volume}
				{480}},\ \bibinfo {pages} {128975} (\bibinfo {year} {2023})}\BibitemShut
		{NoStop}%
		\bibitem [{\citenamefont {Zheng}\ \emph {et~al.}(2023)\citenamefont {Zheng},
			\citenamefont {Kiselev}, \citenamefont {Rybakov}, \citenamefont {Yang},
			\citenamefont {Shi}, \citenamefont {Bl\"{u}gel},\ and\ \citenamefont
			{Dunin-Borkowski}}]{Zheng23}%
		\BibitemOpen
		\bibfield  {author} {\bibinfo {author} {\bibfnamefont {F.}~\bibnamefont
				{Zheng}}, \bibinfo {author} {\bibfnamefont {N.~S.}\ \bibnamefont {Kiselev}},
			\bibinfo {author} {\bibfnamefont {F.~N.}\ \bibnamefont {Rybakov}}, \bibinfo
			{author} {\bibfnamefont {L.}~\bibnamefont {Yang}}, \bibinfo {author}
			{\bibfnamefont {W.}~\bibnamefont {Shi}}, \bibinfo {author} {\bibfnamefont
				{S.}~\bibnamefont {Bl\"{u}gel}},\ and\ \bibinfo {author} {\bibfnamefont
				{R.~E.}\ \bibnamefont {Dunin-Borkowski}},\ }\bibfield  {title} {\bibinfo
			{title} {Hopfion rings in a cubic chiral magnet},\ }\href
		{https://doi.org/10.1038/s41586-023-06658-5} {\bibfield  {journal} {\bibinfo
				{journal} {Nature}\ }\textbf {\bibinfo {volume} {623}},\ \bibinfo {pages}
			{718} (\bibinfo {year} {2023})}\BibitemShut {NoStop}%
		\bibitem [{\citenamefont {Knapman}\ \emph {et~al.}(2025)\citenamefont
			{Knapman}, \citenamefont {Azhar}, \citenamefont {Pignedoli}, \citenamefont
			{Gallard}, \citenamefont {Hertel}, \citenamefont {Leliaert},\ and\
			\citenamefont {Everschor-Sitte}}]{Knapman25}%
		\BibitemOpen
		\bibfield  {author} {\bibinfo {author} {\bibfnamefont {R.}~\bibnamefont
				{Knapman}}, \bibinfo {author} {\bibfnamefont {M.}~\bibnamefont {Azhar}},
			\bibinfo {author} {\bibfnamefont {A.}~\bibnamefont {Pignedoli}}, \bibinfo
			{author} {\bibfnamefont {L.}~\bibnamefont {Gallard}}, \bibinfo {author}
			{\bibfnamefont {R.}~\bibnamefont {Hertel}}, \bibinfo {author} {\bibfnamefont
				{J.}~\bibnamefont {Leliaert}},\ and\ \bibinfo {author} {\bibfnamefont
				{K.}~\bibnamefont {Everschor-Sitte}},\ }\bibfield  {title} {\bibinfo {title}
			{Numerical calculation of the hopf index for three-dimensional magnetic
				textures},\ }\href {https://doi.org/10.1103/physrevb.111.134408} {\bibfield
			{journal} {\bibinfo  {journal} {Physical Review B}\ }\textbf {\bibinfo
				{volume} {111}},\ \bibinfo {pages} {134408} (\bibinfo {year}
			{2025})}\BibitemShut {NoStop}%
		\bibitem [{\citenamefont {Barker}\ and\ \citenamefont
			{Tretiakov}(2016)}]{Barker16}%
		\BibitemOpen
		\bibfield  {author} {\bibinfo {author} {\bibfnamefont {J.}~\bibnamefont
				{Barker}}\ and\ \bibinfo {author} {\bibfnamefont {O.~A.}\ \bibnamefont
				{Tretiakov}},\ }\bibfield  {title} {\bibinfo {title} {Static and dynamical
				properties of antiferromagnetic skyrmions in the presence of applied current
				and temperature},\ }\href {https://doi.org/10.1103/physrevlett.116.147203}
		{\bibfield  {journal} {\bibinfo  {journal} {Physical Review Letters}\
			}\textbf {\bibinfo {volume} {116}},\ \bibinfo {pages} {147203} (\bibinfo
			{year} {2016})}\BibitemShut {NoStop}%
		\bibitem [{\citenamefont {Kovalev}\ and\ \citenamefont
			{Kosevich}(1977)}]{Kovalev77}%
		\BibitemOpen
		\bibfield  {author} {\bibinfo {author} {\bibfnamefont {A.~S.}\ \bibnamefont
				{Kovalev}}\ and\ \bibinfo {author} {\bibfnamefont {A.~M.}\ \bibnamefont
				{Kosevich}},\ }\bibfield  {title} {\bibinfo {title} {Dislocations and domains
				in antiferromagnets (short note)},\ }\href@noop {} {\bibfield  {journal}
			{\bibinfo  {journal} {Low Temperature Physics (Fizika Nizkikh Temperatur)}\
			}\textbf {\bibinfo {volume} {3}},\ \bibinfo {pages} {259} (\bibinfo {year}
			{1977})}\BibitemShut {NoStop}%
		\bibitem [{\citenamefont {Dzyaloshinski{\u{i}}}(1977)}]{Dzyaloshinskiui77}%
		\BibitemOpen
		\bibfield  {author} {\bibinfo {author} {\bibfnamefont {I.~E.}\ \bibnamefont
				{Dzyaloshinski{\u{i}}}},\ }\bibfield  {title} {\bibinfo {title} {Domains and
				dislocations in antferromagnets},\ }\href
		{http://www.jetpletters.ac.ru/ps/1388/article_21067.shtml} {\bibfield
			{journal} {\bibinfo  {journal} {Pis'ma Zh. Exp. Teor. Fiz.}\ }\textbf
			{\bibinfo {volume} {25}},\ \bibinfo {pages} {110} (\bibinfo {year}
			{1977})}\BibitemShut {NoStop}%
		\bibitem [{\citenamefont {Dudko}\ and\ \citenamefont
			{Kovalev}(1998)}]{Dudko98}%
		\BibitemOpen
		\bibfield  {author} {\bibinfo {author} {\bibfnamefont {O.~K.}\ \bibnamefont
				{Dudko}}\ and\ \bibinfo {author} {\bibfnamefont {A.~S.}\ \bibnamefont
				{Kovalev}},\ }\bibfield  {title} {\bibinfo {title} {Magnetostructural
				topological defects in two-dimensional antiferromagnets},\ }\href
		{https://doi.org/10.1063/1.593612} {\bibfield  {journal} {\bibinfo  {journal}
				{Low Temperature Physics}\ }\textbf {\bibinfo {volume} {24}},\ \bibinfo
			{pages} {422} (\bibinfo {year} {1998})}\BibitemShut {NoStop}%
		\bibitem [{\citenamefont {Dudko}\ and\ \citenamefont
			{Kovalev}(2000)}]{Dudko00}%
		\BibitemOpen
		\bibfield  {author} {\bibinfo {author} {\bibfnamefont {O.~K.}\ \bibnamefont
				{Dudko}}\ and\ \bibinfo {author} {\bibfnamefont {A.~S.}\ \bibnamefont
				{Kovalev}},\ }\bibfield  {title} {\bibinfo {title} {Influence of dislocations
				on the magnetic structure of two-dimensional anisotropic antiferromagnets},\
		}\href {https://doi.org/10.1063/1.1289132} {\bibfield  {journal} {\bibinfo
				{journal} {Low Temperature Physics}\ }\textbf {\bibinfo {volume} {26}},\
			\bibinfo {pages} {603} (\bibinfo {year} {2000})}\BibitemShut {NoStop}%
		\bibitem [{\citenamefont {Ivanov}\ and\ \citenamefont
			{Kireev}(2001)}]{Ivanov01b}%
		\BibitemOpen
		\bibfield  {author} {\bibinfo {author} {\bibfnamefont {B.~A.}\ \bibnamefont
				{Ivanov}}\ and\ \bibinfo {author} {\bibfnamefont {V.~E.}\ \bibnamefont
				{Kireev}},\ }\bibfield  {title} {\bibinfo {title} {Spin disclination in a
				layered antiferromagnet with a screw dislocation},\ }\href
		{https://doi.org/10.1134/1.1368712} {\bibfield  {journal} {\bibinfo
				{journal} {Journal of Experimental and Theoretical Physics Letters}\ }\textbf
			{\bibinfo {volume} {73}},\ \bibinfo {pages} {188} (\bibinfo {year}
			{2001})}\BibitemShut {NoStop}%
		\bibitem [{\citenamefont {Galkina}\ \emph {et~al.}(2020)\citenamefont
			{Galkina}, \citenamefont {Kireev}, \citenamefont {Ovcharov}, \citenamefont
			{Khymyn},\ and\ \citenamefont {Ivanov}}]{Galkina20}%
		\BibitemOpen
		\bibfield  {author} {\bibinfo {author} {\bibfnamefont {E.~G.}\ \bibnamefont
				{Galkina}}, \bibinfo {author} {\bibfnamefont {V.~E.}\ \bibnamefont {Kireev}},
			\bibinfo {author} {\bibfnamefont {R.~V.}\ \bibnamefont {Ovcharov}}, \bibinfo
			{author} {\bibfnamefont {R.~S.}\ \bibnamefont {Khymyn}},\ and\ \bibinfo
			{author} {\bibfnamefont {B.~A.}\ \bibnamefont {Ivanov}},\ }\bibfield  {title}
		{\bibinfo {title} {Spin dynamics in antiferromagnets with domain walls and
				disclinations},\ }\href {https://doi.org/10.15407/ujpe65.10.924} {\bibfield
			{journal} {\bibinfo  {journal} {Ukrainian Journal of Physics}\ }\textbf
			{\bibinfo {volume} {65}},\ \bibinfo {pages} {924} (\bibinfo {year}
			{2020})}\BibitemShut {NoStop}%
		\bibitem [{\citenamefont {Kovalev}\ and\ \citenamefont
			{Prilepskii}(2018)}]{Kovalev18}%
		\BibitemOpen
		\bibfield  {author} {\bibinfo {author} {\bibfnamefont {A.~S.}\ \bibnamefont
				{Kovalev}}\ and\ \bibinfo {author} {\bibfnamefont {J.~E.}\ \bibnamefont
				{Prilepskii}},\ }\bibfield  {title} {\bibinfo {title} {Interaction of
				magnetic vortices with defects},\ }\href {https://doi.org/10.1063/1.5041432}
		{\bibfield  {journal} {\bibinfo  {journal} {Low Temperature Physics}\
			}\textbf {\bibinfo {volume} {44}},\ \bibinfo {pages} {663} (\bibinfo {year}
			{2018})}\BibitemShut {NoStop}%
		\bibitem [{\citenamefont {Azhar}\ \emph {et~al.}(2022)\citenamefont {Azhar},
			\citenamefont {Kravchuk},\ and\ \citenamefont {Garst}}]{Azhar22}%
		\BibitemOpen
		\bibfield  {author} {\bibinfo {author} {\bibfnamefont {M.}~\bibnamefont
				{Azhar}}, \bibinfo {author} {\bibfnamefont {V.~P.}\ \bibnamefont
				{Kravchuk}},\ and\ \bibinfo {author} {\bibfnamefont {M.}~\bibnamefont
				{Garst}},\ }\bibfield  {title} {\bibinfo {title} {Screw dislocations in
				chiral magnets},\ }\href {https://doi.org/10.1103/physrevlett.128.157204}
		{\bibfield  {journal} {\bibinfo  {journal} {Physical Review Letters}\
			}\textbf {\bibinfo {volume} {128}},\ \bibinfo {pages} {157204} (\bibinfo
			{year} {2022})}\BibitemShut {NoStop}%
		\bibitem [{\citenamefont {Kravchuk}(2023)}]{Kravchuk23}%
		\BibitemOpen
		\bibfield  {author} {\bibinfo {author} {\bibfnamefont {V.~P.}\ \bibnamefont
				{Kravchuk}},\ }\bibfield  {title} {\bibinfo {title} {Nonlinear dynamics of
				skyrmion strings},\ }\href {https://doi.org/10.1103/physrevb.108.144412}
		{\bibfield  {journal} {\bibinfo  {journal} {Physical Review B}\ }\textbf
			{\bibinfo {volume} {108}},\ \bibinfo {pages} {144412} (\bibinfo {year}
			{2023})}\BibitemShut {NoStop}%
		\bibitem [{\citenamefont {Donnelly}\ \emph {et~al.}(2020)\citenamefont
			{Donnelly}, \citenamefont {Metlov}, \citenamefont {Scagnoli}, \citenamefont
			{Guizar-Sicairos}, \citenamefont {Holler}, \citenamefont {Bingham},
			\citenamefont {Raabe}, \citenamefont {Heyderman}, \citenamefont {Cooper},\
			and\ \citenamefont {Gliga}}]{Donnelly20b}%
		\BibitemOpen
		\bibfield  {author} {\bibinfo {author} {\bibfnamefont {C.}~\bibnamefont
				{Donnelly}}, \bibinfo {author} {\bibfnamefont {K.~L.}\ \bibnamefont
				{Metlov}}, \bibinfo {author} {\bibfnamefont {V.}~\bibnamefont {Scagnoli}},
			\bibinfo {author} {\bibfnamefont {M.}~\bibnamefont {Guizar-Sicairos}},
			\bibinfo {author} {\bibfnamefont {M.}~\bibnamefont {Holler}}, \bibinfo
			{author} {\bibfnamefont {N.~S.}\ \bibnamefont {Bingham}}, \bibinfo {author}
			{\bibfnamefont {J.}~\bibnamefont {Raabe}}, \bibinfo {author} {\bibfnamefont
				{L.~J.}\ \bibnamefont {Heyderman}}, \bibinfo {author} {\bibfnamefont {N.~R.}\
				\bibnamefont {Cooper}},\ and\ \bibinfo {author} {\bibfnamefont
				{S.}~\bibnamefont {Gliga}},\ }\bibfield  {title} {\bibinfo {title}
			{Experimental observation of vortex rings in a bulk magnet},\ }\href
		{https://doi.org/10.1038/s41567-020-01057-3} {\bibfield  {journal} {\bibinfo
				{journal} {Nature Physics}\ }\textbf {\bibinfo {volume} {17}},\ \bibinfo
			{pages} {316} (\bibinfo {year} {2020})}\BibitemShut {NoStop}%
		\bibitem [{\citenamefont {Volkov}\ \emph {et~al.}(2023)\citenamefont {Volkov},
			\citenamefont {Wolf}, \citenamefont {Pylypovskyi}, \citenamefont {K\'{a}kay},
			\citenamefont {Sheka}, \citenamefont {B\"{u}chner}, \citenamefont
			{Fassbender}, \citenamefont {Lubk},\ and\ \citenamefont
			{Makarov}}]{Volkov23a}%
		\BibitemOpen
		\bibfield  {author} {\bibinfo {author} {\bibfnamefont {O.~M.}\ \bibnamefont
				{Volkov}}, \bibinfo {author} {\bibfnamefont {D.}~\bibnamefont {Wolf}},
			\bibinfo {author} {\bibfnamefont {O.~V.}\ \bibnamefont {Pylypovskyi}},
			\bibinfo {author} {\bibfnamefont {A.}~\bibnamefont {K\'{a}kay}}, \bibinfo
			{author} {\bibfnamefont {D.~D.}\ \bibnamefont {Sheka}}, \bibinfo {author}
			{\bibfnamefont {B.}~\bibnamefont {B\"{u}chner}}, \bibinfo {author}
			{\bibfnamefont {J.}~\bibnamefont {Fassbender}}, \bibinfo {author}
			{\bibfnamefont {A.}~\bibnamefont {Lubk}},\ and\ \bibinfo {author}
			{\bibfnamefont {D.}~\bibnamefont {Makarov}},\ }\bibfield  {title} {\bibinfo
			{title} {Chirality coupling in topological magnetic textures with multiple
				magnetochiral parameters},\ }\href
		{https://doi.org/10.1038/s41467-023-37081-z} {\bibfield  {journal} {\bibinfo
				{journal} {Nature Communications}\ }\textbf {\bibinfo {volume} {14}},\
			\bibinfo {pages} {1491} (\bibinfo {year} {2023})}\BibitemShut {NoStop}%
		\bibitem [{Note3()}]{Note3}%
		\BibitemOpen
		\bibinfo {note} {One can show that the same definition $\protect \mathcal {J}
			= \mp \protect \tilde{G}$ is valid a very thin curved magnetic shell with
			the replacement $\partial _x \to g_{11}^{-1/2}\partial _{\eta _1}$, $\partial
			_y \to g_{22}^{-1/2}\partial _{\eta _2}$~\cite {Kravchuk16}.}\BibitemShut
		{Stop}%
		\bibitem [{\citenamefont {Guimar\~{a}es}(2017)}]{Guimaraes17}%
		\BibitemOpen
		\bibfield  {author} {\bibinfo {author} {\bibfnamefont {A.~P.}\ \bibnamefont
				{Guimar\~{a}es}},\ }\href {https://doi.org/10.1007/978-3-319-59409-5} {\emph
			{\bibinfo {title} {Principles of Nanomagnetism}}}\ (\bibinfo  {publisher}
		{Springer-Verlag GmbH},\ \bibinfo {year} {2017})\BibitemShut {NoStop}%
		\bibitem [{\citenamefont {Hubert}\ and\ \citenamefont
			{Sch{\"a}fer}(2009)}]{Hubert09}%
		\BibitemOpen
		\bibfield  {author} {\bibinfo {author} {\bibfnamefont {A.}~\bibnamefont
				{Hubert}}\ and\ \bibinfo {author} {\bibfnamefont {R.}~\bibnamefont
				{Sch{\"a}fer}},\ }\href@noop {} {\emph {\bibinfo {title} {{Magnetic domains:
						The analysis of magnetic microstructures}}}}\ (\bibinfo  {publisher}
		{Springer Berlin Heidelberg},\ \bibinfo {address} {Berlin},\ \bibinfo {year}
		{2009})\BibitemShut {NoStop}%
		\bibitem [{\citenamefont {Stefanita}(2012)}]{Stefanita12}%
		\BibitemOpen
		\bibfield  {author} {\bibinfo {author} {\bibfnamefont {C.-G.}\ \bibnamefont
				{Stefanita}},\ }\href {https://doi.org/10.1007/978-3-642-22977-0} {\emph
			{\bibinfo {title} {Magnetism: Basics and Applications}}}\ (\bibinfo
		{publisher} {Springer Berlin Heidelberg},\ \bibinfo {address} {Berlin,
			Heidelberg},\ \bibinfo {year} {2012})\BibitemShut {NoStop}%
		\bibitem [{\citenamefont {Teresa}\ \emph {et~al.}(2016)\citenamefont {Teresa},
			\citenamefont {Fern{\'{a}}ndez-Pacheco}, \citenamefont {C{\'{o}}rdoba},
			\citenamefont {Serrano-Ram{\'{o}}n}, \citenamefont {Sangiao},\ and\
			\citenamefont {Ibarra}}]{Teresa16}%
		\BibitemOpen
		\bibfield  {author} {\bibinfo {author} {\bibfnamefont {J.~M.~D.}\
				\bibnamefont {Teresa}}, \bibinfo {author} {\bibfnamefont {A.}~\bibnamefont
				{Fern{\'{a}}ndez-Pacheco}}, \bibinfo {author} {\bibfnamefont
				{R.}~\bibnamefont {C{\'{o}}rdoba}}, \bibinfo {author} {\bibfnamefont
				{L.}~\bibnamefont {Serrano-Ram{\'{o}}n}}, \bibinfo {author} {\bibfnamefont
				{S.}~\bibnamefont {Sangiao}},\ and\ \bibinfo {author} {\bibfnamefont {M.~R.}\
				\bibnamefont {Ibarra}},\ }\bibfield  {title} {\bibinfo {title} {Review of
				magnetic nanostructures grown by focused electron beam induced deposition
				({FEBID})},\ }\href {https://doi.org/10.1088/0022-3727/49/24/243003}
		{\bibfield  {journal} {\bibinfo  {journal} {Journal of Physics D: Applied
					Physics}\ }\textbf {\bibinfo {volume} {49}},\ \bibinfo {pages} {243003}
			(\bibinfo {year} {2016})}\BibitemShut {NoStop}%
		\bibitem [{\citenamefont {Fern{\'{a}}ndez-Pacheco}\ \emph
			{et~al.}(2020)\citenamefont {Fern{\'{a}}ndez-Pacheco}, \citenamefont
			{Skoric}, \citenamefont {Teresa}, \citenamefont {Pablo-Navarro},
			\citenamefont {Huth},\ and\ \citenamefont
			{Dobrovolskiy}}]{Fernandez-Pacheco20}%
		\BibitemOpen
		\bibfield  {author} {\bibinfo {author} {\bibfnamefont {A.}~\bibnamefont
				{Fern{\'{a}}ndez-Pacheco}}, \bibinfo {author} {\bibfnamefont
				{L.}~\bibnamefont {Skoric}}, \bibinfo {author} {\bibfnamefont {J.~M.~D.}\
				\bibnamefont {Teresa}}, \bibinfo {author} {\bibfnamefont {J.}~\bibnamefont
				{Pablo-Navarro}}, \bibinfo {author} {\bibfnamefont {M.}~\bibnamefont
				{Huth}},\ and\ \bibinfo {author} {\bibfnamefont {O.~V.}\ \bibnamefont
				{Dobrovolskiy}},\ }\bibfield  {title} {\bibinfo {title} {Writing {3D}
				nanomagnets using focused electron beams},\ }\href
		{https://doi.org/10.3390/ma13173774} {\bibfield  {journal} {\bibinfo
				{journal} {Materials}\ }\textbf {\bibinfo {volume} {13}},\ \bibinfo {pages}
			{3774} (\bibinfo {year} {2020})}\BibitemShut {NoStop}%
		\bibitem [{\citenamefont {Barth}\ \emph {et~al.}(2020)\citenamefont {Barth},
			\citenamefont {Huth},\ and\ \citenamefont {Jungwirth}}]{Barth20}%
		\BibitemOpen
		\bibfield  {author} {\bibinfo {author} {\bibfnamefont {S.}~\bibnamefont
				{Barth}}, \bibinfo {author} {\bibfnamefont {M.}~\bibnamefont {Huth}},\ and\
			\bibinfo {author} {\bibfnamefont {F.}~\bibnamefont {Jungwirth}},\ }\bibfield
		{title} {\bibinfo {title} {Precursors for direct-write nanofabrication with
				electrons},\ }\href {https://doi.org/10.1039/d0tc03689g} {\bibfield
			{journal} {\bibinfo  {journal} {Journal of Materials Chemistry C}\ }\textbf
			{\bibinfo {volume} {8}},\ \bibinfo {pages} {15884} (\bibinfo {year}
			{2020})}\BibitemShut {NoStop}%
		\bibitem [{\citenamefont {Skoric}\ \emph {et~al.}(2020)\citenamefont {Skoric},
			\citenamefont {Sanz-Hern{\'{a}}ndez}, \citenamefont {Meng}, \citenamefont
			{Donnelly}, \citenamefont {Merino-Aceituno},\ and\ \citenamefont
			{Fern{\'{a}}ndez-Pacheco}}]{Skoric20}%
		\BibitemOpen
		\bibfield  {author} {\bibinfo {author} {\bibfnamefont {L.}~\bibnamefont
				{Skoric}}, \bibinfo {author} {\bibfnamefont {D.}~\bibnamefont
				{Sanz-Hern{\'{a}}ndez}}, \bibinfo {author} {\bibfnamefont {F.}~\bibnamefont
				{Meng}}, \bibinfo {author} {\bibfnamefont {C.}~\bibnamefont {Donnelly}},
			\bibinfo {author} {\bibfnamefont {S.}~\bibnamefont {Merino-Aceituno}},\ and\
			\bibinfo {author} {\bibfnamefont {A.}~\bibnamefont
				{Fern{\'{a}}ndez-Pacheco}},\ }\bibfield  {title} {\bibinfo {title}
			{Layer-by-layer growth of complex-shaped three-dimensional nanostructures
				with focused electron beams},\ }\href
		{https://doi.org/10.1021/acs.nanolett.9b03565} {\bibfield  {journal}
			{\bibinfo  {journal} {Nano Letters}\ }\textbf {\bibinfo {volume} {20}},\
			\bibinfo {pages} {184} (\bibinfo {year} {2020})}\BibitemShut {NoStop}%
		\bibitem [{\citenamefont {Skoric}(2020)}]{f3ast_github}%
		\BibitemOpen
		\bibfield  {author} {\bibinfo {author} {\bibfnamefont {L.}~\bibnamefont
				{Skoric}},\ }\href {https://github.com/Skoricius/f3ast} {\bibinfo {title}
			{Febid {3D} algorithm for stream file generation (f3ast)}} (\bibinfo {year}
		{2020}),\ \bibinfo {note} {access date: 2026-03-06}\BibitemShut {NoStop}%
		\bibitem [{\citenamefont {Donnelly}\ \emph
			{et~al.}(2022{\natexlab{a}})\citenamefont {Donnelly}, \citenamefont
			{Hierro-Rodr\'{i}guez}, \citenamefont {Abert}, \citenamefont {Witte},
			\citenamefont {Skoric}, \citenamefont {Sanz-Hern\'{a}ndez}, \citenamefont
			{Finizio}, \citenamefont {Meng}, \citenamefont {McVitie}, \citenamefont
			{Raabe}, \citenamefont {Suess}, \citenamefont {Cowburn},\ and\ \citenamefont
			{Fern\'{a}ndez-Pacheco}}]{Donnelly22}%
		\BibitemOpen
		\bibfield  {author} {\bibinfo {author} {\bibfnamefont {C.}~\bibnamefont
				{Donnelly}}, \bibinfo {author} {\bibfnamefont {A.}~\bibnamefont
				{Hierro-Rodr\'{i}guez}}, \bibinfo {author} {\bibfnamefont {C.}~\bibnamefont
				{Abert}}, \bibinfo {author} {\bibfnamefont {K.}~\bibnamefont {Witte}},
			\bibinfo {author} {\bibfnamefont {L.}~\bibnamefont {Skoric}}, \bibinfo
			{author} {\bibfnamefont {D.}~\bibnamefont {Sanz-Hern\'{a}ndez}}, \bibinfo
			{author} {\bibfnamefont {S.}~\bibnamefont {Finizio}}, \bibinfo {author}
			{\bibfnamefont {F.}~\bibnamefont {Meng}}, \bibinfo {author} {\bibfnamefont
				{S.}~\bibnamefont {McVitie}}, \bibinfo {author} {\bibfnamefont
				{J.}~\bibnamefont {Raabe}}, \bibinfo {author} {\bibfnamefont
				{D.}~\bibnamefont {Suess}}, \bibinfo {author} {\bibfnamefont
				{R.}~\bibnamefont {Cowburn}},\ and\ \bibinfo {author} {\bibfnamefont
				{A.}~\bibnamefont {Fern\'{a}ndez-Pacheco}},\ }\bibfield  {title} {\bibinfo
			{title} {Complex free-space magnetic field textures induced by
				three-dimensional magnetic nanostructures},\ }\href
		{https://doi.org/10.1038/s41565-021-01027-7} {\bibfield  {journal} {\bibinfo
				{journal} {Nature Nanotechnology}\ }\textbf {\bibinfo {volume} {17}},\
			\bibinfo {pages} {136} (\bibinfo {year} {2022}{\natexlab{a}})}\BibitemShut
		{NoStop}%
		\bibitem [{\citenamefont {Ruiz-G\'{o}mez}\ \emph {et~al.}(2025)\citenamefont
			{Ruiz-G\'{o}mez}, \citenamefont {Abert}, \citenamefont
			{Morales-Fern\'{a}ndez}, \citenamefont {Fern\'{a}ndez-Gonz\'{a}lez},
			\citenamefont {Koraltan}, \citenamefont {Danesi}, \citenamefont {Suess},
			\citenamefont {Varela}, \citenamefont {S\'{a}nchez-Santolino}, \citenamefont
			{Bagu\'{e}s}, \citenamefont {Foerster}, \citenamefont {Ni\~{n}o},
			\citenamefont {Mandziak}, \citenamefont {Wilgocka-Ślęzak}, \citenamefont
			{Nita}, \citenamefont {Koenig}, \citenamefont {Seifert}, \citenamefont
			{Hierro-Rodriguez}, \citenamefont {Fern\'{a}ndez-Pacheco},\ and\
			\citenamefont {Donnelly}}]{RuizGomez25}%
		\BibitemOpen
		\bibfield  {author} {\bibinfo {author} {\bibfnamefont {S.}~\bibnamefont
				{Ruiz-G\'{o}mez}}, \bibinfo {author} {\bibfnamefont {C.}~\bibnamefont
				{Abert}}, \bibinfo {author} {\bibfnamefont {P.}~\bibnamefont
				{Morales-Fern\'{a}ndez}}, \bibinfo {author} {\bibfnamefont {C.}~\bibnamefont
				{Fern\'{a}ndez-Gonz\'{a}lez}}, \bibinfo {author} {\bibfnamefont
				{S.}~\bibnamefont {Koraltan}}, \bibinfo {author} {\bibfnamefont
				{L.}~\bibnamefont {Danesi}}, \bibinfo {author} {\bibfnamefont
				{D.}~\bibnamefont {Suess}}, \bibinfo {author} {\bibfnamefont
				{M.}~\bibnamefont {Varela}}, \bibinfo {author} {\bibfnamefont
				{G.}~\bibnamefont {S\'{a}nchez-Santolino}}, \bibinfo {author} {\bibfnamefont
				{N.}~\bibnamefont {Bagu\'{e}s}}, \bibinfo {author} {\bibfnamefont
				{M.}~\bibnamefont {Foerster}}, \bibinfo {author} {\bibfnamefont {M.~A.}\
				\bibnamefont {Ni\~{n}o}}, \bibinfo {author} {\bibfnamefont {A.}~\bibnamefont
				{Mandziak}}, \bibinfo {author} {\bibfnamefont {D.}~\bibnamefont
				{Wilgocka-Ślęzak}}, \bibinfo {author} {\bibfnamefont {P.}~\bibnamefont
				{Nita}}, \bibinfo {author} {\bibfnamefont {M.}~\bibnamefont {Koenig}},
			\bibinfo {author} {\bibfnamefont {S.}~\bibnamefont {Seifert}}, \bibinfo
			{author} {\bibfnamefont {A.}~\bibnamefont {Hierro-Rodriguez}}, \bibinfo
			{author} {\bibfnamefont {A.}~\bibnamefont {Fern\'{a}ndez-Pacheco}},\ and\
			\bibinfo {author} {\bibfnamefont {C.}~\bibnamefont {Donnelly}},\ }\bibfield
		{title} {\bibinfo {title} {Tailoring the energy landscape of a {B}loch point
				domain wall with curvature},\ }\href
		{https://doi.org/10.1038/s41467-025-62705-x} {\bibfield  {journal} {\bibinfo
				{journal} {Nature Communications}\ }\textbf {\bibinfo {volume} {16}},\
			\bibinfo {pages} {7422} (\bibinfo {year} {2025})}\BibitemShut {NoStop}%
		\bibitem [{\citenamefont {Sanz-Hern{\'{a}}ndez}\ \emph
			{et~al.}(2020)\citenamefont {Sanz-Hern{\'{a}}ndez}, \citenamefont
			{Hierro-Rodriguez}, \citenamefont {Donnelly}, \citenamefont {Pablo-Navarro},
			\citenamefont {Sorrentino}, \citenamefont {Pereiro}, \citenamefont
			{Mag{\'{e}}n}, \citenamefont {McVitie}, \citenamefont {de~Teresa},
			\citenamefont {Ferrer}, \citenamefont {Fischer},\ and\ \citenamefont
			{Fern{\'{a}}ndez-Pacheco}}]{Sanz-Hernandez20}%
		\BibitemOpen
		\bibfield  {author} {\bibinfo {author} {\bibfnamefont {D.}~\bibnamefont
				{Sanz-Hern{\'{a}}ndez}}, \bibinfo {author} {\bibfnamefont {A.}~\bibnamefont
				{Hierro-Rodriguez}}, \bibinfo {author} {\bibfnamefont {C.}~\bibnamefont
				{Donnelly}}, \bibinfo {author} {\bibfnamefont {J.}~\bibnamefont
				{Pablo-Navarro}}, \bibinfo {author} {\bibfnamefont {A.}~\bibnamefont
				{Sorrentino}}, \bibinfo {author} {\bibfnamefont {E.}~\bibnamefont {Pereiro}},
			\bibinfo {author} {\bibfnamefont {C.}~\bibnamefont {Mag{\'{e}}n}}, \bibinfo
			{author} {\bibfnamefont {S.}~\bibnamefont {McVitie}}, \bibinfo {author}
			{\bibfnamefont {J.~M.}\ \bibnamefont {de~Teresa}}, \bibinfo {author}
			{\bibfnamefont {S.}~\bibnamefont {Ferrer}}, \bibinfo {author} {\bibfnamefont
				{P.}~\bibnamefont {Fischer}},\ and\ \bibinfo {author} {\bibfnamefont
				{A.}~\bibnamefont {Fern{\'{a}}ndez-Pacheco}},\ }\bibfield  {title} {\bibinfo
			{title} {Artificial double-helix for geometrical control of magnetic
				chirality},\ }\href {https://doi.org/10.1021/acsnano.0c00720} {\bibfield
			{journal} {\bibinfo  {journal} {{ACS} Nano}\ }\textbf {\bibinfo {volume}
				{14}},\ \bibinfo {pages} {8084} (\bibinfo {year} {2020})}\BibitemShut
		{NoStop}%
		\bibitem [{\citenamefont {Fullerton}\ \emph
			{et~al.}(2025{\natexlab{a}})\citenamefont {Fullerton}, \citenamefont {Leo},
			\citenamefont {Jurczyk}, \citenamefont {Donnelly}, \citenamefont
			{Sanz‐Hern\'{a}ndez}, \citenamefont {Skoric}, \citenamefont {Mille},
			\citenamefont {Stanescu}, \citenamefont {MacLaren}, \citenamefont {Belkhou},
			\citenamefont {Hierro‐Rodriguez},\ and\ \citenamefont
			{Fern\'{a}ndez‐Pacheco}}]{Fullerton25a}%
		\BibitemOpen
		\bibfield  {author} {\bibinfo {author} {\bibfnamefont {J.}~\bibnamefont
				{Fullerton}}, \bibinfo {author} {\bibfnamefont {N.}~\bibnamefont {Leo}},
			\bibinfo {author} {\bibfnamefont {J.}~\bibnamefont {Jurczyk}}, \bibinfo
			{author} {\bibfnamefont {C.}~\bibnamefont {Donnelly}}, \bibinfo {author}
			{\bibfnamefont {D.}~\bibnamefont {Sanz‐Hern\'{a}ndez}}, \bibinfo {author}
			{\bibfnamefont {L.}~\bibnamefont {Skoric}}, \bibinfo {author} {\bibfnamefont
				{N.}~\bibnamefont {Mille}}, \bibinfo {author} {\bibfnamefont
				{S.}~\bibnamefont {Stanescu}}, \bibinfo {author} {\bibfnamefont {D.~A.}\
				\bibnamefont {MacLaren}}, \bibinfo {author} {\bibfnamefont {R.}~\bibnamefont
				{Belkhou}}, \bibinfo {author} {\bibfnamefont {A.}~\bibnamefont
				{Hierro‐Rodriguez}},\ and\ \bibinfo {author} {\bibfnamefont
				{A.}~\bibnamefont {Fern\'{a}ndez‐Pacheco}},\ }\bibfield  {title} {\bibinfo
			{title} {Fractional skyrmion tubes in chiral‐interfaced {3D} magnetic
				nanowires},\ }\href {https://doi.org/10.1002/adfm.202501615} {\bibfield
			{journal} {\bibinfo  {journal} {Advanced Functional Materials}\ ,\ \bibinfo
				{pages} {2501615}} (\bibinfo {year} {2025}{\natexlab{a}})}\BibitemShut
		{NoStop}%
		\bibitem [{\citenamefont {Gokita}\ \emph {et~al.}(2026)\citenamefont {Gokita},
			\citenamefont {Jurczyk}, \citenamefont {Leo}, \citenamefont {Koraltan},
			\citenamefont {Anad\'{o}n}, \citenamefont {\'{A}ngel Cascales-Sandoval},
			\citenamefont {Belkhou}, \citenamefont {Abert}, \citenamefont {Suess},
			\citenamefont {Donnelly},\ and\ \citenamefont
			{Fern\'{a}ndez-Pacheco}}]{Gokita26}%
		\BibitemOpen
		\bibfield  {author} {\bibinfo {author} {\bibfnamefont {T.}~\bibnamefont
				{Gokita}}, \bibinfo {author} {\bibfnamefont {J.}~\bibnamefont {Jurczyk}},
			\bibinfo {author} {\bibfnamefont {N.}~\bibnamefont {Leo}}, \bibinfo {author}
			{\bibfnamefont {S.}~\bibnamefont {Koraltan}}, \bibinfo {author}
			{\bibfnamefont {A.}~\bibnamefont {Anad\'{o}n}}, \bibinfo {author}
			{\bibfnamefont {M.}~\bibnamefont {\'{A}ngel Cascales-Sandoval}}, \bibinfo
			{author} {\bibfnamefont {R.}~\bibnamefont {Belkhou}}, \bibinfo {author}
			{\bibfnamefont {C.}~\bibnamefont {Abert}}, \bibinfo {author} {\bibfnamefont
				{D.}~\bibnamefont {Suess}}, \bibinfo {author} {\bibfnamefont
				{C.}~\bibnamefont {Donnelly}},\ and\ \bibinfo {author} {\bibfnamefont
				{A.}~\bibnamefont {Fern\'{a}ndez-Pacheco}},\ }\bibfield  {title} {\bibinfo
			{title} {Magnetization reversal mechanism of double-helix nanowires probed by
				dark-field magneto-optical kerr effect},\ }\bibfield  {journal} {\bibinfo
			{journal} {ArXiv e-prints}\ }\href
		{https://doi.org/10.48550/arXiv.2601.08613} {10.48550/arXiv.2601.08613}
		(\bibinfo {year} {2026})\BibitemShut {NoStop}%
		\bibitem [{\citenamefont {Meng}\ \emph
			{et~al.}(2021{\natexlab{a}})\citenamefont {Meng}, \citenamefont {Donnelly},
			\citenamefont {Skoric}, \citenamefont {Hierro-Rodriguez}, \citenamefont {wei
				Liao},\ and\ \citenamefont {Fern{\'{a}}ndez-Pacheco}}]{Meng21a}%
		\BibitemOpen
		\bibfield  {author} {\bibinfo {author} {\bibfnamefont {F.}~\bibnamefont
				{Meng}}, \bibinfo {author} {\bibfnamefont {C.}~\bibnamefont {Donnelly}},
			\bibinfo {author} {\bibfnamefont {L.}~\bibnamefont {Skoric}}, \bibinfo
			{author} {\bibfnamefont {A.}~\bibnamefont {Hierro-Rodriguez}}, \bibinfo
			{author} {\bibfnamefont {J.}~\bibnamefont {wei Liao}},\ and\ \bibinfo
			{author} {\bibfnamefont {A.}~\bibnamefont {Fern{\'{a}}ndez-Pacheco}},\
		}\bibfield  {title} {\bibinfo {title} {Fabrication of a {3D} nanomagnetic
				circuit with multi-layered materials for applications in spintronics},\
		}\href {https://doi.org/10.3390/mi12080859} {\bibfield  {journal} {\bibinfo
				{journal} {Micromachines}\ }\textbf {\bibinfo {volume} {12}},\ \bibinfo
			{pages} {859} (\bibinfo {year} {2021}{\natexlab{a}})}\BibitemShut {NoStop}%
		\bibitem [{\citenamefont {Volkov}\ \emph {et~al.}(2024)\citenamefont {Volkov},
			\citenamefont {Pylypovskyi}, \citenamefont {Porrati}, \citenamefont
			{Kronast}, \citenamefont {Fernandez-Roldan}, \citenamefont {K\'{a}kay},
			\citenamefont {Kuprava}, \citenamefont {Barth}, \citenamefont {Rybakov},
			\citenamefont {Eriksson}, \citenamefont {Lamb-Camarena}, \citenamefont
			{Makushko}, \citenamefont {Mawass}, \citenamefont {Shakeel}, \citenamefont
			{Dobrovolskiy}, \citenamefont {Huth},\ and\ \citenamefont
			{Makarov}}]{Volkov24a}%
		\BibitemOpen
		\bibfield  {author} {\bibinfo {author} {\bibfnamefont {O.~M.}\ \bibnamefont
				{Volkov}}, \bibinfo {author} {\bibfnamefont {O.~V.}\ \bibnamefont
				{Pylypovskyi}}, \bibinfo {author} {\bibfnamefont {F.}~\bibnamefont
				{Porrati}}, \bibinfo {author} {\bibfnamefont {F.}~\bibnamefont {Kronast}},
			\bibinfo {author} {\bibfnamefont {J.~A.}\ \bibnamefont {Fernandez-Roldan}},
			\bibinfo {author} {\bibfnamefont {A.}~\bibnamefont {K\'{a}kay}}, \bibinfo
			{author} {\bibfnamefont {A.}~\bibnamefont {Kuprava}}, \bibinfo {author}
			{\bibfnamefont {S.}~\bibnamefont {Barth}}, \bibinfo {author} {\bibfnamefont
				{F.~N.}\ \bibnamefont {Rybakov}}, \bibinfo {author} {\bibfnamefont
				{O.}~\bibnamefont {Eriksson}}, \bibinfo {author} {\bibfnamefont
				{S.}~\bibnamefont {Lamb-Camarena}}, \bibinfo {author} {\bibfnamefont
				{P.}~\bibnamefont {Makushko}}, \bibinfo {author} {\bibfnamefont {M.-A.}\
				\bibnamefont {Mawass}}, \bibinfo {author} {\bibfnamefont {S.}~\bibnamefont
				{Shakeel}}, \bibinfo {author} {\bibfnamefont {O.~V.}\ \bibnamefont
				{Dobrovolskiy}}, \bibinfo {author} {\bibfnamefont {M.}~\bibnamefont {Huth}},\
			and\ \bibinfo {author} {\bibfnamefont {D.}~\bibnamefont {Makarov}},\
		}\bibfield  {title} {\bibinfo {title} {Three-dimensional magnetic
				nanotextures with high-order vorticity in soft magnetic wireframes},\ }\href
		{https://doi.org/10.1038/s41467-024-46403-8} {\bibfield  {journal} {\bibinfo
				{journal} {Nature Communications}\ }\textbf {\bibinfo {volume} {15}},\
			\bibinfo {pages} {2193} (\bibinfo {year} {2024})}\BibitemShut {NoStop}%
		\bibitem [{\citenamefont {Sanz-Hern\'{a}ndez}\ and\ \citenamefont
			{Fern\'{a}ndez-Pacheco}(2017)}]{SanzHernandez17}%
		\BibitemOpen
		\bibfield  {author} {\bibinfo {author} {\bibfnamefont {D.}~\bibnamefont
				{Sanz-Hern\'{a}ndez}}\ and\ \bibinfo {author} {\bibfnamefont
				{A.}~\bibnamefont {Fern\'{a}ndez-Pacheco}},\ }\bibfield  {title} {\bibinfo
			{title} {Modelling focused electron beam induced deposition beyond {L}angmuir
				adsorption},\ }\href {https://doi.org/10.3762/bjnano.8.214} {\bibfield
			{journal} {\bibinfo  {journal} {Beilstein Journal of Nanotechnology}\
			}\textbf {\bibinfo {volume} {8}},\ \bibinfo {pages} {2151} (\bibinfo {year}
			{2017})}\BibitemShut {NoStop}%
		\bibitem [{\citenamefont {Hirt}\ \emph {et~al.}(2017)\citenamefont {Hirt},
			\citenamefont {Reiser}, \citenamefont {Spolenak},\ and\ \citenamefont
			{Zambelli}}]{Hirt17}%
		\BibitemOpen
		\bibfield  {author} {\bibinfo {author} {\bibfnamefont {L.}~\bibnamefont
				{Hirt}}, \bibinfo {author} {\bibfnamefont {A.}~\bibnamefont {Reiser}},
			\bibinfo {author} {\bibfnamefont {R.}~\bibnamefont {Spolenak}},\ and\
			\bibinfo {author} {\bibfnamefont {T.}~\bibnamefont {Zambelli}},\ }\bibfield
		{title} {\bibinfo {title} {Additive manufacturing of metal structures at the
				micrometer scale},\ }\href {https://doi.org/10.1002/adma.201604211}
		{\bibfield  {journal} {\bibinfo  {journal} {Advanced Materials}\ }\textbf
			{\bibinfo {volume} {29}},\ \bibinfo {pages} {1604211} (\bibinfo {year}
			{2017})}\BibitemShut {NoStop}%
		\bibitem [{\citenamefont {Askey}\ \emph {et~al.}(2025)\citenamefont {Askey},
			\citenamefont {van~den Berg}, \citenamefont {Gomez}, \citenamefont
			{Donnelly}, \citenamefont {Grunder},\ and\ \citenamefont {Ladak}}]{Askey25}%
		\BibitemOpen
		\bibfield  {author} {\bibinfo {author} {\bibfnamefont {J.}~\bibnamefont
				{Askey}}, \bibinfo {author} {\bibfnamefont {A.}~\bibnamefont {van~den Berg}},
			\bibinfo {author} {\bibfnamefont {S.~R.}\ \bibnamefont {Gomez}}, \bibinfo
			{author} {\bibfnamefont {C.}~\bibnamefont {Donnelly}}, \bibinfo {author}
			{\bibfnamefont {D.}~\bibnamefont {Grunder}},\ and\ \bibinfo {author}
			{\bibfnamefont {S.}~\bibnamefont {Ladak}},\ }\bibfield  {title} {\bibinfo
			{title} {Exploiting two‐photon lithography, deposition, and processing to
				realize complex {3D} magnetic nanostructures},\ }\href
		{https://doi.org/10.1002/adfm.202516383} {\bibfield  {journal} {\bibinfo
				{journal} {Advanced Functional Materials}\ ,\ \bibinfo {pages} {e16383}}
			(\bibinfo {year} {2025})}\BibitemShut {NoStop}%
		\bibitem [{\citenamefont {Sanz-Hern{\'{a}}ndez}\ \emph
			{et~al.}(2018)\citenamefont {Sanz-Hern{\'{a}}ndez}, \citenamefont {Hamans},
			\citenamefont {Osterrieth}, \citenamefont {Liao}, \citenamefont {Skoric},
			\citenamefont {Fowlkes}, \citenamefont {Rack}, \citenamefont {Lippert},
			\citenamefont {Lee}, \citenamefont {Lavrijsen},\ and\ \citenamefont
			{Fern{\'{a}}ndez-Pacheco}}]{Sanz-Hernandez18}%
		\BibitemOpen
		\bibfield  {author} {\bibinfo {author} {\bibfnamefont {D.}~\bibnamefont
				{Sanz-Hern{\'{a}}ndez}}, \bibinfo {author} {\bibfnamefont {R.}~\bibnamefont
				{Hamans}}, \bibinfo {author} {\bibfnamefont {J.}~\bibnamefont {Osterrieth}},
			\bibinfo {author} {\bibfnamefont {J.-W.}\ \bibnamefont {Liao}}, \bibinfo
			{author} {\bibfnamefont {L.}~\bibnamefont {Skoric}}, \bibinfo {author}
			{\bibfnamefont {J.}~\bibnamefont {Fowlkes}}, \bibinfo {author} {\bibfnamefont
				{P.}~\bibnamefont {Rack}}, \bibinfo {author} {\bibfnamefont {A.}~\bibnamefont
				{Lippert}}, \bibinfo {author} {\bibfnamefont {S.}~\bibnamefont {Lee}},
			\bibinfo {author} {\bibfnamefont {R.}~\bibnamefont {Lavrijsen}},\ and\
			\bibinfo {author} {\bibfnamefont {A.}~\bibnamefont
				{Fern{\'{a}}ndez-Pacheco}},\ }\bibfield  {title} {\bibinfo {title}
			{Fabrication of scaffold-based 3d magnetic nanowires for domain wall
				applications},\ }\href {https://doi.org/10.3390/nano8070483} {\bibfield
			{journal} {\bibinfo  {journal} {Nanomaterials}\ }\textbf {\bibinfo {volume}
				{8}},\ \bibinfo {pages} {483} (\bibinfo {year} {2018})}\BibitemShut {NoStop}%
		\bibitem [{\citenamefont {Skoric}\ \emph {et~al.}(2022)\citenamefont {Skoric},
			\citenamefont {Donnelly}, \citenamefont {Hierro-Rodriguez}, \citenamefont
			{Cascales~Sandoval}, \citenamefont {Ruiz-G\'{o}mez}, \citenamefont
			{Foerster}, \citenamefont {Ni\~{n}o}, \citenamefont {Belkhou}, \citenamefont
			{Abert}, \citenamefont {Suess},\ and\ \citenamefont
			{Fern\'{a}ndez-Pacheco}}]{Skoric22}%
		\BibitemOpen
		\bibfield  {author} {\bibinfo {author} {\bibfnamefont {L.}~\bibnamefont
				{Skoric}}, \bibinfo {author} {\bibfnamefont {C.}~\bibnamefont {Donnelly}},
			\bibinfo {author} {\bibfnamefont {A.}~\bibnamefont {Hierro-Rodriguez}},
			\bibinfo {author} {\bibfnamefont {M.~A.}\ \bibnamefont {Cascales~Sandoval}},
			\bibinfo {author} {\bibfnamefont {S.}~\bibnamefont {Ruiz-G\'{o}mez}},
			\bibinfo {author} {\bibfnamefont {M.}~\bibnamefont {Foerster}}, \bibinfo
			{author} {\bibfnamefont {M.~A.}\ \bibnamefont {Ni\~{n}o}}, \bibinfo {author}
			{\bibfnamefont {R.}~\bibnamefont {Belkhou}}, \bibinfo {author} {\bibfnamefont
				{C.}~\bibnamefont {Abert}}, \bibinfo {author} {\bibfnamefont
				{D.}~\bibnamefont {Suess}},\ and\ \bibinfo {author} {\bibfnamefont
				{A.}~\bibnamefont {Fern\'{a}ndez-Pacheco}},\ }\bibfield  {title} {\bibinfo
			{title} {Domain wall automotion in three-dimensional magnetic helical
				interconnectors},\ }\href {https://doi.org/10.1021/acsnano.1c10345}
		{\bibfield  {journal} {\bibinfo  {journal} {ACS Nano}\ }\textbf {\bibinfo
				{volume} {16}},\ \bibinfo {pages} {8860} (\bibinfo {year}
			{2022})}\BibitemShut {NoStop}%
		\bibitem [{\citenamefont {Cui}(2008)}]{Cui08}%
		\BibitemOpen
		\bibfield  {author} {\bibinfo {author} {\bibfnamefont {Z.}~\bibnamefont
				{Cui}},\ }\href {https://doi.org/10.1007/978-0-387-75577-9} {\emph {\bibinfo
				{title} {Nanofabrication}}}\ (\bibinfo  {publisher} {Springer US},\ \bibinfo
		{address} {Boston, MA},\ \bibinfo {year} {2008})\BibitemShut {NoStop}%
		\bibitem [{\citenamefont {Hayat}\ \emph {et~al.}(2011)\citenamefont {Hayat},
			\citenamefont {Nevet}, \citenamefont {Ginzburg},\ and\ \citenamefont
			{Orenstein}}]{Hayat11}%
		\BibitemOpen
		\bibfield  {author} {\bibinfo {author} {\bibfnamefont {A.}~\bibnamefont
				{Hayat}}, \bibinfo {author} {\bibfnamefont {A.}~\bibnamefont {Nevet}},
			\bibinfo {author} {\bibfnamefont {P.}~\bibnamefont {Ginzburg}},\ and\
			\bibinfo {author} {\bibfnamefont {M.}~\bibnamefont {Orenstein}},\ }\bibfield
		{title} {\bibinfo {title} {Applications of two-photon processes in
				semiconductor photonic devices: invited review},\ }\href
		{https://doi.org/10.1088/0268-1242/26/8/083001} {\bibfield  {journal}
			{\bibinfo  {journal} {Semiconductor Science and Technology}\ }\textbf
			{\bibinfo {volume} {26}},\ \bibinfo {pages} {083001} (\bibinfo {year}
			{2011})}\BibitemShut {NoStop}%
		\bibitem [{\citenamefont {Zheng}\ \emph {et~al.}(2016)\citenamefont {Zheng},
			\citenamefont {Smith}, \citenamefont {Jackson}, \citenamefont {Moran},
			\citenamefont {Cui}, \citenamefont {Chen}, \citenamefont {Ye}, \citenamefont
			{Fang}, \citenamefont {Rodriguez}, \citenamefont {Weisgraber},\ and\
			\citenamefont {Spadaccini}}]{Zheng16}%
		\BibitemOpen
		\bibfield  {author} {\bibinfo {author} {\bibfnamefont {X.}~\bibnamefont
				{Zheng}}, \bibinfo {author} {\bibfnamefont {W.}~\bibnamefont {Smith}},
			\bibinfo {author} {\bibfnamefont {J.}~\bibnamefont {Jackson}}, \bibinfo
			{author} {\bibfnamefont {B.}~\bibnamefont {Moran}}, \bibinfo {author}
			{\bibfnamefont {H.}~\bibnamefont {Cui}}, \bibinfo {author} {\bibfnamefont
				{D.}~\bibnamefont {Chen}}, \bibinfo {author} {\bibfnamefont {J.}~\bibnamefont
				{Ye}}, \bibinfo {author} {\bibfnamefont {N.}~\bibnamefont {Fang}}, \bibinfo
			{author} {\bibfnamefont {N.}~\bibnamefont {Rodriguez}}, \bibinfo {author}
			{\bibfnamefont {T.}~\bibnamefont {Weisgraber}},\ and\ \bibinfo {author}
			{\bibfnamefont {C.~M.}\ \bibnamefont {Spadaccini}},\ }\bibfield  {title}
		{\bibinfo {title} {Multiscale metallic metamaterials},\ }\href
		{https://doi.org/10.1038/nmat4694} {\bibfield  {journal} {\bibinfo  {journal}
				{Nature Materials}\ }\textbf {\bibinfo {volume} {15}},\ \bibinfo {pages}
			{1100} (\bibinfo {year} {2016})}\BibitemShut {NoStop}%
		\bibitem [{\citenamefont {Williams}\ \emph {et~al.}(2018)\citenamefont
			{Williams}, \citenamefont {Hunt}, \citenamefont {Boehm}, \citenamefont {May},
			\citenamefont {Taverne}, \citenamefont {Ho}, \citenamefont {Giblin},
			\citenamefont {Read}, \citenamefont {Rarity}, \citenamefont {Allenspach},\
			and\ \citenamefont {Ladak}}]{Williams18}%
		\BibitemOpen
		\bibfield  {author} {\bibinfo {author} {\bibfnamefont {G.}~\bibnamefont
				{Williams}}, \bibinfo {author} {\bibfnamefont {M.}~\bibnamefont {Hunt}},
			\bibinfo {author} {\bibfnamefont {B.}~\bibnamefont {Boehm}}, \bibinfo
			{author} {\bibfnamefont {A.}~\bibnamefont {May}}, \bibinfo {author}
			{\bibfnamefont {M.}~\bibnamefont {Taverne}}, \bibinfo {author} {\bibfnamefont
				{D.}~\bibnamefont {Ho}}, \bibinfo {author} {\bibfnamefont {S.}~\bibnamefont
				{Giblin}}, \bibinfo {author} {\bibfnamefont {D.}~\bibnamefont {Read}},
			\bibinfo {author} {\bibfnamefont {J.}~\bibnamefont {Rarity}}, \bibinfo
			{author} {\bibfnamefont {R.}~\bibnamefont {Allenspach}},\ and\ \bibinfo
			{author} {\bibfnamefont {S.}~\bibnamefont {Ladak}},\ }\bibfield  {title}
		{\bibinfo {title} {Two-photon lithography for {3D} magnetic nanostructure
				fabrication},\ }\href {https://doi.org/10.1007/s12274-017-1694-0} {\bibfield
			{journal} {\bibinfo  {journal} {Nano Research}\ }\textbf {\bibinfo {volume}
				{11}},\ \bibinfo {pages} {845} (\bibinfo {year} {2018})}\BibitemShut
		{NoStop}%
		\bibitem [{\citenamefont {Donnelly}\ \emph {et~al.}(2015)\citenamefont
			{Donnelly}, \citenamefont {Guizar-Sicairos}, \citenamefont {Scagnoli},
			\citenamefont {Holler}, \citenamefont {Huthwelker}, \citenamefont {Menzel},
			\citenamefont {Vartiainen}, \citenamefont {M\"uller}, \citenamefont {Kirk},
			\citenamefont {Gliga}, \citenamefont {Raabe},\ and\ \citenamefont
			{Heyderman}}]{Donnelly15}%
		\BibitemOpen
		\bibfield  {author} {\bibinfo {author} {\bibfnamefont {C.}~\bibnamefont
				{Donnelly}}, \bibinfo {author} {\bibfnamefont {M.}~\bibnamefont
				{Guizar-Sicairos}}, \bibinfo {author} {\bibfnamefont {V.}~\bibnamefont
				{Scagnoli}}, \bibinfo {author} {\bibfnamefont {M.}~\bibnamefont {Holler}},
			\bibinfo {author} {\bibfnamefont {T.}~\bibnamefont {Huthwelker}}, \bibinfo
			{author} {\bibfnamefont {A.}~\bibnamefont {Menzel}}, \bibinfo {author}
			{\bibfnamefont {I.}~\bibnamefont {Vartiainen}}, \bibinfo {author}
			{\bibfnamefont {E.}~\bibnamefont {M\"uller}}, \bibinfo {author}
			{\bibfnamefont {E.}~\bibnamefont {Kirk}}, \bibinfo {author} {\bibfnamefont
				{S.}~\bibnamefont {Gliga}}, \bibinfo {author} {\bibfnamefont
				{J.}~\bibnamefont {Raabe}},\ and\ \bibinfo {author} {\bibfnamefont {L.~J.}\
				\bibnamefont {Heyderman}},\ }\bibfield  {title} {\bibinfo {title}
			{Element-specific {X}-ray phase tomography of {3D} structures at the
				nanoscale},\ }\href {https://doi.org/10.1103/PhysRevLett.114.115501}
		{\bibfield  {journal} {\bibinfo  {journal} {Physical Review Letters}\
			}\textbf {\bibinfo {volume} {114}},\ \bibinfo {pages} {115501} (\bibinfo
			{year} {2015})}\BibitemShut {NoStop}%
		\bibitem [{\citenamefont {May}\ \emph {et~al.}(2019)\citenamefont {May},
			\citenamefont {Hunt}, \citenamefont {Berg}, \citenamefont {Hejazi},\ and\
			\citenamefont {Ladak}}]{May19}%
		\BibitemOpen
		\bibfield  {author} {\bibinfo {author} {\bibfnamefont {A.}~\bibnamefont
				{May}}, \bibinfo {author} {\bibfnamefont {M.}~\bibnamefont {Hunt}}, \bibinfo
			{author} {\bibfnamefont {A.~V.~D.}\ \bibnamefont {Berg}}, \bibinfo {author}
			{\bibfnamefont {A.}~\bibnamefont {Hejazi}},\ and\ \bibinfo {author}
			{\bibfnamefont {S.}~\bibnamefont {Ladak}},\ }\bibfield  {title} {\bibinfo
			{title} {Realisation of a frustrated {3D} magnetic nanowire lattice},\ }\href
		{https://doi.org/10.1038/s42005-018-0104-6} {\bibfield  {journal} {\bibinfo
				{journal} {Communications Physics}\ }\textbf {\bibinfo {volume} {2}},\
			\bibinfo {pages} {13} (\bibinfo {year} {2019})}\BibitemShut {NoStop}%
		\bibitem [{\citenamefont {May}\ \emph {et~al.}(2021)\citenamefont {May},
			\citenamefont {Saccone}, \citenamefont {van~den Berg}, \citenamefont {Askey},
			\citenamefont {Hunt},\ and\ \citenamefont {Ladak}}]{May21}%
		\BibitemOpen
		\bibfield  {author} {\bibinfo {author} {\bibfnamefont {A.}~\bibnamefont
				{May}}, \bibinfo {author} {\bibfnamefont {M.}~\bibnamefont {Saccone}},
			\bibinfo {author} {\bibfnamefont {A.}~\bibnamefont {van~den Berg}}, \bibinfo
			{author} {\bibfnamefont {J.}~\bibnamefont {Askey}}, \bibinfo {author}
			{\bibfnamefont {M.}~\bibnamefont {Hunt}},\ and\ \bibinfo {author}
			{\bibfnamefont {S.}~\bibnamefont {Ladak}},\ }\bibfield  {title} {\bibinfo
			{title} {Magnetic charge propagation upon a {3D} artificial spin-ice},\
		}\href {https://doi.org/10.1038/s41467-021-23480-7} {\bibfield  {journal}
			{\bibinfo  {journal} {Nature Communications}\ }\textbf {\bibinfo {volume}
				{12}},\ \bibinfo {pages} {3217} (\bibinfo {year} {2021})}\BibitemShut
		{NoStop}%
		\bibitem [{\citenamefont {Sahoo}\ \emph {et~al.}(2021)\citenamefont {Sahoo},
			\citenamefont {May}, \citenamefont {van Den~Berg}, \citenamefont {Mondal},
			\citenamefont {Ladak},\ and\ \citenamefont {Barman}}]{Sahoo21}%
		\BibitemOpen
		\bibfield  {author} {\bibinfo {author} {\bibfnamefont {S.}~\bibnamefont
				{Sahoo}}, \bibinfo {author} {\bibfnamefont {A.}~\bibnamefont {May}}, \bibinfo
			{author} {\bibfnamefont {A.}~\bibnamefont {van Den~Berg}}, \bibinfo {author}
			{\bibfnamefont {A.~K.}\ \bibnamefont {Mondal}}, \bibinfo {author}
			{\bibfnamefont {S.}~\bibnamefont {Ladak}},\ and\ \bibinfo {author}
			{\bibfnamefont {A.}~\bibnamefont {Barman}},\ }\bibfield  {title} {\bibinfo
			{title} {Observation of coherent spin waves in a three-dimensional artificial
				spin ice structure},\ }\href {https://doi.org/10.1021/acs.nanolett.1c00650}
		{\bibfield  {journal} {\bibinfo  {journal} {Nano Letters}\ }\textbf {\bibinfo
				{volume} {21}},\ \bibinfo {pages} {4629} (\bibinfo {year}
			{2021})}\BibitemShut {NoStop}%
		\bibitem [{\citenamefont {Pip}\ \emph {et~al.}(2022)\citenamefont {Pip},
			\citenamefont {Treves}, \citenamefont {Massey}, \citenamefont {Finizio},
			\citenamefont {Luo}, \citenamefont {Hrabec}, \citenamefont {Scagnoli},
			\citenamefont {Raabe}, \citenamefont {Philippe}, \citenamefont {Heyderman},\
			and\ \citenamefont {Donnelly}}]{Pip22}%
		\BibitemOpen
		\bibfield  {author} {\bibinfo {author} {\bibfnamefont {P.}~\bibnamefont
				{Pip}}, \bibinfo {author} {\bibfnamefont {S.}~\bibnamefont {Treves}},
			\bibinfo {author} {\bibfnamefont {J.~R.}\ \bibnamefont {Massey}}, \bibinfo
			{author} {\bibfnamefont {S.}~\bibnamefont {Finizio}}, \bibinfo {author}
			{\bibfnamefont {Z.}~\bibnamefont {Luo}}, \bibinfo {author} {\bibfnamefont
				{A.}~\bibnamefont {Hrabec}}, \bibinfo {author} {\bibfnamefont
				{V.}~\bibnamefont {Scagnoli}}, \bibinfo {author} {\bibfnamefont
				{J.}~\bibnamefont {Raabe}}, \bibinfo {author} {\bibfnamefont
				{L.}~\bibnamefont {Philippe}}, \bibinfo {author} {\bibfnamefont {L.~J.}\
				\bibnamefont {Heyderman}},\ and\ \bibinfo {author} {\bibfnamefont
				{C.}~\bibnamefont {Donnelly}},\ }\bibfield  {title} {\bibinfo {title} {X-ray
				imaging of the magnetic configuration of a three-dimensional artificial spin
				ice building block},\ }\href {https://doi.org/10.1063/5.0101797} {\bibfield
			{journal} {\bibinfo  {journal} {APL Materials}\ }\textbf {\bibinfo {volume}
				{10}},\ \bibinfo {pages} {101101} (\bibinfo {year} {2022})}\BibitemShut
		{NoStop}%
		\bibitem [{\citenamefont {Berchialla}\ \emph {et~al.}(2024)\citenamefont
			{Berchialla}, \citenamefont {Macauley},\ and\ \citenamefont
			{Heyderman}}]{Berchialla24}%
		\BibitemOpen
		\bibfield  {author} {\bibinfo {author} {\bibfnamefont {L.}~\bibnamefont
				{Berchialla}}, \bibinfo {author} {\bibfnamefont {G.~M.}\ \bibnamefont
				{Macauley}},\ and\ \bibinfo {author} {\bibfnamefont {L.~J.}\ \bibnamefont
				{Heyderman}},\ }\bibfield  {title} {\bibinfo {title} {Focus on
				three-dimensional artificial spin ice},\ }\href
		{https://doi.org/10.1063/5.0229120} {\bibfield  {journal} {\bibinfo
				{journal} {Applied Physics Letters}\ }\textbf {\bibinfo {volume} {125}},\
			\bibinfo {pages} {220501} (\bibinfo {year} {2024})}\BibitemShut {NoStop}%
		\bibitem [{\citenamefont {van~den Berg}\ \emph {et~al.}(2023)\citenamefont
			{van~den Berg}, \citenamefont {Caruel}, \citenamefont {Hunt},\ and\
			\citenamefont {Ladak}}]{vandenBerg23}%
		\BibitemOpen
		\bibfield  {author} {\bibinfo {author} {\bibfnamefont {A.}~\bibnamefont
				{van~den Berg}}, \bibinfo {author} {\bibfnamefont {M.}~\bibnamefont
				{Caruel}}, \bibinfo {author} {\bibfnamefont {M.}~\bibnamefont {Hunt}},\ and\
			\bibinfo {author} {\bibfnamefont {S.}~\bibnamefont {Ladak}},\ }\bibfield
		{title} {\bibinfo {title} {Combining two-photon lithography with laser
				ablation of sacrificial layers: A route to isolated {3D} magnetic
				nanostructures},\ }\href {https://doi.org/10.1007/s12274-022-4649-z}
		{\bibfield  {journal} {\bibinfo  {journal} {Nano Research}\ }\textbf
			{\bibinfo {volume} {16}},\ \bibinfo {pages} {1441} (\bibinfo {year}
			{2023})}\BibitemShut {NoStop}%
		\bibitem [{\citenamefont {Askey}\ \emph {et~al.}(2024)\citenamefont {Askey},
			\citenamefont {Hunt}, \citenamefont {Payne}, \citenamefont {van~den Berg},
			\citenamefont {Pitsios}, \citenamefont {Hejazi}, \citenamefont {Langbein},\
			and\ \citenamefont {Ladak}}]{Askey24}%
		\BibitemOpen
		\bibfield  {author} {\bibinfo {author} {\bibfnamefont {J.}~\bibnamefont
				{Askey}}, \bibinfo {author} {\bibfnamefont {M.~O.}\ \bibnamefont {Hunt}},
			\bibinfo {author} {\bibfnamefont {L.}~\bibnamefont {Payne}}, \bibinfo
			{author} {\bibfnamefont {A.}~\bibnamefont {van~den Berg}}, \bibinfo {author}
			{\bibfnamefont {I.}~\bibnamefont {Pitsios}}, \bibinfo {author} {\bibfnamefont
				{A.}~\bibnamefont {Hejazi}}, \bibinfo {author} {\bibfnamefont
				{W.}~\bibnamefont {Langbein}},\ and\ \bibinfo {author} {\bibfnamefont
				{S.}~\bibnamefont {Ladak}},\ }\bibfield  {title} {\bibinfo {title} {Direct
				visualization of domain wall pinning in sub-100 nm {3D} magnetic nanowires
				with cross-sectional curvature},\ }\href {https://doi.org/10.1039/d4nr02020k}
		{\bibfield  {journal} {\bibinfo  {journal} {Nanoscale}\ }\textbf {\bibinfo
				{volume} {16}},\ \bibinfo {pages} {17793} (\bibinfo {year}
			{2024})}\BibitemShut {NoStop}%
		\bibitem [{\citenamefont {Grigoriev}\ \emph {et~al.}(2009)\citenamefont
			{Grigoriev}, \citenamefont {Napolskii}, \citenamefont {Grigoryeva},
			\citenamefont {Vasilieva}, \citenamefont {Mistonov}, \citenamefont
			{Chernyshov}, \citenamefont {Petukhov}, \citenamefont {Belov}, \citenamefont
			{Eliseev}, \citenamefont {Lukashin}, \citenamefont {Tretyakov}, \citenamefont
			{Sinitskii},\ and\ \citenamefont {Eckerlebe}}]{Grigoriev09a}%
		\BibitemOpen
		\bibfield  {author} {\bibinfo {author} {\bibfnamefont {S.~V.}\ \bibnamefont
				{Grigoriev}}, \bibinfo {author} {\bibfnamefont {K.~S.}\ \bibnamefont
				{Napolskii}}, \bibinfo {author} {\bibfnamefont {N.~A.}\ \bibnamefont
				{Grigoryeva}}, \bibinfo {author} {\bibfnamefont {A.~V.}\ \bibnamefont
				{Vasilieva}}, \bibinfo {author} {\bibfnamefont {A.~A.}\ \bibnamefont
				{Mistonov}}, \bibinfo {author} {\bibfnamefont {D.~Y.}\ \bibnamefont
				{Chernyshov}}, \bibinfo {author} {\bibfnamefont {A.~V.}\ \bibnamefont
				{Petukhov}}, \bibinfo {author} {\bibfnamefont {D.~V.}\ \bibnamefont {Belov}},
			\bibinfo {author} {\bibfnamefont {A.~A.}\ \bibnamefont {Eliseev}}, \bibinfo
			{author} {\bibfnamefont {A.~V.}\ \bibnamefont {Lukashin}}, \bibinfo {author}
			{\bibfnamefont {Y.~D.}\ \bibnamefont {Tretyakov}}, \bibinfo {author}
			{\bibfnamefont {A.~S.}\ \bibnamefont {Sinitskii}},\ and\ \bibinfo {author}
			{\bibfnamefont {H.}~\bibnamefont {Eckerlebe}},\ }\bibfield  {title} {\bibinfo
			{title} {Structural and magnetic properties of inverse opal photonic crystals
				studied by x-ray diffraction, scanning electron microscopy, and small-angle
				neutron scattering},\ }\href {https://doi.org/10.1103/physrevb.79.045123}
		{\bibfield  {journal} {\bibinfo  {journal} {Physical Review B}\ }\textbf
			{\bibinfo {volume} {79}},\ \bibinfo {pages} {045123} (\bibinfo {year}
			{2009})}\BibitemShut {NoStop}%
		\bibitem [{\citenamefont {Zeissler}(2013)}]{Zeissler13}%
		\BibitemOpen
		\bibfield  {author} {\bibinfo {author} {\bibfnamefont {K.}~\bibnamefont
				{Zeissler}},\ }\emph {\bibinfo {title} {Magnetic and electrical transport
				properties of artificial spin ice}},\ \href {https://doi.org/10.25560/25524}
		{Ph.D. thesis},\ \bibinfo {address} {Imperial College London} (\bibinfo
		{year} {2013})\BibitemShut {NoStop}%
		\bibitem [{\citenamefont {Dubitskiy}\ \emph {et~al.}(2017)\citenamefont
			{Dubitskiy}, \citenamefont {Grigoryeva}, \citenamefont {Mistonov},
			\citenamefont {Valkovskiy}, \citenamefont {Sapoletova},\ and\ \citenamefont
			{Grigoriev}}]{Dubitskiy17}%
		\BibitemOpen
		\bibfield  {author} {\bibinfo {author} {\bibfnamefont {I.~S.}\ \bibnamefont
				{Dubitskiy}}, \bibinfo {author} {\bibfnamefont {N.~A.}\ \bibnamefont
				{Grigoryeva}}, \bibinfo {author} {\bibfnamefont {A.~A.}\ \bibnamefont
				{Mistonov}}, \bibinfo {author} {\bibfnamefont {G.~A.}\ \bibnamefont
				{Valkovskiy}}, \bibinfo {author} {\bibfnamefont {N.~A.}\ \bibnamefont
				{Sapoletova}},\ and\ \bibinfo {author} {\bibfnamefont {S.~V.}\ \bibnamefont
				{Grigoriev}},\ }\bibfield  {title} {\bibinfo {title} {Study of quasi-two- and
				three-dimensional ordered porous structures by means of small-angle {X}-ray
				scattering in the grazing incidence geometry},\ }\href
		{https://doi.org/10.1134/s1063783417120137} {\bibfield  {journal} {\bibinfo
				{journal} {Physics of the Solid State}\ }\textbf {\bibinfo {volume} {59}},\
			\bibinfo {pages} {2464} (\bibinfo {year} {2017})}\BibitemShut {NoStop}%
		\bibitem [{\citenamefont {Rana}\ \emph {et~al.}(2023)\citenamefont {Rana},
			\citenamefont {Liao}, \citenamefont {Iacocca}, \citenamefont {Zou},
			\citenamefont {Pham}, \citenamefont {Lu}, \citenamefont {Subramanian},
			\citenamefont {Lo}, \citenamefont {Ryan}, \citenamefont {Bevis},
			\citenamefont {Karl}, \citenamefont {Glaid}, \citenamefont {Rable},
			\citenamefont {Mahale}, \citenamefont {Hirst}, \citenamefont {Ostler},
			\citenamefont {Liu}, \citenamefont {O’Leary}, \citenamefont {Yu},
			\citenamefont {Bustillo}, \citenamefont {Ohldag}, \citenamefont {Shapiro},
			\citenamefont {Yazdi}, \citenamefont {Mallouk}, \citenamefont {Osher},
			\citenamefont {Kapteyn}, \citenamefont {Crespi}, \citenamefont {Badding},
			\citenamefont {Tserkovnyak}, \citenamefont {Murnane},\ and\ \citenamefont
			{Miao}}]{Rana23}%
		\BibitemOpen
		\bibfield  {author} {\bibinfo {author} {\bibfnamefont {A.}~\bibnamefont
				{Rana}}, \bibinfo {author} {\bibfnamefont {C.-T.}\ \bibnamefont {Liao}},
			\bibinfo {author} {\bibfnamefont {E.}~\bibnamefont {Iacocca}}, \bibinfo
			{author} {\bibfnamefont {J.}~\bibnamefont {Zou}}, \bibinfo {author}
			{\bibfnamefont {M.}~\bibnamefont {Pham}}, \bibinfo {author} {\bibfnamefont
				{X.}~\bibnamefont {Lu}}, \bibinfo {author} {\bibfnamefont {E.-E.~C.}\
				\bibnamefont {Subramanian}}, \bibinfo {author} {\bibfnamefont {Y.~H.}\
				\bibnamefont {Lo}}, \bibinfo {author} {\bibfnamefont {S.~A.}\ \bibnamefont
				{Ryan}}, \bibinfo {author} {\bibfnamefont {C.~S.}\ \bibnamefont {Bevis}},
			\bibinfo {author} {\bibfnamefont {R.~M.}\ \bibnamefont {Karl}}, \bibinfo
			{author} {\bibfnamefont {A.~J.}\ \bibnamefont {Glaid}}, \bibinfo {author}
			{\bibfnamefont {J.}~\bibnamefont {Rable}}, \bibinfo {author} {\bibfnamefont
				{P.}~\bibnamefont {Mahale}}, \bibinfo {author} {\bibfnamefont
				{J.}~\bibnamefont {Hirst}}, \bibinfo {author} {\bibfnamefont
				{T.}~\bibnamefont {Ostler}}, \bibinfo {author} {\bibfnamefont
				{W.}~\bibnamefont {Liu}}, \bibinfo {author} {\bibfnamefont {C.~M.}\
				\bibnamefont {O’Leary}}, \bibinfo {author} {\bibfnamefont {Y.-S.}\
				\bibnamefont {Yu}}, \bibinfo {author} {\bibfnamefont {K.}~\bibnamefont
				{Bustillo}}, \bibinfo {author} {\bibfnamefont {H.}~\bibnamefont {Ohldag}},
			\bibinfo {author} {\bibfnamefont {D.~A.}\ \bibnamefont {Shapiro}}, \bibinfo
			{author} {\bibfnamefont {S.}~\bibnamefont {Yazdi}}, \bibinfo {author}
			{\bibfnamefont {T.~E.}\ \bibnamefont {Mallouk}}, \bibinfo {author}
			{\bibfnamefont {S.~J.}\ \bibnamefont {Osher}}, \bibinfo {author}
			{\bibfnamefont {H.~C.}\ \bibnamefont {Kapteyn}}, \bibinfo {author}
			{\bibfnamefont {V.~H.}\ \bibnamefont {Crespi}}, \bibinfo {author}
			{\bibfnamefont {J.~V.}\ \bibnamefont {Badding}}, \bibinfo {author}
			{\bibfnamefont {Y.}~\bibnamefont {Tserkovnyak}}, \bibinfo {author}
			{\bibfnamefont {M.~M.}\ \bibnamefont {Murnane}},\ and\ \bibinfo {author}
			{\bibfnamefont {J.}~\bibnamefont {Miao}},\ }\bibfield  {title} {\bibinfo
			{title} {Three-dimensional topological magnetic monopoles and their
				interactions in a ferromagnetic meta-lattice},\ }\href
		{https://doi.org/10.1038/s41565-022-01311-0} {\bibfield  {journal} {\bibinfo
				{journal} {Nature Nanotechnology}\ }\textbf {\bibinfo {volume} {18}},\
			\bibinfo {pages} {227} (\bibinfo {year} {2023})}\BibitemShut {NoStop}%
		\bibitem [{\citenamefont {V\'{a}zquez}\ \emph {et~al.}(2005)\citenamefont
			{V\'{a}zquez}, \citenamefont {Pirota}, \citenamefont {Torrej\'{o}n},
			\citenamefont {Navas},\ and\ \citenamefont
			{Hern\'{a}ndez-V\'{e}lez}}]{Vazquez05}%
		\BibitemOpen
		\bibfield  {author} {\bibinfo {author} {\bibfnamefont {M.}~\bibnamefont
				{V\'{a}zquez}}, \bibinfo {author} {\bibfnamefont {K.}~\bibnamefont {Pirota}},
			\bibinfo {author} {\bibfnamefont {J.}~\bibnamefont {Torrej\'{o}n}}, \bibinfo
			{author} {\bibfnamefont {D.}~\bibnamefont {Navas}},\ and\ \bibinfo {author}
			{\bibfnamefont {M.}~\bibnamefont {Hern\'{a}ndez-V\'{e}lez}},\ }\bibfield
		{title} {\bibinfo {title} {Magnetic behaviour of densely packed hexagonal
				arrays of ni nanowires: Influence of geometric characteristics},\ }\href
		{https://doi.org/10.1016/j.jmmm.2005.03.032} {\bibfield  {journal} {\bibinfo
				{journal} {Journal of Magnetism and Magnetic Materials}\ }\textbf {\bibinfo
				{volume} {294}},\ \bibinfo {pages} {174} (\bibinfo {year}
			{2005})}\BibitemShut {NoStop}%
		\bibitem [{\citenamefont {Hauet}\ \emph {et~al.}(2014)\citenamefont {Hauet},
			\citenamefont {Piraux}, \citenamefont {Srivastava}, \citenamefont {Antohe},
			\citenamefont {Lacour}, \citenamefont {Hehn}, \citenamefont {Montaigne},
			\citenamefont {Schwenk}, \citenamefont {Marioni}, \citenamefont {Hug},
			\citenamefont {Hovorka}, \citenamefont {Berger}, \citenamefont {Mangin},\
			and\ \citenamefont {Abreu~Araujo}}]{Hauet14}%
		\BibitemOpen
		\bibfield  {author} {\bibinfo {author} {\bibfnamefont {T.}~\bibnamefont
				{Hauet}}, \bibinfo {author} {\bibfnamefont {L.}~\bibnamefont {Piraux}},
			\bibinfo {author} {\bibfnamefont {S.~K.}\ \bibnamefont {Srivastava}},
			\bibinfo {author} {\bibfnamefont {V.~A.}\ \bibnamefont {Antohe}}, \bibinfo
			{author} {\bibfnamefont {D.}~\bibnamefont {Lacour}}, \bibinfo {author}
			{\bibfnamefont {M.}~\bibnamefont {Hehn}}, \bibinfo {author} {\bibfnamefont
				{F.}~\bibnamefont {Montaigne}}, \bibinfo {author} {\bibfnamefont
				{J.}~\bibnamefont {Schwenk}}, \bibinfo {author} {\bibfnamefont {M.~A.}\
				\bibnamefont {Marioni}}, \bibinfo {author} {\bibfnamefont {H.~J.}\
				\bibnamefont {Hug}}, \bibinfo {author} {\bibfnamefont {O.}~\bibnamefont
				{Hovorka}}, \bibinfo {author} {\bibfnamefont {A.}~\bibnamefont {Berger}},
			\bibinfo {author} {\bibfnamefont {S.}~\bibnamefont {Mangin}},\ and\ \bibinfo
			{author} {\bibfnamefont {F.}~\bibnamefont {Abreu~Araujo}},\ }\bibfield
		{title} {\bibinfo {title} {Reversal mechanism, switching field distribution,
				and dipolar frustrations in {Co/Pt} bit pattern media based on auto-assembled
				anodic alumina hexagonal nanobump arrays},\ }\href
		{https://doi.org/10.1103/physrevb.89.174421} {\bibfield  {journal} {\bibinfo
				{journal} {Physical Review B}\ }\textbf {\bibinfo {volume} {89}},\ \bibinfo
			{pages} {174421} (\bibinfo {year} {2014})}\BibitemShut {NoStop}%
		\bibitem [{\citenamefont {Mart\'{\i}n}\ \emph {et~al.}(2014)\citenamefont
			{Mart\'{\i}n}, \citenamefont {Mart\'{\i}n-Gonz\'{a}lez}, \citenamefont
			{Francisco~Fern\'{a}ndez},\ and\ \citenamefont
			{Caballero-Calero}}]{Martin14}%
		\BibitemOpen
		\bibfield  {author} {\bibinfo {author} {\bibfnamefont {J.}~\bibnamefont
				{Mart\'{\i}n}}, \bibinfo {author} {\bibfnamefont {M.}~\bibnamefont
				{Mart\'{\i}n-Gonz\'{a}lez}}, \bibinfo {author} {\bibfnamefont
				{J.}~\bibnamefont {Francisco~Fern\'{a}ndez}},\ and\ \bibinfo {author}
			{\bibfnamefont {O.}~\bibnamefont {Caballero-Calero}},\ }\bibfield  {title}
		{\bibinfo {title} {Ordered three-dimensional interconnected nanoarchitectures
				in anodic porous alumina},\ }\href {https://doi.org/10.1038/ncomms6130}
		{\bibfield  {journal} {\bibinfo  {journal} {Nature Communications}\ }\textbf
			{\bibinfo {volume} {5}},\ \bibinfo {pages} {5130} (\bibinfo {year}
			{2014})}\BibitemShut {NoStop}%
		\bibitem [{\citenamefont {Dolan}\ \emph {et~al.}(2015)\citenamefont {Dolan},
			\citenamefont {Wilts}, \citenamefont {Vignolini}, \citenamefont {Baumberg},
			\citenamefont {Steiner},\ and\ \citenamefont {Wilkinson}}]{Dolan15}%
		\BibitemOpen
		\bibfield  {author} {\bibinfo {author} {\bibfnamefont {J.~A.}\ \bibnamefont
				{Dolan}}, \bibinfo {author} {\bibfnamefont {B.~D.}\ \bibnamefont {Wilts}},
			\bibinfo {author} {\bibfnamefont {S.}~\bibnamefont {Vignolini}}, \bibinfo
			{author} {\bibfnamefont {J.~J.}\ \bibnamefont {Baumberg}}, \bibinfo {author}
			{\bibfnamefont {U.}~\bibnamefont {Steiner}},\ and\ \bibinfo {author}
			{\bibfnamefont {T.~D.}\ \bibnamefont {Wilkinson}},\ }\bibfield  {title}
		{\bibinfo {title} {Optical properties of gyroid structured materials: From
				photonic crystals to metamaterials},\ }\href
		{https://doi.org/10.1002/adom.201400333} {\bibfield  {journal} {\bibinfo
				{journal} {Advanced Optical Materials}\ }\textbf {\bibinfo {volume} {3}},\
			\bibinfo {pages} {12} (\bibinfo {year} {2015})}\BibitemShut {NoStop}%
		\bibitem [{\citenamefont {Bhattacharya}\ \emph {et~al.}(2022)\citenamefont
			{Bhattacharya}, \citenamefont {Chen}, \citenamefont {Jensen}, \citenamefont
			{Liu}, \citenamefont {Burks}, \citenamefont {Gilbert}, \citenamefont {Zhang},
			\citenamefont {Yin},\ and\ \citenamefont {Liu}}]{Bhattacharya22}%
		\BibitemOpen
		\bibfield  {author} {\bibinfo {author} {\bibfnamefont {D.}~\bibnamefont
				{Bhattacharya}}, \bibinfo {author} {\bibfnamefont {Z.}~\bibnamefont {Chen}},
			\bibinfo {author} {\bibfnamefont {C.~J.}\ \bibnamefont {Jensen}}, \bibinfo
			{author} {\bibfnamefont {C.}~\bibnamefont {Liu}}, \bibinfo {author}
			{\bibfnamefont {E.~C.}\ \bibnamefont {Burks}}, \bibinfo {author}
			{\bibfnamefont {D.~A.}\ \bibnamefont {Gilbert}}, \bibinfo {author}
			{\bibfnamefont {X.}~\bibnamefont {Zhang}}, \bibinfo {author} {\bibfnamefont
				{G.}~\bibnamefont {Yin}},\ and\ \bibinfo {author} {\bibfnamefont
				{K.}~\bibnamefont {Liu}},\ }\bibfield  {title} {\bibinfo {title} {{3D}
				interconnected magnetic nanowire networks as potential integrated multistate
				memristors},\ }\href {https://doi.org/10.1021/acs.nanolett.2c03616}
		{\bibfield  {journal} {\bibinfo  {journal} {Nano Letters}\ }\textbf {\bibinfo
				{volume} {22}},\ \bibinfo {pages} {10010} (\bibinfo {year}
			{2022})}\BibitemShut {NoStop}%
		\bibitem [{\citenamefont {Burks}\ \emph {et~al.}(2020)\citenamefont {Burks},
			\citenamefont {Gilbert}, \citenamefont {Murray}, \citenamefont {Flores},
			\citenamefont {Felter}, \citenamefont {Charnvanichborikarn}, \citenamefont
			{Kucheyev}, \citenamefont {Colvin}, \citenamefont {Yin},\ and\ \citenamefont
			{Liu}}]{Burks20}%
		\BibitemOpen
		\bibfield  {author} {\bibinfo {author} {\bibfnamefont {E.~C.}\ \bibnamefont
				{Burks}}, \bibinfo {author} {\bibfnamefont {D.~A.}\ \bibnamefont {Gilbert}},
			\bibinfo {author} {\bibfnamefont {P.~D.}\ \bibnamefont {Murray}}, \bibinfo
			{author} {\bibfnamefont {C.}~\bibnamefont {Flores}}, \bibinfo {author}
			{\bibfnamefont {T.~E.}\ \bibnamefont {Felter}}, \bibinfo {author}
			{\bibfnamefont {S.}~\bibnamefont {Charnvanichborikarn}}, \bibinfo {author}
			{\bibfnamefont {S.~O.}\ \bibnamefont {Kucheyev}}, \bibinfo {author}
			{\bibfnamefont {J.~D.}\ \bibnamefont {Colvin}}, \bibinfo {author}
			{\bibfnamefont {G.}~\bibnamefont {Yin}},\ and\ \bibinfo {author}
			{\bibfnamefont {K.}~\bibnamefont {Liu}},\ }\bibfield  {title} {\bibinfo
			{title} {3d nanomagnetism in low density interconnected nanowire networks},\
		}\href {https://doi.org/10.1021/acs.nanolett.0c04366} {\bibfield  {journal}
			{\bibinfo  {journal} {Nano Letters}\ }\textbf {\bibinfo {volume} {21}},\
			\bibinfo {pages} {716} (\bibinfo {year} {2020})}\BibitemShut {NoStop}%
		\bibitem [{\citenamefont {Sapozhnikov}\ \emph {et~al.}(2012)\citenamefont
			{Sapozhnikov}, \citenamefont {Ermolaeva}, \citenamefont {Gribkov},
			\citenamefont {Nefedov}, \citenamefont {Karetnikova}, \citenamefont {Gusev},
			\citenamefont {Rogov}, \citenamefont {Troitskii},\ and\ \citenamefont
			{Khokhlova}}]{Sapozhnikov12}%
		\BibitemOpen
		\bibfield  {author} {\bibinfo {author} {\bibfnamefont {M.~V.}\ \bibnamefont
				{Sapozhnikov}}, \bibinfo {author} {\bibfnamefont {O.~L.}\ \bibnamefont
				{Ermolaeva}}, \bibinfo {author} {\bibfnamefont {B.~G.}\ \bibnamefont
				{Gribkov}}, \bibinfo {author} {\bibfnamefont {I.~M.}\ \bibnamefont
				{Nefedov}}, \bibinfo {author} {\bibfnamefont {I.~R.}\ \bibnamefont
				{Karetnikova}}, \bibinfo {author} {\bibfnamefont {S.~A.}\ \bibnamefont
				{Gusev}}, \bibinfo {author} {\bibfnamefont {V.~V.}\ \bibnamefont {Rogov}},
			\bibinfo {author} {\bibfnamefont {B.~B.}\ \bibnamefont {Troitskii}},\ and\
			\bibinfo {author} {\bibfnamefont {L.~V.}\ \bibnamefont {Khokhlova}},\
		}\bibfield  {title} {\bibinfo {title} {Frustrated magnetic vortices in
				hexagonal lattice of magnetic nanocaps},\ }\href
		{https://doi.org/10.1103/PhysRevB.85.054402} {\bibfield  {journal} {\bibinfo
				{journal} {Physical Review B}\ }\textbf {\bibinfo {volume} {85}},\ \bibinfo
			{pages} {054402} (\bibinfo {year} {2012})}\BibitemShut {NoStop}%
		\bibitem [{\citenamefont {Gu}\ \emph {et~al.}(2022)\citenamefont {Gu},
			\citenamefont {Guan}, \citenamefont {Hazra}, \citenamefont {Deniz},
			\citenamefont {Migliorini}, \citenamefont {Zhang},\ and\ \citenamefont
			{Parkin}}]{Gu22}%
		\BibitemOpen
		\bibfield  {author} {\bibinfo {author} {\bibfnamefont {K.}~\bibnamefont
				{Gu}}, \bibinfo {author} {\bibfnamefont {Y.}~\bibnamefont {Guan}}, \bibinfo
			{author} {\bibfnamefont {B.~K.}\ \bibnamefont {Hazra}}, \bibinfo {author}
			{\bibfnamefont {H.}~\bibnamefont {Deniz}}, \bibinfo {author} {\bibfnamefont
				{A.}~\bibnamefont {Migliorini}}, \bibinfo {author} {\bibfnamefont
				{W.}~\bibnamefont {Zhang}},\ and\ \bibinfo {author} {\bibfnamefont
				{S.~S.~P.}\ \bibnamefont {Parkin}},\ }\bibfield  {title} {\bibinfo {title}
			{Three-dimensional racetrack memory devices designed from freestanding
				magnetic heterostructures},\ }\href
		{https://doi.org/10.1038/s41565-022-01213-1} {\bibfield  {journal} {\bibinfo
				{journal} {Nature Nanotechnology}\ }\textbf {\bibinfo {volume} {17}},\
			\bibinfo {pages} {1065} (\bibinfo {year} {2022})}\BibitemShut {NoStop}%
		\bibitem [{\citenamefont {Pip}\ \emph {et~al.}(2020)\citenamefont {Pip},
			\citenamefont {Donnelly}, \citenamefont {D\"{o}beli}, \citenamefont
			{Gunderson}, \citenamefont {Heyderman},\ and\ \citenamefont
			{Philippe}}]{Pip20}%
		\BibitemOpen
		\bibfield  {author} {\bibinfo {author} {\bibfnamefont {P.}~\bibnamefont
				{Pip}}, \bibinfo {author} {\bibfnamefont {C.}~\bibnamefont {Donnelly}},
			\bibinfo {author} {\bibfnamefont {M.}~\bibnamefont {D\"{o}beli}}, \bibinfo
			{author} {\bibfnamefont {C.}~\bibnamefont {Gunderson}}, \bibinfo {author}
			{\bibfnamefont {L.~J.}\ \bibnamefont {Heyderman}},\ and\ \bibinfo {author}
			{\bibfnamefont {L.}~\bibnamefont {Philippe}},\ }\bibfield  {title} {\bibinfo
			{title} {Electroless deposition of {Ni}{\textendash}{Fe} alloys on scaffolds
				for {3D} nanomagnetism},\ }\href {https://doi.org/10.1002/smll.202004099}
		{\bibfield  {journal} {\bibinfo  {journal} {Small}\ }\textbf {\bibinfo
				{volume} {16}},\ \bibinfo {pages} {2004099} (\bibinfo {year}
			{2020})}\BibitemShut {NoStop}%
		\bibitem [{\citenamefont {Xu}\ \emph {et~al.}(2026)\citenamefont {Xu},
			\citenamefont {Deenen}, \citenamefont {Guo}, \citenamefont
			{Morales-Fern\'{a}ndez}, \citenamefont {Wintz}, \citenamefont {Zhakina},
			\citenamefont {Weigand}, \citenamefont {Donnelly},\ and\ \citenamefont
			{Grundler}}]{Xu25b}%
		\BibitemOpen
		\bibfield  {author} {\bibinfo {author} {\bibfnamefont {M.}~\bibnamefont
				{Xu}}, \bibinfo {author} {\bibfnamefont {A.~J.~M.}\ \bibnamefont {Deenen}},
			\bibinfo {author} {\bibfnamefont {H.}~\bibnamefont {Guo}}, \bibinfo {author}
			{\bibfnamefont {P.}~\bibnamefont {Morales-Fern\'{a}ndez}}, \bibinfo {author}
			{\bibfnamefont {S.}~\bibnamefont {Wintz}}, \bibinfo {author} {\bibfnamefont
				{E.}~\bibnamefont {Zhakina}}, \bibinfo {author} {\bibfnamefont
				{M.}~\bibnamefont {Weigand}}, \bibinfo {author} {\bibfnamefont
				{C.}~\bibnamefont {Donnelly}},\ and\ \bibinfo {author} {\bibfnamefont
				{D.}~\bibnamefont {Grundler}},\ }\bibfield  {title} {\bibinfo {title}
			{Geometry-induced spin chirality in a non-chiral ferromagnet at zero field},\
		}\href {https://doi.org/10.1038/s41565-025-02055-3} {\bibfield  {journal}
			{\bibinfo  {journal} {Nature Nanotechnology}\ }\textbf {\bibinfo {volume}
				{21}},\ \bibinfo {pages} {58} (\bibinfo {year} {2026})}\BibitemShut {NoStop}%
		\bibitem [{\citenamefont {Scherer}\ \emph {et~al.}(2014)\citenamefont
			{Scherer}, \citenamefont {Cunha},\ and\ \citenamefont {Steiner}}]{Scherer14}%
		\BibitemOpen
		\bibfield  {author} {\bibinfo {author} {\bibfnamefont {M.~R.~J.}\
				\bibnamefont {Scherer}}, \bibinfo {author} {\bibfnamefont {P.~M.~S.}\
				\bibnamefont {Cunha}},\ and\ \bibinfo {author} {\bibfnamefont
				{U.}~\bibnamefont {Steiner}},\ }\bibfield  {title} {\bibinfo {title}
			{Labyrinth‐induced faceted electrochemical growth},\ }\href
		{https://doi.org/10.1002/adma.201305074} {\bibfield  {journal} {\bibinfo
				{journal} {Advanced Materials}\ }\textbf {\bibinfo {volume} {26}},\ \bibinfo
			{pages} {2403} (\bibinfo {year} {2014})}\BibitemShut {NoStop}%
		\bibitem [{\citenamefont {Birss}(1966)}]{Birss66}%
		\BibitemOpen
		\bibfield  {author} {\bibinfo {author} {\bibfnamefont {R.~R.}\ \bibnamefont
				{Birss}},\ }\href@noop {} {\emph {\bibinfo {title} {Symmetry and
					Magnetism}}},\ \bibinfo {edition} {2nd}\ ed.,\ edited by\ \bibinfo {editor}
		{\bibfnamefont {E.~P.}\ \bibnamefont {Wolhfarth}}\ (\bibinfo  {publisher}
		{North-Holland Publishing Company},\ \bibinfo {address} {Amsterdam},\
		\bibinfo {year} {1966})\BibitemShut {NoStop}%
		\bibitem [{\citenamefont {Nye}(1985)}]{Nye85}%
		\BibitemOpen
		\bibfield  {author} {\bibinfo {author} {\bibfnamefont {J.~F.}\ \bibnamefont
				{Nye}},\ }\href@noop {} {\emph {\bibinfo {title} {Physical Properties of
					Crystals: Their Representation by Tensors and Matrices}}}\ (\bibinfo
		{publisher} {Oxford University Press},\ \bibinfo {address} {Oxford},\
		\bibinfo {year} {1985})\BibitemShut {NoStop}%
		\bibitem [{\citenamefont {Dzyaloshinsky}(1958)}]{Dzyaloshinsky58}%
		\BibitemOpen
		\bibfield  {author} {\bibinfo {author} {\bibfnamefont {I.}~\bibnamefont
				{Dzyaloshinsky}},\ }\bibfield  {title} {\bibinfo {title} {A thermodynamic
				theory of ``weak'' ferromagnetism of antiferromagnetics},\ }\href
		{https://doi.org/10.1016/0022-3697(58)90076-3} {\bibfield  {journal}
			{\bibinfo  {journal} {Journal of Physics and Chemistry of Solids}\ }\textbf
			{\bibinfo {volume} {4}},\ \bibinfo {pages} {241} (\bibinfo {year}
			{1958})}\BibitemShut {NoStop}%
		\bibitem [{\citenamefont {Moriya}(1960)}]{Moriya60}%
		\BibitemOpen
		\bibfield  {author} {\bibinfo {author} {\bibfnamefont {T.}~\bibnamefont
				{Moriya}},\ }\bibfield  {title} {\bibinfo {title} {Anisotropic superexchange
				interaction and weak ferromagnetism},\ }\href
		{https://doi.org/10.1103/PhysRev.120.91} {\bibfield  {journal} {\bibinfo
				{journal} {Physical Review}\ }\textbf {\bibinfo {volume} {120}},\ \bibinfo
			{pages} {91} (\bibinfo {year} {1960})}\BibitemShut {NoStop}%
		\bibitem [{\citenamefont {Bogdanov}\ and\ \citenamefont
			{Yablonski\u{\i}}(1989)}]{Bogdanov89}%
		\BibitemOpen
		\bibfield  {author} {\bibinfo {author} {\bibfnamefont {A.~N.}\ \bibnamefont
				{Bogdanov}}\ and\ \bibinfo {author} {\bibfnamefont {D.~A.}\ \bibnamefont
				{Yablonski\u{\i}}},\ }\bibfield  {title} {\bibinfo {title} {Thermodynamically
				stable ``vortices'' in magnetically ordered crystals. {T}he mixed state of
				magnets},\ }\href {http://jetp.ac.ru/cgi-bin/e/index/e/68/1/p101?a=list}
		{\bibfield  {journal} {\bibinfo  {journal} {Zh. Eksp. Teor. Fiz.}\ }\textbf
			{\bibinfo {volume} {95}},\ \bibinfo {pages} {178} (\bibinfo {year}
			{1989})}\BibitemShut {NoStop}%
		\bibitem [{\citenamefont {M{\"u}hlbauer}\ \emph {et~al.}(2009)\citenamefont
			{M{\"u}hlbauer}, \citenamefont {Binz}, \citenamefont {Jonietz}, \citenamefont
			{Pfleiderer}, \citenamefont {Rosch}, \citenamefont {Neubauer}, \citenamefont
			{Georgii},\ and\ \citenamefont {B{\"o}ni}}]{Muhlbauer09}%
		\BibitemOpen
		\bibfield  {author} {\bibinfo {author} {\bibfnamefont {S.}~\bibnamefont
				{M{\"u}hlbauer}}, \bibinfo {author} {\bibfnamefont {B.}~\bibnamefont {Binz}},
			\bibinfo {author} {\bibfnamefont {F.}~\bibnamefont {Jonietz}}, \bibinfo
			{author} {\bibfnamefont {C.}~\bibnamefont {Pfleiderer}}, \bibinfo {author}
			{\bibfnamefont {A.}~\bibnamefont {Rosch}}, \bibinfo {author} {\bibfnamefont
				{A.}~\bibnamefont {Neubauer}}, \bibinfo {author} {\bibfnamefont
				{R.}~\bibnamefont {Georgii}},\ and\ \bibinfo {author} {\bibfnamefont
				{P.}~\bibnamefont {B{\"o}ni}},\ }\bibfield  {title} {\bibinfo {title}
			{Skyrmion lattice in a chiral magnet},\ }\href
		{https://doi.org/10.1126/science.1166767} {\bibfield  {journal} {\bibinfo
				{journal} {Science}\ }\textbf {\bibinfo {volume} {323}},\ \bibinfo {pages}
			{915} (\bibinfo {year} {2009})}\BibitemShut {NoStop}%
		\bibitem [{\citenamefont {Chacon}\ \emph {et~al.}(2018)\citenamefont {Chacon},
			\citenamefont {Heinen}, \citenamefont {Halder}, \citenamefont {Bauer},
			\citenamefont {Simeth}, \citenamefont {M\"{u}hlbauer}, \citenamefont
			{Berger}, \citenamefont {Garst}, \citenamefont {Rosch},\ and\ \citenamefont
			{Pfleiderer}}]{Chacon18}%
		\BibitemOpen
		\bibfield  {author} {\bibinfo {author} {\bibfnamefont {A.}~\bibnamefont
				{Chacon}}, \bibinfo {author} {\bibfnamefont {L.}~\bibnamefont {Heinen}},
			\bibinfo {author} {\bibfnamefont {M.}~\bibnamefont {Halder}}, \bibinfo
			{author} {\bibfnamefont {A.}~\bibnamefont {Bauer}}, \bibinfo {author}
			{\bibfnamefont {W.}~\bibnamefont {Simeth}}, \bibinfo {author} {\bibfnamefont
				{S.}~\bibnamefont {M\"{u}hlbauer}}, \bibinfo {author} {\bibfnamefont
				{H.}~\bibnamefont {Berger}}, \bibinfo {author} {\bibfnamefont
				{M.}~\bibnamefont {Garst}}, \bibinfo {author} {\bibfnamefont
				{A.}~\bibnamefont {Rosch}},\ and\ \bibinfo {author} {\bibfnamefont
				{C.}~\bibnamefont {Pfleiderer}},\ }\bibfield  {title} {\bibinfo {title}
			{Observation of two independent skyrmion phases in a chiral magnetic
				material},\ }\href {https://doi.org/10.1038/s41567-018-0184-y} {\bibfield
			{journal} {\bibinfo  {journal} {Nature Physics}\ }\textbf {\bibinfo {volume}
				{14}},\ \bibinfo {pages} {936} (\bibinfo {year} {2018})}\BibitemShut
		{NoStop}%
		\bibitem [{\citenamefont {Sutcliffe}(2018)}]{Sutcliffe18}%
		\BibitemOpen
		\bibfield  {author} {\bibinfo {author} {\bibfnamefont {P.}~\bibnamefont
				{Sutcliffe}},\ }\bibfield  {title} {\bibinfo {title} {Hopfions in chiral
				magnets},\ }\href {https://doi.org/10.1088/1751-8121/aad521} {\bibfield
			{journal} {\bibinfo  {journal} {Journal of Physics A: Mathematical and
					Theoretical}\ }\textbf {\bibinfo {volume} {51}},\ \bibinfo {pages} {375401}
			(\bibinfo {year} {2018})}\BibitemShut {NoStop}%
		\bibitem [{\citenamefont {Volkov}\ \emph {et~al.}(2018)\citenamefont {Volkov},
			\citenamefont {Sheka}, \citenamefont {Gaididei}, \citenamefont {Kravchuk},
			\citenamefont {R\"o{\ss}ler}, \citenamefont {Fassbender},\ and\ \citenamefont
			{Makarov}}]{Volkov18}%
		\BibitemOpen
		\bibfield  {author} {\bibinfo {author} {\bibfnamefont {O.~M.}\ \bibnamefont
				{Volkov}}, \bibinfo {author} {\bibfnamefont {D.~D.}\ \bibnamefont {Sheka}},
			\bibinfo {author} {\bibfnamefont {Y.}~\bibnamefont {Gaididei}}, \bibinfo
			{author} {\bibfnamefont {V.~P.}\ \bibnamefont {Kravchuk}}, \bibinfo {author}
			{\bibfnamefont {U.~K.}\ \bibnamefont {R\"o{\ss}ler}}, \bibinfo {author}
			{\bibfnamefont {J.}~\bibnamefont {Fassbender}},\ and\ \bibinfo {author}
			{\bibfnamefont {D.}~\bibnamefont {Makarov}},\ }\bibfield  {title} {\bibinfo
			{title} {{Mesoscale Dzyaloshinskii-Moriya interaction: geometrical tailoring
					of the magnetochirality}},\ }\href
		{https://doi.org/10.1038/s41598-017-18835-4} {\bibfield  {journal} {\bibinfo
				{journal} {Scientific Reports}\ }\textbf {\bibinfo {volume} {8}},\ \bibinfo
			{pages} {866} (\bibinfo {year} {2018})}\BibitemShut {NoStop}%
		\bibitem [{\citenamefont {Huth}\ \emph {et~al.}(2012)\citenamefont {Huth},
			\citenamefont {Porrati}, \citenamefont {Schwalb}, \citenamefont {Winhold},
			\citenamefont {Sachser}, \citenamefont {Dukic}, \citenamefont {Adams},\ and\
			\citenamefont {Fantner}}]{Huth12}%
		\BibitemOpen
		\bibfield  {author} {\bibinfo {author} {\bibfnamefont {M.}~\bibnamefont
				{Huth}}, \bibinfo {author} {\bibfnamefont {F.}~\bibnamefont {Porrati}},
			\bibinfo {author} {\bibfnamefont {C.}~\bibnamefont {Schwalb}}, \bibinfo
			{author} {\bibfnamefont {M.}~\bibnamefont {Winhold}}, \bibinfo {author}
			{\bibfnamefont {R.}~\bibnamefont {Sachser}}, \bibinfo {author} {\bibfnamefont
				{M.}~\bibnamefont {Dukic}}, \bibinfo {author} {\bibfnamefont
				{J.}~\bibnamefont {Adams}},\ and\ \bibinfo {author} {\bibfnamefont
				{G.}~\bibnamefont {Fantner}},\ }\bibfield  {title} {\bibinfo {title} {Focused
				electron beam induced deposition: A perspective},\ }\href
		{https://doi.org/10.3762/bjnano.3.70} {\bibfield  {journal} {\bibinfo
				{journal} {Beilstein Journal of Nanotechnology}\ }\textbf {\bibinfo {volume}
				{3}},\ \bibinfo {pages} {597} (\bibinfo {year} {2012})}\BibitemShut {NoStop}%
		\bibitem [{\citenamefont {Moll}(2018)}]{Moll18}%
		\BibitemOpen
		\bibfield  {author} {\bibinfo {author} {\bibfnamefont {P.~J.}\ \bibnamefont
				{Moll}},\ }\bibfield  {title} {\bibinfo {title} {Focused ion beam
				microstructuring of quantum matter},\ }\href
		{https://doi.org/10.1146/annurev-conmatphys-033117-054021} {\bibfield
			{journal} {\bibinfo  {journal} {Annual Review of Condensed Matter Physics}\
			}\textbf {\bibinfo {volume} {9}},\ \bibinfo {pages} {147} (\bibinfo {year}
			{2018})}\BibitemShut {NoStop}%
		\bibitem [{\citenamefont {Luk’yanchuk}\ \emph {et~al.}(2020)\citenamefont
			{Luk’yanchuk}, \citenamefont {Tikhonov}, \citenamefont {Razumnaya},\ and\
			\citenamefont {Vinokur}}]{Lukyanchuk20}%
		\BibitemOpen
		\bibfield  {author} {\bibinfo {author} {\bibfnamefont {I.}~\bibnamefont
				{Luk’yanchuk}}, \bibinfo {author} {\bibfnamefont {Y.}~\bibnamefont
				{Tikhonov}}, \bibinfo {author} {\bibfnamefont {A.}~\bibnamefont
				{Razumnaya}},\ and\ \bibinfo {author} {\bibfnamefont {V.~M.}\ \bibnamefont
				{Vinokur}},\ }\bibfield  {title} {\bibinfo {title} {Hopfions emerge in
				ferroelectrics},\ }\href {https://doi.org/10.1038/s41467-020-16258-w}
		{\bibfield  {journal} {\bibinfo  {journal} {Nature Communications}\ }\textbf
			{\bibinfo {volume} {11}},\ \bibinfo {pages} {2433} (\bibinfo {year}
			{2020})}\BibitemShut {NoStop}%
		\bibitem [{\citenamefont {Plummer}\ and\ \citenamefont
			{Griffin}(2023)}]{Plummer23_9}%
		\BibitemOpen
		\bibfield  {author} {\bibinfo {author} {\bibfnamefont {J.~D.}\ \bibnamefont
				{Plummer}}\ and\ \bibinfo {author} {\bibfnamefont {P.~B.}\ \bibnamefont
				{Griffin}},\ }\bibinfo {title} {Integrated circuit fabrication: Science and
			technology}\ (\bibinfo  {publisher} {Cambridge University Press},\ \bibinfo
		{year} {2023})\ Chap.\ \bibinfo {chapter} {Etching Technologies and
			Chemical-Mechanical Polishing}, p.\ \bibinfo {pages} {448–502}\BibitemShut
		{NoStop}%
		\bibitem [{\citenamefont {Chen}\ \emph {et~al.}(2017)\citenamefont {Chen},
			\citenamefont {Kim}, \citenamefont {Altieri}, \citenamefont {Chen},\ and\
			\citenamefont {Chang}}]{Chen17c}%
		\BibitemOpen
		\bibfield  {author} {\bibinfo {author} {\bibfnamefont {J.~K.-C.}\
				\bibnamefont {Chen}}, \bibinfo {author} {\bibfnamefont {T.}~\bibnamefont
				{Kim}}, \bibinfo {author} {\bibfnamefont {N.~D.}\ \bibnamefont {Altieri}},
			\bibinfo {author} {\bibfnamefont {E.}~\bibnamefont {Chen}},\ and\ \bibinfo
			{author} {\bibfnamefont {J.~P.}\ \bibnamefont {Chang}},\ }\bibfield  {title}
		{\bibinfo {title} {Ion beam assisted organic chemical vapor etch of magnetic
				thin films},\ }\href {https://doi.org/10.1116/1.4978553} {\bibfield
			{journal} {\bibinfo  {journal} {Journal of Vacuum Science \& Technology A:
					Vacuum, Surfaces, and Films}\ }\textbf {\bibinfo {volume} {35}},\ \bibinfo
			{pages} {031304} (\bibinfo {year} {2017})}\BibitemShut {NoStop}%
		\bibitem [{\citenamefont {Kar}\ \emph {et~al.}(2023)\citenamefont {Kar},
			\citenamefont {Nielsch}, \citenamefont {F\"{a}hler},\ and\ \citenamefont
			{Reith}}]{Kar23}%
		\BibitemOpen
		\bibfield  {author} {\bibinfo {author} {\bibfnamefont {S.}~\bibnamefont
				{Kar}}, \bibinfo {author} {\bibfnamefont {K.}~\bibnamefont {Nielsch}},
			\bibinfo {author} {\bibfnamefont {S.}~\bibnamefont {F\"{a}hler}},\ and\
			\bibinfo {author} {\bibfnamefont {H.}~\bibnamefont {Reith}},\ }\bibfield
		{title} {\bibinfo {title} {Microfabrication approaches on magnetic shape
				memory films},\ }\href {https://doi.org/10.1002/adem.202301133} {\bibfield
			{journal} {\bibinfo  {journal} {Advanced Engineering Materials}\ }\textbf
			{\bibinfo {volume} {25}},\ \bibinfo {pages} {2301133} (\bibinfo {year}
			{2023})}\BibitemShut {NoStop}%
		\bibitem [{\citenamefont {Chumak}\ \emph {et~al.}(2008)\citenamefont {Chumak},
			\citenamefont {Serga}, \citenamefont {Hillebrands},\ and\ \citenamefont
			{Kostylev}}]{Chumak08}%
		\BibitemOpen
		\bibfield  {author} {\bibinfo {author} {\bibfnamefont {A.~V.}\ \bibnamefont
				{Chumak}}, \bibinfo {author} {\bibfnamefont {A.~A.}\ \bibnamefont {Serga}},
			\bibinfo {author} {\bibfnamefont {B.}~\bibnamefont {Hillebrands}},\ and\
			\bibinfo {author} {\bibfnamefont {M.~P.}\ \bibnamefont {Kostylev}},\
		}\bibfield  {title} {\bibinfo {title} {Scattering of backward spin waves in a
				one-dimensional magnonic crystal},\ }\href
		{https://doi.org/10.1063/1.2963027} {\bibfield  {journal} {\bibinfo
				{journal} {Applied Physics Letters}\ }\textbf {\bibinfo {volume} {93}},\
			\bibinfo {pages} {022508} (\bibinfo {year} {2008})}\BibitemShut {NoStop}%
		\bibitem [{\citenamefont {Zhu}\ \emph {et~al.}(2017)\citenamefont {Zhu},
			\citenamefont {Chang}, \citenamefont {Franson}, \citenamefont {Liu},
			\citenamefont {Zhang}, \citenamefont {Johnston-Halperin}, \citenamefont
			{Wu},\ and\ \citenamefont {Tang}}]{Zhu17}%
		\BibitemOpen
		\bibfield  {author} {\bibinfo {author} {\bibfnamefont {N.}~\bibnamefont
				{Zhu}}, \bibinfo {author} {\bibfnamefont {H.}~\bibnamefont {Chang}}, \bibinfo
			{author} {\bibfnamefont {A.}~\bibnamefont {Franson}}, \bibinfo {author}
			{\bibfnamefont {T.}~\bibnamefont {Liu}}, \bibinfo {author} {\bibfnamefont
				{X.}~\bibnamefont {Zhang}}, \bibinfo {author} {\bibfnamefont
				{E.}~\bibnamefont {Johnston-Halperin}}, \bibinfo {author} {\bibfnamefont
				{M.}~\bibnamefont {Wu}},\ and\ \bibinfo {author} {\bibfnamefont {H.~X.}\
				\bibnamefont {Tang}},\ }\bibfield  {title} {\bibinfo {title} {Patterned
				growth of crystalline {Y$_3$Fe$_5$O$_{12}$} nanostructures with engineered
				magnetic shape anisotropy},\ }\href {https://doi.org/10.1063/1.4986474}
		{\bibfield  {journal} {\bibinfo  {journal} {Applied Physics Letters}\
			}\textbf {\bibinfo {volume} {110}},\ \bibinfo {pages} {252401} (\bibinfo
			{year} {2017})}\BibitemShut {NoStop}%
		\bibitem [{\citenamefont {Coburn}\ and\ \citenamefont
			{Winters}(1979)}]{Coburn79}%
		\BibitemOpen
		\bibfield  {author} {\bibinfo {author} {\bibfnamefont {J.~W.}\ \bibnamefont
				{Coburn}}\ and\ \bibinfo {author} {\bibfnamefont {H.~F.}\ \bibnamefont
				{Winters}},\ }\bibfield  {title} {\bibinfo {title} {Ion- and
				electron-assisted gas-surface chemistry -- an important effect in plasma
				etching},\ }\href {https://doi.org/10.1063/1.326355} {\bibfield  {journal}
			{\bibinfo  {journal} {Journal of Applied Physics}\ }\textbf {\bibinfo
				{volume} {50}},\ \bibinfo {pages} {3189} (\bibinfo {year}
			{1979})}\BibitemShut {NoStop}%
		\bibitem [{\citenamefont {Lee}\ \emph {et~al.}(2012)\citenamefont {Lee},
			\citenamefont {Kim},\ and\ \citenamefont {Chung}}]{Lee12}%
		\BibitemOpen
		\bibfield  {author} {\bibinfo {author} {\bibfnamefont {T.~Y.}\ \bibnamefont
				{Lee}}, \bibinfo {author} {\bibfnamefont {E.~H.}\ \bibnamefont {Kim}},\ and\
			\bibinfo {author} {\bibfnamefont {C.~W.}\ \bibnamefont {Chung}},\ }\bibfield
		{title} {\bibinfo {title} {Inductively coupled plasma reactive ion etching of
				{IrMn} magnetic thin films using a {CH$_4$/O$_2$/Ar} gas},\ }\href
		{https://doi.org/10.1016/j.tsf.2011.11.065} {\bibfield  {journal} {\bibinfo
				{journal} {Thin Solid Films}\ }\textbf {\bibinfo {volume} {521}},\ \bibinfo
			{pages} {229} (\bibinfo {year} {2012})}\BibitemShut {NoStop}%
		\bibitem [{\citenamefont {Garay}\ \emph {et~al.}(2015)\citenamefont {Garay},
			\citenamefont {Hwang},\ and\ \citenamefont {Chung}}]{Garay15}%
		\BibitemOpen
		\bibfield  {author} {\bibinfo {author} {\bibfnamefont {A.~A.}\ \bibnamefont
				{Garay}}, \bibinfo {author} {\bibfnamefont {S.~M.}\ \bibnamefont {Hwang}},\
			and\ \bibinfo {author} {\bibfnamefont {C.~W.}\ \bibnamefont {Chung}},\
		}\bibfield  {title} {\bibinfo {title} {Dry etching of {Co$_2$MnSi} magnetic
				thin films using a {CH$_3$OH/Ar} based inductively coupled plasma},\ }\href
		{https://doi.org/10.1016/j.vacuum.2014.09.011} {\bibfield  {journal}
			{\bibinfo  {journal} {Vacuum}\ }\textbf {\bibinfo {volume} {111}},\ \bibinfo
			{pages} {19} (\bibinfo {year} {2015})}\BibitemShut {NoStop}%
		\bibitem [{\citenamefont {Kl{\"a}ui}\ \emph
			{et~al.}(2003{\natexlab{a}})\citenamefont {Kl{\"a}ui}, \citenamefont {Vaz},
			\citenamefont {Lopez-Diaz},\ and\ \citenamefont {Bland}}]{Klaui03}%
		\BibitemOpen
		\bibfield  {author} {\bibinfo {author} {\bibfnamefont {M.}~\bibnamefont
				{Kl{\"a}ui}}, \bibinfo {author} {\bibfnamefont {C.~A.~F.}\ \bibnamefont
				{Vaz}}, \bibinfo {author} {\bibfnamefont {L.}~\bibnamefont {Lopez-Diaz}},\
			and\ \bibinfo {author} {\bibfnamefont {J.~A.~C.}\ \bibnamefont {Bland}},\
		}\bibfield  {title} {\bibinfo {title} {Vortex formation in narrow
				ferromagnetic rings},\ }\href {https://doi.org/10.1088/0953-8984/15/21/201}
		{\bibfield  {journal} {\bibinfo  {journal} {Journal of Physics: Condensed
					Matter}\ }\textbf {\bibinfo {volume} {15}},\ \bibinfo {pages} {R985}
			(\bibinfo {year} {2003}{\natexlab{a}})}\BibitemShut {NoStop}%
		\bibitem [{\citenamefont {Malinauskas}\ \emph {et~al.}(2016)\citenamefont
			{Malinauskas}, \citenamefont {\v{Z}ukauskas}, \citenamefont {Hasegawa},
			\citenamefont {Hayasaki}, \citenamefont {Mizeikis}, \citenamefont
			{Buividas},\ and\ \citenamefont {Juodkazis}}]{Malinauskas16}%
		\BibitemOpen
		\bibfield  {author} {\bibinfo {author} {\bibfnamefont {M.}~\bibnamefont
				{Malinauskas}}, \bibinfo {author} {\bibfnamefont {A.}~\bibnamefont
				{\v{Z}ukauskas}}, \bibinfo {author} {\bibfnamefont {S.}~\bibnamefont
				{Hasegawa}}, \bibinfo {author} {\bibfnamefont {Y.}~\bibnamefont {Hayasaki}},
			\bibinfo {author} {\bibfnamefont {V.}~\bibnamefont {Mizeikis}}, \bibinfo
			{author} {\bibfnamefont {R.}~\bibnamefont {Buividas}},\ and\ \bibinfo
			{author} {\bibfnamefont {S.}~\bibnamefont {Juodkazis}},\ }\bibfield  {title}
		{\bibinfo {title} {Ultrafast laser processing of materials: from science to
				industry},\ }\href {https://doi.org/10.1038/lsa.2016.133} {\bibfield
			{journal} {\bibinfo  {journal} {Light: Science \& Applications}\ }\textbf
			{\bibinfo {volume} {5}},\ \bibinfo {pages} {e16133} (\bibinfo {year}
			{2016})}\BibitemShut {NoStop}%
		\bibitem [{\citenamefont {Gattass}\ and\ \citenamefont
			{Mazur}(2008)}]{Gattass08}%
		\BibitemOpen
		\bibfield  {author} {\bibinfo {author} {\bibfnamefont {R.~R.}\ \bibnamefont
				{Gattass}}\ and\ \bibinfo {author} {\bibfnamefont {E.}~\bibnamefont
				{Mazur}},\ }\bibfield  {title} {\bibinfo {title} {Femtosecond laser
				micromachining in transparent materials},\ }\href
		{https://doi.org/10.1038/nphoton.2008.47} {\bibfield  {journal} {\bibinfo
				{journal} {Nature Photonics}\ }\textbf {\bibinfo {volume} {2}},\ \bibinfo
			{pages} {219} (\bibinfo {year} {2008})}\BibitemShut {NoStop}%
		\bibitem [{\citenamefont {Riddiford}\ \emph {et~al.}(2025)\citenamefont
			{Riddiford}, \citenamefont {Brock}, \citenamefont {Murawska}, \citenamefont
			{Wisser}, \citenamefont {Huang}, \citenamefont {Shepelin}, \citenamefont
			{Nembach}, \citenamefont {Hrabec},\ and\ \citenamefont
			{Heyderman}}]{Riddiford25}%
		\BibitemOpen
		\bibfield  {author} {\bibinfo {author} {\bibfnamefont {L.~J.}\ \bibnamefont
				{Riddiford}}, \bibinfo {author} {\bibfnamefont {J.~A.}\ \bibnamefont
				{Brock}}, \bibinfo {author} {\bibfnamefont {K.}~\bibnamefont {Murawska}},
			\bibinfo {author} {\bibfnamefont {J.}~\bibnamefont {Wisser}}, \bibinfo
			{author} {\bibfnamefont {X.}~\bibnamefont {Huang}}, \bibinfo {author}
			{\bibfnamefont {N.~A.}\ \bibnamefont {Shepelin}}, \bibinfo {author}
			{\bibfnamefont {H.~T.}\ \bibnamefont {Nembach}}, \bibinfo {author}
			{\bibfnamefont {A.}~\bibnamefont {Hrabec}},\ and\ \bibinfo {author}
			{\bibfnamefont {L.~J.}\ \bibnamefont {Heyderman}},\ }\bibfield  {title}
		{\bibinfo {title} {Two-dimensional gradients in magnetic properties created
				with direct-write laser annealing},\ }\href
		{https://doi.org/10.1038/s41467-025-65921-7} {\bibfield  {journal} {\bibinfo
				{journal} {Nature Communications}\ }\textbf {\bibinfo {volume} {16}},\
			\bibinfo {pages} {10979} (\bibinfo {year} {2025})}\BibitemShut {NoStop}%
		\bibitem [{\citenamefont {Sugioka}\ and\ \citenamefont
			{Cheng}(2014)}]{Sugioka14}%
		\BibitemOpen
		\bibfield  {author} {\bibinfo {author} {\bibfnamefont {K.}~\bibnamefont
				{Sugioka}}\ and\ \bibinfo {author} {\bibfnamefont {Y.}~\bibnamefont
				{Cheng}},\ }\bibfield  {title} {\bibinfo {title} {Ultrafast lasers --
				reliable tools for advanced materials processing},\ }\href
		{https://doi.org/10.1038/lsa.2014.30} {\bibfield  {journal} {\bibinfo
				{journal} {Light: Science \& Applications}\ }\textbf {\bibinfo {volume}
				{3}},\ \bibinfo {pages} {e149} (\bibinfo {year} {2014})}\BibitemShut
		{NoStop}%
		\bibitem [{\citenamefont {Finazzi}\ \emph {et~al.}(2013)\citenamefont
			{Finazzi}, \citenamefont {Savoini}, \citenamefont {Khorsand}, \citenamefont
			{Tsukamoto}, \citenamefont {Itoh}, \citenamefont {Du{\`{o}}}, \citenamefont
			{Kirilyuk}, \citenamefont {Rasing},\ and\ \citenamefont {Ezawa}}]{Finazzi13}%
		\BibitemOpen
		\bibfield  {author} {\bibinfo {author} {\bibfnamefont {M.}~\bibnamefont
				{Finazzi}}, \bibinfo {author} {\bibfnamefont {M.}~\bibnamefont {Savoini}},
			\bibinfo {author} {\bibfnamefont {A.~R.}\ \bibnamefont {Khorsand}}, \bibinfo
			{author} {\bibfnamefont {A.}~\bibnamefont {Tsukamoto}}, \bibinfo {author}
			{\bibfnamefont {A.}~\bibnamefont {Itoh}}, \bibinfo {author} {\bibfnamefont
				{L.}~\bibnamefont {Du{\`{o}}}}, \bibinfo {author} {\bibfnamefont
				{A.}~\bibnamefont {Kirilyuk}}, \bibinfo {author} {\bibfnamefont
				{T.}~\bibnamefont {Rasing}},\ and\ \bibinfo {author} {\bibfnamefont
				{M.}~\bibnamefont {Ezawa}},\ }\bibfield  {title} {\bibinfo {title}
			{Laser-induced magnetic nanostructures with tunable topological properties},\
		}\href {https://doi.org/10.1103/physrevlett.110.177205} {\bibfield  {journal}
			{\bibinfo  {journal} {Physical Review Letters}\ }\textbf {\bibinfo {volume}
				{110}},\ \bibinfo {pages} {177205} (\bibinfo {year} {2013})}\BibitemShut
		{NoStop}%
		\bibitem [{\citenamefont {Mangin}\ \emph {et~al.}(2014)\citenamefont {Mangin},
			\citenamefont {Gottwald}, \citenamefont {Lambert}, \citenamefont {Steil},
			\citenamefont {Uhl\'{\i}ř}, \citenamefont {Pang}, \citenamefont {Hehn},
			\citenamefont {Alebrand}, \citenamefont {Cinchetti}, \citenamefont
			{Malinowski}, \citenamefont {Fainman}, \citenamefont {Aeschlimann},\ and\
			\citenamefont {Fullerton}}]{Mangin14}%
		\BibitemOpen
		\bibfield  {author} {\bibinfo {author} {\bibfnamefont {S.}~\bibnamefont
				{Mangin}}, \bibinfo {author} {\bibfnamefont {M.}~\bibnamefont {Gottwald}},
			\bibinfo {author} {\bibfnamefont {C.-H.}\ \bibnamefont {Lambert}}, \bibinfo
			{author} {\bibfnamefont {D.}~\bibnamefont {Steil}}, \bibinfo {author}
			{\bibfnamefont {V.}~\bibnamefont {Uhl\'{\i}ř}}, \bibinfo {author}
			{\bibfnamefont {L.}~\bibnamefont {Pang}}, \bibinfo {author} {\bibfnamefont
				{M.}~\bibnamefont {Hehn}}, \bibinfo {author} {\bibfnamefont {S.}~\bibnamefont
				{Alebrand}}, \bibinfo {author} {\bibfnamefont {M.}~\bibnamefont {Cinchetti}},
			\bibinfo {author} {\bibfnamefont {G.}~\bibnamefont {Malinowski}}, \bibinfo
			{author} {\bibfnamefont {Y.}~\bibnamefont {Fainman}}, \bibinfo {author}
			{\bibfnamefont {M.}~\bibnamefont {Aeschlimann}},\ and\ \bibinfo {author}
			{\bibfnamefont {E.~E.}\ \bibnamefont {Fullerton}},\ }\bibfield  {title}
		{\bibinfo {title} {Engineered materials for all-optical helicity-dependent
				magnetic switching},\ }\href {https://doi.org/10.1038/nmat3864} {\bibfield
			{journal} {\bibinfo  {journal} {Nature Materials}\ }\textbf {\bibinfo
				{volume} {13}},\ \bibinfo {pages} {286} (\bibinfo {year} {2014})}\BibitemShut
		{NoStop}%
		\bibitem [{\citenamefont {Giannuzzi}\ and\ \citenamefont
			{Stevie}(2005)}]{Giannuzzi05}%
		\BibitemOpen
		\bibinfo {editor} {\bibfnamefont {L.~A.}\ \bibnamefont {Giannuzzi}}\ and\
		\bibinfo {editor} {\bibfnamefont {F.~A.}\ \bibnamefont {Stevie}},\ eds.,\
		\href {https://doi.org/10.1007/b101190} {\emph {\bibinfo {title}
				{Introduction to Focused Ion Beams}}}\ (\bibinfo  {publisher} {Springer US},\
		\bibinfo {address} {Boston, MA},\ \bibinfo {year} {2005})\BibitemShut
		{NoStop}%
		\bibitem [{\citenamefont {H\"{o}flich}\ \emph {et~al.}(2023)\citenamefont
			{H\"{o}flich}, \citenamefont {Hobler}, \citenamefont {Allen}, \citenamefont
			{Wirtz}, \citenamefont {Rius}, \citenamefont {McElwee-White}, \citenamefont
			{Krasheninnikov}, \citenamefont {Schmidt}, \citenamefont {Utke},
			\citenamefont {Klingner}, \citenamefont {Osenberg}, \citenamefont
			{C\'{o}rdoba}, \citenamefont {Djurabekova}, \citenamefont {Manke},
			\citenamefont {Moll}, \citenamefont {Manoccio}, \citenamefont {De~Teresa},
			\citenamefont {Bischoff}, \citenamefont {Michler}, \citenamefont {De~Castro},
			\citenamefont {Delobbe}, \citenamefont {Dunne}, \citenamefont {Dobrovolskiy},
			\citenamefont {Frese}, \citenamefont {G\"{o}lzh\"{a}user}, \citenamefont
			{Mazarov}, \citenamefont {Koelle}, \citenamefont {M\"{o}ller}, \citenamefont
			{P\'{e}rez-Murano}, \citenamefont {Philipp}, \citenamefont {Vollnhals},\ and\
			\citenamefont {Hlawacek}}]{Hoflich23}%
		\BibitemOpen
		\bibfield  {author} {\bibinfo {author} {\bibfnamefont {K.}~\bibnamefont
				{H\"{o}flich}}, \bibinfo {author} {\bibfnamefont {G.}~\bibnamefont {Hobler}},
			\bibinfo {author} {\bibfnamefont {F.~I.}\ \bibnamefont {Allen}}, \bibinfo
			{author} {\bibfnamefont {T.}~\bibnamefont {Wirtz}}, \bibinfo {author}
			{\bibfnamefont {G.}~\bibnamefont {Rius}}, \bibinfo {author} {\bibfnamefont
				{L.}~\bibnamefont {McElwee-White}}, \bibinfo {author} {\bibfnamefont {A.~V.}\
				\bibnamefont {Krasheninnikov}}, \bibinfo {author} {\bibfnamefont
				{M.}~\bibnamefont {Schmidt}}, \bibinfo {author} {\bibfnamefont
				{I.}~\bibnamefont {Utke}}, \bibinfo {author} {\bibfnamefont {N.}~\bibnamefont
				{Klingner}}, \bibinfo {author} {\bibfnamefont {M.}~\bibnamefont {Osenberg}},
			\bibinfo {author} {\bibfnamefont {R.}~\bibnamefont {C\'{o}rdoba}}, \bibinfo
			{author} {\bibfnamefont {F.}~\bibnamefont {Djurabekova}}, \bibinfo {author}
			{\bibfnamefont {I.}~\bibnamefont {Manke}}, \bibinfo {author} {\bibfnamefont
				{P.}~\bibnamefont {Moll}}, \bibinfo {author} {\bibfnamefont {M.}~\bibnamefont
				{Manoccio}}, \bibinfo {author} {\bibfnamefont {J.~M.}\ \bibnamefont
				{De~Teresa}}, \bibinfo {author} {\bibfnamefont {L.}~\bibnamefont {Bischoff}},
			\bibinfo {author} {\bibfnamefont {J.}~\bibnamefont {Michler}}, \bibinfo
			{author} {\bibfnamefont {O.}~\bibnamefont {De~Castro}}, \bibinfo {author}
			{\bibfnamefont {A.}~\bibnamefont {Delobbe}}, \bibinfo {author} {\bibfnamefont
				{P.}~\bibnamefont {Dunne}}, \bibinfo {author} {\bibfnamefont {O.~V.}\
				\bibnamefont {Dobrovolskiy}}, \bibinfo {author} {\bibfnamefont
				{N.}~\bibnamefont {Frese}}, \bibinfo {author} {\bibfnamefont
				{A.}~\bibnamefont {G\"{o}lzh\"{a}user}}, \bibinfo {author} {\bibfnamefont
				{P.}~\bibnamefont {Mazarov}}, \bibinfo {author} {\bibfnamefont
				{D.}~\bibnamefont {Koelle}}, \bibinfo {author} {\bibfnamefont
				{W.}~\bibnamefont {M\"{o}ller}}, \bibinfo {author} {\bibfnamefont
				{F.}~\bibnamefont {P\'{e}rez-Murano}}, \bibinfo {author} {\bibfnamefont
				{P.}~\bibnamefont {Philipp}}, \bibinfo {author} {\bibfnamefont
				{F.}~\bibnamefont {Vollnhals}},\ and\ \bibinfo {author} {\bibfnamefont
				{G.}~\bibnamefont {Hlawacek}},\ }\bibfield  {title} {\bibinfo {title}
			{Roadmap for focused ion beam technologies},\ }\href
		{https://doi.org/10.1063/5.0162597} {\bibfield  {journal} {\bibinfo
				{journal} {Applied Physics Reviews}\ }\textbf {\bibinfo {volume} {10}},\
			\bibinfo {pages} {041311} (\bibinfo {year} {2023})}\BibitemShut {NoStop}%
		\bibitem [{\citenamefont {Turnbull}\ \emph {et~al.}(2025)\citenamefont
			{Turnbull}, \citenamefont {Birch}, \citenamefont {Mart\'{\i}nez},
			\citenamefont {Yamamoto}, \citenamefont {Neethirajan}, \citenamefont
			{Ferreira}, \citenamefont {Zhakina}, \citenamefont {Binger}, \citenamefont
			{Choi}, \citenamefont {Belkhou}, \citenamefont {Finizio}, \citenamefont
			{Weigand}, \citenamefont {Suess}, \citenamefont {Mayoh}, \citenamefont
			{Balakrishnan}, \citenamefont {Abert}, \citenamefont {Wintz},\ and\
			\citenamefont {Donnelly}}]{Turnbull25}%
		\BibitemOpen
		\bibfield  {author} {\bibinfo {author} {\bibfnamefont {L.~A.}\ \bibnamefont
				{Turnbull}}, \bibinfo {author} {\bibfnamefont {M.~T.}\ \bibnamefont {Birch}},
			\bibinfo {author} {\bibfnamefont {M.~D.~P.}\ \bibnamefont {Mart\'{\i}nez}},
			\bibinfo {author} {\bibfnamefont {R.}~\bibnamefont {Yamamoto}}, \bibinfo
			{author} {\bibfnamefont {J.}~\bibnamefont {Neethirajan}}, \bibinfo {author}
			{\bibfnamefont {M.~R.}\ \bibnamefont {Ferreira}}, \bibinfo {author}
			{\bibfnamefont {E.}~\bibnamefont {Zhakina}}, \bibinfo {author} {\bibfnamefont
				{H.~J.}\ \bibnamefont {Binger}}, \bibinfo {author} {\bibfnamefont {Y.-G.}\
				\bibnamefont {Choi}}, \bibinfo {author} {\bibfnamefont {R.}~\bibnamefont
				{Belkhou}}, \bibinfo {author} {\bibfnamefont {S.}~\bibnamefont {Finizio}},
			\bibinfo {author} {\bibfnamefont {M.}~\bibnamefont {Weigand}}, \bibinfo
			{author} {\bibfnamefont {D.}~\bibnamefont {Suess}}, \bibinfo {author}
			{\bibfnamefont {D.~A.}\ \bibnamefont {Mayoh}}, \bibinfo {author}
			{\bibfnamefont {G.}~\bibnamefont {Balakrishnan}}, \bibinfo {author}
			{\bibfnamefont {C.}~\bibnamefont {Abert}}, \bibinfo {author} {\bibfnamefont
				{S.}~\bibnamefont {Wintz}},\ and\ \bibinfo {author} {\bibfnamefont
				{C.}~\bibnamefont {Donnelly}},\ }\bibfield  {title} {\bibinfo {title}
			{Interlinking helical spin textures in nanopatterned chiral magnets},\
		}\bibfield  {journal} {\bibinfo  {journal} {ArXiv e-prints}\ }\href
		{https://doi.org/10.48550/arXiv.2511.11372} {10.48550/arXiv.2511.11372}
		(\bibinfo {year} {2025})\BibitemShut {NoStop}%
		\bibitem [{\citenamefont {Birch}\ \emph {et~al.}(2026)\citenamefont {Birch},
			\citenamefont {Fujishiro}, \citenamefont {Belopolski}, \citenamefont {Mogi},
			\citenamefont {Chiew}, \citenamefont {Li}, \citenamefont {Yu}, \citenamefont
			{Nagaosa}, \citenamefont {Kawamura},\ and\ \citenamefont {Tokura}}]{Birch26}%
		\BibitemOpen
		\bibfield  {author} {\bibinfo {author} {\bibfnamefont {M.~T.}\ \bibnamefont
				{Birch}}, \bibinfo {author} {\bibfnamefont {Y.}~\bibnamefont {Fujishiro}},
			\bibinfo {author} {\bibfnamefont {I.}~\bibnamefont {Belopolski}}, \bibinfo
			{author} {\bibfnamefont {M.}~\bibnamefont {Mogi}}, \bibinfo {author}
			{\bibfnamefont {Y.-L.}\ \bibnamefont {Chiew}}, \bibinfo {author}
			{\bibfnamefont {Z.}~\bibnamefont {Li}}, \bibinfo {author} {\bibfnamefont
				{X.}~\bibnamefont {Yu}}, \bibinfo {author} {\bibfnamefont {N.}~\bibnamefont
				{Nagaosa}}, \bibinfo {author} {\bibfnamefont {M.}~\bibnamefont {Kawamura}},\
			and\ \bibinfo {author} {\bibfnamefont {Y.}~\bibnamefont {Tokura}},\
		}\bibfield  {title} {\bibinfo {title} {Nanosculpted {3D} helices of a
				magnetic weyl semimetal with switchable non-reciprocal electron transport},\
		}\bibfield  {journal} {\bibinfo  {journal} {Nature Nanotechnology}\ }\href
		{https://doi.org/10.1038/s41565-025-02104-x} {10.1038/s41565-025-02104-x}
		(\bibinfo {year} {2026})\BibitemShut {NoStop}%
		\bibitem [{\citenamefont {Aramesh}\ \emph {et~al.}(2018)\citenamefont
			{Aramesh}, \citenamefont {Mayamei}, \citenamefont {Wolff},\ and\
			\citenamefont {Ostrikov}}]{Aramesh18}%
		\BibitemOpen
		\bibfield  {author} {\bibinfo {author} {\bibfnamefont {M.}~\bibnamefont
				{Aramesh}}, \bibinfo {author} {\bibfnamefont {Y.}~\bibnamefont {Mayamei}},
			\bibinfo {author} {\bibfnamefont {A.}~\bibnamefont {Wolff}},\ and\ \bibinfo
			{author} {\bibfnamefont {K.}~\bibnamefont {Ostrikov}},\ }\bibfield  {title}
		{\bibinfo {title} {Superplastic nanoscale pore shaping by ion irradiation},\
		}\href {https://doi.org/10.1038/s41467-018-03316-7} {\bibfield  {journal}
			{\bibinfo  {journal} {Nature Communications}\ }\textbf {\bibinfo {volume}
				{9}},\ \bibinfo {pages} {835} (\bibinfo {year} {2018})}\BibitemShut {NoStop}%
		\bibitem [{\citenamefont {Liu}\ \emph {et~al.}(2020{\natexlab{a}})\citenamefont
			{Liu}, \citenamefont {Niu}, \citenamefont {Gu}, \citenamefont {Cabral},
			\citenamefont {Song},\ and\ \citenamefont {Liao}}]{Liu20c}%
		\BibitemOpen
		\bibfield  {author} {\bibinfo {author} {\bibfnamefont {J.}~\bibnamefont
				{Liu}}, \bibinfo {author} {\bibfnamefont {R.}~\bibnamefont {Niu}}, \bibinfo
			{author} {\bibfnamefont {J.}~\bibnamefont {Gu}}, \bibinfo {author}
			{\bibfnamefont {M.}~\bibnamefont {Cabral}}, \bibinfo {author} {\bibfnamefont
				{M.}~\bibnamefont {Song}},\ and\ \bibinfo {author} {\bibfnamefont
				{X.}~\bibnamefont {Liao}},\ }\bibfield  {title} {\bibinfo {title} {Effect of
				ion irradiation introduced by focused ion-beam milling on the mechanical
				behaviour of sub-micron-sized samples},\ }\href
		{https://doi.org/10.1038/s41598-020-66564-y} {\bibfield  {journal} {\bibinfo
				{journal} {Scientific Reports}\ }\textbf {\bibinfo {volume} {10}},\ \bibinfo
			{pages} {10324} (\bibinfo {year} {2020}{\natexlab{a}})}\BibitemShut {NoStop}%
		\bibitem [{\citenamefont {Chappert}\ \emph {et~al.}(1998)\citenamefont
			{Chappert}, \citenamefont {Bernas}, \citenamefont {Ferre}, \citenamefont
			{Kottler}, \citenamefont {Jamet}, \citenamefont {Chen}, \citenamefont
			{Cambril}, \citenamefont {Devolder}, \citenamefont {Rousseaux}, \citenamefont
			{Mathet},\ and\ \citenamefont {Launois}}]{Chappert98}%
		\BibitemOpen
		\bibfield  {author} {\bibinfo {author} {\bibfnamefont {C.}~\bibnamefont
				{Chappert}}, \bibinfo {author} {\bibfnamefont {H.}~\bibnamefont {Bernas}},
			\bibinfo {author} {\bibfnamefont {J.}~\bibnamefont {Ferre}}, \bibinfo
			{author} {\bibfnamefont {V.}~\bibnamefont {Kottler}}, \bibinfo {author}
			{\bibfnamefont {J.-P.}\ \bibnamefont {Jamet}}, \bibinfo {author}
			{\bibfnamefont {Y.}~\bibnamefont {Chen}}, \bibinfo {author} {\bibfnamefont
				{E.}~\bibnamefont {Cambril}}, \bibinfo {author} {\bibfnamefont
				{T.}~\bibnamefont {Devolder}}, \bibinfo {author} {\bibfnamefont
				{F.}~\bibnamefont {Rousseaux}}, \bibinfo {author} {\bibfnamefont
				{V.}~\bibnamefont {Mathet}},\ and\ \bibinfo {author} {\bibfnamefont
				{H.}~\bibnamefont {Launois}},\ }\bibfield  {title} {\bibinfo {title} {Planar
				patterned magnetic media obtained by ion irradiation},\ }\href
		{https://doi.org/10.1126/science.280.5371.1919} {\bibfield  {journal}
			{\bibinfo  {journal} {Science}\ }\textbf {\bibinfo {volume} {280}},\ \bibinfo
			{pages} {1919} (\bibinfo {year} {1998})}\BibitemShut {NoStop}%
		\bibitem [{\citenamefont {Fassbender}\ and\ \citenamefont
			{McCord}(2008)}]{Fassbender08a}%
		\BibitemOpen
		\bibfield  {author} {\bibinfo {author} {\bibfnamefont {J.}~\bibnamefont
				{Fassbender}}\ and\ \bibinfo {author} {\bibfnamefont {J.}~\bibnamefont
				{McCord}},\ }\bibfield  {title} {\bibinfo {title} {Magnetic patterning by
				means of ion irradiation and implantation},\ }\href
		{https://doi.org/10.1016/j.jmmm.2007.07.032} {\bibfield  {journal} {\bibinfo
				{journal} {Journal of Magnetism and Magnetic Materials}\ }\textbf {\bibinfo
				{volume} {320}},\ \bibinfo {pages} {579} (\bibinfo {year}
			{2008})}\BibitemShut {NoStop}%
		\bibitem [{\citenamefont {Cybart}\ \emph {et~al.}(2016)\citenamefont {Cybart},
			\citenamefont {Bali}, \citenamefont {Hlawacek}, \citenamefont {R\"{o}der},\
			and\ \citenamefont {Fassbender}}]{Cybart16}%
		\BibitemOpen
		\bibfield  {author} {\bibinfo {author} {\bibfnamefont {S.~A.}\ \bibnamefont
				{Cybart}}, \bibinfo {author} {\bibfnamefont {R.}~\bibnamefont {Bali}},
			\bibinfo {author} {\bibfnamefont {G.}~\bibnamefont {Hlawacek}}, \bibinfo
			{author} {\bibfnamefont {F.}~\bibnamefont {R\"{o}der}},\ and\ \bibinfo
			{author} {\bibfnamefont {J.}~\bibnamefont {Fassbender}},\ }\bibinfo {title}
		{Focused helium and neon ion beam modification of high-$t_c$ superconductors
			and magnetic materials},\ in\ \href
		{https://doi.org/10.1007/978-3-319-41990-9_17} {\emph {\bibinfo {booktitle}
				{NanoScience and Technology}}}\ (\bibinfo  {publisher} {Springer
			International Publishing},\ \bibinfo {address} {Cham},\ \bibinfo {year}
		{2016})\ pp.\ \bibinfo {pages} {415--445}\BibitemShut {NoStop}%
		\bibitem [{\citenamefont {Buschow}\ and\ \citenamefont
			{de~Boer}(2003)}]{Buschow03}%
		\BibitemOpen
		\bibfield  {author} {\bibinfo {author} {\bibfnamefont {K.~H.~J.}\
				\bibnamefont {Buschow}}\ and\ \bibinfo {author} {\bibfnamefont {F.~R.}\
				\bibnamefont {de~Boer}},\ }\href {https://doi.org/10.1007/b100503} {\emph
			{\bibinfo {title} {Physics of Magnetism and Magnetic Materials}}}\ (\bibinfo
		{publisher} {Springer US},\ \bibinfo {address} {Boston, MA},\ \bibinfo {year}
		{2003})\BibitemShut {NoStop}%
		\bibitem [{\citenamefont {Franco}\ and\ \citenamefont
			{Dodrill}(2021)}]{Franco21a}%
		\BibitemOpen
		\bibinfo {editor} {\bibfnamefont {V.}~\bibnamefont {Franco}}\ and\ \bibinfo
		{editor} {\bibfnamefont {B.}~\bibnamefont {Dodrill}},\ eds.,\ \href
		{https://doi.org/10.1007/978-3-030-70443-8} {\emph {\bibinfo {title}
				{Magnetic Measurement Techniques for Materials Characterization}}}\ (\bibinfo
		{publisher} {Springer International Publishing},\ \bibinfo {address}
		{Cham},\ \bibinfo {year} {2021})\BibitemShut {NoStop}%
		\bibitem [{\citenamefont {Donnelly}\ \emph
			{et~al.}(2022{\natexlab{b}})\citenamefont {Donnelly}, \citenamefont
			{Fischer}, \citenamefont {Kronast}, \citenamefont {Lubk}, \citenamefont
			{Wolf}, \citenamefont {Scagnoli}, \citenamefont {Sch\"{a}fer},\ and\
			\citenamefont {Soldatov}}]{Donnelly22a}%
		\BibitemOpen
		\bibfield  {author} {\bibinfo {author} {\bibfnamefont {C.}~\bibnamefont
				{Donnelly}}, \bibinfo {author} {\bibfnamefont {P.}~\bibnamefont {Fischer}},
			\bibinfo {author} {\bibfnamefont {F.}~\bibnamefont {Kronast}}, \bibinfo
			{author} {\bibfnamefont {A.}~\bibnamefont {Lubk}}, \bibinfo {author}
			{\bibfnamefont {D.}~\bibnamefont {Wolf}}, \bibinfo {author} {\bibfnamefont
				{V.}~\bibnamefont {Scagnoli}}, \bibinfo {author} {\bibfnamefont
				{R.}~\bibnamefont {Sch\"{a}fer}},\ and\ \bibinfo {author} {\bibfnamefont
				{I.}~\bibnamefont {Soldatov}},\ }\bibinfo {title} {Imaging of curved magnetic
			architectures},\ in\ \href {https://doi.org/10.1007/978-3-031-09086-8_6}
		{\emph {\bibinfo {booktitle} {Topics in Applied Physics}}}\ (\bibinfo
		{publisher} {Springer International Publishing},\ \bibinfo {address} {Cham},\
		\bibinfo {year} {2022})\ pp.\ \bibinfo {pages} {269--304}\BibitemShut
		{NoStop}%
		\bibitem [{\citenamefont {Mewes}\ and\ \citenamefont {Mewes}(2021)}]{Mewes21}%
		\BibitemOpen
		\bibfield  {author} {\bibinfo {author} {\bibfnamefont {T.}~\bibnamefont
				{Mewes}}\ and\ \bibinfo {author} {\bibfnamefont {C.~K.~A.}\ \bibnamefont
				{Mewes}},\ }\bibinfo {title} {Ferromagnetic resonance},\ in\ \href
		{https://doi.org/10.1007/978-3-030-70443-8_16} {\emph {\bibinfo {booktitle}
				{Magnetic Measurement Techniques for Materials Characterization}}}\ (\bibinfo
		{publisher} {Springer International Publishing},\ \bibinfo {address}
		{Cham},\ \bibinfo {year} {2021})\ pp.\ \bibinfo {pages}
		{431--452}\BibitemShut {NoStop}%
		\bibitem [{\citenamefont {Barman}\ \emph {et~al.}(2021)\citenamefont {Barman},
			\citenamefont {Gubbiotti}, \citenamefont {Ladak}, \citenamefont {Adeyeye},
			\citenamefont {Krawczyk}, \citenamefont {Gr\"{a}fe}, \citenamefont
			{Adelmann}, \citenamefont {Cotofana}, \citenamefont {Naeemi}, \citenamefont
			{Vasyuchka}, \citenamefont {Hillebrands}, \citenamefont {Nikitov},
			\citenamefont {Yu}, \citenamefont {Grundler}, \citenamefont {Sadovnikov},
			\citenamefont {Grachev}, \citenamefont {Sheshukova}, \citenamefont
			{Duquesne}, \citenamefont {Marangolo}, \citenamefont {Csaba}, \citenamefont
			{Porod}, \citenamefont {Demidov}, \citenamefont {Urazhdin}, \citenamefont
			{Demokritov}, \citenamefont {Albisetti}, \citenamefont {Petti}, \citenamefont
			{Bertacco}, \citenamefont {Schultheiss}, \citenamefont {Kruglyak},
			\citenamefont {Poimanov}, \citenamefont {Sahoo}, \citenamefont {Sinha},
			\citenamefont {Yang}, \citenamefont {Münzenberg}, \citenamefont {Moriyama},
			\citenamefont {Mizukami}, \citenamefont {Landeros}, \citenamefont {Gallardo},
			\citenamefont {Carlotti}, \citenamefont {Kim}, \citenamefont {Stamps},
			\citenamefont {Camley}, \citenamefont {Rana}, \citenamefont {Otani},
			\citenamefont {Yu}, \citenamefont {Yu}, \citenamefont {Bauer}, \citenamefont
			{Back}, \citenamefont {Uhrig}, \citenamefont {Dobrovolskiy}, \citenamefont
			{Budinska}, \citenamefont {Qin}, \citenamefont {van Dijken}, \citenamefont
			{Chumak}, \citenamefont {Khitun}, \citenamefont {Nikonov}, \citenamefont
			{Young}, \citenamefont {Zingsem},\ and\ \citenamefont
			{Winklhofer}}]{Barman21}%
		\BibitemOpen
		\bibfield  {author} {\bibinfo {author} {\bibfnamefont {A.}~\bibnamefont
				{Barman}}, \bibinfo {author} {\bibfnamefont {G.}~\bibnamefont {Gubbiotti}},
			\bibinfo {author} {\bibfnamefont {S.}~\bibnamefont {Ladak}}, \bibinfo
			{author} {\bibfnamefont {A.~O.}\ \bibnamefont {Adeyeye}}, \bibinfo {author}
			{\bibfnamefont {M.}~\bibnamefont {Krawczyk}}, \bibinfo {author}
			{\bibfnamefont {J.}~\bibnamefont {Gr\"{a}fe}}, \bibinfo {author}
			{\bibfnamefont {C.}~\bibnamefont {Adelmann}}, \bibinfo {author}
			{\bibfnamefont {S.}~\bibnamefont {Cotofana}}, \bibinfo {author}
			{\bibfnamefont {A.}~\bibnamefont {Naeemi}}, \bibinfo {author} {\bibfnamefont
				{V.~I.}\ \bibnamefont {Vasyuchka}}, \bibinfo {author} {\bibfnamefont
				{B.}~\bibnamefont {Hillebrands}}, \bibinfo {author} {\bibfnamefont {S.~A.}\
				\bibnamefont {Nikitov}}, \bibinfo {author} {\bibfnamefont {H.}~\bibnamefont
				{Yu}}, \bibinfo {author} {\bibfnamefont {D.}~\bibnamefont {Grundler}},
			\bibinfo {author} {\bibfnamefont {A.~V.}\ \bibnamefont {Sadovnikov}},
			\bibinfo {author} {\bibfnamefont {A.~A.}\ \bibnamefont {Grachev}}, \bibinfo
			{author} {\bibfnamefont {S.~E.}\ \bibnamefont {Sheshukova}}, \bibinfo
			{author} {\bibfnamefont {J.-Y.}\ \bibnamefont {Duquesne}}, \bibinfo {author}
			{\bibfnamefont {M.}~\bibnamefont {Marangolo}}, \bibinfo {author}
			{\bibfnamefont {G.}~\bibnamefont {Csaba}}, \bibinfo {author} {\bibfnamefont
				{W.}~\bibnamefont {Porod}}, \bibinfo {author} {\bibfnamefont {V.~E.}\
				\bibnamefont {Demidov}}, \bibinfo {author} {\bibfnamefont {S.}~\bibnamefont
				{Urazhdin}}, \bibinfo {author} {\bibfnamefont {S.~O.}\ \bibnamefont
				{Demokritov}}, \bibinfo {author} {\bibfnamefont {E.}~\bibnamefont
				{Albisetti}}, \bibinfo {author} {\bibfnamefont {D.}~\bibnamefont {Petti}},
			\bibinfo {author} {\bibfnamefont {R.}~\bibnamefont {Bertacco}}, \bibinfo
			{author} {\bibfnamefont {H.}~\bibnamefont {Schultheiss}}, \bibinfo {author}
			{\bibfnamefont {V.~V.}\ \bibnamefont {Kruglyak}}, \bibinfo {author}
			{\bibfnamefont {V.~D.}\ \bibnamefont {Poimanov}}, \bibinfo {author}
			{\bibfnamefont {S.}~\bibnamefont {Sahoo}}, \bibinfo {author} {\bibfnamefont
				{J.}~\bibnamefont {Sinha}}, \bibinfo {author} {\bibfnamefont
				{H.}~\bibnamefont {Yang}}, \bibinfo {author} {\bibfnamefont {M.}~\bibnamefont
				{Münzenberg}}, \bibinfo {author} {\bibfnamefont {T.}~\bibnamefont
				{Moriyama}}, \bibinfo {author} {\bibfnamefont {S.}~\bibnamefont {Mizukami}},
			\bibinfo {author} {\bibfnamefont {P.}~\bibnamefont {Landeros}}, \bibinfo
			{author} {\bibfnamefont {R.~A.}\ \bibnamefont {Gallardo}}, \bibinfo {author}
			{\bibfnamefont {G.}~\bibnamefont {Carlotti}}, \bibinfo {author}
			{\bibfnamefont {J.-V.}\ \bibnamefont {Kim}}, \bibinfo {author} {\bibfnamefont
				{R.~L.}\ \bibnamefont {Stamps}}, \bibinfo {author} {\bibfnamefont {R.~E.}\
				\bibnamefont {Camley}}, \bibinfo {author} {\bibfnamefont {B.}~\bibnamefont
				{Rana}}, \bibinfo {author} {\bibfnamefont {Y.}~\bibnamefont {Otani}},
			\bibinfo {author} {\bibfnamefont {W.}~\bibnamefont {Yu}}, \bibinfo {author}
			{\bibfnamefont {T.}~\bibnamefont {Yu}}, \bibinfo {author} {\bibfnamefont
				{G.~E.~W.}\ \bibnamefont {Bauer}}, \bibinfo {author} {\bibfnamefont
				{C.}~\bibnamefont {Back}}, \bibinfo {author} {\bibfnamefont {G.~S.}\
				\bibnamefont {Uhrig}}, \bibinfo {author} {\bibfnamefont {O.~V.}\ \bibnamefont
				{Dobrovolskiy}}, \bibinfo {author} {\bibfnamefont {B.}~\bibnamefont
				{Budinska}}, \bibinfo {author} {\bibfnamefont {H.}~\bibnamefont {Qin}},
			\bibinfo {author} {\bibfnamefont {S.}~\bibnamefont {van Dijken}}, \bibinfo
			{author} {\bibfnamefont {A.~V.}\ \bibnamefont {Chumak}}, \bibinfo {author}
			{\bibfnamefont {A.}~\bibnamefont {Khitun}}, \bibinfo {author} {\bibfnamefont
				{D.~E.}\ \bibnamefont {Nikonov}}, \bibinfo {author} {\bibfnamefont {I.~A.}\
				\bibnamefont {Young}}, \bibinfo {author} {\bibfnamefont {B.~W.}\ \bibnamefont
				{Zingsem}},\ and\ \bibinfo {author} {\bibfnamefont {M.}~\bibnamefont
				{Winklhofer}},\ }\bibfield  {title} {\bibinfo {title} {The 2021 magnonics
				roadmap},\ }\href {https://doi.org/10.1088/1361-648x/abec1a} {\bibfield
			{journal} {\bibinfo  {journal} {Journal of Physics: Condensed Matter}\
			}\textbf {\bibinfo {volume} {33}},\ \bibinfo {pages} {413001} (\bibinfo
			{year} {2021})}\BibitemShut {NoStop}%
		\bibitem [{\citenamefont {Kimel}\ \emph {et~al.}(2022)\citenamefont {Kimel},
			\citenamefont {Zvezdin}, \citenamefont {Sharma}, \citenamefont {Shallcross},
			\citenamefont {de~Sousa}, \citenamefont {Garc\'{\i}a-Mart\'{\i}n},
			\citenamefont {Salvan}, \citenamefont {Hamrle}, \citenamefont {Stejskal},
			\citenamefont {McCord}, \citenamefont {Tacchi}, \citenamefont {Carlotti},
			\citenamefont {Gambardella}, \citenamefont {Salis}, \citenamefont
			{M\"{u}nzenberg}, \citenamefont {Schultze}, \citenamefont {Temnov},
			\citenamefont {Bychkov}, \citenamefont {Kotov}, \citenamefont {Maccaferri},
			\citenamefont {Ignatyeva}, \citenamefont {Belotelov}, \citenamefont
			{Donnelly}, \citenamefont {Rodriguez}, \citenamefont {Matsuda}, \citenamefont
			{Ruchon}, \citenamefont {Fanciulli}, \citenamefont {Sacchi}, \citenamefont
			{Du}, \citenamefont {Wang}, \citenamefont {Armitage}, \citenamefont
			{Schubert}, \citenamefont {Darakchieva}, \citenamefont {Liu}, \citenamefont
			{Huang}, \citenamefont {Ding}, \citenamefont {Berger},\ and\ \citenamefont
			{Vavassori}}]{Kimel22}%
		\BibitemOpen
		\bibfield  {author} {\bibinfo {author} {\bibfnamefont {A.}~\bibnamefont
				{Kimel}}, \bibinfo {author} {\bibfnamefont {A.}~\bibnamefont {Zvezdin}},
			\bibinfo {author} {\bibfnamefont {S.}~\bibnamefont {Sharma}}, \bibinfo
			{author} {\bibfnamefont {S.}~\bibnamefont {Shallcross}}, \bibinfo {author}
			{\bibfnamefont {N.}~\bibnamefont {de~Sousa}}, \bibinfo {author}
			{\bibfnamefont {A.}~\bibnamefont {Garc\'{\i}a-Mart\'{\i}n}}, \bibinfo
			{author} {\bibfnamefont {G.}~\bibnamefont {Salvan}}, \bibinfo {author}
			{\bibfnamefont {J.}~\bibnamefont {Hamrle}}, \bibinfo {author} {\bibfnamefont
				{O.}~\bibnamefont {Stejskal}}, \bibinfo {author} {\bibfnamefont
				{J.}~\bibnamefont {McCord}}, \bibinfo {author} {\bibfnamefont
				{S.}~\bibnamefont {Tacchi}}, \bibinfo {author} {\bibfnamefont
				{G.}~\bibnamefont {Carlotti}}, \bibinfo {author} {\bibfnamefont
				{P.}~\bibnamefont {Gambardella}}, \bibinfo {author} {\bibfnamefont
				{G.}~\bibnamefont {Salis}}, \bibinfo {author} {\bibfnamefont
				{M.}~\bibnamefont {M\"{u}nzenberg}}, \bibinfo {author} {\bibfnamefont
				{M.}~\bibnamefont {Schultze}}, \bibinfo {author} {\bibfnamefont
				{V.}~\bibnamefont {Temnov}}, \bibinfo {author} {\bibfnamefont {I.~V.}\
				\bibnamefont {Bychkov}}, \bibinfo {author} {\bibfnamefont {L.~N.}\
				\bibnamefont {Kotov}}, \bibinfo {author} {\bibfnamefont {N.}~\bibnamefont
				{Maccaferri}}, \bibinfo {author} {\bibfnamefont {D.}~\bibnamefont
				{Ignatyeva}}, \bibinfo {author} {\bibfnamefont {V.}~\bibnamefont
				{Belotelov}}, \bibinfo {author} {\bibfnamefont {C.}~\bibnamefont {Donnelly}},
			\bibinfo {author} {\bibfnamefont {A.~H.}\ \bibnamefont {Rodriguez}}, \bibinfo
			{author} {\bibfnamefont {I.}~\bibnamefont {Matsuda}}, \bibinfo {author}
			{\bibfnamefont {T.}~\bibnamefont {Ruchon}}, \bibinfo {author} {\bibfnamefont
				{M.}~\bibnamefont {Fanciulli}}, \bibinfo {author} {\bibfnamefont
				{M.}~\bibnamefont {Sacchi}}, \bibinfo {author} {\bibfnamefont {C.~R.}\
				\bibnamefont {Du}}, \bibinfo {author} {\bibfnamefont {H.}~\bibnamefont
				{Wang}}, \bibinfo {author} {\bibfnamefont {N.~P.}\ \bibnamefont {Armitage}},
			\bibinfo {author} {\bibfnamefont {M.}~\bibnamefont {Schubert}}, \bibinfo
			{author} {\bibfnamefont {V.}~\bibnamefont {Darakchieva}}, \bibinfo {author}
			{\bibfnamefont {B.}~\bibnamefont {Liu}}, \bibinfo {author} {\bibfnamefont
				{Z.}~\bibnamefont {Huang}}, \bibinfo {author} {\bibfnamefont
				{B.}~\bibnamefont {Ding}}, \bibinfo {author} {\bibfnamefont {A.}~\bibnamefont
				{Berger}},\ and\ \bibinfo {author} {\bibfnamefont {P.}~\bibnamefont
				{Vavassori}},\ }\bibfield  {title} {\bibinfo {title} {The 2022 magneto-optics
				roadmap},\ }\href {https://doi.org/10.1088/1361-6463/ac8da0} {\bibfield
			{journal} {\bibinfo  {journal} {Journal of Physics D: Applied Physics}\
			}\textbf {\bibinfo {volume} {55}},\ \bibinfo {pages} {463003} (\bibinfo
			{year} {2022})}\BibitemShut {NoStop}%
		\bibitem [{\citenamefont {Sanz-Hern\'{a}ndez}\ \emph
			{et~al.}(2023)\citenamefont {Sanz-Hern\'{a}ndez}, \citenamefont {Skoric},
			\citenamefont {Cascales-Sandoval},\ and\ \citenamefont
			{Fern\'{a}ndez-Pacheco}}]{SanzHernandez23}%
		\BibitemOpen
		\bibfield  {author} {\bibinfo {author} {\bibfnamefont {D.}~\bibnamefont
				{Sanz-Hern\'{a}ndez}}, \bibinfo {author} {\bibfnamefont {L.}~\bibnamefont
				{Skoric}}, \bibinfo {author} {\bibfnamefont {M.~A.}\ \bibnamefont
				{Cascales-Sandoval}},\ and\ \bibinfo {author} {\bibfnamefont
				{A.}~\bibnamefont {Fern\'{a}ndez-Pacheco}},\ }\bibfield  {title} {\bibinfo
			{title} {Probing {3D} magnetic nanostructures by dark-field magneto-optical
				{K}err effect},\ }\href {https://doi.org/10.1063/5.0132250} {\bibfield
			{journal} {\bibinfo  {journal} {Journal of Applied Physics}\ }\textbf
			{\bibinfo {volume} {133}},\ \bibinfo {pages} {043901} (\bibinfo {year}
			{2023})}\BibitemShut {NoStop}%
		\bibitem [{\citenamefont {Cascales~Sandoval}\ \emph {et~al.}(2025)\citenamefont
			{Cascales~Sandoval}, \citenamefont {Jurczyk}, \citenamefont {Skoric},
			\citenamefont {Sanz-Hern\'{a}ndez}, \citenamefont {Leo}, \citenamefont
			{Kovacs}, \citenamefont {Schrefl}, \citenamefont {Hierro-Rodr\'{\i}guez},\
			and\ \citenamefont {Fern\'{a}ndez-Pacheco}}]{CascalesSandoval25}%
		\BibitemOpen
		\bibfield  {author} {\bibinfo {author} {\bibfnamefont {M.~A.}\ \bibnamefont
				{Cascales~Sandoval}}, \bibinfo {author} {\bibfnamefont {J.}~\bibnamefont
				{Jurczyk}}, \bibinfo {author} {\bibfnamefont {L.}~\bibnamefont {Skoric}},
			\bibinfo {author} {\bibfnamefont {D.}~\bibnamefont {Sanz-Hern\'{a}ndez}},
			\bibinfo {author} {\bibfnamefont {N.}~\bibnamefont {Leo}}, \bibinfo {author}
			{\bibfnamefont {A.}~\bibnamefont {Kovacs}}, \bibinfo {author} {\bibfnamefont
				{T.}~\bibnamefont {Schrefl}}, \bibinfo {author} {\bibfnamefont
				{A.}~\bibnamefont {Hierro-Rodr\'{\i}guez}},\ and\ \bibinfo {author}
			{\bibfnamefont {A.}~\bibnamefont {Fern\'{a}ndez-Pacheco}},\ }\bibfield
		{title} {\bibinfo {title} {Remote-sensing based control of {3D} magnetic
				fields using machine learning for <i>in operando</i> applications},\ }\href
		{https://doi.org/10.1063/5.0249846} {\bibfield  {journal} {\bibinfo
				{journal} {Journal of Applied Physics}\ }\textbf {\bibinfo {volume} {137}},\
			\bibinfo {pages} {113905} (\bibinfo {year} {2025})}\BibitemShut {NoStop}%
		\bibitem [{\citenamefont {Winklhofer}\ and\ \citenamefont
			{Zimanyi}(2006)}]{Winklhofer06}%
		\BibitemOpen
		\bibfield  {author} {\bibinfo {author} {\bibfnamefont {M.}~\bibnamefont
				{Winklhofer}}\ and\ \bibinfo {author} {\bibfnamefont {G.~T.}\ \bibnamefont
				{Zimanyi}},\ }\bibfield  {title} {\bibinfo {title} {Extracting the intrinsic
				switching field distribution in perpendicular media: A comparative
				analysis},\ }\href {https://doi.org/10.1063/1.2176598} {\bibfield  {journal}
			{\bibinfo  {journal} {Journal of Applied Physics}\ }\textbf {\bibinfo
				{volume} {99}},\ \bibinfo {pages} {08E710} (\bibinfo {year}
			{2006})}\BibitemShut {NoStop}%
		\bibitem [{\citenamefont {Dobrot\v{a}}\ and\ \citenamefont
			{Stancu}(2013)}]{Dobrota13}%
		\BibitemOpen
		\bibfield  {author} {\bibinfo {author} {\bibfnamefont {C.-I.}\ \bibnamefont
				{Dobrot\v{a}}}\ and\ \bibinfo {author} {\bibfnamefont {A.}~\bibnamefont
				{Stancu}},\ }\bibfield  {title} {\bibinfo {title} {What does a first-order
				reversal curve diagram really mean? {A} study case: Array of ferromagnetic
				nanowires},\ }\href {https://doi.org/10.1063/1.4789613} {\bibfield  {journal}
			{\bibinfo  {journal} {Journal of Applied Physics}\ }\textbf {\bibinfo
				{volume} {113}},\ \bibinfo {pages} {043928} (\bibinfo {year}
			{2013})}\BibitemShut {NoStop}%
		\bibitem [{\citenamefont {Alikhani}\ \emph {et~al.}(2016)\citenamefont
			{Alikhani}, \citenamefont {Ramazani}, \citenamefont {Almasi~Kashi},
			\citenamefont {Samanifar},\ and\ \citenamefont {Montazer}}]{Alikhani16}%
		\BibitemOpen
		\bibfield  {author} {\bibinfo {author} {\bibfnamefont {M.}~\bibnamefont
				{Alikhani}}, \bibinfo {author} {\bibfnamefont {A.}~\bibnamefont {Ramazani}},
			\bibinfo {author} {\bibfnamefont {M.}~\bibnamefont {Almasi~Kashi}}, \bibinfo
			{author} {\bibfnamefont {S.}~\bibnamefont {Samanifar}},\ and\ \bibinfo
			{author} {\bibfnamefont {A.}~\bibnamefont {Montazer}},\ }\bibfield  {title}
		{\bibinfo {title} {Irreversible evolution of angular-dependent coercivity in
				{Fe$_{80}$Ni$_{20}$} nanowire arrays: Detection of a single vortex state},\
		}\href {https://doi.org/10.1016/j.jmmm.2016.04.035} {\bibfield  {journal}
			{\bibinfo  {journal} {Journal of Magnetism and Magnetic Materials}\ }\textbf
			{\bibinfo {volume} {414}},\ \bibinfo {pages} {158} (\bibinfo {year}
			{2016})}\BibitemShut {NoStop}%
		\bibitem [{\citenamefont {Muscas}\ \emph {et~al.}(2020)\citenamefont {Muscas},
			\citenamefont {Menniti}, \citenamefont {Brucas},\ and\ \citenamefont
			{J\"{o}nsson}}]{Muscas20}%
		\BibitemOpen
		\bibfield  {author} {\bibinfo {author} {\bibfnamefont {G.}~\bibnamefont
				{Muscas}}, \bibinfo {author} {\bibfnamefont {M.}~\bibnamefont {Menniti}},
			\bibinfo {author} {\bibfnamefont {R.}~\bibnamefont {Brucas}},\ and\ \bibinfo
			{author} {\bibfnamefont {P.}~\bibnamefont {J\"{o}nsson}},\ }\bibfield
		{title} {\bibinfo {title} {Mesoscale magnetic rings: Complex magnetization
				reversal uncovered by {FORC}},\ }\href
		{https://doi.org/10.1016/j.jmmm.2020.166559} {\bibfield  {journal} {\bibinfo
				{journal} {Journal of Magnetism and Magnetic Materials}\ }\textbf {\bibinfo
				{volume} {502}},\ \bibinfo {pages} {166559} (\bibinfo {year}
			{2020})}\BibitemShut {NoStop}%
		\bibitem [{\citenamefont {Gr\"{a}fe}\ \emph {et~al.}(2014)\citenamefont
			{Gr\"{a}fe}, \citenamefont {Schmidt}, \citenamefont {Audehm}, \citenamefont
			{Sch\"{u}tz},\ and\ \citenamefont {Goering}}]{Grafe14}%
		\BibitemOpen
		\bibfield  {author} {\bibinfo {author} {\bibfnamefont {J.}~\bibnamefont
				{Gr\"{a}fe}}, \bibinfo {author} {\bibfnamefont {M.}~\bibnamefont {Schmidt}},
			\bibinfo {author} {\bibfnamefont {P.}~\bibnamefont {Audehm}}, \bibinfo
			{author} {\bibfnamefont {G.}~\bibnamefont {Sch\"{u}tz}},\ and\ \bibinfo
			{author} {\bibfnamefont {E.}~\bibnamefont {Goering}},\ }\bibfield  {title}
		{\bibinfo {title} {Application of magneto-optical {K}err effect to
				first-order reversal curve measurements},\ }\href
		{https://doi.org/10.1063/1.4865135} {\bibfield  {journal} {\bibinfo
				{journal} {Review of Scientific Instruments}\ }\textbf {\bibinfo {volume}
				{85}},\ \bibinfo {pages} {023901} (\bibinfo {year} {2014})}\BibitemShut
		{NoStop}%
		\bibitem [{\citenamefont {De~Biasi}(2017)}]{DeBiasi17}%
		\BibitemOpen
		\bibfield  {author} {\bibinfo {author} {\bibfnamefont {E.}~\bibnamefont
				{De~Biasi}},\ }\bibfield  {title} {\bibinfo {title} {Faster modified protocol
				for first order reversal curve measurements},\ }\href
		{https://doi.org/10.1016/j.jmmm.2017.05.005} {\bibfield  {journal} {\bibinfo
				{journal} {Journal of Magnetism and Magnetic Materials}\ }\textbf {\bibinfo
				{volume} {439}},\ \bibinfo {pages} {259} (\bibinfo {year}
			{2017})}\BibitemShut {NoStop}%
		\bibitem [{\citenamefont {Marchiori}\ \emph {et~al.}(2021)\citenamefont
			{Marchiori}, \citenamefont {Ceccarelli}, \citenamefont {Rossi}, \citenamefont
			{Lorenzelli}, \citenamefont {Degen},\ and\ \citenamefont
			{Poggio}}]{Marchiori21}%
		\BibitemOpen
		\bibfield  {author} {\bibinfo {author} {\bibfnamefont {E.}~\bibnamefont
				{Marchiori}}, \bibinfo {author} {\bibfnamefont {L.}~\bibnamefont
				{Ceccarelli}}, \bibinfo {author} {\bibfnamefont {N.}~\bibnamefont {Rossi}},
			\bibinfo {author} {\bibfnamefont {L.}~\bibnamefont {Lorenzelli}}, \bibinfo
			{author} {\bibfnamefont {C.~L.}\ \bibnamefont {Degen}},\ and\ \bibinfo
			{author} {\bibfnamefont {M.}~\bibnamefont {Poggio}},\ }\bibfield  {title}
		{\bibinfo {title} {Nanoscale magnetic field imaging for {2D} materials},\
		}\href {https://doi.org/10.1038/s42254-021-00380-9} {\bibfield  {journal}
			{\bibinfo  {journal} {Nature Reviews Physics}\ }\textbf {\bibinfo {volume}
				{4}},\ \bibinfo {pages} {49} (\bibinfo {year} {2021})}\BibitemShut {NoStop}%
		\bibitem [{\citenamefont {Rugar}\ \emph {et~al.}(1990)\citenamefont {Rugar},
			\citenamefont {Mamin}, \citenamefont {Guethner}, \citenamefont {Lambert},
			\citenamefont {Stern}, \citenamefont {McFadyen},\ and\ \citenamefont
			{Yogi}}]{Rugar90}%
		\BibitemOpen
		\bibfield  {author} {\bibinfo {author} {\bibfnamefont {D.}~\bibnamefont
				{Rugar}}, \bibinfo {author} {\bibfnamefont {H.~J.}\ \bibnamefont {Mamin}},
			\bibinfo {author} {\bibfnamefont {P.}~\bibnamefont {Guethner}}, \bibinfo
			{author} {\bibfnamefont {S.~E.}\ \bibnamefont {Lambert}}, \bibinfo {author}
			{\bibfnamefont {J.~E.}\ \bibnamefont {Stern}}, \bibinfo {author}
			{\bibfnamefont {I.}~\bibnamefont {McFadyen}},\ and\ \bibinfo {author}
			{\bibfnamefont {T.}~\bibnamefont {Yogi}},\ }\bibfield  {title} {\bibinfo
			{title} {Magnetic force microscopy: General principles and application to
				longitudinal recording media},\ }\href {https://doi.org/10.1063/1.346713}
		{\bibfield  {journal} {\bibinfo  {journal} {Journal of Applied Physics}\
			}\textbf {\bibinfo {volume} {68}},\ \bibinfo {pages} {1169} (\bibinfo {year}
			{1990})}\BibitemShut {NoStop}%
		\bibitem [{\citenamefont {Kazakova}\ \emph {et~al.}(2019)\citenamefont
			{Kazakova}, \citenamefont {Puttock}, \citenamefont {Barton}, \citenamefont
			{Corte-Le{\'{o}}n}, \citenamefont {Jaafar}, \citenamefont {Neu},\ and\
			\citenamefont {Asenjo}}]{Kazakova19}%
		\BibitemOpen
		\bibfield  {author} {\bibinfo {author} {\bibfnamefont {O.}~\bibnamefont
				{Kazakova}}, \bibinfo {author} {\bibfnamefont {R.}~\bibnamefont {Puttock}},
			\bibinfo {author} {\bibfnamefont {C.}~\bibnamefont {Barton}}, \bibinfo
			{author} {\bibfnamefont {H.}~\bibnamefont {Corte-Le{\'{o}}n}}, \bibinfo
			{author} {\bibfnamefont {M.}~\bibnamefont {Jaafar}}, \bibinfo {author}
			{\bibfnamefont {V.}~\bibnamefont {Neu}},\ and\ \bibinfo {author}
			{\bibfnamefont {A.}~\bibnamefont {Asenjo}},\ }\bibfield  {title} {\bibinfo
			{title} {Frontiers of magnetic force microscopy},\ }\href
		{https://doi.org/10.1063/1.5050712} {\bibfield  {journal} {\bibinfo
				{journal} {Journal of Applied Physics}\ }\textbf {\bibinfo {volume} {125}},\
			\bibinfo {pages} {060901} (\bibinfo {year} {2019})}\BibitemShut {NoStop}%
		\bibitem [{\citenamefont {H\"{a}berle}\ \emph {et~al.}(2012)\citenamefont
			{H\"{a}berle}, \citenamefont {Haering}, \citenamefont {Pfeifer},
			\citenamefont {Han}, \citenamefont {Achinuq}, \citenamefont {Wiedwald},
			\citenamefont {Herr},\ and\ \citenamefont {Koslowski}}]{Haberle12}%
		\BibitemOpen
		\bibfield  {author} {\bibinfo {author} {\bibfnamefont {T.}~\bibnamefont
				{H\"{a}berle}}, \bibinfo {author} {\bibfnamefont {F.}~\bibnamefont
				{Haering}}, \bibinfo {author} {\bibfnamefont {H.}~\bibnamefont {Pfeifer}},
			\bibinfo {author} {\bibfnamefont {L.}~\bibnamefont {Han}}, \bibinfo {author}
			{\bibfnamefont {B.}~\bibnamefont {Achinuq}}, \bibinfo {author} {\bibfnamefont
				{U.}~\bibnamefont {Wiedwald}}, \bibinfo {author} {\bibfnamefont
				{U.}~\bibnamefont {Herr}},\ and\ \bibinfo {author} {\bibfnamefont
				{B.}~\bibnamefont {Koslowski}},\ }\bibfield  {title} {\bibinfo {title}
			{Towards quantitative magnetic force microscopy: theory and experiment},\
		}\href {https://doi.org/10.1088/1367-2630/14/4/043044} {\bibfield  {journal}
			{\bibinfo  {journal} {New Journal of Physics}\ }\textbf {\bibinfo {volume}
				{14}},\ \bibinfo {pages} {043044} (\bibinfo {year} {2012})}\BibitemShut
		{NoStop}%
		\bibitem [{\citenamefont {Feng}\ \emph {et~al.}(2022)\citenamefont {Feng},
			\citenamefont {Mandru}, \citenamefont {Yıldırım},\ and\ \citenamefont
			{Hug}}]{Feng22}%
		\BibitemOpen
		\bibfield  {author} {\bibinfo {author} {\bibfnamefont {Y.}~\bibnamefont
				{Feng}}, \bibinfo {author} {\bibfnamefont {A.-O.}\ \bibnamefont {Mandru}},
			\bibinfo {author} {\bibfnamefont {O.}~\bibnamefont {Yıldırım}},\ and\
			\bibinfo {author} {\bibfnamefont {H.}~\bibnamefont {Hug}},\ }\bibfield
		{title} {\bibinfo {title} {Quantitative magnetic force microscopy:
				Transfer-function method revisited},\ }\href
		{https://doi.org/10.1103/physrevapplied.18.024016} {\bibfield  {journal}
			{\bibinfo  {journal} {Physical Review Applied}\ }\textbf {\bibinfo {volume}
				{18}},\ \bibinfo {pages} {024016} (\bibinfo {year} {2022})}\BibitemShut
		{NoStop}%
		\bibitem [{\citenamefont {Ulbrich}\ \emph {et~al.}(2006)\citenamefont
			{Ulbrich}, \citenamefont {Makarov}, \citenamefont {Hu}, \citenamefont {Guhr},
			\citenamefont {Suess}, \citenamefont {Schrefl},\ and\ \citenamefont
			{Albrecht}}]{Ulbrich06}%
		\BibitemOpen
		\bibfield  {author} {\bibinfo {author} {\bibfnamefont {T.~C.}\ \bibnamefont
				{Ulbrich}}, \bibinfo {author} {\bibfnamefont {D.}~\bibnamefont {Makarov}},
			\bibinfo {author} {\bibfnamefont {G.}~\bibnamefont {Hu}}, \bibinfo {author}
			{\bibfnamefont {I.~L.}\ \bibnamefont {Guhr}}, \bibinfo {author}
			{\bibfnamefont {D.}~\bibnamefont {Suess}}, \bibinfo {author} {\bibfnamefont
				{T.}~\bibnamefont {Schrefl}},\ and\ \bibinfo {author} {\bibfnamefont
				{M.}~\bibnamefont {Albrecht}},\ }\bibfield  {title} {\bibinfo {title}
			{Magnetization reversal in a novel gradient nanomaterial},\ }\href
		{https://doi.org/10.1103/PhysRevLett.96.077202} {\bibfield  {journal}
			{\bibinfo  {journal} {Physical Review Letters}\ }\textbf {\bibinfo {volume}
				{96}},\ \bibinfo {pages} {077202} (\bibinfo {year} {2006})}\BibitemShut
		{NoStop}%
		\bibitem [{\citenamefont {Baraban}\ \emph {et~al.}(2008)\citenamefont
			{Baraban}, \citenamefont {Makarov}, \citenamefont {Albrecht}, \citenamefont
			{Rivier}, \citenamefont {Leiderer},\ and\ \citenamefont {Erbe}}]{Baraban08}%
		\BibitemOpen
		\bibfield  {author} {\bibinfo {author} {\bibfnamefont {L.}~\bibnamefont
				{Baraban}}, \bibinfo {author} {\bibfnamefont {D.}~\bibnamefont {Makarov}},
			\bibinfo {author} {\bibfnamefont {M.}~\bibnamefont {Albrecht}}, \bibinfo
			{author} {\bibfnamefont {N.}~\bibnamefont {Rivier}}, \bibinfo {author}
			{\bibfnamefont {P.}~\bibnamefont {Leiderer}},\ and\ \bibinfo {author}
			{\bibfnamefont {A.}~\bibnamefont {Erbe}},\ }\bibfield  {title} {\bibinfo
			{title} {Frustration-induced magic number clusters of colloidal magnetic
				particles},\ }\href {https://doi.org/10.1103/physreve.77.031407} {\bibfield
			{journal} {\bibinfo  {journal} {Physical Review E}\ }\textbf {\bibinfo
				{volume} {77}},\ \bibinfo {pages} {031407} (\bibinfo {year}
			{2008})}\BibitemShut {NoStop}%
		\bibitem [{\citenamefont {Jaafar}\ \emph {et~al.}(2011)\citenamefont {Jaafar},
			\citenamefont {Serrano-Ramon}, \citenamefont {Iglesias-Freire}, \citenamefont
			{Fernandez-Pacheco}, \citenamefont {Ibarra}, \citenamefont {De~Teresa},\ and\
			\citenamefont {Asenjo}}]{Jaafar11}%
		\BibitemOpen
		\bibfield  {author} {\bibinfo {author} {\bibfnamefont {M.}~\bibnamefont
				{Jaafar}}, \bibinfo {author} {\bibfnamefont {L.}~\bibnamefont
				{Serrano-Ramon}}, \bibinfo {author} {\bibfnamefont {O.}~\bibnamefont
				{Iglesias-Freire}}, \bibinfo {author} {\bibfnamefont {A.}~\bibnamefont
				{Fernandez-Pacheco}}, \bibinfo {author} {\bibfnamefont {M.~R.}\ \bibnamefont
				{Ibarra}}, \bibinfo {author} {\bibfnamefont {J.}~\bibnamefont {De~Teresa}},\
			and\ \bibinfo {author} {\bibfnamefont {A.}~\bibnamefont {Asenjo}},\
		}\bibfield  {title} {\bibinfo {title} {Hysteresis loops of individual co
				nanostripes measured by magnetic force microscopy},\ }\href
		{https://doi.org/10.1186/1556-276x-6-407} {\bibfield  {journal} {\bibinfo
				{journal} {Nanoscale Research Letters}\ }\textbf {\bibinfo {volume} {6}},\
			\bibinfo {pages} {407} (\bibinfo {year} {2011})}\BibitemShut {NoStop}%
		\bibitem [{\citenamefont {Fern{\'{a}}ndez-Pacheco}\ \emph
			{et~al.}(2013)\citenamefont {Fern{\'{a}}ndez-Pacheco}, \citenamefont
			{Serrano-Ram{\'{o}}n}, \citenamefont {Michalik}, \citenamefont {Ibarra},
			\citenamefont {Teresa}, \citenamefont {O'Brien}, \citenamefont {Petit},
			\citenamefont {Lee},\ and\ \citenamefont {Cowburn}}]{Fernandez-Pacheco13}%
		\BibitemOpen
		\bibfield  {author} {\bibinfo {author} {\bibfnamefont {A.}~\bibnamefont
				{Fern{\'{a}}ndez-Pacheco}}, \bibinfo {author} {\bibfnamefont
				{L.}~\bibnamefont {Serrano-Ram{\'{o}}n}}, \bibinfo {author} {\bibfnamefont
				{J.~M.}\ \bibnamefont {Michalik}}, \bibinfo {author} {\bibfnamefont {M.~R.}\
				\bibnamefont {Ibarra}}, \bibinfo {author} {\bibfnamefont {J.~M.~D.}\
				\bibnamefont {Teresa}}, \bibinfo {author} {\bibfnamefont {L.}~\bibnamefont
				{O'Brien}}, \bibinfo {author} {\bibfnamefont {D.}~\bibnamefont {Petit}},
			\bibinfo {author} {\bibfnamefont {J.}~\bibnamefont {Lee}},\ and\ \bibinfo
			{author} {\bibfnamefont {R.~P.}\ \bibnamefont {Cowburn}},\ }\bibfield
		{title} {\bibinfo {title} {Three dimensional magnetic nanowires grown by
				focused electron-beam induced deposition},\ }\href
		{https://doi.org/10.1038/srep01492} {\bibfield  {journal} {\bibinfo
				{journal} {Scientific Reports}\ }\textbf {\bibinfo {volume} {3}},\ \bibinfo
			{pages} {1492} (\bibinfo {year} {2013})}\BibitemShut {NoStop}%
		\bibitem [{\citenamefont {Fern\'{a}ndez-Gonz\'{a}lez}\ \emph
			{et~al.}(2025)\citenamefont {Fern\'{a}ndez-Gonz\'{a}lez}, \citenamefont
			{\'{A}lvaro G\'{o}mez}, \citenamefont {Fern\'{a}ndez-Garc\'{\i}a},
			\citenamefont {Morales-Fern\'{a}ndez}, \citenamefont {G\'{o}mez-Cruz},
			\citenamefont {Khaliq}, \citenamefont {Donnelly}, \citenamefont {Foerster},
			\citenamefont {Ni\~{n}o}, \citenamefont {Pereiro}, \citenamefont
			{Mart\'{\i}nez}, \citenamefont {Jaafar}, \citenamefont {Aballe},
			\citenamefont {P\'{e}rez},\ and\ \citenamefont
			{Ruiz-G\'{o}mez}}]{FernandezGonzalez25a}%
		\BibitemOpen
		\bibfield  {author} {\bibinfo {author} {\bibfnamefont {C.}~\bibnamefont
				{Fern\'{a}ndez-Gonz\'{a}lez}}, \bibinfo {author} {\bibfnamefont
				{L.}~\bibnamefont {\'{A}lvaro G\'{o}mez}}, \bibinfo {author} {\bibfnamefont
				{L.}~\bibnamefont {Fern\'{a}ndez-Garc\'{\i}a}}, \bibinfo {author}
			{\bibfnamefont {P.}~\bibnamefont {Morales-Fern\'{a}ndez}}, \bibinfo {author}
			{\bibfnamefont {L.}~\bibnamefont {G\'{o}mez-Cruz}}, \bibinfo {author}
			{\bibfnamefont {M.~W.}\ \bibnamefont {Khaliq}}, \bibinfo {author}
			{\bibfnamefont {C.}~\bibnamefont {Donnelly}}, \bibinfo {author}
			{\bibfnamefont {M.}~\bibnamefont {Foerster}}, \bibinfo {author}
			{\bibfnamefont {M.~A.}\ \bibnamefont {Ni\~{n}o}}, \bibinfo {author}
			{\bibfnamefont {E.}~\bibnamefont {Pereiro}}, \bibinfo {author} {\bibfnamefont
				{E.}~\bibnamefont {Mart\'{\i}nez}}, \bibinfo {author} {\bibfnamefont
				{M.}~\bibnamefont {Jaafar}}, \bibinfo {author} {\bibfnamefont
				{L.}~\bibnamefont {Aballe}}, \bibinfo {author} {\bibfnamefont
				{L.}~\bibnamefont {P\'{e}rez}},\ and\ \bibinfo {author} {\bibfnamefont
				{S.}~\bibnamefont {Ruiz-G\'{o}mez}},\ }\bibfield  {title} {\bibinfo {title}
			{Unveiling the {3D} spin texture of nanowires using integrated microscopy
				techniques},\ }\href {https://doi.org/10.1021/acs.nanolett.5c02395}
		{\bibfield  {journal} {\bibinfo  {journal} {Nano Letters}\ }\textbf {\bibinfo
				{volume} {25}},\ \bibinfo {pages} {10648} (\bibinfo {year}
			{2025})}\BibitemShut {NoStop}%
		\bibitem [{\citenamefont {Casola}\ \emph {et~al.}(2018)\citenamefont {Casola},
			\citenamefont {van~der Sar},\ and\ \citenamefont {Yacoby}}]{Casola18}%
		\BibitemOpen
		\bibfield  {author} {\bibinfo {author} {\bibfnamefont {F.}~\bibnamefont
				{Casola}}, \bibinfo {author} {\bibfnamefont {T.}~\bibnamefont {van~der
					Sar}},\ and\ \bibinfo {author} {\bibfnamefont {A.}~\bibnamefont {Yacoby}},\
		}\bibfield  {title} {\bibinfo {title} {Probing condensed matter physics with
				magnetometry based on nitrogen-vacancy centres in diamond},\ }\href
		{https://doi.org/10.1038/natrevmats.2017.88} {\bibfield  {journal} {\bibinfo
				{journal} {Nature Reviews Materials}\ }\textbf {\bibinfo {volume} {3}},\
			\bibinfo {pages} {17088} (\bibinfo {year} {2018})}\BibitemShut {NoStop}%
		\bibitem [{\citenamefont {Wiesendanger}(2009)}]{Wiesendanger09}%
		\BibitemOpen
		\bibfield  {author} {\bibinfo {author} {\bibfnamefont {R.}~\bibnamefont
				{Wiesendanger}},\ }\bibfield  {title} {\bibinfo {title} {Spin mapping at the
				nanoscale and atomic scale},\ }\href
		{https://doi.org/10.1103/RevModPhys.81.1495} {\bibfield  {journal} {\bibinfo
				{journal} {Reviews of Modern Physics}\ }\textbf {\bibinfo {volume} {81}},\
			\bibinfo {pages} {1495} (\bibinfo {year} {2009})}\BibitemShut {NoStop}%
		\bibitem [{\citenamefont {Sebastian}\ \emph {et~al.}(2015)\citenamefont
			{Sebastian}, \citenamefont {Schultheiss}, \citenamefont {Obry}, \citenamefont
			{Hillebrands},\ and\ \citenamefont {Schultheiss}}]{Sebastian15}%
		\BibitemOpen
		\bibfield  {author} {\bibinfo {author} {\bibfnamefont {T.}~\bibnamefont
				{Sebastian}}, \bibinfo {author} {\bibfnamefont {K.}~\bibnamefont
				{Schultheiss}}, \bibinfo {author} {\bibfnamefont {B.}~\bibnamefont {Obry}},
			\bibinfo {author} {\bibfnamefont {B.}~\bibnamefont {Hillebrands}},\ and\
			\bibinfo {author} {\bibfnamefont {H.}~\bibnamefont {Schultheiss}},\
		}\bibfield  {title} {\bibinfo {title} {Micro-focused {B}rillouin light
				scattering: imaging spin waves at the nanoscale},\ }\bibfield  {journal}
		{\bibinfo  {journal} {Frontiers in Physics}\ }\textbf {\bibinfo {volume}
			{3}},\ \href {https://doi.org/10.3389/fphy.2015.00035}
		{10.3389/fphy.2015.00035} (\bibinfo {year} {2015})\BibitemShut {NoStop}%
		\bibitem [{\citenamefont {Gr\"{u}tter}\ and\ \citenamefont
			{Allenspach}(1994)}]{Grutter94}%
		\BibitemOpen
		\bibfield  {author} {\bibinfo {author} {\bibfnamefont {P.}~\bibnamefont
				{Gr\"{u}tter}}\ and\ \bibinfo {author} {\bibfnamefont {R.}~\bibnamefont
				{Allenspach}},\ }\bibfield  {title} {\bibinfo {title} {Can magnetic-force
				microscopy determine micromagnetic structures?},\ }\href
		{https://doi.org/10.1111/j.1365-246x.1994.tb01812.x} {\bibfield  {journal}
			{\bibinfo  {journal} {Geophysical Journal International}\ }\textbf {\bibinfo
				{volume} {116}},\ \bibinfo {pages} {502} (\bibinfo {year}
			{1994})}\BibitemShut {NoStop}%
		\bibitem [{\citenamefont {Blakely}(1995)}]{Blakely95}%
		\BibitemOpen
		\bibfield  {author} {\bibinfo {author} {\bibfnamefont {R.~J.}\ \bibnamefont
				{Blakely}},\ }\href {https://doi.org/10.1017/cbo9780511549816} {\emph
			{\bibinfo {title} {Potential Theory in Gravity and Magnetic Applications}}}\
		(\bibinfo  {publisher} {Cambridge University Press},\ \bibinfo {year}
		{1995})\BibitemShut {NoStop}%
		\bibitem [{\citenamefont {Xu}\ \emph {et~al.}(2023)\citenamefont {Xu},
			\citenamefont {Zhang},\ and\ \citenamefont {Tian}}]{Xu23b}%
		\BibitemOpen
		\bibfield  {author} {\bibinfo {author} {\bibfnamefont {Y.}~\bibnamefont
				{Xu}}, \bibinfo {author} {\bibfnamefont {W.}~\bibnamefont {Zhang}},\ and\
			\bibinfo {author} {\bibfnamefont {C.}~\bibnamefont {Tian}},\ }\bibfield
		{title} {\bibinfo {title} {Recent advances on applications of {NV$^-$}
				magnetometry in condensed matter physics},\ }\href
		{https://doi.org/10.1364/prj.471266} {\bibfield  {journal} {\bibinfo
				{journal} {Photonics Research}\ }\textbf {\bibinfo {volume} {11}},\ \bibinfo
			{pages} {393} (\bibinfo {year} {2023})}\BibitemShut {NoStop}%
		\bibitem [{\citenamefont {Broadway}\ \emph {et~al.}(2025)\citenamefont
			{Broadway}, \citenamefont {Flaks}, \citenamefont {Dubois},\ and\
			\citenamefont {Maletinsky}}]{Broadway25}%
		\BibitemOpen
		\bibfield  {author} {\bibinfo {author} {\bibfnamefont {D.~A.}\ \bibnamefont
				{Broadway}}, \bibinfo {author} {\bibfnamefont {M.}~\bibnamefont {Flaks}},
			\bibinfo {author} {\bibfnamefont {A.~E.}\ \bibnamefont {Dubois}},\ and\
			\bibinfo {author} {\bibfnamefont {P.}~\bibnamefont {Maletinsky}},\ }\bibfield
		{title} {\bibinfo {title} {Reconstruction of nontrivial magnetization
				textures from magnetic field images using neural networks},\ }\href
		{https://doi.org/10.1103/physrevapplied.23.044012} {\bibfield  {journal}
			{\bibinfo  {journal} {Physical Review Applied}\ }\textbf {\bibinfo {volume}
				{23}},\ \bibinfo {pages} {044012} (\bibinfo {year} {2025})}\BibitemShut
		{NoStop}%
		\bibitem [{\citenamefont {Belova}\ \emph {et~al.}(2012)\citenamefont {Belova},
			\citenamefont {Hellwig}, \citenamefont {Dobisz},\ and\ \citenamefont
			{Dan~Dahlberg}}]{Belova12}%
		\BibitemOpen
		\bibfield  {author} {\bibinfo {author} {\bibfnamefont {L.~M.}\ \bibnamefont
				{Belova}}, \bibinfo {author} {\bibfnamefont {O.}~\bibnamefont {Hellwig}},
			\bibinfo {author} {\bibfnamefont {E.}~\bibnamefont {Dobisz}},\ and\ \bibinfo
			{author} {\bibfnamefont {E.}~\bibnamefont {Dan~Dahlberg}},\ }\bibfield
		{title} {\bibinfo {title} {Rapid preparation of electron beam induced
				deposition co magnetic force microscopy tips with 10 nm spatial resolution},\
		}\href {https://doi.org/10.1063/1.4752225} {\bibfield  {journal} {\bibinfo
				{journal} {Review of Scientific Instruments}\ }\textbf {\bibinfo {volume}
				{83}},\ \bibinfo {pages} {093711} (\bibinfo {year} {2012})}\BibitemShut
		{NoStop}%
		\bibitem [{\citenamefont {Gavagnin}\ \emph {et~al.}(2014)\citenamefont
			{Gavagnin}, \citenamefont {Wanzenboeck}, \citenamefont {Wachter},
			\citenamefont {Shawrav}, \citenamefont {Persson}, \citenamefont {Gunnarsson},
			\citenamefont {Svedlindh}, \citenamefont {St\"{o}ger-Pollach},\ and\
			\citenamefont {Bertagnolli}}]{Gavagnin14}%
		\BibitemOpen
		\bibfield  {author} {\bibinfo {author} {\bibfnamefont {M.}~\bibnamefont
				{Gavagnin}}, \bibinfo {author} {\bibfnamefont {H.~D.}\ \bibnamefont
				{Wanzenboeck}}, \bibinfo {author} {\bibfnamefont {S.}~\bibnamefont
				{Wachter}}, \bibinfo {author} {\bibfnamefont {M.~M.}\ \bibnamefont
				{Shawrav}}, \bibinfo {author} {\bibfnamefont {A.}~\bibnamefont {Persson}},
			\bibinfo {author} {\bibfnamefont {K.}~\bibnamefont {Gunnarsson}}, \bibinfo
			{author} {\bibfnamefont {P.}~\bibnamefont {Svedlindh}}, \bibinfo {author}
			{\bibfnamefont {M.}~\bibnamefont {St\"{o}ger-Pollach}},\ and\ \bibinfo
			{author} {\bibfnamefont {E.}~\bibnamefont {Bertagnolli}},\ }\bibfield
		{title} {\bibinfo {title} {Free-standing magnetic nanopillars for {3D}
				nanomagnet logic},\ }\href {https://doi.org/10.1021/am505785t} {\bibfield
			{journal} {\bibinfo  {journal} {ACS Applied Materials \& Interfaces}\
			}\textbf {\bibinfo {volume} {6}},\ \bibinfo {pages} {20254} (\bibinfo {year}
			{2014})}\BibitemShut {NoStop}%
		\bibitem [{\citenamefont {Sato}\ and\ \citenamefont
			{Ishibashi}(2022)}]{Sato22a}%
		\BibitemOpen
		\bibfield  {author} {\bibinfo {author} {\bibfnamefont {K.}~\bibnamefont
				{Sato}}\ and\ \bibinfo {author} {\bibfnamefont {T.}~\bibnamefont
				{Ishibashi}},\ }\bibfield  {title} {\bibinfo {title} {Fundamentals of
				magneto-optical spectroscopy},\ }\href
		{https://doi.org/10.3389/fphy.2022.946515} {\bibfield  {journal} {\bibinfo
				{journal} {Frontiers in Physics}\ }\textbf {\bibinfo {volume} {10}},\
			\bibinfo {pages} {946515} (\bibinfo {year} {2022})}\BibitemShut {NoStop}%
		\bibitem [{\citenamefont {Soldatov}\ \emph {et~al.}(2021)\citenamefont
			{Soldatov}, \citenamefont {Kolesnikova}, \citenamefont {Rodionova},\ and\
			\citenamefont {Sch\"{a}fer}}]{Soldatov21a}%
		\BibitemOpen
		\bibfield  {author} {\bibinfo {author} {\bibfnamefont {I.}~\bibnamefont
				{Soldatov}}, \bibinfo {author} {\bibfnamefont {V.}~\bibnamefont
				{Kolesnikova}}, \bibinfo {author} {\bibfnamefont {V.}~\bibnamefont
				{Rodionova}},\ and\ \bibinfo {author} {\bibfnamefont {R.}~\bibnamefont
				{Sch\"{a}fer}},\ }\bibfield  {title} {\bibinfo {title} {Interpretation of
				{K}err microscopic domain contrast on curved surfaces},\ }\href
		{https://doi.org/10.1109/lmag.2021.3116791} {\bibfield  {journal} {\bibinfo
				{journal} {IEEE Magnetics Letters}\ }\textbf {\bibinfo {volume} {12}},\
			\bibinfo {pages} {1} (\bibinfo {year} {2021})}\BibitemShut {NoStop}%
		\bibitem [{\citenamefont {Richter}\ \emph {et~al.}(2017)\citenamefont
			{Richter}, \citenamefont {Thiaville},\ and\ \citenamefont
			{Varga}}]{Richter17}%
		\BibitemOpen
		\bibfield  {author} {\bibinfo {author} {\bibfnamefont {K.}~\bibnamefont
				{Richter}}, \bibinfo {author} {\bibfnamefont {A.}~\bibnamefont {Thiaville}},\
			and\ \bibinfo {author} {\bibfnamefont {R.}~\bibnamefont {Varga}},\ }\bibfield
		{title} {\bibinfo {title} {Analytical calculation and observation of the
				magnetic contrast in magneto-optical studies of magnetic cylinders},\ }\href
		{https://doi.org/10.1103/physrevb.96.064421} {\bibfield  {journal} {\bibinfo
				{journal} {Physical Review B}\ }\textbf {\bibinfo {volume} {96}},\ \bibinfo
			{pages} {064421} (\bibinfo {year} {2017})}\BibitemShut {NoStop}%
		\bibitem [{\citenamefont {Streubel}\ \emph {et~al.}(2014)\citenamefont
			{Streubel}, \citenamefont {Lee}, \citenamefont {Makarov}, \citenamefont {Im},
			\citenamefont {Karnaushenko}, \citenamefont {Han}, \citenamefont
			{Sch{\"a}fer}, \citenamefont {Fischer}, \citenamefont {Kim},\ and\
			\citenamefont {Schmidt}}]{Streubel14a}%
		\BibitemOpen
		\bibfield  {author} {\bibinfo {author} {\bibfnamefont {R.}~\bibnamefont
				{Streubel}}, \bibinfo {author} {\bibfnamefont {J.}~\bibnamefont {Lee}},
			\bibinfo {author} {\bibfnamefont {D.}~\bibnamefont {Makarov}}, \bibinfo
			{author} {\bibfnamefont {M.-Y.}\ \bibnamefont {Im}}, \bibinfo {author}
			{\bibfnamefont {D.}~\bibnamefont {Karnaushenko}}, \bibinfo {author}
			{\bibfnamefont {L.}~\bibnamefont {Han}}, \bibinfo {author} {\bibfnamefont
				{R.}~\bibnamefont {Sch{\"a}fer}}, \bibinfo {author} {\bibfnamefont
				{P.}~\bibnamefont {Fischer}}, \bibinfo {author} {\bibfnamefont {S.-K.}\
				\bibnamefont {Kim}},\ and\ \bibinfo {author} {\bibfnamefont {O.~G.}\
				\bibnamefont {Schmidt}},\ }\bibfield  {title} {\bibinfo {title} {Magnetic
				microstructure of rolled-up single-layer ferromagnetic nanomembranes},\
		}\href {https://doi.org/10.1002/adma.201303003} {\bibfield  {journal}
			{\bibinfo  {journal} {Advanced Materials}\ }\textbf {\bibinfo {volume}
				{26}},\ \bibinfo {pages} {316} (\bibinfo {year} {2014})}\BibitemShut
		{NoStop}%
		\bibitem [{\citenamefont {Zhao}\ \emph {et~al.}(2026)\citenamefont {Zhao},
			\citenamefont {Rabensteiner}, \citenamefont {\'{A}ngel Cascales-Sandoval},
			\citenamefont {Leo}, \citenamefont {Koraltan},\ and\ \citenamefont
			{Fern\'{a}ndez-Pacheco}}]{Zhao26}%
		\BibitemOpen
		\bibfield  {author} {\bibinfo {author} {\bibfnamefont {L.}~\bibnamefont
				{Zhao}}, \bibinfo {author} {\bibfnamefont {A.}~\bibnamefont {Rabensteiner}},
			\bibinfo {author} {\bibfnamefont {M.}~\bibnamefont {\'{A}ngel
					Cascales-Sandoval}}, \bibinfo {author} {\bibfnamefont {N.}~\bibnamefont
				{Leo}}, \bibinfo {author} {\bibfnamefont {S.}~\bibnamefont {Koraltan}},\ and\
			\bibinfo {author} {\bibfnamefont {A.}~\bibnamefont {Fern\'{a}ndez-Pacheco}},\
		}\bibfield  {title} {\bibinfo {title} {Extended depth of field
				magneto-optical {K}err microscopy for applications in {3D} nanomagnetism},\
		}\bibfield  {journal} {\bibinfo  {journal} {ArXiv e-prints}\ }\href
		{https://doi.org/10.48550/arXiv.2601.08059} {10.48550/arXiv.2601.08059}
		(\bibinfo {year} {2026})\BibitemShut {NoStop}%
		\bibitem [{\citenamefont {Singh}\ \emph {et~al.}(2022)\citenamefont {Singh},
			\citenamefont {Ot\'{a}lora}, \citenamefont {Kang}, \citenamefont {Soldatov},
			\citenamefont {Karnaushenko}, \citenamefont {Becker}, \citenamefont
			{Sch\"{a}fer}, \citenamefont {Karnaushenko}, \citenamefont {Neu},\ and\
			\citenamefont {Schmidt}}]{Singh22}%
		\BibitemOpen
		\bibfield  {author} {\bibinfo {author} {\bibfnamefont {B.}~\bibnamefont
				{Singh}}, \bibinfo {author} {\bibfnamefont {J.~A.}\ \bibnamefont
				{Ot\'{a}lora}}, \bibinfo {author} {\bibfnamefont {T.~H.}\ \bibnamefont
				{Kang}}, \bibinfo {author} {\bibfnamefont {I.}~\bibnamefont {Soldatov}},
			\bibinfo {author} {\bibfnamefont {D.~D.}\ \bibnamefont {Karnaushenko}},
			\bibinfo {author} {\bibfnamefont {C.}~\bibnamefont {Becker}}, \bibinfo
			{author} {\bibfnamefont {R.}~\bibnamefont {Sch\"{a}fer}}, \bibinfo {author}
			{\bibfnamefont {D.}~\bibnamefont {Karnaushenko}}, \bibinfo {author}
			{\bibfnamefont {V.}~\bibnamefont {Neu}},\ and\ \bibinfo {author}
			{\bibfnamefont {O.~G.}\ \bibnamefont {Schmidt}},\ }\bibfield  {title}
		{\bibinfo {title} {Self-assembly as a tool to study microscale curvature and
				strain-dependent magnetic properties},\ }\href
		{https://doi.org/10.1038/s41528-022-00210-7} {\bibfield  {journal} {\bibinfo
				{journal} {npj Flexible Electronics}\ }\textbf {\bibinfo {volume} {6}},\
			\bibinfo {pages} {76} (\bibinfo {year} {2022})}\BibitemShut {NoStop}%
		\bibitem [{\citenamefont {Vahovsky}\ \emph {et~al.}(2019)\citenamefont
			{Vahovsky}, \citenamefont {Varga},\ and\ \citenamefont
			{Richter}}]{Vahovsky19}%
		\BibitemOpen
		\bibfield  {author} {\bibinfo {author} {\bibfnamefont {O.}~\bibnamefont
				{Vahovsky}}, \bibinfo {author} {\bibfnamefont {R.}~\bibnamefont {Varga}},\
			and\ \bibinfo {author} {\bibfnamefont {K.}~\bibnamefont {Richter}},\
		}\bibfield  {title} {\bibinfo {title} {Experimental method for surface domain
				wall shape studies in thin magnetic cylinders},\ }\href
		{https://doi.org/10.1016/j.jmmm.2019.03.015} {\bibfield  {journal} {\bibinfo
				{journal} {Journal of Magnetism and Magnetic Materials}\ }\textbf {\bibinfo
				{volume} {483}},\ \bibinfo {pages} {266} (\bibinfo {year}
			{2019})}\BibitemShut {NoStop}%
		\bibitem [{\citenamefont {Richter}\ \emph {et~al.}(2022)\citenamefont
			{Richter}, \citenamefont {Thiaville}, \citenamefont {Fecova}, \citenamefont
			{Varga},\ and\ \citenamefont {McCord}}]{Richter22}%
		\BibitemOpen
		\bibfield  {author} {\bibinfo {author} {\bibfnamefont {K.}~\bibnamefont
				{Richter}}, \bibinfo {author} {\bibfnamefont {A.}~\bibnamefont {Thiaville}},
			\bibinfo {author} {\bibfnamefont {L.}~\bibnamefont {Fecova}}, \bibinfo
			{author} {\bibfnamefont {R.}~\bibnamefont {Varga}},\ and\ \bibinfo {author}
			{\bibfnamefont {J.}~\bibnamefont {McCord}},\ }\bibfield  {title} {\bibinfo
			{title} {Estimation of a surface magnetization direction of thin cylinders by
				magnetooptical {K}err effect},\ }\href
		{https://doi.org/10.1016/j.jmmm.2022.169752} {\bibfield  {journal} {\bibinfo
				{journal} {Journal of Magnetism and Magnetic Materials}\ }\textbf {\bibinfo
				{volume} {562}},\ \bibinfo {pages} {169752} (\bibinfo {year}
			{2022})}\BibitemShut {NoStop}%
		\bibitem [{\citenamefont {Ye}\ \emph {et~al.}(2013)\citenamefont {Ye},
			\citenamefont {del Real}, \citenamefont {Infante},\ and\ \citenamefont
			{V\'{a}zquez}}]{Ye13a}%
		\BibitemOpen
		\bibfield  {author} {\bibinfo {author} {\bibfnamefont {J.}~\bibnamefont
				{Ye}}, \bibinfo {author} {\bibfnamefont {R.~P.}\ \bibnamefont {del Real}},
			\bibinfo {author} {\bibfnamefont {G.}~\bibnamefont {Infante}},\ and\ \bibinfo
			{author} {\bibfnamefont {M.}~\bibnamefont {V\'{a}zquez}},\ }\bibfield
		{title} {\bibinfo {title} {Local magnetization profile and geometry
				magnetization effects in microwires as determined by magneto-optical {K}err
				effect},\ }\href {https://doi.org/10.1063/1.4776730} {\bibfield  {journal}
			{\bibinfo  {journal} {Journal of Applied Physics}\ }\textbf {\bibinfo
				{volume} {113}},\ \bibinfo {pages} {043904} (\bibinfo {year}
			{2013})}\BibitemShut {NoStop}%
		\bibitem [{\citenamefont {Streubel}\ \emph {et~al.}(2013)\citenamefont
			{Streubel}, \citenamefont {Makarov}, \citenamefont {Lee}, \citenamefont
			{M{\"u}ller}, \citenamefont {Melzer}, \citenamefont {Sch{\"a}fer},
			\citenamefont {Bufon}, \citenamefont {Kim},\ and\ \citenamefont
			{Schmidt}}]{Streubel13b}%
		\BibitemOpen
		\bibfield  {author} {\bibinfo {author} {\bibfnamefont {R.}~\bibnamefont
				{Streubel}}, \bibinfo {author} {\bibfnamefont {D.}~\bibnamefont {Makarov}},
			\bibinfo {author} {\bibfnamefont {J.}~\bibnamefont {Lee}}, \bibinfo {author}
			{\bibfnamefont {C.}~\bibnamefont {M{\"u}ller}}, \bibinfo {author}
			{\bibfnamefont {M.}~\bibnamefont {Melzer}}, \bibinfo {author} {\bibfnamefont
				{R.}~\bibnamefont {Sch{\"a}fer}}, \bibinfo {author} {\bibfnamefont
				{C.~C.~B.}\ \bibnamefont {Bufon}}, \bibinfo {author} {\bibfnamefont {S.-K.}\
				\bibnamefont {Kim}},\ and\ \bibinfo {author} {\bibfnamefont {O.~G.}\
				\bibnamefont {Schmidt}},\ }\bibfield  {title} {\bibinfo {title} {Rolled-up
				permalloy nanomembranes with multiple windings},\ }\href
		{https://doi.org/10.1142/s2010324713400018} {\bibfield  {journal} {\bibinfo
				{journal} {SPIN}\ }\textbf {\bibinfo {volume} {03}},\ \bibinfo {pages}
			{1340001} (\bibinfo {year} {2013})}\BibitemShut {NoStop}%
		\bibitem [{\citenamefont {Wu}\ \emph {et~al.}(2003)\citenamefont {Wu},
			\citenamefont {Won}, \citenamefont {Zhao},\ and\ \citenamefont {Qiu}}]{Wu03}%
		\BibitemOpen
		\bibfield  {author} {\bibinfo {author} {\bibfnamefont {Y.~Z.}\ \bibnamefont
				{Wu}}, \bibinfo {author} {\bibfnamefont {C.}~\bibnamefont {Won}}, \bibinfo
			{author} {\bibfnamefont {H.~W.}\ \bibnamefont {Zhao}},\ and\ \bibinfo
			{author} {\bibfnamefont {Z.~Q.}\ \bibnamefont {Qiu}},\ }\bibfield  {title}
		{\bibinfo {title} {Surface magneto-optic {K}err effect study of {Co} thin
				films grown on double curved {Cu$(0\,0\,1)$}},\ }\href
		{https://doi.org/10.1103/physrevb.67.094409} {\bibfield  {journal} {\bibinfo
				{journal} {Physical Review B}\ }\textbf {\bibinfo {volume} {67}},\ \bibinfo
			{pages} {094409} (\bibinfo {year} {2003})}\BibitemShut {NoStop}%
		\bibitem [{\citenamefont {Bezsmertna}\ \emph {et~al.}(2024)\citenamefont
			{Bezsmertna}, \citenamefont {Xu}, \citenamefont {Pylypovskyi}, \citenamefont
			{Raftrey}, \citenamefont {Sorrentino}, \citenamefont {Fernandez-Roldan},
			\citenamefont {Soldatov}, \citenamefont {Wolf}, \citenamefont {Lubk},
			\citenamefont {Sch\"{a}fer}, \citenamefont {Fischer},\ and\ \citenamefont
			{Makarov}}]{Bezsmertna24}%
		\BibitemOpen
		\bibfield  {author} {\bibinfo {author} {\bibfnamefont {O.}~\bibnamefont
				{Bezsmertna}}, \bibinfo {author} {\bibfnamefont {R.}~\bibnamefont {Xu}},
			\bibinfo {author} {\bibfnamefont {O.}~\bibnamefont {Pylypovskyi}}, \bibinfo
			{author} {\bibfnamefont {D.}~\bibnamefont {Raftrey}}, \bibinfo {author}
			{\bibfnamefont {A.}~\bibnamefont {Sorrentino}}, \bibinfo {author}
			{\bibfnamefont {J.~A.}\ \bibnamefont {Fernandez-Roldan}}, \bibinfo {author}
			{\bibfnamefont {I.}~\bibnamefont {Soldatov}}, \bibinfo {author}
			{\bibfnamefont {D.}~\bibnamefont {Wolf}}, \bibinfo {author} {\bibfnamefont
				{A.}~\bibnamefont {Lubk}}, \bibinfo {author} {\bibfnamefont {R.}~\bibnamefont
				{Sch\"{a}fer}}, \bibinfo {author} {\bibfnamefont {P.}~\bibnamefont
				{Fischer}},\ and\ \bibinfo {author} {\bibfnamefont {D.}~\bibnamefont
				{Makarov}},\ }\bibfield  {title} {\bibinfo {title} {Magnetic solitons in
				hierarchical {3D} magnetic nanoarchitectures of nanoflower shape},\ }\href
		{https://doi.org/10.1021/acs.nanolett.4c04584} {\bibfield  {journal}
			{\bibinfo  {journal} {Nano Letters}\ }\textbf {\bibinfo {volume} {24}},\
			\bibinfo {pages} {15774} (\bibinfo {year} {2024})}\BibitemShut {NoStop}%
		\bibitem [{\citenamefont {R\"{o}ntgen}(1896)}]{Rontgen96}%
		\BibitemOpen
		\bibfield  {author} {\bibinfo {author} {\bibfnamefont {W.~C.}\ \bibnamefont
				{R\"{o}ntgen}},\ }\bibfield  {title} {\bibinfo {title} {On a new kind of
				rays},\ }\href {https://doi.org/10.1126/science.3.59.227} {\bibfield
			{journal} {\bibinfo  {journal} {Science}\ }\textbf {\bibinfo {volume} {3}},\
			\bibinfo {pages} {227} (\bibinfo {year} {1896})}\BibitemShut {NoStop}%
		\bibitem [{\citenamefont {Als-Nielsen}\ and\ \citenamefont
			{McMorrow}(2011)}]{AlsNielsen11}%
		\BibitemOpen
		\bibfield  {author} {\bibinfo {author} {\bibfnamefont {J.}~\bibnamefont
				{Als-Nielsen}}\ and\ \bibinfo {author} {\bibfnamefont {D.}~\bibnamefont
				{McMorrow}},\ }\bibinfo {title} {Elements of modern {X}‐ray physics}\
		(\bibinfo  {publisher} {Wiley},\ \bibinfo {year} {2011})\ Chap.\ \bibinfo
		{chapter} {{X}‐rays and their interaction with matter}, pp.\ \bibinfo
		{pages} {1--28}\BibitemShut {NoStop}%
		\bibitem [{Jae(2016)}]{Jaeschke16}%
		\BibitemOpen
		\href {https://doi.org/10.1007/978-3-319-14394-1} {\bibinfo {title}
			{Synchrotron light sources and free-electron lasers}} (\bibinfo {year}
		{2016})\BibitemShut {NoStop}%
		\bibitem [{\citenamefont {Thole}\ \emph {et~al.}(1985)\citenamefont {Thole},
			\citenamefont {van~der Laan},\ and\ \citenamefont {Sawatzky}}]{Thole85}%
		\BibitemOpen
		\bibfield  {author} {\bibinfo {author} {\bibfnamefont {B.~T.}\ \bibnamefont
				{Thole}}, \bibinfo {author} {\bibfnamefont {G.}~\bibnamefont {van~der
					Laan}},\ and\ \bibinfo {author} {\bibfnamefont {G.~A.}\ \bibnamefont
				{Sawatzky}},\ }\bibfield  {title} {\bibinfo {title} {Strong magnetic
				dichroism predicted in the {M$_{4,5}$} {X}-ray absorption spectra of magnetic
				rare-earth materials},\ }\href {https://doi.org/10.1103/physrevlett.55.2086}
		{\bibfield  {journal} {\bibinfo  {journal} {Physical Review Letters}\
			}\textbf {\bibinfo {volume} {55}},\ \bibinfo {pages} {2086} (\bibinfo {year}
			{1985})}\BibitemShut {NoStop}%
		\bibitem [{\citenamefont {van~der Laan}\ \emph {et~al.}(1986)\citenamefont
			{van~der Laan}, \citenamefont {Thole}, \citenamefont {Sawatzky},
			\citenamefont {Goedkoop}, \citenamefont {Fuggle}, \citenamefont {Esteva},
			\citenamefont {Karnatak}, \citenamefont {Remeika},\ and\ \citenamefont
			{Dabkowska}}]{vanderLaan86}%
		\BibitemOpen
		\bibfield  {author} {\bibinfo {author} {\bibfnamefont {G.}~\bibnamefont
				{van~der Laan}}, \bibinfo {author} {\bibfnamefont {B.~T.}\ \bibnamefont
				{Thole}}, \bibinfo {author} {\bibfnamefont {G.~A.}\ \bibnamefont {Sawatzky}},
			\bibinfo {author} {\bibfnamefont {J.~B.}\ \bibnamefont {Goedkoop}}, \bibinfo
			{author} {\bibfnamefont {J.~C.}\ \bibnamefont {Fuggle}}, \bibinfo {author}
			{\bibfnamefont {J.-M.}\ \bibnamefont {Esteva}}, \bibinfo {author}
			{\bibfnamefont {R.}~\bibnamefont {Karnatak}}, \bibinfo {author}
			{\bibfnamefont {J.~P.}\ \bibnamefont {Remeika}},\ and\ \bibinfo {author}
			{\bibfnamefont {H.~A.}\ \bibnamefont {Dabkowska}},\ }\bibfield  {title}
		{\bibinfo {title} {Experimental proof of magnetic {X}-ray dichroism},\ }\href
		{https://doi.org/10.1103/physrevb.34.6529} {\bibfield  {journal} {\bibinfo
				{journal} {Physical Review B}\ }\textbf {\bibinfo {volume} {34}},\ \bibinfo
			{pages} {6529} (\bibinfo {year} {1986})}\BibitemShut {NoStop}%
		\bibitem [{\citenamefont {Sch{\"u}tz}\ \emph {et~al.}(1987)\citenamefont
			{Sch{\"u}tz}, \citenamefont {Wagner}, \citenamefont {Wilhelm}, \citenamefont
			{Kienle}, \citenamefont {Zeller}, \citenamefont {Frahm},\ and\ \citenamefont
			{Materlik}}]{Schutz87}%
		\BibitemOpen
		\bibfield  {author} {\bibinfo {author} {\bibfnamefont {G.}~\bibnamefont
				{Sch{\"u}tz}}, \bibinfo {author} {\bibfnamefont {W.}~\bibnamefont {Wagner}},
			\bibinfo {author} {\bibfnamefont {W.}~\bibnamefont {Wilhelm}}, \bibinfo
			{author} {\bibfnamefont {P.}~\bibnamefont {Kienle}}, \bibinfo {author}
			{\bibfnamefont {R.}~\bibnamefont {Zeller}}, \bibinfo {author} {\bibfnamefont
				{R.}~\bibnamefont {Frahm}},\ and\ \bibinfo {author} {\bibfnamefont
				{G.}~\bibnamefont {Materlik}},\ }\bibfield  {title} {\bibinfo {title}
			{Absorption of circularly polarized {X} rays in iron},\ }\href
		{https://doi.org/10.1103/physrevlett.58.737} {\bibfield  {journal} {\bibinfo
				{journal} {Physical Review Letters}\ }\textbf {\bibinfo {volume} {58}},\
			\bibinfo {pages} {737} (\bibinfo {year} {1987})}\BibitemShut {NoStop}%
		\bibitem [{\citenamefont {St\"{o}hr}\ \emph {et~al.}(1998)\citenamefont
			{St\"{o}hr}, \citenamefont {Padmore}, \citenamefont {Anders}, \citenamefont
			{Stammler},\ and\ \citenamefont {Scheinfein}}]{Stohr98}%
		\BibitemOpen
		\bibfield  {author} {\bibinfo {author} {\bibfnamefont {J.}~\bibnamefont
				{St\"{o}hr}}, \bibinfo {author} {\bibfnamefont {H.~A.}\ \bibnamefont
				{Padmore}}, \bibinfo {author} {\bibfnamefont {S.}~\bibnamefont {Anders}},
			\bibinfo {author} {\bibfnamefont {T.}~\bibnamefont {Stammler}},\ and\
			\bibinfo {author} {\bibfnamefont {M.~R.}\ \bibnamefont {Scheinfein}},\
		}\bibfield  {title} {\bibinfo {title} {Principles of {X}-ray magnetic
				dichroism spectromicroscopy},\ }\href
		{https://doi.org/10.1142/s0218625x98001638} {\bibfield  {journal} {\bibinfo
				{journal} {Surface Review and Letters}\ }\textbf {\bibinfo {volume} {05}},\
			\bibinfo {pages} {1297} (\bibinfo {year} {1998})}\BibitemShut {NoStop}%
		\bibitem [{\citenamefont {van~der Laan}\ and\ \citenamefont
			{Arenholz}(2009)}]{vanderLaan09}%
		\BibitemOpen
		\bibfield  {author} {\bibinfo {author} {\bibfnamefont {G.}~\bibnamefont
				{van~der Laan}}\ and\ \bibinfo {author} {\bibfnamefont {E.}~\bibnamefont
				{Arenholz}},\ }\bibfield  {title} {\bibinfo {title} {Anisotropic {X}-ray
				magnetic linear dichroism},\ }\href
		{https://doi.org/10.1140/epjst/e2009-00991-x} {\bibfield  {journal} {\bibinfo
				{journal} {The European Physical Journal Special Topics}\ }\textbf {\bibinfo
				{volume} {169}},\ \bibinfo {pages} {187} (\bibinfo {year}
			{2009})}\BibitemShut {NoStop}%
		\bibitem [{\citenamefont {Kuiper}\ \emph {et~al.}(1993)\citenamefont {Kuiper},
			\citenamefont {Searle}, \citenamefont {Rudolf}, \citenamefont {Tjeng},\ and\
			\citenamefont {Chen}}]{Kuiper93}%
		\BibitemOpen
		\bibfield  {author} {\bibinfo {author} {\bibfnamefont {P.}~\bibnamefont
				{Kuiper}}, \bibinfo {author} {\bibfnamefont {B.~G.}\ \bibnamefont {Searle}},
			\bibinfo {author} {\bibfnamefont {P.}~\bibnamefont {Rudolf}}, \bibinfo
			{author} {\bibfnamefont {L.~H.}\ \bibnamefont {Tjeng}},\ and\ \bibinfo
			{author} {\bibfnamefont {C.~T.}\ \bibnamefont {Chen}},\ }\bibfield  {title}
		{\bibinfo {title} {X-ray magnetic dichroism of antiferromagnet {Fe$_2$O$_3$}:
				{T}he orientation of magnetic moments observed by {Fe} $2p$ {X}-ray
				absorption spectroscopy},\ }\href
		{https://doi.org/10.1103/physrevlett.70.1549} {\bibfield  {journal} {\bibinfo
				{journal} {Physical Review Letters}\ }\textbf {\bibinfo {volume} {70}},\
			\bibinfo {pages} {1549} (\bibinfo {year} {1993})}\BibitemShut {NoStop}%
		\bibitem [{\citenamefont {Laboratory}()}]{xray_data_booklet}%
		\BibitemOpen
		\bibfield  {author} {\bibinfo {author} {\bibfnamefont {L.~B.~N.}\
				\bibnamefont {Laboratory}},\ }\href {https://xdb.lbl.gov/} {\bibinfo {title}
			{{X}-ray data booklet. {C}enter for {X}-ray optics and advanced light
				source}},\ \bibinfo {note} {access: 2026-03-17}\BibitemShut {NoStop}%
		\bibitem [{\citenamefont {Thole}\ and\ \citenamefont {van~der
				Laan}(1988)}]{Thole88}%
		\BibitemOpen
		\bibfield  {author} {\bibinfo {author} {\bibfnamefont {B.~T.}\ \bibnamefont
				{Thole}}\ and\ \bibinfo {author} {\bibfnamefont {G.}~\bibnamefont {van~der
					Laan}},\ }\bibfield  {title} {\bibinfo {title} {Branching ratio in {X}-ray
				absorption spectroscopy},\ }\href {https://doi.org/10.1103/physrevb.38.3158}
		{\bibfield  {journal} {\bibinfo  {journal} {Physical Review B}\ }\textbf
			{\bibinfo {volume} {38}},\ \bibinfo {pages} {3158} (\bibinfo {year}
			{1988})}\BibitemShut {NoStop}%
		\bibitem [{\citenamefont {Thole}\ \emph {et~al.}(1992)\citenamefont {Thole},
			\citenamefont {Carra}, \citenamefont {Sette},\ and\ \citenamefont {van~der
				Laan}}]{Thole92}%
		\BibitemOpen
		\bibfield  {author} {\bibinfo {author} {\bibfnamefont {B.~T.}\ \bibnamefont
				{Thole}}, \bibinfo {author} {\bibfnamefont {P.}~\bibnamefont {Carra}},
			\bibinfo {author} {\bibfnamefont {F.}~\bibnamefont {Sette}},\ and\ \bibinfo
			{author} {\bibfnamefont {G.}~\bibnamefont {van~der Laan}},\ }\bibfield
		{title} {\bibinfo {title} {X-ray circular dichroism as a probe of orbital
				magnetization},\ }\href {https://doi.org/10.1103/physrevlett.68.1943}
		{\bibfield  {journal} {\bibinfo  {journal} {Physical Review Letters}\
			}\textbf {\bibinfo {volume} {68}},\ \bibinfo {pages} {1943} (\bibinfo {year}
			{1992})}\BibitemShut {NoStop}%
		\bibitem [{\citenamefont {Locatelli}\ \emph {et~al.}(2025)\citenamefont
			{Locatelli}, \citenamefont {Mentes}, \citenamefont {Brondin}, \citenamefont
			{Cojocariu},\ and\ \citenamefont {Jugovac}}]{Locatelli25}%
		\BibitemOpen
		\bibfield  {author} {\bibinfo {author} {\bibfnamefont {A.}~\bibnamefont
				{Locatelli}}, \bibinfo {author} {\bibfnamefont {T.~O.}\ \bibnamefont
				{Mentes}}, \bibinfo {author} {\bibfnamefont {C.~A.}\ \bibnamefont {Brondin}},
			\bibinfo {author} {\bibfnamefont {I.}~\bibnamefont {Cojocariu}},\ and\
			\bibinfo {author} {\bibfnamefont {M.}~\bibnamefont {Jugovac}},\ }\bibfield
		{title} {\bibinfo {title} {Advancements in {X}-ray photoemission electron
				microscopy applied to thin film magnetism, {2D} materials and molecular
				interfaces},\ }\href {https://doi.org/10.1080/23746149.2025.2549757}
		{\bibfield  {journal} {\bibinfo  {journal} {Advances in Physics: X}\ }\textbf
			{\bibinfo {volume} {10}},\ \bibinfo {pages} {2549757} (\bibinfo {year}
			{2025})}\BibitemShut {NoStop}%
		\bibitem [{\citenamefont {Stohr}\ and\ \citenamefont {Anders}(2000)}]{Stohr00}%
		\BibitemOpen
		\bibfield  {author} {\bibinfo {author} {\bibfnamefont {J.}~\bibnamefont
				{Stohr}}\ and\ \bibinfo {author} {\bibfnamefont {S.}~\bibnamefont {Anders}},\
		}\bibfield  {title} {\bibinfo {title} {X-ray spectro-microscopy of complex
				materials and surfaces},\ }\href {https://doi.org/10.1147/rd.444.0535}
		{\bibfield  {journal} {\bibinfo  {journal} {IBM Journal of Research and
					Development}\ }\textbf {\bibinfo {volume} {44}},\ \bibinfo {pages} {535}
			(\bibinfo {year} {2000})}\BibitemShut {NoStop}%
		\bibitem [{\citenamefont {Kimling}\ \emph {et~al.}(2011)\citenamefont
			{Kimling}, \citenamefont {Kronast}, \citenamefont {Martens}, \citenamefont
			{B{\"o}hnert}, \citenamefont {Martens}, \citenamefont {Herrero-Albillos},
			\citenamefont {Tati-Bismaths}, \citenamefont {Merkt}, \citenamefont
			{Nielsch},\ and\ \citenamefont {Meier}}]{Kimling11}%
		\BibitemOpen
		\bibfield  {author} {\bibinfo {author} {\bibfnamefont {J.}~\bibnamefont
				{Kimling}}, \bibinfo {author} {\bibfnamefont {F.}~\bibnamefont {Kronast}},
			\bibinfo {author} {\bibfnamefont {S.}~\bibnamefont {Martens}}, \bibinfo
			{author} {\bibfnamefont {T.}~\bibnamefont {B{\"o}hnert}}, \bibinfo {author}
			{\bibfnamefont {M.}~\bibnamefont {Martens}}, \bibinfo {author} {\bibfnamefont
				{J.}~\bibnamefont {Herrero-Albillos}}, \bibinfo {author} {\bibfnamefont
				{L.}~\bibnamefont {Tati-Bismaths}}, \bibinfo {author} {\bibfnamefont
				{U.}~\bibnamefont {Merkt}}, \bibinfo {author} {\bibfnamefont
				{K.}~\bibnamefont {Nielsch}},\ and\ \bibinfo {author} {\bibfnamefont
				{G.}~\bibnamefont {Meier}},\ }\bibfield  {title} {\bibinfo {title}
			{Photoemission electron microscopy of three-dimensional magnetization
				configurations in core-shell nanostructures},\ }\href
		{https://doi.org/10.1103/physrevb.84.174406} {\bibfield  {journal} {\bibinfo
				{journal} {Physical Review B}\ }\textbf {\bibinfo {volume} {84}},\ \bibinfo
			{pages} {174406} (\bibinfo {year} {2011})}\BibitemShut {NoStop}%
		\bibitem [{\citenamefont {Da~Col}\ \emph {et~al.}(2014)\citenamefont {Da~Col},
			\citenamefont {Jamet}, \citenamefont {Rougemaille}, \citenamefont
			{Locatelli}, \citenamefont {Mentes}, \citenamefont {Burgos}, \citenamefont
			{Afid}, \citenamefont {Darques}, \citenamefont {Cagnon}, \citenamefont
			{Toussaint},\ and\ \citenamefont {Fruchart}}]{DaCol14}%
		\BibitemOpen
		\bibfield  {author} {\bibinfo {author} {\bibfnamefont {S.}~\bibnamefont
				{Da~Col}}, \bibinfo {author} {\bibfnamefont {S.}~\bibnamefont {Jamet}},
			\bibinfo {author} {\bibfnamefont {N.}~\bibnamefont {Rougemaille}}, \bibinfo
			{author} {\bibfnamefont {A.}~\bibnamefont {Locatelli}}, \bibinfo {author}
			{\bibfnamefont {T.~O.}\ \bibnamefont {Mentes}}, \bibinfo {author}
			{\bibfnamefont {B.~S.}\ \bibnamefont {Burgos}}, \bibinfo {author}
			{\bibfnamefont {R.}~\bibnamefont {Afid}}, \bibinfo {author} {\bibfnamefont
				{M.}~\bibnamefont {Darques}}, \bibinfo {author} {\bibfnamefont
				{L.}~\bibnamefont {Cagnon}}, \bibinfo {author} {\bibfnamefont {J.~C.}\
				\bibnamefont {Toussaint}},\ and\ \bibinfo {author} {\bibfnamefont
				{O.}~\bibnamefont {Fruchart}},\ }\bibfield  {title} {\bibinfo {title}
			{Observation of {B}loch-point domain walls in cylindrical magnetic
				nanowires},\ }\href {https://doi.org/10.1103/physrevb.89.180405} {\bibfield
			{journal} {\bibinfo  {journal} {Physical Review B}\ }\textbf {\bibinfo
				{volume} {89}},\ \bibinfo {pages} {180405(R)} (\bibinfo {year}
			{2014})}\BibitemShut {NoStop}%
		\bibitem [{\citenamefont {Jamet}\ \emph {et~al.}(2015)\citenamefont {Jamet},
			\citenamefont {Col}, \citenamefont {Rougemaille}, \citenamefont {Wartelle},
			\citenamefont {Locatelli}, \citenamefont {Mente{\c{s}}}, \citenamefont
			{Burgos}, \citenamefont {Afid}, \citenamefont {Cagnon}, \citenamefont
			{Bochmann}, \citenamefont {Bachmann}, \citenamefont {Fruchart},\ and\
			\citenamefont {Toussaint}}]{Jamet15a}%
		\BibitemOpen
		\bibfield  {author} {\bibinfo {author} {\bibfnamefont {S.}~\bibnamefont
				{Jamet}}, \bibinfo {author} {\bibfnamefont {S.~D.}\ \bibnamefont {Col}},
			\bibinfo {author} {\bibfnamefont {N.}~\bibnamefont {Rougemaille}}, \bibinfo
			{author} {\bibfnamefont {A.}~\bibnamefont {Wartelle}}, \bibinfo {author}
			{\bibfnamefont {A.}~\bibnamefont {Locatelli}}, \bibinfo {author}
			{\bibfnamefont {T.~O.}\ \bibnamefont {Mente{\c{s}}}}, \bibinfo {author}
			{\bibfnamefont {B.~S.}\ \bibnamefont {Burgos}}, \bibinfo {author}
			{\bibfnamefont {R.}~\bibnamefont {Afid}}, \bibinfo {author} {\bibfnamefont
				{L.}~\bibnamefont {Cagnon}}, \bibinfo {author} {\bibfnamefont
				{S.}~\bibnamefont {Bochmann}}, \bibinfo {author} {\bibfnamefont
				{J.}~\bibnamefont {Bachmann}}, \bibinfo {author} {\bibfnamefont
				{O.}~\bibnamefont {Fruchart}},\ and\ \bibinfo {author} {\bibfnamefont
				{J.~C.}\ \bibnamefont {Toussaint}},\ }\bibfield  {title} {\bibinfo {title}
			{Quantitative analysis of shadow {X}-ray magnetic circular dichroism
				photoemission electron microscopy},\ }\href
		{https://doi.org/10.1103/physrevb.92.144428} {\bibfield  {journal} {\bibinfo
				{journal} {Physical Review B}\ }\textbf {\bibinfo {volume} {92}},\ \bibinfo
			{pages} {144428} (\bibinfo {year} {2015})}\BibitemShut {NoStop}%
		\bibitem [{\citenamefont {Wyss}\ \emph {et~al.}(2017)\citenamefont {Wyss},
			\citenamefont {Mehlin}, \citenamefont {Gross}, \citenamefont {Buchter},
			\citenamefont {Farhan}, \citenamefont {Buzzi}, \citenamefont {Kleibert},
			\citenamefont {T\"{u}t\"{u}nc\"{u}oglu}, \citenamefont {Heimbach},
			\citenamefont {i~Morral}, \citenamefont {Grundler},\ and\ \citenamefont
			{Poggio}}]{Wyss17}%
		\BibitemOpen
		\bibfield  {author} {\bibinfo {author} {\bibfnamefont {M.}~\bibnamefont
				{Wyss}}, \bibinfo {author} {\bibfnamefont {A.}~\bibnamefont {Mehlin}},
			\bibinfo {author} {\bibfnamefont {B.}~\bibnamefont {Gross}}, \bibinfo
			{author} {\bibfnamefont {A.}~\bibnamefont {Buchter}}, \bibinfo {author}
			{\bibfnamefont {A.}~\bibnamefont {Farhan}}, \bibinfo {author} {\bibfnamefont
				{M.}~\bibnamefont {Buzzi}}, \bibinfo {author} {\bibfnamefont
				{A.}~\bibnamefont {Kleibert}}, \bibinfo {author} {\bibfnamefont
				{G.}~\bibnamefont {T\"{u}t\"{u}nc\"{u}oglu}}, \bibinfo {author}
			{\bibfnamefont {F.}~\bibnamefont {Heimbach}}, \bibinfo {author}
			{\bibfnamefont {A.~F.}\ \bibnamefont {i~Morral}}, \bibinfo {author}
			{\bibfnamefont {D.}~\bibnamefont {Grundler}},\ and\ \bibinfo {author}
			{\bibfnamefont {M.}~\bibnamefont {Poggio}},\ }\bibfield  {title} {\bibinfo
			{title} {Imaging magnetic vortex configurations in ferromagnetic nanotubes},\
		}\href {https://doi.org/10.1103/physrevb.96.024423} {\bibfield  {journal}
			{\bibinfo  {journal} {Physical Review B}\ }\textbf {\bibinfo {volume} {96}},\
			\bibinfo {pages} {024423} (\bibinfo {year} {2017})}\BibitemShut {NoStop}%
		\bibitem [{\citenamefont {Streubel}\ \emph {et~al.}(2012)\citenamefont
			{Streubel}, \citenamefont {Thurmer}, \citenamefont {Makarov}, \citenamefont
			{Kronast}, \citenamefont {Kosub}, \citenamefont {Kravchuk}, \citenamefont
			{Sheka}, \citenamefont {Gaididei}, \citenamefont {Sch{\"a}fer},\ and\
			\citenamefont {Schmidt}}]{Streubel12b}%
		\BibitemOpen
		\bibfield  {author} {\bibinfo {author} {\bibfnamefont {R.}~\bibnamefont
				{Streubel}}, \bibinfo {author} {\bibfnamefont {D.~J.}\ \bibnamefont
				{Thurmer}}, \bibinfo {author} {\bibfnamefont {D.}~\bibnamefont {Makarov}},
			\bibinfo {author} {\bibfnamefont {F.}~\bibnamefont {Kronast}}, \bibinfo
			{author} {\bibfnamefont {T.}~\bibnamefont {Kosub}}, \bibinfo {author}
			{\bibfnamefont {V.}~\bibnamefont {Kravchuk}}, \bibinfo {author}
			{\bibfnamefont {D.~D.}\ \bibnamefont {Sheka}}, \bibinfo {author}
			{\bibfnamefont {Y.}~\bibnamefont {Gaididei}}, \bibinfo {author}
			{\bibfnamefont {R.}~\bibnamefont {Sch{\"a}fer}},\ and\ \bibinfo {author}
			{\bibfnamefont {O.~G.}\ \bibnamefont {Schmidt}},\ }\bibfield  {title}
		{\bibinfo {title} {Magnetically capped rolled-up nanomembranes},\ }\href
		{https://doi.org/10.1021/nl301147h} {\bibfield  {journal} {\bibinfo
				{journal} {Nano Letters}\ }\textbf {\bibinfo {volume} {12}},\ \bibinfo
			{pages} {3961} (\bibinfo {year} {2012})}\BibitemShut {NoStop}%
		\bibitem [{\citenamefont {Fischer}\ \emph {et~al.}(2001)\citenamefont
			{Fischer}, \citenamefont {Eim{\"u}ller}, \citenamefont {Sch{\"u}tz},
			\citenamefont {Denbeaux}, \citenamefont {Pearson}, \citenamefont {Johnson},
			\citenamefont {Attwood}, \citenamefont {Tsunashima}, \citenamefont
			{Kumazawa}, \citenamefont {Takagi}, \citenamefont {K{\"o}hler},\ and\
			\citenamefont {Bayreuther}}]{Fischer01}%
		\BibitemOpen
		\bibfield  {author} {\bibinfo {author} {\bibfnamefont {P.}~\bibnamefont
				{Fischer}}, \bibinfo {author} {\bibfnamefont {T.}~\bibnamefont
				{Eim{\"u}ller}}, \bibinfo {author} {\bibfnamefont {G.}~\bibnamefont
				{Sch{\"u}tz}}, \bibinfo {author} {\bibfnamefont {G.}~\bibnamefont
				{Denbeaux}}, \bibinfo {author} {\bibfnamefont {A.}~\bibnamefont {Pearson}},
			\bibinfo {author} {\bibfnamefont {L.}~\bibnamefont {Johnson}}, \bibinfo
			{author} {\bibfnamefont {D.}~\bibnamefont {Attwood}}, \bibinfo {author}
			{\bibfnamefont {S.}~\bibnamefont {Tsunashima}}, \bibinfo {author}
			{\bibfnamefont {M.}~\bibnamefont {Kumazawa}}, \bibinfo {author}
			{\bibfnamefont {N.}~\bibnamefont {Takagi}}, \bibinfo {author} {\bibfnamefont
				{M.}~\bibnamefont {K{\"o}hler}},\ and\ \bibinfo {author} {\bibfnamefont
				{G.}~\bibnamefont {Bayreuther}},\ }\bibfield  {title} {\bibinfo {title}
			{Element-specific imaging of magnetic domains at 25~{nm} spatial resolution
				using soft {X}-ray microscopy},\ }\href {https://doi.org/10.1063/1.1351840}
		{\bibfield  {journal} {\bibinfo  {journal} {Review of Scientific
					Instruments}\ }\textbf {\bibinfo {volume} {72}},\ \bibinfo {pages} {2322}
			(\bibinfo {year} {2001})}\BibitemShut {NoStop}%
		\bibitem [{\citenamefont {Chao}\ \emph {et~al.}(2012)\citenamefont {Chao},
			\citenamefont {Fischer}, \citenamefont {Tyliszczak}, \citenamefont {Rekawa},
			\citenamefont {Anderson},\ and\ \citenamefont {Naulleau}}]{Chao12}%
		\BibitemOpen
		\bibfield  {author} {\bibinfo {author} {\bibfnamefont {W.}~\bibnamefont
				{Chao}}, \bibinfo {author} {\bibfnamefont {P.}~\bibnamefont {Fischer}},
			\bibinfo {author} {\bibfnamefont {T.}~\bibnamefont {Tyliszczak}}, \bibinfo
			{author} {\bibfnamefont {S.}~\bibnamefont {Rekawa}}, \bibinfo {author}
			{\bibfnamefont {E.}~\bibnamefont {Anderson}},\ and\ \bibinfo {author}
			{\bibfnamefont {P.}~\bibnamefont {Naulleau}},\ }\bibfield  {title} {\bibinfo
			{title} {Real space soft {X}-ray imaging at 10\,nm spatial resolution},\
		}\href {https://doi.org/10.1364/oe.20.009777} {\bibfield  {journal} {\bibinfo
				{journal} {Optics Express}\ }\textbf {\bibinfo {volume} {20}},\ \bibinfo
			{pages} {9777} (\bibinfo {year} {2012})}\BibitemShut {NoStop}%
		\bibitem [{\citenamefont {Finizio}\ \emph {et~al.}(2022)\citenamefont
			{Finizio}, \citenamefont {Donnelly}, \citenamefont {Mayr}, \citenamefont
			{Hrabec},\ and\ \citenamefont {Raabe}}]{Finizio22}%
		\BibitemOpen
		\bibfield  {author} {\bibinfo {author} {\bibfnamefont {S.}~\bibnamefont
				{Finizio}}, \bibinfo {author} {\bibfnamefont {C.}~\bibnamefont {Donnelly}},
			\bibinfo {author} {\bibfnamefont {S.}~\bibnamefont {Mayr}}, \bibinfo {author}
			{\bibfnamefont {A.}~\bibnamefont {Hrabec}},\ and\ \bibinfo {author}
			{\bibfnamefont {J.}~\bibnamefont {Raabe}},\ }\bibfield  {title} {\bibinfo
			{title} {Three-dimensional vortex gyration dynamics unraveled by
				time-resolved soft x-ray laminography with freely selectable excitation
				frequencies},\ }\href {https://doi.org/10.1021/acs.nanolett.1c04662}
		{\bibfield  {journal} {\bibinfo  {journal} {Nano Letters}\ }\textbf {\bibinfo
				{volume} {22}},\ \bibinfo {pages} {1971} (\bibinfo {year}
			{2022})}\BibitemShut {NoStop}%
		\bibitem [{\citenamefont {Chushkin}\ and\ \citenamefont
			{Zontone}(2025)}]{Chushkin25}%
		\BibitemOpen
		\bibfield  {author} {\bibinfo {author} {\bibfnamefont {Y.}~\bibnamefont
				{Chushkin}}\ and\ \bibinfo {author} {\bibfnamefont {F.}~\bibnamefont
				{Zontone}},\ }\bibfield  {title} {\bibinfo {title} {Prospects for coherent
				{X}-ray diffraction imaging at fourth-generation synchrotron sources},\
		}\href {https://doi.org/10.1107/s2052252525001526} {\bibfield  {journal}
			{\bibinfo  {journal} {IUCrJ}\ }\textbf {\bibinfo {volume} {12}},\ \bibinfo
			{pages} {280} (\bibinfo {year} {2025})}\BibitemShut {NoStop}%
		\bibitem [{\citenamefont {Gabor}(1948)}]{Gabor48}%
		\BibitemOpen
		\bibfield  {author} {\bibinfo {author} {\bibfnamefont {D.}~\bibnamefont
				{Gabor}},\ }\bibfield  {title} {\bibinfo {title} {A new microscopic
				principle},\ }\href {https://doi.org/10.1038/161777a0} {\bibfield  {journal}
			{\bibinfo  {journal} {Nature}\ }\textbf {\bibinfo {volume} {161}},\ \bibinfo
			{pages} {777} (\bibinfo {year} {1948})}\BibitemShut {NoStop}%
		\bibitem [{\citenamefont {Guizar-Sicairos}\ and\ \citenamefont
			{Fienup}(2008)}]{GuizarSicairos08}%
		\BibitemOpen
		\bibfield  {author} {\bibinfo {author} {\bibfnamefont {M.}~\bibnamefont
				{Guizar-Sicairos}}\ and\ \bibinfo {author} {\bibfnamefont {J.~R.}\
				\bibnamefont {Fienup}},\ }\bibfield  {title} {\bibinfo {title} {Direct image
				reconstruction from a fourier intensity pattern using {HERALDO}},\ }\href
		{https://doi.org/10.1364/ol.33.002668} {\bibfield  {journal} {\bibinfo
				{journal} {Optics Letters}\ }\textbf {\bibinfo {volume} {33}},\ \bibinfo
			{pages} {2668} (\bibinfo {year} {2008})}\BibitemShut {NoStop}%
		\bibitem [{\citenamefont {Pfeiffer}(2018)}]{Pfeiffer18}%
		\BibitemOpen
		\bibfield  {author} {\bibinfo {author} {\bibfnamefont {F.}~\bibnamefont
				{Pfeiffer}},\ }\bibfield  {title} {\bibinfo {title} {X-ray ptychography},\
		}\href {https://doi.org/10.1038/s41566-017-0072-5} {\bibfield  {journal}
			{\bibinfo  {journal} {Nature Photonics}\ }\textbf {\bibinfo {volume} {12}},\
			\bibinfo {pages} {9} (\bibinfo {year} {2018})}\BibitemShut {NoStop}%
		\bibitem [{\citenamefont {Favre-Nicolin}\ \emph {et~al.}(2020)\citenamefont
			{Favre-Nicolin}, \citenamefont {Girard}, \citenamefont {Leake}, \citenamefont
			{Carnis}, \citenamefont {Chushkin}, \citenamefont {Kieffer}, \citenamefont
			{Paleo},\ and\ \citenamefont {Richard}}]{FavreNicolin20}%
		\BibitemOpen
		\bibfield  {author} {\bibinfo {author} {\bibfnamefont {V.}~\bibnamefont
				{Favre-Nicolin}}, \bibinfo {author} {\bibfnamefont {G.}~\bibnamefont
				{Girard}}, \bibinfo {author} {\bibfnamefont {S.}~\bibnamefont {Leake}},
			\bibinfo {author} {\bibfnamefont {J.}~\bibnamefont {Carnis}}, \bibinfo
			{author} {\bibfnamefont {Y.}~\bibnamefont {Chushkin}}, \bibinfo {author}
			{\bibfnamefont {J.}~\bibnamefont {Kieffer}}, \bibinfo {author} {\bibfnamefont
				{P.}~\bibnamefont {Paleo}},\ and\ \bibinfo {author} {\bibfnamefont {M.-I.}\
				\bibnamefont {Richard}},\ }\bibfield  {title} {\bibinfo {title}
			{\textit{PyNX}: high-performance computing toolkit for coherent {X}-ray
				imaging based on operators},\ }\href
		{https://doi.org/10.1107/s1600576720010985} {\bibfield  {journal} {\bibinfo
				{journal} {Journal of Applied Crystallography}\ }\textbf {\bibinfo {volume}
				{53}},\ \bibinfo {pages} {1404} (\bibinfo {year} {2020})}\BibitemShut
		{NoStop}%
		\bibitem [{\citenamefont {Neethirajan}\ \emph {et~al.}(2024)\citenamefont
			{Neethirajan}, \citenamefont {Daurer}, \citenamefont {Mart\'{\i}nez},
			\citenamefont {Hrabec}, \citenamefont {Turnbull}, \citenamefont {Yamamoto},
			\citenamefont {Ferreira}, \citenamefont {\v{S}tefan\v{c}i\v{c}},
			\citenamefont {Mayoh}, \citenamefont {Balakrishnan}, \citenamefont {Pei},
			\citenamefont {Xue}, \citenamefont {Chang}, \citenamefont {Ringe},
			\citenamefont {Harrison}, \citenamefont {Valencia}, \citenamefont {Kazemian},
			\citenamefont {Kaulich},\ and\ \citenamefont {Donnelly}}]{Neethirajan24}%
		\BibitemOpen
		\bibfield  {author} {\bibinfo {author} {\bibfnamefont {J.}~\bibnamefont
				{Neethirajan}}, \bibinfo {author} {\bibfnamefont {B.~J.}\ \bibnamefont
				{Daurer}}, \bibinfo {author} {\bibfnamefont {M.~D.~P.}\ \bibnamefont
				{Mart\'{\i}nez}}, \bibinfo {author} {\bibfnamefont {A.}~\bibnamefont
				{Hrabec}}, \bibinfo {author} {\bibfnamefont {L.}~\bibnamefont {Turnbull}},
			\bibinfo {author} {\bibfnamefont {R.}~\bibnamefont {Yamamoto}}, \bibinfo
			{author} {\bibfnamefont {M.~R.}\ \bibnamefont {Ferreira}}, \bibinfo {author}
			{\bibfnamefont {A.}~\bibnamefont {\v{S}tefan\v{c}i\v{c}}}, \bibinfo {author}
			{\bibfnamefont {D.~A.}\ \bibnamefont {Mayoh}}, \bibinfo {author}
			{\bibfnamefont {G.}~\bibnamefont {Balakrishnan}}, \bibinfo {author}
			{\bibfnamefont {Z.}~\bibnamefont {Pei}}, \bibinfo {author} {\bibfnamefont
				{P.}~\bibnamefont {Xue}}, \bibinfo {author} {\bibfnamefont {L.}~\bibnamefont
				{Chang}}, \bibinfo {author} {\bibfnamefont {E.}~\bibnamefont {Ringe}},
			\bibinfo {author} {\bibfnamefont {R.}~\bibnamefont {Harrison}}, \bibinfo
			{author} {\bibfnamefont {S.}~\bibnamefont {Valencia}}, \bibinfo {author}
			{\bibfnamefont {M.}~\bibnamefont {Kazemian}}, \bibinfo {author}
			{\bibfnamefont {B.}~\bibnamefont {Kaulich}},\ and\ \bibinfo {author}
			{\bibfnamefont {C.}~\bibnamefont {Donnelly}},\ }\bibfield  {title} {\bibinfo
			{title} {Soft {X}-ray phase nanomicroscopy of micrometer-thick magnets},\
		}\href {https://doi.org/10.1103/physrevx.14.031028} {\bibfield  {journal}
			{\bibinfo  {journal} {Physical Review X}\ }\textbf {\bibinfo {volume} {14}},\
			\bibinfo {pages} {031028} (\bibinfo {year} {2024})}\BibitemShut {NoStop}%
		\bibitem [{\citenamefont {Donnelly}\ \emph {et~al.}(2017)\citenamefont
			{Donnelly}, \citenamefont {Guizar-Sicairos}, \citenamefont {Scagnoli},
			\citenamefont {Gliga}, \citenamefont {Holler}, \citenamefont {Raabe},\ and\
			\citenamefont {Heyderman}}]{Donnelly17}%
		\BibitemOpen
		\bibfield  {author} {\bibinfo {author} {\bibfnamefont {C.}~\bibnamefont
				{Donnelly}}, \bibinfo {author} {\bibfnamefont {M.}~\bibnamefont
				{Guizar-Sicairos}}, \bibinfo {author} {\bibfnamefont {V.}~\bibnamefont
				{Scagnoli}}, \bibinfo {author} {\bibfnamefont {S.}~\bibnamefont {Gliga}},
			\bibinfo {author} {\bibfnamefont {M.}~\bibnamefont {Holler}}, \bibinfo
			{author} {\bibfnamefont {J.}~\bibnamefont {Raabe}},\ and\ \bibinfo {author}
			{\bibfnamefont {L.~J.}\ \bibnamefont {Heyderman}},\ }\bibfield  {title}
		{\bibinfo {title} {Three-dimensional magnetization structures revealed with
				{X}-ray vector nanotomography},\ }\href {https://doi.org/10.1038/nature23006}
		{\bibfield  {journal} {\bibinfo  {journal} {Nature}\ }\textbf {\bibinfo
				{volume} {547}},\ \bibinfo {pages} {328} (\bibinfo {year}
			{2017})}\BibitemShut {NoStop}%
		\bibitem [{\citenamefont {Donnelly}\ \emph {et~al.}(2018)\citenamefont
			{Donnelly}, \citenamefont {Gliga}, \citenamefont {Scagnoli}, \citenamefont
			{Holler}, \citenamefont {Raabe}, \citenamefont {Heyderman},\ and\
			\citenamefont {Guizar-Sicairos}}]{Donnelly18}%
		\BibitemOpen
		\bibfield  {author} {\bibinfo {author} {\bibfnamefont {C.}~\bibnamefont
				{Donnelly}}, \bibinfo {author} {\bibfnamefont {S.}~\bibnamefont {Gliga}},
			\bibinfo {author} {\bibfnamefont {V.}~\bibnamefont {Scagnoli}}, \bibinfo
			{author} {\bibfnamefont {M.}~\bibnamefont {Holler}}, \bibinfo {author}
			{\bibfnamefont {J.}~\bibnamefont {Raabe}}, \bibinfo {author} {\bibfnamefont
				{L.~J.}\ \bibnamefont {Heyderman}},\ and\ \bibinfo {author} {\bibfnamefont
				{M.}~\bibnamefont {Guizar-Sicairos}},\ }\bibfield  {title} {\bibinfo {title}
			{Tomographic reconstruction of a three-dimensional magnetization vector
				field},\ }\href {https://doi.org/10.1088/1367-2630/aad35a} {\bibfield
			{journal} {\bibinfo  {journal} {New Journal of Physics}\ }\textbf {\bibinfo
				{volume} {20}},\ \bibinfo {pages} {083009} (\bibinfo {year}
			{2018})}\BibitemShut {NoStop}%
		\bibitem [{\citenamefont {Witte}\ \emph {et~al.}(2020)\citenamefont {Witte},
			\citenamefont {Späth}, \citenamefont {Finizio}, \citenamefont {Donnelly},
			\citenamefont {Watts}, \citenamefont {Sarafimov}, \citenamefont {Odstrcil},
			\citenamefont {Guizar-Sicairos}, \citenamefont {Holler}, \citenamefont
			{Fink},\ and\ \citenamefont {Raabe}}]{Witte20}%
		\BibitemOpen
		\bibfield  {author} {\bibinfo {author} {\bibfnamefont {K.}~\bibnamefont
				{Witte}}, \bibinfo {author} {\bibfnamefont {A.}~\bibnamefont {Späth}},
			\bibinfo {author} {\bibfnamefont {S.}~\bibnamefont {Finizio}}, \bibinfo
			{author} {\bibfnamefont {C.}~\bibnamefont {Donnelly}}, \bibinfo {author}
			{\bibfnamefont {B.}~\bibnamefont {Watts}}, \bibinfo {author} {\bibfnamefont
				{B.}~\bibnamefont {Sarafimov}}, \bibinfo {author} {\bibfnamefont
				{M.}~\bibnamefont {Odstrcil}}, \bibinfo {author} {\bibfnamefont
				{M.}~\bibnamefont {Guizar-Sicairos}}, \bibinfo {author} {\bibfnamefont
				{M.}~\bibnamefont {Holler}}, \bibinfo {author} {\bibfnamefont {R.~H.}\
				\bibnamefont {Fink}},\ and\ \bibinfo {author} {\bibfnamefont
				{J.}~\bibnamefont {Raabe}},\ }\bibfield  {title} {\bibinfo {title} {From {2D}
				{STXM} to {3D} imaging: Soft {X-ray} laminography of thin specimens},\ }\href
		{https://doi.org/10.1021/acs.nanolett.9b04782} {\bibfield  {journal}
			{\bibinfo  {journal} {Nano Letters}\ }\textbf {\bibinfo {volume} {20}},\
			\bibinfo {pages} {1305} (\bibinfo {year} {2020})}\BibitemShut {NoStop}%
		\bibitem [{\citenamefont {Di~Pietro~Mart\'{\i}nez}\ \emph
			{et~al.}(2023)\citenamefont {Di~Pietro~Mart\'{\i}nez}, \citenamefont
			{Wartelle}, \citenamefont {Herrero~Mart\'{\i}nez}, \citenamefont {Fettar},
			\citenamefont {Blondelle}, \citenamefont {Motte}, \citenamefont {Donnelly},
			\citenamefont {Turnbull}, \citenamefont {Ogrin}, \citenamefont {van~der
				Laan}, \citenamefont {Popescu}, \citenamefont {Jaouen}, \citenamefont
			{Yakhou-Harris},\ and\ \citenamefont {Beutier}}]{DiPietroMartinez23}%
		\BibitemOpen
		\bibfield  {author} {\bibinfo {author} {\bibfnamefont {M.}~\bibnamefont
				{Di~Pietro~Mart\'{\i}nez}}, \bibinfo {author} {\bibfnamefont
				{A.}~\bibnamefont {Wartelle}}, \bibinfo {author} {\bibfnamefont
				{C.}~\bibnamefont {Herrero~Mart\'{\i}nez}}, \bibinfo {author} {\bibfnamefont
				{F.}~\bibnamefont {Fettar}}, \bibinfo {author} {\bibfnamefont
				{F.}~\bibnamefont {Blondelle}}, \bibinfo {author} {\bibfnamefont {J.-F.}\
				\bibnamefont {Motte}}, \bibinfo {author} {\bibfnamefont {C.}~\bibnamefont
				{Donnelly}}, \bibinfo {author} {\bibfnamefont {L.}~\bibnamefont {Turnbull}},
			\bibinfo {author} {\bibfnamefont {F.}~\bibnamefont {Ogrin}}, \bibinfo
			{author} {\bibfnamefont {G.}~\bibnamefont {van~der Laan}}, \bibinfo {author}
			{\bibfnamefont {H.}~\bibnamefont {Popescu}}, \bibinfo {author} {\bibfnamefont
				{N.}~\bibnamefont {Jaouen}}, \bibinfo {author} {\bibfnamefont
				{F.}~\bibnamefont {Yakhou-Harris}},\ and\ \bibinfo {author} {\bibfnamefont
				{G.}~\bibnamefont {Beutier}},\ }\bibfield  {title} {\bibinfo {title}
			{Three-dimensional tomographic imaging of the magnetization vector field
				using {F}ourier transform holography},\ }\href
		{https://doi.org/10.1103/physrevb.107.094425} {\bibfield  {journal} {\bibinfo
				{journal} {Physical Review B}\ }\textbf {\bibinfo {volume} {107}},\ \bibinfo
			{pages} {094425} (\bibinfo {year} {2023})}\BibitemShut {NoStop}%
		\bibitem [{\citenamefont {Di~Pietro~Mart\'{\i}nez}\ \emph
			{et~al.}(2025)\citenamefont {Di~Pietro~Mart\'{\i}nez}, \citenamefont
			{Turnbull}, \citenamefont {Neethirajan}, \citenamefont {Birch}, \citenamefont
			{Finizio}, \citenamefont {Raabe}, \citenamefont {Lesne}, \citenamefont
			{Markou}, \citenamefont {V\'{e}lez}, \citenamefont {Hierro-Rodr\'{\i}guez},
			\citenamefont {Salvalaglio},\ and\ \citenamefont
			{Donnelly}}]{DiPietroMartinez25}%
		\BibitemOpen
		\bibfield  {author} {\bibinfo {author} {\bibfnamefont {M.}~\bibnamefont
				{Di~Pietro~Mart\'{\i}nez}}, \bibinfo {author} {\bibfnamefont {L.~A.}\
				\bibnamefont {Turnbull}}, \bibinfo {author} {\bibfnamefont {J.}~\bibnamefont
				{Neethirajan}}, \bibinfo {author} {\bibfnamefont {M.}~\bibnamefont {Birch}},
			\bibinfo {author} {\bibfnamefont {S.}~\bibnamefont {Finizio}}, \bibinfo
			{author} {\bibfnamefont {J.}~\bibnamefont {Raabe}}, \bibinfo {author}
			{\bibfnamefont {E.}~\bibnamefont {Lesne}}, \bibinfo {author} {\bibfnamefont
				{A.}~\bibnamefont {Markou}}, \bibinfo {author} {\bibfnamefont
				{M.}~\bibnamefont {V\'{e}lez}}, \bibinfo {author} {\bibfnamefont
				{A.}~\bibnamefont {Hierro-Rodr\'{\i}guez}}, \bibinfo {author} {\bibfnamefont
				{M.}~\bibnamefont {Salvalaglio}},\ and\ \bibinfo {author} {\bibfnamefont
				{C.}~\bibnamefont {Donnelly}},\ }\bibfield  {title} {\bibinfo {title}
			{Unidirectional motion of topological defects mediating continuous rotation
				processes},\ }\href {https://doi.org/10.1038/s44306-025-00111-1} {\bibfield
			{journal} {\bibinfo  {journal} {npj Spintronics}\ }\textbf {\bibinfo {volume}
				{3}},\ \bibinfo {pages} {47} (\bibinfo {year} {2025})}\BibitemShut {NoStop}%
		\bibitem [{\citenamefont {Williams}\ and\ \citenamefont
			{Carter}(2009)}]{Williams09}%
		\BibitemOpen
		\bibfield  {author} {\bibinfo {author} {\bibfnamefont {D.~B.}\ \bibnamefont
				{Williams}}\ and\ \bibinfo {author} {\bibfnamefont {C.~B.}\ \bibnamefont
				{Carter}},\ }\href {https://doi.org/10.1007/978-0-387-76501-3} {\emph
			{\bibinfo {title} {Transmission Electron Microscopy}}}\ (\bibinfo
		{publisher} {Springer US},\ \bibinfo {address} {Boston, MA},\ \bibinfo {year}
		{2009})\BibitemShut {NoStop}%
		\bibitem [{\citenamefont {Pennycook}\ and\ \citenamefont
			{Nellist}(2011)}]{Pennycook11}%
		\BibitemOpen
		\bibinfo {editor} {\bibfnamefont {S.~J.}\ \bibnamefont {Pennycook}}\ and\
		\bibinfo {editor} {\bibfnamefont {P.~D.}\ \bibnamefont {Nellist}},\ eds.,\
		\href {https://doi.org/10.1007/978-1-4419-7200-2} {\emph {\bibinfo {title}
				{Scanning Transmission Electron Microscopy}}}\ (\bibinfo  {publisher}
		{Springer New York},\ \bibinfo {address} {New York, NY},\ \bibinfo {year}
		{2011})\BibitemShut {NoStop}%
		\bibitem [{\citenamefont {Chapman}\ and\ \citenamefont
			{Scheinfein}(1999)}]{Chapman99}%
		\BibitemOpen
		\bibfield  {author} {\bibinfo {author} {\bibfnamefont {J.}~\bibnamefont
				{Chapman}}\ and\ \bibinfo {author} {\bibfnamefont {M.}~\bibnamefont
				{Scheinfein}},\ }\bibfield  {title} {\bibinfo {title} {Transmission electron
				microscopies of magnetic microstructures},\ }\href
		{https://doi.org/10.1016/s0304-8853(99)00317-0} {\bibfield  {journal}
			{\bibinfo  {journal} {Journal of Magnetism and Magnetic Materials}\ }\textbf
			{\bibinfo {volume} {200}},\ \bibinfo {pages} {729} (\bibinfo {year}
			{1999})}\BibitemShut {NoStop}%
		\bibitem [{\citenamefont {Loudon}\ \emph {et~al.}(2019)\citenamefont {Loudon},
			\citenamefont {Twitchett‐Harrison}, \citenamefont
			{Cort\'{e}s‐Ortu\~{n}o}, \citenamefont {Birch}, \citenamefont {Turnbull},
			\citenamefont {\v{S}tefan\v{c}i\v{c}}, \citenamefont {Ogrin}, \citenamefont
			{Burgos‐Parra}, \citenamefont {Bukin}, \citenamefont {Laurenson},
			\citenamefont {Popescu}, \citenamefont {Beg}, \citenamefont {Hovorka},
			\citenamefont {Fangohr}, \citenamefont {Midgley}, \citenamefont
			{Balakrishnan},\ and\ \citenamefont {Hatton}}]{Loudon19}%
		\BibitemOpen
		\bibfield  {author} {\bibinfo {author} {\bibfnamefont {J.~C.}\ \bibnamefont
				{Loudon}}, \bibinfo {author} {\bibfnamefont {A.~C.}\ \bibnamefont
				{Twitchett‐Harrison}}, \bibinfo {author} {\bibfnamefont {D.}~\bibnamefont
				{Cort\'{e}s‐Ortu\~{n}o}}, \bibinfo {author} {\bibfnamefont {M.~T.}\
				\bibnamefont {Birch}}, \bibinfo {author} {\bibfnamefont {L.~A.}\ \bibnamefont
				{Turnbull}}, \bibinfo {author} {\bibfnamefont {A.}~\bibnamefont
				{\v{S}tefan\v{c}i\v{c}}}, \bibinfo {author} {\bibfnamefont {F.~Y.}\
				\bibnamefont {Ogrin}}, \bibinfo {author} {\bibfnamefont {E.~O.}\ \bibnamefont
				{Burgos‐Parra}}, \bibinfo {author} {\bibfnamefont {N.}~\bibnamefont
				{Bukin}}, \bibinfo {author} {\bibfnamefont {A.}~\bibnamefont {Laurenson}},
			\bibinfo {author} {\bibfnamefont {H.}~\bibnamefont {Popescu}}, \bibinfo
			{author} {\bibfnamefont {M.}~\bibnamefont {Beg}}, \bibinfo {author}
			{\bibfnamefont {O.}~\bibnamefont {Hovorka}}, \bibinfo {author} {\bibfnamefont
				{H.}~\bibnamefont {Fangohr}}, \bibinfo {author} {\bibfnamefont {P.~A.}\
				\bibnamefont {Midgley}}, \bibinfo {author} {\bibfnamefont {G.}~\bibnamefont
				{Balakrishnan}},\ and\ \bibinfo {author} {\bibfnamefont {P.~D.}\ \bibnamefont
				{Hatton}},\ }\bibfield  {title} {\bibinfo {title} {Do images of biskyrmions
				show type‐{II} bubbles?},\ }\href {https://doi.org/10.1002/adma.201806598}
		{\bibfield  {journal} {\bibinfo  {journal} {Advanced Materials}\ }\textbf
			{\bibinfo {volume} {31}},\ \bibinfo {pages} {1806598} (\bibinfo {year}
			{2019})}\BibitemShut {NoStop}%
		\bibitem [{\citenamefont {Kov{\'{a}}cs}\ and\ \citenamefont
			{Dunin-Borkowski}(2018)}]{Kovacs18}%
		\BibitemOpen
		\bibfield  {author} {\bibinfo {author} {\bibfnamefont {A.}~\bibnamefont
				{Kov{\'{a}}cs}}\ and\ \bibinfo {author} {\bibfnamefont {R.~E.}\ \bibnamefont
				{Dunin-Borkowski}},\ }\bibfield  {title} {\bibinfo {title} {Magnetic imaging
				of nanostructures using off-axis electron holography},\ }in\ \href
		{https://doi.org/10.1016/bs.hmm.2018.09.001} {\emph {\bibinfo {booktitle}
				{Handbook of Magnetic Materials}}}\ (\bibinfo  {publisher} {Elsevier},\
		\bibinfo {year} {2018})\ pp.\ \bibinfo {pages} {59--153}\BibitemShut
		{NoStop}%
		\bibitem [{\citenamefont {McVitie}\ \emph {et~al.}(2015)\citenamefont
			{McVitie}, \citenamefont {McGrouther}, \citenamefont {McFadzean},
			\citenamefont {MacLaren}, \citenamefont {O'Shea},\ and\ \citenamefont
			{Benitez}}]{McVitie15}%
		\BibitemOpen
		\bibfield  {author} {\bibinfo {author} {\bibfnamefont {S.}~\bibnamefont
				{McVitie}}, \bibinfo {author} {\bibfnamefont {D.}~\bibnamefont {McGrouther}},
			\bibinfo {author} {\bibfnamefont {S.}~\bibnamefont {McFadzean}}, \bibinfo
			{author} {\bibfnamefont {D.}~\bibnamefont {MacLaren}}, \bibinfo {author}
			{\bibfnamefont {K.}~\bibnamefont {O'Shea}},\ and\ \bibinfo {author}
			{\bibfnamefont {M.}~\bibnamefont {Benitez}},\ }\bibfield  {title} {\bibinfo
			{title} {Aberration corrected lorentz scanning transmission electron
				microscopy},\ }\href {https://doi.org/10.1016/j.ultramic.2015.01.003}
		{\bibfield  {journal} {\bibinfo  {journal} {Ultramicroscopy}\ }\textbf
			{\bibinfo {volume} {152}},\ \bibinfo {pages} {57} (\bibinfo {year}
			{2015})}\BibitemShut {NoStop}%
		\bibitem [{\citenamefont {Hassan}\ \emph {et~al.}(2024)\citenamefont {Hassan},
			\citenamefont {Koraltan}, \citenamefont {Ullrich}, \citenamefont {Bruckner},
			\citenamefont {Serha}, \citenamefont {Levchenko}, \citenamefont {Varvaro},
			\citenamefont {Kiselev}, \citenamefont {Heigl}, \citenamefont {Abert},
			\citenamefont {Suess},\ and\ \citenamefont {Albrecht}}]{Hassan24}%
		\BibitemOpen
		\bibfield  {author} {\bibinfo {author} {\bibfnamefont {M.}~\bibnamefont
				{Hassan}}, \bibinfo {author} {\bibfnamefont {S.}~\bibnamefont {Koraltan}},
			\bibinfo {author} {\bibfnamefont {A.}~\bibnamefont {Ullrich}}, \bibinfo
			{author} {\bibfnamefont {F.}~\bibnamefont {Bruckner}}, \bibinfo {author}
			{\bibfnamefont {R.~O.}\ \bibnamefont {Serha}}, \bibinfo {author}
			{\bibfnamefont {K.~V.}\ \bibnamefont {Levchenko}}, \bibinfo {author}
			{\bibfnamefont {G.}~\bibnamefont {Varvaro}}, \bibinfo {author} {\bibfnamefont
				{N.~S.}\ \bibnamefont {Kiselev}}, \bibinfo {author} {\bibfnamefont
				{M.}~\bibnamefont {Heigl}}, \bibinfo {author} {\bibfnamefont
				{C.}~\bibnamefont {Abert}}, \bibinfo {author} {\bibfnamefont
				{D.}~\bibnamefont {Suess}},\ and\ \bibinfo {author} {\bibfnamefont
				{M.}~\bibnamefont {Albrecht}},\ }\bibfield  {title} {\bibinfo {title}
			{Dipolar skyrmions and antiskyrmions of arbitrary topological charge at room
				temperature},\ }\href {https://doi.org/10.1038/s41567-023-02358-z} {\bibfield
			{journal} {\bibinfo  {journal} {Nature Physics}\ }\textbf {\bibinfo {volume}
				{20}},\ \bibinfo {pages} {615} (\bibinfo {year} {2024})}\BibitemShut
		{NoStop}%
		\bibitem [{\citenamefont {Fullerton}\ and\ \citenamefont
			{Phatak}(2025)}]{Fullerton25}%
		\BibitemOpen
		\bibfield  {author} {\bibinfo {author} {\bibfnamefont {J.}~\bibnamefont
				{Fullerton}}\ and\ \bibinfo {author} {\bibfnamefont {C.}~\bibnamefont
				{Phatak}},\ }\bibfield  {title} {\bibinfo {title} {Design and control of
				three-dimensional topological magnetic fields using interwoven helical
				nanostructures},\ }\href {https://doi.org/10.1021/acs.nanolett.4c06152}
		{\bibfield  {journal} {\bibinfo  {journal} {Nano Letters}\ }\textbf {\bibinfo
				{volume} {25}},\ \bibinfo {pages} {5148} (\bibinfo {year}
			{2025})}\BibitemShut {NoStop}%
		\bibitem [{\citenamefont {Dugato}\ \emph {et~al.}(2025)\citenamefont {Dugato},
			\citenamefont {Jalil}, \citenamefont {Cardias}, \citenamefont {Albuquerque},
			\citenamefont {Costa}, \citenamefont {Almeida}, \citenamefont {Fallon},
			\citenamefont {Kov\'{a}cs}, \citenamefont {McVitie}, \citenamefont
			{Dunin-Borkowski},\ and\ \citenamefont {Garcia}}]{Dugato25}%
		\BibitemOpen
		\bibfield  {author} {\bibinfo {author} {\bibfnamefont {D.~A.}\ \bibnamefont
				{Dugato}}, \bibinfo {author} {\bibfnamefont {W.~B.~F.}\ \bibnamefont
				{Jalil}}, \bibinfo {author} {\bibfnamefont {R.}~\bibnamefont {Cardias}},
			\bibinfo {author} {\bibfnamefont {M.}~\bibnamefont {Albuquerque}}, \bibinfo
			{author} {\bibfnamefont {M.}~\bibnamefont {Costa}}, \bibinfo {author}
			{\bibfnamefont {T.~P.}\ \bibnamefont {Almeida}}, \bibinfo {author}
			{\bibfnamefont {K.}~\bibnamefont {Fallon}}, \bibinfo {author} {\bibfnamefont
				{A.}~\bibnamefont {Kov\'{a}cs}}, \bibinfo {author} {\bibfnamefont
				{S.}~\bibnamefont {McVitie}}, \bibinfo {author} {\bibfnamefont {R.~E.}\
				\bibnamefont {Dunin-Borkowski}},\ and\ \bibinfo {author} {\bibfnamefont
				{F.}~\bibnamefont {Garcia}},\ }\bibfield  {title} {\bibinfo {title} {Curved
				nanomagnets: An archetype for the skyrmionic states at ambient conditions},\
		}\href {https://doi.org/10.1021/acs.nanolett.5c00773} {\bibfield  {journal}
			{\bibinfo  {journal} {Nano Letters}\ }\textbf {\bibinfo {volume} {25}},\
			\bibinfo {pages} {8901} (\bibinfo {year} {2025})}\BibitemShut {NoStop}%
		\bibitem [{\citenamefont {Wolf}\ \emph {et~al.}(2022)\citenamefont {Wolf},
			\citenamefont {Schneider}, \citenamefont {R\"{o}{\ss}ler}, \citenamefont
			{Kov\'{a}cs}, \citenamefont {Schmidt}, \citenamefont {Dunin-Borkowski},
			\citenamefont {B\"{u}chner}, \citenamefont {Rellinghaus},\ and\ \citenamefont
			{Lubk}}]{Wolf22}%
		\BibitemOpen
		\bibfield  {author} {\bibinfo {author} {\bibfnamefont {D.}~\bibnamefont
				{Wolf}}, \bibinfo {author} {\bibfnamefont {S.}~\bibnamefont {Schneider}},
			\bibinfo {author} {\bibfnamefont {U.~K.}\ \bibnamefont {R\"{o}{\ss}ler}},
			\bibinfo {author} {\bibfnamefont {A.}~\bibnamefont {Kov\'{a}cs}}, \bibinfo
			{author} {\bibfnamefont {M.}~\bibnamefont {Schmidt}}, \bibinfo {author}
			{\bibfnamefont {R.~E.}\ \bibnamefont {Dunin-Borkowski}}, \bibinfo {author}
			{\bibfnamefont {B.}~\bibnamefont {B\"{u}chner}}, \bibinfo {author}
			{\bibfnamefont {B.}~\bibnamefont {Rellinghaus}},\ and\ \bibinfo {author}
			{\bibfnamefont {A.}~\bibnamefont {Lubk}},\ }\bibfield  {title} {\bibinfo
			{title} {Unveiling the three-dimensional magnetic texture of skyrmion
				tubes},\ }\href {https://doi.org/10.1038/s41565-021-01031-x} {\bibfield
			{journal} {\bibinfo  {journal} {Nature Nanotechnology}\ }\textbf {\bibinfo
				{volume} {17}},\ \bibinfo {pages} {250} (\bibinfo {year} {2022})}\BibitemShut
		{NoStop}%
		\bibitem [{\citenamefont {Yu}\ \emph {et~al.}(2024)\citenamefont {Yu},
			\citenamefont {Nakanishi}, \citenamefont {Chiew}, \citenamefont {Liu},
			\citenamefont {Nakajima}, \citenamefont {Kanazawa}, \citenamefont {Karube},
			\citenamefont {Taguchi}, \citenamefont {Nagaosa},\ and\ \citenamefont
			{Tokura}}]{Yu24}%
		\BibitemOpen
		\bibfield  {author} {\bibinfo {author} {\bibfnamefont {X.}~\bibnamefont
				{Yu}}, \bibinfo {author} {\bibfnamefont {N.}~\bibnamefont {Nakanishi}},
			\bibinfo {author} {\bibfnamefont {Y.-L.}\ \bibnamefont {Chiew}}, \bibinfo
			{author} {\bibfnamefont {Y.}~\bibnamefont {Liu}}, \bibinfo {author}
			{\bibfnamefont {K.}~\bibnamefont {Nakajima}}, \bibinfo {author}
			{\bibfnamefont {N.}~\bibnamefont {Kanazawa}}, \bibinfo {author}
			{\bibfnamefont {K.}~\bibnamefont {Karube}}, \bibinfo {author} {\bibfnamefont
				{Y.}~\bibnamefont {Taguchi}}, \bibinfo {author} {\bibfnamefont
				{N.}~\bibnamefont {Nagaosa}},\ and\ \bibinfo {author} {\bibfnamefont
				{Y.}~\bibnamefont {Tokura}},\ }\bibfield  {title} {\bibinfo {title} {{3D}
				skyrmion strings and their melting dynamics revealed via scalar-field
				electron tomography},\ }\href {https://doi.org/10.1038/s43246-024-00512-5}
		{\bibfield  {journal} {\bibinfo  {journal} {Communications Materials}\
			}\textbf {\bibinfo {volume} {5}},\ \bibinfo {pages} {80} (\bibinfo {year}
			{2024})}\BibitemShut {NoStop}%
		\bibitem [{OOM()}]{OOMMFb}%
		\BibitemOpen
		\href {http://math.nist.gov/oommf/} {\bibinfo {title} {The {O}bject
				{O}riented {M}icro{M}agnetic {F}ramework}},\ \bibinfo {note} {developed by M.
			J. Donahue and D. Porter mainly, from NIST}\BibitemShut {NoStop}%
		\bibitem [{\citenamefont {Donahue}\ and\ \citenamefont
			{Porter}(1999)}]{Donahue99}%
		\BibitemOpen
		\bibfield  {author} {\bibinfo {author} {\bibfnamefont {M.~J.}\ \bibnamefont
				{Donahue}}\ and\ \bibinfo {author} {\bibfnamefont {D.~G.}\ \bibnamefont
				{Porter}},\ }\href
		{http://math.nist.gov/~MDonahue/pubs/abstracts.html#Donahue199909} {\emph
			{\bibinfo {title} {{OOMMF} User's Guide, Version 1.0}}},\ \bibinfo {type}
		{Tech. Rep.}\ (\bibinfo  {institution} {Interagency Report NISTIR 6376},\
		\bibinfo {year} {1999})\BibitemShut {NoStop}%
		\bibitem [{\citenamefont {Vansteenkiste}\ \emph {et~al.}(2014)\citenamefont
			{Vansteenkiste}, \citenamefont {Leliaert}, \citenamefont {Dvornik},
			\citenamefont {Helsen}, \citenamefont {Garcia-Sanchez},\ and\ \citenamefont
			{Van~Waeyenberge}}]{Vansteenkiste14}%
		\BibitemOpen
		\bibfield  {author} {\bibinfo {author} {\bibfnamefont {A.}~\bibnamefont
				{Vansteenkiste}}, \bibinfo {author} {\bibfnamefont {J.}~\bibnamefont
				{Leliaert}}, \bibinfo {author} {\bibfnamefont {M.}~\bibnamefont {Dvornik}},
			\bibinfo {author} {\bibfnamefont {M.}~\bibnamefont {Helsen}}, \bibinfo
			{author} {\bibfnamefont {F.}~\bibnamefont {Garcia-Sanchez}},\ and\ \bibinfo
			{author} {\bibfnamefont {B.}~\bibnamefont {Van~Waeyenberge}},\ }\bibfield
		{title} {\bibinfo {title} {The design and verification of {MuMax3}},\ }\href
		{https://doi.org/10.1063/1.4899186} {\bibfield  {journal} {\bibinfo
				{journal} {AIP Advances}\ }\textbf {\bibinfo {volume} {4}},\ \bibinfo {pages}
			{107133} (\bibinfo {year} {2014})}\BibitemShut {NoStop}%
		\bibitem [{\citenamefont {Moreels}\ \emph {et~al.}(2026)\citenamefont
			{Moreels}, \citenamefont {Lateur}, \citenamefont {De~Gusem}, \citenamefont
			{Mulkers}, \citenamefont {Maes}, \citenamefont {Milo\v{s}ević},
			\citenamefont {Leliaert},\ and\ \citenamefont {Van~Waeyenberge}}]{Moreels26}%
		\BibitemOpen
		\bibfield  {author} {\bibinfo {author} {\bibfnamefont {L.}~\bibnamefont
				{Moreels}}, \bibinfo {author} {\bibfnamefont {I.}~\bibnamefont {Lateur}},
			\bibinfo {author} {\bibfnamefont {D.}~\bibnamefont {De~Gusem}}, \bibinfo
			{author} {\bibfnamefont {J.}~\bibnamefont {Mulkers}}, \bibinfo {author}
			{\bibfnamefont {J.}~\bibnamefont {Maes}}, \bibinfo {author} {\bibfnamefont
				{M.~V.}\ \bibnamefont {Milo\v{s}ević}}, \bibinfo {author} {\bibfnamefont
				{J.}~\bibnamefont {Leliaert}},\ and\ \bibinfo {author} {\bibfnamefont
				{B.}~\bibnamefont {Van~Waeyenberge}},\ }\bibfield  {title} {\bibinfo {title}
			{{mumax+}: extensible {GPU}-accelerated micromagnetics and beyond},\ }\href
		{https://doi.org/10.1038/s41524-025-01893-y} {\bibfield  {journal} {\bibinfo
				{journal} {npj Computational Materials}\ }\textbf {\bibinfo {volume} {12}},\
			\bibinfo {pages} {71} (\bibinfo {year} {2026})}\BibitemShut {NoStop}%
		\bibitem [{mag()}]{magpar}%
		\BibitemOpen
		\href {http://www.magpar.net} {\bibinfo {title} {{MAGPAR} finite element
				micromagnetics package}},\ \bibinfo {note} {developed by Werner
			Scholz.}\BibitemShut {Stop}%
		\bibitem [{\citenamefont {Scholz}\ \emph {et~al.}(2003)\citenamefont {Scholz},
			\citenamefont {Fidler}, \citenamefont {Schrefl}, \citenamefont {Suess},
			\citenamefont {Dittrich}, \citenamefont {Forster},\ and\ \citenamefont
			{Tsiantos}}]{Scholz03}%
		\BibitemOpen
		\bibfield  {author} {\bibinfo {author} {\bibfnamefont {W.}~\bibnamefont
				{Scholz}}, \bibinfo {author} {\bibfnamefont {J.}~\bibnamefont {Fidler}},
			\bibinfo {author} {\bibfnamefont {T.}~\bibnamefont {Schrefl}}, \bibinfo
			{author} {\bibfnamefont {D.}~\bibnamefont {Suess}}, \bibinfo {author}
			{\bibfnamefont {R.}~\bibnamefont {Dittrich}}, \bibinfo {author}
			{\bibfnamefont {H.}~\bibnamefont {Forster}},\ and\ \bibinfo {author}
			{\bibfnamefont {V.}~\bibnamefont {Tsiantos}},\ }\bibfield  {title} {\bibinfo
			{title} {Scalable parallel micromagnetic solvers for magnetic
				nanostructures},\ }\href {https://doi.org/10.1016/S0927-0256(03)00119-8}
		{\bibfield  {journal} {\bibinfo  {journal} {Computational Materials Science}\
			}\textbf {\bibinfo {volume} {28}},\ \bibinfo {pages} {366} (\bibinfo {year}
			{2003})},\ \bibinfo {note} {{P}roceedings of the {S}ymposium on Software
			Development for Process and Materials Design}\BibitemShut {NoStop}%
		\bibitem [{\citenamefont {Fischbacher}\ \emph {et~al.}(2007)\citenamefont
			{Fischbacher}, \citenamefont {Franchin}, \citenamefont {Bordignon},\ and\
			\citenamefont {Fangohr}}]{Fischbacher07}%
		\BibitemOpen
		\bibfield  {author} {\bibinfo {author} {\bibfnamefont {T.}~\bibnamefont
				{Fischbacher}}, \bibinfo {author} {\bibfnamefont {M.}~\bibnamefont
				{Franchin}}, \bibinfo {author} {\bibfnamefont {G.}~\bibnamefont
				{Bordignon}},\ and\ \bibinfo {author} {\bibfnamefont {H.}~\bibnamefont
				{Fangohr}},\ }\bibfield  {title} {\bibinfo {title} {A systematic approach to
				multiphysics extensions of finite-element-based micromagnetic simulations:
				Nmag},\ }\href {https://doi.org/10.1109/tmag.2007.893843} {\bibfield
			{journal} {\bibinfo  {journal} {IEEE Transactions on Magnetics}\ }\textbf
			{\bibinfo {volume} {43}},\ \bibinfo {pages} {2896} (\bibinfo {year}
			{2007})}\BibitemShut {NoStop}%
		\bibitem [{\citenamefont {Hertel}(2023)}]{Hertel23}%
		\BibitemOpen
		\bibfield  {author} {\bibinfo {author} {\bibfnamefont {R.}~\bibnamefont
				{Hertel}},\ }\href {https://github.com/R-Hertel/tetmag} {\bibinfo {title}
			{{tetmag}: a finite-element software for general-purpose micromagnetic
				simulations}} (\bibinfo {year} {2023}),\ \bibinfo {note} {accessed:
			2026-03-13}\BibitemShut {NoStop}%
		\bibitem [{\citenamefont {Abert}\ \emph {et~al.}(2013)\citenamefont {Abert},
			\citenamefont {Exl}, \citenamefont {Bruckner}, \citenamefont {Drews},\ and\
			\citenamefont {Suess}}]{Abert13a}%
		\BibitemOpen
		\bibfield  {author} {\bibinfo {author} {\bibfnamefont {C.}~\bibnamefont
				{Abert}}, \bibinfo {author} {\bibfnamefont {L.}~\bibnamefont {Exl}}, \bibinfo
			{author} {\bibfnamefont {F.}~\bibnamefont {Bruckner}}, \bibinfo {author}
			{\bibfnamefont {A.}~\bibnamefont {Drews}},\ and\ \bibinfo {author}
			{\bibfnamefont {D.}~\bibnamefont {Suess}},\ }\bibfield  {title} {\bibinfo
			{title} {magnum.fe: A micromagnetic finite-element simulation code based on
				{FEniCS}},\ }\href {https://doi.org/10.1016/j.jmmm.2013.05.051} {\bibfield
			{journal} {\bibinfo  {journal} {Journal of Magnetism and Magnetic Materials}\
			}\textbf {\bibinfo {volume} {345}},\ \bibinfo {pages} {29} (\bibinfo {year}
			{2013})}\BibitemShut {NoStop}%
		\bibitem [{\citenamefont {Carvalho-Santos}\ \emph {et~al.}(2010)\citenamefont
			{Carvalho-Santos}, \citenamefont {Moura-Melo},\ and\ \citenamefont
			{Pereira}}]{Carvalho-Santos10}%
		\BibitemOpen
		\bibfield  {author} {\bibinfo {author} {\bibfnamefont {V.~L.}\ \bibnamefont
				{Carvalho-Santos}}, \bibinfo {author} {\bibfnamefont {W.~A.}\ \bibnamefont
				{Moura-Melo}},\ and\ \bibinfo {author} {\bibfnamefont {A.~R.}\ \bibnamefont
				{Pereira}},\ }\bibfield  {title} {\bibinfo {title} {Miniaturization of
				vortex-comprising system using ferromagnetic nanotori},\ }\href
		{https://doi.org/10.1063/1.3487924} {\bibfield  {journal} {\bibinfo
				{journal} {Journal of Applied Physics}\ }\textbf {\bibinfo {volume} {108}},\
			\bibinfo {pages} {094310} (\bibinfo {year} {2010})}\BibitemShut {NoStop}%
		\bibitem [{\citenamefont {Corona}\ \emph {et~al.}(2023)\citenamefont {Corona},
			\citenamefont {Saavedra}, \citenamefont {Castillo-Sepulveda}, \citenamefont
			{Escrig}, \citenamefont {Altbir},\ and\ \citenamefont
			{Carvalho-Santos}}]{Corona23}%
		\BibitemOpen
		\bibfield  {author} {\bibinfo {author} {\bibfnamefont {R.~M.}\ \bibnamefont
				{Corona}}, \bibinfo {author} {\bibfnamefont {E.}~\bibnamefont {Saavedra}},
			\bibinfo {author} {\bibfnamefont {S.}~\bibnamefont {Castillo-Sepulveda}},
			\bibinfo {author} {\bibfnamefont {J.}~\bibnamefont {Escrig}}, \bibinfo
			{author} {\bibfnamefont {D.}~\bibnamefont {Altbir}},\ and\ \bibinfo {author}
			{\bibfnamefont {V.~L.}\ \bibnamefont {Carvalho-Santos}},\ }\bibfield  {title}
		{\bibinfo {title} {Curvature-induced stabilization and field-driven dynamics
				of magnetic hopfions in toroidal nanorings},\ }\href
		{https://doi.org/10.1088/1361-6528/acb557} {\bibfield  {journal} {\bibinfo
				{journal} {Nanotechnology}\ }\textbf {\bibinfo {volume} {34}},\ \bibinfo
			{pages} {165702} (\bibinfo {year} {2023})}\BibitemShut {NoStop}%
		\bibitem [{\citenamefont {Teixeira}\ \emph {et~al.}(2019)\citenamefont
			{Teixeira}, \citenamefont {Castillo-Sep{\'{u}}lveda}, \citenamefont
			{Vojkovic}, \citenamefont {Fonseca}, \citenamefont {Altbir}, \citenamefont
			{N{\'{u}}{\~{n}}ez},\ and\ \citenamefont {Carvalho-Santos}}]{Teixeira19}%
		\BibitemOpen
		\bibfield  {author} {\bibinfo {author} {\bibfnamefont {A.}~\bibnamefont
				{Teixeira}}, \bibinfo {author} {\bibfnamefont {S.}~\bibnamefont
				{Castillo-Sep{\'{u}}lveda}}, \bibinfo {author} {\bibfnamefont
				{S.}~\bibnamefont {Vojkovic}}, \bibinfo {author} {\bibfnamefont
				{J.}~\bibnamefont {Fonseca}}, \bibinfo {author} {\bibfnamefont
				{D.}~\bibnamefont {Altbir}}, \bibinfo {author} {\bibfnamefont
				{{\'{A}}.}~\bibnamefont {N{\'{u}}{\~{n}}ez}},\ and\ \bibinfo {author}
			{\bibfnamefont {V.}~\bibnamefont {Carvalho-Santos}},\ }\bibfield  {title}
		{\bibinfo {title} {Analysis on the stability of in-surface magnetic
				configurations in toroidal nanoshells},\ }\href
		{https://doi.org/10.1016/j.jmmm.2019.01.093} {\bibfield  {journal} {\bibinfo
				{journal} {Journal of Magnetism and Magnetic Materials}\ }\textbf {\bibinfo
				{volume} {478}},\ \bibinfo {pages} {253} (\bibinfo {year}
			{2019})}\BibitemShut {NoStop}%
		\bibitem [{\citenamefont {O'Neill}(2006)}]{ONeill06}%
		\BibitemOpen
		\bibfield  {author} {\bibinfo {author} {\bibfnamefont {B.}~\bibnamefont
				{O'Neill}},\ }\bibinfo {title} {Shape operators},\ in\ \href
		{https://doi.org/10.1016/b978-0-12-088735-4.50009-2} {\emph {\bibinfo
				{booktitle} {Elementary Differential Geometry}}}\ (\bibinfo  {publisher}
		{Elsevier},\ \bibinfo {year} {2006})\ pp.\ \bibinfo {pages}
		{202--262}\BibitemShut {NoStop}%
		\bibitem [{\citenamefont {Kovalchuk}\ and\ \citenamefont
			{Chudzi\'{n}ski}(2023)}]{Kovalchuk23}%
		\BibitemOpen
		\bibfield  {author} {\bibinfo {author} {\bibfnamefont {V.}~\bibnamefont
				{Kovalchuk}}\ and\ \bibinfo {author} {\bibfnamefont {P.}~\bibnamefont
				{Chudzi\'{n}ski}},\ }\bibfield  {title} {\bibinfo {title} {Unifying approach
				to {H}eisenberg models for spin distributions on curved surfaces of
				revolution with isothermal (conformal) metrics},\ }\href
		{https://doi.org/10.1002/mma.8587} {\bibfield  {journal} {\bibinfo  {journal}
				{Mathematical Methods in the Applied Sciences}\ }\textbf {\bibinfo {volume}
				{46}},\ \bibinfo {pages} {1443} (\bibinfo {year} {2023})}\BibitemShut
		{NoStop}%
		\bibitem [{\citenamefont {Saxena}\ \emph {et~al.}(1998)\citenamefont {Saxena},
			\citenamefont {Dandoloff},\ and\ \citenamefont {Lookman}}]{Saxena98a}%
		\BibitemOpen
		\bibfield  {author} {\bibinfo {author} {\bibfnamefont {A.}~\bibnamefont
				{Saxena}}, \bibinfo {author} {\bibfnamefont {R.}~\bibnamefont {Dandoloff}},\
			and\ \bibinfo {author} {\bibfnamefont {T.}~\bibnamefont {Lookman}},\
		}\bibfield  {title} {\bibinfo {title} {Deformable curved magnetic surfaces},\
		}\href {https://doi.org/10.1016/S0378-4371(98)00378-1} {\bibfield  {journal}
			{\bibinfo  {journal} {Physica A: Statistical and Theoretical Physics}\
			}\textbf {\bibinfo {volume} {261}},\ \bibinfo {pages} {13} (\bibinfo {year}
			{1998})}\BibitemShut {NoStop}%
		\bibitem [{\citenamefont {Kravchuk}\ \emph {et~al.}(2012)\citenamefont
			{Kravchuk}, \citenamefont {Sheka}, \citenamefont {Streubel}, \citenamefont
			{Makarov}, \citenamefont {Schmidt},\ and\ \citenamefont
			{Gaididei}}]{Kravchuk12a}%
		\BibitemOpen
		\bibfield  {author} {\bibinfo {author} {\bibfnamefont {V.~P.}\ \bibnamefont
				{Kravchuk}}, \bibinfo {author} {\bibfnamefont {D.~D.}\ \bibnamefont {Sheka}},
			\bibinfo {author} {\bibfnamefont {R.}~\bibnamefont {Streubel}}, \bibinfo
			{author} {\bibfnamefont {D.}~\bibnamefont {Makarov}}, \bibinfo {author}
			{\bibfnamefont {O.~G.}\ \bibnamefont {Schmidt}},\ and\ \bibinfo {author}
			{\bibfnamefont {Y.}~\bibnamefont {Gaididei}},\ }\bibfield  {title} {\bibinfo
			{title} {Out-of-surface vortices in spherical shells},\ }\href
		{https://doi.org/10.1103/PhysRevB.85.144433} {\bibfield  {journal} {\bibinfo
				{journal} {Physical Review B}\ }\textbf {\bibinfo {volume} {85}},\ \bibinfo
			{pages} {144433} (\bibinfo {year} {2012})}\BibitemShut {NoStop}%
		\bibitem [{\citenamefont {Dandoloff}\ and\ \citenamefont
			{Saxena}(2011)}]{Dandoloff11}%
		\BibitemOpen
		\bibfield  {author} {\bibinfo {author} {\bibfnamefont {R.}~\bibnamefont
				{Dandoloff}}\ and\ \bibinfo {author} {\bibfnamefont {A.}~\bibnamefont
				{Saxena}},\ }\bibfield  {title} {\bibinfo {title} {Heisenberg spins on a
				bilayer connected by a neck and other geometries with a characteristic length
				scale},\ }\href {https://doi.org/10.1088/1751-8113/44/4/045203} {\bibfield
			{journal} {\bibinfo  {journal} {Journal of Physics A: Mathematical and
					Theoretical}\ }\textbf {\bibinfo {volume} {44}},\ \bibinfo {pages} {045203}
			(\bibinfo {year} {2011})}\BibitemShut {NoStop}%
		\bibitem [{\citenamefont {Bogomol'nyi}(1976)}]{Bogomolnyi76}%
		\BibitemOpen
		\bibfield  {author} {\bibinfo {author} {\bibfnamefont {E.~B.}\ \bibnamefont
				{Bogomol'nyi}},\ }\bibfield  {title} {\bibinfo {title} {The stability of
				classical solutions},\ }\href {https://www.osti.gov/biblio/7309001}
		{\bibfield  {journal} {\bibinfo  {journal} {Sov. J. Nucl. Phys. [Yad.Fiz. 24
					(1976) 861-870]}\ }\textbf {\bibinfo {volume} {24}},\ \bibinfo {pages} {449}
			(\bibinfo {year} {1976})}\BibitemShut {NoStop}%
		\bibitem [{\citenamefont {Belavin}\ and\ \citenamefont
			{Polyakov}(1975)}]{Belavin75}%
		\BibitemOpen
		\bibfield  {author} {\bibinfo {author} {\bibfnamefont {A.~A.}\ \bibnamefont
				{Belavin}}\ and\ \bibinfo {author} {\bibfnamefont {A.~M.}\ \bibnamefont
				{Polyakov}},\ }\bibfield  {title} {\bibinfo {title} {Metastable states of a
				{2D} isotropic ferromagnet},\ }\href
		{http://www.jetpletters.ac.ru/ps/1529/article_23383.shtml} {\bibfield
			{journal} {\bibinfo  {journal} {JETP Letters}\ }\textbf {\bibinfo {volume}
				{22}},\ \bibinfo {pages} {245} (\bibinfo {year} {1975})}\BibitemShut
		{NoStop}%
		\bibitem [{\citenamefont {Gradshteyn}\ and\ \citenamefont
			{Ryzhik}(2007)}]{Gradshteyn07}%
		\BibitemOpen
		\bibfield  {author} {\bibinfo {author} {\bibfnamefont {I.}~\bibnamefont
				{Gradshteyn}}\ and\ \bibinfo {author} {\bibfnamefont {I.}~\bibnamefont
				{Ryzhik}},\ }\href
		{https://www.sciencedirect.com/book/edited-volume/9780123736376/table-of-integrals-series-and-products}
		{\emph {\bibinfo {title} {Table of Integrals, Series, and Products}}},\
		\bibinfo {edition} {7th}\ ed.,\ edited by\ \bibinfo {editor} {\bibfnamefont
			{A.}~\bibnamefont {Jeffrey}}\ and\ \bibinfo {editor} {\bibfnamefont
			{D.}~\bibnamefont {Zwillinger}}\ (\bibinfo  {publisher} {Academic Press},\
		\bibinfo {year} {2007})\BibitemShut {NoStop}%
		\bibitem [{\citenamefont {Aharoni}(1996)}]{Aharoni96}%
		\BibitemOpen
		\bibfield  {author} {\bibinfo {author} {\bibfnamefont {A.}~\bibnamefont
				{Aharoni}},\ }\href@noop {} {\emph {\bibinfo {title} {Introduction to the
					theory of {F}erromagnetism}}}\ (\bibinfo  {publisher} {Oxford University
			Press},\ \bibinfo {year} {1996})\BibitemShut {NoStop}%
		\bibitem [{\citenamefont {Guslienko}\ \emph {et~al.}(2001)\citenamefont
			{Guslienko}, \citenamefont {Novosad}, \citenamefont {Otani}, \citenamefont
			{Shima},\ and\ \citenamefont {Fukamichi}}]{Guslienko01b}%
		\BibitemOpen
		\bibfield  {author} {\bibinfo {author} {\bibfnamefont {K.~Y.}\ \bibnamefont
				{Guslienko}}, \bibinfo {author} {\bibfnamefont {V.}~\bibnamefont {Novosad}},
			\bibinfo {author} {\bibfnamefont {Y.}~\bibnamefont {Otani}}, \bibinfo
			{author} {\bibfnamefont {H.}~\bibnamefont {Shima}},\ and\ \bibinfo {author}
			{\bibfnamefont {K.}~\bibnamefont {Fukamichi}},\ }\bibfield  {title} {\bibinfo
			{title} {Field evolution of magnetic vortex state in ferromagnetic disks},\
		}\href {https://doi.org/10.1063/1.1377850} {\bibfield  {journal} {\bibinfo
				{journal} {Applied Physics Letters}\ }\textbf {\bibinfo {volume} {78}},\
			\bibinfo {pages} {3848} (\bibinfo {year} {2001})}\BibitemShut {NoStop}%
		\bibitem [{\citenamefont {Shibata}\ \emph {et~al.}(2003)\citenamefont
			{Shibata}, \citenamefont {Shigeto},\ and\ \citenamefont
			{Otani}}]{Shibata03a}%
		\BibitemOpen
		\bibfield  {author} {\bibinfo {author} {\bibfnamefont {J.}~\bibnamefont
				{Shibata}}, \bibinfo {author} {\bibfnamefont {K.}~\bibnamefont {Shigeto}},\
			and\ \bibinfo {author} {\bibfnamefont {Y.}~\bibnamefont {Otani}},\ }\bibfield
		{title} {\bibinfo {title} {Dynamics of magnetostatically coupled vortices in
				magnetic nanodisks},\ }\href {https://doi.org/10.1103/physrevb.67.224404}
		{\bibfield  {journal} {\bibinfo  {journal} {Physical Review B}\ }\textbf
			{\bibinfo {volume} {67}},\ \bibinfo {pages} {224404} (\bibinfo {year}
			{2003})}\BibitemShut {NoStop}%
		\bibitem [{\citenamefont {Hertel}\ and\ \citenamefont
			{Kirschner}(2004{\natexlab{a}})}]{Hertel04}%
		\BibitemOpen
		\bibfield  {author} {\bibinfo {author} {\bibfnamefont {R.}~\bibnamefont
				{Hertel}}\ and\ \bibinfo {author} {\bibfnamefont {J.}~\bibnamefont
				{Kirschner}},\ }\bibfield  {title} {\bibinfo {title} {Resonant modes of
				vortex structures in soft--magnetic nanodiscs},\ }\href
		{https://doi.org/10.1016/j.jmmm.2003.12.678} {\bibfield  {journal} {\bibinfo
				{journal} {Journal of Magnetism and Magnetic Materials}\ }\textbf {\bibinfo
				{volume} {272-276}},\ \bibinfo {pages} {655} (\bibinfo {year}
			{2004}{\natexlab{a}})}\BibitemShut {NoStop}%
		\bibitem [{\citenamefont {Pereira}\ \emph {et~al.}(2007)\citenamefont
			{Pereira}, \citenamefont {Moura}, \citenamefont {Moura-Melo}, \citenamefont
			{Carneiro}, \citenamefont {Leonel},\ and\ \citenamefont {Coura}}]{Pereira07}%
		\BibitemOpen
		\bibfield  {author} {\bibinfo {author} {\bibfnamefont {A.~R.}\ \bibnamefont
				{Pereira}}, \bibinfo {author} {\bibfnamefont {A.~R.}\ \bibnamefont {Moura}},
			\bibinfo {author} {\bibfnamefont {W.~A.}\ \bibnamefont {Moura-Melo}},
			\bibinfo {author} {\bibfnamefont {D.~F.}\ \bibnamefont {Carneiro}}, \bibinfo
			{author} {\bibfnamefont {S.~A.}\ \bibnamefont {Leonel}},\ and\ \bibinfo
			{author} {\bibfnamefont {P.~Z.}\ \bibnamefont {Coura}},\ }\bibfield  {title}
		{\bibinfo {title} {Analytical and monte carlo study of two antidots in
				magnetic nanodisks with vortex-like magnetization},\ }\href
		{https://doi.org/10.1063/1.2432312} {\bibfield  {journal} {\bibinfo
				{journal} {Journal of Applied Physics}\ }\textbf {\bibinfo {volume} {101}},\
			\bibinfo {pages} {034310} (\bibinfo {year} {2007})}\BibitemShut {NoStop}%
		\bibitem [{\citenamefont {Goll}\ \emph {et~al.}(2004)\citenamefont {Goll},
			\citenamefont {Berkowitz},\ and\ \citenamefont {Bertram}}]{Goll04}%
		\BibitemOpen
		\bibfield  {author} {\bibinfo {author} {\bibfnamefont {D.}~\bibnamefont
				{Goll}}, \bibinfo {author} {\bibfnamefont {A.~E.}\ \bibnamefont
				{Berkowitz}},\ and\ \bibinfo {author} {\bibfnamefont {H.~N.}\ \bibnamefont
				{Bertram}},\ }\bibfield  {title} {\bibinfo {title} {Critical sizes for
				ferromagnetic spherical hollow nanoparticles},\ }\href
		{https://doi.org/10.1103/physrevb.70.184432} {\bibfield  {journal} {\bibinfo
				{journal} {Physical Review B}\ }\textbf {\bibinfo {volume} {70}},\ \bibinfo
			{pages} {184432} (\bibinfo {year} {2004})}\BibitemShut {NoStop}%
		\bibitem [{\citenamefont {Beleggia}\ \emph {et~al.}(2006)\citenamefont
			{Beleggia}, \citenamefont {Lau}, \citenamefont {Schofield}, \citenamefont
			{Zhu}, \citenamefont {Tandon},\ and\ \citenamefont {De~Graef}}]{Beleggia06}%
		\BibitemOpen
		\bibfield  {author} {\bibinfo {author} {\bibfnamefont {M.}~\bibnamefont
				{Beleggia}}, \bibinfo {author} {\bibfnamefont {J.}~\bibnamefont {Lau}},
			\bibinfo {author} {\bibfnamefont {M.}~\bibnamefont {Schofield}}, \bibinfo
			{author} {\bibfnamefont {Y.}~\bibnamefont {Zhu}}, \bibinfo {author}
			{\bibfnamefont {S.}~\bibnamefont {Tandon}},\ and\ \bibinfo {author}
			{\bibfnamefont {M.}~\bibnamefont {De~Graef}},\ }\bibfield  {title} {\bibinfo
			{title} {Phase diagram for magnetic nano-rings},\ }\href
		{https://doi.org/10.1016/j.jmmm.2005.06.024} {\bibfield  {journal} {\bibinfo
				{journal} {Journal of Magnetism and Magnetic Materials}\ }\textbf {\bibinfo
				{volume} {301}},\ \bibinfo {pages} {131} (\bibinfo {year}
			{2006})}\BibitemShut {NoStop}%
		\bibitem [{\citenamefont {Vaz}\ \emph {et~al.}(2007)\citenamefont {Vaz},
			\citenamefont {Hayward}, \citenamefont {Llandro}, \citenamefont {Schackert},
			\citenamefont {Morecroft}, \citenamefont {Bland}, \citenamefont {Kl\"{a}ui},
			\citenamefont {Laufenberg}, \citenamefont {Backes}, \citenamefont
			{R\"{u}diger}, \citenamefont {Casta\~{n}o}, \citenamefont {Ross},
			\citenamefont {Heyderman}, \citenamefont {Nolting}, \citenamefont
			{Locatelli}, \citenamefont {Faini}, \citenamefont {Cherifi},\ and\
			\citenamefont {Wernsdorfer}}]{Vaz07}%
		\BibitemOpen
		\bibfield  {author} {\bibinfo {author} {\bibfnamefont {C.~A.~F.}\
				\bibnamefont {Vaz}}, \bibinfo {author} {\bibfnamefont {T.~J.}\ \bibnamefont
				{Hayward}}, \bibinfo {author} {\bibfnamefont {J.}~\bibnamefont {Llandro}},
			\bibinfo {author} {\bibfnamefont {F.}~\bibnamefont {Schackert}}, \bibinfo
			{author} {\bibfnamefont {D.}~\bibnamefont {Morecroft}}, \bibinfo {author}
			{\bibfnamefont {J.~A.~C.}\ \bibnamefont {Bland}}, \bibinfo {author}
			{\bibfnamefont {M.}~\bibnamefont {Kl\"{a}ui}}, \bibinfo {author}
			{\bibfnamefont {M.}~\bibnamefont {Laufenberg}}, \bibinfo {author}
			{\bibfnamefont {D.}~\bibnamefont {Backes}}, \bibinfo {author} {\bibfnamefont
				{U.}~\bibnamefont {R\"{u}diger}}, \bibinfo {author} {\bibfnamefont {F.~J.}\
				\bibnamefont {Casta\~{n}o}}, \bibinfo {author} {\bibfnamefont {C.~A.}\
				\bibnamefont {Ross}}, \bibinfo {author} {\bibfnamefont {L.~J.}\ \bibnamefont
				{Heyderman}}, \bibinfo {author} {\bibfnamefont {F.}~\bibnamefont {Nolting}},
			\bibinfo {author} {\bibfnamefont {A.}~\bibnamefont {Locatelli}}, \bibinfo
			{author} {\bibfnamefont {G.}~\bibnamefont {Faini}}, \bibinfo {author}
			{\bibfnamefont {S.}~\bibnamefont {Cherifi}},\ and\ \bibinfo {author}
			{\bibfnamefont {W.}~\bibnamefont {Wernsdorfer}},\ }\bibfield  {title}
		{\bibinfo {title} {Ferromagnetic nanorings},\ }\href
		{https://doi.org/10.1088/0953-8984/19/25/255207} {\bibfield  {journal}
			{\bibinfo  {journal} {Journal of Physics: Condensed Matter}\ }\textbf
			{\bibinfo {volume} {19}},\ \bibinfo {pages} {255207} (\bibinfo {year}
			{2007})}\BibitemShut {NoStop}%
		\bibitem [{\citenamefont {Kravchuk}\ \emph {et~al.}(2007)\citenamefont
			{Kravchuk}, \citenamefont {Sheka},\ and\ \citenamefont
			{Gaididei}}]{Kravchuk07}%
		\BibitemOpen
		\bibfield  {author} {\bibinfo {author} {\bibfnamefont {V.~P.}\ \bibnamefont
				{Kravchuk}}, \bibinfo {author} {\bibfnamefont {D.~D.}\ \bibnamefont
				{Sheka}},\ and\ \bibinfo {author} {\bibfnamefont {Y.~B.}\ \bibnamefont
				{Gaididei}},\ }\bibfield  {title} {\bibinfo {title} {Equilibrium
				magnetisation structures in ferromagnetic nanorings},\ }\href
		{https://doi.org/10.1016/j.jmmm.2006.07.035} {\bibfield  {journal} {\bibinfo
				{journal} {Journal of Magnetism and Magnetic Materials}\ }\textbf {\bibinfo
				{volume} {310}},\ \bibinfo {pages} {116} (\bibinfo {year}
			{2007})}\BibitemShut {NoStop}%
		\bibitem [{\citenamefont {Landeros}\ \emph {et~al.}(2006)\citenamefont
			{Landeros}, \citenamefont {Escrig}, \citenamefont {Altbir}, \citenamefont
			{Bahiana},\ and\ \citenamefont {d'Albuquerque~e Castro}}]{Landeros06}%
		\BibitemOpen
		\bibfield  {author} {\bibinfo {author} {\bibfnamefont {P.}~\bibnamefont
				{Landeros}}, \bibinfo {author} {\bibfnamefont {J.}~\bibnamefont {Escrig}},
			\bibinfo {author} {\bibfnamefont {D.}~\bibnamefont {Altbir}}, \bibinfo
			{author} {\bibfnamefont {M.}~\bibnamefont {Bahiana}},\ and\ \bibinfo {author}
			{\bibfnamefont {J.}~\bibnamefont {d'Albuquerque~e Castro}},\ }\bibfield
		{title} {\bibinfo {title} {Stability of magnetic configurations in
				nanorings},\ }\href {https://doi.org/10.1063/1.2218997} {\bibfield  {journal}
			{\bibinfo  {journal} {Journal of Applied Physics}\ }\textbf {\bibinfo
				{volume} {100}},\ \bibinfo {pages} {044311} (\bibinfo {year}
			{2006})}\BibitemShut {NoStop}%
		\bibitem [{\citenamefont {Kl\"{a}ui}\ \emph {et~al.}(2001)\citenamefont
			{Kl\"{a}ui}, \citenamefont {Rothman}, \citenamefont {Lopez-Diaz},
			\citenamefont {Vaz}, \citenamefont {Bland},\ and\ \citenamefont
			{Cui}}]{Klaui01}%
		\BibitemOpen
		\bibfield  {author} {\bibinfo {author} {\bibfnamefont {M.}~\bibnamefont
				{Kl\"{a}ui}}, \bibinfo {author} {\bibfnamefont {J.}~\bibnamefont {Rothman}},
			\bibinfo {author} {\bibfnamefont {L.}~\bibnamefont {Lopez-Diaz}}, \bibinfo
			{author} {\bibfnamefont {C.~A.~F.}\ \bibnamefont {Vaz}}, \bibinfo {author}
			{\bibfnamefont {J.~A.~C.}\ \bibnamefont {Bland}},\ and\ \bibinfo {author}
			{\bibfnamefont {Z.}~\bibnamefont {Cui}},\ }\bibfield  {title} {\bibinfo
			{title} {Vortex circulation control in mesoscopic ring magnets},\ }\href
		{https://doi.org/10.1063/1.1361282} {\bibfield  {journal} {\bibinfo
				{journal} {Applied Physics Letters}\ }\textbf {\bibinfo {volume} {78}},\
			\bibinfo {pages} {3268} (\bibinfo {year} {2001})}\BibitemShut {NoStop}%
		\bibitem [{\citenamefont {Kl\"{a}ui}\ \emph {et~al.}(2002)\citenamefont
			{Kl\"{a}ui}, \citenamefont {Vaz}, \citenamefont {Bland}, \citenamefont
			{Wernsdorfer}, \citenamefont {Faini},\ and\ \citenamefont
			{Cambril}}]{Klaui02}%
		\BibitemOpen
		\bibfield  {author} {\bibinfo {author} {\bibfnamefont {M.}~\bibnamefont
				{Kl\"{a}ui}}, \bibinfo {author} {\bibfnamefont {C.~A.~F.}\ \bibnamefont
				{Vaz}}, \bibinfo {author} {\bibfnamefont {J.~A.~C.}\ \bibnamefont {Bland}},
			\bibinfo {author} {\bibfnamefont {W.}~\bibnamefont {Wernsdorfer}}, \bibinfo
			{author} {\bibfnamefont {G.}~\bibnamefont {Faini}},\ and\ \bibinfo {author}
			{\bibfnamefont {E.}~\bibnamefont {Cambril}},\ }\bibfield  {title} {\bibinfo
			{title} {Controlled magnetic switching in single narrow rings probed by
				magnetoresistance measurements},\ }\href {https://doi.org/10.1063/1.1490626}
		{\bibfield  {journal} {\bibinfo  {journal} {Applied Physics Letters}\
			}\textbf {\bibinfo {volume} {81}},\ \bibinfo {pages} {108} (\bibinfo {year}
			{2002})}\BibitemShut {NoStop}%
		\bibitem [{\citenamefont {Kl{\"a}ui}\ \emph
			{et~al.}(2003{\natexlab{b}})\citenamefont {Kl{\"a}ui}, \citenamefont {Vaz},
			\citenamefont {Rothman}, \citenamefont {Bland}, \citenamefont {Wernsdorfer},
			\citenamefont {Faini},\ and\ \citenamefont {Cambril}}]{Klaui03b}%
		\BibitemOpen
		\bibfield  {author} {\bibinfo {author} {\bibfnamefont {M.}~\bibnamefont
				{Kl{\"a}ui}}, \bibinfo {author} {\bibfnamefont {C.~A.~F.}\ \bibnamefont
				{Vaz}}, \bibinfo {author} {\bibfnamefont {J.}~\bibnamefont {Rothman}},
			\bibinfo {author} {\bibfnamefont {J.~A.~C.}\ \bibnamefont {Bland}}, \bibinfo
			{author} {\bibfnamefont {W.}~\bibnamefont {Wernsdorfer}}, \bibinfo {author}
			{\bibfnamefont {G.}~\bibnamefont {Faini}},\ and\ \bibinfo {author}
			{\bibfnamefont {E.}~\bibnamefont {Cambril}},\ }\bibfield  {title} {\bibinfo
			{title} {Domain wall pinning in narrow ferromagnetic ring structures probed
				by magnetoresistance measurements},\ }\href
		{https://doi.org/10.1103/PhysRevLett.90.097202} {\bibfield  {journal}
			{\bibinfo  {journal} {Physical Review Letters}\ }\textbf {\bibinfo {volume}
				{90}},\ \bibinfo {pages} {097202} (\bibinfo {year}
			{2003}{\natexlab{b}})}\BibitemShut {NoStop}%
		\bibitem [{\citenamefont {Kl\"{a}ui}\ \emph
			{et~al.}(2003{\natexlab{a}})\citenamefont {Kl\"{a}ui}, \citenamefont {Vaz},
			\citenamefont {Bland}, \citenamefont {Wernsdorfer}, \citenamefont {Faini},\
			and\ \citenamefont {Cambril}}]{Klaui03a}%
		\BibitemOpen
		\bibfield  {author} {\bibinfo {author} {\bibfnamefont {M.}~\bibnamefont
				{Kl\"{a}ui}}, \bibinfo {author} {\bibfnamefont {C.~A.~F.}\ \bibnamefont
				{Vaz}}, \bibinfo {author} {\bibfnamefont {J.~A.~C.}\ \bibnamefont {Bland}},
			\bibinfo {author} {\bibfnamefont {W.}~\bibnamefont {Wernsdorfer}}, \bibinfo
			{author} {\bibfnamefont {G.}~\bibnamefont {Faini}},\ and\ \bibinfo {author}
			{\bibfnamefont {E.}~\bibnamefont {Cambril}},\ }\bibfield  {title} {\bibinfo
			{title} {Domain wall pinning and controlled magnetic switching in narrow
				ferromagnetic ring structures with notches (invited)},\ }\href
		{https://doi.org/10.1063/1.1557758} {\bibfield  {journal} {\bibinfo
				{journal} {Journal of Applied Physics}\ }\textbf {\bibinfo {volume} {93}},\
			\bibinfo {pages} {7885} (\bibinfo {year} {2003}{\natexlab{a}})}\BibitemShut
		{NoStop}%
		\bibitem [{\citenamefont {Zhu}\ \emph {et~al.}(2006)\citenamefont {Zhu},
			\citenamefont {Chern}, \citenamefont {Tchernyshyov}, \citenamefont {Zhu},
			\citenamefont {Zhu},\ and\ \citenamefont {Chien}}]{Zhu06a}%
		\BibitemOpen
		\bibfield  {author} {\bibinfo {author} {\bibfnamefont {F.~Q.}\ \bibnamefont
				{Zhu}}, \bibinfo {author} {\bibfnamefont {G.~W.}\ \bibnamefont {Chern}},
			\bibinfo {author} {\bibfnamefont {O.}~\bibnamefont {Tchernyshyov}}, \bibinfo
			{author} {\bibfnamefont {X.~C.}\ \bibnamefont {Zhu}}, \bibinfo {author}
			{\bibfnamefont {J.~G.}\ \bibnamefont {Zhu}},\ and\ \bibinfo {author}
			{\bibfnamefont {C.~L.}\ \bibnamefont {Chien}},\ }\bibfield  {title} {\bibinfo
			{title} {Magnetic bistability and controllable reversal of asymmetric
				ferromagnetic nanorings},\ }\href
		{https://doi.org/10.1103/physrevlett.96.027205} {\bibfield  {journal}
			{\bibinfo  {journal} {Physical Review Letters}\ }\textbf {\bibinfo {volume}
				{96}},\ \bibinfo {pages} {027205} (\bibinfo {year} {2006})}\BibitemShut
		{NoStop}%
		\bibitem [{\citenamefont {Kl\"{a}ui}\ \emph
			{et~al.}(2004{\natexlab{a}})\citenamefont {Kl\"{a}ui}, \citenamefont {Vaz},
			\citenamefont {Bland}, \citenamefont {Heyderman}, \citenamefont {David},
			\citenamefont {Sinnecker},\ and\ \citenamefont {Guimarães}}]{Klaui04b}%
		\BibitemOpen
		\bibfield  {author} {\bibinfo {author} {\bibfnamefont {M.}~\bibnamefont
				{Kl\"{a}ui}}, \bibinfo {author} {\bibfnamefont {C.~A.~F.}\ \bibnamefont
				{Vaz}}, \bibinfo {author} {\bibfnamefont {J.~A.~C.}\ \bibnamefont {Bland}},
			\bibinfo {author} {\bibfnamefont {L.~J.}\ \bibnamefont {Heyderman}}, \bibinfo
			{author} {\bibfnamefont {C.}~\bibnamefont {David}}, \bibinfo {author}
			{\bibfnamefont {E.~H. C.~P.}\ \bibnamefont {Sinnecker}},\ and\ \bibinfo
			{author} {\bibfnamefont {A.~P.}\ \bibnamefont {Guimarães}},\ }\bibfield
		{title} {\bibinfo {title} {Multistep switching phase diagram of ferromagnetic
				ring structures},\ }\href {https://doi.org/10.1063/1.1689491} {\bibfield
			{journal} {\bibinfo  {journal} {Journal of Applied Physics}\ }\textbf
			{\bibinfo {volume} {95}},\ \bibinfo {pages} {6639} (\bibinfo {year}
			{2004}{\natexlab{a}})}\BibitemShut {NoStop}%
		\bibitem [{\citenamefont {Vojkovic}\ \emph {et~al.}(2016)\citenamefont
			{Vojkovic}, \citenamefont {Nunez}, \citenamefont {Altbir},\ and\
			\citenamefont {Carvalho-Santos}}]{Vojkovic16}%
		\BibitemOpen
		\bibfield  {author} {\bibinfo {author} {\bibfnamefont {S.}~\bibnamefont
				{Vojkovic}}, \bibinfo {author} {\bibfnamefont {A.~S.}\ \bibnamefont {Nunez}},
			\bibinfo {author} {\bibfnamefont {D.}~\bibnamefont {Altbir}},\ and\ \bibinfo
			{author} {\bibfnamefont {V.~L.}\ \bibnamefont {Carvalho-Santos}},\ }\bibfield
		{title} {\bibinfo {title} {Magnetization ground state and reversal modes of
				magnetic nanotori},\ }\href {https://doi.org/10.1063/1.4958740} {\bibfield
			{journal} {\bibinfo  {journal} {Journal of Applied Physics}\ }\textbf
			{\bibinfo {volume} {120}},\ \bibinfo {pages} {033901} (\bibinfo {year}
			{2016})}\BibitemShut {NoStop}%
		\bibitem [{\citenamefont {Lewis}\ \emph {et~al.}(2020)\citenamefont {Lewis},
			\citenamefont {Loudon}, \citenamefont {Tovey}, \citenamefont {Chen},
			\citenamefont {Roberts}, \citenamefont {Harrison}, \citenamefont {Midgley},\
			and\ \citenamefont {Ringe}}]{Lewis20}%
		\BibitemOpen
		\bibfield  {author} {\bibinfo {author} {\bibfnamefont {G.~R.}\ \bibnamefont
				{Lewis}}, \bibinfo {author} {\bibfnamefont {J.~C.}\ \bibnamefont {Loudon}},
			\bibinfo {author} {\bibfnamefont {R.}~\bibnamefont {Tovey}}, \bibinfo
			{author} {\bibfnamefont {Y.-H.}\ \bibnamefont {Chen}}, \bibinfo {author}
			{\bibfnamefont {A.~P.}\ \bibnamefont {Roberts}}, \bibinfo {author}
			{\bibfnamefont {R.~J.}\ \bibnamefont {Harrison}}, \bibinfo {author}
			{\bibfnamefont {P.~A.}\ \bibnamefont {Midgley}},\ and\ \bibinfo {author}
			{\bibfnamefont {E.}~\bibnamefont {Ringe}},\ }\bibfield  {title} {\bibinfo
			{title} {Magnetic vortex states in toroidal iron oxide nanoparticles:
				Combining micromagnetics with tomography},\ }\href
		{https://doi.org/10.1021/acs.nanolett.0c02795} {\bibfield  {journal}
			{\bibinfo  {journal} {Nano Letters}\ }\textbf {\bibinfo {volume} {20}},\
			\bibinfo {pages} {7405} (\bibinfo {year} {2020})}\BibitemShut {NoStop}%
		\bibitem [{\citenamefont {Morse}\ and\ \citenamefont
			{Feshbach}(1953)}]{Morse53}%
		\BibitemOpen
		\bibfield  {author} {\bibinfo {author} {\bibfnamefont {P.~M.}\ \bibnamefont
				{Morse}}\ and\ \bibinfo {author} {\bibfnamefont {H.}~\bibnamefont
				{Feshbach}},\ }\href@noop {} {\emph {\bibinfo {title} {Methods of theoretical
					physics, part {I} and {II}}}}\ (\bibinfo  {publisher} {McGraw-Hill Science},\
		\bibinfo {address} {New York, NY, USA},\ \bibinfo {year} {1953})\BibitemShut
		{NoStop}%
		\bibitem [{\citenamefont {Gaididei}\ \emph {et~al.}(2014)\citenamefont
			{Gaididei}, \citenamefont {Kravchuk},\ and\ \citenamefont
			{Sheka}}]{Gaididei14}%
		\BibitemOpen
		\bibfield  {author} {\bibinfo {author} {\bibfnamefont {Y.}~\bibnamefont
				{Gaididei}}, \bibinfo {author} {\bibfnamefont {V.~P.}\ \bibnamefont
				{Kravchuk}},\ and\ \bibinfo {author} {\bibfnamefont {D.~D.}\ \bibnamefont
				{Sheka}},\ }\bibfield  {title} {\bibinfo {title} {Curvature effects in thin
				magnetic shells},\ }\href {https://doi.org/10.1103/PhysRevLett.112.257203}
		{\bibfield  {journal} {\bibinfo  {journal} {Physical Review Letters}\
			}\textbf {\bibinfo {volume} {112}},\ \bibinfo {pages} {257203} (\bibinfo
			{year} {2014})}\BibitemShut {NoStop}%
		\bibitem [{\citenamefont {Vojkovic}\ \emph {et~al.}(2017)\citenamefont
			{Vojkovic}, \citenamefont {Carvalho-Santos}, \citenamefont {Fonseca},\ and\
			\citenamefont {Nunez}}]{Vojkovic17}%
		\BibitemOpen
		\bibfield  {author} {\bibinfo {author} {\bibfnamefont {S.}~\bibnamefont
				{Vojkovic}}, \bibinfo {author} {\bibfnamefont {V.~L.}\ \bibnamefont
				{Carvalho-Santos}}, \bibinfo {author} {\bibfnamefont {J.~M.}\ \bibnamefont
				{Fonseca}},\ and\ \bibinfo {author} {\bibfnamefont {A.~S.}\ \bibnamefont
				{Nunez}},\ }\bibfield  {title} {\bibinfo {title} {Vortex-antivortex pairs
				induced by curvature in toroidal nanomagnets},\ }\href
		{https://doi.org/10.1063/1.4977983} {\bibfield  {journal} {\bibinfo
				{journal} {Journal of Applied Physics}\ }\textbf {\bibinfo {volume} {121}},\
			\bibinfo {pages} {113906} (\bibinfo {year} {2017})}\BibitemShut {NoStop}%
		\bibitem [{\citenamefont {Castillo-Sep{\'{u}}lveda}\ \emph
			{et~al.}(2021)\citenamefont {Castillo-Sep{\'{u}}lveda}, \citenamefont
			{Cacilhas}, \citenamefont {Carvalho-Santos}, \citenamefont {Corona},\ and\
			\citenamefont {Altbir}}]{Castillo-Sepulveda21}%
		\BibitemOpen
		\bibfield  {author} {\bibinfo {author} {\bibfnamefont {S.}~\bibnamefont
				{Castillo-Sep{\'{u}}lveda}}, \bibinfo {author} {\bibfnamefont
				{R.}~\bibnamefont {Cacilhas}}, \bibinfo {author} {\bibfnamefont {V.~L.}\
				\bibnamefont {Carvalho-Santos}}, \bibinfo {author} {\bibfnamefont {R.~M.}\
				\bibnamefont {Corona}},\ and\ \bibinfo {author} {\bibfnamefont
				{D.}~\bibnamefont {Altbir}},\ }\bibfield  {title} {\bibinfo {title} {Magnetic
				hopfions in toroidal nanostructures driven by an {O}ersted magnetic field},\
		}\href {https://doi.org/10.1103/physrevb.104.184406} {\bibfield  {journal}
			{\bibinfo  {journal} {Physical Review B}\ }\textbf {\bibinfo {volume}
				{104}},\ \bibinfo {pages} {184406} (\bibinfo {year} {2021})}\BibitemShut
		{NoStop}%
		\bibitem [{\citenamefont {McCord}\ and\ \citenamefont
			{Sch\"{a}fer}(2009)}]{McCord09}%
		\BibitemOpen
		\bibfield  {author} {\bibinfo {author} {\bibfnamefont {J.}~\bibnamefont
				{McCord}}\ and\ \bibinfo {author} {\bibfnamefont {R.}~\bibnamefont
				{Sch\"{a}fer}},\ }\bibfield  {title} {\bibinfo {title} {Domain wall
				asymmetries in {Ni$_{81}$Fe$_{19}$/NiO}: proof of variable anisotropies in
				exchange bias systems},\ }\href
		{https://doi.org/10.1088/1367-2630/11/8/083016} {\bibfield  {journal}
			{\bibinfo  {journal} {New Journal of Physics}\ }\textbf {\bibinfo {volume}
				{11}},\ \bibinfo {pages} {083016} (\bibinfo {year} {2009})}\BibitemShut
		{NoStop}%
		\bibitem [{\citenamefont {Middelhoek}(1963)}]{Middelhoek63}%
		\BibitemOpen
		\bibfield  {author} {\bibinfo {author} {\bibfnamefont {S.}~\bibnamefont
				{Middelhoek}},\ }\bibfield  {title} {\bibinfo {title} {Domain walls in thin
				{Ni--Fe} films},\ }\href {https://doi.org/10.1063/1.1729367} {\bibfield
			{journal} {\bibinfo  {journal} {Journal of Applied Physics}\ }\textbf
			{\bibinfo {volume} {34}},\ \bibinfo {pages} {1054} (\bibinfo {year}
			{1963})}\BibitemShut {NoStop}%
		\bibitem [{\citenamefont {Carvalho-Santos}\ \emph
			{et~al.}(2015{\natexlab{b}})\citenamefont {Carvalho-Santos}, \citenamefont
			{Elias}, \citenamefont {Fonseca},\ and\ \citenamefont
			{Altbir}}]{CarvalhoSantos15a}%
		\BibitemOpen
		\bibfield  {author} {\bibinfo {author} {\bibfnamefont {V.~L.}\ \bibnamefont
				{Carvalho-Santos}}, \bibinfo {author} {\bibfnamefont {R.~G.}\ \bibnamefont
				{Elias}}, \bibinfo {author} {\bibfnamefont {J.~M.}\ \bibnamefont {Fonseca}},\
			and\ \bibinfo {author} {\bibfnamefont {D.}~\bibnamefont {Altbir}},\
		}\bibfield  {title} {\bibinfo {title} {Curvature-induced changes in the
				magnetic energy of vortices and skyrmions in paraboloidal nanoparticles},\
		}\href {https://doi.org/10.1063/1.4918565} {\bibfield  {journal} {\bibinfo
				{journal} {Journal of Applied Physics}\ }\textbf {\bibinfo {volume} {117}},\
			\bibinfo {pages} {17E518} (\bibinfo {year} {2015}{\natexlab{b}})}\BibitemShut
		{NoStop}%
		\bibitem [{\citenamefont {Carvalho-Santos}\ \emph {et~al.}(2020)\citenamefont
			{Carvalho-Santos}, \citenamefont {Corona}, \citenamefont {Altbir},\ and\
			\citenamefont {Castillo-Sep\'ulveda}}]{Carvalho-Santos20}%
		\BibitemOpen
		\bibfield  {author} {\bibinfo {author} {\bibfnamefont {V.~L.}\ \bibnamefont
				{Carvalho-Santos}}, \bibinfo {author} {\bibfnamefont {R.~M.}\ \bibnamefont
				{Corona}}, \bibinfo {author} {\bibfnamefont {D.}~\bibnamefont {Altbir}},\
			and\ \bibinfo {author} {\bibfnamefont {S.}~\bibnamefont
				{Castillo-Sep\'ulveda}},\ }\bibfield  {title} {\bibinfo {title} {Shifts in
				the skyrmion stabilization due to curvature effects in dome- and
				antidome-shaped surfaces},\ }\href
		{https://doi.org/10.1103/PhysRevB.102.024444} {\bibfield  {journal} {\bibinfo
				{journal} {Physical Review B}\ }\textbf {\bibinfo {volume} {102}},\ \bibinfo
			{pages} {024444} (\bibinfo {year} {2020})}\BibitemShut {NoStop}%
		\bibitem [{\citenamefont {El{\'{\i}}as}\ \emph {et~al.}(2019)\citenamefont
			{El{\'{\i}}as}, \citenamefont {Vidal-Silva},\ and\ \citenamefont
			{Carvalho-Santos}}]{Elias19}%
		\BibitemOpen
		\bibfield  {author} {\bibinfo {author} {\bibfnamefont {R.~G.}\ \bibnamefont
				{El{\'{\i}}as}}, \bibinfo {author} {\bibfnamefont {N.}~\bibnamefont
				{Vidal-Silva}},\ and\ \bibinfo {author} {\bibfnamefont {V.~L.}\ \bibnamefont
				{Carvalho-Santos}},\ }\bibfield  {title} {\bibinfo {title} {Winding number
				selection on merons by gaussian curvature's sign},\ }\href
		{https://doi.org/10.1038/s41598-019-50395-7} {\bibfield  {journal} {\bibinfo
				{journal} {Scientific Reports}\ }\textbf {\bibinfo {volume} {9}},\ \bibinfo
			{pages} {14309} (\bibinfo {year} {2019})}\BibitemShut {NoStop}%
		\bibitem [{\citenamefont {Makarov}\ \emph {et~al.}(2022)\citenamefont
			{Makarov}, \citenamefont {Volkov}, \citenamefont {K\'{a}kay}, \citenamefont
			{Pylypovskyi}, \citenamefont {Budinsk\'{a}},\ and\ \citenamefont
			{Dobrovolskiy}}]{Makarov22}%
		\BibitemOpen
		\bibfield  {author} {\bibinfo {author} {\bibfnamefont {D.}~\bibnamefont
				{Makarov}}, \bibinfo {author} {\bibfnamefont {O.~M.}\ \bibnamefont {Volkov}},
			\bibinfo {author} {\bibfnamefont {A.}~\bibnamefont {K\'{a}kay}}, \bibinfo
			{author} {\bibfnamefont {O.~V.}\ \bibnamefont {Pylypovskyi}}, \bibinfo
			{author} {\bibfnamefont {B.}~\bibnamefont {Budinsk\'{a}}},\ and\ \bibinfo
			{author} {\bibfnamefont {O.~V.}\ \bibnamefont {Dobrovolskiy}},\ }\bibfield
		{title} {\bibinfo {title} {New dimension in magnetism and superconductivity:
				{3D} and curvilinear nanoarchitectures},\ }\href
		{https://doi.org/10.1002/adma.202101758} {\bibfield  {journal} {\bibinfo
				{journal} {Advanced Materials}\ }\textbf {\bibinfo {volume} {34}},\ \bibinfo
			{pages} {2101758} (\bibinfo {year} {2022})}\BibitemShut {NoStop}%
		\bibitem [{\citenamefont {Faddeev}\ and\ \citenamefont
			{Niemi}(1997)}]{Faddeev97}%
		\BibitemOpen
		\bibfield  {author} {\bibinfo {author} {\bibfnamefont {L.}~\bibnamefont
				{Faddeev}}\ and\ \bibinfo {author} {\bibfnamefont {A.~J.}\ \bibnamefont
				{Niemi}},\ }\bibfield  {title} {\bibinfo {title} {Stable knot-like structures
				in classical field theory},\ }\href {https://doi.org/10.1038/387058a0}
		{\bibfield  {journal} {\bibinfo  {journal} {Nature}\ }\textbf {\bibinfo
				{volume} {387}},\ \bibinfo {pages} {58} (\bibinfo {year} {1997})}\BibitemShut
		{NoStop}%
		\bibitem [{\citenamefont {Sutcliffe}(2017)}]{Sutcliffe17}%
		\BibitemOpen
		\bibfield  {author} {\bibinfo {author} {\bibfnamefont {P.}~\bibnamefont
				{Sutcliffe}},\ }\bibfield  {title} {\bibinfo {title} {Skyrmion knots in
				frustrated magnets},\ }\href {https://doi.org/10.1103/physrevlett.118.247203}
		{\bibfield  {journal} {\bibinfo  {journal} {Physical Review Letters}\
			}\textbf {\bibinfo {volume} {118}},\ \bibinfo {pages} {247203} (\bibinfo
			{year} {2017})}\BibitemShut {NoStop}%
		\bibitem [{\citenamefont {Rybakov}\ \emph {et~al.}(2022)\citenamefont
			{Rybakov}, \citenamefont {Kiselev}, \citenamefont {Borisov}, \citenamefont
			{D\"{o}ring}, \citenamefont {Melcher},\ and\ \citenamefont
			{Bl\"{u}gel}}]{Rybakov22}%
		\BibitemOpen
		\bibfield  {author} {\bibinfo {author} {\bibfnamefont {F.~N.}\ \bibnamefont
				{Rybakov}}, \bibinfo {author} {\bibfnamefont {N.~S.}\ \bibnamefont
				{Kiselev}}, \bibinfo {author} {\bibfnamefont {A.~B.}\ \bibnamefont
				{Borisov}}, \bibinfo {author} {\bibfnamefont {L.}~\bibnamefont {D\"{o}ring}},
			\bibinfo {author} {\bibfnamefont {C.}~\bibnamefont {Melcher}},\ and\ \bibinfo
			{author} {\bibfnamefont {S.}~\bibnamefont {Bl\"{u}gel}},\ }\bibfield  {title}
		{\bibinfo {title} {Magnetic hopfions in solids},\ }\href
		{https://doi.org/10.1063/5.0099942} {\bibfield  {journal} {\bibinfo
				{journal} {APL Materials}\ }\textbf {\bibinfo {volume} {10}},\ \bibinfo
			{pages} {111113} (\bibinfo {year} {2022})}\BibitemShut {NoStop}%
		\bibitem [{\citenamefont {Guslienko}(2025)}]{Guslienko25}%
		\BibitemOpen
		\bibfield  {author} {\bibinfo {author} {\bibfnamefont {K.~Y.}\ \bibnamefont
				{Guslienko}},\ }\bibfield  {title} {\bibinfo {title} {Magnetic hopfions as
				local rotations of the uniform magnetization background},\ }\href
		{https://doi.org/10.1002/pssr.202500022} {\bibfield  {journal} {\bibinfo
				{journal} {physica status solidi (RRL) – Rapid Research Letters}\ ,\
				\bibinfo {pages} {2500022}} (\bibinfo {year} {2025})}\BibitemShut {NoStop}%
		\bibitem [{\citenamefont {Brevis}\ \emph {et~al.}(2025)\citenamefont {Brevis},
			\citenamefont {K\"{o}rber}, \citenamefont {Mimica-Figari}, \citenamefont
			{Gallardo}, \citenamefont {K\'{a}kay},\ and\ \citenamefont
			{Landeros}}]{Brevis25}%
		\BibitemOpen
		\bibfield  {author} {\bibinfo {author} {\bibfnamefont {F.}~\bibnamefont
				{Brevis}}, \bibinfo {author} {\bibfnamefont {L.}~\bibnamefont {K\"{o}rber}},
			\bibinfo {author} {\bibfnamefont {B.}~\bibnamefont {Mimica-Figari}}, \bibinfo
			{author} {\bibfnamefont {R.}~\bibnamefont {Gallardo}}, \bibinfo {author}
			{\bibfnamefont {A.}~\bibnamefont {K\'{a}kay}},\ and\ \bibinfo {author}
			{\bibfnamefont {P.}~\bibnamefont {Landeros}},\ }\bibfield  {title} {\bibinfo
			{title} {Toroidal moments in confined nanomagnets and their impact on
				magnonics},\ }\href {https://doi.org/10.1103/2pzb-n2q2} {\bibfield  {journal}
			{\bibinfo  {journal} {Physical Review Applied}\ }\textbf {\bibinfo {volume}
				{24}},\ \bibinfo {pages} {024058} (\bibinfo {year} {2025})}\BibitemShut
		{NoStop}%
		\bibitem [{\citenamefont {Saji}\ \emph {et~al.}(2023)\citenamefont {Saji},
			\citenamefont {Troncoso}, \citenamefont {Carvalho-Santos}, \citenamefont
			{Altbir},\ and\ \citenamefont {Nunez}}]{Saji23}%
		\BibitemOpen
		\bibfield  {author} {\bibinfo {author} {\bibfnamefont {C.}~\bibnamefont
				{Saji}}, \bibinfo {author} {\bibfnamefont {R.~E.}\ \bibnamefont {Troncoso}},
			\bibinfo {author} {\bibfnamefont {V.~L.}\ \bibnamefont {Carvalho-Santos}},
			\bibinfo {author} {\bibfnamefont {D.}~\bibnamefont {Altbir}},\ and\ \bibinfo
			{author} {\bibfnamefont {A.~S.}\ \bibnamefont {Nunez}},\ }\bibfield  {title}
		{\bibinfo {title} {Hopfion-driven magnonic hall effect and magnonic
				focusing},\ }\href {https://doi.org/10.1103/physrevlett.131.166702}
		{\bibfield  {journal} {\bibinfo  {journal} {Physical Review Letters}\
			}\textbf {\bibinfo {volume} {131}},\ \bibinfo {pages} {166702} (\bibinfo
			{year} {2023})}\BibitemShut {NoStop}%
		\bibitem [{\citenamefont {Bo}\ \emph {et~al.}(2021)\citenamefont {Bo},
			\citenamefont {Ji}, \citenamefont {Hu}, \citenamefont {Zhao}, \citenamefont
			{Li}, \citenamefont {Zhang},\ and\ \citenamefont {Zhang}}]{Bo21}%
		\BibitemOpen
		\bibfield  {author} {\bibinfo {author} {\bibfnamefont {L.}~\bibnamefont
				{Bo}}, \bibinfo {author} {\bibfnamefont {L.}~\bibnamefont {Ji}}, \bibinfo
			{author} {\bibfnamefont {C.}~\bibnamefont {Hu}}, \bibinfo {author}
			{\bibfnamefont {R.}~\bibnamefont {Zhao}}, \bibinfo {author} {\bibfnamefont
				{Y.}~\bibnamefont {Li}}, \bibinfo {author} {\bibfnamefont {J.}~\bibnamefont
				{Zhang}},\ and\ \bibinfo {author} {\bibfnamefont {X.}~\bibnamefont {Zhang}},\
		}\bibfield  {title} {\bibinfo {title} {Spin excitation spectrum of a magnetic
				hopfion},\ }\href {https://doi.org/10.1063/5.0072349} {\bibfield  {journal}
			{\bibinfo  {journal} {Applied Physics Letters}\ }\textbf {\bibinfo {volume}
				{119}},\ \bibinfo {pages} {212408} (\bibinfo {year} {2021})}\BibitemShut
		{NoStop}%
		\bibitem [{\citenamefont {Raftrey}\ and\ \citenamefont
			{Fischer}(2021)}]{Raftrey21}%
		\BibitemOpen
		\bibfield  {author} {\bibinfo {author} {\bibfnamefont {D.}~\bibnamefont
				{Raftrey}}\ and\ \bibinfo {author} {\bibfnamefont {P.}~\bibnamefont
				{Fischer}},\ }\bibfield  {title} {\bibinfo {title} {Field-driven dynamics of
				magnetic hopfions},\ }\href {https://doi.org/10.1103/physrevlett.127.257201}
		{\bibfield  {journal} {\bibinfo  {journal} {Physical Review Letters}\
			}\textbf {\bibinfo {volume} {127}},\ \bibinfo {pages} {257201} (\bibinfo
			{year} {2021})}\BibitemShut {NoStop}%
		\bibitem [{\citenamefont {Wu}\ and\ \citenamefont {Lan}(2025)}]{Wu25}%
		\BibitemOpen
		\bibfield  {author} {\bibinfo {author} {\bibfnamefont {H.}~\bibnamefont
				{Wu}}\ and\ \bibinfo {author} {\bibfnamefont {J.}~\bibnamefont {Lan}},\
		}\bibfield  {title} {\bibinfo {title} {Spin-wave vortex as a topological
				probe of magnetic textures},\ }\href {https://doi.org/10.1103/bvvv-kk4j}
		{\bibfield  {journal} {\bibinfo  {journal} {Physical Review B}\ }\textbf
			{\bibinfo {volume} {112}},\ \bibinfo {pages} {064410} (\bibinfo {year}
			{2025})}\BibitemShut {NoStop}%
		\bibitem [{\citenamefont {Liu}\ \emph {et~al.}(2018)\citenamefont {Liu},
			\citenamefont {Lake},\ and\ \citenamefont {Zang}}]{Liu18a}%
		\BibitemOpen
		\bibfield  {author} {\bibinfo {author} {\bibfnamefont {Y.}~\bibnamefont
				{Liu}}, \bibinfo {author} {\bibfnamefont {R.~K.}\ \bibnamefont {Lake}},\ and\
			\bibinfo {author} {\bibfnamefont {J.}~\bibnamefont {Zang}},\ }\bibfield
		{title} {\bibinfo {title} {Binding a hopfion in a chiral magnet nanodisk},\
		}\href {https://doi.org/10.1103/physrevb.98.174437} {\bibfield  {journal}
			{\bibinfo  {journal} {Physical Review B}\ }\textbf {\bibinfo {volume} {98}},\
			\bibinfo {pages} {174437} (\bibinfo {year} {2018})}\BibitemShut {NoStop}%
		\bibitem [{\citenamefont {G\"{o}bel}\ \emph {et~al.}(2020)\citenamefont
			{G\"{o}bel}, \citenamefont {Akosa}, \citenamefont {Tatara},\ and\
			\citenamefont {Mertig}}]{Gobel20a}%
		\BibitemOpen
		\bibfield  {author} {\bibinfo {author} {\bibfnamefont {B.}~\bibnamefont
				{G\"{o}bel}}, \bibinfo {author} {\bibfnamefont {C.~A.}\ \bibnamefont
				{Akosa}}, \bibinfo {author} {\bibfnamefont {G.}~\bibnamefont {Tatara}},\ and\
			\bibinfo {author} {\bibfnamefont {I.}~\bibnamefont {Mertig}},\ }\bibfield
		{title} {\bibinfo {title} {Topological {H}all signatures of magnetic
				hopfions},\ }\href {https://doi.org/10.1103/physrevresearch.2.013315}
		{\bibfield  {journal} {\bibinfo  {journal} {Physical Review Research}\
			}\textbf {\bibinfo {volume} {2}},\ \bibinfo {pages} {013315} (\bibinfo {year}
			{2020})}\BibitemShut {NoStop}%
		\bibitem [{\citenamefont {Liu}\ \emph {et~al.}(2020{\natexlab{b}})\citenamefont
			{Liu}, \citenamefont {Hou}, \citenamefont {Han},\ and\ \citenamefont
			{Zang}}]{Liu20a}%
		\BibitemOpen
		\bibfield  {author} {\bibinfo {author} {\bibfnamefont {Y.}~\bibnamefont
				{Liu}}, \bibinfo {author} {\bibfnamefont {W.}~\bibnamefont {Hou}}, \bibinfo
			{author} {\bibfnamefont {X.}~\bibnamefont {Han}},\ and\ \bibinfo {author}
			{\bibfnamefont {J.}~\bibnamefont {Zang}},\ }\bibfield  {title} {\bibinfo
			{title} {Three-dimensional dynamics of a magnetic hopfion driven by spin
				transfer torque},\ }\href {https://doi.org/10.1103/PhysRevLett.124.127204}
		{\bibfield  {journal} {\bibinfo  {journal} {Physical Review Letters}\
			}\textbf {\bibinfo {volume} {124}},\ \bibinfo {pages} {127204} (\bibinfo
			{year} {2020}{\natexlab{b}})}\BibitemShut {NoStop}%
		\bibitem [{\citenamefont {Liu}\ \emph {et~al.}(2022)\citenamefont {Liu},
			\citenamefont {Watanabe},\ and\ \citenamefont {Nagaosa}}]{Liu22d}%
		\BibitemOpen
		\bibfield  {author} {\bibinfo {author} {\bibfnamefont {Y.}~\bibnamefont
				{Liu}}, \bibinfo {author} {\bibfnamefont {H.}~\bibnamefont {Watanabe}},\ and\
			\bibinfo {author} {\bibfnamefont {N.}~\bibnamefont {Nagaosa}},\ }\bibfield
		{title} {\bibinfo {title} {Emergent magnetomultipoles and nonlinear responses
				of a magnetic hopfion},\ }\href
		{https://doi.org/10.1103/physrevlett.129.267201} {\bibfield  {journal}
			{\bibinfo  {journal} {Physical Review Letters}\ }\textbf {\bibinfo {volume}
				{129}},\ \bibinfo {pages} {267201} (\bibinfo {year} {2022})}\BibitemShut
		{NoStop}%
		\bibitem [{\citenamefont {Khodzhaev}\ and\ \citenamefont
			{Turgut}(2022)}]{Khodzhaev22}%
		\BibitemOpen
		\bibfield  {author} {\bibinfo {author} {\bibfnamefont {Z.}~\bibnamefont
				{Khodzhaev}}\ and\ \bibinfo {author} {\bibfnamefont {E.}~\bibnamefont
				{Turgut}},\ }\bibfield  {title} {\bibinfo {title} {Hopfion dynamics in chiral
				magnets},\ }\href {https://doi.org/10.1088/1361-648x/ac533d} {\bibfield
			{journal} {\bibinfo  {journal} {Journal of Physics: Condensed Matter}\
			}\textbf {\bibinfo {volume} {34}},\ \bibinfo {pages} {225805} (\bibinfo
			{year} {2022})}\BibitemShut {NoStop}%
		\bibitem [{\citenamefont {Zheng}\ \emph {et~al.}(2017)\citenamefont {Zheng},
			\citenamefont {Li}, \citenamefont {Wang}, \citenamefont {Song}, \citenamefont
			{Jin}, \citenamefont {Wei}, \citenamefont {Kov\'acs}, \citenamefont {Zang},
			\citenamefont {Tian}, \citenamefont {Zhang}, \citenamefont {Du},\ and\
			\citenamefont {Dunin-Borkowski}}]{Zheng17a}%
		\BibitemOpen
		\bibfield  {author} {\bibinfo {author} {\bibfnamefont {F.}~\bibnamefont
				{Zheng}}, \bibinfo {author} {\bibfnamefont {H.}~\bibnamefont {Li}}, \bibinfo
			{author} {\bibfnamefont {S.}~\bibnamefont {Wang}}, \bibinfo {author}
			{\bibfnamefont {D.}~\bibnamefont {Song}}, \bibinfo {author} {\bibfnamefont
				{C.}~\bibnamefont {Jin}}, \bibinfo {author} {\bibfnamefont {W.}~\bibnamefont
				{Wei}}, \bibinfo {author} {\bibfnamefont {A.}~\bibnamefont {Kov\'acs}},
			\bibinfo {author} {\bibfnamefont {J.}~\bibnamefont {Zang}}, \bibinfo {author}
			{\bibfnamefont {M.}~\bibnamefont {Tian}}, \bibinfo {author} {\bibfnamefont
				{Y.}~\bibnamefont {Zhang}}, \bibinfo {author} {\bibfnamefont
				{H.}~\bibnamefont {Du}},\ and\ \bibinfo {author} {\bibfnamefont {R.~E.}\
				\bibnamefont {Dunin-Borkowski}},\ }\bibfield  {title} {\bibinfo {title}
			{Direct imaging of a zero-field target skyrmion and its polarity switch in a
				chiral magnetic nanodisk},\ }\href
		{https://doi.org/10.1103/physrevlett.119.197205} {\bibfield  {journal}
			{\bibinfo  {journal} {Physical Review Letters}\ }\textbf {\bibinfo {volume}
				{119}},\ \bibinfo {pages} {197205} (\bibinfo {year} {2017})}\BibitemShut
		{NoStop}%
		\bibitem [{\citenamefont {Ackerman}\ and\ \citenamefont
			{Smalyukh}(2017)}]{Ackerman17}%
		\BibitemOpen
		\bibfield  {author} {\bibinfo {author} {\bibfnamefont {P.~J.}\ \bibnamefont
				{Ackerman}}\ and\ \bibinfo {author} {\bibfnamefont {I.~I.}\ \bibnamefont
				{Smalyukh}},\ }\bibfield  {title} {\bibinfo {title} {Diversity of knot
				solitons in liquid crystals manifested by linking of preimages in torons and
				hopfions},\ }\href {https://doi.org/10.1103/physrevx.7.011006} {\bibfield
			{journal} {\bibinfo  {journal} {Physical Review X}\ }\textbf {\bibinfo
				{volume} {7}},\ \bibinfo {pages} {011006} (\bibinfo {year}
			{2017})}\BibitemShut {NoStop}%
		\bibitem [{\citenamefont {Tai}\ and\ \citenamefont {Smalyukh}(2020)}]{Tai20}%
		\BibitemOpen
		\bibfield  {author} {\bibinfo {author} {\bibfnamefont {J.-S.~B.}\
				\bibnamefont {Tai}}\ and\ \bibinfo {author} {\bibfnamefont {I.~I.}\
				\bibnamefont {Smalyukh}},\ }\bibfield  {title} {\bibinfo {title} {Surface
				anchoring as a control parameter for stabilizing torons, skyrmions, twisted
				walls, fingers, and their hybrids in chiral nematics},\ }\href
		{https://doi.org/10.1103/physreve.101.042702} {\bibfield  {journal} {\bibinfo
				{journal} {Physical Review E}\ }\textbf {\bibinfo {volume} {101}},\ \bibinfo
			{pages} {042702} (\bibinfo {year} {2020})}\BibitemShut {NoStop}%
		\bibitem [{\citenamefont {Hu}\ \emph {et~al.}(2021)\citenamefont {Hu},
			\citenamefont {Lyu}, \citenamefont {Tang}, \citenamefont {Kong},
			\citenamefont {Du},\ and\ \citenamefont {Wang}}]{Hu21}%
		\BibitemOpen
		\bibfield  {author} {\bibinfo {author} {\bibfnamefont {Q.}~\bibnamefont
				{Hu}}, \bibinfo {author} {\bibfnamefont {B.}~\bibnamefont {Lyu}}, \bibinfo
			{author} {\bibfnamefont {J.}~\bibnamefont {Tang}}, \bibinfo {author}
			{\bibfnamefont {L.}~\bibnamefont {Kong}}, \bibinfo {author} {\bibfnamefont
				{H.}~\bibnamefont {Du}},\ and\ \bibinfo {author} {\bibfnamefont
				{W.}~\bibnamefont {Wang}},\ }\bibfield  {title} {\bibinfo {title}
			{Unidirectional current-driven toron motion in a cylindrical nanowire},\
		}\href {https://doi.org/10.1063/5.0033239} {\bibfield  {journal} {\bibinfo
				{journal} {Applied Physics Letters}\ }\textbf {\bibinfo {volume} {118}},\
			\bibinfo {pages} {022404} (\bibinfo {year} {2021})}\BibitemShut {NoStop}%
		\bibitem [{\citenamefont {Li}\ \emph {et~al.}(2022)\citenamefont {Li},
			\citenamefont {Xia}, \citenamefont {Shen}, \citenamefont {Zhang},
			\citenamefont {Ezawa},\ and\ \citenamefont {Zhou}}]{Li22d}%
		\BibitemOpen
		\bibfield  {author} {\bibinfo {author} {\bibfnamefont {S.}~\bibnamefont
				{Li}}, \bibinfo {author} {\bibfnamefont {J.}~\bibnamefont {Xia}}, \bibinfo
			{author} {\bibfnamefont {L.}~\bibnamefont {Shen}}, \bibinfo {author}
			{\bibfnamefont {X.}~\bibnamefont {Zhang}}, \bibinfo {author} {\bibfnamefont
				{M.}~\bibnamefont {Ezawa}},\ and\ \bibinfo {author} {\bibfnamefont
				{Y.}~\bibnamefont {Zhou}},\ }\bibfield  {title} {\bibinfo {title} {Mutual
				conversion between a magnetic {N}\'{e}el hopfion and a {N}\'{e}el toron},\
		}\href {https://doi.org/10.1103/physrevb.105.174407} {\bibfield  {journal}
			{\bibinfo  {journal} {Physical Review B}\ }\textbf {\bibinfo {volume}
				{105}},\ \bibinfo {pages} {174407} (\bibinfo {year} {2022})}\BibitemShut
		{NoStop}%
		\bibitem [{\citenamefont {Saavedra}\ \emph
			{et~al.}(2024{\natexlab{a}})\citenamefont {Saavedra}, \citenamefont {Corona},
			\citenamefont {Castillo-Sep\'{u}lveda}, \citenamefont {Laroze}, \citenamefont
			{Altbir},\ and\ \citenamefont {Carvalho-Santos}}]{Saavedra24a}%
		\BibitemOpen
		\bibfield  {author} {\bibinfo {author} {\bibfnamefont {E.}~\bibnamefont
				{Saavedra}}, \bibinfo {author} {\bibfnamefont {R.~M.}\ \bibnamefont
				{Corona}}, \bibinfo {author} {\bibfnamefont {S.}~\bibnamefont
				{Castillo-Sep\'{u}lveda}}, \bibinfo {author} {\bibfnamefont {D.}~\bibnamefont
				{Laroze}}, \bibinfo {author} {\bibfnamefont {D.}~\bibnamefont {Altbir}},\
			and\ \bibinfo {author} {\bibfnamefont {V.~L.}\ \bibnamefont
				{Carvalho-Santos}},\ }\bibfield  {title} {\bibinfo {title} {Micromagnetic
				study of the dynamics of toron chains},\ }\href
		{https://doi.org/10.1063/5.0216480} {\bibfield  {journal} {\bibinfo
				{journal} {Applied Physics Letters}\ }\textbf {\bibinfo {volume} {125}},\
			\bibinfo {pages} {182401} (\bibinfo {year} {2024}{\natexlab{a}})}\BibitemShut
		{NoStop}%
		\bibitem [{\citenamefont {Kuchkin}\ \emph {et~al.}(2025)\citenamefont
			{Kuchkin}, \citenamefont {Kiselev}, \citenamefont {Haller}, \citenamefont
			{Li\v{s}\v{c}\'{a}k}, \citenamefont {Michels},\ and\ \citenamefont
			{Schmidt}}]{Kuchkin25}%
		\BibitemOpen
		\bibfield  {author} {\bibinfo {author} {\bibfnamefont {V.~M.}\ \bibnamefont
				{Kuchkin}}, \bibinfo {author} {\bibfnamefont {N.~S.}\ \bibnamefont
				{Kiselev}}, \bibinfo {author} {\bibfnamefont {A.}~\bibnamefont {Haller}},
			\bibinfo {author} {\bibfnamefont {v.}~\bibnamefont {Li\v{s}\v{c}\'{a}k}},
			\bibinfo {author} {\bibfnamefont {A.}~\bibnamefont {Michels}},\ and\ \bibinfo
			{author} {\bibfnamefont {T.~L.}\ \bibnamefont {Schmidt}},\ }\bibfield
		{title} {\bibinfo {title} {Stability and nucleation of dipole strings in
				uniaxial chiral magnets},\ }\href
		{https://doi.org/10.1103/physrevb.111.174410} {\bibfield  {journal} {\bibinfo
				{journal} {Physical Review B}\ }\textbf {\bibinfo {volume} {111}},\ \bibinfo
			{pages} {174410} (\bibinfo {year} {2025})}\BibitemShut {NoStop}%
		\bibitem [{\citenamefont {Liu}\ and\ \citenamefont {Nagaosa}(2024)}]{Liu24d}%
		\BibitemOpen
		\bibfield  {author} {\bibinfo {author} {\bibfnamefont {Y.}~\bibnamefont
				{Liu}}\ and\ \bibinfo {author} {\bibfnamefont {N.}~\bibnamefont {Nagaosa}},\
		}\bibfield  {title} {\bibinfo {title} {Current-induced creation of
				topological vortex rings in a magnetic nanocylinder},\ }\href
		{https://doi.org/10.1103/physrevlett.132.126701} {\bibfield  {journal}
			{\bibinfo  {journal} {Physical Review Letters}\ }\textbf {\bibinfo {volume}
				{132}},\ \bibinfo {pages} {126701} (\bibinfo {year} {2024})}\BibitemShut
		{NoStop}%
		\bibitem [{\citenamefont {Vidal-Silva}\ \emph {et~al.}(2019)\citenamefont
			{Vidal-Silva}, \citenamefont {Riveros}, \citenamefont {Tejo}, \citenamefont
			{Escrig},\ and\ \citenamefont {Altbir}}]{VidalSilva19}%
		\BibitemOpen
		\bibfield  {author} {\bibinfo {author} {\bibfnamefont {N.}~\bibnamefont
				{Vidal-Silva}}, \bibinfo {author} {\bibfnamefont {A.}~\bibnamefont
				{Riveros}}, \bibinfo {author} {\bibfnamefont {F.}~\bibnamefont {Tejo}},
			\bibinfo {author} {\bibfnamefont {J.}~\bibnamefont {Escrig}},\ and\ \bibinfo
			{author} {\bibfnamefont {D.}~\bibnamefont {Altbir}},\ }\bibfield  {title}
		{\bibinfo {title} {Controlling the nucleation and annihilation of skyrmions
				with magnetostatic interactions},\ }\href {https://doi.org/10.1063/1.5099898}
		{\bibfield  {journal} {\bibinfo  {journal} {Applied Physics Letters}\
			}\textbf {\bibinfo {volume} {115}},\ \bibinfo {pages} {082405} (\bibinfo
			{year} {2019})}\BibitemShut {NoStop}%
		\bibitem [{\citenamefont {Saavedra}\ \emph
			{et~al.}(2024{\natexlab{b}})\citenamefont {Saavedra}, \citenamefont {Altbir},
			\citenamefont {Escrig}, \citenamefont {Castillo-Sep\'{u}lveda}, \citenamefont
			{Corona},\ and\ \citenamefont {Carvalho-Santos}}]{Saavedra24b}%
		\BibitemOpen
		\bibfield  {author} {\bibinfo {author} {\bibfnamefont {E.}~\bibnamefont
				{Saavedra}}, \bibinfo {author} {\bibfnamefont {D.}~\bibnamefont {Altbir}},
			\bibinfo {author} {\bibfnamefont {J.}~\bibnamefont {Escrig}}, \bibinfo
			{author} {\bibfnamefont {S.}~\bibnamefont {Castillo-Sep\'{u}lveda}}, \bibinfo
			{author} {\bibfnamefont {R.}~\bibnamefont {Corona}},\ and\ \bibinfo {author}
			{\bibfnamefont {V.}~\bibnamefont {Carvalho-Santos}},\ }\bibfield  {title}
		{\bibinfo {title} {Exploring hopfion stability and dynamics in ring-like
				structures},\ }\href {https://doi.org/10.1016/j.rinp.2024.107835} {\bibfield
			{journal} {\bibinfo  {journal} {Results in Physics}\ }\textbf {\bibinfo
				{volume} {62}},\ \bibinfo {pages} {107835} (\bibinfo {year}
			{2024}{\natexlab{b}})}\BibitemShut {NoStop}%
		\bibitem [{\citenamefont {Tai}\ and\ \citenamefont {Smalyukh}(2018)}]{Tai18}%
		\BibitemOpen
		\bibfield  {author} {\bibinfo {author} {\bibfnamefont {J.-S.~B.}\
				\bibnamefont {Tai}}\ and\ \bibinfo {author} {\bibfnamefont {I.~I.}\
				\bibnamefont {Smalyukh}},\ }\bibfield  {title} {\bibinfo {title} {Static hopf
				solitons and knotted emergent fields in solid-state noncentrosymmetric
				magnetic nanostructures},\ }\href
		{https://doi.org/10.1103/physrevlett.121.187201} {\bibfield  {journal}
			{\bibinfo  {journal} {Physical Review Letters}\ }\textbf {\bibinfo {volume}
				{121}},\ \bibinfo {pages} {187201} (\bibinfo {year} {2018})}\BibitemShut
		{NoStop}%
		\bibitem [{\citenamefont {Pathak}\ and\ \citenamefont
			{Hertel}(2021)}]{Pathak21}%
		\BibitemOpen
		\bibfield  {author} {\bibinfo {author} {\bibfnamefont {S.~A.}\ \bibnamefont
				{Pathak}}\ and\ \bibinfo {author} {\bibfnamefont {R.}~\bibnamefont
				{Hertel}},\ }\bibfield  {title} {\bibinfo {title} {Three-dimensional chiral
				magnetization structures in {FeGe} nanospheres},\ }\href
		{https://doi.org/10.1103/physrevb.103.104414} {\bibfield  {journal} {\bibinfo
				{journal} {Physical Review B}\ }\textbf {\bibinfo {volume} {103}},\ \bibinfo
			{pages} {104414} (\bibinfo {year} {2021})}\BibitemShut {NoStop}%
		\bibitem [{\citenamefont {Wilczek}\ and\ \citenamefont
			{Zee}(1983)}]{Wilczek83}%
		\BibitemOpen
		\bibfield  {author} {\bibinfo {author} {\bibfnamefont {F.}~\bibnamefont
				{Wilczek}}\ and\ \bibinfo {author} {\bibfnamefont {A.}~\bibnamefont {Zee}},\
		}\bibfield  {title} {\bibinfo {title} {Linking numbers, spin, and statistics
				of solitons},\ }\href {https://doi.org/10.1103/PhysRevLett.51.2250}
		{\bibfield  {journal} {\bibinfo  {journal} {Physical Review Letters}\
			}\textbf {\bibinfo {volume} {51}},\ \bibinfo {pages} {2250} (\bibinfo {year}
			{1983})}\BibitemShut {NoStop}%
		\bibitem [{\citenamefont {Rothman}\ \emph {et~al.}(2001)\citenamefont
			{Rothman}, \citenamefont {Kl{\"a}ui}, \citenamefont {Lopez-Diaz},
			\citenamefont {Vaz}, \citenamefont {Bleloch}, \citenamefont {Bland},
			\citenamefont {Cui},\ and\ \citenamefont {Speaks}}]{Rothman01}%
		\BibitemOpen
		\bibfield  {author} {\bibinfo {author} {\bibfnamefont {J.}~\bibnamefont
				{Rothman}}, \bibinfo {author} {\bibfnamefont {M.}~\bibnamefont {Kl{\"a}ui}},
			\bibinfo {author} {\bibfnamefont {L.}~\bibnamefont {Lopez-Diaz}}, \bibinfo
			{author} {\bibfnamefont {C.~A.~F.}\ \bibnamefont {Vaz}}, \bibinfo {author}
			{\bibfnamefont {A.}~\bibnamefont {Bleloch}}, \bibinfo {author} {\bibfnamefont
				{J.~A.~C.}\ \bibnamefont {Bland}}, \bibinfo {author} {\bibfnamefont
				{Z.}~\bibnamefont {Cui}},\ and\ \bibinfo {author} {\bibfnamefont
				{R.}~\bibnamefont {Speaks}},\ }\bibfield  {title} {\bibinfo {title}
			{Observation of a bi-domain state and nucleation free switching in mesoscopic
				ring magnets},\ }\href {https://doi.org/10.1103/physrevlett.86.1098}
		{\bibfield  {journal} {\bibinfo  {journal} {Physical Review Letters}\
			}\textbf {\bibinfo {volume} {86}},\ \bibinfo {pages} {1098} (\bibinfo {year}
			{2001})}\BibitemShut {NoStop}%
		\bibitem [{\citenamefont {Nee}(1996)}]{Nee96}%
		\BibitemOpen
		\bibfield  {author} {\bibinfo {author} {\bibfnamefont {S.-M.~F.}\
				\bibnamefont {Nee}},\ }\bibfield  {title} {\bibinfo {title} {Polarization of
				specular reflection and near-specular scattering by a rough surface},\ }\href
		{https://doi.org/10.1364/ao.35.003570} {\bibfield  {journal} {\bibinfo
				{journal} {Applied Optics}\ }\textbf {\bibinfo {volume} {35}},\ \bibinfo
			{pages} {3570} (\bibinfo {year} {1996})}\BibitemShut {NoStop}%
		\bibitem [{\citenamefont {Kl\"{a}ui}\ \emph
			{et~al.}(2004{\natexlab{b}})\citenamefont {Kl\"{a}ui}, \citenamefont {Vaz},
			\citenamefont {Bland}, \citenamefont {Sinnecker}, \citenamefont {Guimarães},
			\citenamefont {Wernsdorfer}, \citenamefont {Faini}, \citenamefont {Cambril},
			\citenamefont {Heyderman},\ and\ \citenamefont {David}}]{Klaui04}%
		\BibitemOpen
		\bibfield  {author} {\bibinfo {author} {\bibfnamefont {M.}~\bibnamefont
				{Kl\"{a}ui}}, \bibinfo {author} {\bibfnamefont {C.~A.~F.}\ \bibnamefont
				{Vaz}}, \bibinfo {author} {\bibfnamefont {J.~A.~C.}\ \bibnamefont {Bland}},
			\bibinfo {author} {\bibfnamefont {E.~H. C.~P.}\ \bibnamefont {Sinnecker}},
			\bibinfo {author} {\bibfnamefont {A.~P.}\ \bibnamefont {Guimarães}},
			\bibinfo {author} {\bibfnamefont {W.}~\bibnamefont {Wernsdorfer}}, \bibinfo
			{author} {\bibfnamefont {G.}~\bibnamefont {Faini}}, \bibinfo {author}
			{\bibfnamefont {E.}~\bibnamefont {Cambril}}, \bibinfo {author} {\bibfnamefont
				{L.~J.}\ \bibnamefont {Heyderman}},\ and\ \bibinfo {author} {\bibfnamefont
				{C.}~\bibnamefont {David}},\ }\bibfield  {title} {\bibinfo {title} {Switching
				processes and switching reproducibility in ferromagnetic ring structures},\
		}\href {https://doi.org/10.1063/1.1640451} {\bibfield  {journal} {\bibinfo
				{journal} {Applied Physics Letters}\ }\textbf {\bibinfo {volume} {84}},\
			\bibinfo {pages} {951} (\bibinfo {year} {2004}{\natexlab{b}})}\BibitemShut
		{NoStop}%
		\bibitem [{\citenamefont {Yoo}\ \emph {et~al.}(2003)\citenamefont {Yoo},
			\citenamefont {Kl\"{a}ui}, \citenamefont {Vaz}, \citenamefont {Heyderman},\
			and\ \citenamefont {Bland}}]{Yoo03}%
		\BibitemOpen
		\bibfield  {author} {\bibinfo {author} {\bibfnamefont {Y.~G.}\ \bibnamefont
				{Yoo}}, \bibinfo {author} {\bibfnamefont {M.}~\bibnamefont {Kl\"{a}ui}},
			\bibinfo {author} {\bibfnamefont {C.~A.~F.}\ \bibnamefont {Vaz}}, \bibinfo
			{author} {\bibfnamefont {L.~J.}\ \bibnamefont {Heyderman}},\ and\ \bibinfo
			{author} {\bibfnamefont {J.~A.~C.}\ \bibnamefont {Bland}},\ }\bibfield
		{title} {\bibinfo {title} {Switching field phase diagram of {Co} nanoring
				magnets},\ }\href {https://doi.org/10.1063/1.1568167} {\bibfield  {journal}
			{\bibinfo  {journal} {Applied Physics Letters}\ }\textbf {\bibinfo {volume}
				{82}},\ \bibinfo {pages} {2470} (\bibinfo {year} {2003})}\BibitemShut
		{NoStop}%
		\bibitem [{\citenamefont {Kl\"{a}ui}\ \emph
			{et~al.}(2005{\natexlab{a}})\citenamefont {Kl\"{a}ui}, \citenamefont {Vaz},
			\citenamefont {Bland}, \citenamefont {Wernsdorfer}, \citenamefont {Faini},
			\citenamefont {Cambril}, \citenamefont {Heyderman}, \citenamefont {Nolting},\
			and\ \citenamefont {R\"{u}diger}}]{Klaui05b}%
		\BibitemOpen
		\bibfield  {author} {\bibinfo {author} {\bibfnamefont {M.}~\bibnamefont
				{Kl\"{a}ui}}, \bibinfo {author} {\bibfnamefont {C.~A.~F.}\ \bibnamefont
				{Vaz}}, \bibinfo {author} {\bibfnamefont {J.~A.~C.}\ \bibnamefont {Bland}},
			\bibinfo {author} {\bibfnamefont {W.}~\bibnamefont {Wernsdorfer}}, \bibinfo
			{author} {\bibfnamefont {G.}~\bibnamefont {Faini}}, \bibinfo {author}
			{\bibfnamefont {E.}~\bibnamefont {Cambril}}, \bibinfo {author} {\bibfnamefont
				{L.~J.}\ \bibnamefont {Heyderman}}, \bibinfo {author} {\bibfnamefont
				{F.}~\bibnamefont {Nolting}},\ and\ \bibinfo {author} {\bibfnamefont
				{U.}~\bibnamefont {R\"{u}diger}},\ }\bibfield  {title} {\bibinfo {title}
			{Controlled and reproducible domain wall displacement by current pulses
				injected into ferromagnetic ring structures},\ }\href
		{https://doi.org/10.1103/PhysRevLett.94.106601} {\bibfield  {journal}
			{\bibinfo  {journal} {Physical Review Letters}\ }\textbf {\bibinfo {volume}
				{94}},\ \bibinfo {pages} {106601} (\bibinfo {year}
			{2005}{\natexlab{a}})}\BibitemShut {NoStop}%
		\bibitem [{\citenamefont {Vidamour}\ \emph {et~al.}(2023)\citenamefont
			{Vidamour}, \citenamefont {Swindells}, \citenamefont {Venkat}, \citenamefont
			{Manneschi}, \citenamefont {Fry}, \citenamefont {Welbourne}, \citenamefont
			{Rowan-Robinson}, \citenamefont {Backes}, \citenamefont {Maccherozzi},
			\citenamefont {Dhesi}, \citenamefont {Vasilaki}, \citenamefont {Allwood},\
			and\ \citenamefont {Hayward}}]{Vidamour23}%
		\BibitemOpen
		\bibfield  {author} {\bibinfo {author} {\bibfnamefont {I.~T.}\ \bibnamefont
				{Vidamour}}, \bibinfo {author} {\bibfnamefont {C.}~\bibnamefont {Swindells}},
			\bibinfo {author} {\bibfnamefont {G.}~\bibnamefont {Venkat}}, \bibinfo
			{author} {\bibfnamefont {L.}~\bibnamefont {Manneschi}}, \bibinfo {author}
			{\bibfnamefont {P.~W.}\ \bibnamefont {Fry}}, \bibinfo {author} {\bibfnamefont
				{A.}~\bibnamefont {Welbourne}}, \bibinfo {author} {\bibfnamefont {R.~M.}\
				\bibnamefont {Rowan-Robinson}}, \bibinfo {author} {\bibfnamefont
				{D.}~\bibnamefont {Backes}}, \bibinfo {author} {\bibfnamefont
				{F.}~\bibnamefont {Maccherozzi}}, \bibinfo {author} {\bibfnamefont {S.~S.}\
				\bibnamefont {Dhesi}}, \bibinfo {author} {\bibfnamefont {E.}~\bibnamefont
				{Vasilaki}}, \bibinfo {author} {\bibfnamefont {D.~A.}\ \bibnamefont
				{Allwood}},\ and\ \bibinfo {author} {\bibfnamefont {T.~J.}\ \bibnamefont
				{Hayward}},\ }\bibfield  {title} {\bibinfo {title} {Reconfigurable reservoir
				computing in a magnetic metamaterial},\ }\href
		{https://doi.org/10.1038/s42005-023-01352-4} {\bibfield  {journal} {\bibinfo
				{journal} {Communications Physics}\ }\textbf {\bibinfo {volume} {6}},\
			\bibinfo {pages} {230} (\bibinfo {year} {2023})}\BibitemShut {NoStop}%
		\bibitem [{\citenamefont {Laufenberg}\ \emph {et~al.}(2006)\citenamefont
			{Laufenberg}, \citenamefont {Backes}, \citenamefont {B{\"{u}}hrer},
			\citenamefont {Bedau}, \citenamefont {Kl{\"{a}}ui}, \citenamefont
			{R{\"{u}}diger}, \citenamefont {Vaz}, \citenamefont {Bland}, \citenamefont
			{Heyderman}, \citenamefont {Nolting}, \citenamefont {Cherifi}, \citenamefont
			{Locatelli}, \citenamefont {Belkhou}, \citenamefont {Heun},\ and\
			\citenamefont {Bauer}}]{Laufenberg06}%
		\BibitemOpen
		\bibfield  {author} {\bibinfo {author} {\bibfnamefont {M.}~\bibnamefont
				{Laufenberg}}, \bibinfo {author} {\bibfnamefont {D.}~\bibnamefont {Backes}},
			\bibinfo {author} {\bibfnamefont {W.}~\bibnamefont {B{\"{u}}hrer}}, \bibinfo
			{author} {\bibfnamefont {D.}~\bibnamefont {Bedau}}, \bibinfo {author}
			{\bibfnamefont {M.}~\bibnamefont {Kl{\"{a}}ui}}, \bibinfo {author}
			{\bibfnamefont {U.}~\bibnamefont {R{\"{u}}diger}}, \bibinfo {author}
			{\bibfnamefont {C.~A.~F.}\ \bibnamefont {Vaz}}, \bibinfo {author}
			{\bibfnamefont {J.~A.~C.}\ \bibnamefont {Bland}}, \bibinfo {author}
			{\bibfnamefont {L.~J.}\ \bibnamefont {Heyderman}}, \bibinfo {author}
			{\bibfnamefont {F.}~\bibnamefont {Nolting}}, \bibinfo {author} {\bibfnamefont
				{S.}~\bibnamefont {Cherifi}}, \bibinfo {author} {\bibfnamefont
				{A.}~\bibnamefont {Locatelli}}, \bibinfo {author} {\bibfnamefont
				{R.}~\bibnamefont {Belkhou}}, \bibinfo {author} {\bibfnamefont
				{S.}~\bibnamefont {Heun}},\ and\ \bibinfo {author} {\bibfnamefont
				{E.}~\bibnamefont {Bauer}},\ }\bibfield  {title} {\bibinfo {title}
			{Observation of thermally activated domain wall transformations},\ }\href
		{https://doi.org/10.1063/1.2168677} {\bibfield  {journal} {\bibinfo
				{journal} {Applied Physics Letters}\ }\textbf {\bibinfo {volume} {88}},\
			\bibinfo {pages} {052507} (\bibinfo {year} {2006})}\BibitemShut {NoStop}%
		\bibitem [{\citenamefont {Kl\"{a}ui}\ \emph
			{et~al.}(2003{\natexlab{b}})\citenamefont {Kl\"{a}ui}, \citenamefont {Bland},
			\citenamefont {Monchesky}, \citenamefont {Unguris}, \citenamefont {Bauer},
			\citenamefont {Cherifi}, \citenamefont {Heun}, \citenamefont {Locatelli},
			\citenamefont {Heyderman},\ and\ \citenamefont {Cui}}]{Klaui03c}%
		\BibitemOpen
		\bibfield  {author} {\bibinfo {author} {\bibfnamefont {M.}~\bibnamefont
				{Kl\"{a}ui}}, \bibinfo {author} {\bibfnamefont {J.~A.~C.}\ \bibnamefont
				{Bland}}, \bibinfo {author} {\bibfnamefont {T.~L.}\ \bibnamefont
				{Monchesky}}, \bibinfo {author} {\bibfnamefont {J.}~\bibnamefont {Unguris}},
			\bibinfo {author} {\bibfnamefont {E.}~\bibnamefont {Bauer}}, \bibinfo
			{author} {\bibfnamefont {S.}~\bibnamefont {Cherifi}}, \bibinfo {author}
			{\bibfnamefont {S.}~\bibnamefont {Heun}}, \bibinfo {author} {\bibfnamefont
				{A.}~\bibnamefont {Locatelli}}, \bibinfo {author} {\bibfnamefont {L.~J.}\
				\bibnamefont {Heyderman}},\ and\ \bibinfo {author} {\bibfnamefont
				{Z.}~\bibnamefont {Cui}},\ }\bibfield  {title} {\bibinfo {title} {Direct
				observation of spin configurations and classification of switching processes
				in mesoscopic ferromagnetic rings},\ }\href
		{https://doi.org/10.1103/physrevb.68.134426} {\bibfield  {journal} {\bibinfo
				{journal} {Physical Review B}\ }\textbf {\bibinfo {volume} {68}},\ \bibinfo
			{pages} {134426} (\bibinfo {year} {2003}{\natexlab{b}})}\BibitemShut
		{NoStop}%
		\bibitem [{\citenamefont {Kl\"{a}ui}\ \emph
			{et~al.}(2004{\natexlab{c}})\citenamefont {Kl\"{a}ui}, \citenamefont {Vaz},
			\citenamefont {Bland}, \citenamefont {Heyderman}, \citenamefont {Nolting},
			\citenamefont {Pavlovska}, \citenamefont {Bauer}, \citenamefont {Cherifi},
			\citenamefont {Heun},\ and\ \citenamefont {Locatelli}}]{Klaui04a}%
		\BibitemOpen
		\bibfield  {author} {\bibinfo {author} {\bibfnamefont {M.}~\bibnamefont
				{Kl\"{a}ui}}, \bibinfo {author} {\bibfnamefont {C.~A.~F.}\ \bibnamefont
				{Vaz}}, \bibinfo {author} {\bibfnamefont {J.~A.~C.}\ \bibnamefont {Bland}},
			\bibinfo {author} {\bibfnamefont {L.~J.}\ \bibnamefont {Heyderman}}, \bibinfo
			{author} {\bibfnamefont {F.}~\bibnamefont {Nolting}}, \bibinfo {author}
			{\bibfnamefont {A.}~\bibnamefont {Pavlovska}}, \bibinfo {author}
			{\bibfnamefont {E.}~\bibnamefont {Bauer}}, \bibinfo {author} {\bibfnamefont
				{S.}~\bibnamefont {Cherifi}}, \bibinfo {author} {\bibfnamefont
				{S.}~\bibnamefont {Heun}},\ and\ \bibinfo {author} {\bibfnamefont
				{A.}~\bibnamefont {Locatelli}},\ }\bibfield  {title} {\bibinfo {title}
			{Head-to-head domain-wall phase diagram in mesoscopic ring magnets},\ }\href
		{https://doi.org/10.1063/1.1829800} {\bibfield  {journal} {\bibinfo
				{journal} {Applied Physics Letters}\ }\textbf {\bibinfo {volume} {85}},\
			\bibinfo {pages} {5637} (\bibinfo {year} {2004}{\natexlab{c}})}\BibitemShut
		{NoStop}%
		\bibitem [{\citenamefont {Kl\"{a}ui}\ \emph
			{et~al.}(2003{\natexlab{c}})\citenamefont {Kl\"{a}ui}, \citenamefont {Vaz},
			\citenamefont {Bland}, \citenamefont {Wernsdorfer}, \citenamefont {Faini},
			\citenamefont {Cambril},\ and\ \citenamefont {Heyderman}}]{Klaui03d}%
		\BibitemOpen
		\bibfield  {author} {\bibinfo {author} {\bibfnamefont {M.}~\bibnamefont
				{Kl\"{a}ui}}, \bibinfo {author} {\bibfnamefont {C.~A.~F.}\ \bibnamefont
				{Vaz}}, \bibinfo {author} {\bibfnamefont {J.~A.~C.}\ \bibnamefont {Bland}},
			\bibinfo {author} {\bibfnamefont {W.}~\bibnamefont {Wernsdorfer}}, \bibinfo
			{author} {\bibfnamefont {G.}~\bibnamefont {Faini}}, \bibinfo {author}
			{\bibfnamefont {E.}~\bibnamefont {Cambril}},\ and\ \bibinfo {author}
			{\bibfnamefont {L.~J.}\ \bibnamefont {Heyderman}},\ }\bibfield  {title}
		{\bibinfo {title} {Domain wall motion induced by spin polarized currents in
				ferromagnetic ring structures},\ }\href {https://doi.org/10.1063/1.1588736}
		{\bibfield  {journal} {\bibinfo  {journal} {Applied Physics Letters}\
			}\textbf {\bibinfo {volume} {83}},\ \bibinfo {pages} {105} (\bibinfo {year}
			{2003}{\natexlab{c}})}\BibitemShut {NoStop}%
		\bibitem [{\citenamefont {Kl\"{a}ui}\ \emph
			{et~al.}(2005{\natexlab{b}})\citenamefont {Kl\"{a}ui}, \citenamefont
			{Jubert}, \citenamefont {Allenspach}, \citenamefont {Bischof}, \citenamefont
			{Bland}, \citenamefont {Faini}, \citenamefont {R\"{u}diger}, \citenamefont
			{Vaz}, \citenamefont {Vila},\ and\ \citenamefont {Vouille}}]{Klaui05}%
		\BibitemOpen
		\bibfield  {author} {\bibinfo {author} {\bibfnamefont {M.}~\bibnamefont
				{Kl\"{a}ui}}, \bibinfo {author} {\bibfnamefont {P.-O.}\ \bibnamefont
				{Jubert}}, \bibinfo {author} {\bibfnamefont {R.}~\bibnamefont {Allenspach}},
			\bibinfo {author} {\bibfnamefont {A.}~\bibnamefont {Bischof}}, \bibinfo
			{author} {\bibfnamefont {J.~A.~C.}\ \bibnamefont {Bland}}, \bibinfo {author}
			{\bibfnamefont {G.}~\bibnamefont {Faini}}, \bibinfo {author} {\bibfnamefont
				{U.}~\bibnamefont {R\"{u}diger}}, \bibinfo {author} {\bibfnamefont
				{C.~A.~F.}\ \bibnamefont {Vaz}}, \bibinfo {author} {\bibfnamefont
				{L.}~\bibnamefont {Vila}},\ and\ \bibinfo {author} {\bibfnamefont
				{C.}~\bibnamefont {Vouille}},\ }\bibfield  {title} {\bibinfo {title} {Direct
				observation of domain-wall configurations transformed by spin currents},\
		}\href {https://doi.org/10.1103/PhysRevLett.95.026601} {\bibfield  {journal}
			{\bibinfo  {journal} {Physical Review Letters}\ }\textbf {\bibinfo {volume}
				{95}},\ \bibinfo {pages} {026601} (\bibinfo {year}
			{2005}{\natexlab{b}})}\BibitemShut {NoStop}%
		\bibitem [{\citenamefont {Neudecker}\ \emph {et~al.}(2006)\citenamefont
			{Neudecker}, \citenamefont {Klaui}, \citenamefont {Perzlmaier}, \citenamefont
			{Backes}, \citenamefont {Heyderman}, \citenamefont {Vaz}, \citenamefont
			{Bland}, \citenamefont {Rudiger},\ and\ \citenamefont {Back}}]{Neudecker06}%
		\BibitemOpen
		\bibfield  {author} {\bibinfo {author} {\bibfnamefont {I.}~\bibnamefont
				{Neudecker}}, \bibinfo {author} {\bibfnamefont {M.}~\bibnamefont {Klaui}},
			\bibinfo {author} {\bibfnamefont {K.}~\bibnamefont {Perzlmaier}}, \bibinfo
			{author} {\bibfnamefont {D.}~\bibnamefont {Backes}}, \bibinfo {author}
			{\bibfnamefont {L.~J.}\ \bibnamefont {Heyderman}}, \bibinfo {author}
			{\bibfnamefont {C.~A.~F.}\ \bibnamefont {Vaz}}, \bibinfo {author}
			{\bibfnamefont {J.~A.~C.}\ \bibnamefont {Bland}}, \bibinfo {author}
			{\bibfnamefont {U.}~\bibnamefont {Rudiger}},\ and\ \bibinfo {author}
			{\bibfnamefont {C.~H.}\ \bibnamefont {Back}},\ }\bibfield  {title} {\bibinfo
			{title} {Spatially resolved dynamic eigenmode spectrum of {Co} rings},\
		}\href {https://doi.org/10.1103/physrevlett.96.057207} {\bibfield  {journal}
			{\bibinfo  {journal} {Physical Review Letters}\ }\textbf {\bibinfo {volume}
				{96}},\ \bibinfo {pages} {057207} (\bibinfo {year} {2006})}\BibitemShut
		{NoStop}%
		\bibitem [{\citenamefont {Heins}\ \emph {et~al.}(2026)\citenamefont {Heins},
			\citenamefont {K\'{a}kay}, \citenamefont {Kim}, \citenamefont {Hlawacek},
			\citenamefont {Fassbender}, \citenamefont {Schultheiss},\ and\ \citenamefont
			{Schultheiss}}]{Heins26a}%
		\BibitemOpen
		\bibfield  {author} {\bibinfo {author} {\bibfnamefont {C.}~\bibnamefont
				{Heins}}, \bibinfo {author} {\bibfnamefont {A.}~\bibnamefont {K\'{a}kay}},
			\bibinfo {author} {\bibfnamefont {J.-V.}\ \bibnamefont {Kim}}, \bibinfo
			{author} {\bibfnamefont {G.}~\bibnamefont {Hlawacek}}, \bibinfo {author}
			{\bibfnamefont {J.}~\bibnamefont {Fassbender}}, \bibinfo {author}
			{\bibfnamefont {K.}~\bibnamefont {Schultheiss}},\ and\ \bibinfo {author}
			{\bibfnamefont {H.}~\bibnamefont {Schultheiss}},\ }\bibfield  {title}
		{\bibinfo {title} {Control of magnon frequency combs in magnetic rings},\
		}\href {https://doi.org/10.1063/5.0311482} {\bibfield  {journal} {\bibinfo
				{journal} {Applied Physics Letters}\ }\textbf {\bibinfo {volume} {128}},\
			\bibinfo {pages} {082402} (\bibinfo {year} {2026})}\BibitemShut {NoStop}%
		\bibitem [{\citenamefont {Jia}\ \emph {et~al.}(2008)\citenamefont {Jia},
			\citenamefont {Sun}, \citenamefont {Luo}, \citenamefont {Han}, \citenamefont
			{Heyderman}, \citenamefont {Yan}, \citenamefont {Yan}, \citenamefont {Zheng},
			\citenamefont {Zhang}, \citenamefont {Takano}, \citenamefont {Hayashi},
			\citenamefont {Eltschka}, \citenamefont {Kläui}, \citenamefont {Rüdiger},
			\citenamefont {Kasama}, \citenamefont {Cervera-Gontard}, \citenamefont
			{Dunin-Borkowski}, \citenamefont {Tzvetkov},\ and\ \citenamefont
			{Raabe}}]{Jia08}%
		\BibitemOpen
		\bibfield  {author} {\bibinfo {author} {\bibfnamefont {C.-J.}\ \bibnamefont
				{Jia}}, \bibinfo {author} {\bibfnamefont {L.-D.}\ \bibnamefont {Sun}},
			\bibinfo {author} {\bibfnamefont {F.}~\bibnamefont {Luo}}, \bibinfo {author}
			{\bibfnamefont {X.-D.}\ \bibnamefont {Han}}, \bibinfo {author} {\bibfnamefont
				{L.~J.}\ \bibnamefont {Heyderman}}, \bibinfo {author} {\bibfnamefont {Z.-G.}\
				\bibnamefont {Yan}}, \bibinfo {author} {\bibfnamefont {C.-H.}\ \bibnamefont
				{Yan}}, \bibinfo {author} {\bibfnamefont {K.}~\bibnamefont {Zheng}}, \bibinfo
			{author} {\bibfnamefont {Z.}~\bibnamefont {Zhang}}, \bibinfo {author}
			{\bibfnamefont {M.}~\bibnamefont {Takano}}, \bibinfo {author} {\bibfnamefont
				{N.}~\bibnamefont {Hayashi}}, \bibinfo {author} {\bibfnamefont
				{M.}~\bibnamefont {Eltschka}}, \bibinfo {author} {\bibfnamefont
				{M.}~\bibnamefont {Kläui}}, \bibinfo {author} {\bibfnamefont
				{U.}~\bibnamefont {Rüdiger}}, \bibinfo {author} {\bibfnamefont
				{T.}~\bibnamefont {Kasama}}, \bibinfo {author} {\bibfnamefont
				{L.}~\bibnamefont {Cervera-Gontard}}, \bibinfo {author} {\bibfnamefont
				{R.~E.}\ \bibnamefont {Dunin-Borkowski}}, \bibinfo {author} {\bibfnamefont
				{G.}~\bibnamefont {Tzvetkov}},\ and\ \bibinfo {author} {\bibfnamefont
				{J.}~\bibnamefont {Raabe}},\ }\bibfield  {title} {\bibinfo {title}
			{Large-scale synthesis of single-crystalline iron oxide magnetic nanorings},\
		}\href {https://doi.org/10.1021/ja805152t} {\bibfield  {journal} {\bibinfo
				{journal} {Journal of the American Chemical Society}\ }\textbf {\bibinfo
				{volume} {130}},\ \bibinfo {pages} {16968} (\bibinfo {year}
			{2008})}\BibitemShut {NoStop}%
		\bibitem [{\citenamefont {Eltschka}\ \emph {et~al.}(2008)\citenamefont
			{Eltschka}, \citenamefont {Kl\"{a}ui}, \citenamefont {R\"{u}diger},
			\citenamefont {Kasama}, \citenamefont {Cervera-Gontard}, \citenamefont
			{Dunin-Borkowski}, \citenamefont {Luo}, \citenamefont {Heyderman},
			\citenamefont {Jia}, \citenamefont {Sun},\ and\ \citenamefont
			{Yan}}]{Eltschka08}%
		\BibitemOpen
		\bibfield  {author} {\bibinfo {author} {\bibfnamefont {M.}~\bibnamefont
				{Eltschka}}, \bibinfo {author} {\bibfnamefont {M.}~\bibnamefont {Kl\"{a}ui}},
			\bibinfo {author} {\bibfnamefont {U.}~\bibnamefont {R\"{u}diger}}, \bibinfo
			{author} {\bibfnamefont {T.}~\bibnamefont {Kasama}}, \bibinfo {author}
			{\bibfnamefont {L.}~\bibnamefont {Cervera-Gontard}}, \bibinfo {author}
			{\bibfnamefont {R.~E.}\ \bibnamefont {Dunin-Borkowski}}, \bibinfo {author}
			{\bibfnamefont {F.}~\bibnamefont {Luo}}, \bibinfo {author} {\bibfnamefont
				{L.~J.}\ \bibnamefont {Heyderman}}, \bibinfo {author} {\bibfnamefont {C.-J.}\
				\bibnamefont {Jia}}, \bibinfo {author} {\bibfnamefont {L.-D.}\ \bibnamefont
				{Sun}},\ and\ \bibinfo {author} {\bibfnamefont {C.-H.}\ \bibnamefont {Yan}},\
		}\bibfield  {title} {\bibinfo {title} {Correlation between magnetic spin
				structure and the three-dimensional geometry in chemically synthesized
				nanoscale magnetite rings},\ }\href {https://doi.org/10.1063/1.2936989}
		{\bibfield  {journal} {\bibinfo  {journal} {Applied Physics Letters}\
			}\textbf {\bibinfo {volume} {92}},\ \bibinfo {pages} {222508} (\bibinfo
			{year} {2008})}\BibitemShut {NoStop}%
		\bibitem [{\citenamefont {Lebech}\ \emph {et~al.}(1989)\citenamefont {Lebech},
			\citenamefont {Bernhard},\ and\ \citenamefont {Freltoft}}]{Lebech89}%
		\BibitemOpen
		\bibfield  {author} {\bibinfo {author} {\bibfnamefont {B.}~\bibnamefont
				{Lebech}}, \bibinfo {author} {\bibfnamefont {J.}~\bibnamefont {Bernhard}},\
			and\ \bibinfo {author} {\bibfnamefont {T.}~\bibnamefont {Freltoft}},\
		}\bibfield  {title} {\bibinfo {title} {Magnetic structures of cubic {FeGe}
				studied by small-angle neutron scattering},\ }\href
		{https://doi.org/10.1088/0953-8984/1/35/010} {\bibfield  {journal} {\bibinfo
				{journal} {Journal of Physics: Condensed Matter}\ }\textbf {\bibinfo {volume}
				{1}},\ \bibinfo {pages} {6105} (\bibinfo {year} {1989})}\BibitemShut
		{NoStop}%
		\bibitem [{\citenamefont {Sloika}\ \emph {et~al.}(2017)\citenamefont {Sloika},
			\citenamefont {Sheka}, \citenamefont {Kravchuk}, \citenamefont
			{Pylypovskyi},\ and\ \citenamefont {Gaididei}}]{Sloika17}%
		\BibitemOpen
		\bibfield  {author} {\bibinfo {author} {\bibfnamefont {M.~I.}\ \bibnamefont
				{Sloika}}, \bibinfo {author} {\bibfnamefont {D.~D.}\ \bibnamefont {Sheka}},
			\bibinfo {author} {\bibfnamefont {V.~P.}\ \bibnamefont {Kravchuk}}, \bibinfo
			{author} {\bibfnamefont {O.~V.}\ \bibnamefont {Pylypovskyi}},\ and\ \bibinfo
			{author} {\bibfnamefont {Y.}~\bibnamefont {Gaididei}},\ }\bibfield  {title}
		{\bibinfo {title} {Geometry induced phase transitions in magnetic spherical
				shell},\ }\href {https://doi.org/10.1016/j.jmmm.2017.07.036} {\bibfield
			{journal} {\bibinfo  {journal} {Journal of Magnetism and Magnetic Materials}\
			}\textbf {\bibinfo {volume} {443}},\ \bibinfo {pages} {404} (\bibinfo {year}
			{2017})}\BibitemShut {NoStop}%
		\bibitem [{\citenamefont {Sloika}\ \emph {et~al.}(2022)\citenamefont {Sloika},
			\citenamefont {Gaididei}, \citenamefont {Kravchuk}, \citenamefont
			{Pylypovskyi}, \citenamefont {Makarov},\ and\ \citenamefont
			{Sheka}}]{Sloika22}%
		\BibitemOpen
		\bibfield  {author} {\bibinfo {author} {\bibfnamefont {M.~I.}\ \bibnamefont
				{Sloika}}, \bibinfo {author} {\bibfnamefont {Y.}~\bibnamefont {Gaididei}},
			\bibinfo {author} {\bibfnamefont {V.~P.}\ \bibnamefont {Kravchuk}}, \bibinfo
			{author} {\bibfnamefont {O.~V.}\ \bibnamefont {Pylypovskyi}}, \bibinfo
			{author} {\bibfnamefont {D.}~\bibnamefont {Makarov}},\ and\ \bibinfo {author}
			{\bibfnamefont {D.~D.}\ \bibnamefont {Sheka}},\ }\bibfield  {title} {\bibinfo
			{title} {Impact of curvature-induced {D}zyaloshinskii-{M}oriya interaction on
				magnetic vortex texture in spherical caps},\ }\href
		{https://doi.org/10.1063/10.0014596} {\bibfield  {journal} {\bibinfo
				{journal} {Low Temperature Physics}\ }\textbf {\bibinfo {volume} {48}},\
			\bibinfo {pages} {956} (\bibinfo {year} {2022})}\BibitemShut {NoStop}%
		\bibitem [{\citenamefont {Sloika}\ \emph {et~al.}(2025)\citenamefont {Sloika},
			\citenamefont {Gaididei}, \citenamefont {Kravchuk}, \citenamefont
			{Pylypovskyi}, \citenamefont {Makarov},\ and\ \citenamefont
			{Sheka}}]{Sloika25}%
		\BibitemOpen
		\bibfield  {author} {\bibinfo {author} {\bibfnamefont {M.~I.}\ \bibnamefont
				{Sloika}}, \bibinfo {author} {\bibfnamefont {Y.}~\bibnamefont {Gaididei}},
			\bibinfo {author} {\bibfnamefont {V.~P.}\ \bibnamefont {Kravchuk}}, \bibinfo
			{author} {\bibfnamefont {O.~V.}\ \bibnamefont {Pylypovskyi}}, \bibinfo
			{author} {\bibfnamefont {D.}~\bibnamefont {Makarov}},\ and\ \bibinfo {author}
			{\bibfnamefont {D.~D.}\ \bibnamefont {Sheka}},\ }\bibfield  {title} {\bibinfo
			{title} {Magnetic vortex dynamics on a spherical cap},\ }\href
		{https://doi.org/10.1063/10.0036747} {\bibfield  {journal} {\bibinfo
				{journal} {Low Temperature Physics}\ }\textbf {\bibinfo {volume} {51}},\
			\bibinfo {pages} {700} (\bibinfo {year} {2025})}\BibitemShut {NoStop}%
		\bibitem [{\citenamefont {Sayed}\ \emph {et~al.}(2018)\citenamefont {Sayed},
			\citenamefont {Yaacoub}, \citenamefont {Labaye}, \citenamefont {Hassan},
			\citenamefont {Singh}, \citenamefont {Kumar}, \citenamefont {Greneche},
			\citenamefont {Mathieu}, \citenamefont {Hadjipanayis}, \citenamefont
			{Agostinelli},\ and\ \citenamefont {Peddis}}]{Sayed18a}%
		\BibitemOpen
		\bibfield  {author} {\bibinfo {author} {\bibfnamefont {F.}~\bibnamefont
				{Sayed}}, \bibinfo {author} {\bibfnamefont {N.}~\bibnamefont {Yaacoub}},
			\bibinfo {author} {\bibfnamefont {Y.}~\bibnamefont {Labaye}}, \bibinfo
			{author} {\bibfnamefont {R.~S.}\ \bibnamefont {Hassan}}, \bibinfo {author}
			{\bibfnamefont {G.}~\bibnamefont {Singh}}, \bibinfo {author} {\bibfnamefont
				{P.~A.}\ \bibnamefont {Kumar}}, \bibinfo {author} {\bibfnamefont {J.~M.}\
				\bibnamefont {Greneche}}, \bibinfo {author} {\bibfnamefont {R.}~\bibnamefont
				{Mathieu}}, \bibinfo {author} {\bibfnamefont {G.~C.}\ \bibnamefont
				{Hadjipanayis}}, \bibinfo {author} {\bibfnamefont {E.}~\bibnamefont
				{Agostinelli}},\ and\ \bibinfo {author} {\bibfnamefont {D.}~\bibnamefont
				{Peddis}},\ }\bibfield  {title} {\bibinfo {title} {Surface effects in
				ultrathin iron oxide hollow nanoparticles: Exploring magnetic disorder at the
				nanoscale},\ }\href {https://doi.org/10.1021/acs.jpcc.8b00300} {\bibfield
			{journal} {\bibinfo  {journal} {The Journal of Physical Chemistry C}\
			}\textbf {\bibinfo {volume} {122}},\ \bibinfo {pages} {7516} (\bibinfo {year}
			{2018})}\BibitemShut {NoStop}%
		\bibitem [{\citenamefont {Guo}\ \emph {et~al.}(2024)\citenamefont {Guo},
			\citenamefont {Zhang}, \citenamefont {Cai}, \citenamefont {Ren},\ and\
			\citenamefont {Liu}}]{Guo24}%
		\BibitemOpen
		\bibfield  {author} {\bibinfo {author} {\bibfnamefont {J.}~\bibnamefont
				{Guo}}, \bibinfo {author} {\bibfnamefont {X.}~\bibnamefont {Zhang}}, \bibinfo
			{author} {\bibfnamefont {N.}~\bibnamefont {Cai}}, \bibinfo {author}
			{\bibfnamefont {R.}~\bibnamefont {Ren}},\ and\ \bibinfo {author}
			{\bibfnamefont {Y.}~\bibnamefont {Liu}},\ }\bibfield  {title} {\bibinfo
			{title} {Stability and evolution of skyrmionium and skyrmions in a spherical
				shell},\ }\href {https://doi.org/10.1088/1402-4896/ad7919} {\bibfield
			{journal} {\bibinfo  {journal} {Physica Scripta}\ }\textbf {\bibinfo {volume}
				{99}},\ \bibinfo {pages} {105994} (\bibinfo {year} {2024})}\BibitemShut
		{NoStop}%
		\bibitem [{\citenamefont {Fullerton}\ \emph
			{et~al.}(2025{\natexlab{b}})\citenamefont {Fullerton}, \citenamefont
			{Koenig},\ and\ \citenamefont {Phatak}}]{Fullerton25b}%
		\BibitemOpen
		\bibfield  {author} {\bibinfo {author} {\bibfnamefont {J.}~\bibnamefont
				{Fullerton}}, \bibinfo {author} {\bibfnamefont {J.}~\bibnamefont {Koenig}},\
			and\ \bibinfo {author} {\bibfnamefont {C.}~\bibnamefont {Phatak}},\
		}\bibfield  {title} {\bibinfo {title} {Magnetic solitons and
				thickness‐dependent magnetization reversal in interconnected helical
				nanowire arrays},\ }\href {https://doi.org/10.1002/apxr.202500135} {\bibfield
			{journal} {\bibinfo  {journal} {Advanced Physics Research}\ ,\ \bibinfo
				{pages} {e00135}} (\bibinfo {year} {2025}{\natexlab{b}})}\BibitemShut
		{NoStop}%
		\bibitem [{\citenamefont {Ternovsky}\ \emph {et~al.}(2001)\citenamefont
			{Ternovsky}, \citenamefont {Luk'yanchuk},\ and\ \citenamefont
			{Wang}}]{Ternovsky01}%
		\BibitemOpen
		\bibfield  {author} {\bibinfo {author} {\bibfnamefont {V.}~\bibnamefont
				{Ternovsky}}, \bibinfo {author} {\bibfnamefont {B.}~\bibnamefont
				{Luk'yanchuk}},\ and\ \bibinfo {author} {\bibfnamefont {J.}~\bibnamefont
				{Wang}},\ }\bibfield  {title} {\bibinfo {title} {Remanent states of small
				ferromagnetic cylinder},\ }\href {https://doi.org/10.1134/1.1397750}
		{\bibfield  {journal} {\bibinfo  {journal} {JETP Letters}\ }\textbf {\bibinfo
				{volume} {73}},\ \bibinfo {pages} {661} (\bibinfo {year} {2001})}\BibitemShut
		{NoStop}%
		\bibitem [{\citenamefont {Hertel}\ and\ \citenamefont
			{Kirschner}(2004{\natexlab{b}})}]{Hertel04a}%
		\BibitemOpen
		\bibfield  {author} {\bibinfo {author} {\bibfnamefont {R.}~\bibnamefont
				{Hertel}}\ and\ \bibinfo {author} {\bibfnamefont {J.}~\bibnamefont
				{Kirschner}},\ }\bibfield  {title} {\bibinfo {title} {Magnetic drops in a
				soft-magnetic cylinder},\ }\href {https://doi.org/10.1016/j.jmmm.2004.02.032}
		{\bibfield  {journal} {\bibinfo  {journal} {Journal of Magnetism and Magnetic
					Materials}\ }\textbf {\bibinfo {volume} {278}},\ \bibinfo {pages} {L291}
			(\bibinfo {year} {2004}{\natexlab{b}})}\BibitemShut {NoStop}%
		\bibitem [{\citenamefont {Cheenikundil}\ and\ \citenamefont
			{Hertel}(2021)}]{Cheenikundil21}%
		\BibitemOpen
		\bibfield  {author} {\bibinfo {author} {\bibfnamefont {R.}~\bibnamefont
				{Cheenikundil}}\ and\ \bibinfo {author} {\bibfnamefont {R.}~\bibnamefont
				{Hertel}},\ }\bibfield  {title} {\bibinfo {title} {Switchable magnetic
				frustration in buckyball nanoarchitectures},\ }\href
		{https://doi.org/10.1063/5.0048936} {\bibfield  {journal} {\bibinfo
				{journal} {Applied Physics Letters}\ }\textbf {\bibinfo {volume} {118}},\
			\bibinfo {pages} {212403} (\bibinfo {year} {2021})}\BibitemShut {NoStop}%
		\bibitem [{\citenamefont {Cheenikundil}\ \emph {et~al.}(2022)\citenamefont
			{Cheenikundil}, \citenamefont {Bauer}, \citenamefont {Goharyan},
			\citenamefont {d’Aquino},\ and\ \citenamefont {Hertel}}]{Cheenikundil22}%
		\BibitemOpen
		\bibfield  {author} {\bibinfo {author} {\bibfnamefont {R.}~\bibnamefont
				{Cheenikundil}}, \bibinfo {author} {\bibfnamefont {J.}~\bibnamefont {Bauer}},
			\bibinfo {author} {\bibfnamefont {M.}~\bibnamefont {Goharyan}}, \bibinfo
			{author} {\bibfnamefont {M.}~\bibnamefont {d’Aquino}},\ and\ \bibinfo
			{author} {\bibfnamefont {R.}~\bibnamefont {Hertel}},\ }\bibfield  {title}
		{\bibinfo {title} {High-frequency modes in a magnetic buckyball
				nanoarchitecture},\ }\href {https://doi.org/10.1063/5.0097695} {\bibfield
			{journal} {\bibinfo  {journal} {APL Materials}\ }\textbf {\bibinfo {volume}
				{10}},\ \bibinfo {pages} {081106} (\bibinfo {year} {2022})}\BibitemShut
		{NoStop}%
		\bibitem [{\citenamefont {Morales‐Fern\'{a}ndez}\ \emph
			{et~al.}(2026)\citenamefont {Morales‐Fern\'{a}ndez}, \citenamefont
			{Konstantinos‐Douveas}, \citenamefont {Abert}, \citenamefont {Zhakina},
			\citenamefont {Ruiz‐G\'{o}mez}, \citenamefont
			{Fern\'{a}ndez‐Gonz\'{a}lez}, \citenamefont {Turnbull}, \citenamefont
			{Wintz}, \citenamefont {K\"{o}nig}, \citenamefont {Finizio}, \citenamefont
			{Weigand}, \citenamefont {Leo}, \citenamefont {Suess}, \citenamefont
			{Hierro‐Rodr\'{\i}guez}, \citenamefont {Fern\'{a}ndez‐Pacheco},\ and\
			\citenamefont {Donnelly}}]{MoralesFernandez26}%
		\BibitemOpen
		\bibfield  {author} {\bibinfo {author} {\bibfnamefont {P.}~\bibnamefont
				{Morales‐Fern\'{a}ndez}}, \bibinfo {author} {\bibfnamefont
				{I.}~\bibnamefont {Konstantinos‐Douveas}}, \bibinfo {author} {\bibfnamefont
				{C.}~\bibnamefont {Abert}}, \bibinfo {author} {\bibfnamefont
				{E.}~\bibnamefont {Zhakina}}, \bibinfo {author} {\bibfnamefont
				{S.}~\bibnamefont {Ruiz‐G\'{o}mez}}, \bibinfo {author} {\bibfnamefont
				{C.}~\bibnamefont {Fern\'{a}ndez‐Gonz\'{a}lez}}, \bibinfo {author}
			{\bibfnamefont {L.~A.}\ \bibnamefont {Turnbull}}, \bibinfo {author}
			{\bibfnamefont {S.}~\bibnamefont {Wintz}}, \bibinfo {author} {\bibfnamefont
				{M.}~\bibnamefont {K\"{o}nig}}, \bibinfo {author} {\bibfnamefont
				{S.}~\bibnamefont {Finizio}}, \bibinfo {author} {\bibfnamefont
				{M.}~\bibnamefont {Weigand}}, \bibinfo {author} {\bibfnamefont
				{N.}~\bibnamefont {Leo}}, \bibinfo {author} {\bibfnamefont {D.}~\bibnamefont
				{Suess}}, \bibinfo {author} {\bibfnamefont {A.}~\bibnamefont
				{Hierro‐Rodr\'{\i}guez}}, \bibinfo {author} {\bibfnamefont
				{A.}~\bibnamefont {Fern\'{a}ndez‐Pacheco}},\ and\ \bibinfo {author}
			{\bibfnamefont {C.}~\bibnamefont {Donnelly}},\ }\bibfield  {title} {\bibinfo
			{title} {Resonant domain wall dynamics in a three‐dimensional magnetic nano
				double helix},\ }\href {https://doi.org/10.1002/adma.202600004} {\bibfield
			{journal} {\bibinfo  {journal} {Advanced Materials}\ ,\ \bibinfo {pages}
				{e00004}} (\bibinfo {year} {2026})}\BibitemShut {NoStop}%
		\bibitem [{\citenamefont {Llandro}\ \emph {et~al.}(2020)\citenamefont
			{Llandro}, \citenamefont {Love}, \citenamefont {Kov{\'{a}}cs}, \citenamefont
			{Caron}, \citenamefont {Vyas}, \citenamefont {K{\'{a}}kay}, \citenamefont
			{Salikhov}, \citenamefont {Lenz}, \citenamefont {Fassbender}, \citenamefont
			{Scherer}, \citenamefont {Cimorra}, \citenamefont {Steiner}, \citenamefont
			{Barnes}, \citenamefont {Dunin-Borkowski}, \citenamefont {Fukami},\ and\
			\citenamefont {Ohno}}]{Llandro20}%
		\BibitemOpen
		\bibfield  {author} {\bibinfo {author} {\bibfnamefont {J.}~\bibnamefont
				{Llandro}}, \bibinfo {author} {\bibfnamefont {D.~M.}\ \bibnamefont {Love}},
			\bibinfo {author} {\bibfnamefont {A.}~\bibnamefont {Kov{\'{a}}cs}}, \bibinfo
			{author} {\bibfnamefont {J.}~\bibnamefont {Caron}}, \bibinfo {author}
			{\bibfnamefont {K.~N.}\ \bibnamefont {Vyas}}, \bibinfo {author}
			{\bibfnamefont {A.}~\bibnamefont {K{\'{a}}kay}}, \bibinfo {author}
			{\bibfnamefont {R.}~\bibnamefont {Salikhov}}, \bibinfo {author}
			{\bibfnamefont {K.}~\bibnamefont {Lenz}}, \bibinfo {author} {\bibfnamefont
				{J.}~\bibnamefont {Fassbender}}, \bibinfo {author} {\bibfnamefont {M.~R.~J.}\
				\bibnamefont {Scherer}}, \bibinfo {author} {\bibfnamefont {C.}~\bibnamefont
				{Cimorra}}, \bibinfo {author} {\bibfnamefont {U.}~\bibnamefont {Steiner}},
			\bibinfo {author} {\bibfnamefont {C.~H.~W.}\ \bibnamefont {Barnes}}, \bibinfo
			{author} {\bibfnamefont {R.~E.}\ \bibnamefont {Dunin-Borkowski}}, \bibinfo
			{author} {\bibfnamefont {S.}~\bibnamefont {Fukami}},\ and\ \bibinfo {author}
			{\bibfnamefont {H.}~\bibnamefont {Ohno}},\ }\bibfield  {title} {\bibinfo
			{title} {Visualizing magnetic structure in {3D} nanoscale {Ni{\textendash}Fe}
				gyroid networks},\ }\href {https://doi.org/10.1021/acs.nanolett.0c00578}
		{\bibfield  {journal} {\bibinfo  {journal} {Nano Letters}\ }\textbf {\bibinfo
				{volume} {20}},\ \bibinfo {pages} {3642} (\bibinfo {year}
			{2020})}\BibitemShut {NoStop}%
		\bibitem [{\citenamefont {Koshikawa}\ \emph {et~al.}(2023)\citenamefont
			{Koshikawa}, \citenamefont {Llandro}, \citenamefont {Ohzeki}, \citenamefont
			{Fukami}, \citenamefont {Ohno},\ and\ \citenamefont {Leo}}]{Koshikawa23}%
		\BibitemOpen
		\bibfield  {author} {\bibinfo {author} {\bibfnamefont {A.~S.}\ \bibnamefont
				{Koshikawa}}, \bibinfo {author} {\bibfnamefont {J.}~\bibnamefont {Llandro}},
			\bibinfo {author} {\bibfnamefont {M.}~\bibnamefont {Ohzeki}}, \bibinfo
			{author} {\bibfnamefont {S.}~\bibnamefont {Fukami}}, \bibinfo {author}
			{\bibfnamefont {H.}~\bibnamefont {Ohno}},\ and\ \bibinfo {author}
			{\bibfnamefont {N.}~\bibnamefont {Leo}},\ }\bibfield  {title} {\bibinfo
			{title} {Magnetic order in nanoscale gyroid networks},\ }\href
		{https://doi.org/10.1103/physrevb.108.024414} {\bibfield  {journal} {\bibinfo
				{journal} {Physical Review B}\ }\textbf {\bibinfo {volume} {108}},\ \bibinfo
			{pages} {024414} (\bibinfo {year} {2023})}\BibitemShut {NoStop}%
		\bibitem [{\citenamefont {Wu}\ \emph {et~al.}(2015)\citenamefont {Wu},
			\citenamefont {Zhang},\ and\ \citenamefont {Zhang}}]{Wu15a}%
		\BibitemOpen
		\bibfield  {author} {\bibinfo {author} {\bibfnamefont {L.}~\bibnamefont
				{Wu}}, \bibinfo {author} {\bibfnamefont {W.}~\bibnamefont {Zhang}},\ and\
			\bibinfo {author} {\bibfnamefont {D.}~\bibnamefont {Zhang}},\ }\bibfield
		{title} {\bibinfo {title} {Engineering gyroid‐structured functional
				materials via templates discovered in nature and in the lab},\ }\href
		{https://doi.org/10.1002/smll.201500812} {\bibfield  {journal} {\bibinfo
				{journal} {Small}\ }\textbf {\bibinfo {volume} {11}},\ \bibinfo {pages}
			{5004} (\bibinfo {year} {2015})}\BibitemShut {NoStop}%
		\bibitem [{\citenamefont {Dolan}\ \emph {et~al.}(2018)\citenamefont {Dolan},
			\citenamefont {Korzeb}, \citenamefont {Dehmel}, \citenamefont {G\"{o}del},
			\citenamefont {Stefik}, \citenamefont {Wiesner}, \citenamefont {Wilkinson},
			\citenamefont {Baumberg}, \citenamefont {Wilts}, \citenamefont {Steiner},\
			and\ \citenamefont {Gunkel}}]{Dolan18}%
		\BibitemOpen
		\bibfield  {author} {\bibinfo {author} {\bibfnamefont {J.~A.}\ \bibnamefont
				{Dolan}}, \bibinfo {author} {\bibfnamefont {K.}~\bibnamefont {Korzeb}},
			\bibinfo {author} {\bibfnamefont {R.}~\bibnamefont {Dehmel}}, \bibinfo
			{author} {\bibfnamefont {K.~C.}\ \bibnamefont {G\"{o}del}}, \bibinfo {author}
			{\bibfnamefont {M.}~\bibnamefont {Stefik}}, \bibinfo {author} {\bibfnamefont
				{U.}~\bibnamefont {Wiesner}}, \bibinfo {author} {\bibfnamefont {T.~D.}\
				\bibnamefont {Wilkinson}}, \bibinfo {author} {\bibfnamefont {J.~J.}\
				\bibnamefont {Baumberg}}, \bibinfo {author} {\bibfnamefont {B.~D.}\
				\bibnamefont {Wilts}}, \bibinfo {author} {\bibfnamefont {U.}~\bibnamefont
				{Steiner}},\ and\ \bibinfo {author} {\bibfnamefont {I.}~\bibnamefont
				{Gunkel}},\ }\bibfield  {title} {\bibinfo {title} {Controlling
				self‐assembly in gyroid terpolymer films by solvent vapor annealing},\
		}\href {https://doi.org/10.1002/smll.201802401} {\bibfield  {journal}
			{\bibinfo  {journal} {Small}\ }\textbf {\bibinfo {volume} {14}},\ \bibinfo
			{pages} {1802401} (\bibinfo {year} {2018})}\BibitemShut {NoStop}%
		\bibitem [{\citenamefont {Taguchi}\ \emph {et~al.}(2001)\citenamefont
			{Taguchi}, \citenamefont {Oohara}, \citenamefont {Yoshizawa}, \citenamefont
			{Nagaosa},\ and\ \citenamefont {Tokura}}]{Taguchi01}%
		\BibitemOpen
		\bibfield  {author} {\bibinfo {author} {\bibfnamefont {Y.}~\bibnamefont
				{Taguchi}}, \bibinfo {author} {\bibfnamefont {Y.}~\bibnamefont {Oohara}},
			\bibinfo {author} {\bibfnamefont {H.}~\bibnamefont {Yoshizawa}}, \bibinfo
			{author} {\bibfnamefont {N.}~\bibnamefont {Nagaosa}},\ and\ \bibinfo {author}
			{\bibfnamefont {Y.}~\bibnamefont {Tokura}},\ }\bibfield  {title} {\bibinfo
			{title} {Spin chirality, {B}erry phase, and anomalous {H}all effect in a
				frustrated ferromagnet},\ }\href {https://doi.org/10.1126/science.1058161}
		{\bibfield  {journal} {\bibinfo  {journal} {Science}\ }\textbf {\bibinfo
				{volume} {291}},\ \bibinfo {pages} {2573} (\bibinfo {year}
			{2001})}\BibitemShut {NoStop}%
		\bibitem [{\citenamefont {Park}\ \emph {et~al.}(2008)\citenamefont {Park},
			\citenamefont {Onoda}, \citenamefont {Nagaosa},\ and\ \citenamefont
			{Han}}]{Park08}%
		\BibitemOpen
		\bibfield  {author} {\bibinfo {author} {\bibfnamefont {J.-H.}\ \bibnamefont
				{Park}}, \bibinfo {author} {\bibfnamefont {S.}~\bibnamefont {Onoda}},
			\bibinfo {author} {\bibfnamefont {N.}~\bibnamefont {Nagaosa}},\ and\ \bibinfo
			{author} {\bibfnamefont {J.~H.}\ \bibnamefont {Han}},\ }\bibfield  {title}
		{\bibinfo {title} {Nematic and chiral order for planar spins on a triangular
				lattice},\ }\href {https://doi.org/10.1103/physrevlett.101.167202} {\bibfield
			{journal} {\bibinfo  {journal} {Physical Review Letters}\ }\textbf {\bibinfo
				{volume} {101}},\ \bibinfo {pages} {167202} (\bibinfo {year}
			{2008})}\BibitemShut {NoStop}%
		\bibitem [{\citenamefont {Wannier}(1950)}]{Wannier50}%
		\BibitemOpen
		\bibfield  {author} {\bibinfo {author} {\bibfnamefont {G.~H.}\ \bibnamefont
				{Wannier}},\ }\bibfield  {title} {\bibinfo {title} {Antiferromagnetism. {T}he
				triangular {I}sing net},\ }\href {https://doi.org/10.1103/physrev.79.357}
		{\bibfield  {journal} {\bibinfo  {journal} {Physical Review}\ }\textbf
			{\bibinfo {volume} {79}},\ \bibinfo {pages} {357} (\bibinfo {year}
			{1950})}\BibitemShut {NoStop}%
		\bibitem [{\citenamefont {Diep}(2013)}]{Diep13}%
		\BibitemOpen
		\bibinfo {editor} {\bibfnamefont {H.~T.}\ \bibnamefont {Diep}},\ ed.,\ \href
		{https://doi.org/10.1142/8676} {\emph {\bibinfo {title} {Frustrated Spin
					Systems (2nd Edition)}}}\ (\bibinfo  {publisher} {World Scientific Publishing
			Company},\ \bibinfo {year} {2013})\BibitemShut {NoStop}%
		\bibitem [{\citenamefont {Lacroix}\ \emph {et~al.}(2011)\citenamefont
			{Lacroix}, \citenamefont {Mendels},\ and\ \citenamefont {Mila}}]{Lacroix11}%
		\BibitemOpen
		\bibinfo {editor} {\bibfnamefont {C.}~\bibnamefont {Lacroix}}, \bibinfo
		{editor} {\bibfnamefont {P.}~\bibnamefont {Mendels}},\ and\ \bibinfo {editor}
		{\bibfnamefont {F.}~\bibnamefont {Mila}},\ eds.,\ \href
		{https://doi.org/10.1007/978-3-642-10589-0} {\emph {\bibinfo {title}
				{Introduction to Frustrated Magnetism}}}\ (\bibinfo  {publisher} {Springer
			Berlin Heidelberg},\ \bibinfo {address} {Berlin, Heidelberg},\ \bibinfo
		{year} {2011})\BibitemShut {NoStop}%
		\bibitem [{\citenamefont {Skj{\ae}rv{\o}}\ \emph {et~al.}(2019)\citenamefont
			{Skj{\ae}rv{\o}}, \citenamefont {Marrows}, \citenamefont {Stamps},\ and\
			\citenamefont {Heyderman}}]{Skjaervo19}%
		\BibitemOpen
		\bibfield  {author} {\bibinfo {author} {\bibfnamefont {S.~H.}\ \bibnamefont
				{Skj{\ae}rv{\o}}}, \bibinfo {author} {\bibfnamefont {C.~H.}\ \bibnamefont
				{Marrows}}, \bibinfo {author} {\bibfnamefont {R.~L.}\ \bibnamefont
				{Stamps}},\ and\ \bibinfo {author} {\bibfnamefont {L.~J.}\ \bibnamefont
				{Heyderman}},\ }\bibfield  {title} {\bibinfo {title} {Advances in artificial
				spin ice},\ }\href {https://doi.org/10.1038/s42254-019-0118-3} {\bibfield
			{journal} {\bibinfo  {journal} {Nature Reviews Physics}\ }\textbf {\bibinfo
				{volume} {2}},\ \bibinfo {pages} {13} (\bibinfo {year} {2019})}\BibitemShut
		{NoStop}%
		\bibitem [{\citenamefont {Rikken}\ \emph {et~al.}(2001)\citenamefont {Rikken},
			\citenamefont {F\"{o}lling},\ and\ \citenamefont {Wyder}}]{Rikken01}%
		\BibitemOpen
		\bibfield  {author} {\bibinfo {author} {\bibfnamefont {G.~L. J.~A.}\
				\bibnamefont {Rikken}}, \bibinfo {author} {\bibfnamefont {J.}~\bibnamefont
				{F\"{o}lling}},\ and\ \bibinfo {author} {\bibfnamefont {P.}~\bibnamefont
				{Wyder}},\ }\bibfield  {title} {\bibinfo {title} {Electrical magnetochiral
				anisotropy},\ }\href {https://doi.org/10.1103/physrevlett.87.236602}
		{\bibfield  {journal} {\bibinfo  {journal} {Physical Review Letters}\
			}\textbf {\bibinfo {volume} {87}},\ \bibinfo {pages} {236602} (\bibinfo
			{year} {2001})}\BibitemShut {NoStop}%
		\bibitem [{\citenamefont {Seki}\ \emph {et~al.}(2016)\citenamefont {Seki},
			\citenamefont {Okamura}, \citenamefont {Kondou}, \citenamefont {Shibata},
			\citenamefont {Kubota}, \citenamefont {Takagi}, \citenamefont {Kagawa},
			\citenamefont {Kawasaki}, \citenamefont {Tatara}, \citenamefont {Otani},\
			and\ \citenamefont {Tokura}}]{Seki16a}%
		\BibitemOpen
		\bibfield  {author} {\bibinfo {author} {\bibfnamefont {S.}~\bibnamefont
				{Seki}}, \bibinfo {author} {\bibfnamefont {Y.}~\bibnamefont {Okamura}},
			\bibinfo {author} {\bibfnamefont {K.}~\bibnamefont {Kondou}}, \bibinfo
			{author} {\bibfnamefont {K.}~\bibnamefont {Shibata}}, \bibinfo {author}
			{\bibfnamefont {M.}~\bibnamefont {Kubota}}, \bibinfo {author} {\bibfnamefont
				{R.}~\bibnamefont {Takagi}}, \bibinfo {author} {\bibfnamefont
				{F.}~\bibnamefont {Kagawa}}, \bibinfo {author} {\bibfnamefont
				{M.}~\bibnamefont {Kawasaki}}, \bibinfo {author} {\bibfnamefont
				{G.}~\bibnamefont {Tatara}}, \bibinfo {author} {\bibfnamefont
				{Y.}~\bibnamefont {Otani}},\ and\ \bibinfo {author} {\bibfnamefont
				{Y.}~\bibnamefont {Tokura}},\ }\bibfield  {title} {\bibinfo {title}
			{Magnetochiral nonreciprocity of volume spin wave propagation in
				chiral-lattice ferromagnets},\ }\href
		{https://doi.org/10.1103/physrevb.93.235131} {\bibfield  {journal} {\bibinfo
				{journal} {Physical Review B}\ }\textbf {\bibinfo {volume} {93}},\ \bibinfo
			{pages} {235131} (\bibinfo {year} {2016})}\BibitemShut {NoStop}%
		\bibitem [{\citenamefont {Tokura}\ and\ \citenamefont
			{Nagaosa}(2018)}]{Tokura18}%
		\BibitemOpen
		\bibfield  {author} {\bibinfo {author} {\bibfnamefont {Y.}~\bibnamefont
				{Tokura}}\ and\ \bibinfo {author} {\bibfnamefont {N.}~\bibnamefont
				{Nagaosa}},\ }\bibfield  {title} {\bibinfo {title} {Nonreciprocal responses
				from non-centrosymmetric quantum materials},\ }\href
		{https://doi.org/10.1038/s41467-018-05759-4} {\bibfield  {journal} {\bibinfo
				{journal} {Nature Communications}\ }\textbf {\bibinfo {volume} {9}},\
			\bibinfo {pages} {3740} (\bibinfo {year} {2018})}\BibitemShut {NoStop}%
		\bibitem [{\citenamefont {Barman}\ \emph {et~al.}(2020)\citenamefont {Barman},
			\citenamefont {Mondal}, \citenamefont {Sahoo},\ and\ \citenamefont
			{De}}]{Barman20}%
		\BibitemOpen
		\bibfield  {author} {\bibinfo {author} {\bibfnamefont {A.}~\bibnamefont
				{Barman}}, \bibinfo {author} {\bibfnamefont {S.}~\bibnamefont {Mondal}},
			\bibinfo {author} {\bibfnamefont {S.}~\bibnamefont {Sahoo}},\ and\ \bibinfo
			{author} {\bibfnamefont {A.}~\bibnamefont {De}},\ }\bibfield  {title}
		{\bibinfo {title} {Magnetization dynamics of nanoscale magnetic materials: A
				perspective},\ }\href {https://doi.org/10.1063/5.0023993} {\bibfield
			{journal} {\bibinfo  {journal} {Journal of Applied Physics}\ }\textbf
			{\bibinfo {volume} {128}},\ \bibinfo {pages} {170901} (\bibinfo {year}
			{2020})}\BibitemShut {NoStop}%
		\bibitem [{\citenamefont {Atzori}\ \emph {et~al.}(2021)\citenamefont {Atzori},
			\citenamefont {Train}, \citenamefont {Hillard}, \citenamefont {Avarvari},\
			and\ \citenamefont {Rikken}}]{Atzori21}%
		\BibitemOpen
		\bibfield  {author} {\bibinfo {author} {\bibfnamefont {M.}~\bibnamefont
				{Atzori}}, \bibinfo {author} {\bibfnamefont {C.}~\bibnamefont {Train}},
			\bibinfo {author} {\bibfnamefont {E.~A.}\ \bibnamefont {Hillard}}, \bibinfo
			{author} {\bibfnamefont {N.}~\bibnamefont {Avarvari}},\ and\ \bibinfo
			{author} {\bibfnamefont {G.~L. J.~A.}\ \bibnamefont {Rikken}},\ }\bibfield
		{title} {\bibinfo {title} {Magneto‐chiral anisotropy: From fundamentals to
				perspectives},\ }\href {https://doi.org/10.1002/chir.23361} {\bibfield
			{journal} {\bibinfo  {journal} {Chirality}\ }\textbf {\bibinfo {volume}
				{33}},\ \bibinfo {pages} {844} (\bibinfo {year} {2021})}\BibitemShut
		{NoStop}%
		\bibitem [{\citenamefont {Pignard}\ \emph {et~al.}(2000)\citenamefont
			{Pignard}, \citenamefont {Goglio}, \citenamefont {Radulescu}, \citenamefont
			{Piraux}, \citenamefont {Dubois}, \citenamefont {Decl\'{e}my},\ and\
			\citenamefont {Duvail}}]{Pignard00}%
		\BibitemOpen
		\bibfield  {author} {\bibinfo {author} {\bibfnamefont {S.}~\bibnamefont
				{Pignard}}, \bibinfo {author} {\bibfnamefont {G.}~\bibnamefont {Goglio}},
			\bibinfo {author} {\bibfnamefont {A.}~\bibnamefont {Radulescu}}, \bibinfo
			{author} {\bibfnamefont {L.}~\bibnamefont {Piraux}}, \bibinfo {author}
			{\bibfnamefont {S.}~\bibnamefont {Dubois}}, \bibinfo {author} {\bibfnamefont
				{A.}~\bibnamefont {Decl\'{e}my}},\ and\ \bibinfo {author} {\bibfnamefont
				{J.~L.}\ \bibnamefont {Duvail}},\ }\bibfield  {title} {\bibinfo {title}
			{Study of the magnetization reversal in individual nickel nanowires},\ }\href
		{https://doi.org/10.1063/1.371947} {\bibfield  {journal} {\bibinfo  {journal}
				{Journal of Applied Physics}\ }\textbf {\bibinfo {volume} {87}},\ \bibinfo
			{pages} {824} (\bibinfo {year} {2000})}\BibitemShut {NoStop}%
		\bibitem [{\citenamefont {Wang}\ \emph {et~al.}(2005)\citenamefont {Wang},
			\citenamefont {Adeyeye}, \citenamefont {Singh}, \citenamefont {Huang},\ and\
			\citenamefont {Wu}}]{Wang05b}%
		\BibitemOpen
		\bibfield  {author} {\bibinfo {author} {\bibfnamefont {C.~C.}\ \bibnamefont
				{Wang}}, \bibinfo {author} {\bibfnamefont {A.~O.}\ \bibnamefont {Adeyeye}},
			\bibinfo {author} {\bibfnamefont {N.}~\bibnamefont {Singh}}, \bibinfo
			{author} {\bibfnamefont {Y.~S.}\ \bibnamefont {Huang}},\ and\ \bibinfo
			{author} {\bibfnamefont {Y.~H.}\ \bibnamefont {Wu}},\ }\bibfield  {title}
		{\bibinfo {title} {Magnetoresistance behavior of nanoscale antidot arrays},\
		}\href {https://doi.org/10.1103/physrevb.72.174426} {\bibfield  {journal}
			{\bibinfo  {journal} {Physical Review B}\ }\textbf {\bibinfo {volume} {72}},\
			\bibinfo {pages} {174426} (\bibinfo {year} {2005})}\BibitemShut {NoStop}%
		\bibitem [{\citenamefont {Branford}\ \emph {et~al.}(2012)\citenamefont
			{Branford}, \citenamefont {Ladak}, \citenamefont {Read}, \citenamefont
			{Zeissler},\ and\ \citenamefont {Cohen}}]{Branford12}%
		\BibitemOpen
		\bibfield  {author} {\bibinfo {author} {\bibfnamefont {W.~R.}\ \bibnamefont
				{Branford}}, \bibinfo {author} {\bibfnamefont {S.}~\bibnamefont {Ladak}},
			\bibinfo {author} {\bibfnamefont {D.~E.}\ \bibnamefont {Read}}, \bibinfo
			{author} {\bibfnamefont {K.}~\bibnamefont {Zeissler}},\ and\ \bibinfo
			{author} {\bibfnamefont {L.~F.}\ \bibnamefont {Cohen}},\ }\bibfield  {title}
		{\bibinfo {title} {Emerging chirality in artificial spin ice},\ }\href
		{https://doi.org/10.1126/science.1211379} {\bibfield  {journal} {\bibinfo
				{journal} {Science}\ }\textbf {\bibinfo {volume} {335}},\ \bibinfo {pages}
			{1597} (\bibinfo {year} {2012})}\BibitemShut {NoStop}%
		\bibitem [{\citenamefont {Le}\ \emph {et~al.}(2017)\citenamefont {Le},
			\citenamefont {Park}, \citenamefont {Sklenar}, \citenamefont {Chern},
			\citenamefont {Nisoli}, \citenamefont {Watts}, \citenamefont {Manno},
			\citenamefont {Rench}, \citenamefont {Samarth}, \citenamefont {Leighton},\
			and\ \citenamefont {Schiffer}}]{Le17}%
		\BibitemOpen
		\bibfield  {author} {\bibinfo {author} {\bibfnamefont {B.~L.}\ \bibnamefont
				{Le}}, \bibinfo {author} {\bibfnamefont {J.}~\bibnamefont {Park}}, \bibinfo
			{author} {\bibfnamefont {J.}~\bibnamefont {Sklenar}}, \bibinfo {author}
			{\bibfnamefont {G.-W.}\ \bibnamefont {Chern}}, \bibinfo {author}
			{\bibfnamefont {C.}~\bibnamefont {Nisoli}}, \bibinfo {author} {\bibfnamefont
				{J.~D.}\ \bibnamefont {Watts}}, \bibinfo {author} {\bibfnamefont
				{M.}~\bibnamefont {Manno}}, \bibinfo {author} {\bibfnamefont {D.~W.}\
				\bibnamefont {Rench}}, \bibinfo {author} {\bibfnamefont {N.}~\bibnamefont
				{Samarth}}, \bibinfo {author} {\bibfnamefont {C.}~\bibnamefont {Leighton}},\
			and\ \bibinfo {author} {\bibfnamefont {P.}~\bibnamefont {Schiffer}},\
		}\bibfield  {title} {\bibinfo {title} {Understanding magnetotransport
				signatures in networks of connected permalloy nanowires},\ }\href
		{https://doi.org/10.1103/physrevb.95.060405} {\bibfield  {journal} {\bibinfo
				{journal} {Physical Review B}\ }\textbf {\bibinfo {volume} {95}},\ \bibinfo
			{pages} {060405} (\bibinfo {year} {2017})}\BibitemShut {NoStop}%
		\bibitem [{\citenamefont {Meng}\ \emph
			{et~al.}(2021{\natexlab{b}})\citenamefont {Meng}, \citenamefont {Donnelly},
			\citenamefont {Abert}, \citenamefont {Skoric}, \citenamefont {Holmes},
			\citenamefont {Xiao}, \citenamefont {Liao}, \citenamefont {Newton},
			\citenamefont {Barnes}, \citenamefont {Sanz-Hern{\'{a}}ndez}, \citenamefont
			{Hierro-Rodriguez}, \citenamefont {Suess}, \citenamefont {Cowburn},\ and\
			\citenamefont {Fern{\'{a}}ndez-Pacheco}}]{Meng21}%
		\BibitemOpen
		\bibfield  {author} {\bibinfo {author} {\bibfnamefont {F.}~\bibnamefont
				{Meng}}, \bibinfo {author} {\bibfnamefont {C.}~\bibnamefont {Donnelly}},
			\bibinfo {author} {\bibfnamefont {C.}~\bibnamefont {Abert}}, \bibinfo
			{author} {\bibfnamefont {L.}~\bibnamefont {Skoric}}, \bibinfo {author}
			{\bibfnamefont {S.}~\bibnamefont {Holmes}}, \bibinfo {author} {\bibfnamefont
				{Z.}~\bibnamefont {Xiao}}, \bibinfo {author} {\bibfnamefont {J.-W.}\
				\bibnamefont {Liao}}, \bibinfo {author} {\bibfnamefont {P.~J.}\ \bibnamefont
				{Newton}}, \bibinfo {author} {\bibfnamefont {C.~H.}\ \bibnamefont {Barnes}},
			\bibinfo {author} {\bibfnamefont {D.}~\bibnamefont {Sanz-Hern{\'{a}}ndez}},
			\bibinfo {author} {\bibfnamefont {A.}~\bibnamefont {Hierro-Rodriguez}},
			\bibinfo {author} {\bibfnamefont {D.}~\bibnamefont {Suess}}, \bibinfo
			{author} {\bibfnamefont {R.~P.}\ \bibnamefont {Cowburn}},\ and\ \bibinfo
			{author} {\bibfnamefont {A.}~\bibnamefont {Fern{\'{a}}ndez-Pacheco}},\
		}\bibfield  {title} {\bibinfo {title} {Non-planar geometrical effects on the
				magnetoelectrical signal in a three-dimensional nanomagnetic circuit},\
		}\href {https://doi.org/10.1021/acsnano.0c10272} {\bibfield  {journal}
			{\bibinfo  {journal} {ACS Nano}\ }\textbf {\bibinfo {volume} {15}},\ \bibinfo
			{pages} {6765} (\bibinfo {year} {2021}{\natexlab{b}})}\BibitemShut {NoStop}%
		\bibitem [{\citenamefont {Pylypovskyi}\ \emph {et~al.}(2015)\citenamefont
			{Pylypovskyi}, \citenamefont {Kravchuk}, \citenamefont {Sheka}, \citenamefont
			{Makarov}, \citenamefont {Schmidt},\ and\ \citenamefont
			{Gaididei}}]{Pylypovskyi15b}%
		\BibitemOpen
		\bibfield  {author} {\bibinfo {author} {\bibfnamefont {O.~V.}\ \bibnamefont
				{Pylypovskyi}}, \bibinfo {author} {\bibfnamefont {V.~P.}\ \bibnamefont
				{Kravchuk}}, \bibinfo {author} {\bibfnamefont {D.~D.}\ \bibnamefont {Sheka}},
			\bibinfo {author} {\bibfnamefont {D.}~\bibnamefont {Makarov}}, \bibinfo
			{author} {\bibfnamefont {O.~G.}\ \bibnamefont {Schmidt}},\ and\ \bibinfo
			{author} {\bibfnamefont {Y.}~\bibnamefont {Gaididei}},\ }\bibfield  {title}
		{\bibinfo {title} {Coupling of chiralities in spin and physical spaces: {T}he
				{M}\"obius ring as a case study},\ }\href
		{https://doi.org/10.1103/PhysRevLett.114.197204} {\bibfield  {journal}
			{\bibinfo  {journal} {Physical Review Letters}\ }\textbf {\bibinfo {volume}
				{114}},\ \bibinfo {pages} {197204} (\bibinfo {year} {2015})}\BibitemShut
		{NoStop}%
		\bibitem [{\citenamefont {Lu}\ and\ \citenamefont {Wu}(2001)}]{Lu01}%
		\BibitemOpen
		\bibfield  {author} {\bibinfo {author} {\bibfnamefont {W.~T.}\ \bibnamefont
				{Lu}}\ and\ \bibinfo {author} {\bibfnamefont {F.~Y.}\ \bibnamefont {Wu}},\
		}\bibfield  {title} {\bibinfo {title} {Ising model on nonorientable surfaces:
				Exact solution for the m{\"o}bius strip and the klein bottle},\ }\href
		{https://doi.org/10.1103/PhysRevE.63.026107} {\bibfield  {journal} {\bibinfo
				{journal} {Physical Review E}\ }\textbf {\bibinfo {volume} {63}},\ \bibinfo
			{pages} {026107} (\bibinfo {year} {2001})}\BibitemShut {NoStop}%
		\bibitem [{\citenamefont {Kaneda}\ and\ \citenamefont
			{Okabe}(2001)}]{Kaneda01}%
		\BibitemOpen
		\bibfield  {author} {\bibinfo {author} {\bibfnamefont {K.}~\bibnamefont
				{Kaneda}}\ and\ \bibinfo {author} {\bibfnamefont {Y.}~\bibnamefont {Okabe}},\
		}\bibfield  {title} {\bibinfo {title} {Finite-size scaling for the {I}sing
				model on the {M}{\"o}bius strip and the {K}lein bottle},\ }\href
		{https://doi.org/10.1103/physrevlett.86.2134} {\bibfield  {journal} {\bibinfo
				{journal} {Physical Review Letters}\ }\textbf {\bibinfo {volume} {86}},\
			\bibinfo {pages} {2134} (\bibinfo {year} {2001})}\BibitemShut {NoStop}%
		\bibitem [{Note4()}]{Note4}%
		\BibitemOpen
		\bibinfo {note} {Projective plane is the generalization of a usual plane that
			contains a line (boundary) at the infinity instead of the single point
			associated with all infinitely distant points.}\BibitemShut {Stop}%
		\bibitem [{\citenamefont {Vasilyev}\ \emph {et~al.}(2019)\citenamefont
			{Vasilyev}, \citenamefont {Macio{\l}ek},\ and\ \citenamefont
			{Dietrich}}]{Vasilyev19}%
		\BibitemOpen
		\bibfield  {author} {\bibinfo {author} {\bibfnamefont {O.~A.}\ \bibnamefont
				{Vasilyev}}, \bibinfo {author} {\bibfnamefont {A.}~\bibnamefont
				{Macio{\l}ek}},\ and\ \bibinfo {author} {\bibfnamefont {S.}~\bibnamefont
				{Dietrich}},\ }\bibfield  {title} {\bibinfo {title} {Criticality senses
				topology},\ }\href {https://doi.org/10.1209/0295-5075/128/20002} {\bibfield
			{journal} {\bibinfo  {journal} {{EPL} (Europhysics Letters)}\ }\textbf
			{\bibinfo {volume} {128}},\ \bibinfo {pages} {20002} (\bibinfo {year}
			{2019})}\BibitemShut {NoStop}%
		\bibitem [{\citenamefont {Yoneya}\ \emph {et~al.}(2008)\citenamefont {Yoneya},
			\citenamefont {Kuboki},\ and\ \citenamefont {Hayashi}}]{Yoneya08}%
		\BibitemOpen
		\bibfield  {author} {\bibinfo {author} {\bibfnamefont {M.}~\bibnamefont
				{Yoneya}}, \bibinfo {author} {\bibfnamefont {K.}~\bibnamefont {Kuboki}},\
			and\ \bibinfo {author} {\bibfnamefont {M.}~\bibnamefont {Hayashi}},\
		}\bibfield  {title} {\bibinfo {title} {{D}omain-wall structure of a classical
				{H}eisenberg ferromagnet on a {M}{\"o}bius strip},\ }\href
		{https://doi.org/10.1103/physrevb.78.064419} {\bibfield  {journal} {\bibinfo
				{journal} {Physical Review B}\ }\textbf {\bibinfo {volume} {78}},\ \bibinfo
			{pages} {064419} (\bibinfo {year} {2008})}\BibitemShut {NoStop}%
		\bibitem [{\citenamefont {Hatcher}(2002)}]{Hatcher02}%
		\BibitemOpen
		\bibfield  {author} {\bibinfo {author} {\bibfnamefont {A.}~\bibnamefont
				{Hatcher}},\ }\href@noop {} {\emph {\bibinfo {title} {Algebraic Topology}}}\
		(\bibinfo  {publisher} {Cambridge University Press},\ \bibinfo {year}
		{2002})\BibitemShut {NoStop}%
		\bibitem [{\citenamefont {Saji}\ \emph {et~al.}(2025)\citenamefont {Saji},
			\citenamefont {Castro}, \citenamefont {Carvalho-Santos}, \citenamefont
			{Saavedra}, \citenamefont {Nunez},\ and\ \citenamefont {Troncoso}}]{Saji25a}%
		\BibitemOpen
		\bibfield  {author} {\bibinfo {author} {\bibfnamefont {C.}~\bibnamefont
				{Saji}}, \bibinfo {author} {\bibfnamefont {M.~A.}\ \bibnamefont {Castro}},
			\bibinfo {author} {\bibfnamefont {V.~L.}\ \bibnamefont {Carvalho-Santos}},
			\bibinfo {author} {\bibfnamefont {E.}~\bibnamefont {Saavedra}}, \bibinfo
			{author} {\bibfnamefont {A.~S.}\ \bibnamefont {Nunez}},\ and\ \bibinfo
			{author} {\bibfnamefont {R.~E.}\ \bibnamefont {Troncoso}},\ }\bibfield
		{title} {\bibinfo {title} {Magnetic topological textures in nonorientable
				surfaces},\ }\bibfield  {journal} {\bibinfo  {journal} {ArXiv e-prints}\
		}\href {https://doi.org/10.48550/arXiv.2512.12853}
		{10.48550/arXiv.2512.12853} (\bibinfo {year} {2025})\BibitemShut {NoStop}%
		\bibitem [{\citenamefont {K\"{o}rber}(2023)}]{Korber23}%
		\BibitemOpen
		\bibfield  {author} {\bibinfo {author} {\bibfnamefont {L.}~\bibnamefont
				{K\"{o}rber}},\ }\emph {\bibinfo {title} {Spin waves in curved magnetic
				shells}},\ \href {https://doi.org/10.25368/2023.131} {Ph.D. thesis},\
		\bibinfo {address} {Dresden} (\bibinfo {year} {2023})\BibitemShut {NoStop}%
		\bibitem [{\citenamefont {Thonikkadavan}\ \emph {et~al.}(2026)\citenamefont
			{Thonikkadavan}, \citenamefont {d’Aquino},\ and\ \citenamefont
			{Hertel}}]{Thonikkadavan26}%
		\BibitemOpen
		\bibfield  {author} {\bibinfo {author} {\bibfnamefont {A.}~\bibnamefont
				{Thonikkadavan}}, \bibinfo {author} {\bibfnamefont {M.}~\bibnamefont
				{d’Aquino}},\ and\ \bibinfo {author} {\bibfnamefont {R.}~\bibnamefont
				{Hertel}},\ }\bibfield  {title} {\bibinfo {title} {Rotating spin wave modes
				in nanoscale {M}\"{o}bius strips},\ }\href
		{https://doi.org/10.1038/s44306-025-00118-8} {\bibfield  {journal} {\bibinfo
				{journal} {npj Spintronics}\ }\textbf {\bibinfo {volume} {4}},\ \bibinfo
			{pages} {4} (\bibinfo {year} {2026})}\BibitemShut {NoStop}%
		\bibitem [{\citenamefont {d’Aquino}\ and\ \citenamefont
			{Hertel}(2025)}]{dAquino25}%
		\BibitemOpen
		\bibfield  {author} {\bibinfo {author} {\bibfnamefont {M.}~\bibnamefont
				{d’Aquino}}\ and\ \bibinfo {author} {\bibfnamefont {R.}~\bibnamefont
				{Hertel}},\ }\bibfield  {title} {\bibinfo {title} {Nonreciprocal inertial
				spin-wave dynamics in twisted magnetic nanostrips},\ }\href
		{https://doi.org/10.1103/gvwf-fsz6} {\bibfield  {journal} {\bibinfo
				{journal} {Physical Review Letters}\ }\textbf {\bibinfo {volume} {135}},\
			\bibinfo {pages} {216705} (\bibinfo {year} {2025})}\BibitemShut {NoStop}%
		\bibitem [{\citenamefont {Ciornei}\ \emph {et~al.}(2011)\citenamefont
			{Ciornei}, \citenamefont {Rub\'{\i}},\ and\ \citenamefont
			{Wegrowe}}]{Ciornei11}%
		\BibitemOpen
		\bibfield  {author} {\bibinfo {author} {\bibfnamefont {M.-C.}\ \bibnamefont
				{Ciornei}}, \bibinfo {author} {\bibfnamefont {J.~M.}\ \bibnamefont
				{Rub\'{\i}}},\ and\ \bibinfo {author} {\bibfnamefont {J.-E.}\ \bibnamefont
				{Wegrowe}},\ }\bibfield  {title} {\bibinfo {title} {Magnetization dynamics in
				the inertial regime: Nutation predicted at short time scales},\ }\href
		{https://doi.org/10.1103/physrevb.83.020410} {\bibfield  {journal} {\bibinfo
				{journal} {Physical Review B}\ }\textbf {\bibinfo {volume} {83}},\ \bibinfo
			{pages} {020410} (\bibinfo {year} {2011})}\BibitemShut {NoStop}%
		\bibitem [{\citenamefont {Mondal}\ \emph {et~al.}(2017)\citenamefont {Mondal},
			\citenamefont {Berritta}, \citenamefont {Nandy},\ and\ \citenamefont
			{Oppeneer}}]{Mondal17}%
		\BibitemOpen
		\bibfield  {author} {\bibinfo {author} {\bibfnamefont {R.}~\bibnamefont
				{Mondal}}, \bibinfo {author} {\bibfnamefont {M.}~\bibnamefont {Berritta}},
			\bibinfo {author} {\bibfnamefont {A.~K.}\ \bibnamefont {Nandy}},\ and\
			\bibinfo {author} {\bibfnamefont {P.~M.}\ \bibnamefont {Oppeneer}},\
		}\bibfield  {title} {\bibinfo {title} {Relativistic theory of magnetic
				inertia in ultrafast spin dynamics},\ }\href
		{https://doi.org/10.1103/PhysRevB.96.024425} {\bibfield  {journal} {\bibinfo
				{journal} {Physical Review B}\ }\textbf {\bibinfo {volume} {96}},\ \bibinfo
			{pages} {024425} (\bibinfo {year} {2017})}\BibitemShut {NoStop}%
		\bibitem [{\citenamefont {Neeraj}\ \emph {et~al.}(2021)\citenamefont {Neeraj},
			\citenamefont {Awari}, \citenamefont {Kovalev}, \citenamefont {Polley},
			\citenamefont {Zhou~Hagstr\"{o}m}, \citenamefont {Arekapudi}, \citenamefont
			{Semisalova}, \citenamefont {Lenz}, \citenamefont {Green}, \citenamefont
			{Deinert}, \citenamefont {Ilyakov}, \citenamefont {Chen}, \citenamefont
			{Bawatna}, \citenamefont {Scalera}, \citenamefont {d’Aquino}, \citenamefont
			{Serpico}, \citenamefont {Hellwig}, \citenamefont {Wegrowe}, \citenamefont
			{Gensch},\ and\ \citenamefont {Bonetti}}]{Neeraj21}%
		\BibitemOpen
		\bibfield  {author} {\bibinfo {author} {\bibfnamefont {K.}~\bibnamefont
				{Neeraj}}, \bibinfo {author} {\bibfnamefont {N.}~\bibnamefont {Awari}},
			\bibinfo {author} {\bibfnamefont {S.}~\bibnamefont {Kovalev}}, \bibinfo
			{author} {\bibfnamefont {D.}~\bibnamefont {Polley}}, \bibinfo {author}
			{\bibfnamefont {N.}~\bibnamefont {Zhou~Hagstr\"{o}m}}, \bibinfo {author}
			{\bibfnamefont {S.~S. P.~K.}\ \bibnamefont {Arekapudi}}, \bibinfo {author}
			{\bibfnamefont {A.}~\bibnamefont {Semisalova}}, \bibinfo {author}
			{\bibfnamefont {K.}~\bibnamefont {Lenz}}, \bibinfo {author} {\bibfnamefont
				{B.}~\bibnamefont {Green}}, \bibinfo {author} {\bibfnamefont {J.-C.}\
				\bibnamefont {Deinert}}, \bibinfo {author} {\bibfnamefont {I.}~\bibnamefont
				{Ilyakov}}, \bibinfo {author} {\bibfnamefont {M.}~\bibnamefont {Chen}},
			\bibinfo {author} {\bibfnamefont {M.}~\bibnamefont {Bawatna}}, \bibinfo
			{author} {\bibfnamefont {V.}~\bibnamefont {Scalera}}, \bibinfo {author}
			{\bibfnamefont {M.}~\bibnamefont {d’Aquino}}, \bibinfo {author}
			{\bibfnamefont {C.}~\bibnamefont {Serpico}}, \bibinfo {author} {\bibfnamefont
				{O.}~\bibnamefont {Hellwig}}, \bibinfo {author} {\bibfnamefont {J.-E.}\
				\bibnamefont {Wegrowe}}, \bibinfo {author} {\bibfnamefont {M.}~\bibnamefont
				{Gensch}},\ and\ \bibinfo {author} {\bibfnamefont {S.}~\bibnamefont
				{Bonetti}},\ }\bibfield  {title} {\bibinfo {title} {Inertial spin dynamics in
				ferromagnets},\ }\href {https://doi.org/10.1038/s41567-020-01040-y}
		{\bibfield  {journal} {\bibinfo  {journal} {Nature Physics}\ }\textbf
			{\bibinfo {volume} {17}},\ \bibinfo {pages} {245} (\bibinfo {year}
			{2021})}\BibitemShut {NoStop}%
		\bibitem [{\citenamefont {Quarenta}\ \emph {et~al.}(2024)\citenamefont
			{Quarenta}, \citenamefont {Tharmalingam}, \citenamefont {Ludwig},
			\citenamefont {Yuan}, \citenamefont {Karwacki}, \citenamefont {Verstraten},\
			and\ \citenamefont {Duine}}]{Quarenta24}%
		\BibitemOpen
		\bibfield  {author} {\bibinfo {author} {\bibfnamefont {M.~G.}\ \bibnamefont
				{Quarenta}}, \bibinfo {author} {\bibfnamefont {M.}~\bibnamefont
				{Tharmalingam}}, \bibinfo {author} {\bibfnamefont {T.}~\bibnamefont
				{Ludwig}}, \bibinfo {author} {\bibfnamefont {H.~Y.}\ \bibnamefont {Yuan}},
			\bibinfo {author} {\bibfnamefont {L.}~\bibnamefont {Karwacki}}, \bibinfo
			{author} {\bibfnamefont {R.~C.}\ \bibnamefont {Verstraten}},\ and\ \bibinfo
			{author} {\bibfnamefont {R.~A.}\ \bibnamefont {Duine}},\ }\bibfield  {title}
		{\bibinfo {title} {Bath-induced spin inertia},\ }\href
		{https://doi.org/10.1103/physrevlett.133.136701} {\bibfield  {journal}
			{\bibinfo  {journal} {Physical Review Letters}\ }\textbf {\bibinfo {volume}
				{133}},\ \bibinfo {pages} {136701} (\bibinfo {year} {2024})}\BibitemShut
		{NoStop}%
		\bibitem [{\citenamefont {Reyes-Osorio}\ and\ \citenamefont
			{Nikolić}(2025)}]{ReyesOsorio25}%
		\BibitemOpen
		\bibfield  {author} {\bibinfo {author} {\bibfnamefont {F.}~\bibnamefont
				{Reyes-Osorio}}\ and\ \bibinfo {author} {\bibfnamefont {B.~K.}\ \bibnamefont
				{Nikolić}},\ }\bibfield  {title} {\bibinfo {title} {Optically induced
				magnetic inertia and magnons from non-markovian extension of the
				{L}andau-lifshitz-gilbert equation},\ }\href
		{https://doi.org/10.1103/sl4k-pcvq} {\bibfield  {journal} {\bibinfo
				{journal} {Physical Review Letters}\ }\textbf {\bibinfo {volume} {135}},\
			\bibinfo {pages} {246701} (\bibinfo {year} {2025})}\BibitemShut {NoStop}%
		\bibitem [{\citenamefont {K\"{o}rber}\ \emph {et~al.}(2021)\citenamefont
			{K\"{o}rber}, \citenamefont {Zimmermann}, \citenamefont {Wintz},
			\citenamefont {Finizio}, \citenamefont {Kronseder}, \citenamefont {Bougeard},
			\citenamefont {Dirnberger}, \citenamefont {Weigand}, \citenamefont {Raabe},
			\citenamefont {Ot\'{a}lora}, \citenamefont {Schultheiss}, \citenamefont
			{Josten}, \citenamefont {Lindner}, \citenamefont {K\'{e}zsm\'{a}rki},
			\citenamefont {Back},\ and\ \citenamefont {K\'{a}kay}}]{Korber21a}%
		\BibitemOpen
		\bibfield  {author} {\bibinfo {author} {\bibfnamefont {L.}~\bibnamefont
				{K\"{o}rber}}, \bibinfo {author} {\bibfnamefont {M.}~\bibnamefont
				{Zimmermann}}, \bibinfo {author} {\bibfnamefont {S.}~\bibnamefont {Wintz}},
			\bibinfo {author} {\bibfnamefont {S.}~\bibnamefont {Finizio}}, \bibinfo
			{author} {\bibfnamefont {M.}~\bibnamefont {Kronseder}}, \bibinfo {author}
			{\bibfnamefont {D.}~\bibnamefont {Bougeard}}, \bibinfo {author}
			{\bibfnamefont {F.}~\bibnamefont {Dirnberger}}, \bibinfo {author}
			{\bibfnamefont {M.}~\bibnamefont {Weigand}}, \bibinfo {author} {\bibfnamefont
				{J.}~\bibnamefont {Raabe}}, \bibinfo {author} {\bibfnamefont {J.~A.}\
				\bibnamefont {Ot\'{a}lora}}, \bibinfo {author} {\bibfnamefont
				{H.}~\bibnamefont {Schultheiss}}, \bibinfo {author} {\bibfnamefont
				{E.}~\bibnamefont {Josten}}, \bibinfo {author} {\bibfnamefont
				{J.}~\bibnamefont {Lindner}}, \bibinfo {author} {\bibfnamefont
				{I.}~\bibnamefont {K\'{e}zsm\'{a}rki}}, \bibinfo {author} {\bibfnamefont
				{C.~H.}\ \bibnamefont {Back}},\ and\ \bibinfo {author} {\bibfnamefont
				{A.}~\bibnamefont {K\'{a}kay}},\ }\bibfield  {title} {\bibinfo {title}
			{Symmetry and curvature effects on spin waves in vortex-state hexagonal
				nanotubes},\ }\href {https://doi.org/10.1103/physrevb.104.184429} {\bibfield
			{journal} {\bibinfo  {journal} {Physical Review B}\ }\textbf {\bibinfo
				{volume} {104}},\ \bibinfo {pages} {184429} (\bibinfo {year}
			{2021})}\BibitemShut {NoStop}%
		\bibitem [{\citenamefont {Dhungana}\ and\ \citenamefont
			{Pati}(2014)}]{Dhungana14}%
		\BibitemOpen
		\bibfield  {author} {\bibinfo {author} {\bibfnamefont {K.}~\bibnamefont
				{Dhungana}}\ and\ \bibinfo {author} {\bibfnamefont {R.}~\bibnamefont
				{Pati}},\ }\bibfield  {title} {\bibinfo {title} {Boron nitride nanotubes for
				spintronics},\ }\href {https://doi.org/10.3390/s140917655} {\bibfield
			{journal} {\bibinfo  {journal} {Sensors}\ }\textbf {\bibinfo {volume} {14}},\
			\bibinfo {pages} {17655} (\bibinfo {year} {2014})}\BibitemShut {NoStop}%
		\bibitem [{\citenamefont {Ot\'{a}lora}\ \emph {et~al.}(2019)\citenamefont
			{Ot\'{a}lora}, \citenamefont {Schultheiss},\ and\ \citenamefont
			{K\'{a}kay}}]{Otalora19}%
		\BibitemOpen
		\bibfield  {author} {\bibinfo {author} {\bibfnamefont {J.~A.}\ \bibnamefont
				{Ot\'{a}lora}}, \bibinfo {author} {\bibfnamefont {H.}~\bibnamefont
				{Schultheiss}},\ and\ \bibinfo {author} {\bibfnamefont {A.}~\bibnamefont
				{K\'{a}kay}},\ }\bibinfo {title} {Spin waves in nanotubes: Impact of
			curvature on transport properties},\ in\ \href
		{https://doi.org/10.1201/9780429299155-10} {\emph {\bibinfo {booktitle}
				{Three-Dimensional Magnonics}}}\ (\bibinfo  {publisher} {Jenny Stanford
			Publishing},\ \bibinfo {year} {2019})\ pp.\ \bibinfo {pages}
		{315--345}\BibitemShut {NoStop}%
		\bibitem [{\citenamefont {Proenca}\ \emph {et~al.}(2020)\citenamefont
			{Proenca}, \citenamefont {Sousa}, \citenamefont {Ventura},\ and\
			\citenamefont {Ara\'{u}jo}}]{Proenca20}%
		\BibitemOpen
		\bibfield  {author} {\bibinfo {author} {\bibfnamefont {M.~P.}\ \bibnamefont
				{Proenca}}, \bibinfo {author} {\bibfnamefont {C.~T.}\ \bibnamefont {Sousa}},
			\bibinfo {author} {\bibfnamefont {J.}~\bibnamefont {Ventura}},\ and\ \bibinfo
			{author} {\bibfnamefont {J.~P.}\ \bibnamefont {Ara\'{u}jo}},\ }\bibinfo
		{title} {Cylindrical magnetic nanotubes: Synthesis, magnetism and
			applications},\ in\ \href
		{https://doi.org/10.1016/b978-0-08-102832-2.00006-2} {\emph {\bibinfo
				{booktitle} {Magnetic Nano- and Microwires}}}\ (\bibinfo  {publisher}
		{Elsevier},\ \bibinfo {year} {2020})\ pp.\ \bibinfo {pages}
		{135--184}\BibitemShut {NoStop}%
		\bibitem [{\citenamefont {Shigeto}\ \emph {et~al.}(2002)\citenamefont
			{Shigeto}, \citenamefont {Okuno}, \citenamefont {Mibu}, \citenamefont
			{Shinjo},\ and\ \citenamefont {Ono}}]{Shigeto02}%
		\BibitemOpen
		\bibfield  {author} {\bibinfo {author} {\bibfnamefont {K.}~\bibnamefont
				{Shigeto}}, \bibinfo {author} {\bibfnamefont {T.}~\bibnamefont {Okuno}},
			\bibinfo {author} {\bibfnamefont {K.}~\bibnamefont {Mibu}}, \bibinfo {author}
			{\bibfnamefont {T.}~\bibnamefont {Shinjo}},\ and\ \bibinfo {author}
			{\bibfnamefont {T.}~\bibnamefont {Ono}},\ }\bibfield  {title} {\bibinfo
			{title} {Magnetic force microscopy observation of antivortex core with
				perpendicular magnetization in patterned thin film of permalloy},\ }\href
		{https://doi.org/10.1063/1.1483386} {\bibfield  {journal} {\bibinfo
				{journal} {Applied Physics Letters}\ }\textbf {\bibinfo {volume} {80}},\
			\bibinfo {pages} {4190} (\bibinfo {year} {2002})}\BibitemShut {NoStop}%
		\bibitem [{\citenamefont {Wang}\ and\ \citenamefont {Campbell}(2007)}]{Wang07}%
		\BibitemOpen
		\bibfield  {author} {\bibinfo {author} {\bibfnamefont {H.}~\bibnamefont
				{Wang}}\ and\ \bibinfo {author} {\bibfnamefont {C.~E.}\ \bibnamefont
				{Campbell}},\ }\bibfield  {title} {\bibinfo {title} {Spin dynamics of a
				magnetic antivortex: Micromagnetic simulations},\ }\href
		{https://doi.org/10.1103/PhysRevB.76.220407} {\bibfield  {journal} {\bibinfo
				{journal} {Physical Review B}\ }\textbf {\bibinfo {volume} {76}},\ \bibinfo
			{pages} {220407} (\bibinfo {year} {2007})}\BibitemShut {NoStop}%
		\bibitem [{\citenamefont {Gliga}\ \emph {et~al.}(2008)\citenamefont {Gliga},
			\citenamefont {Yan}, \citenamefont {Hertel},\ and\ \citenamefont
			{Schneider}}]{Gliga08}%
		\BibitemOpen
		\bibfield  {author} {\bibinfo {author} {\bibfnamefont {S.}~\bibnamefont
				{Gliga}}, \bibinfo {author} {\bibfnamefont {M.}~\bibnamefont {Yan}}, \bibinfo
			{author} {\bibfnamefont {R.}~\bibnamefont {Hertel}},\ and\ \bibinfo {author}
			{\bibfnamefont {C.~M.}\ \bibnamefont {Schneider}},\ }\bibfield  {title}
		{\bibinfo {title} {Ultrafast dynamics of a magnetic antivortex: Micromagnetic
				simulations},\ }\href {https://doi.org/10.1103/PhysRevB.77.060404} {\bibfield
			{journal} {\bibinfo  {journal} {Physical Review B}\ }\textbf {\bibinfo
				{volume} {77}},\ \bibinfo {pages} {060404} (\bibinfo {year}
			{2008})}\BibitemShut {NoStop}%
		\bibitem [{\citenamefont {Li}\ \emph {et~al.}(2014)\citenamefont {Li},
			\citenamefont {Tan}, \citenamefont {Moon}, \citenamefont {Doran},
			\citenamefont {Marcus}, \citenamefont {Young}, \citenamefont {Arenholz},
			\citenamefont {Ma}, \citenamefont {Yang}, \citenamefont {Hwang},\ and\
			\citenamefont {et~al.}}]{Li14}%
		\BibitemOpen
		\bibfield  {author} {\bibinfo {author} {\bibfnamefont {J.}~\bibnamefont
				{Li}}, \bibinfo {author} {\bibfnamefont {A.}~\bibnamefont {Tan}}, \bibinfo
			{author} {\bibfnamefont {K.~W.}\ \bibnamefont {Moon}}, \bibinfo {author}
			{\bibfnamefont {A.}~\bibnamefont {Doran}}, \bibinfo {author} {\bibfnamefont
				{M.~A.}\ \bibnamefont {Marcus}}, \bibinfo {author} {\bibfnamefont {A.~T.}\
				\bibnamefont {Young}}, \bibinfo {author} {\bibfnamefont {E.}~\bibnamefont
				{Arenholz}}, \bibinfo {author} {\bibfnamefont {S.}~\bibnamefont {Ma}},
			\bibinfo {author} {\bibfnamefont {R.~F.}\ \bibnamefont {Yang}}, \bibinfo
			{author} {\bibfnamefont {C.}~\bibnamefont {Hwang}},\ and\ \bibinfo {author}
			{\bibnamefont {et~al.}},\ }\bibfield  {title} {\bibinfo {title} {Stabilizing
				a magnetic vortex/antivortex array in single crystalline fe/ag(001)
				microstructures},\ }\href {https://doi.org/10.1063/1.4887345} {\bibfield
			{journal} {\bibinfo  {journal} {Applied Physics Letters}\ }\textbf {\bibinfo
				{volume} {104}},\ \bibinfo {pages} {262409} (\bibinfo {year}
			{2014})}\BibitemShut {NoStop}%
		\bibitem [{\citenamefont {Tey}\ \emph {et~al.}(2022)\citenamefont {Tey},
			\citenamefont {Chen}, \citenamefont {Soumyanarayanan},\ and\ \citenamefont
			{Ho}}]{Tey22}%
		\BibitemOpen
		\bibfield  {author} {\bibinfo {author} {\bibfnamefont {M.~S.~N.}\
				\bibnamefont {Tey}}, \bibinfo {author} {\bibfnamefont {X.}~\bibnamefont
				{Chen}}, \bibinfo {author} {\bibfnamefont {A.}~\bibnamefont
				{Soumyanarayanan}},\ and\ \bibinfo {author} {\bibfnamefont {P.}~\bibnamefont
				{Ho}},\ }\bibfield  {title} {\bibinfo {title} {Chiral spin textures for
				next-generation memory and unconventional computing},\ }\href
		{https://doi.org/10.1021/acsaelm.2c01023} {\bibfield  {journal} {\bibinfo
				{journal} {ACS Applied Electronic Materials}\ }\textbf {\bibinfo {volume}
				{4}},\ \bibinfo {pages} {5088} (\bibinfo {year} {2022})}\BibitemShut
		{NoStop}%
		\bibitem [{\citenamefont {M\"uller}\ \emph {et~al.}(2019)\citenamefont
			{M\"uller}, \citenamefont {Hoffmann}, \citenamefont {Di\ss{}elkamp},
			\citenamefont {Sch\"urhoff}, \citenamefont {Mavros}, \citenamefont
			{Sallermann}, \citenamefont {Kiselev}, \citenamefont {J\'onsson},\ and\
			\citenamefont {Bl\"ugel}}]{Muller19}%
		\BibitemOpen
		\bibfield  {author} {\bibinfo {author} {\bibfnamefont {G.~P.}\ \bibnamefont
				{M\"uller}}, \bibinfo {author} {\bibfnamefont {M.}~\bibnamefont {Hoffmann}},
			\bibinfo {author} {\bibfnamefont {C.}~\bibnamefont {Di\ss{}elkamp}}, \bibinfo
			{author} {\bibfnamefont {D.}~\bibnamefont {Sch\"urhoff}}, \bibinfo {author}
			{\bibfnamefont {S.}~\bibnamefont {Mavros}}, \bibinfo {author} {\bibfnamefont
				{M.}~\bibnamefont {Sallermann}}, \bibinfo {author} {\bibfnamefont {N.~S.}\
				\bibnamefont {Kiselev}}, \bibinfo {author} {\bibfnamefont {H.}~\bibnamefont
				{J\'onsson}},\ and\ \bibinfo {author} {\bibfnamefont {S.}~\bibnamefont
				{Bl\"ugel}},\ }\bibfield  {title} {\bibinfo {title} {Spirit: Multifunctional
				framework for atomistic spin simulations},\ }\href
		{https://doi.org/10.1103/PhysRevB.99.224414} {\bibfield  {journal} {\bibinfo
				{journal} {Physical Review B}\ }\textbf {\bibinfo {volume} {99}},\ \bibinfo
			{pages} {224414} (\bibinfo {year} {2019})}\BibitemShut {NoStop}%
		\bibitem [{\citenamefont {Kovacs}\ \emph {et~al.}(2022)\citenamefont {Kovacs},
			\citenamefont {Exl}, \citenamefont {Kornell}, \citenamefont {Fischbacher},
			\citenamefont {Hovorka}, \citenamefont {Gusenbauer}, \citenamefont {Breth},
			\citenamefont {Oezelt}, \citenamefont {Yano}, \citenamefont {Sakuma},
			\citenamefont {Kinoshita}, \citenamefont {Shoji}, \citenamefont {Kato},\ and\
			\citenamefont {Schrefl}}]{Kovacs22}%
		\BibitemOpen
		\bibfield  {author} {\bibinfo {author} {\bibfnamefont {A.}~\bibnamefont
				{Kovacs}}, \bibinfo {author} {\bibfnamefont {L.}~\bibnamefont {Exl}},
			\bibinfo {author} {\bibfnamefont {A.}~\bibnamefont {Kornell}}, \bibinfo
			{author} {\bibfnamefont {J.}~\bibnamefont {Fischbacher}}, \bibinfo {author}
			{\bibfnamefont {M.}~\bibnamefont {Hovorka}}, \bibinfo {author} {\bibfnamefont
				{M.}~\bibnamefont {Gusenbauer}}, \bibinfo {author} {\bibfnamefont
				{L.}~\bibnamefont {Breth}}, \bibinfo {author} {\bibfnamefont
				{H.}~\bibnamefont {Oezelt}}, \bibinfo {author} {\bibfnamefont
				{M.}~\bibnamefont {Yano}}, \bibinfo {author} {\bibfnamefont {N.}~\bibnamefont
				{Sakuma}}, \bibinfo {author} {\bibfnamefont {A.}~\bibnamefont {Kinoshita}},
			\bibinfo {author} {\bibfnamefont {T.}~\bibnamefont {Shoji}}, \bibinfo
			{author} {\bibfnamefont {A.}~\bibnamefont {Kato}},\ and\ \bibinfo {author}
			{\bibfnamefont {T.}~\bibnamefont {Schrefl}},\ }\bibfield  {title} {\bibinfo
			{title} {Conditional physics informed neural networks},\ }\href
		{https://doi.org/10.1016/j.cnsns.2021.106041} {\bibfield  {journal} {\bibinfo
				{journal} {Communications in Nonlinear Science and Numerical Simulation}\
			}\textbf {\bibinfo {volume} {104}},\ \bibinfo {pages} {106041} (\bibinfo
			{year} {2022})}\BibitemShut {NoStop}%
		\bibitem [{\citenamefont {Abert}\ \emph {et~al.}(2025)\citenamefont {Abert},
			\citenamefont {Bruckner}, \citenamefont {Voronov}, \citenamefont {Lang},
			\citenamefont {Pathak}, \citenamefont {Holt}, \citenamefont {Kraft},
			\citenamefont {Allayarov}, \citenamefont {Flauger}, \citenamefont {Koraltan},
			\citenamefont {Schrefl}, \citenamefont {Chumak}, \citenamefont {Fangohr},\
			and\ \citenamefont {Suess}}]{Abert25}%
		\BibitemOpen
		\bibfield  {author} {\bibinfo {author} {\bibfnamefont {C.}~\bibnamefont
				{Abert}}, \bibinfo {author} {\bibfnamefont {F.}~\bibnamefont {Bruckner}},
			\bibinfo {author} {\bibfnamefont {A.}~\bibnamefont {Voronov}}, \bibinfo
			{author} {\bibfnamefont {M.}~\bibnamefont {Lang}}, \bibinfo {author}
			{\bibfnamefont {S.~A.}\ \bibnamefont {Pathak}}, \bibinfo {author}
			{\bibfnamefont {S.}~\bibnamefont {Holt}}, \bibinfo {author} {\bibfnamefont
				{R.}~\bibnamefont {Kraft}}, \bibinfo {author} {\bibfnamefont
				{R.}~\bibnamefont {Allayarov}}, \bibinfo {author} {\bibfnamefont
				{P.}~\bibnamefont {Flauger}}, \bibinfo {author} {\bibfnamefont
				{S.}~\bibnamefont {Koraltan}}, \bibinfo {author} {\bibfnamefont
				{T.}~\bibnamefont {Schrefl}}, \bibinfo {author} {\bibfnamefont
				{A.}~\bibnamefont {Chumak}}, \bibinfo {author} {\bibfnamefont
				{H.}~\bibnamefont {Fangohr}},\ and\ \bibinfo {author} {\bibfnamefont
				{D.}~\bibnamefont {Suess}},\ }\bibfield  {title} {\bibinfo {title}
			{Neuralmag: an open-source nodal finite-difference code for inverse
				micromagnetics},\ }\href {https://doi.org/10.1038/s41524-025-01688-1}
		{\bibfield  {journal} {\bibinfo  {journal} {npj Computational Materials}\
			}\textbf {\bibinfo {volume} {11}},\ \bibinfo {pages} {193} (\bibinfo {year}
			{2025})}\BibitemShut {NoStop}%
		\bibitem [{\citenamefont {Pylypovskyi}\ \emph {et~al.}(2020)\citenamefont
			{Pylypovskyi}, \citenamefont {Kononenko}, \citenamefont {Yershov},
			\citenamefont {R{\"{o}}{\ss}ler}, \citenamefont {Tomilo}, \citenamefont
			{Fassbender}, \citenamefont {van~den Brink}, \citenamefont {Makarov},\ and\
			\citenamefont {Sheka}}]{Pylypovskyi20}%
		\BibitemOpen
		\bibfield  {author} {\bibinfo {author} {\bibfnamefont {O.~V.}\ \bibnamefont
				{Pylypovskyi}}, \bibinfo {author} {\bibfnamefont {D.~Y.}\ \bibnamefont
				{Kononenko}}, \bibinfo {author} {\bibfnamefont {K.~V.}\ \bibnamefont
				{Yershov}}, \bibinfo {author} {\bibfnamefont {U.~K.}\ \bibnamefont
				{R{\"{o}}{\ss}ler}}, \bibinfo {author} {\bibfnamefont {A.~V.}\ \bibnamefont
				{Tomilo}}, \bibinfo {author} {\bibfnamefont {J.}~\bibnamefont {Fassbender}},
			\bibinfo {author} {\bibfnamefont {J.}~\bibnamefont {van~den Brink}}, \bibinfo
			{author} {\bibfnamefont {D.}~\bibnamefont {Makarov}},\ and\ \bibinfo {author}
			{\bibfnamefont {D.~D.}\ \bibnamefont {Sheka}},\ }\bibfield  {title} {\bibinfo
			{title} {Curvilinear one-dimensional antiferromagnets},\ }\href
		{https://doi.org/10.1021/acs.nanolett.0c03246} {\bibfield  {journal}
			{\bibinfo  {journal} {Nano Letters}\ }\textbf {\bibinfo {volume} {20}},\
			\bibinfo {pages} {8157} (\bibinfo {year} {2020})}\BibitemShut {NoStop}%
		\bibitem [{\citenamefont {Borysenko}\ \emph {et~al.}(2022)\citenamefont
			{Borysenko}, \citenamefont {Sheka}, \citenamefont {Fassbender}, \citenamefont
			{van~den Brink}, \citenamefont {Makarov},\ and\ \citenamefont
			{Pylypovskyi}}]{Borysenko22}%
		\BibitemOpen
		\bibfield  {author} {\bibinfo {author} {\bibfnamefont {Y.~A.}\ \bibnamefont
				{Borysenko}}, \bibinfo {author} {\bibfnamefont {D.~D.}\ \bibnamefont
				{Sheka}}, \bibinfo {author} {\bibfnamefont {J.}~\bibnamefont {Fassbender}},
			\bibinfo {author} {\bibfnamefont {J.}~\bibnamefont {van~den Brink}}, \bibinfo
			{author} {\bibfnamefont {D.}~\bibnamefont {Makarov}},\ and\ \bibinfo {author}
			{\bibfnamefont {O.~V.}\ \bibnamefont {Pylypovskyi}},\ }\bibfield  {title}
		{\bibinfo {title} {Field-induced spin reorientation transitions in
				antiferromagnetic ring-shaped spin chains},\ }\href
		{https://doi.org/10.1103/physrevb.106.174426} {\bibfield  {journal} {\bibinfo
				{journal} {Physical Review B}\ }\textbf {\bibinfo {volume} {106}},\ \bibinfo
			{pages} {174426} (\bibinfo {year} {2022})}\BibitemShut {NoStop}%
		\bibitem [{\citenamefont {Yershov}(2022)}]{Yershov22a}%
		\BibitemOpen
		\bibfield  {author} {\bibinfo {author} {\bibfnamefont {K.~V.}\ \bibnamefont
				{Yershov}},\ }\bibfield  {title} {\bibinfo {title} {Dynamics of domain walls
				in curved antiferromagnetic wires},\ }\href
		{https://doi.org/10.1103/physrevb.105.064407} {\bibfield  {journal} {\bibinfo
				{journal} {Physical Review B}\ }\textbf {\bibinfo {volume} {105}},\ \bibinfo
			{pages} {064407} (\bibinfo {year} {2022})}\BibitemShut {NoStop}%
		\bibitem [{\citenamefont {Das~Gupta}\ and\ \citenamefont
			{Maiti}(2022)}]{DasGupta22}%
		\BibitemOpen
		\bibfield  {author} {\bibinfo {author} {\bibfnamefont {D.}~\bibnamefont
				{Das~Gupta}}\ and\ \bibinfo {author} {\bibfnamefont {S.~K.}\ \bibnamefont
				{Maiti}},\ }\bibfield  {title} {\bibinfo {title} {Antiferromagnetic helix as
				an efficient spin polarizer: Interplay between electric field and
				higher-order hopping},\ }\href {https://doi.org/10.1103/physrevb.106.125420}
		{\bibfield  {journal} {\bibinfo  {journal} {Physical Review B}\ }\textbf
			{\bibinfo {volume} {106}},\ \bibinfo {pages} {125420} (\bibinfo {year}
			{2022})}\BibitemShut {NoStop}%
		\bibitem [{\citenamefont {Yershov}\ \emph {et~al.}(2022)\citenamefont
			{Yershov}, \citenamefont {K\'{a}kay},\ and\ \citenamefont
			{Kravchuk}}]{Yershov22}%
		\BibitemOpen
		\bibfield  {author} {\bibinfo {author} {\bibfnamefont {K.~V.}\ \bibnamefont
				{Yershov}}, \bibinfo {author} {\bibfnamefont {A.}~\bibnamefont {K\'{a}kay}},\
			and\ \bibinfo {author} {\bibfnamefont {V.~P.}\ \bibnamefont {Kravchuk}},\
		}\bibfield  {title} {\bibinfo {title} {Curvature-induced drift and
				deformation of magnetic skyrmions: Comparison of the ferromagnetic and
				antiferromagnetic cases},\ }\href
		{https://doi.org/10.1103/physrevb.105.054425} {\bibfield  {journal} {\bibinfo
				{journal} {Physical Review B}\ }\textbf {\bibinfo {volume} {105}},\ \bibinfo
			{pages} {054425} (\bibinfo {year} {2022})}\BibitemShut {NoStop}%
		\bibitem [{\citenamefont {Yershov}\ \emph {et~al.}(2025)\citenamefont
			{Yershov}, \citenamefont {Gomonay}, \citenamefont {Sinova}, \citenamefont
			{van~den Brink},\ and\ \citenamefont {Kravchuk}}]{Yershov25}%
		\BibitemOpen
		\bibfield  {author} {\bibinfo {author} {\bibfnamefont {K.~V.}\ \bibnamefont
				{Yershov}}, \bibinfo {author} {\bibfnamefont {O.}~\bibnamefont {Gomonay}},
			\bibinfo {author} {\bibfnamefont {J.}~\bibnamefont {Sinova}}, \bibinfo
			{author} {\bibfnamefont {J.}~\bibnamefont {van~den Brink}},\ and\ \bibinfo
			{author} {\bibfnamefont {V.~P.}\ \bibnamefont {Kravchuk}},\ }\bibfield
		{title} {\bibinfo {title} {Curvature-induced magnetization of altermagnetic
				films},\ }\href {https://doi.org/10.1103/physrevlett.134.116701} {\bibfield
			{journal} {\bibinfo  {journal} {Physical Review Letters}\ }\textbf {\bibinfo
				{volume} {134}},\ \bibinfo {pages} {116701} (\bibinfo {year}
			{2025})}\BibitemShut {NoStop}%
		\bibitem [{\citenamefont {Ivanov}(2019)}]{Ivanov19}%
		\BibitemOpen
		\bibfield  {author} {\bibinfo {author} {\bibfnamefont {B.~A.}\ \bibnamefont
				{Ivanov}},\ }\bibfield  {title} {\bibinfo {title} {Ultrafast spin dynamics
				and spintronics for ferrimagnets close to the spin compensation point
				(review)},\ }\href {https://doi.org/10.1063/1.5121265} {\bibfield  {journal}
			{\bibinfo  {journal} {Low Temperature Physics}\ }\textbf {\bibinfo {volume}
				{45}},\ \bibinfo {pages} {935} (\bibinfo {year} {2019})}\BibitemShut
		{NoStop}%
		\bibitem [{\citenamefont {Salamone}\ \emph {et~al.}(2021)\citenamefont
			{Salamone}, \citenamefont {Svendsen}, \citenamefont {Amundsen},\ and\
			\citenamefont {Jacobsen}}]{Salamone21}%
		\BibitemOpen
		\bibfield  {author} {\bibinfo {author} {\bibfnamefont {T.}~\bibnamefont
				{Salamone}}, \bibinfo {author} {\bibfnamefont {M.~B.~M.}\ \bibnamefont
				{Svendsen}}, \bibinfo {author} {\bibfnamefont {M.}~\bibnamefont {Amundsen}},\
			and\ \bibinfo {author} {\bibfnamefont {S.}~\bibnamefont {Jacobsen}},\
		}\bibfield  {title} {\bibinfo {title} {Curvature-induced long-range
				supercurrents in diffusive superconductor-ferromagnet-superconductor
				josephson junctions with a dynamic $0-\pi$ transition},\ }\href
		{https://doi.org/10.1103/physrevb.104.l060505} {\bibfield  {journal}
			{\bibinfo  {journal} {Physical Review B}\ }\textbf {\bibinfo {volume}
				{104}},\ \bibinfo {pages} {L060505} (\bibinfo {year} {2021})}\BibitemShut
		{NoStop}%
		\bibitem [{\citenamefont {Johan~Skarpeid}\ \emph {et~al.}(2024)\citenamefont
			{Johan~Skarpeid}, \citenamefont {Hugdal}, \citenamefont {Salamone},
			\citenamefont {Amundsen},\ and\ \citenamefont {Jacobsen}}]{JohanSkarpeid24}%
		\BibitemOpen
		\bibfield  {author} {\bibinfo {author} {\bibfnamefont {A.}~\bibnamefont
				{Johan~Skarpeid}}, \bibinfo {author} {\bibfnamefont {H.~G.}\ \bibnamefont
				{Hugdal}}, \bibinfo {author} {\bibfnamefont {T.}~\bibnamefont {Salamone}},
			\bibinfo {author} {\bibfnamefont {M.}~\bibnamefont {Amundsen}},\ and\
			\bibinfo {author} {\bibfnamefont {S.~H.}\ \bibnamefont {Jacobsen}},\
		}\bibfield  {title} {\bibinfo {title} {Non-constant geometric curvature for
				tailored spin–orbit coupling and chirality in superconductor-magnet
				heterostructures},\ }\href {https://doi.org/10.1088/1361-648x/ad2e23}
		{\bibfield  {journal} {\bibinfo  {journal} {Journal of Physics: Condensed
					Matter}\ }\textbf {\bibinfo {volume} {36}},\ \bibinfo {pages} {235302}
			(\bibinfo {year} {2024})}\BibitemShut {NoStop}%
		\bibitem [{\citenamefont {Heinrich}\ \emph {et~al.}(2025)\citenamefont
			{Heinrich}, \citenamefont {Hugdal}, \citenamefont {Amundsen},\ and\
			\citenamefont {Jacobsen}}]{Heinrich25}%
		\BibitemOpen
		\bibfield  {author} {\bibinfo {author} {\bibfnamefont {K.~S.}\ \bibnamefont
				{Heinrich}}, \bibinfo {author} {\bibfnamefont {H.~G.}\ \bibnamefont
				{Hugdal}}, \bibinfo {author} {\bibfnamefont {M.}~\bibnamefont {Amundsen}},\
			and\ \bibinfo {author} {\bibfnamefont {S.~H.}\ \bibnamefont {Jacobsen}},\
		}\bibfield  {title} {\bibinfo {title} {Bending-strain effects in conventional
				superconductors and superconducting junctions},\ }\href
		{https://doi.org/10.1038/s44306-025-00084-1} {\bibfield  {journal} {\bibinfo
				{journal} {npj Spintronics}\ }\textbf {\bibinfo {volume} {3}},\ \bibinfo
			{pages} {18} (\bibinfo {year} {2025})}\BibitemShut {NoStop}%
		\bibitem [{\citenamefont {Edstr\"{o}m}\ \emph {et~al.}(2022)\citenamefont
			{Edstr\"{o}m}, \citenamefont {Amoroso}, \citenamefont {Picozzi},
			\citenamefont {Barone},\ and\ \citenamefont {Stengel}}]{Edstrom22}%
		\BibitemOpen
		\bibfield  {author} {\bibinfo {author} {\bibfnamefont {A.}~\bibnamefont
				{Edstr\"{o}m}}, \bibinfo {author} {\bibfnamefont {D.}~\bibnamefont
				{Amoroso}}, \bibinfo {author} {\bibfnamefont {S.}~\bibnamefont {Picozzi}},
			\bibinfo {author} {\bibfnamefont {P.}~\bibnamefont {Barone}},\ and\ \bibinfo
			{author} {\bibfnamefont {M.}~\bibnamefont {Stengel}},\ }\bibfield  {title}
		{\bibinfo {title} {Curved magnetism in {CrI$_3$}},\ }\href
		{https://doi.org/10.1103/physrevlett.128.177202} {\bibfield  {journal}
			{\bibinfo  {journal} {Physical Review Letters}\ }\textbf {\bibinfo {volume}
				{128}},\ \bibinfo {pages} {177202} (\bibinfo {year} {2022})}\BibitemShut
		{NoStop}%
		\bibitem [{\citenamefont {Gao}\ \emph {et~al.}(2026)\citenamefont {Gao},
			\citenamefont {Wang}, \citenamefont {Lyu}, \citenamefont {Liu}, \citenamefont
			{Zhu}, \citenamefont {Liu}, \citenamefont {Li}, \citenamefont {Ji},
			\citenamefont {Zhang}, \citenamefont {Gu}, \citenamefont {Ma}, \citenamefont
			{Cao},\ and\ \citenamefont {Lu}}]{Gao26}%
		\BibitemOpen
		\bibfield  {author} {\bibinfo {author} {\bibfnamefont {L.}~\bibnamefont
				{Gao}}, \bibinfo {author} {\bibfnamefont {Y.}~\bibnamefont {Wang}}, \bibinfo
			{author} {\bibfnamefont {X.}~\bibnamefont {Lyu}}, \bibinfo {author}
			{\bibfnamefont {P.}~\bibnamefont {Liu}}, \bibinfo {author} {\bibfnamefont
				{M.}~\bibnamefont {Zhu}}, \bibinfo {author} {\bibfnamefont {J.}~\bibnamefont
				{Liu}}, \bibinfo {author} {\bibfnamefont {M.}~\bibnamefont {Li}}, \bibinfo
			{author} {\bibfnamefont {A.}~\bibnamefont {Ji}}, \bibinfo {author}
			{\bibfnamefont {Q.}~\bibnamefont {Zhang}}, \bibinfo {author} {\bibfnamefont
				{L.}~\bibnamefont {Gu}}, \bibinfo {author} {\bibfnamefont {L.}~\bibnamefont
				{Ma}}, \bibinfo {author} {\bibfnamefont {Z.}~\bibnamefont {Cao}},\ and\
			\bibinfo {author} {\bibfnamefont {N.}~\bibnamefont {Lu}},\ }\bibfield
		{title} {\bibinfo {title} {Strain‐assembled crystalline {SrRuO$_3$}
				microtube and emergent curvilinear magnetism},\ }\href
		{https://doi.org/10.1002/advs.202522085} {\bibfield  {journal} {\bibinfo
				{journal} {Advanced Science}\ ,\ \bibinfo {pages} {e22085}} (\bibinfo {year}
			{2026})}\BibitemShut {NoStop}%
		\bibitem [{\citenamefont {Kosevich}(2004)}]{Kosevich04a}%
		\BibitemOpen
		\bibfield  {author} {\bibinfo {author} {\bibfnamefont {A.~M.}\ \bibnamefont
				{Kosevich}},\ }\bibfield  {title} {\bibinfo {title} {Topology and solid-state
				physics (review)},\ }\href {https://doi.org/10.1063/1.1645161} {\bibfield
			{journal} {\bibinfo  {journal} {Low Temperature Physics}\ }\textbf {\bibinfo
				{volume} {30}},\ \bibinfo {pages} {97} (\bibinfo {year} {2004})}\BibitemShut
		{NoStop}%
		\bibitem [{\citenamefont {Cador}\ \emph {et~al.}(2004)\citenamefont {Cador},
			\citenamefont {Gatteschi}, \citenamefont {Sessoli}, \citenamefont {Larsen},
			\citenamefont {Overgaard}, \citenamefont {Barra}, \citenamefont {Teat},
			\citenamefont {Timco},\ and\ \citenamefont {Winpenny}}]{Cador04}%
		\BibitemOpen
		\bibfield  {author} {\bibinfo {author} {\bibfnamefont {O.}~\bibnamefont
				{Cador}}, \bibinfo {author} {\bibfnamefont {D.}~\bibnamefont {Gatteschi}},
			\bibinfo {author} {\bibfnamefont {R.}~\bibnamefont {Sessoli}}, \bibinfo
			{author} {\bibfnamefont {F.~K.}\ \bibnamefont {Larsen}}, \bibinfo {author}
			{\bibfnamefont {J.}~\bibnamefont {Overgaard}}, \bibinfo {author}
			{\bibfnamefont {A.-L.}\ \bibnamefont {Barra}}, \bibinfo {author}
			{\bibfnamefont {S.~J.}\ \bibnamefont {Teat}}, \bibinfo {author}
			{\bibfnamefont {G.~A.}\ \bibnamefont {Timco}},\ and\ \bibinfo {author}
			{\bibfnamefont {R.~E.~P.}\ \bibnamefont {Winpenny}},\ }\bibfield  {title}
		{\bibinfo {title} {The magnetic {M}{\"{o}}bius strip: Synthesis, structure,
				and magnetic studies of odd-numbered antiferromagnetically coupled wheels},\
		}\href {https://doi.org/10.1002/anie.200460211} {\bibfield  {journal}
			{\bibinfo  {journal} {Angewandte Chemie International Edition}\ }\textbf
			{\bibinfo {volume} {43}},\ \bibinfo {pages} {5196} (\bibinfo {year}
			{2004})}\BibitemShut {NoStop}%
		\bibitem [{\citenamefont {Murray}(2022)}]{Murray22}%
		\BibitemOpen
		\bibinfo {editor} {\bibfnamefont {K.}~\bibnamefont {Murray}},\ ed.,\ \href
		{https://doi.org/10.1007/978-3-031-11709-1} {\emph {\bibinfo {title} {Single
					Molecule Toroics}}}\ (\bibinfo  {publisher} {Springer International
			Publishing},\ \bibinfo {address} {Cham},\ \bibinfo {year} {2022})\BibitemShut
		{NoStop}%
	\end{thebibliography}
\end{document}